\documentclass[10pt]{article}
\usepackage[lmargin=1in,rmargin=1in,tmargin=1in,bmargin=1in]{geometry}
\usepackage{amsmath,amsthm,amsfonts,amssymb,comment,slashed,mathtools}
\usepackage[utf8]{inputenc}
\usepackage[T1]{fontenc}  
\usepackage[english]{babel}

\usepackage{mathtools}

\usepackage{authblk}

\usepackage{csquotes}
\usepackage{comment}
\usepackage{graphicx} 
\graphicspath{ {./images/} }
\usepackage{amsmath}
\usepackage{amssymb}
\usepackage{amsthm}
\usepackage{float}
\usepackage{hyperref}
\hypersetup{
	colorlinks=true,
	linkcolor=black,}

\usepackage{tikz}
\usepackage{pgfplots}
\pgfplotsset{compat=1.15}
\usepackage{mathrsfs}
\usetikzlibrary{arrows}
\usetikzlibrary{decorations}

\numberwithin{equation}{section}
\mathtoolsset{showonlyrefs,showmanualtags}

\title{Shell crossing and the dynamical formation of thin shells \\ in the collapse of self-gravitating relativistic dust}
\author{Matt Smith\thanks{ms2652@cam.ac.uk}}
\affil{University of Cambridge, Department of Applied Mathematics and Theoretical Physics}
\date{\today}

\newtheorem{theorem}{Theorem}
\newtheorem{lemma}{Lemma}[section]
\newtheorem{proposition}{Proposition}[section]
\newtheorem{corollary}{Corollary}[section]
\newtheorem{conjecture}{Conjecture}
\newtheorem{definition}{Definition}[section]

\newtheorem{introtheorem}{Theorem}

\theoremstyle{remark}
\newtheorem{remark}{Remark}[section]

\newtheorem*{note}{Note}

\newcommand{\QGHa}{\mathscr{Q}_{GH, 0}(\tau_{\Gamma, 0}, \tau_{\chi_b, 0})}
\newcommand{\QGHb}{\mathscr{Q}_{GH, 1}(\underline{s})}
\newcommand{\R}{\mathbb{R}}
\newcommand{\B}{[0,\Delta_5] \times \tilde{\mathcal{B}}}
\newcommand{\uu}[1]{\underline{\underline{#1}}}
\newcommand{\metric}{(\mathfrak{D}_{\chi_b},d_k)}

\begin{document}
	
	\maketitle
	
	\begin{abstract}
		In this paper, we show that shell crossing in the collapse of self-gravitating relativistic dust in spherical symmetry can be understood as dynamical formation of a thin shell of dust. More precisely, we consider the spherically symmetric Einstein-dust system in double null coordinates (E-d), for which first singularities with area radius bounded away from zero are known as `shell crossing' singularities. We show that there exists a weak solution of (E-d), in which a thin shell of dust forms, extending beyond any `non-degenerate' first shell crossing singularity of a smooth `collapsing' solution of (E-d). This local extension is unique in the class of piecewise smooth solutions of (E-d) satisfying a `determinism condition'. 
		
		We then consider the physical problem of smooth data for (E-d) on an initial hypersurface diffeomorphic to $\R^3$, containing a collapsing dust cloud. We show that, for generic such initial data, there exists a maximal globally hyperbolic piecewise smooth development satisfying a `propagation condition' (stronger than the determinism condition). The development may contain an arbitrary (finite) number of thin shells which can moreover merge. All shell crossing singularities are `resolved' and the area radius extends continuously to zero at any first singularity of the maximal development. We note that, for an open set of initial data in the smooth topology, as shown by Christodoulou \cite{C1}, a Cauchy horizon emanates from a central `shell-focussing' singularity.
	\end{abstract}
	
	\tableofcontents
	
	\section{Introduction}
	
	In their celebrated 1939 paper, Oppenheimer and Snyder \cite{OS} studied the collapse of self-gravitating relativistic dust, restricted to the special case of a \emph{homogeneous} dust cloud. Their work, significantly ahead of its time, would later be understood as the first example of black hole formation in gravitational collapse, as a solution of the equations of general relativity. In particular, it was one of the motivations for Penrose \cite{P} to formulate the so-called \emph{cosmic censorship} conjectures. In a twist of irony, the spherically symmetric Einstein-dust system, of which the `Oppenheimer--Snyder spacetimes' are a solution, exhibits a dramatric violation of cosmic censorship. The first such claim \cite{Y}, made already in 1974, showed that so-called `shell crossing' singularities, formed in the collapse of an \emph{inhomogeneous} dust cloud, can be \emph{globally naked}. Furthermore, in 1984 Christodoulou \cite{C1} studied a family of inhomogeneous dust clouds that do not form shell crossing singularities, for which he showed that the central shell-focussing singularity is in fact locally naked, and can moreover be globally naked under suitable conditions. 
	
	We will be primarily concerned here with the issue of shell crossing singularities. In classical solutions of the Einstein-dust system, `shells' of dust (the fluid flow lines of a sphere) locally foliate spacetime: shell crossing singularities form when the density of this foliation diverges. It has long been expected that the issue of shell crossing singularities can be `resolved', primarily due to the (only result thus far) of an \emph{extension} past a first shell crossing singularity: the 1967 work of Papapetrou and Hamoui \cite{PH}. They showed that spacetime can be \emph{extended} past the (highly non-generic) case of a \emph{degenerate} shell crossing singularity, where a dust cloud `crosses' through a spacelike codimension-2 sphere. We term this a \emph{collisionless} extension. The crucial insight was a change of coordinates adapted to the structure of the first singularity, away from the usual `comoving coordinates' (first introduced by Lema\^itre \cite{L} in 1933) in which one can essentially write down the general classical solution of Einstein-dust in spherical symmetry. Indeed, comoving coordinates themselves are particularly ill-suited at shell crossing singularities: the spacetime metric in comoving coordinates degenerates, not even extending continuously as a Lorentzian metric.
	
	In this paper, we consider a physically distinct extension to the collisionless extension of Papapetrou and Hamoui. We make essential use of \emph{double null} coordinates $(u,v)$, which are particularly well-suited to the study of non-smooth solutions of the Einstein-dust system. Indeed, in double null coordinates the metric components extend in $C^1$ to `non-degenerate' first shell crossing singularities. Moreover, we can define `piecewise smooth' solutions of the Einstein-dust system in double null coordinates, which consist of a thin shell of dust (first introduced by Israel \cite{I}) with (locally) classical solutions of Einstein-dust in the interior and exterior. Our first main result of the paper is that spacetime can be extended beyond non-degenerate first shell crossing singularities with a dynamically forming thin shell of dust, which we refer to as `thin shell development at shell crossing'.
	\begin{introtheorem}[rough formulation] \label{thm:1}
		Let $(\mathcal{Q},r,\Omega^2,\rho,\mathbf{u}^\mu)$ be a smooth collapsing\footnote{By collapsing, we mean that the velocity of the dust is ingoing: $\mathbf{u}(r) = \mathbf{u}^\mu \partial_\mu r < 0$. Note that we also impose that $r, \rho >0$.} solution of the Einstein-dust system in double null coordinates (E-d), with a non-degenerate first shell crossing singularity $p \in \R^{1+1}_{u,v}$ (as in Figure \ref{MGHDsubset}). Then there exists a piecewise smooth solution of (E-d), $(\tilde{\mathcal{Q}},\tilde{r},\tilde{\Omega}^2,\tilde{\rho},\tilde{\mathbf{u}}^\mu)$, \underline{extending} $(\mathcal{Q},r,\Omega^2,\rho,\mathbf{u}^\mu)$, \underline{containing a thin shell of dust}, $\Sigma$, \underline{forming at $p$}. The metric components $\tilde{r},\tilde{\Omega}^2$ are $C^1_{u,v}$ at $p$, $C^{0,1}_{u,v}$ at the thin shell to the future of $p$, and at least $C^2_{u,v}$ otherwise. The extension is \underline{unique} in the class of piecewise smooth solutions of (E-d) satisfying a `\underline{determinism condition}'.
		\begin{figure}[H]
			\centering
			\begin{minipage}{0.4\linewidth}
				\centering
				\def\svgwidth{0.7\linewidth}   
\begingroup%
  \makeatletter%
  \providecommand\color[2][]{%
    \errmessage{(Inkscape) Color is used for the text in Inkscape, but the package 'color.sty' is not loaded}%
    \renewcommand\color[2][]{}%
  }%
  \providecommand\transparent[1]{%
    \errmessage{(Inkscape) Transparency is used (non-zero) for the text in Inkscape, but the package 'transparent.sty' is not loaded}%
    \renewcommand\transparent[1]{}%
  }%
  \providecommand\rotatebox[2]{#2}%
  \newcommand*\fsize{\dimexpr\f@size pt\relax}%
  \newcommand*\lineheight[1]{\fontsize{\fsize}{#1\fsize}\selectfont}%
  \ifx\svgwidth\undefined%
    \setlength{\unitlength}{401.59259466bp}%
    \ifx\svgscale\undefined%
      \relax%
    \else%
      \setlength{\unitlength}{\unitlength * \real{\svgscale}}%
    \fi%
  \else%
    \setlength{\unitlength}{\svgwidth}%
  \fi%
  \global\let\svgwidth\undefined%
  \global\let\svgscale\undefined%
  \makeatother%
  \begin{picture}(1,0.99989865)%
    \lineheight{1}%
    \setlength\tabcolsep{0pt}%
    \put(0,0){\includegraphics[width=\unitlength,page=1]{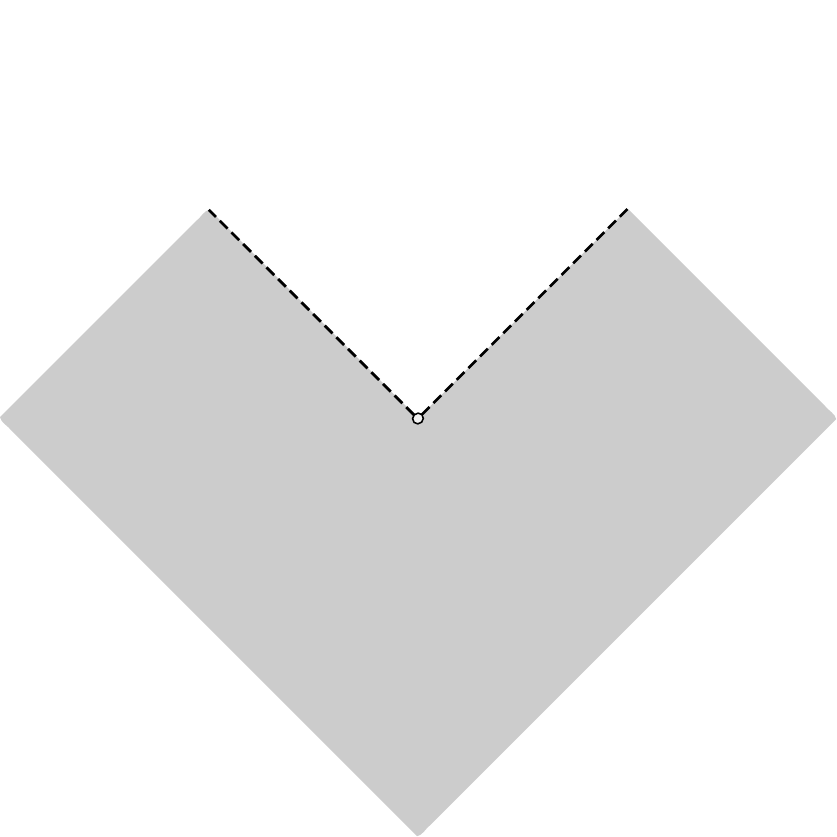}}%
    \put(0.37605147,0.67624102){\rotatebox{-45}{\makebox(0,0)[lt]{\lineheight{2.04999995}\smash{\begin{tabular}[t]{l}$$\end{tabular}}}}}%
    \put(0.51446408,0.47093298){\color[rgb]{0,0,0}\makebox(0,0)[lt]{\lineheight{2.04999995}\smash{\begin{tabular}[t]{l}$p$\end{tabular}}}}%
    \put(0,0){\includegraphics[width=\unitlength,page=2]{MGHD1a.pdf}}%
  \end{picture}%
\endgroup%

				\caption{A smooth solution of the Einstein-dust system with a non-degenerate first shell crossing singularity.} \label{MGHDsubset}
			\end{minipage}\hspace{5em}
			\begin{minipage}{0.4\linewidth}
				\centering
				\def\svgwidth{0.7\linewidth}  
\begingroup%
  \makeatletter%
  \providecommand\color[2][]{%
    \errmessage{(Inkscape) Color is used for the text in Inkscape, but the package 'color.sty' is not loaded}%
    \renewcommand\color[2][]{}%
  }%
  \providecommand\transparent[1]{%
    \errmessage{(Inkscape) Transparency is used (non-zero) for the text in Inkscape, but the package 'transparent.sty' is not loaded}%
    \renewcommand\transparent[1]{}%
  }%
  \providecommand\rotatebox[2]{#2}%
  \newcommand*\fsize{\dimexpr\f@size pt\relax}%
  \newcommand*\lineheight[1]{\fontsize{\fsize}{#1\fsize}\selectfont}%
  \ifx\svgwidth\undefined%
    \setlength{\unitlength}{401.59259466bp}%
    \ifx\svgscale\undefined%
      \relax%
    \else%
      \setlength{\unitlength}{\unitlength * \real{\svgscale}}%
    \fi%
  \else%
    \setlength{\unitlength}{\svgwidth}%
  \fi%
  \global\let\svgwidth\undefined%
  \global\let\svgscale\undefined%
  \makeatother%
  \begin{picture}(1,0.99989865)%
    \lineheight{1}%
    \setlength\tabcolsep{0pt}%
    \put(0,0){\includegraphics[width=\unitlength,page=1]{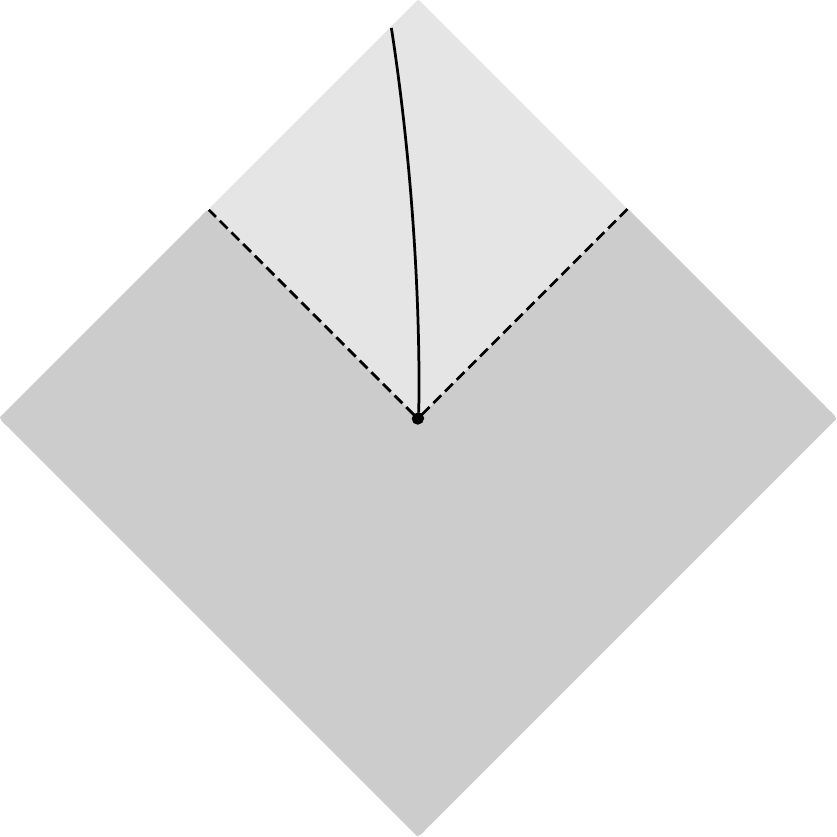}}%
    \put(0.37605147,0.67624102){\rotatebox{-45}{\makebox(0,0)[lt]{\lineheight{2.04999995}\smash{\begin{tabular}[t]{l}$$\end{tabular}}}}}%
    \put(0.50718687,0.75019796){\color[rgb]{0,0,0}\makebox(0,0)[lt]{\lineheight{2.04999995}\smash{\begin{tabular}[t]{l}$\Sigma$\end{tabular}}}}%
    \put(0.51446408,0.47093298){\color[rgb]{0,0,0}\makebox(0,0)[lt]{\lineheight{2.04999995}\smash{\begin{tabular}[t]{l}$p$\end{tabular}}}}%
    \put(0,0){\includegraphics[width=\unitlength,page=2]{MGHD2a.pdf}}%
  \end{picture}%
\endgroup%

				\caption{The unique piecewise smooth extension of the spacetime in Figure \ref{MGHDsubset}, satisfying the determinism condition.} \label{extension}
			\end{minipage}
		\end{figure}
	\end{introtheorem}
	For an explanation of the setup and proof of the theorem, see already Section~\ref{formationexp}. See already Theorem~\ref{devthm} in Section~\ref{statementdevthm} for the precise statement.
	
	One can interpret this extension as dust `sticking' upon collision, as opposed to the collisionless extension of Papapetrou and Hamoui, see Section \ref{prev} for more on the collisionless extension, and Section \ref{future} for the interpretation of the physically distinct extensions.

	We then consider the physical problem of smooth initial data for the Einstein-dust system containing a compactly supported collapsing dust cloud, as initiated by Oppenheimer and Snyder. It is possible for smooth collapsing dust clouds to form infinitely many first shell crossing singularities, however, we show that \emph{generic} dust clouds only form finitely many first shell crossing singularities, and moreover that they are \emph{non-degenerate}. It is therefore possible, using Theorem \ref{thm:1}, to extend as a piecewise smooth solution of (E-d) with \emph{finitely many thin shells}. In order to have any hope of proving uniqueness of such solutions, in particular at thin shell \emph{collisions}, we need to introduce yet another condition, the `propagation condition': at any point on, or at the limit of, a thin shell, the solution to the future must locally consist of a single thin shell (which must also satisfy the determinism condition). This implies that thin shells `stick' and form a single shell to the future at any collision.\footnote{One expects that such a condition is the natural choice for which piecewise smooth solutions of Einstein-dust can be approximated by solutions of Einstein--Euler in the pressureless limit, see already Section \ref{future}.} Equipped with this framework, our second main result is the construction of a unique maximal piecewise smooth development of generic collapsing dust clouds, where we completely `resolve' the issue of shell crossing singularities.
	\begin{introtheorem}[rough formulation] \label{MGHD}
		Let $\mathfrak{D}_{\chi_b}$ be the moduli space of smooth initial data for (E-d) consisting of a collapsing dust cloud of fixed compact support, posed on an initial hypersurface $\mathcal{D}_0$ diffeomorphic to $\R^3$. There exists a \underline{generic}\footnote{By which we mean open and dense in a suitably defined metric space on $\mathfrak{D}_{\chi_b}$, for which a Cauchy stability result holds.} subset $\mathfrak{D}_{\chi_b}^* \subset \mathfrak{D}_{\chi_b}$, such that there exists a unique maximal globally hyperbolic piecewise smooth development satisfying the \underline{propagation condition}, $(\mathcal{Q}_{MGHD}, r, \Omega^2, \rho, \mathbf{u}^\mu)$, of initial data in $\mathfrak{D}_{\chi_b}^*$ posed on $\mathcal{D}_0$. At any first singularity of $\mathcal{Q}_{MGHD}$, the area radius $r$ extends continuously to zero. Finitely many ($N_{ts} \in \mathbb{N}-1$) thin shells form in $\mathcal{Q}_{MGHD}$. If $N_{ts} > 0$, then we denote the thin shells by $\Sigma_i \subset \mathcal{Q}_{MGHD}$ for $1 \leq i \leq N_{ts} \in \mathbb{N}$, with points of formation $\partial^- \Sigma_i \in \mathcal{Q}_{MGHD}$ which correspond to either \underline{non-degenerate} first shell crossing singularities or \underline{thin shell collisions}. 
		
		Moreover, the generic subset $\mathfrak{D}_{\chi_b}^* \subset \mathfrak{D}_{\chi_b}$ can be expressed as the following disjoint union:
		\begin{equation} \label{union}
			\mathfrak{D}_{\chi_b}^* = \mathfrak{D}^*_{\chi_b, \mathcal{O} \in \partial^+\Sigma} \cup \mathfrak{D}^*_{\chi_b, \mathcal{S} \neq \emptyset \neq \mathcal{CH}} \cup \mathfrak{D}^*_{\chi_b, \mathcal{S} = \emptyset \neq \mathcal{CH}},
		\end{equation} 
		where each of $\mathfrak{D}^*_{\chi_b, \mathcal{O} \in \partial^+\Sigma},  \mathfrak{D}^*_{\chi_b, \mathcal{S} \neq \emptyset \neq \mathcal{CH}}, \mathfrak{D}^*_{\chi_b, \mathcal{S} \neq \emptyset \neq \mathcal{CH}}$ have non-empty interior (in the smooth topology on initial data). The Penrose diagram of $\mathcal{Q}_{MGHD}$ for initial data in each respective subset of \eqref{union} have the following boundary decompositions:
		\begin{figure} [H] 
			\centering
			\def\svgwidth{0.75\linewidth}
			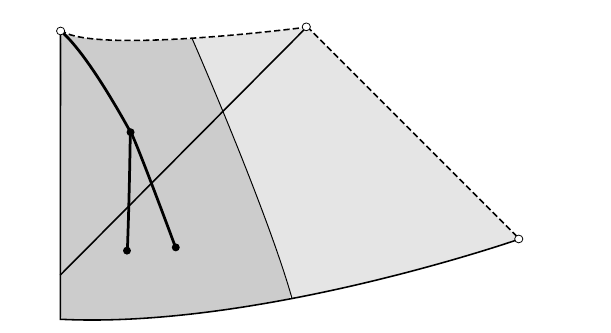 
			\caption{Penrose diagram of $\mathcal{Q}_{MGHD}$ for initial data in $\mathfrak{D}^*_{\chi_b, \mathcal{O} \in \partial^+\Sigma}$ posed on $\mathcal{D}_0$. In this particular example $N_{ts} = 3$, with two thin shells forming at shell crossing singularities and one forming at a thin shell collision. In general, for initial data in $\mathfrak{D}^*_{\chi_b, \mathcal{O} \in \partial^+\Sigma}$, $N_{ts} \geq 1$.}
			\label{fig:O-ts}
		\end{figure}
		\begin{figure}[H]
			\centering
			\begin{minipage}{0.52\linewidth}
				\centering
				\def\svgwidth{\linewidth}
				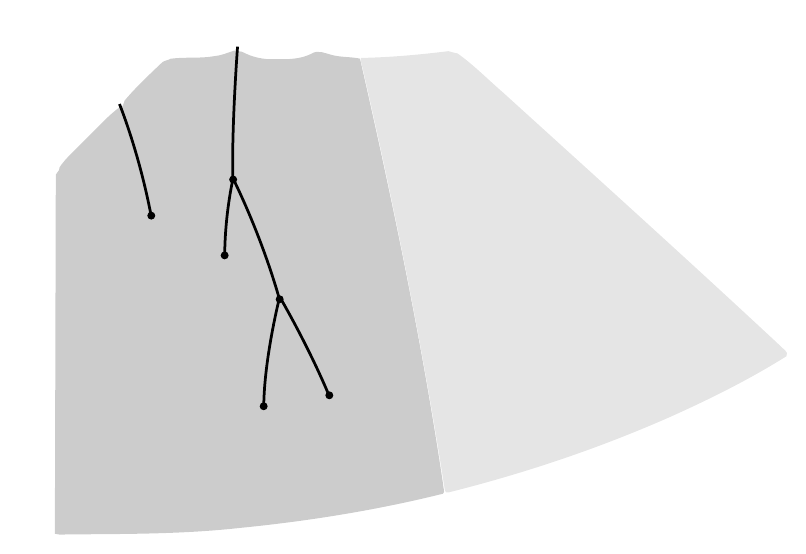
				\caption{Penrose diagram of $\mathcal{Q}_{MGHD}$ for initial data in $\mathfrak{D}^*_{\chi_b, \mathcal{S} \neq \emptyset \neq \mathcal{CH}}$ posed on $\mathcal{D}_0$. In this particular example, $N_{ts} = 7$. In general, for initial data in $\mathfrak{D}^*_{\chi_b, \mathcal{S} \neq \emptyset \neq \mathcal{CH}}$, $N_{ts} \geq 0$.}
				\label{fig:loc-naked}
			\end{minipage}\hfill
			\begin{minipage}{0.43\linewidth}
				\centering
				\def\svgwidth{\linewidth}
				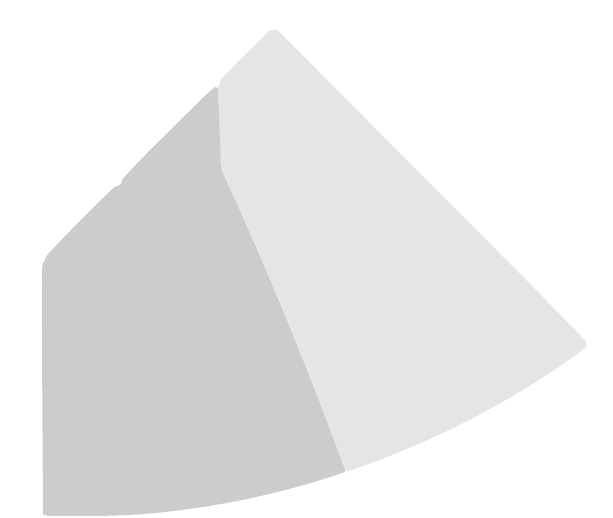
				\caption{Penrose diagram of $\mathcal{Q}_{MGHD}$ for initial data in $\mathfrak{D}^*_{\chi_b, \mathcal{S} = \emptyset \neq \mathcal{CH}}$ posed on $\mathcal{D}_0$. In this particular example, $N_{ts} = 2$. In general, for initial data in $\mathfrak{D}^*_{\chi_b, \mathcal{S} = \emptyset \neq \mathcal{CH}}$, $N_{ts} \geq 0$. }
				\label{fig:glob-naked}
			\end{minipage}
		\end{figure}
		Here\footnote{For the precise definitions of the boundary components, see the precise statement in Section \ref{sec:thm2}.}, $\Gamma$ is the centre of spherical symmetry, $\mathcal{O}$ is the first singularity at the centre, $\mathcal{S}$ is an achronal curve to which the area radius extends continuously to zero, $i^0$ is spacelike infinity, $i^+$ is future timelike infinity, $\mathcal{I}^+$ is future null infinity. In Figures \ref{fig:loc-naked} and \ref{fig:glob-naked}, $\mathcal{CH}$ is a Cauchy horizon emanating from $\mathcal{O}$ (a central shell-focussing singularity). In Figure \ref{fig:glob-naked}, $i^\Box$ is the future endpoint of $\mathcal{CH}$/future limit point of $\mathcal{I}^+$: either $\mathcal{I}^+$ is complete\footnote{In the sense of Christodoulou \cite{C4}.} and $i^\Box = i^+$, or $\mathcal{I}^+$ is incomplete and $i^\Box = i^{naked}$ (in which case $\mathcal{O}$ is a `globally' naked singularity).
		
		In each figure, the dust cloud (the darker shaded region) is bounded by $\mathcal{B}_d$, a timelike connected (possibly piecewise, see for example Figure \ref{fig:glob-naked}) $C^1$ curve, and the lighter shaded region is isometric to Schwarzschild with metric $g_{M}$ for some $M>0$.	For initial data in $\mathfrak{D}^*_{\chi_b, \mathcal{O} \in \partial^+\Sigma}$, as depicted in Figure \ref{fig:O-ts}, at least one thin shell must form in the maximal piecewise smooth development and $\mathcal{O}$ is the future limit point of at least one thin shell. Let $\partial^+\Sigma$ be the set of future limit points of thin shells, and $\partial^-\Sigma$ the set of points of formation of thin shells, then:
		\begin{equation}
			\partial^+\Sigma \subset \partial^-\Sigma \cup \mathcal{O} \cup \mathcal{CH} \cup \mathcal{S}.
		\end{equation}
	\end{introtheorem} 
	For an overview of the main ideas in the setup and proof of the theorem, see already Section \ref{intromaxdev}. See already Theorem \ref{MGHDprecise} in Section \ref{sec:thm2} for the precise statement of the theorem. 
	
	Note that Theorem \ref{MGHD} does not include the exact Oppenheimer--Snyder (O--S) spacetimes, since they are not smooth due to a jump discontinuity in the energy density $\rho$ at the boundary of the dust cloud.\footnote{The optimal regularity of the O--S metric at the boundary of the dust cloud is $C^{1,1}$, which is attained in double null coordinates.} One can consider instead initial data $\Psi_{OS} \in \mathfrak{D}_{\chi_b}$ that coincides with O--S data for area radius $0 \leq r \leq \chi_a$, with $0 < \chi_a < \chi_b$ (where $r = \chi_b$ is the area radius of the support of the dust cloud), and such that no shell crossing singularities form in evolution of the initial data.\footnote{The Penrose diagram of the (smooth) maximal globally hyperbolic development of the data $\Psi_{OS}$ posed on $\mathcal{D}_0$ is therefore the same as that of the O--S spacetimes; see already the last part of Section \ref{gen-data} for the existence of data $\Psi_{OS} \in \mathfrak{D}_{\chi_b}^*$ satisfying the assumptions.} The structure of any such data in a neighbourhood of the centre implies that $\Psi_{OS} \notin \mathfrak{D}_{\chi_b}^*$. In the smooth topology on $\mathfrak{D}_{\chi_b}$, $\Psi_{OS}$ is in the closure of both $\mathfrak{D}^*_{\chi_b, \mathcal{O} \in \partial^+\Sigma}$ and $\mathfrak{D}^*_{\chi_b, \mathcal{S} \neq \emptyset \neq \mathcal{CH}}$. Indeed, the Oppenheimer--Snyder spacetimes (with a smoothed boundary) are unstable to \emph{both} formation of a locally naked shell-focussing singularity at the centre \emph{and} formation of (arbitrarily many) thin shells. (For an explanation, see already Sections \ref{comovingintro} and \ref{gen-data}.)
	
	Theorem \ref{MGHD} implies in particular that the cosmic censorship conjectures are violated for maximal piecewise smooth developments of \emph{smooth} initial data for collapsing dust clouds. One could have hoped that the resolution of shell crossing could somehow have `saved' cosmic censorship, but the spacetimes constructed by Christodoulou in \cite{C1} are \emph{stable} against formation of shell crossing singularities, in the smooth topology on data (see already Section \ref{gen-data}). Indeed, the fact that $\mathfrak{D}_{\chi_b, \mathcal{S} \neq \emptyset \neq \mathcal{CH}}^*$, $\mathfrak{D}^*_{\chi_b, \mathcal{S} = \emptyset \neq \mathcal{CH}}$ have non-empty interior follows already from \cite{C1}. Moreover, the \emph{blueshift} of the (at least locally) naked singularities in Figures \ref{fig:loc-naked}, \ref{fig:glob-naked} is \emph{finite} (see already \eqref{blueshift} and Proposition \ref{prop:NS}), so one cannot directly implement the instability mechanism of Christodoulou \cite{C5} for the coupled Einstein-dust-scalar field system. It is however possible that in a moduli space of \emph{non-smooth} initial data for the Einstein-dust system, where Christodoulou's naked singularities are not necessarily stable, that \emph{generic} initial data could have maximal piecewise smooth development with Penrose diagram as in Figure \ref{fig:O-ts}. This will be addressed in upcoming work. Note that, in order to prove a \emph{strong} cosmic censorship result, one would have to show moreover a $C^{0,1}$ \emph{inextendibility} result\footnote{Note that the metric for piecewise smooth solutions of Einstein-dust in double null coordinates is only $C^{0,1}$ at thin shells (and this is the optimal regularity), hence the requirement of a $C^{0,1}$ inextendibility result.} at $\mathcal{S}$. Sbierski's result \cite{S1} on the $C^0$ inextendibility of Schwarzschild implies that the metric of any maximal piecewise smooth development with Penrose diagram as in Figure \ref{fig:O-ts} is $C^0$ inextendible at the future boundary of the lighter shaded (vacuum) region. It is (to the best of the author's knowledge) an open problem to prove a generic $C^{0,1}$ inextendibility result for the subset of $\mathcal{S}$ bounding the dust region.\footnote{Another result of Sbierski \cite{S2} implies that the Oppenheimer--Snyder spacetimes are in fact $C^{0,1}$ inextendible.} 
	
	\subsection{The main ideas of the setup and proof of Theorem \ref{thm:1}} \label{formationexp}
	
	In the following sections we give an overview of the problem of thin shell formation and development\footnote{Here, the terminology of `formation' and `development' is borrowed from the study of shocks in the compressible Euler equations. See for example \cite{CL}.} in Einstein-dust, which will serve as a guide to the setup and proof of Theorem \ref{thm:1}.
	
	In Section \ref{comovingintro}, we introduce comoving coordinates for smooth solutions of Einstein-dust. We show that any smooth solutions in comoving coordinates are \emph{uniquely determined} by integrating along fluid flow lines (integral curves of the fluid velocity). We're led to define the notion of the `maximal dust development' (MDD), a singular boundary of which corresponds to \emph{shell crossing}. 
	
	In Section \ref{dn-intro}, we introduce the Einstein-dust system in double null coordinates, along with the definition of a non-degenerate first shell crossing singularity. We briefly discuss the construction of double null coordinates on a solution of the Einstein-dust system in comoving coordinates, as solutions of the \emph{eikonal equation}. 
	
	We then introduce piecewise smooth solutions of (E-d) in Section \ref{thinshells}, which we construct as solutions of the \emph{thin shell ODE}, using again the eikonal equation to construct double null coordinates. This is exactly the same method that we use to construct a solution of the \emph{thin shell development problem}, which we discuss in Section \ref{introdev}. The proof reduces to an existence/uniqueness statement for the thin shell ODE posed on a maximal dust development, with certain initial conditions. The ODE has a singular point, and we show existence/uniqueness by expanding in higher order terms which we show successively are \emph{bounded}, ultimately expanding the ODE into the form of a regular singular point, which we show has a unique solution using a contraction mapping argument.
	
	\subsubsection{Einstein-dust in comoving coordinates and the `maximal dust development'} \label{comovingintro}
	
	\textbf{Einstein-dust system in comoving coordinates.} The Einstein-dust system consists of the Einstein equations
	\begin{equation}
		R_{\mu \nu} - \frac{1}{2} R g_{\mu \nu} = 8\pi T_{\mu \nu},
	\end{equation}
	with energy-momentum tensor corresponding to \emph{pressureless} fluid,
	\begin{equation}
		T_{\mu \nu} = \rho \mathbf{u}_\mu \mathbf{u}_\nu,
	\end{equation}
	where $\rho$ is the energy density, and $\mathbf{u}^a$ is the fluid velocity, a timelike vector field normalised by $g(\mathbf{u},\mathbf{u}) = -1$. On any smooth solution of the spherically symmetric Einstein-dust system, one can define functions $\tau,\chi$, so-called \emph{comoving coordinates}, where $\tau$ is proper time along fluid flow lines (the integral curves of $\mathbf{u}^a$), and $\chi$ `labels' the fluid flow lines (see Lemma \ref{comovinglemma}): 
	\begin{equation} \label{gaugeprop}
		\mathbf{u}(\tau) = \mathbf{u}^\mu \partial_\mu \tau = 1, \qquad \mathbf{u}(\chi) = 0.
	\end{equation}
	In comoving coordinates, the metric, energy density and fluid velocity take the respective forms:
	\begin{equation} \label{int:metric-rho-u}
		g = -d\tau^2 + \frac{(\partial_\chi r(\tau,\chi))^2}{1 + f(\chi)} d\chi^2 + r(\tau,\chi)^2 g_{\mathbb{S}^2}, \qquad \rho(\tau,\chi) = \frac{m'(\chi)}{ (r^2 \partial_\chi r)(\tau,\chi)}, \qquad \mathbf{u} = \partial_\tau,
	\end{equation}
	where $m,f$ are functions determined by initial data ($m$ is the Hawking mass), satisfying $\partial_\tau m = 0 = \partial_\tau f$. We consider only \emph{collapsing} dust, with $\mathbf{u}(r) < 0$ (when $r > 0$), where the function $r(\tau,\chi)$ is determined by the following ODE along fluid flow lines:
	\begin{equation} \label{vel}
		\mathbf{u}(r) = \partial_\tau r(\tau,\chi) = - \left( \frac{2m(\chi)}{r(\tau,\chi)} + f(\chi)\right)^\frac{1}{2}.
	\end{equation}
	We observe from \eqref{int:metric-rho-u} that the energy density can blow up if $r \to 0$ or $\partial_\chi r \to 0$. The term \underline{shell crossing} in the literature refers to the singularities for which, upon integrating \eqref{vel}, there exists a (locally defined) function $\tau_c(\chi)$ such that:
	\begin{equation} \label{shell-crossing}
		\lim_{\tau \to \tau_c(\chi)} \partial_\chi r(\tau,\chi) = 0, \qquad \liminf_{\tau \to \tau_c(\chi)} r(\tau,\chi) > 0,
	\end{equation}
	where from \eqref{int:metric-rho-u} we see that the metric, in comoving coordinates, does not even extend continuously, as a Lorentzian metric, to shell crossing singularities!\footnote{This can be expected since comoving coordinates are defined with respect to the foliation of spacetime by the fluid flow lines, and shell crossing is exactly the density of this foliation $\sim (\partial_\chi r)^{-1}$ blowing up.} 
	
	\textbf{Maximal dust development.} In the construction of comoving coordinates, we've seen that the \emph{gauge} is uniquely determined along fluid flow lines, \eqref{gaugeprop}, and moreover the solution in comoving coordinates is uniquely determined by integrating \eqref{vel} in $\tau$ for fixed $\chi$, along with $\partial_\tau m = 0 = \partial_\tau f$. The fluid flow lines are timelike, and therefore smooth solutions of Einstein-dust \emph{in spherical symmetry} are uniquely determined possibly \emph{beyond} the usual maximal globally hyperbolic development, in comoving coordinates. Note that this is only possible because the gravity degrees of freedom of the Einstein equations in spherical symmetry are pure gauge, cf.~Birkhoff's Theorem.
	
	We therefore define the `maximal dust development' (MDD) of smooth solutions of Einstein-dust in comoving coordinates (which will be essential to understanding piecewise smooth solutions of Einstein-dust in double null coordinates) as follows. Suppose that we have suitably defined smooth initial data (with $r > 0$)\footnote{See Section \ref{gen-data} for a note on the inclusion of the centre.} posed on a curve in the plane $\R^2_{\tau,\chi}$, e.g.~$\{\tau = \tau_{\mathscr{D}}(\chi)\}$, for $\tau_{\mathscr{D}}(\chi) \in C^\infty([\chi_1,\chi_2])$, then we define the MDD, $\mathscr{Q}_{MDD} \subset \R^2_{\tau,\chi}$, to be the maximal subset of $\{ \chi \in [\chi_1,\chi_2], \tau \geq \tau_{\mathscr{D}}(\chi) \} \subset \R^{2}_{\tau,\chi}$ such that for the solution $r(\tau,\chi)$ of \eqref{vel} with the given initial data: $r(\tau,\chi), \partial_\chi r(\tau,\chi) > 0$ for all $(\tau,\chi) \in \mathscr{Q}_{MDD}$.
	
	It will be useful to introduce the functions that characterise the MDD: one can always define the function $\tau_s \in C^\infty([\chi_1,\chi_2])$, which satisfies $r(\tau_s(\chi),\chi) = 0$. However, if the set $\mathscr{C}_\chi := \{ \chi \in [\chi_1,\chi_2] : \tau_s'(\chi) < 0 \}$ is non-empty, then there exists $\tau_c \in C^\infty(\mathscr{C}_\chi)$, with $\tau_c(\chi) < \tau_s(\chi)$, such that \eqref{shell-crossing} holds (so that $\mathscr{C}_\chi$ is the set of fluid flows that terminate at a shell crossing singularity). We then define the \emph{continuous} function 
	\begin{equation}
		\tau_{MDD}(\chi) := \begin{cases}
			\tau_s(\chi) : \chi \in \mathscr{S}_\chi := ([\chi_1,\chi_2]) \setminus \mathscr{C}_\chi, \\ \tau_c(\chi) : \chi \in \mathscr{C}_\chi,
		\end{cases}
	\end{equation}
	which denotes the maximal time of existence along each fluid flow line, and the MDD is given by
	\begin{equation}
		\mathscr{Q}_{MDD} := \{ (\tau,\chi) : \tau_{\mathscr{D}}(\chi) \leq \tau < \tau_{MDD}(\chi), \ \chi \in [\chi_1,\chi_2]\}.
	\end{equation}
	The \emph{singular boundary} of the maximal dust development $\mathscr{Q}_{MDD}$ (as induced in $\R^2_{\tau,\chi}$), consists of $\mathscr{S} \cup \mathscr{C}$, where $\mathscr{S} := \{ (\tau_s(\chi),\chi) : \chi \in \mathscr{S}_\chi\}$ and the `shell crossing singular boundary': $\mathscr{C} := \{ (\tau_c(\chi),\chi) : \chi \in \mathscr{C}_\chi\}$.
	
	Note that, for smooth initial data, the function $r(\tau,\chi)$ (and obviously $m(\chi),f(\chi)$) extend in $C^\infty$ to $\mathscr{C}$. See Section \ref{mdd} for the precise definitions and the structure of the MDD, in particular Proposition \ref{mddprop}. 
	
	The Oppenheimer--Snyder spacetimes are (a subset of)\footnote{In the original paper \cite{OS}, O--S consider the special case of a \emph{marginally bound} (homogeneous) dust cloud, for which $f=0$ (for $f$ as in \eqref{int:metric-rho-u}, \eqref{vel}).} the (highly non-generic) family of \emph{homogeneous} dust clouds, for which $\tau_s(\chi) = const$. Since $\tau_s'(\chi) = 0$, there are no shell crossing singularities, but arbitrarily small perturbations of initial data lead to shell crossing singularities (and, separately, locally naked shell-focusing singularities at the centre). See already Section \ref{gen-data} for a discussion of the metric space structure on initial data, and in particular the last part for a discussion of the (smoothed) Oppenheimer--Snyder spacetimes and Christodoulou's locally naked singularities in the moduli space of initial data. 
	
	In any homogeneous dust cloud, the foliation of spacetime by fluid flow lines \emph{diverges} at $\mathscr{S}$ (away from the centre): $\partial_\chi r \sim r$. This contrasts with the `generic' structure of $\mathscr{S}$, $\tau_s'(\chi) > 0$, for which $\partial_\chi r \sim r^{-\frac{1}{2}}$. (See Section \ref{firstsingrzero}, in particular Lemma \ref{lem:eik-S} and \eqref{pchir-exp}.) 
	
	\subsubsection{Einstein-dust in double null coordinates (E-d), non-degenerate first shell crossing singularities and the eikonal equation} \label{dn-intro}
	
	\textbf{Einstein-dust in double null coordinates and first shell crossing singularities.} We've seen in Section \ref{comovingintro} that by constructing comoving coordinates we can completely understand \emph{smooth} solutions of Einstein-dust in spherical symmetry. In order to understand \emph{non-smooth} solutions however, we introduce double null coordinates, which we'll see extend in `optimal' regularity to \emph{first shell crossing singularities}, which we define. Indeed, double null coordinates are defined to satisfy the `eikonal equation':
	\begin{equation} \label{int:eik}
		g^{-1}(du,du) = 0 = g^{-1}(dv,dv),
	\end{equation}
	with $g^{-1}(du,dv) < 0$, so that the metric is given by
	\begin{equation} \label{dn-metric}
		g = -\Omega^2 dudv + r^2 g_{\mathbb{S}^2}.
	\end{equation}
	The spherically symmetric Einstein-dust system in double null coordinates, which we refer to as (E-d), takes the form:
	\begin{align}
		\partial_u \partial_v r = \Omega^2 \left( \pi r \rho - \frac{m}{2r^2}\right),& \qquad
		\partial_u \partial_v \log{\Omega^2} =  \Omega^2 \left( \frac{m}{r^3} - 2\pi \rho\right), \label{int:wave} \\
		\partial_u \left( -\frac{\partial_u r}{\Omega^2} \right) = \pi r\rho (\mathbf{u}^v)^2 \Omega^2,& \qquad \partial_v \left( -\frac{\partial_v r}{\Omega^2} \right) = \pi r\rho (\mathbf{u}^u)^2 \Omega^2, \label{int:Raych}
	\end{align}
	where $m$ is the Hawking mass, and normalisation of the fluid velocity is equivalent to $\Omega^2 \mathbf{u}^u \mathbf{u}^v = 1$. The system (E-d) consists of wave equations  \eqref{int:wave} for the functions $r, \log \Omega^2$, and the so-called Raychaudhuri equations \eqref{int:Raych}. We consider smooth solutions of (E-d) defined on a domain $\mathcal{Q} \subset \R^{1+1}_{u,v}$, with positive functions $r, \Omega^2, \rho, \mathbf{u}^\mu \in C^\infty(\mathcal{Q})$, solving \eqref{int:wave}, \eqref{int:Raych}. We moreover restrict to \emph{collapsing} solutions, which, defining first the vector field $\mathbf{u}_* := \mathbf{u}^v \partial_v - \mathbf{u}^u \partial_u$, satisfy the inequalities\footnote{Note that if these inequalities hold on initial data, they hold on any globally hyperbolic development of the data.}: $\mathbf{u}(r) < 0$ and $\mathbf{u}_*(r) > 0$. We define a \underline{first shell crossing singularity} (Definition \ref{firstscsing}) to be a point $(u_c,v_c) \in \R^{1+1}_{u,v}$ such that there exist $u_0 < u_c, v_0 < v_c$ with
	\begin{equation}
		\mathcal{R}' :=  ([u_0,u_c] \times [v_0,v_c]) \setminus \{ (u_c,v_c) \} \subset \mathcal{Q},
	\end{equation}
	and: $\inf_{\mathcal{R}'} r > 0$, $\sup_{\mathcal{R}'} \rho = \infty$. (Note that if instead $\sup_{\mathcal{R}'} \rho < \infty$, then we show that the solution extends smoothly to a neighbourhood of $(u_c,v_c)$, see Section \ref{sec:extp}.) 
	
\begin{figure}[h] 
		\centering
		\def\svgwidth{0.7\linewidth}
		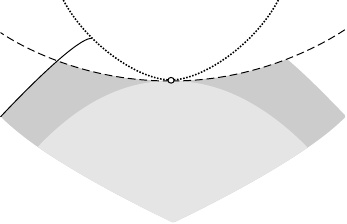 
		\caption{A subset of $\mathscr{Q}_{MDD} \subset \R^2_{\tau,\chi}$, in a neighbourhood of a non-degenerate minima of $\tau_c(\chi)$. The long-dashed lines represent future-directed ingoing/outgoing null curves from $(\tau_c(\chi_0,\chi_0))$, while the solid (thick) lines also represent ingoing and outgoing null curves. The union of the light shaded region, $\mathscr{R}'$, with the dark shaded regions $\mathscr{R}^\pm$ is the subset of $\mathscr{Q}_{MDD}$ corresponding to the shaded region in Figure \ref{MGHDsubset}.}
		\label{fig:4}
	\end{figure}

	\textbf{The eikonal equation and null curves on the maximal dust development.} It turns out to be useful to not analyse the system (E-d) directly, but to instead define comoving coordinates on $\mathcal{R}'$: $\tau, \chi \in C^\infty(\mathcal{R}')$, and transform to a solution of the Einstein-dust system in comoving coordinates, which we denote $(\mathscr{R}',r,\rho,m,f)$, with $\mathscr{R}' := \{ (\tau(u,v),\chi(u,v)) : (u,v) \in \mathcal{R}' \}$. Let $\mathscr{Q}_{MDD}$ be the maximal dust development extending $\mathscr{R}'$. The double null coordinates are given by functions $u, v \in C^\infty(\mathscr{R}')$ which satisfy the \emph{eikonal equation}, in comoving coordinates this is equivalent to (see Section \ref{eikonalsection}):
	\begin{equation} \label{int:eik-com}
		\partial_\chi u(\tau,\chi) = - \frac{\partial_\chi r(\tau,\chi)}{(1+f(\chi))^\frac{1}{2}} \partial_\tau u(\tau,\chi), \qquad \partial_\chi v(\tau,\chi) = \frac{\partial_\chi r(\tau,\chi)}{(1+f(\chi))^\frac{1}{2}} \partial_\tau v(\tau,\chi).
	\end{equation}
	By definition of the null coordinates (or e.g.~by the method of characteristics), $u(\tau,\chi)$ is constant on outgoing null curves $\tau = \tau_+(\chi)$, $v(\tau,\chi)$ is constant on ingoing null curves, $\tau = \tau_-(\chi)$, which satisfy
	\begin{equation} \label{int:null}
		\tau_\pm'(\chi) = \pm \frac{\partial_\chi r(\tau,\chi)}{(1+f(\chi))^\frac{1}{2}}.
	\end{equation}
	The advantage of studying the eikonal equation on the maximal dust development $\mathscr{Q}_{MDD}$ is that solutions of \eqref{int:eik-com} (on their respective domains of definition) extend smoothly to $\mathscr{C}$ (see Proposition \ref{eikprop}). Then, simply by writing down the transformation back to double null coordinates (see Lemma \ref{dntransf}), we can read off the regularity of the functions $r, \Omega^2, \rho, \mathbf{u}^\mu$. Indeed, we show in Proposition \ref{firstscprop}, using $\mathbf{u}^u(u,v) = \partial_\tau u(\tau(u,v),\chi(u,v))$, $\mathbf{u}^v(u,v) = \partial_\tau v(\tau(u,v),\chi(u,v))$  (and the fact that the inequality $\partial_\tau u, \partial_\tau v > 0$ extends to $\mathscr{C}$, see Proposition \ref{eikprop}), that $\mathbf{u}^u, \mathbf{u}^v, \Omega^2, m \sim 1$ at any first shell crossing singularity $(u_c,v_c)$, and the following integrals converge:
	\begin{equation}
		\int_{u_0}^{u_c} \rho(u,v_c) du < \infty, \qquad \int_{v_0}^{v_c} \rho(u_c,v) dv < \infty.
	\end{equation}
	\textbf{Non-degenerate first shell crossing singularities.} This motivates the definition: a first shell crossing singularity is \underline{non-degenerate} if there exists $\beta < 1$ such that $\rho(u,v_c) \lesssim (u_c - u)^{-\beta}, \ \rho(u_c,v) \lesssim (v_c - v)^{-\beta}$ (Definition \ref{nondegensc}). At any non-degenerate first shell crossing singularity (see Lemma \ref{nondeglemma}), we show that $\tau(u,v), \chi(u,v)$ extend continuously to $(u_c,v_c)$ and moreover that $\tau_c(\chi)$ has a `non-degenerate' minima at $\chi_0 := \chi(u_c,v_c)$. Indeed, there exists $n \in \mathbb{N}$ such that:
	\begin{equation} \label{non-deg-minima}
		\tau_c^{(2n)}(\chi_0) > 0, \ \tau_c^{l}(\chi_0) = 0 \text{ for all } 1 \leq l < 2n.
	\end{equation}
	In Proposition \ref{regext}, using \eqref{non-deg-minima}, and $\partial_\chi r \sim (\tau_c(\chi) - \tau)$ (which follows from \eqref{vel}), we show that there exist future-directed ingoing/outgoing null curves emanating from $(\tau_c(\chi_0),\chi_0)$ in $\mathscr{Q}_{MDD}$, and therefore that the smooth solution of (E-d) $(\mathcal{Q}, r, \Omega^2, \rho, \mathbf{u}^\mu)$ uniquely extends to include a region as depicted in Figure \ref{MGHDsubset}, with $p = (u_c,v_c)$. This defines solutions of the eikonal equation on further regions $\mathscr{R}^\pm \subset \mathscr{Q}_{MDD}$, depicted in Figure \ref{fig:4}.
	
	We note the following domain of dependence property for solutions of the eikonal equation (see Proposition \ref{eikprop}): $u(\tau,\chi)$ is uniquely determined along outgoing null curves, $v(\tau,\chi)$ is uniquely determined on ingoing null curves. As demonstrated in Figure \ref{fig:4}, for any non-degenerate shell crossing singularity, there exists $\chi_- < \chi_0$ such that for all $\chi \in [\chi_-, \chi_0)$, $u(\tau,\chi)$ is uniquely determined up to $(\tau_c(\chi),\chi)$ along outgoing null curves, and similarly there exists $\chi_+ > \chi_0$ such that for $\chi \in (\chi_0, \chi_+]$, $v(\tau,\chi)$ is uniquely determined up to $(\tau_c(\chi),\chi)$ along ingoing null curves. This will be important for the thin shell development problem.
	
	We show moreover in Proposition \ref{regext}, again using the transformation to double null coordinates from solutions of the eikonal equation in comoving coordinates, that $r,\Omega^2$ extend in $C^1_{u,v}$ to $(u_c,v_c)$ for any non-degenerate first shell crossing singularity, and in fact $(r\Omega)^2$ extends in $C^2_{u,v}$. One can therefore consider double null coordinates to be `optimally regular' at first shell crossing singularites: curvature blows up, so second derivatives of the metric components necessarily blow up, and the optimal regularity is continuous first derivatives of the metric components.
		
	\subsubsection{Piecewise smooth solutions of (E-d), thin shells of dust, the `thin shell ODE' and the determinism condition} \label{thinshells}
	
	One of the advantages of studying the Einstein-dust system in double null coordinates is the ability to define \emph{piecewise smooth} solutions; comoving coordinates can only be constructed when spacetime is foliated by fluid flow lines, they will be essential however in understanding the dynamics of piecewise smooth solutions. 
	
	\textbf{Piecewise smooth solutions of (E-d) and thin shells of dust.} Let $\mathcal{Q} := [0,u_1) \times [0,v_1) \subset \R^{1,1}_{u,v}$, with metric $g_\mathcal{Q} = -\Omega^2 du dv$. We define weak solutions of (E-d) by multiplying \eqref{int:wave}, \eqref{int:Raych} by a $C^1_{c}$ test function and integrating by parts once on the left hand side. (See Section \ref{pwisesection} for the precise definitions of the setup discussed here.) For example, the wave equation for $r(u,v)$ is satisfied in `weak integrated form' if
	\begin{equation}
		\int_0^{u_1} \int_0^{v_1} \partial_v r \partial_u \varphi dudv + \int_0^{v_1} \varphi(0,v)\partial_v r(0,v) dv = \int_\mathcal{Q} \varphi\left(\frac{m}{r^2} - 2\pi r\rho \right),
	\end{equation}
	\begin{equation}
		\int_0^{u_1} \int_0^{v_1} \partial_u r \partial_v \varphi dudv + \int_0^{u_1} \varphi(u,0)\partial_u r(u,0) du = \int_\mathcal{Q} \varphi\left(\frac{m}{r^2} - 2\pi r\rho \right),
	\end{equation}
	for all $\varphi \in C^1_c(\mathcal{Q})$. Now, let $\Sigma$ be a timelike curve defined by the functions $u_\Sigma,v_\Sigma \in C^1([0,s_1]) \cap C^2([0,s_1))$:
	\begin{equation}
		\Sigma = \{ (u_\Sigma(s),v_\Sigma(s)) : s \in [0,s_1)\},
	\end{equation}
	where $s$ is proper time along $\Sigma$, so that $\Omega^2(u_\Sigma(s),v_\Sigma(s)) \dot{u}_\Sigma(s) \dot{v}_\Sigma(s) = 1$.
	We suppose that $\mathcal{Q}\setminus\Sigma$ has two connected components: an interior $\mathcal{Q}^-$ and exterior $\mathcal{Q}^+$, where $r, \Omega \in C^{0,1}_{u,v}(\mathcal{Q})$, and $r \in C^2(\mathcal{Q}^\pm)$, $\Omega^2 \in C^1(\mathcal{Q^\pm})$, with extension in $C^2$/$C^1$ respectively to $\Sigma$ from $\mathcal{Q}^\pm$. We then consider the following piecewise smooth distribution of dust $(\rho,\mathbf{u}^\mu)$, with delta function support for $\rho$ on $\Sigma$. We suppose that, for $(u,v) \in \mathcal{Q}^\pm$: $\rho(u,v) = \mathring{\rho}(u,v)$, $\mathbf{u}^\mu(u,v) = \mathring{\mathbf{u}}^\mu(u,v)$, where $\mathring{\rho}, \mathring{\mathbf{u}}^\mu \in C^1(\mathcal{Q}^\pm)$ with extension in $C^1$ to $\Sigma$ from $\mathcal{Q}^\pm$. Let $\mathbf{u}(u_\Sigma(s),v_\Sigma(s)) := \dot{u}_\Sigma(s) \partial_u + \dot{v}_\Sigma(s) \partial_v$, and suppose that exists $\sigma \in C^0([0,s_1)) \cap C^1((0,s_1))$ such that, for any $\varphi \in C^1(\mathcal{Q})$,
	\begin{equation}
		\int_{\mathcal{Q}} \varphi(u,v) \rho = \int_{0}^{s_1} \varphi(u_\Sigma(s),v_\Sigma(s)) \sigma(s) ds + \int_{\mathcal{Q}^+} \varphi(u,v) \mathring{\rho} + \int_{\mathcal{Q}^-} \varphi(u,v) \mathring{\rho}.
	\end{equation}
	It is straightforward to show, using integration by parts (see Lemma \ref{thinshellequiv}), that this construction is a weak solution of (E-d) if the Einstein equations are solved classically\footnote{We actually assume additional regularity for $\Omega^2$: $\partial_u \log\Omega^2$ can be differentiated in $v$; $\partial_v \log\Omega^2$ can be differentiated in $u$; so that the wave equation for $\log \Omega^2$ is satisfied continuously.} on $\mathcal{Q}^\pm$, and the following jump conditions on $\Sigma$ are satisfied:
	\begin{align}
		\dot{u}_{\Sigma}(s) [\partial_u r](s) = 2\pi \sigma(s) r(s) &, \qquad \dot{v}_{\Sigma}(s) [\partial_v r](s) = -2\pi\sigma(s) r(s), \label{rjump} \\ \dot{u}_{\Sigma}(s) [\partial_u \log\Omega^2](s) = -4\pi\sigma(s) &, \qquad \dot{v}_{\Sigma}(s) [\partial_v \log\Omega^2](s) = 4\pi\sigma(s), 
	\end{align}
	(where we've defined e.g.~$r(s) := r(u_\Sigma(s),v_\Sigma(s))$). This is equivalent to $\Sigma \times \mathbb{S}^2$ being a \emph{thin shell of dust} in the spacetime $(\mathcal{M} = \mathcal{Q} \times \mathbb{S}^2,g = g_{\mathcal{Q}} + r^2 g_{\mathbb{S}^2})$ (which we show in Lemma \ref{thinshellequiv}): the induced metric on $\Sigma \times \mathbb{S}^2$ is $h= - ds^2 + r(s)^2 g_{\mathbb{S}^2}$ , and the extrinsic curvature $K_{i j}^\pm(s)$ of $\Sigma$ as a boundary of $\mathcal{M}^\pm$ satisfies:
	\begin{equation} \label{thinshellcond}
		[K_{ss}](s) = - 4\pi \sigma(s), \qquad [K_{A B}](s) = -  4 \pi \sigma(s) r(s)^2 (g_{\mathbb{S}^2})_{A B},
	\end{equation}
	so that the `surface energy momentum tensor' of $\Sigma \times \mathbb{S}^2$ is that of a thin shell of dust (see the start of Section \ref{pwise} or \cite{I}, \cite{Po} for the definition). 
	
\begin{figure}[htbp]
		\centering
		\begin{minipage}{0.4\linewidth}
			\centering
			\def\svgwidth{\linewidth}
\begingroup%
  \makeatletter%
  \providecommand\color[2][]{%
    \errmessage{(Inkscape) Color is used for the text in Inkscape, but the package 'color.sty' is not loaded}%
    \renewcommand\color[2][]{}%
  }%
  \providecommand\transparent[1]{%
    \errmessage{(Inkscape) Transparency is used (non-zero) for the text in Inkscape, but the package 'transparent.sty' is not loaded}%
    \renewcommand\transparent[1]{}%
  }%
  \providecommand\rotatebox[2]{#2}%
  \newcommand*\fsize{\dimexpr\f@size pt\relax}%
  \newcommand*\lineheight[1]{\fontsize{\fsize}{#1\fsize}\selectfont}%
  \ifx\svgwidth\undefined%
    \setlength{\unitlength}{210.18082746bp}%
    \ifx\svgscale\undefined%
      \relax%
    \else%
      \setlength{\unitlength}{\unitlength * \real{\svgscale}}%
    \fi%
  \else%
    \setlength{\unitlength}{\svgwidth}%
  \fi%
  \global\let\svgwidth\undefined%
  \global\let\svgscale\undefined%
  \makeatother%
  \begin{picture}(1,1.11503381)%
    \lineheight{1}%
    \setlength\tabcolsep{0pt}%
    \put(0,0){\includegraphics[width=\unitlength,page=1]{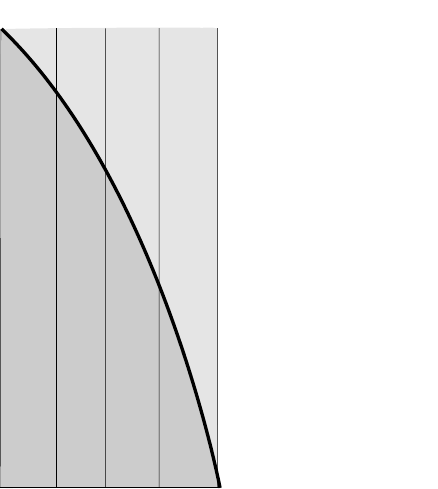}}%
    \put(0.2674835,0.70847035){\color[rgb]{0,0,0}\makebox(0,0)[lt]{\lineheight{2.04999995}\smash{\begin{tabular}[t]{l}$\Sigma^-$\end{tabular}}}}%
    \put(0,0){\includegraphics[width=\unitlength,page=2]{thinshells_MDD.pdf}}%
    \put(0.79703793,0.69766655){\color[rgb]{0,0,0}\makebox(0,0)[lt]{\lineheight{2.04999995}\smash{\begin{tabular}[t]{l}$\Sigma^+$\end{tabular}}}}%
    \put(0.19607969,1.07040152){\color[rgb]{0,0,0}\makebox(0,0)[lt]{\lineheight{2.04999995}\smash{\begin{tabular}[t]{l}$\mathscr{Q}_{MDD}^-$\end{tabular}}}}%
    \put(0.75630351,0.997562){\color[rgb]{0,0,0}\makebox(0,0)[lt]{\lineheight{2.04999995}\smash{\begin{tabular}[t]{l}$\mathscr{Q}_{MDD}^+$\end{tabular}}}}%
  \end{picture}%
\endgroup%
 
			\caption{The exterior/interior curves of a solution of the thin shell ODE on (subsets of) maximal dust developments $\mathscr{Q}_{MDD}^\pm$.}
			\label{fig:thinshells_MDD}
		\end{minipage}\hspace{5em}
		\begin{minipage}{0.48\linewidth}
			\centering
			\def\svgwidth{0.8\linewidth}  
\begingroup%
  \makeatletter%
  \providecommand\color[2][]{%
    \errmessage{(Inkscape) Color is used for the text in Inkscape, but the package 'color.sty' is not loaded}%
    \renewcommand\color[2][]{}%
  }%
  \providecommand\transparent[1]{%
    \errmessage{(Inkscape) Transparency is used (non-zero) for the text in Inkscape, but the package 'transparent.sty' is not loaded}%
    \renewcommand\transparent[1]{}%
  }%
  \providecommand\rotatebox[2]{#2}%
  \newcommand*\fsize{\dimexpr\f@size pt\relax}%
  \newcommand*\lineheight[1]{\fontsize{\fsize}{#1\fsize}\selectfont}%
  \ifx\svgwidth\undefined%
    \setlength{\unitlength}{400.8794349bp}%
    \ifx\svgscale\undefined%
      \relax%
    \else%
      \setlength{\unitlength}{\unitlength * \real{\svgscale}}%
    \fi%
  \else%
    \setlength{\unitlength}{\svgwidth}%
  \fi%
  \global\let\svgwidth\undefined%
  \global\let\svgscale\undefined%
  \makeatother%
  \begin{picture}(1,1)%
    \lineheight{1}%
    \setlength\tabcolsep{0pt}%
    \put(0,0){\includegraphics[width=\unitlength,page=1]{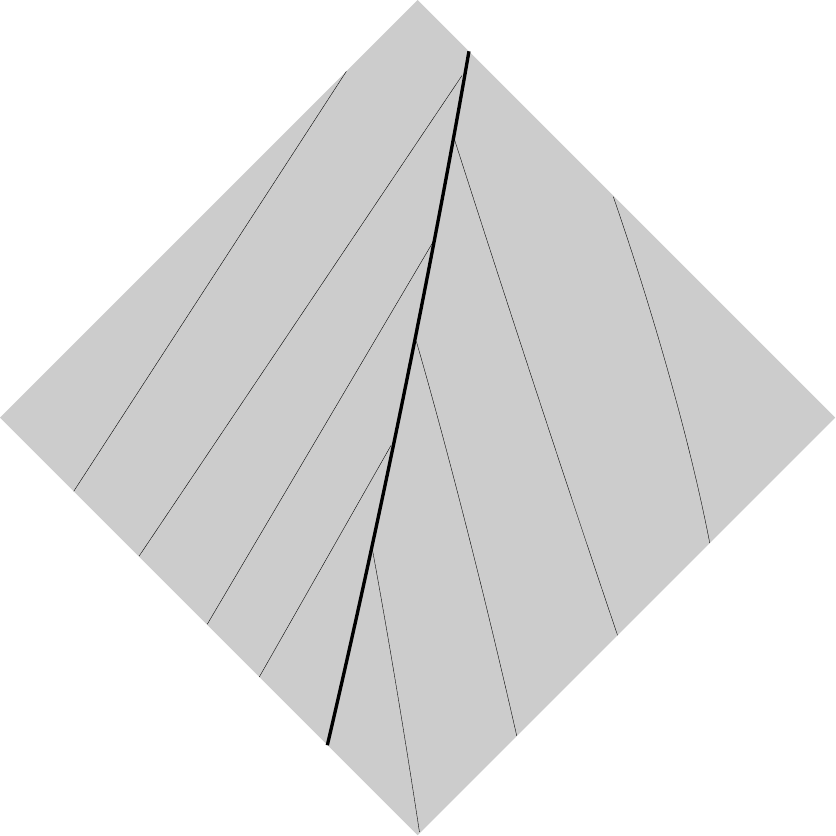}}%
    \put(0.47009794,0.73215578){\color[rgb]{0,0,0}\makebox(0,0)[lt]{\lineheight{2.04999995}\smash{\begin{tabular}[t]{l}$\Sigma$\end{tabular}}}}%
    \put(0.80882589,0.85602053){\color[rgb]{0,0,0}\makebox(0,0)[lt]{\lineheight{2.04999995}\smash{\begin{tabular}[t]{l}$\mathcal{Q}^+$\end{tabular}}}}%
    \put(0,0){\includegraphics[width=\unitlength,page=2]{thinshell_Q.pdf}}%
    \put(0.1363947,0.85056914){\color[rgb]{0,0,0}\makebox(0,0)[lt]{\lineheight{2.04999995}\smash{\begin{tabular}[t]{l}$\mathcal{Q}^-$\end{tabular}}}}%
    \put(0,0){\includegraphics[width=\unitlength,page=3]{thinshell_Q.pdf}}%
  \end{picture}%
\endgroup%

			\caption{The piecewise smooth solution constructed by solving the eikonal equation on the dark shaded regions in Figure \ref{fig:thinshells_MDD}, with the boundary conditions \eqref{eik-BCs} at $\Sigma^\pm$, and transforming to double null coordinates.} \label{fig:thinshell_in_Q}
		\end{minipage}
	\end{figure}

	\textbf{The thin shell ODE and constructing piecewise smooth solutions of (E-d).} We now suppose that the exterior/interior spacetimes $\mathcal{Q}^\pm$ are actually \emph{smooth} in comoving coordinates, corresponding to a subset of the maximal dust developments $(\mathscr{Q}_{MDD}^\pm,r^\pm,\rho^\pm,m^\pm,f^\pm)$. The evolution of $\Sigma$ (in comoving coordinates) then reduces to an \emph{ODE} (see Proposition \ref{tsodeprop}) defined on $\mathscr{Q}_{MDD}^\pm$: there exists a function $\epsilon(s)$ defined along $\Sigma$ (depending on the derivatives of $r$ in the normal directions, see \eqref{epsdef}, \eqref{eps}), and functions $\mathcal{R}, \mathcal{M}^\pm, \mathcal{E}$ depending on $(\mathscr{Q}_{MDD}^\pm,r^\pm,\rho^\pm,m^\pm,f^\pm)$ (see Definition \ref{def:ODE}), such that:
	\begin{equation} \label{thin-shell-ode}
		\dot{r}(s) = \mathcal{R}(r(s),m^+(s),m^-(s),\epsilon(s)), \ \dot{m}^\pm(s) = \mathcal{M}^\pm(r(s),m^+(s),m^-(s),\epsilon(s)), \ \dot{\epsilon}(s) = \mathcal{E}(r(s),m^+(s),m^-(s),\epsilon(s)),
	\end{equation}
	where $r(s)$ is the area radius of the thin shell, and $m^\pm(s)$ is the Hawking mass along $\Sigma$ in $\mathcal{Q}^\pm$ respectively. We refer to \eqref{thin-shell-ode} as the \emph{thin shell ODE}. The functions $r(s), m^\pm(s)$ uniquely determine comoving coordinates $(\tau^\pm(s),\chi^\pm(s))$ in $\mathscr{Q}_{MDD}^\pm$, and we define:
	\begin{equation} \label{Sigma-pm}
		\Sigma^\pm := \{ (\tau^\pm(s),\chi^\pm(s)) : s \in [0,s_1) \},
	\end{equation}
	which we refer to as the exterior/interior curves of the thin shell $\Sigma$. Importantly, we can \emph{construct} piecewise smooth solutions given $(\tau^\pm(s),\chi^\pm(s))$ determined by solutions of the thin shell ODE, by solving for double null coordinates with boundary conditions on $\Sigma^\pm$ (see Proposition \ref{prop:tsode-eik}). Suppose that $u \in C^2(\mathscr{Q}_{MDD}^-)$, $v \in C^2(\mathscr{Q}_{MDD}^+)$ are solutions of the respective eikonal equations \eqref{int:eik-com}. We then pose the boundary conditions:
	\begin{equation} \label{eik-BCs}
		u(\tau^+(s),\chi^+(s)) = u(\tau^-(s),\chi^-(s)), \qquad v(\tau^-(s),\chi^-(s)) = v(\tau^+(s),\chi^+(s)), 
	\end{equation}
	and solve for $u \in C^2(\mathscr{Q}_{MDD}^+)$, $v \in C^2(\mathscr{Q}_{MDD}^-)$. It then follows that the transformation to double null coordinates in the interior of $\Sigma^-$, and in the exterior of $\Sigma^+$, defines a piecewise smooth solution of (E-d), with thin shell $\Sigma$ given by $u_{\Sigma}(s) := u(\tau^\pm(s),\chi^\pm(s))$, $v_{\Sigma}(s) := v(\tau^\pm(s),\chi^\pm(s))$, see Figures \ref{fig:thinshells_MDD} and \ref{fig:thinshell_in_Q}.
	
	\textbf{The determinism condition.} For any piecewise smooth solution, the interior/exterior dust spacetimes $\mathcal{Q}^\pm$ can only be uniquely determined by initial data if the fluid flow lines from any point $q \in \Sigma$ (which exist by the regularity assumption), lie in the \emph{causal past} of $q$. We therefore impose the \underline{determinism condition}:
	\begin{equation} \label{int:det-cond0}
		(\mathbf{u}^v)^+(s) < \dot{v}(s) < (\mathbf{u}^v)^-(s), \text{ for } s > 0,
	\end{equation}
	so that moreover the fluid flow lines are not tangent to $\Sigma$. Note that we allow this to degenerate at the \emph{formation point} of a thin shell (which we always normalise to be at $s=0$). The inequalities \eqref{int:det-cond0} are equivalent to the following, in comoving coordinates (see Lemma \ref{detcondlemma}):
	\begin{equation} \label{int:det-cond}
		\dot{\chi}^-(s) < 0, \qquad \dot{\chi}^+(s) > 0,
	\end{equation}
	which we impose on solutions of the thin shell ODE. This is illustrated in Figures \ref{fig:thinshells_MDD}, \ref{fig:thinshell_in_Q}, where the determinism condition \eqref{int:det-cond0} is satisfied for $\Sigma$ in Figure \ref{fig:thinshell_in_Q}, and \eqref{int:det-cond} holds for $\Sigma^\pm$ in Figure \ref{fig:thinshells_MDD}.
	
	\subsubsection{Thin shell development at non-degenerate first shell crossing singularities} \label{introdev}
	
	In this section, we discuss the proof of the local thin shell development result, Theorem \ref{thm:1}. The proof consists of two parts: we first use the geometric construction already discussed in Section \ref{thinshells} to reduce the existence/uniqueness of a piecewise smooth extension satisfying the determinism condition to an existence/uniqueness result for the thin shell ODE \eqref{thin-shell-ode} satisfying certain initial conditions, and then prove the necessary ODE result. In this section, we first discuss the geometric arguments that imply existence and uniqueness from the ODE result (Proposition \ref{equivprop}), and then given an outline of the proof of the ODE result (Proposition \ref{thinshellODEprop}). 
	
	\textbf{Existence of a piecewise smooth extension (a solution to the thin shell development problem).} We've discussed that we can construct piecewise smooth solutions of (E-d) by solving the eikonal equation on maximal dust developments with boundary conditions on a solution of the thin shell ODE \eqref{thin-shell-ode}. This is precisely the construction we use for the thin shell development problem, now with a solution of the thin shell ODE with initial conditions at the point $(\tau_c(\chi_0),\chi_0) \in \mathscr{C}$ corresponding to a non-degenerate first shell crossing singularity. 
	
\begin{figure} [h] 
		\centering
		\def\svgwidth{0.7\linewidth}
		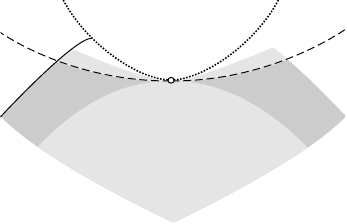 
		\caption{The exterior/interior curves $\Sigma^\pm$ of a solution of the thin shell ODE on a maximal dust development as in Figure \ref{fig:4}, with initial conditions at a non-degenerate first shell crossing singularity.}
		\label{fig:7}
	\end{figure}

	Indeed, given a non-degenerate first shell crossing singularity as depicted in Figure \ref{MGHDsubset}, we've discussed in Section \ref{dn-intro} that the corresponding \emph{uniquely determined} maximal dust development $(\mathscr{Q}_{MDD},r,\rho,m,f)$ in a neighbourhood of the first shell crossing singularity has the form of Figure \ref{fig:4}. Let $(\tau_c(\chi_0),\chi_0) \in \mathscr{C}$ be the non-degenerate shell crossing singularity (as a boundary point of the MDD), and \emph{suppose} that there exists a solution of the thin shell ODE \eqref{thin-shell-ode}, $(r(s),m^+(s),m^-(s),\epsilon(s))$, on $\mathscr{Q}_{MDD}$, with initial conditions
	\begin{equation} \label{ts-ICs}
		r(0) = r(\tau_c(\chi_0),\chi_0), \qquad m^\pm(0) = m(\chi_0), \qquad \epsilon(0) = \epsilon_0,
	\end{equation}
	(where $\epsilon_0$ must be a fixed number depending on $f(\chi_0)$, which we'll come back to) and moreover that $r(s),m^\pm(s)$ correspond to comoving coordinates in the maximal dust development:
	\begin{equation} \label{det-cond}
		(\tau^\pm(s),\chi^\pm(s)) \in \mathscr{Q}_{MDD} \text{ satisfying } \dot{\chi}^+(s) > 0, \ \dot{\chi}^-(s) < 0 \text{ for } s > 0.
	\end{equation}  
	See Figure \ref{fig:7} for $\Sigma^\pm$ (as in \eqref{Sigma-pm}) for such a solution of the thin shell ODE \eqref{thin-shell-ode}.
	
	As discussed at the end of Section \ref{dn-intro}, $u \in C^\infty(\mathscr{R}^-)$ extends uniquely to the region $\widetilde{\mathscr{R}}^- \subset \mathscr{Q}_{MDD}$ labelled in Figure \ref{fig:7}, and $v \in C^\infty(\mathscr{R}^+)$ extends uniquely to the region $\widetilde{\mathscr{R}}^+ \subset \mathscr{Q}_{MDD}$. We then use the boundary conditions as in \eqref{eik-BCs}: there exist unique $\tilde{u} \in C^\infty(\widetilde{\mathscr{R}}^+)$, $\tilde{v} \in C^\infty(\widetilde{\mathscr{R}}^-)$ solving \eqref{int:eik-com} and satisfying the boundary conditions
	\begin{equation} \label{ts-BCs}
		\tilde{u}(\tau^+(s),\chi^+(s)) = u(\tau^-(s),\chi^-(s)), \qquad \tilde{v}(\tau^-(s),\chi^-(s)) = v(\tau^+(s),\chi^+(s)).
	\end{equation}
	Now, the transformation to double null coordinates from the regions $\widetilde{\mathscr{R}}^\pm$ will define a piecewise smooth solution of (E-d), but we need to consider the regularity of the constructed solution of the eikonal equation across the ingoing/outgoing null curves from $(\tau_c(\chi_0),\chi_0)$ (the long dashed lines on Figure \ref{fig:7}). This corresponds to showing, in double null coordinates, that the solution is sufficiently regular across $\{ u = u_c, v > v_c\}$, $\{ u > u_c, v = v_c\}$, i.e.~the `Cauchy horizons' of the first shell crossing singularity in Figure \ref{MGHDsubset}.
	
	Recall (from Section \ref{dn-intro}) that solutions of the eikonal equation extend smoothly to $\mathscr{C}$. In this case, $u \in C^\infty(\widetilde{\mathscr{R}}^- \cup \mathscr{R}^- \cup \mathscr{R}' \cup \mathscr{R}^+)$ and $v \in C^\infty(\widetilde{\mathscr{R}}^+ \cup \mathscr{R}^+ \cup \mathscr{R}' \cup \mathscr{R}^-)$ extend smoothly to $(\tau_c(\chi_0),\chi_0)$. If we show, for example, that:
	\begin{equation} \label{eik-ctty}
		\partial_\tau^k\tilde{u}(\tau^+(s),\chi^+(s)) \to \partial_\tau^k u(\tau_c(\chi_0),\chi_0) \text{ as } s \to 0,
	\end{equation}
	then it follows that $u, \tilde{u}$ together define a $C^k$ solution of the eikonal equation on $\mathscr{R}^+ \cup \widetilde{\mathscr{R}}^+$. This condition for $k=1$ (which is equivalent to requiring \emph{continuity of the metric}\footnote{In terms of solutions of the eikonal equation, $\Omega^2$ is given by: $\Omega^2 = (\partial_\tau u \cdot \partial_\tau v)^{-1}$.} in double null coordinates) imposes the unique initial condition for $\epsilon(s)$: $\partial_\tau\tilde{u}(\tau^+(s),\chi^+(s)) \to \partial_\tau u(\tau_c(\chi_0),\chi_0)$ as $s \to 0$ if and only if $\epsilon(0) = (1+f(\chi_0))^\frac{1}{2} =: \epsilon_0$. This is equivalent to imposing that the thin shell must be initially tangent to the dust shell\footnote{Note that the term `shell' refers to the hypersurface in the `upstairs' manifold corresponding to a fluid flow line in the quotient manifold.} that first shell crosses. It can then be shown that $\Sigma^\pm$ satisfying the thin shell ODE \eqref{thin-shell-ode} imposes that \eqref{eik-ctty} also holds for $k=2,3$, and similarly for $v, \tilde{v}$. (See in particular Lemma \ref{eiklemma}.)
	
	The transformation to double null coordinates then indeed defines a piecewise smooth solution, as in Figure \ref{extension}. The essential difficulty in the proof of existence of a solution to the thin shell development problem is then the existence of a solution to the thin shell ODE \eqref{thin-shell-ode} with initial conditions \eqref{ts-ICs}, where the ODE is \emph{singular} since $\partial_\chi r(\tau_c(\chi_0),\chi_0) = 0$ (so the energy density $\rho$ is infinite), and the mass of the thin shell is initially zero, $m^+(0) - m^-(0) = 0$. See the discussion at the end of this section for an outline of the proof of existence (and uniqueness) of solutions of the thin shell ODE \eqref{thin-shell-ode} with initial conditions \eqref{ts-ICs}.
	
	\textbf{Uniqueness of the piecewise smooth extension.} Suppose that there exists a piecewise smooth extension as in Figure \ref{extension}. The determinism condition implies that the past fluid flow lines from any point $p \in \Sigma$ must enter the region on which the smooth solution is uniquely determined prior to shell crossing. We can then \emph{extend} the definition of comoving coordinates to the regions either side of the thin shell. It follows that these regions must embed into the maximal dust development $\mathscr{Q}_{MDD}$, and moreover that the thin shell satisfies the thin shell ODE \eqref{thin-shell-ode}, and the comoving coordinates along the thin shell satisfy \eqref{det-cond}. The exterior/interior curves in the maximal dust development for any such thin shell must therefore be as depicted in Figure \ref{fig:7}. 
	
	The supposed piecewise smooth extension defines solutions of the eikonal equation in the regions $\widetilde{\mathscr{R}}^\pm$, and the same continuity condition (as discussed for the existence argument) as $s \to 0$ must apply. Continuity of the metric across the `Cauchy horizons' of the first shell crossing singularity then imply that the solution of the thin shell ODE must satisfy the initial condition $\epsilon(0) = (1+f(\chi_0))^\frac{1}{2}$. The uniqueness argument then reduces to a uniqueness statement for solutions of the thin shell ODE \eqref{thin-shell-ode} with initial conditions \eqref{ts-ICs}. 
	
	\textbf{Outline of the proof of existence/uniqueness for solutions to the thin shell ODE \eqref{thin-shell-ode} with initial data \eqref{ts-ICs} at a non-degenerate first shell crossing singularity.} As noted already, the thin shell ODE \eqref{thin-shell-ode} has a singular point at $s=0$, for the initial conditions \eqref{ts-ICs} with $\epsilon_0 = (1+f(\chi_0))^\frac{1}{2}$. For example, $\mathcal{E}$ (as in \eqref{thin-shell-ode}, see \eqref{def:E} for the definition) has an overall factor of $(m^+(s) - m^-(s))^{-1}$, and $\mathcal{M}^\pm$ depends on the energy density in the exterior/interior respectively, which blow up: as $s\to0$,  $\rho^\pm(s) \to \infty$. Crucially, we need to show existence/uniqueness of solutions to the thin shell ODE \eqref{thin-shell-ode} that \emph{in addition} satisfy the determinism condition \eqref{det-cond}. Indeed, the thin shell ODE \eqref{thin-shell-ode} itself is only well-defined for solutions satisfying the determinism condition. It turns out that we can exploit the singular point structure of the ODE \emph{and} the structure of the determinism condition by reparametrising the ODE in terms of $\Delta(s) := m^+(s) - m^-(s)$ (see the start of Section \ref{ODEdet}). The idea is then that we can effectively Taylor expand (to second order) the solutions of the ODE \eqref{thin-shell-ode} about $\Delta=0$, where the combination of the ODE \eqref{thin-shell-ode} and determinism condition \eqref{det-cond} fixes the leading order terms and implies that the `second order terms' are \emph{bounded} (by constants depending only on the structure of the non-degenerate first shell crossing singularity) for $\Delta \in (0,\Delta_4]$, for some sufficiently small $\Delta_4 > 0$ (see the second part of Section \ref{ODEdet}, and Section \ref{firstorder}). We can then further restrict to $\Delta \in (0,\Delta_5]$ for $0 < \Delta_5 \leq \Delta_4$ (again independently of any solution of the ODE), and we show that the determinism condition \eqref{det-cond} is satisfied for any bounded solution of the ODE for the second order terms (the first part of Section \ref{secondorder}). We then expand the ODE for the second order terms to see that it has a regular singular point at $s=0$ (the second part of Section \ref{secondorder}). The proof is then closed with a contraction mapping argument at the level of the second order terms, where we show that there is a unique bounded solution of the ODE (Section \ref{contraction}). 
	
	We will now give a more detailed outline of the steps of the proof. We first note the following convention for notation, used in Sections \ref{ODEdet}, \ref{firstorder}: underlined quantities denote solutions of the thin shell ODE parametrised by $\Delta$ (or functions of such solutions), \eqref{drdDelta}, \eqref{detadDelta}, \eqref{depsdDelta}, which moreover satisfy the determinism condition \eqref{MDDbd}, \eqref{detcondDelta}, and some additional inequalities \eqref{sign}. Here, the system \eqref{drdDelta}, \eqref{detadDelta}, \eqref{depsdDelta} is an ODE for $(\underline{r}(\Delta), \underline{\eta}(\Delta), \underline{\epsilon}(\Delta))$, where we define (in Section \ref{ODEdet}):
	\begin{equation}
		\underline{\eta}(\Delta) := \frac{m(\chi_0) - \underline{m}^-(\Delta)}{\Delta}.
	\end{equation}
	By the determinism condition (recall $\Delta = \underline{m}^+(\Delta) - \underline{m}^-(\Delta)$, and $\underline{m}^+(\Delta) > m(\chi_0) > \underline{m}^-(\Delta)$ for $\Delta > 0$), we have $\underline{\eta}(\Delta) \in [0,1]$ for $\Delta > 0$ (where $\Delta=0$ is the point of formation of the thin shell). The condition \eqref{detcondDelta} is exactly the determinism condition, i.e.~the second half of \eqref{det-cond}, and \eqref{MDDbd} is equivalent to the first half of \eqref{det-cond} for sufficiently small $\Delta$. Indeed, let $r \in C^\infty(\mathscr{Q}_{MDD})$ be the solution of the collapsing Einstein-dust system \eqref{vel}. We define (choosing the Hawking mass as the comoving coordinate, $\chi = m$, with e.g.~$m_0 := m(\chi_0)$), $r_c(m) := r(\tau_c(m),m)$, so that the first half of \eqref{det-cond} (for $\Delta$ sufficiently small) is equivalent to \eqref{MDDbd}: 
	\begin{equation} \label{int:MDDbd}
		\underline{r}(\Delta) > r_c(\underline{m}^\pm(\Delta)). 
	\end{equation}
	
	The \underline{first step} of the proof (the second half of Section \ref{ODEdet}) consists of choosing suitable `first order' terms (e.g.~in a Taylor series of $\underline{r}(\Delta), \underline{\epsilon}(\Delta)$), which are \emph{bounded}. By the assumption that $(\tau_c(m_0), m_0)$ (where we've used $\chi = m$) is a \emph{non-degenerate} first shell crossing singularity, \eqref{non-deg-minima} holds, which implies $r_c^{(2n)}(m_0) < 0$, with all lower derivatives vanishing (where we only require that $n \in \mathbb{N}$). Note that we can always bound $|\underline{m}^\pm(\Delta) - m_0| \leq \Delta$, and the leading order term in $r_c(\underline{m}^\pm(\Delta))$ is $\sim \Delta^{2n}$, depending on $\underline{\eta}(\Delta)$. We define (see Definition \ref{phi-alpha-def}):
	\begin{equation}
		\underline{\phi}(\Delta) := (r_{2n} \Delta^{2n})^{-1} (r_0 - \underline{r}(\Delta)), \qquad \underline{\alpha}(\Delta) := \Delta^{-1} ( (1+f_0)^{-\frac{1}{2}}\underline{\epsilon}(\Delta) - 1),
	\end{equation}
	where $r_{2n} > 0$ is a constant depending in particular on $r_c^{(2n)}(m_0) < 0$ (see \eqref{rc:nondeg-exp}) and \eqref{int:MDDbd} implies that $\underline{\phi}$ is bounded, while the determinism condition \eqref{detcondDelta} implies that $\underline{\alpha}$ is bounded, for sufficiently small $\Delta$, see Lemma \ref{proof1}. Note that these bounds follow \emph{only} from the determinism condition and the initial conditions, and are independent of actually solving the ODE \eqref{drdDelta}--\eqref{depsdDelta}. In summary for this part, we've shown that the `first order terms':
	\begin{equation}
		\underline{\theta}(\Delta) := (\underline{\alpha}(\Delta), \underline{\eta}(\Delta), \underline{\phi}(\Delta)),
	\end{equation}
	are bounded for sufficiently small $\Delta$, i.e.~there exists a closed, bounded, subset $\mathcal{B} \subset \R^3$ such that $\underline{\theta}(\Delta) \in \mathcal{B}$ for all $\Delta \in (0,\Delta_1]$. (This statement is contained in the statement of Lemma \ref{proof1}.)
	
	In the \underline{second step} of the proof (Section \ref{firstorder}) we show that $\underline{\theta}(\Delta)$ in fact \emph{extends continuously} to $\Delta = 0$, with $\underline{\theta}(0) = \theta_0$ uniquely determined by the structure of the non-degenerate first shell crossing singularity. Moreover, we show that the second order terms, i.e.~$\Delta^{-1}(\underline{\theta}(\Delta) - \theta_0)$, are \emph{bounded}. We show this by exploiting the regular singular point structure of the ODE \eqref{drdDelta}--\eqref{depsdDelta} for solutions satisfying the inequalities \eqref{MDDbd}, \eqref{sign}, \eqref{detcondDelta}. 
	
	Indeed, in Lemma \ref{lem:beta} we identify a linear combination of $\underline{\alpha}, \underline{\eta}$, which we denote $\underline{\beta}$ (see \eqref{def:beta}), that we show satisfies $\underline{\beta}(0) = \beta_0$ and $\underline{\tilde{\beta}}(\Delta) := \Delta^{-1}(\underline{\beta} - \beta_0)$ is bounded. We prove this by deriving a \emph{decoupled} ODE for $\underline{\beta}$ of the form (see \eqref{dbeta}): 
	\begin{equation} \label{int:beta-ode}
		\frac{d\underline{\beta}}{d\Delta} + \frac{2(\underline{\beta} - \beta_0)}{\Delta} = B(\Delta,\underline{\theta}),
	\end{equation}
	where $\beta_0 \in \R$ is a constant (see the statement of Lemma \ref{lem:beta}) and $B(\Delta,\underline{\theta})$ is \emph{bounded}, where we use in particular the determinism condition bounds \eqref{MDDbd}, \eqref{detcondDelta}, see \eqref{def:N} and the bound below. Note that we make essential use here of $\underline{\beta}(\Delta)$ being bounded, i.e.~$\underline{\theta}(\Delta) \in \mathcal{B}$, in order to prove the result after arriving at the ODE \eqref{int:beta-ode}.
	
	Using the bound for $\tilde{\underline{\beta}}(\Delta)$, we show in Lemma \ref{lem:eta} that the ODE for $\underline{\eta}$ effectively decouples from $\underline{\phi}$ at leading order, and the regular singular point structure allows us to show that (since $\underline{\eta}$ is \emph{bounded}): $\underline{\eta}(0) = \frac{1}{2}$ and $\underline{\tilde{\eta}}(\Delta) := \Delta^{-1}(\underline{\eta}(\Delta) - \frac{1}{2})$ is bounded. Finally, expanding now in terms of $\tilde{\eta}$, this simplifies the ODE for $\underline{\phi}(\Delta)$ to a regular singular point, and we show in Lemma \ref{lem:phi} that there exists a constant $\phi_0 \in \R$ (see the statement of Lemma \ref{lem:phi}) such that $\underline{\phi}(0) = \phi_0$ and $\underline{\tilde{\phi}}(\Delta) := \Delta^{-1}(\underline{\phi}(\Delta) - \phi_0)$ is bounded. Together, the results in Section \ref{firstorder} show that (as noted in Remark \ref{rem:firstorder}):
	\begin{equation}
		\tilde{\underline{\theta}}(\Delta) := (\underline{\tilde{\beta}}(\Delta), \underline{\tilde{\eta}}(\Delta), \underline{\tilde{\phi}}(\Delta)) \in \tilde{\mathcal{B}},
	\end{equation}
	for $\Delta \in (0,\Delta_4]$ ($\Delta_4$ is defined in Definition \ref{def:a+Delta4}), where $\tilde{\mathcal{B}} \subset \R^3$ is closed and bounded (see Definition \ref{def:tilde-theta}).
	
	In the \underline{third (and final) step} (Sections \ref{secondorder} and \ref{contraction}) we first expand the thin shell ODE \eqref{drdDelta}, \eqref{detadDelta}, \eqref{depsdDelta} in terms of $\underline{\tilde{\theta}}(\Delta)$. Importantly, the fixed initial values $\underline{\tilde{\theta}}(0)$ imply that, if we expand any functions $r(\Delta),\eta(\Delta),\epsilon(\Delta)$ to second order (as we've done for solutions of the thin shell ODE \eqref{drdDelta}, \eqref{detadDelta}, \eqref{depsdDelta} in Section \ref{firstorder}), and impose that the second order terms satisfy $\tilde{\underline{\theta}}(\Delta) \in \tilde{\mathcal{B}}$ for $\Delta \in (0,\Delta_5]$ (where $\Delta_5 \in (0,\Delta_4]$, see \eqref{def:Delta5}) then the conditions \eqref{MDDbd} and \eqref{detcondDelta} are satisfied, see Remark \ref{rem:cond}. This shows that the ODE with determinism condition system is equivalent to the ODE for the \emph{bounded} second order terms. Indeed, in Lemma \ref{lem:finalode} at the start of Section \ref{contraction}, we show that $(r(s),m^+(s),m^-(s),\epsilon(s))$ is a solution of the thin shell ODE \eqref{thin-shell-ode} satisfying the determinism condition \eqref{det-cond} if and only if the second order terms $\tilde{\theta}(\Delta)$, as defined above (see \eqref{Psi2} and Definitions \ref{phi-alpha-def}, \ref{def:Theta1}, \ref{def:tilde-eta}, \ref{def:tilde-phi}), satisfy the ODE:
	\begin{equation} \label{int:ode1}
		\frac{d\tilde{\beta}}{d\Delta} + \frac{3(\tilde{\beta} - \tilde{\beta}_0)}{\Delta} = P_1(\Delta,\tilde{\theta}),
	\end{equation}
	\begin{equation} \label{int:ode2}
		\frac{d\tilde{\eta}}{d\Delta} + \frac{(2n+3)\tilde{\eta} - c_{\beta} \tilde{\beta} - c_{\eta}}{\Delta} = N_2(\Delta,\tilde{\theta}),
	\end{equation}
	\begin{equation} \label{int:ode3}
		\frac{d\tilde{\phi}}{d\Delta} + \frac{(2n+2)\tilde{\phi}}{\Delta} = P_3(\Delta,\tilde{\theta}),
	\end{equation}
	and $\tilde{\theta}(\Delta) \in \tilde{\mathcal{B}}$ for $\Delta \in (0,\Delta_5]$. Here, we use the expansions of the relevant functions composing the ODE in terms of $(\Delta,\tilde{\theta}) \in [0,\Delta_5] \times \tilde{\mathcal{B}}$ from Section \ref{secondorder} to see that $P_1, N_2, P_3 \in C^\infty([0,\Delta_5] \times \tilde{\mathcal{B}})$, and $\tilde{\beta}_0, c_\beta, c_\eta$ are constants determined in Section \ref{secondorder}. We finally conclude the proof by diagonalising the regular singular point of the system \eqref{int:ode1}, \eqref{int:ode2}, \eqref{int:ode3}, see Lemma \ref{lem:cont}, and we then use a contraction mapping argument to show that there is a unique bounded solution of \eqref{int:ode1}--\eqref{int:ode3}, see Lemma \ref{lem:contraction}.
 
	\subsection{The main ideas of the setup and proof of Theorem \ref{MGHD}} \label{intromaxdev}
	
	In this section we discuss the statement and proof of the second main result of the paper, Theorem \ref{MGHD}: the complete resolution of shell crossing singularities for generic collapsing dust clouds. We first define piecewise smooth solutions of (E-d) with finitely many thin shells in Section \ref{sec:int-pwise-ext}. We discuss the propagation condition and show that it implies an extension principle: if certain regularity assumptions and inequalities are satisfied at any `first singularity', then there is a unique local piecewise smooth extension, determined by a solution of the thin shell ODE \eqref{thin-shell-ode}. 
	
	In Section \ref{gen-data}, we briefly discuss smooth initial data for (E-d), which is equivalent to initial data in comoving coordinates along with a choice of double null gauge. We then define a family of metrics on the space of initial data, and define the set $\mathfrak{D}_{\chi_b}^*$, before giving a brief overview of the proof that $\mathfrak{D}_{\chi_b}^*$ is `generic' (open and dense) in $\mathfrak{D}_{\chi_b}$.
	
	In Section \ref{sec:int-ode}, we give a quick overview of the analysis of solutions of the thin shell ODE, which is essential for constructing the maximal piecewise smooth development for generic data.
	
	We finally give an overview of the construction of the maximal piecewise smooth development in Section \ref{sec:col}, in which we show in particular that there is always a unique extension beyond thin shell collisions, where the thin shell that forms must satisfy the determinism condition. 
	
	\subsubsection{Piecewise smooth solutions with finitely many thin shells, the propagation condition and an extension principle} \label{sec:int-pwise-ext}
	
	We define piecewise smooth solutions with finitely many thin shells (see Definition \ref{pwisefinite}) as the immediate generalisation of the definition discussed in Section \ref{thinshells}. We however restrict that thin shells can only form at shell crossing singularities and at the limit point of other thin shells.\footnote{It should be possible to show that, given the propagation condition, this extra `weak-strong uniqueness' assumption is unneccesary.} This implies that smooth solutions remain smooth until the (possible) formation of a first shell crossing singularity. We moreover impose the \underline{propagation condition} (see Section \ref{propcond} for the precise definition), as discussed briefly already in the introduction, which consists of the following: for any $(u_0,v_0) \in \mathcal{Q}$ that lies on or at the limit point of a thin shell, we require that there exists $u_1 > u_0$, $v_1 > v_0$ such that the restriction to $[u_0, u_1) \times [v_0, v_1)$ is a piecewise smooth solution with a single thin shell, satisfying the determinism condition; moreover, we impose that the metric components and the fluid variables extend (in respective regularity) to $(u_0,v_0)$ from either side of the thin shell. Note that this additional regularity condition is not necessary in order to get uniqueness for the thin shell development problem, but it is satisfied by the unique solution satisfying the determinism condition, see the statement of Theorem \ref{devthm} and Corollary \ref{devcor}.
	
	We note that imposing some condition\footnote{In the field of hyperbolic conservation laws, closely related to this problem, cf.~Section \ref{pressurelessEuler}, such uniqueness conditions take the form of `entropy conditions'.} for uniqueness is necessary, indeed, otherwise thin shells can `split' at any time into multiple thin shells, or form multiple thin shells at collisions (see Frauendiener and Klein's \cite{F} for a discussion of this problem). One can consider the extension beyond first shell crossing singularities as piecewise smooth solutions as a `sticky dust' model, where dust shells stick together upon collision, forming a thin shell. The propagation condition then further imposes that thin shells stick together upon collision. (See Section \ref{future} for conjectures that this condition arises in the pressureless limit of Einstein--Euler.) `Sticky' dust models are well studied in the literature for solutions of the pressureless Euler equations, see Section \ref{pressurelessEuler}. 
	
\begin{figure}[h]
		\centering
		\begin{minipage}{0.4\linewidth}
			\centering
			\def\svgwidth{\linewidth}
\begingroup%
  \makeatletter%
  \providecommand\color[2][]{%
    \errmessage{(Inkscape) Color is used for the text in Inkscape, but the package 'color.sty' is not loaded}%
    \renewcommand\color[2][]{}%
  }%
  \providecommand\transparent[1]{%
    \errmessage{(Inkscape) Transparency is used (non-zero) for the text in Inkscape, but the package 'transparent.sty' is not loaded}%
    \renewcommand\transparent[1]{}%
  }%
  \providecommand\rotatebox[2]{#2}%
  \newcommand*\fsize{\dimexpr\f@size pt\relax}%
  \newcommand*\lineheight[1]{\fontsize{\fsize}{#1\fsize}\selectfont}%
  \ifx\svgwidth\undefined%
    \setlength{\unitlength}{402.00190026bp}%
    \ifx\svgscale\undefined%
      \relax%
    \else%
      \setlength{\unitlength}{\unitlength * \real{\svgscale}}%
    \fi%
  \else%
    \setlength{\unitlength}{\svgwidth}%
  \fi%
  \global\let\svgwidth\undefined%
  \global\let\svgscale\undefined%
  \makeatother%
  \begin{picture}(1,1.01034609)%
    \lineheight{1}%
    \setlength\tabcolsep{0pt}%
    \put(0,0){\includegraphics[width=\unitlength,page=1]{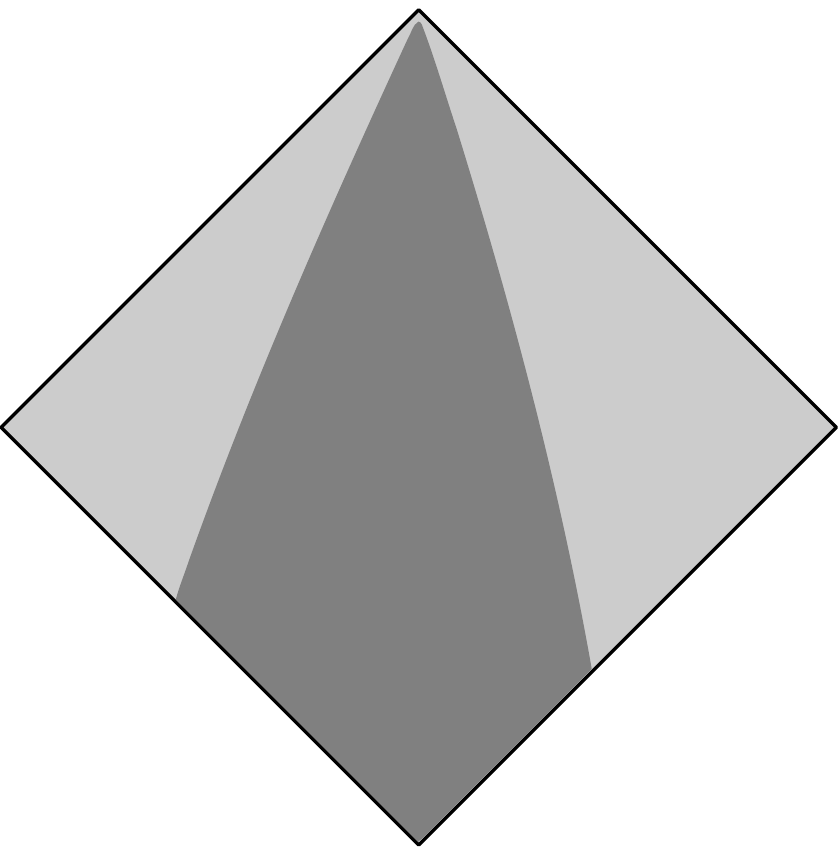}}%
    \put(0.69807047,0.06786733){\color[rgb]{0,0,0}\makebox(0,0)[lt]{\lineheight{2.04999995}\smash{\begin{tabular}[t]{l}$\widetilde{\mathcal{R}}_{ts}$\end{tabular}}}}%
    \put(0,0){\includegraphics[width=\unitlength,page=2]{extp.pdf}}%
    \put(0.08872357,0.81615859){\color[rgb]{0,0,0}\makebox(0,0)[lt]{\lineheight{2.04999995}\smash{\begin{tabular}[t]{l}$\widetilde{\mathcal{R}}^-$\end{tabular}}}}%
    \put(0.80236575,0.81616994){\color[rgb]{0,0,0}\makebox(0,0)[lt]{\lineheight{2.04999995}\smash{\begin{tabular}[t]{l}$\widetilde{\mathcal{R}}^+$\end{tabular}}}}%
    \put(0,0){\includegraphics[width=\unitlength,page=3]{extp.pdf}}%
    \put(0.52669837,1.00006902){\color[rgb]{0,0,0}\makebox(0,0)[lt]{\lineheight{2.04999995}\smash{\begin{tabular}[t]{l}$(u_1,v_1)$\end{tabular}}}}%
    \put(0,0){\includegraphics[width=\unitlength,page=4]{extp.pdf}}%
  \end{picture}%
\endgroup%
 
			\caption{The subset $\mathcal{R}' \subset \mathcal{Q}$ for which we prove an extension principle. The subset $\widetilde{\mathcal{R}}_{ts}$ contains finitely many thin shells, at least one with future limit point at $(u_1,v_1)$.} \label{fig:ext}
		\end{minipage}\hspace{5em}
		\begin{minipage}{0.4\linewidth}
			\centering
			\def\svgwidth{\linewidth}
			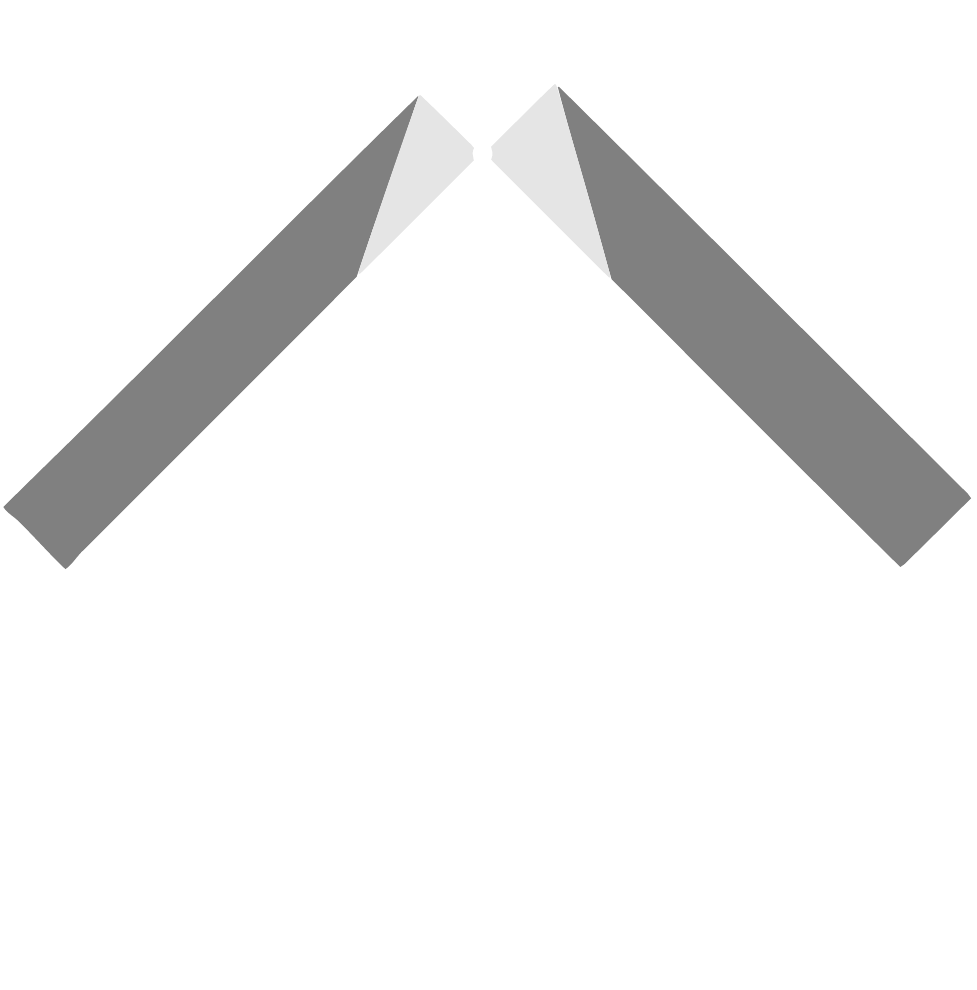 
			\caption{The unique local extension satisfying the propagation condition, with a thin shell forming at $(u_1,v_1)$.} 
			\label{fig:extp_ext}
		\end{minipage}
	\end{figure}

	We show in Proposition \ref{prop:pwise-extension} (Section \ref{extp}) that the propagation condition imposes that there is a unique piecewise smooth extension for certain `first singularities' that can arise in evolution. Indeed, let $(\mathcal{Q},r,\Omega^2,\rho,\mathbf{u}^\mu)$ be a piecewise smooth solution satisfying the propagation condition, and suppose that $\mathcal{R}' \subset \mathcal{Q}$ can be decomposed as $\mathcal{R}' = \widetilde{\mathcal{R}}^+ \cup \widetilde{\mathcal{R}}_{ts} \cup \widetilde{\mathcal{R}}^-$, with $(u_1,v_1) = \overline{\mathcal{R}'} \setminus \mathcal{R}'$ the limit point of a thin shell in $\widetilde{\mathcal{R}}_{ts}$, as in Figure \ref{fig:ext}.
	We assume that the restrictions to $\mathcal{R}^\pm$ are classical solutions of the Einstein equations, which are moreover \emph{smooth} in comoving coordinates, and $r, \Omega^2$ extend in $C^1$, $\rho, \mathbf{u}^\mu$ extend continuously, to $(u_c,v_c)$ from the null curves $\{ u = u_1\}$, $\{ v=v_1\}$, while we assume also that $r, \Omega^2$ extend continuously to $(u_1,v_1)$ from $\mathcal{R}'$, with $r(u_1,v_1), \Omega^2(u_1, v_1) > 0$. We include the case in which $\rho$ extends continuously to infinity, in which case we require the non-degeneracy condition along the respective null curve(s) (i.e.~we allow formation of a non-degenerate first shell crossing singularity on a thin shell, or at a thin shell collision). 
	
	We then let e.g.~$(\partial_u r)^\pm_1$ denote the respective limits from $\{ v=v_1\}$, $\{ u = u_1\}$, and $[\partial_u r]_1 := (\partial_u r)^+_1 - (\partial_u r)^-_1$. We assume various conditions on the limits at $(u_1,v_1)$ from the interior and exterior (see the statement of Proposition \ref{prop:pwise-extension} for the details, e.g.~\eqref{firstbd}), in particular we require that the following inequalities (\eqref{mindet} and \eqref{posdet}) hold:
	\begin{equation} \label{ext-det-cond}
		\frac{(\mathbf{u}^v)_1^+}{(\mathbf{u}^u)_1^+} < \frac{[\partial_u r]_1}{- [\partial_v r]_1} < \frac{(\mathbf{u}^v)_1^-}{(\mathbf{u}^u)_1^-}. 
	\end{equation}
	The main concept of the proof of Proposition \ref{prop:pwise-extension} is then essentially identical to that of thin shell development. We show that we can locally uniquely extend the classical solutions from $\widetilde{\mathcal{R}}^\pm$ to $\widetilde{\mathcal{R}}_{ext}^\pm$, the `lightly shaded regions' in Figure \ref{fig:extp_ext}, where the solution is uniquely determined in comoving coordinates as a subset of the maximal dust development extending $\widetilde{\mathcal{R}}^\pm$. The double null gauge is determined from $\widetilde{\mathcal{R}}^\pm_{ts}$, as labelled in Figure \ref{fig:extp_ext} (which must locally be subsets of $\mathcal{Q}$ by compactness, and may contain finitely many thin shells). The unique piecewise smooth extension is then determined by the existence of a unique solution of the thin shell ODE in the interior/exterior maximal dust developments, with initial data (\eqref{extdata1}, \eqref{extdata2}) determined by the jump conditions \eqref{rjump} holding at $(u_1,v_1)$. In particular, the initial velocity in double null coordinates of the thin shell forming at $(u_0,v_0)$  \eqref{initvel} is uniquely determined, and the condition \eqref{ext-det-cond} implies that this velocity satisfies the determinism condition \eqref{detcondzero} (as depicted in Figure \ref{fig:extp_ext}). 
	
	The existence/uniqueness result for the thin shell ODE is immediate by Picard--Lindel\"of in the case that $\rho$ is bounded along the null curves $\{u=u_1\}, \{v=v_1\}$. In the case where $\rho$ extends continuously to infinity, the thin shell ODE \eqref{thin-shell-ode} is mildly singular, and we show in Lemma \ref{lem:ode} by expanding to first order that the system has a regular singular point, and hence a unique solution. Here, it is important that the determinism condition is satisfied by continuity for any solution of the thin shell ODE \eqref{thin-shell-ode} attaining the induced initial data \eqref{extdata1}, \eqref{extdata2}, see the proof of Lemma \ref{lem:ode}. (Note that the initial mass $m^+(0) - m^-(0)$ of the solution of the thin shell ODE \eqref{thin-shell-ode} in Lemma \ref{lem:ode} is \emph{positive}. The proof is much simpler in this case than that of the thin shell development problem, where the initial mass necessarily vanishes.)
	
	\subsubsection{Generic smooth initial data for (E-d) and Christodoulou's naked singularities} \label{gen-data}
	
	In this section we briefly discuss the setup of smooth initial data for (E-d) and then introduce the generic subset of initial data used in Theorem \ref{MGHD}, along with a brief discussion of the proof of genericity. In the last part, we describe how the choice of generic data relates to the geometry of the singular boundary of the maximal dust development, including the possibility of (locally) naked shell-focussing singularities first discovered by Christodoulou \cite{C1}.
	
	\textbf{The space of smooth initial data for (E-d) with fixed compact support.} We pose smooth initial data for (E-d) on a simultaneous curve, which is equivalent to posing data on the curve $\{ \tau = 0\}$ in comoving coordinates with a gauge choice of initial data for the eikonal equation (and a gauge choice in the vacuum exterior, see Definition \ref{E-ddata}), see Section \ref{smoothdataEd} for the full setup. Smooth developments of initial data then exactly correspond to the transformation to double null coordinates of solutions of the eikonal equation on (globally hyperbolic subsets of)\footnote{The maximal dust development has a timelike exterior boundary, the boundary of support of the dust $\{\chi=\chi_b\}$, so subsets of the MDD can only be thought of as globally hyperbolic after including the vacuum exterior.} the maximal dust development. 
	
	We fix the compact support of initial data: we make the gauge choice $r(0,\chi) = \chi$, and restrict to $\chi \in [0,\chi_b]$, where $\{ \chi = 0\}$ is the \emph{centre} of spherical symmetry, and $\{ \chi = \chi_b \}$ is the boundary of the dust cloud. In order for solutions to be regular at the centre of symmetry (in suitably defined Cartesian coordinates, see Section \ref{initial-data}), the space of initial data in comoving coordinates reduces to even functions $\tilde{m}, \tilde{f} \in C^\infty([0,\chi_b])$ (where $\tilde{m}(\chi) := \chi^{-3} m(\chi)$, $\tilde{f}(\chi) := \chi^{-2} f(\chi)$, for $m, f$ as introduced in Section \ref{comovingintro}, see e.g.~\eqref{int:metric-rho-u}, \eqref{vel}), which are rescaled to extend smoothly to the centre, and moreover must satisfy various inequalities\footnote{The inequalities follow, for example, from imposing $\mathbf{u}(r) < 0$ for $r> 0$, and positive energy density $\rho(0,\chi) > 0$ for $\chi \in [0,\chi_b)$.}. We call the moduli space of $(\tilde{m},\tilde{f})$ satisfying these conditions $\mathfrak{D}_{\chi_b}$ (see Definition \ref{D}).
	
	The maximal dust development of data $(\tilde{m},\tilde{f}) \in \mathfrak{D}_{\chi_b}$ is characterised by the functions $\tau_s(\tilde{m},\tilde{f}) \in C^\infty([0,\chi_b])$, $\tau_c(\tilde{m},\tilde{f}) \in C^\infty(\mathscr{C}_\chi(\tilde{m},\tilde{f}))$ (see Proposition \ref{prop:mddcentre}), as introduced in Section \ref{comovingintro} (where we've added the parametrisation by $(\tilde{m},\tilde{f})$ to note the dependence on initial data). The function $\tau_s(\tilde{m},\tilde{f})$ can be written explicitly in terms of $\tilde{m}, \tilde{f}$ (see Proposition \ref{prop:mddcentre}, \eqref{taus}):
	\begin{equation} \label{int:taus}
		\tau_s(\chi) = \int_0^1 \frac{R^\frac{1}{2}}{(2\tilde{m}(\chi) + \tilde{f}(\chi)R)^\frac{1}{2}} dR \in C^\infty([0,\chi_b]),
	\end{equation}
	while $\tau_c(\tilde{m},\tilde{f})$ (only defined on $\mathscr{C}_\chi(\tilde{m},\tilde{f})$, i.e.~when $\tau_s(\tilde{m},\tilde{f})'(\chi) < 0$) is determined implicitly by an expression involving $\tilde{m}$, $\tilde{f}$ and their first derivatives (see Proposition \ref{mddprop}). For $(\underline{\tilde{m}},\underline{\tilde{f}}), (\tilde{m},\tilde{f}) \in \mathfrak{D}_{\chi_b}$, we define the following metric, for $k \in \mathbb{N}$:
	\begin{equation} \label{int:metric}
		d_k((\underline{\tilde{m}},\underline{\tilde{f}}), (\tilde{m},\tilde{f})) := \| \underline{\tilde{m}} - \tilde{m}\|_{C^k([0,\chi_b])} + \| \underline{\tilde{f}} - \tilde{f}\|_{C^k([0,\chi_b])}.
	\end{equation}
	In Section \ref{top}, we show that a $C^k$ Cauchy stability result holds with this metric on the space of initial data, for smooth solutions of (E-d). 
	
	\textbf{Generic smooth initial data in comoving coordinates.} Defining the quantity:
	\begin{equation} \label{N2-def}
		\mathcal{N}_2(\tau_s)(\chi) := |\tau_s'(\chi)| + |\tau_s''(\chi)|,
	\end{equation}
	we then teleologically define $\mathfrak{D}_{\chi_b}^* \subset \mathfrak{D}_{\chi_b}$ to be the set of all $(\tilde{m},\tilde{f}) \in \mathfrak{D}_{\chi_b}$ such that:
	\begin{equation} \label{N2}
		\min_{\chi \in [0,\chi_b]} \mathcal{N}_2(\tau_s(\tilde{m},\tilde{f}))(\chi) > 0, \qquad \inf_{\chi \in \mathscr{C}_\chi(\tilde{m},\tilde{f})} \mathcal{N}_2(\tau_c(\tilde{m},\tilde{f}))(\chi) > 0.
	\end{equation}
	We show in Section \ref{genericsection} (Proposition \ref{genericdata}) that $\mathfrak{D}_{\chi_b}^*$ is \emph{open and dense} in the metric space $\metric$ for $k \geq 3$. Note the significance of the choice of $\mathfrak{D}_{\chi_b}^*$ is that all local extrema of $\tau_s$, $\tau_c$ are parabolic minima/maxima, and there can only be \emph{finitely many}.
	
	It is relatively straightforward to see that the condition on $\tau_s(\tilde{m},\tilde{f})$ in \eqref{N2} holds on an open and dense set in $(\mathfrak{D}_{\chi_b},d_k)$ (for $k \geq 2$). We use the bound (which follows from \eqref{int:taus})
	\begin{equation} \label{taus-bdd}
		\| \tau_s(\underline{\tilde{m}},\underline{\tilde{f}}) - \tau_s(\tilde{m},\tilde{f}) \|_{C^k([0,\chi_b])} \lesssim d_k((\underline{\tilde{m}},\underline{\tilde{f}}), (\tilde{m},\tilde{f})),
	\end{equation}
	to see that the condition on $\tau_s$ is open for $k \geq 2$. Density follows from the fact that even functions with \eqref{N2-def} everywhere positive (i.e.~even Morse functions)\footnote{Note that in Morse theory, functions with $\mathcal{N}_2(f) > 0$, i.e.~functions with only non-degenerate critical points, are known as \emph{Morse functions}.} are dense in $C^k([0,\chi_b])$ for all $k \in \mathbb{N}$ (this is a basic result in Morse theory, see Lemma \ref{lem:dense}), and we can directly construct $(\tilde{m},\tilde{f})$ such that $\tau_s(\tilde{m},\tilde{f})$ is any given smooth even function. See Lemma \ref{Dsgen} for the precise statement and proof. 
	
	There is an added difficulty in showing that the $\tau_c$ part of \eqref{N2} holds on an open set: the domain of definition of $\tau_c$, $\mathscr{C}_{\chi}(\tilde{m},\tilde{f})$, \emph{depends} on data. We crucially use that $\min_{\chi \in [0,\chi_b]} \mathcal{N}_2(\tau_s(\tilde{m},\tilde{f}))(\chi) > 0$ on open sets, and for such data $\mathscr{C}_{\chi}(\tilde{m},\tilde{f})$ is composed of \emph{finitely many disjoint open intervals}. Moreover, at any endpoint $\chi = \chi_i > 0$ of any such interval, we have $\tau_s'(\chi_i) = 0$, $|\tau_s''(\chi_i)| > 0$. We use this structure to show `orbital stability' for the endpoints of each interval, and to control $|\tau_c(\underline{\tilde{m}},\underline{\tilde{f}})'(\chi)| \sim 1$ in a fixed neighbourhood of each endpoint, see Lemma \ref{scprop1}.
	
	There is also an additional difficulty in the density argument (again in Lemma \ref{genericsc}), since $\tau_c(\tilde{m},\tilde{f})$ is determined implicitly by $(\tilde{m},\tilde{f})$ and their first derivatives. We fix $\tilde{m}$, and construct nearby data satisfying \eqref{N2} by rearranging the definition of $\tau_c$ into an ODE for $\tilde{f}$. We then crucially use \emph{Cauchy stability} of this ODE to construct $\underline{\tilde{f}}$ arbitrarily close to $\tilde{f}$, so that $(\tilde{m},\underline{\tilde{f}}) \in \mathfrak{D}_{\chi_b}^*$ and $\tau_c(\tilde{m},\underline{\tilde{f}})$ satisfies \eqref{N2}.
	
\begin{figure} [h] 
		\centering
		\def\svgwidth{0.65\linewidth}
		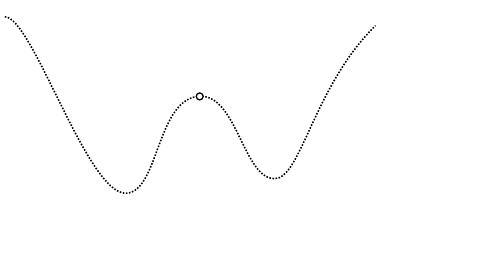 
		\caption{The maximal dust development of initial data in $\mathfrak{D}_{\chi_b}^*$ with $\tau_s''(0)<0$ and two local minima of $\tau_c(\chi)$.}
		\label{fig:MDD-sc}
	\end{figure}

	\textbf{Decomposition of the singular boundary of the maximal dust development for generic initial data.} For $(\tilde{m},\tilde{f}) \in \mathfrak{D}_{\chi_b}^*$, we have a finite decomposition of the shell crossing singular boundary (see Corollary \ref{structuregendata}), $\mathscr{C}$, into components $\mathscr{C}^-_i$ where $\tau_c'(\chi) < 0$, $\mathscr{C}^+_i$ where $\tau_c'(\chi) > 0$, and minima/maxima $\partial^-\mathscr{C}_i$, $\partial^+\mathscr{C}_i$. The set $\mathscr{S}$ also has a decompostion into finitely many connected components. See Figure \ref{fig:MDD-sc} for an example.
	
	\textbf{Christodoulou's (locally) naked singularities, the dichotomy with shell crossing `up to the centre' on the maximal dust development and the Oppenheimer--Snyder spacetimes.} The nature of the singular boundary of the maximal dust development in a neighbourhood of the centre $\Gamma = \{ \chi = 0\}$ is determined by the function $\tau_s(\chi)$. Indeed, if $\tau_s''(0)<0$ then there is `shell crossing up to the centre', as shown in Figure \ref{fig:MDD-sc}. 
	
	We've seen already, in the context of shell crossing (i.e.~in Sections \ref{comovingintro}, \ref{dn-intro}), that the MDD can be uniquely determined \emph{beyond} the usual maximal globally hyperbolic development, as a solution of the Einstein-dust system in comoving coordinates. Indeed, this `unphysical' construction is crucial in constructing the piecewise smooth solution of (E-d) extending beyond shell crossing (see Section \ref{introdev}), in which a \emph{physical} subset of the MDD is identified in the construction (i.e.~the shaded regions in Figure \ref{fig:7}). 
	
	If $\tau_s''(0) > 0$, then dust shells in the maximal dust development sufficiently close to the centre must terminate at a component of $\mathscr{S}$. As first shown by Christodoulou \cite{C1} in the case of inhomogeneous dust clouds collapsing from rest\footnote{In this case, the condition $\tau_s''(0) > 0$ is equivalent to $\rho_0''(0) < 0$, where $\rho_0(\chi)$ is the initial energy density.}, in this case there exists an \emph{outgoing null geodesic} emanating from the first singularity at the centre $(\tau_s(0),0)$, in the MDD, see Proposition \ref{prop:NS} for a proof. (There is in fact an infinite family of null geodesics emanating from $(\tau_s(0),0)$ in the MDD.) In the case in which there is no shell crossing, i.e.~$\tau_s'(\chi) \geq 0$ everywhere, then the maximal globally hyperbolic development has Penrose diagram as in one of Figures \ref{fig:loc-naked}, \ref{fig:glob-naked}, depending on global properties of the dust cloud (see \cite{C1} for the proof that globally naked singularities, i.e.~as in Figure \ref{fig:glob-naked}, exist). One can think of the Cauchy horizon in Figures \ref{fig:loc-naked}, \ref{fig:glob-naked} as the `earliest' outgoing null geodesic which emanates from $(\tau_s(0),0)$ in the maximal dust development. 
	
	By the definition of the generic subset of initial data $\mathfrak{D}_{\chi_b}^*$, we must have \emph{either} $\tau_s''(0) > 0$ \emph{or} $\tau_s''(0) < 0$. As defined in the introduction, we say $\Psi_{OS} \in \mathfrak{D}_{\chi_b}$ is `smoothed' Oppenheimer--Snyder initial data if there exists $0 < \chi_a < \chi_b$ and $\tilde{m}_{OS} \in C^\infty([0,\chi_b]), \tilde{m}_0 > 0$ such that: $\Psi_{OS} = (\tilde{m}_{OS},0) \in \mathfrak{D}_{\chi_b}$, $\tilde{m}_{OS}(\chi) = \tilde{m}_0 \text{ for all } 0 \leq \chi \leq \chi_a$ and $\tau_s(\tilde{m}_{OS},0)'(\chi) \geq 0$ for all $\chi \in [\chi_a, \chi_b]$. From \eqref{int:taus}, we see that, if $\tilde{f} = 0$ (i.e.~\emph{marginally bound} dust): $\tau_s(\tilde{m},0)(\chi) = \frac{2}{3} (2\tilde{m}(\chi))^{-\frac{1}{2}}$, so we must have $\tilde{m}_{OS}'(\chi) \leq 0$ for $\chi \in [\chi_a, \chi_b]$. This is satisfied if the initial energy density $\rho_{OS, 0}(\chi) \in C^\infty([0,\chi_b])$ is non-increasing for $\chi \in [\chi_a, \chi_b]$ (while $\rho_{OS, 0}(\chi) = \frac{3 \tilde{m}_0}{4\pi} \text{ for all } 0 \leq \chi \leq \chi_a$). Note that $\tau_s(\tilde{m}_{OS},0)(\chi) = const$ for $0 \leq \chi \leq \chi_a$ implies that $\Psi_{OS} \notin \mathfrak{D}_{\chi_b}^*$. By the density argument already discussed in this section (Lemma \ref{Dsgen}), one can construct arbitrarily close (in $d_k$ for any $k \in \mathbb{N}$) initial data in $\mathfrak{D}_{\chi_b}^*$ to $\Psi_{OS}$ with arbitrarily many first shell crossing singularities.\footnote{One could construct such data so that all first shell crossing singularities form on a simultaneous curve of the dust (i.e.~$\tau = const$), so that arbitrarily many thin shells indeed form in the maximal piecewise smooth development.} Moreover, one can show by explicitly constructing a converging sequence (see Lemma \ref{lem:dense-nosc}) that $\Psi_{OS}$ lies in the closure of $\mathfrak{D}_{\chi_b, \mathcal{S} \neq \emptyset \neq \mathcal{CH}}$. ($\Psi_{OS}$ cannot lie in the closure of $\mathfrak{D}_{\chi_b, \mathcal{S} = \emptyset \neq \mathcal{CH}}^*$ in the metric space $(\mathfrak{D}_{\chi_b},d_k)$ for any $k \in \mathbb{N}$ by Cauchy stability.)

	\subsubsection{The thin shell ODE on the maximal dust development of generic initial data} \label{sec:int-ode}
	
\begin{figure} [h] 
		\centering
		\def\svgwidth{0.65\linewidth}
		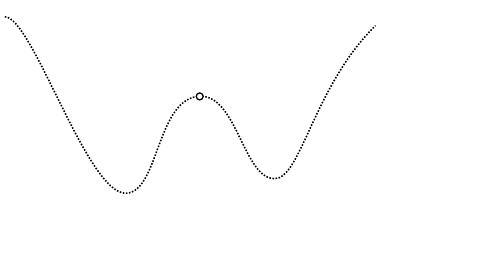 
		\caption{Maximal solution of the thin shell ODE on a maximal dust development as in Figure \ref{fig:MDD-sc}. The bounds \eqref{det-cond-bd} imply that the limit point $\Sigma^+$ cannot be in $\mathscr{C}_1^+ \cup \partial^+\mathscr{C}_1$, and the limit point of $\Sigma^-$ cannot be in $\mathscr{C}_1^-$.}
		\label{fig:max-ts-ode}
	\end{figure}

	By the propagation condition, as discussed in Section \ref{sec:int-pwise-ext}, the evolution of thin shells in piecewise smooth solutions is determined by solutions of the thin shell ODE and then the transformation to double null coordinates, constructed as solutions of the eikonal equation with boundary conditions on the thin shells. It is therefore crucial to understand how solutions of the thin shell ODE behave on any given maximal dust development (of generic smooth initial data). In Section \ref{ODEMDD}, we show that (Proposition \ref{odemddprop} and in particular Proposition \ref{maximal}): given initial data for the thin shell ODE on $\mathscr{Q}_{MDD}$, satisfying the determinism condition, the \emph{maximal} solution extends until either of $\Sigma^\pm$ reach the \emph{singular} boundary of $\mathscr{Q}_{MDD}$, and moreover that the determinism condition remains satisfied in evolution. This follows from analysis of the thin shell ODE \eqref{thin-shell-ode}, where in particular we show that (see \eqref{detplus}, \eqref{detmin} in the statement of Proposition \ref{maximal}):
	\begin{equation} \label{det-cond-bd}
		\partial_\chi r(\tau^+(s),\chi^+(s)) \dot{\chi}^+(s) \sim 1, \qquad \partial_\chi r(\tau^-(s),\chi^-(s)) \dot{\chi}^-(s) \sim - 1,
	\end{equation}
	which holds \emph{with extension} to the shell crossing singular boundary $\mathscr{C}$. We further show (in Lemma \ref{lem:detbd}) that the bound \eqref{det-cond-bd} implies $\Sigma^-$ cannot have a limit point in $\mathscr{C}_i^-$ or $\partial^+\mathscr{C}_i$, similarly $\Sigma^+$ cannot have a limit point in $\mathscr{C}_i^+$ or $\partial^+\mathscr{C}_i$. See Figure \ref{fig:max-ts-ode} for an example. This analysis is essential for showing that the maximal piecewise smooth development of generic data forms only finitely many thin shells, see the discussion of the construction in the following section.
	
	\subsubsection{Thin shell collisions and the maximal piecewise smooth development} \label{sec:col}
	
	Using the setup discussed thus far, we prove in Proposition \ref{prop:MGHPSD} that, given generic initial data $(\tilde{m},\tilde{f}) \in \mathfrak{D}_{\chi_b}$ (along with a gauge choice for the vacuum exterior of the dust cloud, see Definition \ref{E-ddata}), there exists a unique maximal globally hyperbolic piecewise smooth development (MGHPSD) attaining the initial data and satisfying the propagation condition. It follows from the construction that $r$ is not bounded away from zero at any first singularity of the MGHPSD, and in Section \ref{firstsingrzero} we complete the proof of (the precise statement of) Theorem \ref{MGHD}, where in particular we show that $r$ extends continuously to zero at any first singularity.
	 
	See the end of this section for a discussion of the first singularity at the centre of the MGHPSD and the boundary decomposition, for generic initial data. We now outline the steps of the proof of Proposition \ref{prop:MGHPSD}, using the initial data with maximal dust development as in Figure \ref{fig:MDD-sc} as an example. 
	
	`Step 0' in the proof is to construct the maximal `smooth' globally hyperbolic development of the initial data, and show that it has only finitely many first singularities with $r$ bounded away from zero, which moreover must all be \emph{non-degenerate} first shell crossing singularities. We do this by solving the eikonal equation on the maximal dust development $\mathscr{Q}_{MDD}$ (of any given generic initial data $(\tilde{m},\tilde{f}) \in \mathfrak{D}_{\chi_b}^*$), with boundary conditions at $\{ \chi = \chi_b\}$ determined by the gauge choice in the vacuum exterior, and the boundary condition $u(\tau,0) = v(\tau,0)$ on $\{\chi=0\}$ (see \eqref{eikIBCs}). We define a family of subsets of the MDD for which such solutions of the eikonal equation are uniquely defined (see Definition \ref{def:QGH0}), and show that there is a unique maximal such subset, which we denote $\mathscr{Q}_{MGH, 0} \subset \mathscr{Q}_{MDD}$. The set $\mathscr{Q}_{MGH, 0}$ can be characterised as the union of all points $(\tau,\chi) \in \mathscr{Q}_{MDD}$ such that there is an outgoing null curve from initial data or the centre $\Gamma$ to $(\tau,\chi)$ \emph{and} there is an ingoing null curve from initial data or the boundary of the dust cloud $\{\chi=\chi_b\}$ to $(\tau,\chi)$. 
	
	We then define a smooth globally hyperbolic solution of (E-d), $(\mathcal{Q}_0,r,\Omega^2,\rho,\mathbf{u}^\mu)$, to consist of the transformation to double null coordinates of the solution of the collapsing Einstein-dust system on $\mathscr{Q}_{MGH, 0}$ with the uniquely determined vacuum solution in the exterior of the dust cloud, see \eqref{Q0}. By definition of $\mathscr{Q}_{MGH, 0}$, $(\mathcal{Q}_0,r,\Omega^2,\rho,\mathbf{u}^\mu)$ is then the maximal `smooth' globally hyperbolic development. Any first singularity of $\mathcal{Q}_0$ with $r$ bounded away from zero must correspond to a point on the shell crossing boundary $\mathscr{C} \subset \overline{\mathscr{Q}_{MDD}} \setminus \mathscr{Q}_{MDD}$. It follows from the ODE for null curves on the MDD \eqref{int:null} that outgoing null curves only meet the shell crossing boundary at $\partial^-\mathscr{C} \cup \mathscr{C}^-$ (see Section \ref{gen-data} for the decomposition of $\mathscr{C}$) and ingoing null curves only meet the shell crossing boundary at $\partial^-\mathscr{C} \cup \mathscr{C}^+$ (see \eqref{null-bd1} and \eqref{null-bd2}). This implies that the only possible first singularities of $\mathcal{Q}_0$ with area radius bounded away from zero are \emph{non-degenerate} first shell crossing singularities, which in comoving coordinates correspond to points in $\partial^-\mathscr{C}$ (see \eqref{fsing}). By the choice of $\mathfrak{D}_{\chi_b}^*$, as discussed in Section \ref{gen-data}, $\partial^-\mathscr{C}$ is \emph{finite}.
	
	In the next step (`Step 1a' in the proof of Proposition \ref{prop:MGHPSD}), we use Theorem \ref{thm:1} to extend beyond $\mathcal{Q}_0$ as a globally hyperbolic piecewise smooth solution of (E-d). In particular, we construct the maximal such extension, \emph{restricted} to thin shells formed at first shell crossing singularities of $\mathcal{Q}_0$. We do this similarly to the previous step by constructing directly the corresponding maximal subset of $\mathscr{Q}_{MDD}$, now (partially) bounded to the future by solutions of the thin shell ODE. Indeed, we order the $n_1 \in \mathbb{N}$ first singularities by their comoving coordinate $0 < \chi_{f, 0, i} < \chi_{f, 0 , i+1} < \chi_b$, and let $(r_{0, i}, m_{0, i}^+, m_{0, i}^-,\epsilon_{0, i}) \in C^0([0,s_{*, 0, i}]) \cap C^\infty((0,s_{*, 0 ,i}))$ be the \emph{maximal} unique solution (which exists by the proof of Theorem \ref{thm:1}, see Section \ref{introdev}) of the thin shell ODE \eqref{thin-shell-ode} on $\overline{\mathscr{Q}_{MDD}}$ with initial conditions as in the proof of Theorem \ref{devthm}, \eqref{ts-formation-ICs}. 
	 
	We \emph{restrict} these solutions of the thin shell ODE (see the first half of Definition \ref{def:QGH1}) to proper time $0 < s_i \leq s_{*, 0 ,i}$, notated by $\Sigma^\pm_{0, i, s_i} \subset \mathscr{Q}_{MDD} \cup \partial^-\mathscr{C}$ (see \eqref{Sigma-ts}), such that all past-directed outgoing null curves from each thin shell interior $\Sigma_{0, i, s_i}^-$ meet the exterior of the neighbouring thin shell $\Sigma_{0, i-1, s_{i-1}}^+$ (or the centre if $i=1$) or $\mathscr{Q}_{MGH, 0}$, and similarly all past-directed ingoing null curves from each thin shell exterior $\Sigma_{0, i, s_i}^-$ meet the interior of the  neighbouring thin shell $\Sigma_{0, i+1, s_{i+1}}^-$ (or the boundary of the dust cloud) or $\mathscr{Q}_{MGH, 0}$. We then define (see the second half of Definition \ref{def:QGH1}), for any $s_i > 0$ satisfying this condition, $\mathscr{Q}_{GH, 1}(s_0, ..., s_{n_1 + 1})$ to be the union of $\mathscr{Q}_{MGH, 0}$ with the set of all points that have past-directed null curves which intersect neighbouring thin shell interior/exteriors (or the centre and the boundary of the dust cloud). 
	
\begin{figure} [h] 
		\centering
		\def\svgwidth{0.65\linewidth}
		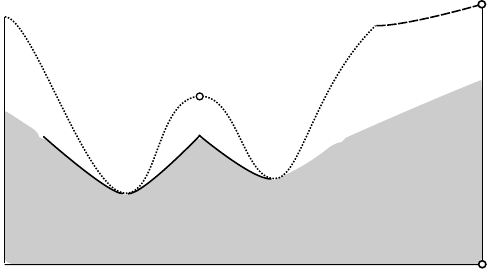 
		\caption{A collision of two solutions of the thin shell ODE on the maximal dust development as in Figure \ref{fig:MDD-sc}. The curves $\Sigma_1^\pm$ correspond to a subset of $\Sigma_1$ as in Figure \ref{fig:max-ts-ode}.\protect\footnotemark~The dark shaded region is $\mathscr{Q}_{MGH, 1}$ in the proof of Proposition \ref{prop:MGHPSD}.}
		\label{fig:9}
	\end{figure}

	It follows from Definition \ref{def:QGH1} that solutions of the eikonal equation with boundary conditions on the thin shell interior/exteriors, i.e.~\eqref{eik-BCs} (\eqref{SigmaBCs} in the proof), and the initial conditions/boundary conditions at the centre/vacuum boundary \eqref{eikIBCs} are uniquely determined on $\mathscr{Q}_{GH, 1}(\underline{s})$, and we show that there is a unique maximal choice of $\underline{s} = (s_0, ..., s_{n_1 + 1})$ which we denote $\underline{s_1}$, and define $\mathscr{Q}_{MGH, 1} := \mathscr{Q}_{GH, 1}(\underline{s_1})$. See for example Figure \ref{fig:9}. We then transform to double null coordinates on $\mathscr{Q}_{MGH, 1}$ and attach the vacuum exterior, defining $(\mathcal{Q}_1, r, \Omega^2, \rho, \mathbf{u}^\mu)$, a piecewise smooth solution of (E-d) satisfying the propagation condition and crucially the unique globally hyperbolic piecewise smooth development of initial data restricted to $\mathcal{Q}_1$, satisfying the propagation condition. 
	
	\textbf{Thin shell collisions.} In `Step 1b' of the proof of Proposition \ref{prop:MGHPSD}, we show that $\mathcal{Q}_1$ has finitely many first singularities with $r$ bounded away from zero. Indeed, using crucially the ODE analysis discussed in Section \ref{sec:int-ode} (i.e.~Proposition \ref{odemddprop}), and the already discussed properties of null curves, any first singularity of $\mathcal{Q}_1$ is a either a shell crossing singularity or a \emph{thin shell collision}. In the construction of $\mathscr{Q}_{MGH, 1}$, thin shell collisions arise when the limit point of the exterior of one thin shell is the limit point of the interior of the neighbouring thin shell (see Definition \ref{def:collision} for the precise definition; Figure \ref{fig:9} for an example). By the analysis of the thin shell ODE (Proposition \ref{odemddprop}), this limit point must lie in $\mathscr{Q}_{MDD} \cup \partial^-\mathscr{C}$.
	
	\footnotetext{For clarity, we've restricted here the notation used in the proof of Proposition \ref{prop:MGHPSD}: $\Sigma_i^\pm = \Sigma_{0, i, (s_1)_i}^\pm$.} 
	
\begin{figure}[H]
		\centering
		\begin{minipage}{0.43\linewidth}
			\centering
			\def\svgwidth{0.7\linewidth}
\begingroup%
  \makeatletter%
  \providecommand\color[2][]{%
    \errmessage{(Inkscape) Color is used for the text in Inkscape, but the package 'color.sty' is not loaded}%
    \renewcommand\color[2][]{}%
  }%
  \providecommand\transparent[1]{%
    \errmessage{(Inkscape) Transparency is used (non-zero) for the text in Inkscape, but the package 'transparent.sty' is not loaded}%
    \renewcommand\transparent[1]{}%
  }%
  \providecommand\rotatebox[2]{#2}%
  \newcommand*\fsize{\dimexpr\f@size pt\relax}%
  \newcommand*\lineheight[1]{\fontsize{\fsize}{#1\fsize}\selectfont}%
  \ifx\svgwidth\undefined%
    \setlength{\unitlength}{201.34329969bp}%
    \ifx\svgscale\undefined%
      \relax%
    \else%
      \setlength{\unitlength}{\unitlength * \real{\svgscale}}%
    \fi%
  \else%
    \setlength{\unitlength}{\svgwidth}%
  \fi%
  \global\let\svgwidth\undefined%
  \global\let\svgscale\undefined%
  \makeatother%
  \begin{picture}(1,1.00823751)%
    \lineheight{1}%
    \setlength\tabcolsep{0pt}%
    \put(0,0){\includegraphics[width=\unitlength,page=1]{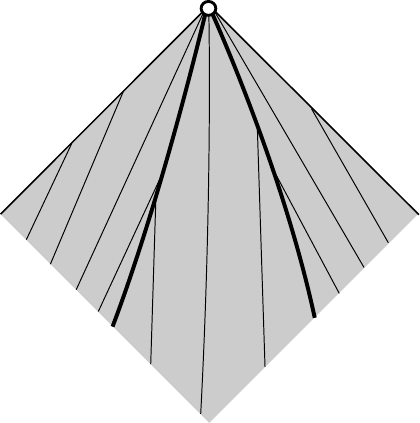}}%
    \put(0.64982931,0.6540731){\color[rgb]{0,0,0}\makebox(0,0)[lt]{\lineheight{2.04999995}\smash{\begin{tabular}[t]{l}$\Sigma_2$\end{tabular}}}}%
    \put(0.41514641,0.61827728){\color[rgb]{0,0,0}\makebox(0,0)[lt]{\lineheight{2.04999995}\smash{\begin{tabular}[t]{l}$\Sigma_1$\end{tabular}}}}%
  \end{picture}%
\endgroup%
 
			\caption{The thin shell collision in the transformation to double null coordinates of the dark shaded region in Figure \ref{fig:9}.}
			\label{fig:10}
		\end{minipage}\hspace{5em}
		\begin{minipage}{0.43\linewidth}
			\centering
			\def\svgwidth{0.7\linewidth}
\begingroup%
  \makeatletter%
  \providecommand\color[2][]{%
    \errmessage{(Inkscape) Color is used for the text in Inkscape, but the package 'color.sty' is not loaded}%
    \renewcommand\color[2][]{}%
  }%
  \providecommand\transparent[1]{%
    \errmessage{(Inkscape) Transparency is used (non-zero) for the text in Inkscape, but the package 'transparent.sty' is not loaded}%
    \renewcommand\transparent[1]{}%
  }%
  \providecommand\rotatebox[2]{#2}%
  \newcommand*\fsize{\dimexpr\f@size pt\relax}%
  \newcommand*\lineheight[1]{\fontsize{\fsize}{#1\fsize}\selectfont}%
  \ifx\svgwidth\undefined%
    \setlength{\unitlength}{240.94488189bp}%
    \ifx\svgscale\undefined%
      \relax%
    \else%
      \setlength{\unitlength}{\unitlength * \real{\svgscale}}%
    \fi%
  \else%
    \setlength{\unitlength}{\svgwidth}%
  \fi%
  \global\let\svgwidth\undefined%
  \global\let\svgscale\undefined%
  \makeatother%
  \begin{picture}(1,1.00000009)%
    \lineheight{1}%
    \setlength\tabcolsep{0pt}%
    \put(0,0){\includegraphics[width=\unitlength,page=1]{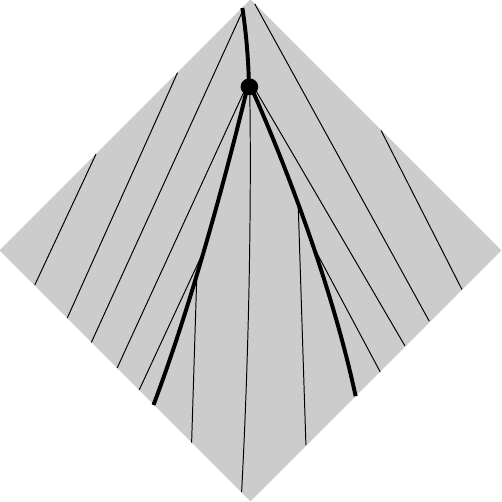}}%
    \put(0.62451562,0.54786642){\color[rgb]{0,0,0}\makebox(0,0)[lt]{\lineheight{2.04999995}\smash{\begin{tabular}[t]{l}$\Sigma_2$\end{tabular}}}}%
    \put(0.42840509,0.51795399){\color[rgb]{0,0,0}\makebox(0,0)[lt]{\lineheight{2.04999995}\smash{\begin{tabular}[t]{l}$\Sigma_1$\end{tabular}}}}%
    \put(0.50983658,0.88250576){\color[rgb]{0,0,0}\makebox(0,0)[lt]{\lineheight{2.04999995}\smash{\begin{tabular}[t]{l}$\Sigma_3$\end{tabular}}}}%
  \end{picture}%
\endgroup%
 
			\caption{The unique local piecewise smooth extension satisfying the propagation condition of the thin shell collision in Figure \ref{fig:10}.}
			\label{fig:11} 
		\end{minipage}
	\end{figure}

\begin{figure} [h] 
		\centering
		\def\svgwidth{0.65\linewidth}
		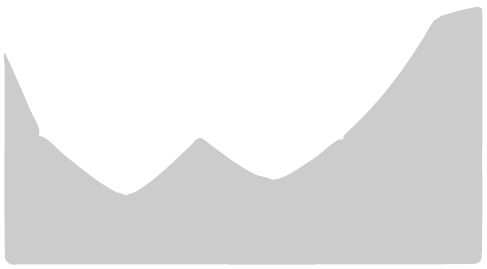 
		\caption{The solutions of the thin shell ODE on the MDD, extending 
			Figure \ref{fig:9} with the thin shell $\Sigma_3$ formed at the collision of $\Sigma_1$, $\Sigma_2$. The `dust region' of the maximal piecewise smooth development $\mathcal{Q}_{MGHD}$ is exactly the transformation to double null coordinates on the dark shaded region (which is $\mathscr{Q}_{MGH}$ in the notation of Proposition \ref{prop:MGHPSD}). The Penrose diagram of $\mathcal{Q}_{MGHD}$ in this case is given by Figure \ref{fig:O-ts}.}
		\label{fig:ts-form-at-collision}
	\end{figure}

	Now, we show that any thin shell collision satisfies the assumptions of the extension principle discussed in Section \ref{sec:int-pwise-ext}. Indeed, see Figure \ref{fig:10} for the `first singularity' (of $\mathcal{Q}_1$) corresponding to the collision of $\Sigma_1$, $\Sigma_2$ in the transformation to double null coordinates of Figure \ref{fig:9}.
	The conditions of the extension principle stated in Proposition \ref{prop:pwise-extension} (other than \eqref{ext-det-cond}, that is, \eqref{mindet} and \eqref{posdet} in Proposition \ref{prop:pwise-extension}) are essentially satisfied immediately by the properties of solutions of the eikonal equation on the maximal dust development. See for example \eqref{deriv-sum}, which implies some of the conditions of \eqref{firstbd} in the statement of Proposition \ref{prop:pwise-extension}. The condition \eqref{ext-det-cond} follows from the colliding thin shells satisfying the determinism condition, and the extension of \eqref{det-cond-bd} to the limit points, see \eqref{nu-bd+}, \eqref{nu-bd-} and the following argument. Hence, there exists a unique piecewise smooth extension (see Figures \ref{fig:11}, \ref{fig:ts-form-at-collision}), given by a solution of the thin shell ODE on the maximal dust development with initial data determined by the extension principle (see Proposition \ref{prop:pwise-extension}, and \eqref{eps-extp}, \eqref{init-extp}).
	
	We then repeat this process, extending $(\mathcal{Q}_1, r, \Omega^2, \rho, \mathbf{u}^\mu)$ with a thin shell forming at each first singularity with $r$ bounded away from zero. At each step, we restrict to thin shells that form at the first singularities of the previous development (and thin shells that can be continued from the previous development). There can only be finitely many steps: there are finitely many first shell crossing singularities by the choice of the generic initial data (i.e.~$\partial^-\mathscr{C}$ is finite), and there can only be finitely many thin shell collisions, so only finitely many thin shells can form. We let $\mathscr{Q}_{MGH} \subset \mathscr{Q}_{MDD}$ correspond to the dust region of the maximal piecewise smooth development. Note that the future boundary of $\mathscr{Q}_{MGH}$ is `acoustically spacelike' (see Section \ref{mdd}), and consists of subsets of $\mathscr{S}$, a possible Cauchy horizon emanating from $(\tau_s(0),0)$ if $\tau_s''(0) > 0$, and the interior/exterior of thin shells. 

	\textbf{The first singularity at the centre of the maximal piecewise smooth development of generic initial data, and the boundary decomposition.} In Section \ref{firstsingrzero}, we show that any first singularity of the maximal piecewise smooth development has area radius extending continuously to zero, and give a full decomposition of the singular boundary of the Penrose diagram of the maximal globally hyperbolic piecewise smooth development. In particular, we show in Lemma \ref{lem:O-limit-thinshell} that if $\tau_s''(0) < 0$, then the first singularity at the centre is formed by a collapsing thin shell, i.e.~$\mathcal{O} \in \partial^+ \Sigma$. This follows from the analysis of the ODE discussed in Section \ref{sec:int-ode}: the interior of a solution to the thin shell ODE, $\Sigma^-_i$, cannot have a limit point in a component of $\mathscr{C}^-$ or $\partial^+ \mathscr{C}$, see Proposition \ref{odemddprop}. There must therefore exist a solution of the thin shell ODE bounding $\mathscr{Q}_{MGH}$ with $\chi_i^-(s) \to 0$. This implies that $\partial^+ \Sigma_i = \mathcal{O}$ in the maximal piecewise smooth development. Moreover, in the case $\tau_s''(0) < 0$, Proposition \ref{odemddprop} implies that in the maximal piecewise smooth development: $\sup_{(u,v)\in\Gamma} \rho(u,v) < \infty$ (see Theorem \ref{MGHDprecise}, Section \ref{sec:thm2}). See Figure \ref{fig:ts-form-at-collision} (and Figure \ref{fig:O-ts}) for an example. 
	
	If $\tau_s''(0) > 0$, then either a thin shell forms the first singularity at the centre (i.e.~a thin shell collapses before the energy density at the centre of the dust cloud blows up), or a Cauchy horizon emanates from the first singularity, as discussed in Section \ref{gen-data}, and shown in Proposition \ref{prop:NS}. This gives the dichotomy \eqref{union} of the possible local structure of the first singularity at the centre. For the detailed structure of the singular boundary of the maximal piecewise smooth development, see Theorem \ref{MGHDprecise}, Section \ref{sec:thm2}. In Lemma \ref{lem:eik-S}, we show that solutions of the eikonal equation extend smoothly to subsets of $\mathscr{S}$ with $\tau_s'(\chi) > 0$. The subset of $\mathscr{S}$ (away from the centre) which bounds $\mathscr{Q}_{MGH}$ satisfies this condition, and, after transforming to double null coordinates, corresponds to the \emph{achronal} singular boundary $\mathcal{S}$ of the maximal piecewise smooth development. In particular, the connected components of $\mathcal{S} \setminus \partial^+ \Sigma$ are $C^1$ spacelike, see Lemma \ref{lem:eik-S} and Proposition \ref{prop:O-in-limit-ts} for the proof. Note that if the exterior curve $\Sigma_i^+$ of a solution of the thin shell ODE bounding $\mathscr{Q}_{MGH}$ meets the boundary of the dust cloud $\{\chi=\chi_b\}$, then the interior curve $\Sigma_i^-$ continues smoothly. The jump conditions for the thin shell provide boundary conditions for the eikonal equation, when transforming to double null coordinates, see \eqref{eikBC-vacuum}. In this case, $\mathcal{B}_d$ is piecewise $C^1$, see for example Figure \ref{fig:glob-naked}.
	
	\subsection{Outline of the paper}
	
	\textbf{Section \ref{dust}.} We construct comoving coordinates $(\tau,\chi)$ for smooth solutions of Einstein-dust in spherical symmetry. We then define the `maximal dust development' (MDD) of smooth compactly supported initial data in comoving coordinates, and give a decomposition of its singular boundary in $\R^2_{\tau,\chi}$. We then introduce the Einstein-dust system in double null coordinates (E-d), which we understand by studying the eikonal equation on solutions of the Einstein-dust system in comoving coordinates. We show an extension principle for smooth solutions of (E-d) with bounded energy density, and then study the geometry of first shell crossing singularities. 
	 
 	\textbf{Section \ref{pwise}.} Using the original definition of Lanczos \cite{Lanczos} for `delta function support' on a curve, we define piecewise smooth solutions directly on the quotient manifold $\mathcal{Q}$, which we show are equivalent to Israel's formulation of thin shells in terms of extrinsic curvature on the `upstairs' manifold $\mathcal{M} = \mathcal{Q} \times \mathbb{S}^2$. We then show that, under suitable conditions, piecewise smooth solutions are equivalent to solutions of the thin shell ODE on maximal dust developments, with double null coordinates again constructed as solutions of the eikonal equation. We define the determinism condition in Section \ref{detcondsec}.
	
	\textbf{Section \ref{development}.} We give the precise statement of Theorem \ref{thm:1}, and state the existence/uniqueness result for the thin shell ODE (Proposition \ref{thinshellODEprop}), which we then show is equivalent to the precise statement, in Proposition \ref{equiv}. We then prove the ODE result, Proposition \ref{thinshellODEprop}.
	
	\textbf{Section \ref{maxdev}.} We first consider solutions of the Einstein-dust system including the centre of spherical symmetry, and define the moduli space of smooth initial data for dust clouds including the centre in Section \ref{smoothdataEd}. We then define piecewise smooth solutions with multiple thin shells in Section \ref{finiteshells}, and define the `propagation condition' in Section \ref{propcond}, which we show in Section \ref{extp} uniquely determines piecewise smooth solutions under certain conditions. We then give the precise statement of Theorem \ref{MGHD} in Section \ref{sec:thm2}, and prove the result in Section \ref{MGHDproof}.

	\subsection{Previous work on shell crossing} \label{prev}
	
	As discussed in the introduction, the first study of shell crossing was carried out by Papapetrou and Hamoui \cite{PH} in 1967, for a special case of a \emph{degenerate} first shell crossing singularity.\footnote{In the notation introduced in Section \ref{comovingintro}, the degenerate case considered by Papapetrou and Hamoui is $\tau_c(\chi) = \tau_0$ for $\chi \in [\chi_1,\chi_2]$, which implies that the dust shells with $\chi \in [\chi_1,\chi_2]$ collide at a spacelike codimension-2 sphere.} They realised that there was a unique continuation such that all of the dust shells \emph{crossed}\footnote{This may be the origin of the slightly misleading term `shell crossing'.}, which they identified by choosing coordinates so that the metric is $C^{0,1}$ at the collision sphere. Crucially, in their extension, the spacetime to the future of the collision sphere is again a \emph{single dust} region. 
	
	Clarke and O'Donnell \cite{CD} considered the generic caustic (as opposed to the degenerate case of \cite{PH}) formed in a proposed collisionless extension at a non-degenerate first shell crossing singularity, and showed that it is `self-consistent'. In this case, there is a \emph{three-dust} region, as opposed to the single-dust regions of \cite{PH}. Bick \cite{David} has recently given the first (local) construction of a caustic curve as a weak solution of the Einstein(-multi)-dust system, bounding a two-dust region. (It remains an open problem to show the existence and uniqueness of a local collisionless extension at a non-degenerate first shell crossing singularity.)
	
	Nolan \cite{N}, in 2004, considered the Einstein-dust system in comoving coordinates as a hyperbolic conservation law. In particular, for marginally bound dust, he showed that after a change of coordinates the equation of motion is a scalar conservation law for the Hawking mass, analogous to Burgers' equation. He identifies however a fundamental non-uniqueness in the weak solutions that he constructs, which satisfy the standard Rankine--Hugoniot relations for conservation laws. One can interpret Nolan's spacetimes as solutions of the Einstein equations with thin shells \cite{Tegai}, but the matter on the thin shells does not satisfy a prescribed equation of state, and they are in particular not thin shells of dust. The spacetimes are therefore, for example, when changed into double null coordinates, not weak solutions of the Einstein-dust system (see Section \ref{pwisesection}, in particular Definition \ref{pwisesmooth}). 
	
	In \cite{FM1}, the authors show that Nolan's shocks become null (and then spacelike) as they collapse, and comment that this arises because of the use of comoving coordinates in constructing the weak solutions, in particular that the comoving coordinate $\tau$ satisfies $\tau^+(s)=\tau^-(s)$. See also the recent \cite{FM2} for a discussion of extending spacetime through shell crossing singularities using thin shells in loop quantum gravity, and \cite{Bobula} for numerical results. 
	
	\subsubsection{Delta shocks in pressureless Euler} \label{pressurelessEuler}
	
	The Einstein-dust system can be formulated as the Einstein--Euler system with equation of state $p=0$, and the fluid variables $(\rho,\mathbf{u}^\mu)$ satisfy the `relativistic pressureless Euler equations'. The pressureless Euler equations (outside of relativity) have been understood to admit `measure solutions' (including piecewise smooth solutions analogous to our framework), since the seminal paper of Bouchut \cite{B} for pressureless Euler in one spatial dimension. Bouchut showed in particular that there exist measure solutions containing a curve on which the density has delta function support, termed a `delta shock'. Existence and uniqueness has been shown in suitable classes of measure solutions, see for example E, Rykov and Sinai \cite{ERS} for a construction of weak solutions, and Bouchut and James \cite{BF} for an existence and uniqueness result. These measure solutions can be described as an `adhesion' or `sticky dust' model for pressureless Euler: particles stick upon collision, conserving mass and momentum. Such models have their roots in Zeldovich's study on large density perturbations in the cosmological setting \cite{Z}, and have been used in the study of large scale structure formation in the universe, see e.g.~\cite{GSS}. 
	
	It is straightforward to see from the pressureless Euler equations in 1D (and indeed in spherical symmetry) that smooth solutions can form singularities in finite time, where the density (and the density of the foliation by fluid flow lines) diverges. This is analogous to the phenomenon of shell crossing in the Einstein-dust system. The pressureless Euler equations in 1D are:
	\begin{align}
		\partial_t \rho + \partial_x(\rho u) &= 0 \label{pE1} \\ \partial_t (\rho u) + \partial_x (\rho u^2) &= 0, \label{pE2}
	\end{align}
	where $\rho$ is the fluid density and $u$ is the fluid velocity. Combining \eqref{pE1} and \eqref{pE2}, we see that the fluid velocity $u$ satisfies \emph{Burgers' equation}:
	\begin{equation} \label{Burgers}
			\partial_t u + u \partial_x u = 0,
	\end{equation}
	Therefore, given initial data with $u'(0,x) < 0$, we get shock formation in Burgers' equation, and one can see (by introducing a Lagrangrian coordinate $\chi(t,x)$ solving $\partial_t \chi + u \partial_x \chi = 0$, analogous to the \emph{comoving} coordinate used for solutions of the Einstein-dust system in spherical symmetry) that $\partial_\chi x \to 0$ at shock formation, and hence that the fluid density $\rho$ blows up. An `extension' beyond a `shock formation' first singularity as a measure solution of \eqref{pE1}, \eqref{pE2} would correspond to a solution of the `delta shock development problem'. Note that measure solutions are defined by integrating \eqref{pE1}, \eqref{pE2}, \emph{not} Burgers' equation \eqref{Burgers}. Indeed, delta shocks are \emph{not} Burgers' equation shocks and satisfy distinct Rankine--Hugoniot conditions.
	
	Given a non-degenerate first shell crossing singularity, as in Figure \ref{MGHDsubset}, there exist \emph{non-unique} smooth extensions of double null coordinates, beyond the `Cauchy horizons' of the first singularity, defined on the maximal dust development, see Lemma \ref{smoothext} in Section \ref{regsc}. Such a choice of coordinates can be extended to the shell crossing singular boundary, and the corresponding solution of the Einstein-dust system in double null coordinates then does not embed into $\R^{1+1}_{u,v}$. The smooth structure of such a spacetime in the neighbourhood of the first shell crossing singularity is comparable to that of pressureless Euler/Burgers' equation: spacetime is bounded by two (timelike) caustic curves (corresponding to the shell crossing singular boundary) forming at the first singularity. Projecting onto $\R^{1+1}_{u,v}$, the caustics envelope a region of crossing characteristics (fluid flow lines).
	
	The fundamental difference of pressureless Euler to the coupled Einstein-dust system is the existence of `physical' $(t,x)$ coordinates, in which weak solutions are defined directly. Moreover, the exact structure of the 1D pressureless Euler equations allow one to show existence and uniqueness for more general measure solutions than `piecewise smooth', see e.g.~\cite{BF}. In general relativity, there are of course no such `physical' coordinates. Comoving coordinates are essential for understanding the structure of smooth solutions, and we can define the `singular boundary' of the maximal dust development, which is analogous to the structure of shock formation for Burgers' equation in Langrangian coordinates. However, piecewise smooth solutions of Einstein-dust in spherical symmetry cannot be defined directly in `global' comoving coordinates, hence the use of double null coordinates. The solution of the thin shell development problem follows from \emph{teleologically} identifying double null coordinates along solutions of the `thin shell ODE' on the maximal dust development; it remains to be seen whether weak solutions more general than `piecewise smooth' can be defined for the Einstein-dust system.

	\subsection{Further problems and future work} \label{future}
	
	In upcoming work, Feh\'ert\'oi-Nagy \cite{Lili} shows that, in the singular limit, solutions of Einstein--Vlasov converge to solutions of Einstein--dust on compact sets (see also \cite{AR} for solutions of Einstein--Vlasov converging to the Oppenheimer--Snyder spacetimes). One would expect that the choice of \emph{collisionless} dust at shell crossing, i.e.~the case considered by Papapetrou and Hamoui \cite{PH}, Clarke and O'Donnell \cite{CD} and Bick \cite{David} could be understood as the singular limit of Einstein--Vlasov.
	
	On the other hand, the choice considered in this paper, that thin shells of dust form at shell crossing, should correspond to the pressureless limit of Einstein--Euler. The pressureless limit of the Riemann problem in 1D compressible Euler was studied by Chen and Liu \cite{Ch}. They show that, given equation of state $p(\rho) = \varepsilon \rho^\gamma$, and piecewise constant data such that the solution contains two shocks, the solution converges in measure to the corresponding delta shock solution of the pressureless Euler equations, as $\varepsilon \to 0$, where $\rho \sim \varepsilon^{-\frac{1}{\gamma}}, p(\rho) \sim 1$ between the shocks. Peng and Wang \cite{PengWang} have shown that the pressureless limit for continuous solutions of 1D Euler converges to the corresponding continuous solution of pressureless Euler. We note also that Guo, Had\v{z}i\'{c} and Jang \cite{GHJ2} have constructed collapsing solutions of the Euler--Poisson system with leading order singular behaviour described by a solution of the pressureless Euler--Poisson system. 
	
	It is an interesting problem, already for the Euler equations, whether `delta shock formation', or indeed `delta shock development', can be understood in the pressureless limit.\footnote{Note that the existing BV theory for 1D Euler breaks down in the pressureless limit, indeed the total variation of fixed initial data diverges in the pressureless limit. This is of course expected since the pressureless Euler equations are ill-posed in BV.} The heuristic idea is as follows: fix initial data for pressureless Euler that forms a delta shock at $(t_*,x_*)$, then consider the analogous initial data with e.g.~equation of state $p = \varepsilon \rho^\gamma$, for some $\gamma > 1$. One would expect, for $\varepsilon > 0$ sufficiently small, formation of \emph{two shocks} close to $(t_*,x_*)$, such that, after shock development, the shocks are \emph{subsonic} with respect to the fluid region between them, and \emph{supersonic} with respect to the respective left/right exterior regions. It appears to be a difficult problem to show that the shocks remain close and approximate the delta shock for $t \sim 1$. 
	
	It would then be extremely interesting to understand the pressureless limit in the spherically symmetric Einstein--Euler system, in particular whether thin shells of dust can be approximated by pairs of shocks in weak solutions of Einstein--Euler, and whether one can understand shock formation/development, from smooth initial data, in the pressureless limit of Einstein--Euler. Weak solutions of the spherically symmetric Einstein--Euler system with bounded variation have been studied by Burtscher and LeFloch \cite{Burtscher-LeFloch}, but to the best of the author's knowledge, there are no results on \emph{shock formation} from smooth initial data in the \emph{spherically symmetric} Einstein--Euler system, let alone shock development. There do exist results on shock formation for Einstein--Euler that are not spherically symmetric: see \cite{Rendall} for shock formation in plane symmetric spacetimes, and upcoming work of Anderson and Luk \cite{John-Jonathan} on stable shock formation for the Einstein--Euler system. 
	
	We make the following conjectures for the Einstein--Euler system in spherical symmetry with equation of state
	\begin{equation} \label{eos}
		p = \varepsilon \rho.
	\end{equation}
	Motivated by \cite{Ch}, we expect thin shells of dust to be approximated in the pressureless limit by solutions of Einstein--Euler containing pairs of shocks.
	
	\begin{conjecture}[Approximation of dust spacetimes containing a thin shell by piecewise smooth solutions of Einstein--Euler] \label{thinshell-approx}
		Given a piecewise smooth solution of the spherically symmetric Einstein-dust system in double null coordinates, containing a thin shell of dust, then there exists $\varepsilon_*>0$, and weak solutions of the spherically symmetric Einstein--Euler system with equation of state \eqref{eos}, depending on $\varepsilon \in (0,\varepsilon_*)$, and containing two shocks, which converge as $\varepsilon \to 0$, in an appropriate sense, to the piecewise smooth solution of Einstein-dust.
	\end{conjecture}
	
	Given an (affirmative) resolution of the above conjecture, a (much harder) next step would be to study the \emph{formation} of thin shells in the pressureless limit:
	
	\begin{conjecture} [Pressureless limit of Einstein--Euler to thin shell development] \label{thinshelldev-approx}
		Choose smooth initial data for the spherically symmetric Einstein--Euler system with equation of state \eqref{eos}, in double null coordinates, depending continuously on $\varepsilon$. Suppose that the classical development of the dust initial data, i.e.~the data for $\varepsilon=0$, forms a non-degenerate first shell crossing singularity, and let $(\mathcal{Q}_0,r_0,\Omega^2_0,\rho_0,\mathbf{u}^\mu_0)$ be a piecewise smooth globally hyperbolic development of the initial data, containing a thin shell forming at the first shell crossing singularity, which exists for proper time $s \in [0,s_*]$ and satisfies the (thin shell) \underline{determinism condition}. Then there exists $\varepsilon_* > 0$ such that: for $\varepsilon \in (0,\varepsilon_*]$ there exist weak solutions of the Einstein--Euler system with equation of state \eqref{eos}, $(\mathcal{Q}_{0}, r_{\varepsilon}, \Omega^2_{\varepsilon}, \rho_{\varepsilon}, \mathbf{u}^\mu_{\varepsilon})$, attaining the corresponding initial data, and which form two shocks that exist for proper time $\geq s_*$ and satisfy the determinism condition\footnote{This is the determinism condition for shocks in Einstein--Euler: any shock is \emph{acoustically spacelike}/supersonic with respect to the `state ahead' and \emph{acoustically timelike}/subsonic with respect to the `state behind', see \cite{CL}.}. In the limit $\varepsilon \to 0$, then $(\mathcal{Q}_{0}, r_{\varepsilon}, \Omega^2_{\varepsilon}, \rho_{\varepsilon}, \mathbf{u}^\mu_{\varepsilon})$ converges in an appropriate sense to $(\mathcal{Q}_0,r_0,\Omega^2_0,\rho_0,\mathbf{u}^\mu_0)$.
	\end{conjecture}
	
	One can also ask whether the \emph{global} structure of collapsing dust clouds with multiple thin shells formed at shell crossing, in particular the structure of \emph{thin shell collisions}, can be realised in the pressureless limit. 
	
	\begin{conjecture}[Approximation of thin shell collisions by piecewise smooth solutions of Einstein--Euler] \label{collision-approx}
		Given a piecewise smooth solution of the spherically symmetric Einstein-dust system in double null coordinates satisfying the \underline{propagation condition}, containing a thin shell collision, then there exists $\varepsilon_*>0$, and weak solutions of the spherically symmetric Einstein--Euler system with equation of state \eqref{eos}, depending on $\varepsilon \in (0,\varepsilon_*)$, which converge as $\varepsilon \to 0$, in an appropriate sense, to the piecewise smooth solution of Einstein-dust.
	\end{conjecture}
	
	The above conjectures would imply that the propagation condition for piecewise smooth solutions of Einstein-dust (which we impose in order to prove uniqueness) arises naturally in the pressureless limit of Einstein--Euler.
	
	It is a separate interesting question to understand what happens at the centre in the pressureless limit, particularly because of the existence of Christodoulou-type naked singularities in dust \cite{C1}, and the existence of self-similar naked singularities in Einstein--Euler, which has been shown by Guo, Had\v{z}i\'{c} and Jang \cite{GHJ} for equation of state $p = \varepsilon \rho$ with $\varepsilon$ sufficiently small. 

	\subsection{Acknowledgements}
	
	The author would like to thank Mihalis Dafermos for valuable guidance and advice in writing this paper. The author would also like to thank David Bick, Lili Feh\'ert\'oi-Nagy, Istvan Kadar, Christoph Kehle, Warren Li, Harvey Reall, M\'onica Tapia del Moral, Rita Teixeira da Costa, Ryan Unger and Claude Warnick for helpful discussions. This work was supported by the Science and Technology Facilities Council.
	
	\subsection*{AI declaration}
	
	None of the mathematical results or broader mathematical ideas of this paper were generated by AI. The exposition of the results/ideas in this paper was not AI generated. AI was used as a tool to search and understand the literature. 
	
	\section{Smooth solutions of Einstein-dust in spherical symmetry} \label{dust}
	
	In this section, we first consider smooth, spherically symmetric, solutions of the Einstein-dust system in comoving coordinates $(\tau,\chi)$, in which the Einstein-dust equations reduce to a remarkably simple system. The characteristics of this system are the fluid flow lines of the dust, and we therefore introduce a notion of `acoustic hyperbolicity', using which we show the existence (and uniqueness) of a `maximal acoustically globally hyperbolic development', which we term the `maximal dust development'. The embedding of the maximal dust development into $\R^2_{\tau,\chi}$ defines the (acoustic) singular boundary of the quotient manifold: either corresponding to \emph{shell crossing}, or \emph{collapse} (the area radius extends continuously to zero). We identify conditions on initial data that characterise components of this singular boundary being empty/non-empty.

	We then study the Einstein-dust system in spherical symmetry in double null coordinates (E-d). In Section \ref{eikonalsection} we construct double null coordinates on dust spacetimes in comoving coordinates as solutions of the \emph{eikonal equation}. In Section \ref{CIVP}, we use existence and uniqueness of smooth solutions to Einstein-dust in comoving coordinates, as well as that of the eikonal equation, to show local existence and uniqueness of $C^\infty$ solutions of (E-d) with characteristic initial data. We then prove an extension principle for $C^\infty$ collapsing solutions of (E-d) away from the centre in Section \ref{sec:extp}. We're led to the definition of a first shell crossing singularity in Section \ref{firstscsec}, where we define also \emph{non-degenerate} first shell crossing singularities, and show that they correspond to minima of the function $\tau_c(\chi)$ defining the singular boundary in comoving coordinates. 
	
	In Section \ref{regsc}, we make essential use of the eikonal equation to show that the metric components $r, \Omega$ extend in $C^1_{u,v}$ to non-degenerate first shell crossing singularities. We also show that we can \emph{non-uniquely} smoothly extend the solution of (E-d) beyond the Cauchy horizon of a non-degenerate first shell crossing singularity: the extension is a non-unique choice of double null coordinates on the MDD, and is necessarily not globally hyperbolic. We show that any such extension \emph{does not embed} into $\R^{1+1}_{u,v}$, reminiscent of, for example, the failure of smooth solutions of Burgers' equation to embed into $\R^2_{t, x}$ in a neighbourhood of shock formation.
	
	\subsection{Comoving coordinates} \label{comoving}
	
	Let $(\mathcal{M},g)$ be a smooth connected, four-dimensional, time orientable, Lorenztian manifold. We say that $(\mathcal{M},g)$ is \emph{spherically symmetric} if the group $SO(3)$ acts as an isometry on $(\mathcal{M},g)$, and $Q := \mathcal{M} / SO(3)$ (the `quotient manifold') inherits the structure of a $1 + 1$ dimensional Lorentzian manifold, with boundary $\Gamma \subset Q$ which we refer to as the `centre' (corresponding to the set of fixed points of the action of $SO(3)$ on $\mathcal{M}$). We require moreover that the metric $g$ on $\mathcal{M}$ takes the form 
	\begin{equation}
		g = g_{Q} + r^2 g_{\mathbb{S}^2},
	\end{equation}
	where $g_{Q}$ is the induced metric on $Q$, $g_{\mathbb{S}^2}$ is the round metric on the sphere, and $r$ (the `area radius') is a non-negative function on $Q$, positive on $\mathring{Q} := Q \setminus \Gamma$, and zero on $\Gamma$. We now assume that $(\mathcal{M},g)$ is in fact a spherically symmetric smooth solution of the Einstein-dust system:
	\begin{equation*}
		R_{\mu \nu} - \frac{1}{2} R g_{\mu \nu} = 8\pi T_{\mu \nu},
	\end{equation*}
	where the energy-momentum tensor $T_{\mu \nu}$ is that of a pressureless perfect fluid,
	\begin{equation*}
		T_{\mu \nu} = \rho \mathbf{u}_\mu \mathbf{u}_\nu.
	\end{equation*}
	Here, $\rho$ is a smooth function on $\mathcal{M}$, which we call the energy density, and $\mathbf{u}^a$ a smooth vector field (defined on the support of $\rho$) satisfying $g(\mathbf{u},\mathbf{u}) = -1$, which we call the fluid velocity. We time-orient the spacetime $(\mathcal{M},g)$ by imposing that the fluid velocity $\mathbf{u}^a$ is future-directed. Conservation of the energy momentum tensor, $\nabla_\mu T^{\mu \nu} = 0$, implies that
	\begin{align}
		\nabla_\mu (\rho \mathbf{u}^\mu) &= 0, \label{masscons} \\ \rho \mathbf{u}^\mu \nabla_\mu \mathbf{u}^\nu &= 0. \label{geodesic}
	\end{align}
	 After taking the quotient of the spherically symmetric action of $SO(3)$ on $(\mathcal{M},g)$, $\rho$ descends to a smooth function on $(Q, g_{Q})$, and $\mathbf{u}^a$ descends to a smooth vector field on $(Q, g_{Q})$. We call the integral curves of $\mathbf{u}^a$ in $Q$ `\emph{fluid flow lines}' (which for $r>0$ correspond to timelike hypersurfaces in $\mathcal{M}$). We define the Hawking mass $m$ by
	 \begin{equation} \label{mdef0}
	 	1 - \frac{2m}{r} = g(\nabla r,\nabla r).
	 \end{equation}
	 The remarkably simple evolution of \emph{smooth} dust in spherical symmetry is demonstrated in \emph{comoving coordinates}:
	 \begin{lemma} \label{comovinglemma}
	 	Let $(Q,g,\rho,\mathbf{u})$ be a smooth solution of Einstein-dust in spherical symmetry, such that the manifold $(Q,g)$ is simply connected and orientable, and suppose that $r, \rho$ are positive functions on $Q$, so that in particular $\mathbf{u}^a$ is a global vector field, then there exists a global spacelike vector field $\mathbf{u}_*^a$ satisfying:
	 	\begin{equation}
	 		g(\mathbf{u}_*,\mathbf{u}) = 0, \qquad g(\mathbf{u}_*,\mathbf{u}_*) = 1.
	 	\end{equation}
	 	If moreover $\mathbf{u}_*(r)^2 > 0$, then we choose $\mathbf{u}_*$ to be `outward-directed': $\mathbf{u}_*(r) > 0$, and there exist global `comoving coordinates', $(\tau,\chi)$, for which $\mathbf{u} = \partial_\tau$, the Einstein equations reduce to the following system:
	 	\begin{align}
	 		\partial_\tau m &= 0, \label{mtau} \\ \partial_\chi m &= 4\pi r^2 \rho \partial_\chi r, \label{mchi} \\ \partial_\tau^2 r &= -\frac{m}{r^2}, \label{acc}
	 	\end{align}
	 	so that $m$ is a function of $\chi$ only, $m(\chi)$, and 
	 	\begin{equation} \label{f}
	 		(\partial_\tau r)^2 = \frac{2m(\chi)}{r} + f(\chi),
	 	\end{equation}
	 	\begin{equation} \label{velstar}
	 		\mathbf{u}_*(r)^2 = e^{-2\omega} (\partial_\chi r)^2 = 1 - \frac{2m}{r} + (\partial_\tau r)^2 = 1 + f(\chi) > 0,
	 	\end{equation}
	 	so that $\mathbf{u}_* = \frac{(1+f(\chi))^\frac{1}{2}}{\partial_\chi r(\tau,\chi)} \partial_\chi$. The metric is given by
	 	\begin{equation} \label{metric}
	 		g = -d\tau^2 + \frac{(\partial_\chi r(\tau,\chi))^2}{1 + f(\chi)} d\chi^2 + r(\tau,\chi)^2 g_{\mathbb{S}^2},
	 	\end{equation}
	 	and the energy density is given by
	 	\begin{equation} \label{rho}
	 		\rho(\tau,\chi) = \frac{\partial_\chi m(\chi)}{4\pi r(\tau,\chi)^2 \partial_\chi r(\tau,\chi)}.
	 	\end{equation}
	 \end{lemma}
	 
	 \begin{remark}
	 	We refer to the integral curves of $\mathbf{u}_*$, i.e.~curves of constant $\tau$, as `simultaneous curves'.
	 \end{remark}
	
	\begin{proof}
		We follow \cite{C2}. The existence of $\mathbf{u}_*$ follows immediately from the orientability of $Q$. We define the $1$-form 
		\begin{equation} \label{beta}
			\beta_\mu := - g_{\mu \nu} \mathbf{u}^\nu,
		\end{equation}
		which satisfies $d\beta = 0$ by the geodesic equation \eqref{geodesic} and $g(\mathbf{u},\mathbf{u}) = 1$. Now, $Q$ being simply connected, there exists a function $\tau : Q \to \R$ such that $\beta = d\tau$, which is uniquely defined up to a constant term. We have $\beta(\mathbf{u}_*) := \beta_\mu \mathbf{u}_*^\mu = 0$, so $\mathbf{u}_*(\tau) = \mathbf{u}_*^\mu \partial_\mu \tau = 0$. We also have $\beta(\mathbf{u})=1$, so that $\mathbf{u}(\tau)=\mathbf{u}^\mu \partial_\mu \tau =1$. Given any $p \in Q$, we can construct comoving coordinates in a neighbourhood of $p$ as follows. We gauge fix $\tau(p) = 0$, and let $\mathcal{D} \subset \mathcal{Q}$ be the simultaneous curve containing $p$, so that $\mathcal{D} = \{\tau=0\}$. Choose a smooth function $\mathring{\chi} : \mathcal{D} \to \R$ such that $\mathbf{u}_*(\mathring{\chi}) > 0$ on $\mathcal{D}$. We extend $\mathring{\chi}$ to a smooth function $\chi$ on (a subset of) $\mathcal{Q}$ satisfying $\mathbf{u}(\chi) = 0$, so that the level sets of $\chi$ are fluid flow lines. The pair $(\tau,\chi)$ are `comoving coordinates', and we can choose functions $\psi, \omega$ on the subset of $\mathcal{Q}$ on which $\chi$ is defined, such that
		\begin{equation*}
			\mathbf{u} = e^{-\psi} \partial_\tau, \qquad \mathbf{u}_* = e^{-\omega} \partial_\chi,
		\end{equation*}
		where we note that $\mathbf{u}(\tau)=1$ from above so in fact $\psi \equiv 0$. The metric then takes the form
		\begin{equation*}
			g = - d\tau^2 + e^{2\omega} d\chi^2 + r(\tau,\chi)^2 g_{\mathbb{S}^2}.
		\end{equation*}
		The definition of the Hawking mass \eqref{mdef0} in comoving coordinates is
		\begin{equation} \label{defm}
			1 - \frac{2m}{r} = e^{-2\omega} (\partial_\chi r)^2 - (\partial_\tau r)^2.
		\end{equation}
		One can then compute (see \cite{C2}) that the Einstein-dust equations in comoving coordinates reduce to the system \eqref{mtau}, \eqref{mchi}, \eqref{acc}. By \eqref{mtau}, the Hawking mass is a function of $\chi$ only, and we can therefore integrate \eqref{acc} to get \eqref{f}. By the definition of the Hawking mass \eqref{defm} we then get \eqref{velstar}, and therefore the metric is given by \eqref{metric}. Finally, \eqref{rho} follows from \eqref{mchi}.
		
		We have thus far only defined $\chi$ locally in a neighbourhood of each fluid flow line. To see that we can choose a global $\chi$, note that $\mathbf{u}(m) = 0$, and $\mathbf{u}_*(m) = 4\pi r^2 \rho \mathbf{u}_*(r) > 0$, independent of a choice of coordinates. We can therefore always choose $\chi = m$, and $(\tau,m)$ are global comoving coordinates.
	\end{proof}
	
	Given a choice of comoving coordinates, the system therefore reduces to determining $r(\tau,\chi)$, which follows from integating \eqref{f} given suitable initial data. The characteristics of the system are the fluid flow lines: the solution at any point is completely determined by integrating from initial data along its unique fluid flow line. Note in particular that the fluid flow lines are the characteristics of the \emph{gauge choice}: given a choice of comoving coordinates at a point $p \in \mathring{Q}$, then $\mathbf{u}(\chi) = 0$, $\mathbf{u}(\tau) = 1$, so the gauge is determined along the fluid flow line.
	
	We will consider only \emph{collapsing} dust spacetimes: 
	\begin{definition} 
		We say $(Q,g_Q,\rho,\mathbf{u})$ is a \emph{collapsing} spherically symmetric dust spacetime if $\mathbf{u}(r) = \mathbf{u}^\mu \partial_\mu r < 0$ on $\mathring{Q}$, and $\mathbf{u}_*(r) = \mathbf{u}_*^\mu \partial_\mu r > 0$ on $Q$.
	\end{definition}
	We note that 
	\begin{equation}
		\mathbf{u}_*(r) =  e^{-\omega} \partial_\chi r = (1 + f(\chi))^\frac{1}{2},
	\end{equation}
	and therefore $\mathbf{u}(\mathbf{u}_*(r)) = \partial_\tau (1+f(\chi))^\frac{1}{2} = 0$. If $\mathbf{u}_*(r) > 0$ on initial data, then $\mathbf{u}_*(r) > 0$ on every fluid flow line extending the data. Further, from \eqref{velstar}, we see that if $\frac{2m}{r} < 1$, then we must have $\mathbf{u}_*(r)^2 > 0$. Therefore, any collapse from untrapped (orientable) initial data must have $\mathbf{u}_*(r) > 0$. For any collapsing dust spacetime, we have
	\begin{equation} \label{Edcollapse}
		\partial_\tau r(\tau,\chi) = -\left( \frac{2m(\chi)}{r} + f(\chi) \right)^\frac{1}{2}
	\end{equation} 
	on $\mathring{Q}$. In this section, we'll consider solutions of the Einstein-dust system with $r>0$, so that $\Gamma = \emptyset$. See Section \ref{initial-data} for the inclusion of the centre. We assume that we can construct global comoving coordinates as in Lemma \ref{comovinglemma}, and we define:
	\begin{definition} \label{colldust}
		A $C^\infty$ solution of the collapsing Einstein-dust system, in comoving coordinates, consists of a domain $\mathscr{Q} \subset \R^2_{\tau,\chi}$ with functions $r, \rho, m, f \in C^\infty_{\tau,\chi}(\mathscr{Q})$, such that $r, \rho, m, (1+f), \partial_\chi r > 0$, and the functions satisfy \eqref{mtau}, \eqref{rho}, \eqref{Edcollapse}.
	\end{definition}
	
	\begin{remark} \label{Tremark}
		Given any $C^\infty$ solution of the collapsing Einstein-dust system, the implicit function theorem and $\mathbf{u}(r) = \partial_\tau r < 0$ implies that there exists a smooth function $T \in C^\infty( \{ (r(\tau,\chi),\chi) : (\tau,\chi) \in \mathscr{Q} \})$ such that for all $(\tau,\chi) \in \mathscr{Q}$:
		\begin{equation} \label{Tdef0}
			\tau = T(r(\tau,\chi), \chi).
		\end{equation}
	\end{remark}
	
	\begin{definition} \label{data}
		Smooth initial data for the collapsing Einstein-dust system consists of $\mathscr{D} := \{ (\tau_{\mathscr{D}}(\chi), \chi) : \chi \in [\chi_1,\chi_2] \}$, where $\tau_{\mathscr{D}}(\chi)$ is a smooth function, along with functions $r_{\mathscr{D}}, m_{\mathscr{D}}, \mathbf{u}(r)_{\mathscr{D}} \in C^\infty([\chi_1,\chi_2])$, such that: $r_{\mathscr{D}}, m_{\mathscr{D}} > 0$, $\mathbf{u}(r)_{\mathscr{D}} < 0$ and for $\chi \in [\chi_1,\chi_2]$:
		\begin{equation}
			m_{\mathscr{D}}'(\chi) > 0, \qquad (\partial_\chi r)_{\mathscr{D}}(\chi) := r_{\mathscr{D}}'(\chi) - \mathbf{u}(r)_{\mathscr{D}}(\chi) \cdot \tau_{\mathscr{D}}'(\chi) > 0, \label{pchirD}
		\end{equation}
		\begin{equation} 
			 [\mathbf{u}_*(r)_{\mathscr{D}}]^2 := 1 - \frac{2m_{\mathscr{D}}}{r_{\mathscr{D}}} + [\mathbf{u}(r)_{\mathscr{D}}]^2 > 0.
		\end{equation}
	\end{definition}
	
	\subsubsection{Acoustic global hyperbolicity and the maximal dust development (MDD)} \label{mdd}
	
	In the previous section, we saw that solutions in comoving coordinates are uniquely determined by integrating along fluid flow lines. We formalise this here as the notion of `acoustic global hyperbolicity'.\footnote{The term `acoustic' is used because one can think of Einstein-dust as the $p = 0$ case of the Einstein--Euler equations, describing self-gravitating relativistic fluids. For solutions of the Einstein--Euler system in spherical symmetry, one can define an `acoustic' metric, a Lorentzian metric for which the null cone is referred to as the `sound cone'. In the dust case, the assumption of vanishing pressure implies that the speed of sound vanishes: the sound cone collapses to a line and the acoustic metric becomes degenerate, given by (up to a conformal factor):
		\begin{equation}
			h = d\chi^2.
		\end{equation}
	The fluid flow lines are \emph{acoustically null}, and any vector not proportional to $\mathbf{u}^a$ is \emph{acoustically spacelike}.}	
	\begin{definition}
		An acoustic Cauchy surface $\mathscr{D} \subset \mathscr{Q}$ is a curve such that $\rho > 0$ on $\mathscr{D}$, and no two points on $\mathscr{D}$ are connected by a fluid flow line. The acoustic future domain of dependence of $\mathscr{D}$, which we denote $D_{ac}^+(\mathscr{D})$, is the subset of $\mathscr{Q}$ such that for any $p \in \mathscr{Q}$, the past-inextendible fluid flow line intersects $\mathcal{D}$.
	\end{definition}
	\begin{definition}
		The spacetime $(\mathscr{Q},g_{\mathscr{Q}},\rho,\mathbf{u})$ is (future) \emph{acoustically globally hyperbolic} if there exists an acoustic Cauchy surface $\mathscr{D} \subset \mathscr{Q}$ such that $\mathscr{Q} = D_{ac}^+(\mathscr{D})$.
	\end{definition}
	Acoustic globally hyperbolic spacetimes are then uniquely determined by data on an acoustic Cauchy surface. We define:
	\begin{definition} \label{MAGHD}
		A $C^\infty$ collapsing acoustically globally hyperbolic development of smooth initial data on $\mathscr{D}$ consists of a domain $\mathscr{Q} \subset \{ (\tau,\chi) : \chi \in [\chi_1,\chi_2], \tau \geq \tau_{\mathscr{D}}(\chi) \} \subset \R^2_{\tau,\chi}$ such that:
		\begin{equation}
			(\tau,\chi) \in \mathscr{Q} \iff \forall \tilde{\tau} \in [\tau_{\mathscr{D}}, \tau], (\tilde{\tau}, \chi) \in \mathscr{Q},
		\end{equation}
		along with functions $r, \rho, m, f \in C^\infty(\mathscr{Q})$ such that $(\mathscr{Q}, r, \rho, m, f)$ is a $C^\infty$ solution of the collapsing Einstein-dust system extending the initial data on $\mathscr{D}$, and $\mathscr{D}$ is an acoustic Cauchy surface for $\mathscr{Q}$.
	\end{definition}
	Note that $\mathscr{Q}$ is defined here as a manifold-with-boundary, where the boundary is: $\mathscr{D} \cup \mathscr{B}_{\chi_1} \cup \mathscr{B}_{\chi_2}$, for
	\begin{equation}
		\mathscr{B}_{\chi_i} := \mathscr{Q} \cap \{ \chi = \chi_i \}.
	\end{equation}
	analogously to the consideration of Penrose diagrams for spherically symmetric solutions of the Einstein equations foliated by a choice of double null coordinates, we can consider the \emph{induced boundary} $\mathscr{B}^+$ of $\mathscr{Q}$ by the ambient $\R^2_{\tau,\chi}$: 
	\begin{equation}
		\mathscr{B}^+ := \partial\mathscr{Q} \setminus \{ \mathscr{D} \cup \mathscr{B}_{\chi_1} \cup \mathscr{B}_{\chi_2} \} \subset \R^{2}_{\tau,\chi},
	\end{equation}
	where $\partial \mathscr{Q} := \overline{\mathscr{Q}} \setminus \mathscr{Q} \subset \R^2_{\tau, \chi}$. We define the \emph{singular boundary} $\mathscr{B}^+_s$ of $\mathscr{Q}$ to be:
	\begin{equation}
		\mathscr{B}^+_s := \{ p \in \mathscr{B}^+ : \lim_{\mathscr{Q} \ni (\tau,\chi) \to p} \rho(\tau,\chi) = \infty \}.
	\end{equation}
	Clearly, $\rho \to \infty$ is a sufficient condition to be a (classical!) singularity.
	\begin{definition}
		We define the subset $\mathscr{S} \subset \mathscr{B}^+_s$ by:
		\begin{equation}
			\mathscr{S} := \{ p \in \mathscr{B}^+_s : \lim_{\mathscr{Q} \ni (\tau,\chi) \to p} r(\tau,\chi) = 0\}.
		\end{equation}
	\end{definition}
	\begin{definition}
		The \underline{shell crossing} subset of the singular boundary, $\mathscr{C} \subset \mathscr{B}_s^+$, is given by
		\begin{equation}
			\mathscr{C} := \{ p \in \mathscr{B}^+_s : \liminf_{\mathscr{Q} \ni (\tau,\chi) \to p} r(\tau,\chi) > 0 \}.
		\end{equation}
	\end{definition}
	\begin{note}
		This definition of the shell crossing singular boundary corresponds exactly to points where $\partial_\chi r \to 0$ with $\liminf r > 0$ (which follows from \eqref{rho}).
	\end{note}
	
	We now consider the \emph{maximal} acoustic development of smooth solutions to the Einstein-dust system, given initial data as in Definition \ref{data}. We term such a development the `maximal dust development' (MDD). In the following proposition, we show that the singular boundary of the MDD is exactly $\mathscr{S} \cup \mathscr{C}$: any first singularity of the collapsing Einstein-dust system (with $\rho > 0$) must have $\rho \to \infty$.
	\begin{proposition} \label{mddprop}
		Given initial data $(\mathscr{D}, r_{\mathscr{D}},m_{\mathscr{D}}, \mathbf{u}(r)_{\mathscr{D}})$, there exists a unique maximal $C^\infty$ collapsing acoustically globally hyperbolic development (Definition \ref{MAGHD}) attaining the data, the \underline{maximal dust development} (MDD): 
		\begin{equation}
			(\mathscr{Q}_{MDD}, r,\rho,m,f)(\mathscr{D}, r_{\mathscr{D}},m_{\mathscr{D}}, \mathbf{u}(r)_{\mathscr{D}}).
		\end{equation}
		\textbf{Decomposition of the singular boundary.} Defining
		\begin{equation} \label{tausdef}
			\tau_s(\chi) := \tau_{\mathcal{D}}(\chi) + \int_0^{r_{\mathcal{D}}(\chi)} \left( \frac{2m(\chi)}{r} + f(\chi) \right)^{-\frac{1}{2}} dr,
		\end{equation}
		then $\tau_s : [\chi_1,\chi_2] \to \R$ is a smooth function. Let
		\begin{equation} 
			\mathscr{S}_{\chi} := \{ \chi \in [\chi_1,\chi_2] : \tau_s(\chi) \geq 0\}, \qquad \mathscr{C}_{\chi} := \{ \chi \in [\chi_1,\chi_2] : \tau_s(\chi) < 0 \},
		\end{equation}
		then there exists a smooth function $\tau_c : \mathscr{C}_{\chi} \to \R$, depending on initial data, with $\tau_c(\chi) < \tau_s(\chi)$ for all $\chi \in \mathscr{C}_\chi$, such that: the function
		\begin{equation} 
			\tau_{MDD}(\chi) := \begin{cases}
				\tau_s(\chi) &: \chi \in \mathscr{S}_{\chi} \\ \tau_c(\chi) &: \chi \in \mathscr{C}_{\chi}
			\end{cases}
		\end{equation}
		is continuous on $[\chi_1,\chi_2]$, and the underlying manifold, $\mathscr{Q}_{MDD}$, is given by:
		\begin{equation}
			\mathscr{Q}_{MDD} = \{ (\tau,\chi) : \tau_{\mathcal{D}}(\chi) \leq \tau < \tau_{MDD}(\chi) , \ \chi \in [\chi_1,\chi_2] \},
		\end{equation}
		with $\mathscr{B}^+ = \mathscr{B}^+_s$, and the singular boundary decomposes as
		\begin{equation}
			\mathscr{B}_s = \{ (\tau_{MDD} (\chi),\chi) : \chi \in [\chi_1,\chi_2] \} = \mathscr{S} \cup \mathscr{C},
		\end{equation}
		where
		\begin{equation}
			\mathscr{S} = \{ (\tau_s(\chi),\chi) : \chi \in \mathscr{S}_{\chi}\}, \qquad \mathscr{C} = \{ (\tau_c(\chi), \chi) : \chi \in \mathscr{C}_{\chi} \}.
		\end{equation}
		\textbf{Structure near $\mathscr{C}$.} If $\mathscr{C}_\chi \neq \emptyset$, then the function $r(\tau,\chi)$ extends smoothly to (a positive function on) $\mathscr{C}$:
		\begin{equation}
			r_c(\chi) := r(\tau_c(\chi),\chi) > 0,
		\end{equation}
		is a smooth function for $\chi \in \mathscr{C}_\chi$, and moreover, $\forall \chi \in \mathscr{C}_{\chi}$:
		\begin{equation} \label{diff-pos}
			\frac{2m'(\chi)}{r_c(\chi)} + f'(\chi) > 0,
		\end{equation}
		so that
		\begin{equation} \label{pchiu}
			\partial_\chi (\mathbf{u}(r))(\tau_c(\chi),\chi) = \frac{1}{2 \mathbf{u}(r)(\tau_c(\chi),\chi)} \left( \frac{2m'(\chi)}{r_c(\chi)} + f'(\chi)\right) < 0,
		\end{equation}
		and we can express:
		\begin{equation} \label{pchirA}
			\partial_\chi r = \frac{1}{2} \frac{\frac{2m'(\chi)}{r_c(\chi)}+ f'(\chi)}{\frac{2m(\chi)}{r} + f(\chi)} (r - r_c(\chi)) \left[ 1 + (r-r_c(\chi)) A(r,\chi) \right],
		\end{equation}
		where, given any compact subset $\mathscr{K}_{\chi} \subset \mathscr{C}_\chi$, $A(r,\chi)$ is smooth for $\{ (r,\chi) : r > 0, \chi \in \mathscr{K}_\chi \}$. 
	\end{proposition}
	
	\begin{remark}
		While the MDD is defined to be acoustically globally hyperbolic, it is not necessarily globally hyperbolic with respect to the causal structure of the spacetime metric $g$. Indeed, if $\mathscr{C}_\chi \neq \emptyset$ then the MDD is \emph{necessarily not} globally hyperbolic. 
	\end{remark}

	\begin{remark}
		The metric \eqref{metric} degenerates at $\mathscr{C}$, where $\partial_\chi r \to 0$.
	\end{remark}
	
	\begin{proof}
		\textbf{Decomposition of the singular boundary.} Integrating \eqref{Edcollapse}, we get $\tau = T(r(\tau,\chi),\chi)$, where
		\begin{equation} \label{Trchi}
			T(r,\chi) = \tau_{\mathcal{D}}(\chi) + \int_{r}^{r_{\mathcal{D}}(\chi)} \left( \frac{2m(\chi)}{\tilde{r}} + f(\chi)\right)^{-\frac{1}{2}}  d\tilde{r},
		\end{equation}
		and $\tau_s(\chi) = T(0,\chi)$, which is clearly smooth since $\frac{2m(\chi)}{r} + f(\chi) > 0$ for $r\leq r_{\mathcal{D}}(\chi)$. By the implicit function, we get $r(\tau,\chi) \in (0,r_{\mathcal{D}}(\chi)]$ a smooth function of $(\tau,\chi)$, for $\tau \in [\tau_{\mathcal{D}}(\chi), \tau_s(\chi))$. Differentiating \eqref{f} with respect to $\chi$, and using \eqref{acc}, we get
		\begin{equation} \label{dtdchir}
			\partial_\tau \left(\frac{\partial_\chi r}{\partial_\tau r}\right) =\frac{1}{2} \cdot \frac{\frac{2m'(\chi)}{r} + f'(\chi)}{\frac{2m(\chi)}{r} + f(\chi)}.
		\end{equation}
		Integrating in $\tau$ from $\tau_\mathscr{D}(\chi)$, we get
		\begin{equation} \label{pchir}
			\partial_\chi r (\tau,\chi) = (\partial_\chi r)_{\mathscr{D}}(\chi) \left( \frac{\frac{2m(\chi)}{r(\tau,\chi)} + f(\chi)}{\frac{2m(\chi)}{r_\mathcal{D}(\chi)} + f(\chi)}\right)^\frac{1}{2} \left( 1 - H(\tau,\chi) \right)
		\end{equation}
		where 
		\begin{equation} \label{H}
			H(\tau,\chi) :=  \frac{1}{2 (\partial_\chi r)_{\mathscr{D}}(\chi)} \left( \frac{2m(\chi)}{r_\mathscr{D}(\chi)} + f(\chi) \right)^{\frac{1}{2}} \int_{r(\tau,\chi)}^{r_\mathscr{D}(\chi)} \frac{\frac{2m'(\chi)}{\tilde{r}} + f'(\chi)}{\left(\frac{2m(\chi)}{\tilde{r}} + f(\chi)\right)^\frac{3}{2}} d\tilde{r},
		\end{equation}
		is smooth for $\tau < \tau_s(\chi)$, with $H(\tau_{\mathscr{D}}(\chi),\chi) = 0$, and $\partial_\chi r (\tau,\chi) = 0 \iff H(\tau,\chi)  = 1$. Differentiating the definition of $\tau_s(\chi)$, \eqref{tausdef}, we get:
		\begin{align} \label{tausp}
			\tau_s'(\chi) &= \tau_{\mathscr{D}}'(\chi) + r_{\mathscr{D}}'(\chi) \left( \frac{2m(\chi)}{r_{\mathscr{D}}(\chi)} + f(\chi) \right)^{-\frac{1}{2}} - \frac{1}{2} \int_0^{r_{\mathscr{D}}(\chi)} \frac{\frac{2m'(\chi)}{r} + f'(\chi)}{\left(\frac{2m(\chi)}{r} + f(\chi)\right)^\frac{3}{2}} dr.
		\end{align}
		Using \eqref{pchirD}, and \eqref{H} with $\tau = \tau_s(\chi)$, we obtain the expression
		\begin{equation} \label{Hs}
			H_s(\chi) := H(\tau_s(\chi),\chi) = 1 - \frac{\left( \frac{2m(\chi)}{r_{\mathscr{D}}(\chi)} + f(\chi) \right)^{\frac{1}{2}}}{(\partial_\chi r)_{\mathscr{D}}(\chi)} \tau_s'(\chi).
		\end{equation}
		Now, suppose that $\exists \tau_c < \tau_s(\chi)$ (so that $\lim_{\tau \to \tau_c} r(\tau,\chi) > 0$) such that $\partial_\chi r(\tau,\chi) > 0$ for $\tau < \tau_c$, and
		\begin{equation}
			\liminf_{\tau \uparrow \tau_c} \partial_\chi r(\tau,\chi) = 0,
		\end{equation}
		which by \eqref{pchir} is equivalent to
		\begin{equation} \label{limsupH}
			\limsup_{\tau\uparrow\tau_c} H(\tau,\chi) = 1.
		\end{equation}
		Differentiating \eqref{H} with respect to $\tau$:
		\begin{equation} \label{ptauH}
			\partial_\tau H(\tau,\chi) =  \frac{1}{2 (\partial_\chi r)_{\mathscr{D}}(\chi)} \left( \frac{2m(\chi)}{r_\mathscr{D}(\chi)} + f(\chi) \right)^{\frac{1}{2}} \frac{\frac{2m'(\chi)}{r(\tau,\chi)} + f'(\chi)}{\frac{2m(\chi)}{r(\tau,\chi)} + f(\chi)},
		\end{equation}
		where we note that
		\begin{equation} \label{monotone}
			\partial_\tau \left( \frac{2m'(\chi)}{r(\tau,\chi)} + f'(\chi) \right) = \frac{2m'(\chi)}{r(\tau,\chi)^2} \left( \frac{2m(\chi)}{r(\tau,\chi)} + f(\chi) \right)^\frac{1}{2} \geq 0.
		\end{equation}
		By the mean value theorem, if $H(\tau_0,\chi)>0$ for anuhiy $\tau_0>\tau_{\mathscr{D}(\chi)}$, then $\exists \tau_1 \in (\tau_{\mathscr{D}(\chi)},\tau_0)$ such that $\partial_\tau H(\tau_1,\chi) > 0$, so by \eqref{ptauH} and \eqref{monotone} we must have $\partial_\tau H(\tau,\chi) > 0 \ \forall \tau \in [\tau_1,\tau_s(\chi))$. Equation \eqref{limsupH} then implies that $H(\tau_c,\chi) = 1$, so we do have $\partial_\chi r (\tau,\chi) \rightarrow 0$ as $\tau \rightarrow \tau_c$. We have also $H(\tau_s,\chi) > 1$, so \eqref{Hs} implies that $\partial_\chi r(\tau,\chi) \to 0$ as $\tau \to \tau_c < \tau_s(\chi)$ if and only if $\tau_s'(\chi) < 0$. We define $\tau_c : \mathscr{C}_\chi \rightarrow \R$ to be the unique solution of $H(\tau_c(\chi),\chi) = 1$. By the above, we have $\partial_\tau H(\tau_c(\chi),\chi) > 0$, so by the implicit function theorem $\tau_c(\chi)$ is smooth. Defining also $r_c(\chi) := r(\tau_c(\chi),\chi)$, $\partial_\tau H(\tau_c(\chi),\chi) > 0$ implies that
		\begin{equation}
			\frac{2m'(\chi)}{r_c(\chi)} + f'(\chi) > 0,
		\end{equation}
		and \eqref{pchiu} follows from differentiating \eqref{Edcollapse} and using $\partial_\chi r(\tau_c(\chi),\chi) = 0$. If $\tau_s'(\chi) \geq 0$, then $H_s(\chi) \leq 1$, so we have $\partial_\chi r(\tau,\chi) > 0$ for $\tau \in [\tau_{\mathscr{D}}(\chi), \tau_s(\chi))$, and $r(\tau,\chi) \to 0$ as $\tau \to \tau_s(\chi)$. The function $\tau_{MDD}(\chi)$ is continuous by continuity of $H_s(\chi)$. Uniqueness of the MDD follows from uniqueness of solutions of the ODEs \eqref{Edcollapse} and \eqref{dtdchir} for the given initial data.
		
		\textbf{Derivation of the expression \eqref{pchirA} with $A(r,\chi)$ smooth.} We integrate \eqref{dtdchir} from $\tilde{r} = r_c(\chi)$:
		\begin{equation}
			\partial_\chi r(\tau,\chi) = \frac{1}{2} \left( \frac{2m(\chi)}{r(\tau,\chi)} + f(\chi) \right)^\frac{1}{2} \int_{r_c(\chi)}^{r(\tau,\chi)} \frac{\frac{2m'(\chi)}{\tilde{r}} + f'(\chi)}{\left( \frac{2m(\chi)}{\tilde{r}} + f(\chi)\right)^\frac{3}{2}} d\tilde{r},
		\end{equation}
		which is smooth for $r(\tau,\chi) > 0$, $\chi \in \mathscr{K}_\chi$. We make the change of variables $\tilde{r} = r_c(\chi) + \tilde{s} (r-r_c(\chi))$, so that
		\begin{equation}
			\partial_\chi r (\tau(r,\chi),\chi) = \frac{1}{2} \left( \frac{2m(\chi)}{r} + f(\chi) \right)^\frac{1}{2} (r-r_c(\chi)) \int_0^1 \frac{\frac{2m'(\chi)}{r_c(\chi) + \tilde{s}(r-r_c(\chi))} + f'(\chi)}{\left( \frac{2m(\chi)}{r_c(\chi) + \tilde{s}(r-r_c(\chi))} + f(\chi)\right)^\frac{3}{2}} d\tilde{s}.
		\end{equation}
		By definition of $\tilde{r}$, we can write
		\begin{equation}
			\frac{2m'(\chi)}{\tilde{r}} = \frac{2m'(\chi)}{r_c(\chi)} - \frac{2m'(\chi)}{r_c(\chi)} \frac{\tilde{s}(r - r_c(\chi))}{r_c(\chi) + \tilde{s}(r - r_c(\chi))},
		\end{equation}
		so that 
		\begin{equation}
			\frac{\frac{2m'}{\tilde{r}} + f'}{\frac{2m'}{\tilde{r_c}} + f'} = 1 - (r-r_c(\chi))A_1(r,\chi;\tilde{s}),
		\end{equation}
		\begin{equation}
			A_1(r,\chi;\tilde{s}) := \frac{\frac{2m'(\chi)}{r_c(\chi)}}{\frac{2m'(\chi)}{r_c(\chi)}+f'(\chi)}\frac{\tilde{s}}{r_c(\chi) + \tilde{s}(r-r_c(\chi))},
		\end{equation}
		where $A_1(r,\chi;s)$ is smooth and well-defined for $\tilde{s}\in[0,1]$. Similarly for 
		\begin{align}
			\left( \frac{2m(\chi)}{\tilde{r}} +f(\chi) \right)^{-\frac{3}{2}} &= \left( \frac{2m(\chi)}{r} +f(\chi) \right)^{-\frac{3}{2}} \left( 1 + (r-r_c(\chi)) A_2(r,\chi;s) \right)^{-\frac{3}{2}}\\  &=: \left( \frac{2m(\chi)}{r} +f(\chi) \right)^{-\frac{3}{2}} (1 - (r-r_c(\chi)) A_3(r,\chi;\tilde{s}))
		\end{align}
		where we've defined
		\begin{equation}
			A_2(r,\chi;s) := \frac{\frac{2m(\chi)}{r}}{\frac{2m(\chi)}{r} +f(\chi)} \frac{1 - \tilde{s}}{r - (1-\tilde{s})(r-r_c(\chi))}.
		\end{equation}
		Defining $y_{-3/2} \in C^\infty((-1,\infty),\R)$ by 
		\begin{equation}
			y_{-3/2}(x) := (1+x)^{-\frac{3}{2}} \frac{3+3x + x^2}{(1+x)^\frac{3}{2} + 1},
		\end{equation}
		so that $(1+x)^{-\frac{3}{2}} = 1 - xy_{-3/2}(x)$, we have
		\begin{equation}
			A_3(r,\chi;s) := -A_2(r,\chi;s) y_{-3/2}((r-r_c)A_2(r,\chi;s)),
		\end{equation}
		 smooth for $\chi \in \mathscr{K}_\chi$, $r > 0$, and $\tilde{s} \in [0,1]$. We finally define
		\begin{equation}
			A(r,\chi) := - \int_0^1 \left(A_1(r,\chi;\tilde{s}) + A_3(r,\chi;\tilde{s})(1 - (r-r_c(\chi))A_1(\chi,r;\tilde{s}))\right) d\tilde{s},
		\end{equation}
		which satisfies the assumption of the proposition.
	\end{proof}

	\subsection{Einstein-dust in double null coordinates (E-d)} \label{dnmdd}

	We now consider the Einstein-dust system in double null coordinates $(u,v)$, which are defined to be solutions of the eikonal equation
	\begin{equation} \label{eikonaleq}
		g^{-1}(du,du) = 0 = g^{-1}(dv,dv),
	\end{equation}
	with $g^{-1}(du,dv) < 0$. In double null coordinates the metric takes the form
	\begin{equation} \label{metric-dn}
		g = -\Omega^2 du dv + r^2 g_{\mathbb{S}^2},
	\end{equation}
	and the Einstein-dust system: $G_{\mu \nu} = 8\pi T_{\mu \nu}$, with $T_{\mu \nu} = \rho \mathbf{u}_\mu \mathbf{u}_\nu$, $g(\mathbf{u},\mathbf{u}) = -1$, is equivalent to the following system of equations:
	\begin{align}
		\partial_u \partial_v r &= \Omega^2 \left( \pi r \rho - \frac{m}{2r^2}\right), \label{rwave} \\
		\partial_u \partial_v \log{\Omega^2} &= \Omega^2 \left(\frac{m}{r^3} - 2 \pi \rho \right), \label{Ruv} \\
		\partial_u \left( -\frac{\partial_u r}{\Omega^2} \right) &= \pi r\rho (\mathbf{u}^v)^2 \Omega^2,  \label{Rayu} \\ \partial_v \left( -\frac{\partial_v r}{\Omega^2} \right) &= \pi r\rho (\mathbf{u}^u)^2 \Omega^2, \label{Rayv}
	\end{align}
	where 
	\begin{equation} \label{fluidvel}
		\Omega^2 \mathbf{u}^u \mathbf{u}^v = 1,
	\end{equation}
	and the Hawking mass $m$ is defined by:
	\begin{equation} \label{m}
		1 - \frac{2m}{r} := g(\nabla r, \nabla r) = - \frac{4\partial_u r \partial_v r}{\Omega^2}.
	\end{equation}
	We refer to the full system \eqref{rwave}-\eqref{Rayv} with \eqref{fluidvel} and \eqref{m} as (E-d). In double null coordinates, mass conservation, \eqref{masscons}, takes the form
	\begin{equation}
		(\mathbf{u}^u \partial_u + \mathbf{u}^v \partial_v)\left( \log((r\Omega)^2 \rho)\right) = - (\partial_u \mathbf{u}^u + \partial_v \mathbf{u}^v),  \label{masscons2}
	\end{equation}
	and the geodedic equation, \eqref{geodesic}, becomes
	\begin{align} 
		\partial_u \left( \frac{1}{\mathbf{u}^v}\right) = \partial_v \left( \frac{1}{\mathbf{u}^u}\right). \label{geo} 
	\end{align}
	We orient the $(u,v)$ coordinates by imposing that $\partial_u, \partial_v$ are future-directed, and $\partial_v$ is `outgoing' while $\partial_u$ is `ingoing'. By $\mathbf{u}^a$ being future-directed, and
	\begin{equation}
		g(\partial_u, \mathbf{u}) = - \Omega^2 \mathbf{u}^v < 0, \qquad g(\partial_v, \mathbf{u}) = - \Omega^2 \mathbf{u}^u < 0, \label{fd}
	\end{equation}
	$\partial_u, \partial_v$ being future-directed is equivalent to $\mathbf{u}^u, \mathbf{u}^v > 0$. Since we assume that $\mathbf{u}_*$ is outward-directed (where $\mathbf{u}_*$ is as defined in Lemma \ref{comovinglemma}), the outgoing/ingoing condition is equivalent to $g(\partial_v, \mathbf{u}_*) > 0$, $g(\partial_u, \mathbf{u}_*) < 0$, and therefore
	\begin{equation} \label{ustar}
		\mathbf{u}_* =  \mathbf{u}^v \partial_v - \mathbf{u}^u \partial_u.
	\end{equation}
	We make the following definitions for solutions of (E-d) that do not include the centre. See Section \ref{smoothdataEd}, in particular Definition \ref{def:E-d-centre} in Section \ref{localex} for the definition of $C^\infty$ solutions of (E-d) including the centre.
	\begin{definition} \label{kcollapse}
		Given $k \in \mathbb{N}$, a $C^k$ solution of (E-d) consists of a domain $\mathcal{Q} \subset \R^{1 + 1}_{u,v}$ and functions $r, m \in C^{k+1}(\mathcal{Q})$, $\Omega^2, \rho \in C^k(\mathcal{Q})$, with $r, \Omega^2, m > 0$, $\rho \geq 0$, along with positive functions $\mathbf{u}^u, \mathbf{u}^v \in C^k(\text{supp}(\rho))$.
		
		The functions are requrired to satisfy \eqref{rwave}--\eqref{m}. In the case $k=1$ we require in addition that \eqref{geo}, \eqref{masscons2} are satisfied and for all $(u,v) \in \mathcal{Q}$: $\partial_u \log\Omega^2(u,v)$ is differentiable in $v$, $\partial_v \log\Omega^2(u,v)$ is differentiable in $u$, and $\partial_v(\partial_u \log\Omega^2(u,v)) = \partial_u(\partial_v \log\Omega^2(u,v))$, so that \eqref{Ruv} is satisfied classically. 
	\end{definition}
	\begin{definition} \label{collapsesol}
		A $C^\infty$ solution of (E-d) consists of a domain $\mathcal{Q} \subset \R^{1 + 1}_{u,v}$ and positive functions $r, m, \Omega^2, \rho, \mathbf{u}^u, \mathbf{u}^v \in C^\infty(\mathcal{Q})$, satisfying \eqref{rwave}--\eqref{m}.
	\end{definition}
	\begin{definition}
		A $C^k$ or $C^\infty$ solution of (E-d) $(\mathcal{Q},r,\Omega^2,\rho,\mathbf{u}^\mu)$ is \underline{collapsing} if $\mathbf{u}(r) < 0, \mathbf{u}_*(r) > 0$ for all $(u,v) \in \text{supp}(\rho)$.
	\end{definition}
	\begin{remark} \label{open-future}
		We assume that the manifolds for all $C^k$, $C^\infty$ solutions of (E-d) are `open to the future': for all $(u,v) \in \mathcal{Q}$, there exists $(\tilde{u},\tilde{v}) \in I^+(u,v)$ such that $J^+(u,v) \cap I^-(\tilde{u},\tilde{v}) \in \mathcal{Q}$.
	\end{remark}
	\begin{remark}
		We will only consider solutions of (E-d) with $\rho > 0$ in this section. Note that given smooth collapsing initial data with $\rho > 0$, any globally hyperbolic development of the data will have $\rho > 0$.
	\end{remark}
	Recalling the discussion in Section \ref{comoving}, we note that the conditions $\mathbf{u}(r) < 0$ on $\mathcal{Q} \setminus \Gamma$ and $\mathbf{u}_*(r) > 0$ on $\mathcal{Q}$ follow from assuming the same bounds on initial data, since $\mathbf{u}(\mathbf{u}(r)) = -\frac{m}{r^2} < 0$, and $\mathbf{u}(\mathbf{u}_*(r)) = 0$. In particular, if initial data is collapsing and untrapped, then we can always choose the orientation of $\mathbf{u}_*$ such that $\mathbf{u}_*(r) > 0$, since
	\begin{equation}
		\mathbf{u}_*(r)^2 = 1 - \frac{2m}{r} + \mathbf{u}(r)^2 > 0,
	\end{equation}
	for $2m < r$. Note that on any $C^k$ collapsing solution of (E-d), we have
	\begin{equation}
		\partial_u r = \frac{1}{2 \mathbf{u}^u} (\mathbf{u}(r) - \mathbf{u}_*(r)) < 0.
	\end{equation}
	
	\subsubsection{The eikonal equation in comoving coordinates} \label{eikonalsection}
	
	Let $(\mathscr{Q},r,\rho,m,f)$ be a solution of the collapsing Einstein-dust system in comoving coordinates. We now construct double null coordinates on (a subset of) $\mathscr{Q}$, and consider the transformation to $C^\infty$ collapsing solutions of (E-d). This will enable us to completely understand the geometry of shell crossing in double null coordinates.
	
	We use the expression for the metric in comoving coordinates \eqref{metric}, along with the conditions \eqref{fd} and \eqref{ustar}, to see that $u(\tau,\chi)$, $v(\tau,\chi)$ solving the eikonal equation \eqref{eikonaleq} is equivalent to:
	\begin{align}
		\partial_\chi u(\tau,\chi) &= - \frac{\partial_\chi r(\tau,\chi)}{(1+f(\chi))^\frac{1}{2}} \partial_\tau u(\tau,\chi) \label{eiku} \\ \partial_\chi v(\tau,\chi) &= \frac{\partial_\chi r(\tau,\chi)}{(1+f(\chi))^\frac{1}{2}} \partial_\tau v(\tau,\chi). \label{eikv}
	\end{align}
	The characteristics of \eqref{eiku} are outgoing null curves: given $\tau_+(\chi)$ satisfying
	\begin{equation} \label{outnull}
		\tau_+'(\chi) = \frac{\partial_\chi r(\tau_+(\chi),\chi)}{(1+f(\chi))^\frac{1}{2}},
	\end{equation}
	then $u(\tau_+(\tilde{\chi}),\tilde{\chi}) = u(\tau_+(\chi),\chi)$. The characteristics of \eqref{eikv} are ingoing null curves: given $\tau_-(\chi)$ satisfying
	\begin{equation} \label{innull}
		\tau_-'(\chi) = - \frac{\partial_\chi r(\tau_-(\chi),\chi)}{(1+f(\chi))^\frac{1}{2}},
	\end{equation}
	then $v(\tau_+(\tilde{\chi}),\tilde{\chi}) = v(\tau_+(\chi),\chi)$. We therefore define the following subsets of $\mathscr{Q} \cup \mathscr{C}$, which will correspond to the domain on which there are unique solutions to the eikonal equations \eqref{eiku}, \eqref{eikv} given data on $\mathscr{D}$.
	\begin{definition} \label{eikdomdef}
		Let $(\mathscr{Q}, r, \rho, m, f)$ be a $C^\infty$ collapsing dust spacetime. For $\mathscr{X}, \mathscr{Y} \subset \mathscr{Q} \cup \mathscr{C}$, we define the unions of the outgoing/ingoing future-directed null curves from $\mathscr{X}$ in $\mathscr{Y}$:
		\begin{align}
			\mathscr{U}_+(\mathscr{X},\mathscr{Y}) := \{ (\tau,\chi) : \exists \chi_0 \leq \chi, \ \tau_+(\tilde{\chi}) \in C^1([\chi_0, \chi]) &\text{ solving } \eqref{outnull}, \text{ with }  (\tau_+(\tilde{\chi}),\tilde{\chi}) \in \mathscr{Y} \text{ for } \tilde{\chi} \in (\chi_0,\chi], \\ &\tau = \tau_+(\chi) \geq \tau_+(\chi_0), \text{ and } (\tau_+(\chi_0),\chi_0) \in \mathscr{X} \}, \\ \mathscr{V}_+(\mathscr{X},\mathscr{Y}) := \{ (\tau,\chi) : \exists  \chi_0 \geq \chi, \ \tau_-(\tilde{\chi}) \in C^1([\chi, \chi_0]) &\text{ solving } \eqref{innull}, \text{ with }  (\tau_-(\tilde{\chi}),\tilde{\chi}) \in \mathscr{Y} \text{ for } \tilde{\chi} \in [\chi,\chi_0), \\ &\tau = \tau_-(\chi) \geq \tau_-(\chi_0), \text{ and } (\tau_-(\chi_0),\chi_0) \in \mathscr{X} \}.
		\end{align}
		The sets $\mathscr{U}_-(\mathscr{X},\mathscr{Y})$, $\mathscr{V}_-(\mathscr{X},\mathscr{Y})$ are defined analogously for past-directed null curves from $\mathscr{D}$. 
	\end{definition}
	\begin{note}
		By definition, we always have $\mathscr{X} \subset \mathscr{U}_\pm(\mathscr{X},\mathscr{Y}), \mathscr{V}_\pm(\mathscr{X},\mathscr{Y})$.
	\end{note}
	\begin{remark}
		Note that we define $\mathscr{X}$, $\mathscr{Y}$ as subsets of $\mathscr{Q} \cup \mathscr{C}$ since $r(\tau,\chi)$ extends smoothly to $\mathscr{C}$, null curves \eqref{outnull}, \eqref{innull} extend smoothly to $\mathscr{C}$ from $\mathscr{Q}$, and we'll show in the following proposition that solutions of the eikonal equation extend smoothly to $\mathscr{C}$.
	\end{remark}
	\begin{definition} \label{nullvec}
		We define also null vector fields on $(\mathscr{Q},r,\rho,m,f)$:
		\begin{equation}
			L := \frac{\partial_\chi r(\tau,\chi)}{(1+f(\chi))^\frac{1}{2}} \partial_\tau + \partial_\chi, \qquad \underline{L} := \frac{\partial_\chi r(\tau,\chi)}{(1+f(\chi))^\frac{1}{2}} \partial_\tau - \partial_\chi.
		\end{equation}
	\end{definition}
	Note that $L, \underline{L}$ are normalised so that they extend smoothly to the shell crossing subset of the singular boundary, $\mathscr{C}$. The following proposition is standard method of characteristics to show that there exist unique $C^k$ solutions to the eikonal equation, with additional regularity in the null directions.
	\begin{proposition} \label{eikprop}
		Let $(\mathscr{Q}, r, \rho, m, f)$ be a $C^\infty$ solution of the collapsing Einstein-dust system with curve $\mathscr{D}$:
		\begin{equation}
			\mathscr{D} := \{ (\tau_{\mathscr{D}}(\chi),\chi) : \chi \in [\chi_1,\chi_2]\} \subset \mathscr{Q},
		\end{equation}
		given by $\tau_{\mathscr{D}} \in C^k([\chi_1,\chi_2])$, for some $k \in \mathbb{N} + 1$, satisfying
		\begin{equation} \label{outbd}
			\tau_{\mathscr{D}}'(\chi) \neq \frac{\partial_\chi r(\tau_{\mathscr{D}}(\chi),\chi)}{(1+f(\chi))^\frac{1}{2}},
		\end{equation}
		and let $u_{\mathscr{D}}(\chi) \in C^k([\chi_1,\chi_2])$, then there exists a unique $u(\tau,\chi) \in C^k(\mathscr{U}_+(\mathscr{D},\mathscr{Q} \cup \mathscr{C}))$ solving \eqref{eiku} and attaining the data on $\mathscr{D}$: $u(\tau_{\mathscr{D}}(\chi),\chi) = u_{\mathscr{D}}(\chi)$. Moreover, for all $1 \leq j \leq k$: $L^m(\partial_\tau^j u) \in C^0(\mathscr{U}_+(\mathscr{D},\mathscr{Q} \cup \mathscr{C}))$ for all $m \in \mathbb{N}$,
		where in particular
		\begin{equation} \label{Lptauu}
			L(\log \partial_\tau u) = (1+f)^{-\frac{1}{2}} \partial_\chi \left[\left( \frac{2m}{r} + f \right)^\frac{1}{2}\right], \qquad (\partial_\tau u) L\left( \frac{\partial_\tau^2 u}{(\partial_\tau u)^2}\right) = (1+f)^{-\frac{1}{2}} \partial_\chi \left( \frac{m}{r^2}\right).
		\end{equation}
		If $\partial_\tau u >0$ on $\mathscr{D}$ then $\partial_\tau u(\tau,\chi) > 0$ for all $(\tau,\chi) \in \mathscr{U}_+(\mathscr{D},\mathscr{Q} \cup \mathscr{C})$. analogously, if 
		\begin{equation} \label{inbd}
			\tau_{\mathscr{D}}'(\chi) \neq -\frac{\partial_\chi r(\tau_{\mathscr{D}}(\chi),\chi)}{(1+f(\chi))^\frac{1}{2}},
		\end{equation}
		and $v_{\mathscr{D}}(\chi) \in C^k([\chi_1,\chi_2])$, then there exists a unique $v(\tau,\chi) \in C^k(\mathscr{V}_+(\mathscr{D},\mathscr{Q} \cup \mathscr{C}))$ solving \eqref{eikv} and satisfying $v(\tau_{\mathscr{D}}(\chi),\chi) = v_{\mathscr{D}}(\chi)$. Moreover, for all $1 \leq j \leq k$: $\underline{L}^m(\partial_\tau^j v) \in C^0(\mathscr{V}_+(\mathscr{D},\mathscr{Q} \cup \mathscr{C}))$ for all $m \in \mathbb{N}$,
		where in particular
		\begin{equation} \label{Lptauv}
			\underline{L}(\log \partial_\tau v) = (1+f)^{-\frac{1}{2}} \partial_\chi \left[\left( \frac{2m}{r} + f \right)^\frac{1}{2}\right], \qquad (\partial_\tau v) \underline{L}\left( \frac{\partial_\tau^2 v}{(\partial_\tau v)^2}\right) = (1+f)^{-\frac{1}{2}} \partial_\chi \left( \frac{m}{r^2}\right).
		\end{equation}
		If $\partial_\tau v > 0$ on $\mathscr{D}$ then $\partial_\tau v(\tau,\chi) > 0$ for all $(\tau,\chi) \in \mathscr{V}_+(\mathscr{D},\mathscr{Q} \cup \mathscr{C})$.
	\end{proposition}
	
	\begin{proof}
		We prove the case of $u(\tau,\chi)$ solving \eqref{eiku}, the proof for \eqref{eikv} is identical up to minus signs. 
		
		Let $F(\tau,\chi) := (1+f(\chi))^{-\frac{1}{2}} \partial_\chi r(\tau,\chi)$, so that $F \in C^\infty(\mathscr{Q} \cup \mathscr{C})$ by Proposition \ref{mddprop}. For each $(\tau,\chi) \in \mathscr{U}_+(\mathscr{D}, \mathscr{Q} \cup \mathscr{C})$, we define the `flow map' $T(\xi; \tau,\chi)$ satisfying
		\begin{equation} \label{flow}
			\partial_\xi T(\xi; \tau,\chi) = F(T(\xi; \tau,\chi), \xi),
		\end{equation}
		and $T(\chi ; \tau,\chi) = \tau$. Standard ODE theory and $F \in C^\infty(\mathscr{Q} \cup \mathscr{C})$ implies that $T(\xi;\tau,\chi)$ is a smooth function for all $(\tau,\chi) \in \mathscr{Q} \cup \mathscr{C}$, and all $\xi$ such that $(T(\xi; \tau,\chi),\xi) \in \mathscr{Q} \cup \mathscr{C}$. Now, for all $(\tau,\chi) \in \mathscr{U}_+(\mathscr{D},\mathscr{Q}\cup\mathscr{C})$, there exists $\chi_0 \in [\chi_1,\chi_2]$ such that 
		\begin{equation} \label{chi0}
			T(\chi_0;\tau,\chi) = \tau_{\mathscr{D}}(\chi_0).
		\end{equation}
		The assumption \eqref{outbd} implies, via the implicit function theorem, that there exists a unique function $\chi_0(\tau,\chi)$ satisfying \eqref{chi0} such that $T(\xi;\tau,\chi) \in \mathscr{U}_+(\mathscr{D},\mathscr{Q} \cup \mathscr{C})$ for all $\xi \in [\chi_0(\tau,\chi), \chi]$, and moreover: $\chi_0(\tau,\chi) \in C^\infty(\mathscr{U}_+(\mathscr{D},\mathscr{Q} \cup \mathscr{C}))$. The unique solution of the eikonal equation given data $u_{\mathscr{D}} \in C^k([\chi_1,\chi_2])$ is then:
		\begin{equation}
			u(\tau,\chi) = u_{\mathscr{D}}(\chi_0(\tau,\chi)) \in C^k(\mathscr{U}_+(\mathscr{D},\mathscr{Q}\cup \mathscr{C})).
		\end{equation}
		Differentiating in $\tau$, and using $u_{\mathscr{D}}'(\chi) = \partial_\tau u(\tau_{\mathscr{D}}(\chi),\chi) (\tau_{\mathscr{D}}'(\chi) - F(\tau_{\mathscr{D}}(\chi),\chi))$, we get
		\begin{equation} \label{ptauu1}
			\partial_\tau u(\tau,\chi) = \partial_\tau u(\tau_{\mathscr{D}}(\chi_0(\tau,\chi)),\chi_0(\tau,\chi)) \cdot \partial_\tau T(\chi_0(\tau,\chi); \tau,\chi).
		\end{equation}
		Noting that $L(\chi_0) = 0$, then $L^m(\partial_\tau u)$ can only depend on $\partial_\tau u(\tau_{\mathscr{D}}(\chi_0(\tau,\chi)),\chi_0(\tau,\chi))$ for all $m \in \mathbb{N}$, and so is continuous on $\mathscr{U}_+(\mathscr{D},\mathscr{Q} \cup \mathscr{C})$, and similarly for all $1\leq j \leq k$: $L^m(\partial_\tau^j u)$ can only depend on $\partial_\tau^l u(\tau_{\mathscr{D}}(\chi_0(\tau,\chi)),\chi_0(\tau,\chi))$ for $1 \leq l \leq j$.
		
		The identities \eqref{Lptauu} follow by direct computation from \eqref{eiku} (or from \eqref{ptauu1} if the data is only assumed to be $C^2$.)
	\end{proof}
	
	We now consider the transformation from comoving to double null coordinates, given a solution of the eikonal equation.
	\begin{lemma} \label{dntransf}
		Given $(\mathscr{Q},r,\rho,m,f)$ a $C^\infty$ solution of the collapsing Einstein-dust system, then for any $k \in \mathbb{N}+1$, and an open subset $\mathscr{W} \subset \mathscr{Q}$, along with $u(\tau,\chi), v(\tau,\chi) \in C^k(\mathscr{W})$ solutions of the eikonal equation \eqref{eiku}, \eqref{eikv} respectively, with $\partial_\tau u$, $\partial_\tau v > 0$, then defining:
		\begin{equation}
			\mathcal{Q} := \{ (u(\tau,\chi),v(\tau,\chi)) : (\tau,\chi) \in \mathscr{W} \},
		\end{equation}
		the map $(u(\tau,\chi),v(\tau,\chi)) : \mathscr{W} \to \mathcal{Q}$ is a bijection, with inverse $(\tau,\chi) \in C^k_{u,v}(\mathcal{Q})^2$. The functions $(\tau(u,v),\chi(u,v))$ satisfy the `inverted' eikonal equation:
		\begin{align}
			\partial_v \chi(u,v) &= \frac{(1+f(\chi(u,v)))^\frac{1}{2}}{\partial_\chi r(\tau(u,v),\chi(u,v))} \partial_v \tau(u,v), \label{inveikv} \\ \partial_u \chi(u,v) &= -\frac{(1+f(\chi(u,v)))^\frac{1}{2}}{\partial_\chi r(\tau(u,v),\chi(u,v))} \partial_u \tau(u,v), \label{inveiku}
		\end{align}
		and the coordinate derivatives in terms of $L, \underline{L}$ (Definition \ref{nullvec}) are:
		\begin{equation} \label{L}
			\partial_v = \frac{(1+f)^\frac{1}{2}}{\partial_\chi r} \frac{L}{2 \partial_\tau v}, \qquad \partial_u = \frac{(1+f)^\frac{1}{2}}{\partial_\chi r} \frac{\underline{L}}{2 \partial_\tau u},
		\end{equation}
		The area radius and energy density are given by
		\begin{equation} \label{rrho}
			r(u,v) = r(\tau(u,v),\chi(u,v)), \qquad \rho(u,v) = \rho(\tau(u,v),\chi(u,v)),
		\end{equation}
		the components of the fluid velocity, and $\Omega^2$, satisfy:
		\begin{equation} \label{velcomp}
			\mathbf{u}^u(u,v) = \partial_\tau u(\tau(u,v),\chi(u,v)), \qquad \mathbf{u}^v(u,v) = \partial_\tau v(\tau(u,v),\chi(u,v)),
		\end{equation}
		\begin{equation} 
			 \Omega^2(u,v) = \left[ \partial_\tau u(\tau(u,v),\chi(u,v)) \partial_\tau v(\tau(u,v),\chi(u,v))\right]^{-1}, \label{omega}
		\end{equation}
		so that:
		\begin{equation}\label{uvtau}
			\partial_u \tau = \frac{1}{2\mathbf{u}^u}, \qquad \partial_v \tau = \frac{1}{2\mathbf{u}^v},
		\end{equation}
		\begin{equation} \label{tauwave}
			\partial_u \partial_v \tau = \frac{\Omega^2}{4 \mathbf{u}(r)} \left(4\pi r \rho - \frac{m}{r^2} + \mathbf{u}_*(\mathbf{u}_*(r))\right).
		\end{equation}
		The first derivatives of $r, \Omega^2, m$ are given by:
		\begin{equation}
			\partial_u r = - \left((1+f)^\frac{1}{2} + \left( \frac{2m}{r} + f \right)^\frac{1}{2} \right) \frac{1}{2\partial_\tau u}, \qquad \partial_v r = \left( (1+f)^\frac{1}{2} - \left( \frac{2m}{r} + f \right)^\frac{1}{2} \right)  \frac{1}{2\partial_\tau v}, \label{pr}
		\end{equation}
		\begin{equation}
			\partial_u \log \Omega^2 = - \frac{\partial_\tau^2 u}{(\partial_\tau u)^2}, \qquad \partial_v \log \Omega^2 = - \frac{\partial_\tau^2 v}{(\partial_\tau v)^2}, \label{pomega}
		\end{equation}
		\begin{equation}
			\partial_u m = - (1+f)^\frac{1}{2} \frac{2\pi r^2 \rho}{\partial_\tau u}, \qquad \partial_v m = (1+f)^\frac{1}{2} \frac{2\pi r^2 \rho}{\partial_\tau v}, \label{pm}
		\end{equation}
		and $(\mathcal{Q},r,\Omega^2,\rho,\mathbf{u}^\mu)$ is a $C^k$ collapsing solution of (E-d).
	\end{lemma}
	
	\begin{remark}
		All derivatives of $r,\Omega^2,\rho,\mathbf{u}^\mu$ can be computed explicitly in terms of the solution of the collapsing Einstein-dust, and the solution of the eikonal equation, using \eqref{L} and \eqref{rrho}, \eqref{velcomp}, \eqref{omega}. 
	\end{remark}
	
	\begin{remark}
		The equality \eqref{tauwave} (which we'll show in the proof follows from \eqref{Lptauu}, \eqref{Lptauv}) is equivalent to mass conservation \eqref{masscons2}, given \eqref{uvtau}. This can be seen from the identity (equivalent to \eqref{masscons2} with \eqref{uvtau})
		\begin{equation}
			\mathbf{u}(\log(r^2\rho)) = - 4 \Omega^2 \partial_u \partial_v \tau, 
		\end{equation}
		along with $\mathbf{u}(r^2 \rho \partial_\chi r) = 0$. The geodesic equation \eqref{geo} is equivalent to $\partial_u \partial_v \tau = \partial_v \partial_u \tau$.
	\end{remark}
	
	\begin{proof}
		Given $u(\tau,\chi), v(\tau,\chi)$, we can locally invert for $(\tau(u,v),\chi(u,v))$ in a neighbourhood of any $(\tilde{\tau},\tilde{\chi}) \in \mathscr{W}$, since the Jacobian determinant $\mathcal{J}(\tilde{\tau},\tilde{\chi})$ is positive on $\mathscr{Q}$:
		\begin{equation}
			\mathcal{J}(\tilde{\tau},\tilde{\chi}) := \begin{vmatrix}
				\partial_\tau u & \partial_\chi u \\ \partial_\tau v & \partial_\chi v 
			\end{vmatrix}(\tilde{\tau},\tilde{\chi}) = \frac{2\partial_\chi r(\tilde{\tau},\tilde{\chi})}{(1+f(\tilde{\chi})^\frac{1}{2})} \partial_\tau u(\tilde{\tau},\tilde{\chi}) \partial_\tau v(\tilde{\tau},\tilde{\chi}) > 0,
		\end{equation}
		so the map is locally invertible, and since $\partial_\tau u, \partial_\tau v, \partial_\chi v > 0$, $\partial_\chi u < 0$, the map is certainly injective (and surjective by definition of $\mathcal{Q}$), hence globally invertible.
		
		The relations \eqref{inveikv}, \eqref{inveiku}, \eqref{L} follow from inverting the Jacobian matrix for $(u(\tau,\chi),v(\tau,\chi))$. The area radius and energy density transform as scalars, and \eqref{velcomp}, \eqref{omega} follow from the the respective coordinate transformations, where \eqref{velcomp} then implies \eqref{uvtau}. The expressions \eqref{pr}, \eqref{pomega}, \eqref{pm} follow from differentiating with \eqref{L}, while \eqref{tauwave} follows from either of \eqref{Lptauu}, \eqref{Lptauv}. 
		
		We have from \eqref{rrho} that for $C^k$ solutions of the eikonal equation we have $r, \rho \in C^k_{u, v}(\mathcal{Q})$, and similarly from \eqref{velcomp},  \eqref{omega} we have $\mathbf{u}^\mu, \Omega^2 \in C^{k-1}_{u, v}(\mathcal{Q})$, while from \eqref{pm} we then see that $m \in C^k_{u, v}(\mathcal{Q})$. 
		
		For $k > 2$, this is sufficient regularity to compute directly that (E-d) is satisfied, while $\mathbf{u}(r)<0$, $\mathbf{u}_*(r) > 0$ so the solution is `collapsing'. For $k=2$, the additional regularity in the respective null directions of the solutions of the eikonal equation implies that $\partial_u \log\Omega^2$, $\partial_v \log\Omega^2$ are continuously differentiable in $v, u$ respectively. Moreover, from \eqref{Lptauu}, \eqref{Lptauv} and \eqref{L}, \eqref{pomega} we see that:
		\begin{equation}
			\partial_u (\partial_v \log \Omega^2) = \partial_v (\partial_u \log \Omega^2) = - \frac{1}{2} \Omega^2 (\partial_\chi r)^{-1} \partial_\chi \left( \frac{m}{r^2}\right) = \Omega^2 \left(\frac{m}{r^3} - 2\pi \rho \right),
		\end{equation}
		and so $(\mathcal{Q},r,\Omega^2,\rho,\mathbf{u}^\mu)$ is a $C^1$ (collapsing) solution of (E-d).
	\end{proof}
	
	\subsubsection{Characteristic initial value problem} \label{CIVP}
	
	In this section, we use the results of the previous sections to show local existence and uniqueness of solutions to the characteristic initial value problem. We consider data on a bifurcate null hypersurface $C_{v_0, v_1} \cup \underline{C}_{u_0, u_1} \subset \R^{1+1}_{u,v}$, defined for $u_0 < u_1, v_0 < v_1$ by:
	\begin{equation}
		C_{v_0, v_1} := \{ u= u_0 \} \cap \{ v_0 \leq v_1 \}, \qquad \underline{C}_{u_0, u_1} := \{ v = v_0 \} \cap \{ u_0 \leq u_1 \}.
	\end{equation}
	We define the notion of a smooth function on a bifurcate null hypersurface:
	\begin{definition}
		A continuous function $f : C_{v_0, v_1} \cup \underline{C}_{u_0, u_1} \to \R$ is \emph{smooth} on the bifurcate null hypersurface, $f \in C^\infty_{u,v}(C_{v_0, v_1} \cup \underline{C}_{u_0, u_1})$, if and only if
		\begin{equation}
			f(u_0,v) \in C^\infty_v([v_0,v_1]), \qquad f(u,v_0) \in C^\infty_u([u_0,u_1]).
		\end{equation}
	\end{definition}
	\begin{definition}
		A $C^\infty$ bifurcate characteristic initial data set on $C_{v_0, v_1} \cup \underline{C}_{u_0, u_1}$ consists of functions $r, \Omega^2, \rho, \mathbf{u}^u \in C^\infty(C_{v_0, v_1} \cup \underline{C}_{u_0, u_1})$, with $r, \Omega^2, \rho, \mathbf{u}^u > 0$, and the data is required to satisfy Raychaudhuri's equations \eqref{Rayu}, \eqref{Rayv}.
	\end{definition}

	\begin{lemma} \label{smoothtransverse}
		For any $C^\infty$ solution of (E-d) extending the data on $C_{v_0, v_1} \cup \underline{C}_{u_0, u_1}$, we can determine all initial transverse derivatives directly from the data on $C_{v_0, v_1} \cup \underline{C}_{u_0, u_1}$.
	\end{lemma}
	
	\begin{proof}
		We show that we can determine all $v$ derivatives of $r,\Omega^2, \rho, \mathbf{u}^\mu$ on $\underline{C}_{u_0, u_1}$ by integrating the evolution equations. The procedure for determining $u$ derivatives on $C_{v_0,v_1}$ is identical. Using the definition of the Hawking mass \eqref{m}, we can write the wave equation for $r$, \eqref{rwave}, as:
		\begin{equation} \label{rwave1}
			\partial_u (r \partial_v r) = \Omega^2 \left( \pi r^2 \rho - \frac{1}{4} \right),
		\end{equation}
		so integrating from $(u_0,v_0)$ along $\underline{C}_{u_0, u_1}$, we get $\partial_v r(u,v_0)$. We obtain $m(u,v_0)$, so we can integrate \eqref{Ruv} for $\partial_v \log \Omega^2(u,v_0)$. Now, for any smooth solution, $\rho, \mathbf{u}^u$ solve the dust equations of motion \eqref{geo} and \eqref{masscons2}: we can read off $\partial_v \mathbf{u}^u(u,v_0)$ from \eqref{geo}, given first derivatives of $\Omega^2$ along with $\partial_u \mathbf{u}^u(u,v_0)$, and $\partial_v \rho(u,v_0)$ from \eqref{masscons2}. We can now differentiate \eqref{rwave1} with respect to $v$, integrate the ODE for $\partial_v^2 r$ (or equivalently read off $\partial_v^2 r$ from Raychaudhuri's equation). The procedure then iterates: differentiating \eqref{rwave} $(k-1)$ times, we determine $\partial_v^k r(u,v_0)$ in terms of $\leq (k-1)$ derivatives of $r, \Omega, \rho$, then (differentiate in $v$ and) integrate \eqref{Ruv} for $\partial_v^k\log\Omega^2(u,v_0)$. We then differentiate \eqref{geo} and read off $\partial_v^{k} \mathbf{u}^u(u,v_0)$ in terms of lower derivatives, and differentiate \eqref{masscons2} to read off $\partial_v^k \rho(u,v_0)$ in terms of $\partial_v^{k} \mathbf{u}^u(u,v_0)$ and lower derivatives. 
	\end{proof}
	We can now make the following definition for initial data data sets, corresponding to the notion of a `collapsing' solution of (E-d).
	\begin{definition}
		A $C^\infty$ bifurcate characteristic data set is `collapsing' if $\mathbf{u}(r) < 0$, $\mathbf{u}_*(r) > 0$.
	\end{definition}
	By defining comoving coordinates and solving the eikonal equation for $u(\tau,\chi)$, $v(\tau,\chi)$, we have local existence and uniqueness of $C^\infty$ solutions. 
	\begin{proposition} \label{CIVPprop}
		Given $C^\infty$ collapsing data on a bifurcate characteristic hypersurface, there exists a globally hyperbolic domain $\mathcal{Q} \subset \R^{1+1}_{u, v}$ with past Cauchy surface $C_{v_0, v_1} \cup \underline{C}_{u_0, u_1}$, such that there exists a unique $C^\infty$ solution on $\mathcal{Q}$ extending the data on $C_{v_0, v_1} \cup \underline{C}_{u_0, u_1}$.
	\end{proposition} 

	\begin{proof}
		We can define comoving coordinates direcly on $C_{v_0, v_1} \cup \underline{C}_{u_0, u_1}$. For any $C^\infty$ solution of (E-d), we have \eqref{uvtau}, so we can integrate for $\tau(u,v_0)$, $\tau(u_0,v)$, given a gauge choice, i.e.~$\tau(u_0,v_0) = 0$. Now, $\rho > 0$ on $C_{v_0, v_1} \cup \underline{C}_{u_0, u_1}$, so we can choose $\chi = m$, which satisfies
		\begin{equation}
			\partial_u m = - \frac{2\pi r^2 \rho}{\mathbf{u}^u} \mathbf{u}_*(r) <0, \qquad \partial_v m = \frac{2\pi r^2 \rho}{\mathbf{u}^v} \mathbf{u}_*(r) > 0. \label{dn:pm}
		\end{equation}
		Using this construction of comoving coordinates, we define
		\begin{equation}
			\mathscr{D}_+ := \{ (\tau(u_0,v), m(u_0,v)) : v \in [v_0,v_1] \}, \ \mathscr{D}_- := \{ (\tau(u,v_0), m(u,v_0)) : u \in [u_0,u_1] \}.
		\end{equation}
		The smooth data defined on $C_{v_0, v_1} \cup \underline{C}_{u_0, u_1}$ induces smooth data in comoving coordinates on $\mathscr{D}_\pm$. We now show that, given any separate developments of the induced data on $\mathscr{D}_\pm$, the union is smooth across $\{ m = m(u_0,v_0)\}$. Clearly $m \in C^\infty(C_{v_0, v_1} \cup \underline{C}_{u_0, u_1})$ by Lemma \ref{smoothtransverse}, and so is $f := \mathbf{u}_*(r)^2 - 1$. Therefore all $\tau$ derivatives of $r(\tau,m)$ are continuous, it is sufficient to show that $\partial_\chi^k r(\tau,m) \in C^0(C^\infty(C_{v_0, v_1} \cup \underline{C}_{u_0, u_1}))$ for all $k \in \mathbb{N}$. For the gauge choice $\chi = m$, we have $\partial_m r = (4\pi r^2 \rho)^{-1}$, and therefore
		\begin{equation}
			\partial_m = \frac{\mathbf{u}_*}{4\pi r^2 \rho \mathbf{u}_*(r)},
		\end{equation}
		so we need $(\mathbf{u}_*)^k(r), (\mathbf{u}_*)^k(\rho) \in C^0(C_{v_0, v_1} \cup \underline{C}_{u_0, u_1})$ for all $k \in \mathbb{N}$, which follows from Lemma \ref{smoothtransverse}. 
		
		Now, we have smooth induced data for the eikonal equation on the respective characteristics: $u_{\mathscr{D}^-}(\chi)$, $v_{\mathscr{D}^+}(\chi)$, for which there is a unique smooth solution on $\mathscr{Q}(\mathscr{D}_+) \cup \mathscr{Q}(\mathscr{D}_-)$, and therefore by the proof of Lemma \ref{dntransf}, we construct a $C^\infty$ collapsing solution of (E-d) extending the data on $C_{v_0, v_1} \cup \underline{C}_{u_0, u_1}$.
		
		For uniqueness, suppose that $\mathcal{Q}$ is a globally hyperbolic domain with past Cauchy surface $C_{v_0, v_1} \cup \underline{C}_{u_0, u_1}$, and there exists a $C^\infty$ collapsing solution (as in Definition \ref{collapsesol}) on $\mathcal{Q}$, extending the bifurcate characteristic initial data. Now, $\mathcal{Q}$ is simply connected, so we can use the argument from Section \ref{comoving} to construct $\tau$: the one-form $\beta_\mu := - g_{\mu \nu} \mathbf{u}^\nu$ satisfies $d\beta = 0$, so there exists a smooth function $\tau : \mathcal{Q} \to \R$, where we gauge fix $\tau(u_0,v_0) = 0$. In $(\tau,m)$ coordinates, the solution must embed into an acoustic globally hyperbolic development of $\mathscr{D}^- \cup \mathscr{D}^+$, which is uniquely determined, and then $(u,v)$ are uniquely determined as solutions of the eikonal equation given the induced data on $\mathscr{D}^- \cup \mathscr{D}^+$.
	\end{proof}
	
	\begin{remark} \label{bifurcatedata}
		Given a bifurcate characteristic hypersurface $C_{v_0, v_1} \cup \underline{C}_{u_0, u_1}$ in a $C^\infty$ collapsing solution of (E-d), with a smooth choice of comoving coordinates ($\tau(u,v), \chi(u,v))$, and
		\begin{equation}
			\mathscr{D}_+ := \{ (\tau(u_0,v), \chi(u_0,v)) : v \in [v_0,v_1] \}, \qquad \mathscr{D}_- := \{ (\tau(u,v_0), \chi(u,v_0)) : u \in [u_0,u_1] \},
		\end{equation}
		then by Proposition \ref{CIVPprop} we can consider the induced data on $\mathscr{D} := \mathscr{D}^+ \cup \mathscr{D}^-$: $(\mathscr{D}, r_{\mathscr{D}}, m_{\mathscr{D}}, \mathbf{u}(r)_{\mathscr{D}})$ as smooth initial data for Einstein-dust in comoving coordinates (i.e.~extending Definition \ref{data} to including a piecewise $C^\infty$ initial curve). We can therefore use the definition of the maximal dust development of Proposition \ref{mddprop} for such induced data.
	\end{remark}
	
	\subsubsection{Extension principle away from the centre} \label{sec:extp}
	
	Using the understanding of the eikonal equation from Section \ref{eikonalsection}, we can use Proposition \ref{mddprop} to completely understand the formation of singularities to (E-d) away from the centre, in the interior of the support of the dust. We show in the following proposition that any first singularity away from the centre, with $\inf r > 0$, must have $\sup \rho = \infty$. 
	
	We first introduce some notation. By definition of $C^\infty$ (and $C^k$) solutions of (E-d), we have $\mathcal{Q} \subset \R^{1+1}_{u,v}$, while the metric $g$ given by \eqref{metric-dn} is conformal to the Minkowski metric $\eta_{\mu \nu}$, so causality in $(\mathcal{Q},g)$ coincides with the causality of the ambient $(\R^{1+1}_{u,v},\eta_{\mu \nu})$, where $\eta_{\mu \nu}$ is the flat Minkowski metric. We accordingly define the causal/chronological future/past ($J^\pm, I^\pm$ respectively) as subsets of $\R^{1+1}_{u,v}$:
	\begin{definition}
		For any $(u_0,v_0) \in \R^{1+1}_{u,v}$, we define:
		\begin{align}
			&J^+(u_0,v_0) = \{ (u,v) \in \R^{1+1}_{u,v} : u  \geq u_0, v \geq v_0\}, \qquad J^-(u_0,v_0) = \{ (u,v) \in \R^{1+1}_{u,v} : u  \leq u_0, v \leq v_0\}, \\ &I^+(u_0,v_0) = \{ (u,v) \in \R^{1+1}_{u,v} : u  > u_0, v > v_0\}, \qquad I^-(u_0,v_0) = \{ (u,v) \in \R^{1+1}_{u,v} : u  < u_0, v < v_0\}.
		\end{align}
	\end{definition}
	We can now state the extension principle for $C^\infty$ collapsing solutions of (E-d) away from the centre:
	\begin{proposition} \label{extensionprinciple}
		Let $(\mathcal{Q},r,\Omega^2,\rho,\mathbf{u}^u)$ be a globally hyperbolic $C^\infty$ collapsing solution of (E-d). Suppose that there exists $(u_1,v_1) \in \R^{1+1}_{u,v}$, and $(u_0,v_0) \in I^-(u_1,v_1)$, such that for:
		\begin{equation}
			\mathcal{R} := ( [u_0, u_1] \times [v_0, v_1] ) \setminus \{ (u_1,v_1)\},
		\end{equation}
		we have $\mathcal{R} \subset \mathcal{Q} \setminus \Gamma$, with $\sup_{(u,v) \in \mathcal{R}} \rho(u,v) < \infty$, and $\inf_{(u,v) \in \mathcal{R}} r(u,v) > 0$; then the solution extends smoothly to a neighbourhood of $(u_1,v_1) \in \mathcal{Q}$.
	\end{proposition}
	\begin{proof}
		The domain $\mathcal{R}$ is simply connected and orientable, so we can define comoving coordinates $(\tau(u,v),m(u,v))$ by Lemma \ref{comovinglemma}, where $m(u,v)$ is the Hawking mass. Let $(\mathscr{Q}_{MDD},r,\rho,m,f)$ be the maximal dust development of Proposition \ref{mddprop} for the induced data in comoving coordinates on $C_{v_0,v_1} \cup \underline{C}_{u_0,u_1}$. Defining $\mathscr{R} := \{ (\tau(u,v),\chi(u,v)) : (u,v) \in \mathcal{R} \}$, we must have $\mathscr{R} \subset \mathscr{Q}_{MDD}$. 
		
		We first observe that $\rho \sim 1$ on $\mathcal{R}$. Indeed, one can see from Proposition \ref{mddprop} that we must have $\inf_{(\tau,\chi) \in \overline{\mathscr{R}}} \rho(\tau,\chi) > 0$, and therefore $\inf_{(u,v) \in \mathcal{R}} \rho(u,v) > 0$ (where we're using the implicit assumption that $\rho > 0$ for $C^\infty$ solutions of (E-d) away from the centre). We have $r \sim 1$ by assumption, which implies also that $\mathbf{u}(r) \sim -1$ (since the solution is `collapsing').
		
		Given any $(\tilde{u},\tilde{v}) \in \mathcal{R}$, we integrate \eqref{Lptauu} on $\{ u = \tilde{u} \}$, for $v \in [v_0,\tilde{v}]$, where we parametrise the integral in terms of $m(\tilde{u},v)$. Letting $v_{\tilde{u}}(m)$ be the inverse of $m(\tilde{u},v)$ for fixed $\tilde{u}$, with $v_{\tilde{u}}'(m) = [\partial_v m(\tilde{u}, v_{\tilde{u}}(m))]^{-1} > 0$, then: 
		\begin{equation} \label{int-eik}
			\mathbf{u}^u(\tilde{u},\tilde{v}) = \mathbf{u}^u(\tilde{u},v_0) \exp\left[ \int_{m(\tilde{u},v_0)}^{m(\tilde{u},\tilde{v})}  \frac{r(\tilde{u}, v_{\tilde{u}}(\tilde{m}))^{-1} + \frac{1}{2}f'(\tilde{m}) - \tilde{m} \cdot [8\pi(r^4 \rho)(\tilde{u}, v_{\tilde{u}}(\tilde{m}))]^{-1} }{(1 + f(\tilde{m}))^{\frac{1}{2}} \left( \frac{2\tilde{m}}{r(\tilde{u}, v_{\tilde{u}}(\tilde{m}))} + f(\tilde{m})\right)^{\frac{1}{2}} } d\tilde{m} \right],
		\end{equation}
		where we've used \eqref{velcomp} to equate $\mathbf{u}^u(\tilde{u},\tilde{v}) = \partial_\tau u(\tilde{u},\tilde{v}),m(\tilde{u},\tilde{v}))$. We have already $r \sim 1$, $\rho \sim 1$, and monotonicity of $m(u,v)$: $\partial_u m(u,v) < 0$, $\partial_v m(u,v) > 0$, implies that $m \sim 1$ on $\mathcal{R}'$, while we have $f(m) \in C^\infty(m(v_0,u_1),m(u_0,v_1))$, so $1 + f(m), |f'(m)| \sim 1$. We therefore have $\mathbf{u}^u(\tilde{u},\tilde{v}) \sim \mathbf{u}^u(\tilde{u},v_0) \sim 1$, and we can analogously integrate \eqref{Lptauv} along $\{ v=\tilde{v}\}$ to see that $\mathbf{u}^v \sim 1$. From \eqref{uvtau}, we have $\partial_u \tau, \partial_v \tau \sim 1$, so $\tau(u,v)$ extends continuously to $(u_1,v_1)$, and $\rho \sim 1$ implies that $m(u,v)$ also extends continuously, and $(\tau_1,m_1) := (\tau(u_1,v_1),m(u_1,v_1)) \in \mathscr{Q}_{MDD}$. The solution of the Einstein-dust system in comoving coordinates therefore extends smoothly to a neighbourhood of $(\tau_1,m_1)$, and moreover extends smoothly (and uniquely) to a `future' neighbourhood of $\overline{\mathscr{R}}$. By Remark \ref{open-future}, we can choose $\delta>0$ such that $C_{v_0,v_1 + \delta} \cup \underline{C}_{u_0, u_1 + \delta} \subset \mathcal{Q}$, and existence/uniqueness of $C^\infty$ solutions of the eikonal equation, Proposition \ref{eikprop}, along with Lemma \ref{dntransf}, imply that $[u_0, u_1 + \delta] \times [v_0, v_1 + \delta] \subset \mathcal{Q}$ for $\delta > 0$ sufficiently small.
	\end{proof}
	
	\subsubsection{First shell crossing singularities and non-degeneracy} \label{firstscsec}
	
	Any first singularity with $\inf r > 0$ must have $\sup \rho = \infty$, which we refer to as a first \emph{shell crossing} singularity:
	\begin{definition} \label{firstscsing}
		Let $(\mathcal{Q},r,\Omega^2,\rho,\mathbf{u}^u)$ be a $C^\infty$ collapsing solution of (E-d). A point $(u_c,v_c) \in \R^{1+1}_{u,v}$ is a \underline{first shell crossing singularity} if there exists $(u_0,v_0) \in I^-((u_c,v_c))$ such that
		\begin{equation}
			\mathcal{R} := ([u_0,u_c] \times [v_0,v_c]) \setminus \{ (u_c,v_c) \} \subset \mathcal{Q},
		\end{equation}
		with $\sup_{(u,v) \in \mathcal{R}} \rho(u,v) = \infty$ and $\inf_{(u,v) \in \mathcal{R}} r(u,v) > 0$.
	\end{definition}
	
	\begin{lemma} \label{firstscprop}
		Let $(\mathcal{Q},r,\Omega^2,\rho,\mathbf{u}^\mu)$ be a $C^\infty$ solution of (E-d), and suppose that $(u_c,v_c) \in \R^{1+1}_{u,v}$ is a first shell crossing singularity. Then $r,\Omega^2$ extend in $C^{0,1}$ to $(u_c,v_c)$, and $\mathbf{u}^\mu, m \sim 1$ on $\mathcal{R}$, with
		\begin{equation} \label{mlimit}
			\lim_{v \to v_c} m(u_c,v) \leq \lim_{u \to u_c} m(u,v_c),
		\end{equation}
		and $m$ extends continuously to $(u_c,v_c)$ if and only if $\lim_{v \to v_c} m(u_c,v) = \lim_{u \to _1} m(u,v_1)$, while $\rho^{-1}$ extends continuously to zero at $(u_c,v_c)$, with
		\begin{equation}
			\int_{u_0}^{u_c} \rho(u,v_c) du < \infty, \qquad \int_{v_0}^{v_c} \rho(u_c,v) dv < \infty.
		\end{equation}
	\end{lemma}
	\begin{proof}
		We proceed analogously to the proof of Proposition \ref{extensionprinciple}: we define $(\tau(u,v),m(u,v))$ comoving coordinates on $\mathcal{R}$ using Lemma \ref{comovinglemma}, and we have $r, m, -\mathbf{u}(r), \mathbf{u}_*(r) \sim 1$, $\rho \gtrsim 1$, while \eqref{mlimit} follows from $\partial_v m > 0$, $\partial_u m <0$. In addition, using $\rho \gtrsim 1$, $r \sim 1$, \eqref{int-eik} implies that $\mathbf{u}^u \sim 1$, and the analogous equation for $\mathbf{u}^v$ implies that $\mathbf{u}^v \sim 1$. Integrating \eqref{dn:pm} along $\{v=v_c\}$, we see that:
		\begin{equation}
			\int_{u_0}^{u_c} \rho(u,v_c) du = \int_{u_0}^{u_c} \frac{ (-\mathbf{u}^u  \partial_u m)}{2\pi r^2 \mathbf{u}_*(r)} du \lesssim m(u_0,v_c) - m(v_c,u_0) < \infty,
		\end{equation}
		and we get $\int_{v_0}^{v_c} \rho(v,u_c) dv < \infty$ analogously. From $\mathbf{u}^u, \mathbf{u}^v \sim 1$ and \eqref{uvtau} we get $\tau(u,v)$ extending in $C_{u,v}^{0,1}$ to $(u_c,v_c)$, where we define $\tau_* := \tau(u_c,v_c)$, and \eqref{pr} implies that $r$ extends to $C^{0,1}$ to $(u_c,v_c)$. Integrating the second equation in \eqref{Lptauu}, \eqref{Lptauv} (equivalent to \eqref{Ruv}) along the respective null curves, and using \eqref{pomega}, we see that $\partial_u \log \Omega^2, \partial_v \log \Omega^2 \sim 1$, so $\Omega^2$ extends in $C^{0,1}$ to $(u_c,v_c)$.
		
		Let $(\mathscr{Q}_{MDD},r,\rho,m,f)$ be the maximal dust development of the induced data in comoving coordinates on $C_{v_0,v_c} \cup \underline{C}_{u_0,u_c}$. We have $\mathscr{R} := \{ (\tau(u,v),m(u,v)) : (u,v) \in \mathcal{R} \} \subset \mathscr{Q}_{MDD}$, and $\inf_{(\tau,\chi) \in \mathscr{R}} r(\tau,\chi) > 0$ implies that $\overline{\mathscr{R}} \subset \mathscr{Q}_{MDD} \cup \mathscr{C}$. By Proposition \ref{mddprop}, we have $r(\tau,\chi) \in C^\infty(\mathscr{Q}_{MDD} \cup \mathscr{C})$. If $m_c^- := \lim_{v \to v_1} m(u_1,v) < \lim_{u \to _1} m(u,v_1) =: m_c^+$, then: $\overline{\mathscr{R}} \setminus \mathscr{R} = \{ (\tau_*, m) : m \in [m_c^-,m_c^+] \} \subset \mathscr{Q}_{MDD} \cup \mathscr{C}$. By continuity of $r(u,v)$ on $\overline{\mathcal{R}}$: $r(\tilde{\tau},m_c^+) - r(\tilde{\tau},m_c^-) = \int_{m_c^-}^{m_c^+} \partial_m r(\tilde{\tau}, m) dm \to 0$ as $\tilde{\tau} \to \tau_*$, so we must have $\partial_m r(\tau_*,m) = 0$ for $m \in [m_c^-,m_c^+]$ so $\overline{\mathscr{R}} \setminus \mathscr{R} \subset \mathscr{C}$ and $\rho^{-1}$ extends continuously to zero at $(u_c,v_c)$.
		
		If $m_c^- = m_c^+ =: m_*$, then $\sup_{(u,v) \in \mathcal{R}} \rho(u,v) = \infty$ implies that we must have $\partial_m r(\tau_*,m_*) = 0$ and therefore $\rho^{-1}$ extends continuously to zero.
	\end{proof}
 	 \begin{remark}
 	 	In the `degenerate' case studied by Papapetrou and Hamoui \cite{PH}, the function $\tau_c(m)$ is a constant on an interval, and the metric components are only $C^{0,1}$ at the first shell crossing singularity. 
 	 \end{remark}
 	 We avoid such \emph{degenerate} cases:
	\begin{definition} \label{nondegensc}
		A first shell crossing singularity is \underline{non-degenerate} if there exists $\beta < 1$ such that
		\begin{equation}
			\rho(u,v_c) \lesssim (u_c - u)^{-\beta}, \qquad \rho(u_c,v) \lesssim (v_c - v)^{-\beta},
		\end{equation}
		as $u \to u_c$, $v \to v_c$ respectively. 
	\end{definition}
	
	\begin{remark}
		We'll see in Section \ref{genericsection} that suitably defined generic smooth data in comoving coordinates satisfies this non-degeneracy condition at any first shell crossing singularity (and in fact $\beta = \frac{2}{3}$ generically).
	\end{remark}
	
	The importance of the non-degeneracy condition is that it characterises the structure of the function $\tau_c(m)$ describing the singular boundary due to shell crossing in comoving coordinates: $\tau_c(m)$ must be a local minima with a non-vanishing derivative:
	\begin{lemma} \label{nondeglemma}
		Let $(\mathcal{Q},r,\Omega^2,\rho,\mathbf{u}^\mu)$ be a $C^\infty$ collapsing solution of (E-d) with a non-degenerate first shell crossing singularity $(u_c,v_c) \in \R^{1+1}_{u,v}$, with $(u_0,v_0) \in I^-(u_c,v_c)$ such that:
		\begin{equation}
			\mathcal{R} := ([u_0,u_c] \times [v_0,v_c]) \setminus \{ (u_c,v_c) \} \subset \mathcal{Q},
		\end{equation}
		then by Lemma \ref{comovinglemma}, there exist comoving coordinates $(\tau(u,v), m(u,v))$ on $\mathcal{R}$ (where $m$ is the Hawking mass). Let $(\mathscr{Q}_{MDD},r,\rho,m,f)$ be the maximal dust development (of Proposition \ref{mddprop}) for the induced data on 
		\begin{equation}
			\mathscr{D} := \{ (\tau(u,v),m(u,v)) : (u,v) \in C_{v_0,v_c} \cup \underline{C}_{u_0,u_c} \},
		\end{equation}
		then $m(u,v) \in C^0(\overline{\mathcal{R}})$, and letting $m_* := m(u_c,v_c)$, there exists $\delta > 0$ such that the function $\tau_c(m)$ has a local minima at $m_* := m(u_c,v_c)$: $[m_* - \delta, m_* + \delta]\subset \mathscr{C}_\chi$; $\tau_c'(m) < 0$ for $m \in [m_* - \delta, m_*)$, $\tau_c'(m) > 0$ for $m \in (m_*, m_* +\delta]$ and there exists $n \in \mathbb{N}$ such that $\tau_c^{(2n)}(m_*) > 0$, and $\tau_c^{(k)}(m_*) = 0$ for $1 \leq k < 2n$.
		
		Noting that $r_c'(m) = - \left( \frac{2m}{r_c(m)} + f(m)\right)^\frac{1}{2} \tau_c'(m)$, we have $r_c^{(k)}(m_*) = 0$ for $1 \leq k < 2n$ and $r_c^{(2n)}(m_*) < 0$. 
	\end{lemma}
	\begin{proof}
		We have $\partial_v m = \frac{2\pi r^2 \mathbf{u}_*(r)}{\mathbf{u}^v} \cdot \rho \sim \rho$, where we've used $\mathbf{u}^v, r, \mathbf{u}_*(r) \sim 1$ on $\mathcal{R}$. Let $m_c^- := \lim_{v \to v_c} m(u_c, v), \ m_c^+ := \lim_{u \to u_c} m(u, v_c)$. Given $\beta < 1$ such that $\rho \lesssim (v_c - v)^{-\beta}$, by the assumption of non-degeneracy, then we have $\partial_v m \lesssim (v_c - v)^{-\beta}$, and integrating from $v_c$: $m_c^- - m(u_c,v) \lesssim (v_c - v)^{1-\beta}$. By Proposition \ref{mddprop}, we have $\partial_\tau(\partial_m r) \sim -1$, so $\partial_m r \sim \tau_c(m) - \tau$. Using $\rho^{-1} \gtrsim (v_c - v)^{\beta}$, we get
		\begin{equation}
			\tau_c(m(u_c,v)) - \tau(u_c,v) \gtrsim (v_c - v)^\beta .
		\end{equation}
		Now, we have $\partial_v \tau = \frac{1}{2\mathbf{u}^v} \sim 1$, so $\tau(u_c,v) - \tau(u_c,v_c) \sim - (v_c - v)$, and therefore
		\begin{equation} \label{minbd}
			\tau_c(m(u_c,v)) - \tau(u_c,v_c) \gtrsim (v_c - v)^\beta\left[ 1 - (v_c - v)^{1-\beta} \right] \sim (v_c - v)^\beta \gtrsim (m_c^- - m(u_c,v))^\frac{\beta}{1-\beta}.
		\end{equation}
		Now, $\tau_c(m)$ is a smooth function on $\mathscr{C}_{\chi} \subset [m_c^-,m_c^+]$, so \eqref{minbd} implies that there exists $k \in \mathbb{N}$ such that $\tau_c^{(k)}(m_c^-) > 0$, therefore we must have $m_c^- = m_c^+ =: m_*$, so $m(u,v)$ extends continuously to $(u_c,v_c)$. Doing the same analysis as above on the ingoing null curve $\{ v= v_c\}$, we get
		\begin{equation} \label{minbd2}
			\tau_c(m(u,v_c)) - \tau(u_c,v_c) \gtrsim (v_c - v)^\beta\left[ 1 - (v_c - v)^{1-\beta} \right] \sim (v_c - v)^\beta \gtrsim (m_* - m(u,v_c))^\frac{\beta}{1-\beta},
		\end{equation}
		so $\tau_c(m)$ has a local minima at $m_*$, and since $\tau_c(m)$ is smooth, there must exist $n \leq \frac{\beta}{2(1-\beta)}$ such that $\tau_c^{(2n)}(m_*) > 0$, and $\tau_c^{(k)}(m_*) = 0$ for $1 \leq k < 2n$.
	\end{proof}

	\subsubsection{Regularity of metric components at a non-degenerate first shell crossing singularity} \label{regsc}
	
	We can now prove the fundamental proposition for the structure of non-degenerate first shell crossing singularities, which will be essential for the proof of Theorem \ref{thm:1}.
	\begin{proposition} \label{regext}
		Let $(\mathcal{Q},r,\Omega^2,\rho,\mathbf{u}^\mu)$ be a $C^\infty$ collapsing solution of (E-d), and suppose that $(u_c,v_c) \in\R$ is a non-degenerate first shell crossing singularity, with $(u_0,v_0) \in I^-(u_c,v_c)$ such that:
		\begin{equation}
			\mathcal{R} := ([u_0,u_c] \times [v_0,v_c]) \setminus \{ (u_c,v_c) \} \subset \mathcal{Q},
		\end{equation}
		then the metric components satisfy $r,\Omega^2 \in C^1(\overline{\mathcal{R}})$ and $\mathbf{u}^u,\mathbf{u}^v \in C^0(\overline{\mathcal{R}})$, $\log (r\Omega)^2 \in C^2(\overline{\mathcal{R}})$. Moreover, given $\delta' > 0$ such that $\{ u= u_0, v \in [v_0, v_c + \delta'] \}, \{ v = v_0, u \in [u_0, u_c + \delta'] \} \subset \mathcal{Q}$, then there exists $0 < \delta \leq \delta'$ such that: ($(\mathcal{Q}, r, \Omega^2, \rho, \mathbf{u}^u)$ can be uniquely extended to include)
		\begin{equation}
			\mathcal{R}' := \mathcal{R} \cup ([u_0,u_c) \times [v_c, v_c + \delta]) \cup ([v_0,v_c) \times [u_c, u_c + \delta]) \subset \mathcal{Q},
		\end{equation}
		and the functions $r, \Omega^2, \rho, \mathbf{u}^\mu$ extend smoothly to $\overline{\mathcal{R}'} \setminus \{ (u_c,v_c)\}$:
		\begin{equation}
			r, \Omega^2, \rho, \mathbf{u}^\mu \in C^\infty_{u,v}\left(\overline{\mathcal{R}'} \setminus \{ (u_c,v_c)\} \right).
		\end{equation}
	\end{proposition}
	
	\begin{proof}
		By Lemma \ref{comovinglemma}, we can choose smooth comoving coordinates $\tau(u,v), m(u,v)$ on $\mathcal{R}$. Let $(\mathscr{Q}_{MDD},r,\rho,m,f)$ be the maximal dust development of the induced data on $C_{v_0, v_c} \cup \underline{C}_{v_0, v_c}$. By Lemma \ref{nondeglemma}, $m(u,v)$ extends continuously to $(u_c,v_c)$, so let $m_* := m(u_c,v_c)$, $\tau_* := \tau(u_c,v_c)$. Let $\mathscr{R} := \{ (\tau(u,v),m(u,v)) : (u,v) \in \mathcal{R}\} \subset \mathscr{Q}_{MDD}$, and let $(u(\tau,m), v(\tau,m))$ be the solutions of the eikonal equation on $\mathscr{R}$ that are the inverse of $(\tau(u,v), m(u,v))$. From Proposition \ref{eikprop}, $u(\tau,m)$, $v(\tau,m)$ extend smoothly to $(\tau_*,m_*) \in \mathscr{C}$, with $\partial_\tau u (\tau_*, m_*), \partial_\tau v (\tau_*, m_*) > 0$. Continuity of $\tau(u,v), m(u,v)$ on $\overline{\mathcal{R}}$ implies that all derivatives $\partial_\tau^k u(\tau(u,v),m(u,v))$, $\partial_\tau^k v(\tau(u,v),m(u,v))$ are continuous as functions of $(u,v)$ on $\overline{\mathcal{R}}$. From the expressions for first derivatives of $r, \Omega^2$: \eqref{pr}, \eqref{pomega} we see that $r, \Omega^2 \in C^1(\overline{\mathcal{R}})$. From (E-d), second derivatives of $r$ blow up when $\rho \to \infty$. Note that we can combine \eqref{rwave} and \eqref{Ruv} to get
		\begin{equation} \label{wavelog}
			\partial_u \partial_v \log (r\Omega)^2 = -\frac{2\partial_u r \partial_v r}{r^2} = \frac{\Omega^2}{2r^2}\left( 1 - \frac{2m}{r} \right),
		\end{equation}
		which extends continuously to the shell crossing boundary. We compute the second derivatives of $\log (r\Omega)^2$ directly by differentiating $\partial_u \log(r\Omega)^2$, $\partial_v \log(r\Omega)^2$ using the eikonal equation:
		\begin{align}  \label{pusqrom}
			\partial_u^2 \log(r\Omega)^2 &= - \frac{1}{2\partial_\tau u}\left[ 2\partial_\tau \left( \frac{\partial_\tau^2 u}{(\partial_\tau u)^2}\right) + \frac{3m}{2r^3 \partial_\tau u} + 2\partial_u r \left( \frac{\partial_\tau^2 u}{r\partial_\tau u} - \frac{ (1+f)^\frac{1}{2}}{r^2} \right) \right], \\ \partial_v^2 \log(r\Omega)^2 &= - \frac{1}{2\partial_\tau v} \left[ 2\partial_\tau \left( \frac{\partial_\tau^2 v}{(\partial_\tau v)^2}\right) +\frac{3m}{2r^3 \partial_\tau v} + 2\partial_v r \left( \frac{\partial_\tau^2 v}{r\partial_\tau v} + \frac{ (1+f)^\frac{1}{2}}{r^2} \right) \right], \label{pvsqrom}
		\end{align}
		so we see that  $\log(r\Omega)^2$ indeed extends in $C^2_{u,v}$ to $(u_c,v_c)$.
		
		We now consider null curves on the maximal dust development, in order to show that we can choose $\delta$ sufficiently small so that $\mathcal{R}' \subset \mathcal{Q}$. The curves $\{ u= u_c, v < v_c\}$, $\{ u < u_c, v = v_c\}$ are the unique ingoing/outgoing null curves with future limit point $(\tau_*,m_*) \in \mathscr{C}$. To see existence and uniqueness, note that by Lemma \ref{nondeglemma}, there exists $\delta_1 > 0$ such that $\tau_c(m) > \tau_*$ for $\chi \in \{[m_*- \delta_1, m_* + \delta_1]\} \setminus \{m_*\}$. The ODEs for null curves on $\mathscr{Q}_{MDD}$, \eqref{outnull}, \eqref{innull}, are of the form
		\begin{equation} \label{nullODE}
			\tau_+'(m) = F(\chi, \tau_+(m)), \qquad \tau_-'(m) = - F(\chi,\tau_-(m)),
		\end{equation}
		where $F(\tau,m) = (1+f(m))^{-\frac{1}{2}} \partial_m r(\tau,m)$ is smooth and non-negative on $\mathscr{Q}_{MDD} \cup \mathscr{C}$. Existence and uniqueness for past-directed null curves from $(\tau_*,m_*)$ follows essentially by Picard-Lindel\"of, since the ODEs clearly map into the domain $\{ \tau \leq \tau_*\}$. In order to show that $\mathcal{R}' \subset \mathcal{Q}$, it is sufficient to show that there exist unique future-directed ingoing/outgoing null curves from $(\tau_*, m_*)$ in $\mathscr{Q}_{MDD} \cup \{(\tau_*, m_*)\}$. Uniqueness of solutions of the eikonal equation then implies that we can uniquely extend the solution to include $\mathcal{R}'$, for some $\delta > 0$. Any future-directed outgoing null curve from $(\tau_*,m_*)$ in $\mathcal{Q}_{MDD}$ must have $\tau_+(m) < \tau_c(m)$ for $m \in (m_*, m_*+\delta_1)$, for some $\delta_1 > 0$. Bounding $\partial_m r(\tau_+(m),m)$ using \eqref{pchiu}:
		\begin{equation}
			\partial_m r \lesssim (\tau_c(m) - \tau_+(m)) \leq \tau_c(m) - \tau_* \lesssim (m - m_*)^{2n},
		\end{equation}
		for some $n \in \mathbb{N}$ by Lemma \ref{nondeglemma}, possibly shrinking $\delta_1 > 0$ so that we can apply Taylor's theorem on $[m_*, m_* + \delta_1]$. But we then have, for any solution of \eqref{nullODE}:
		\begin{equation}
			\tau_+(m) - \tau_* = \int_{m_*}^m F(\tau_+(x),x) dx \lesssim (m - m_*)^{2n+1}.
		\end{equation}
		We can therefore define a contraction map for sufficiently small $\delta_1 > 0$, and show that there is a unique outgoing null curve from $(\tau_*,m_*)$. The argument is identical for the ingoing null curve. Now, for $0 < \delta \leq \delta'$, we consider:
		\begin{align}
			\mathscr{D}_{\delta, -} &:= \{ (\tau(u,v),m(u,v)) : v = v_0, u \in [u_0,u_c + \delta]\}, \\ \mathscr{D}_{\delta, +} &:= \{ (\tau(u,v),m(u,v)) : u = u_0, v \in [v_0,v_c + \delta]\}.
		\end{align}
		For sufficiently small $\delta > 0$, the outgoing null curve from $(\tau(u_c+\delta,v_0),m(u_c+\delta,v_0))$ in $\mathscr{Q}_{MDD}$ intersects the unique ingoing null curve from $(\tau_*,m_*)$, and also the ingoing null curve from $(\tau(u_0,v_c+\delta),m(u_0,v_c+\delta))$ in $\mathscr{Q}_{MDD}$ intersects the outgoin null curve from $(\tau_*,m_*)$. Since the maximal dust development is acoustically globally hyperbolic by definition, $\mathcal{Q}$ extends uniquely to include $\mathcal{R}'$ for such a $\delta > 0$. Moreover, denoting the future-directed null curves from $(\tau_*,m_*)$ by:
		\begin{align}
			\mathscr{CH}_{\delta, -} &:= \{ (\tau(u,v),m(u,v)) : v = v_c, u \in (u_c,u_c + \delta]\}, \\ \mathscr{CH}_{\delta, +} &:= \{ (\tau(u,v),m(u,v)) : u = u_c, v \in (v_c,v_c + \delta]\},
		\end{align}
		then $\partial_\tau^k u(\tau,m)$, $\partial_\tau^k v(\tau,m)$ extend continuously to $\mathscr{CH}_{\delta, \pm}$ by Proposition \ref{eikprop}, and $\partial_\chi r(\tau,m) > 0$ for $(\tau,m) \in \mathscr{CH}_{\delta, \pm}$ so the transformation to double null coordinates is smooth and we have $r, \Omega^2, \rho, \mathbf{u}^\mu$ extending in $C^\infty$ to $\overline{\mathcal{R}'} \setminus \{(u_c,v_c)\}$.
	\end{proof}

	(The last part of this section is an independent observation and is not necessary for any further result in the paper.) We now show that we can make a \emph{non-unique} smooth extension of double null coordinates beyond the Cauchy horizon of the (non-degenerate) first shell crossing singularity. The non-uniqueness is purely in the \emph{gauge}, the solution can be uniquely smoothly extended in \emph{comoving coordinates}, cf.~Proposition \ref{mddprop}. Any such extension is however \emph{unphysical}: it is necessarily not globally hyperbolic with respect to spacetime causal curves (despite being `acoustically globally hyperbolic'). We show in addition that any such smooth extension in double null coordinates cannot embed into $\R^{1+1}_{u,v}$.
	\begin{lemma} \label{smoothext}
		Let $(\mathcal{Q},r,\Omega^2,\rho,\mathbf{u}^\mu)$ be a globally hyperbolic $C^\infty$ collapsing solution of (E-d), and suppose that $(u_c,v_c) \in \R$ is a non-degenerate first shell crossing singularity, so that, by Proposition \ref{regext}, there exists a unique smooth extension of $(\mathcal{Q},r,\Omega^2,\rho,\mathbf{u}^\mu)$ and $(u_0,v_0) \in I^-(u_c,v_c), \delta > 0$ such that 
		\begin{equation}
			\mathcal{R}' := \mathcal{R} \cup ([u_0,u_c) \times [v_c, v_c + \delta]) \cup ([v_0,v_c) \times [u_c, u_c + \delta]) \subset \mathcal{Q}.
		\end{equation}
		For $\delta>0$ sufficiently small, there exist $C^\infty$ functions $v_{\mathcal{C}^-}(u) : (u_c, u_c + \delta] \to \R$, $u_{\mathcal{C}^+}(v) : (v_c, v_c + \delta] \to \R$, with $v_{\mathcal{C}^-}'(u), u_{\mathcal{C}^+}'(v) > 0$ such that: defining 
		\begin{equation}
			\mathcal{R}^- := \{ (u,v) : u_c \leq u \leq u_c+\delta, \ v > v_{\mathcal{C}^-}(u) \}, \qquad \mathcal{R}^+ := \{ (u,v) : v_c \leq v \leq v_c+\delta, \ u < u_{\mathcal{C}^+}(v) \},
		\end{equation}
		there exist \underline{non-unique} smooth extensions of $(\mathcal{Q},r,\Omega^2,\rho,\mathbf{u}^\mu)$ to $\mathcal{R}' \cup \mathcal{R}^-$ and to $\mathcal{R}' \cup \mathcal{R}^+$, which are necessarily \underline{not globally hyperbolic}. Any such extensions fail to jointly embed into $\R^{1 + 1}_{u,v}$: $\mathcal{R}^+ \cap \mathcal{R}^- \neq \emptyset$. The (timelike) singular boundaries of $\mathcal{R}^\pm$ correspond to \underline{shell crossing}: $\rho(u,v) \to \infty$ as $(u,v) \to (u,v_{\mathcal{C}^-}(u))$ in $\mathcal{R}^-$, and as $(u,v) \to (u_{\mathcal{C}^+}(v),v)$ in $\mathcal{R}^+$.
	\end{lemma}
	
	\begin{proof}
		By Lemma \ref{comovinglemma}, we can choose smooth comoving coordinates $\tau(u,v), m(u,v)$ on $\mathcal{R}'$, with $m$ the Hawking mass. Let $\mathscr{Q}_{MDD}$ be the maximal dust development of the induced data on $C_{v_0, v_c} \cup \underline{C}_{v_0, v_c}$. Proposition \ref{regext} implies that $(\tau(u,v),m(u,v))$ extend smoothly to the 'Cauchy horizons':
		\begin{equation}
			 \mathcal{CH}^- := \{ u \in (u_c, u_c + \delta], \ v = v_c \}, \qquad \mathcal{CH}^+ := \{ u = u_c, \ v \in (v_c,v_c+\delta] \},
		\end{equation} 
		which correspond in comoving coordinates to $\mathscr{D}^\pm := \{ (\tau(u,v),m(u,v)) : (u,v) \in \mathcal{CH}^\pm \}$. Consider the induced data on $\mathscr{D}^-$ in comoving coordinates, and let $\mathscr{Q}^- := D^+_{ac}(\mathscr{D}^-) \subset \mathscr{Q}_{MDD}$. The solution of the eikonal equation \eqref{eiku}, $u(\tau,m)$, corresponding to the choice of comoving coordinates on $\mathcal{R}'$, extends smoothly, and \emph{uniquely}, across $\mathscr{D}^-$ to $\mathscr{U}_+(\mathscr{D}^-, \mathscr{Q}^- \cup \mathscr{C})$. The domain $\mathscr{U}_+(\mathscr{D}^-, \mathscr{Q}^- \cup \mathscr{C})$ is bounded by the outgoing null curve $\{ (\tau,\chi) : u(\tau,\chi) = u_c + \delta\} \subset \mathscr{Q}^-$, and by the singular boundary due to shell crossing: $\mathscr{C} \cap \mathscr{U}_+(\mathscr{D}^-, \mathscr{Q}^- \cup \mathscr{C})$. (Here, we take $\delta>0$ sufficiently small so that $[m(u_c + \delta,v_c),m_*] \subset \mathscr{C}_{\chi}$ (where $m_* := m(u_c,v_c)$), which is possible by Lemma \ref{nondeglemma}.)
		
		Now, from Proposition \ref{regext}, we have that $v(\tau,m)$ extends in $C^\infty$ to $\mathscr{D}^-$. We can therefore \emph{non-uniquely} choose a smooth extension of $v(\tau,m)$ to $\mathscr{U}_+(\mathscr{D}^-, \mathscr{Q}^- \cup \mathscr{C})$, e.g.~consider the induced data on the outgoing null curve $\{ u(\tau,m) = u_c + \delta\}$, choose a smooth extension of the data, then solve the eikonal equation for $v(\tau,m)$ in $\mathscr{U}_+(\mathscr{D}^-, \mathscr{Q}^- \cup \mathscr{C})$. We define, for $m$ such that $(\tau_c(m),m) \in \mathscr{C} \cap \mathscr{U}_+(\mathscr{D}^-, \mathscr{Q}^- \cup \mathscr{C})$, 
		\begin{equation} 
			u_{\mathscr{C}}(m) := u(\tau_c(m),m), \qquad v_{\mathscr{C}}(m) := v(\tau_c(m),m).
		\end{equation}
		Using the eikonal equation \eqref{eiku}, we have
		\begin{equation}\label{ucp}
			u_{\mathscr{C}}'(m) = \partial_\tau u(\tau_c(m),m) \left( \tau_c'(m) - \frac{\partial_\chi r(\tau_c(m),m)}{(1+f(m))^\frac{1}{2}}\right) = \partial_\tau u(\tau_c(m),m)\tau_c'(m),
		\end{equation}
		where we've used $\partial_m r(\tau_c(m),m) = 0$, by definition of $\tau_c(m)$. Similarly using \eqref{eikv}:
		\begin{equation}\label{vcp}
			v_{\mathscr{C}}'(m) = \partial_\tau v(\tau_c(m),m) \tau_c'(m). 
		\end{equation}
		By Proposition \ref{eikprop}, $\partial_\tau u(\tau_c(m),m), \partial_\tau v(\tau_c(m),m) > 0$, and by Lemma \ref{nondeglemma}: we can again choose $\delta > 0$ sufficiently small such that $\tau_c'(m) > 0$ for all $(\tau_c(m),m) \in \mathscr{C} \cap \mathscr{U}_+(\mathscr{D}^-, \mathscr{Q}^- \cup \mathscr{C})$. We therefore have $u_{\mathscr{C}}'(m), v_{\mathscr{C}}'(m) > 0$, so we can invert $u_{\mathscr{C}}(m)$ and define: $v_{\mathcal{C}^-}(u) := v_{\mathscr{C}}( u_{\mathscr{C}}^{-1}(u))$, for $u \in (u_c, u_c + \delta]$, which is $C^\infty$ on its domain by the smooth extension of solutions of the eikonal equation to $\mathscr{C}$. We've constructed a smooth extension to $\mathcal{R}^-$ as defined in the statement of the lemma, and by construction $\{ v = v_{\mathcal{C}^-}(u)\}$ is a timelike singular boundary corresponding to shell crossing. The existence of smooth extensions to $\mathcal{R}^+$ follows analogously. 
		
		We now observe the failure of smooth extensions to embed into $\R^{1+1}_{u,v}$. Indeed, suppose that we are given smooth extensions to both $\mathcal{R}^-$ and $\mathcal{R}^+$. Let $u_{\mathcal{C}^-}(v)$ be the smooth inverse of $v_{\mathcal{C}^-}(u)$, defined on $(v_c, v_{\mathcal{C}^-}(u_c + \delta)]$. We now consider the behaviour of $u_{\mathcal{C}^-}(v)$ in a neighbourhood of $v = v_c$. From \eqref{ucp} and \eqref{vcp}, we have
		\begin{equation}
			u_{\mathcal{C}^-}'(v_{\mathscr{C}}(m)) = \frac{\partial_\tau u(\tau_c(m),m)}{\partial_\tau v(\tau_c(m),m)} > 0.
		\end{equation}
		For the second derivative, we differentiate the eikonal equations \eqref{eiku},\eqref{eikv} and evaluate at $(\tau_c(m),m) \in \mathscr{C}$ to get
		\begin{align}
			\partial_\tau \partial_m u(\tau_c(m),m) &=  (1+f(m))^{-\frac{1}{2}}\partial_\chi \left( \frac{2m}{r_c(m)} + f(m)\right)^\frac{1}{2} \partial_\tau u(\tau_c(m),m), \\ \partial_\tau \partial_m v(\tau_c(m),m) &=  - (1+f(m))^{-\frac{1}{2}}\partial_m \left( \frac{2m}{r_c(m)} + f(m)\right)^\frac{1}{2} \partial_\tau v(\tau_c(m),m),
		\end{align}
		and we can then compute
		\begin{equation}
			u_{\mathcal{C}^-}''(v_{\mathscr{C}}(m)) = \frac{2 \partial_\tau u}{(\partial_\tau v)^2} \frac{\partial_m \left( \frac{2m}{r_c(m)} + f(m)\right)^\frac{1}{2}}{(1+f(m))^{\frac{1}{2}} \tau_c'(m)} = \frac{2 \partial_\tau u}{(\partial_\tau v)^2} \left[ \frac{(\tau_c'(m))^{-1} \left(\frac{2}{r_c(m)} + f'(m)\right)}{(1+f(m))^\frac{1}{2} (\frac{2m}{r_c(m)} + f(m))^\frac{1}{2}} - (1+f(m))^{-\frac{1}{2}} \frac{2m}{r_c(m)^2}\right],
		\end{equation}
		so that $u_{\mathcal{C}^-}''(v_{\mathscr{C}}(m)) \sim (\tau_c'(m))^{-1} \sim (m - m_*)^{1 - 2n} < 0$, for $m_* := m(u_c,v_c)$, $n \in \mathbb{N}$ as in the statement of Lemma \ref{nondeglemma}. From \eqref{vcp}, we have
		\begin{equation}
			v_{\mathscr{C}}'(m) \sim (m - m_*)^{2n-1},
		\end{equation}
		so that $v_{\mathscr{C}}(m) - v_c \sim (m - m_*)^{2n}$, and therefore
		\begin{equation}
			u_{\mathcal{C}^-}''(v) \sim - (v - v_c)^{\frac{1}{2n} - 1},
		\end{equation}
		for $v \in (v_c, v_{\mathcal{C}^-}(u_c + \delta)]$. The identical calculation for the singular boundary of $\mathcal{R}^+$ gives:
		\begin{equation}
			u_{\mathcal{C}^+}''(v) \sim (v - v_c)^{\frac{1}{2n} - 1},
		\end{equation}
		where we have $u_{\mathcal{C}^+}'(v_c) = u_{\mathcal{C}^-}'(v_c)$. Therefore, $u_{\mathcal{C}^+}(v) > u_{\mathcal{C}^-}(v)$ for $v > v_c$ sufficiently close to $v_c$: $\mathcal{R}^+ \cap \mathcal{R}^- \neq \emptyset$. 
	\end{proof}
	
	\section{Piecewise smooth solutions of (E-d) and thin shells of dust} \label{pwise}
	
	In this section, we consider the class of `piecewise smooth' solutions to the Einstein-dust system in double null coordinates: weak (measure-valued) solutions that are smooth either side of a timelike hypersurface. We show that any sufficiently regular timelike hypersurface on which the energy density of dust has delta function support is a thin shell of dust. The notion of thin shells being coordinate independent, we can understand the propagation of thin shells solely in comoving coordinates, as a solution of the \emph{thin shell ODE}. Piecewise smooth solutions are then uniquely determined as long as they satisfy the `determinism condition': the fluid flow lines of the smooth dust either side of the thin shell must intersect initial data. 
	
	Thin shells were first studied by Lanczos \cite{Lanczos} in 1924, with Israel \cite{I} later formulating the definition in terms of extrinsic curvature. The standard setup is as follows. Let $(\mathcal{M},g)$ be a spacetime with timelike hypersurface $\Sigma$ such that $\mathcal{M}\setminus\Sigma$ has two connected components, which we denote by $\mathcal{M}^\pm$, with metric $g^\pm$. 
	
	Choose coordinates on $\Sigma$ with we denote by $\{i,j\}$. Let $h^\pm_{i j}$ be the induced metric on $\Sigma$ by $(\mathcal{M}^\pm,g^\pm)$, and $K^\pm_{i j}$ the corresponding extrinsic curvature. We are interested in the case in which $\Sigma$ is a \emph{singular hypersurface}. Note that in order to have a continous metric in spacetime coordinates in a neighbourhood of $\Sigma$, we must have 
	\begin{equation}
		h^+_{i j} = h^-_{i j} =: h_{i j}.
	\end{equation}
	The insight of Israel was that a `jump' in extrinsic curvature, by which we mean
	\begin{equation}
		\gamma_{i j} := K^+_{i j} - K^-_{i j} \neq 0,
	\end{equation}
	corresponds to matter with `delta function support' on the hypersurface $\Sigma$. One can construct Gaussian normal coordinates on $\Sigma$ (see \cite{Po}) to see that the Einstein equations are solved in a distributional sense: letting $l$ be the geodesic distance from $\Sigma$, then the `surface energy momentum tensor' $S_{\mu \nu}$ is defined by
	\begin{equation}
		T_{\mu \nu} = \Theta(l)T^+_{\mu \nu} + \Theta(-l)T^-_{\mu \nu} + \delta(l)S_{\mu \nu},
	\end{equation}
	where $\Theta(l)$ is the Heaviside step function, and $\delta(l)$ the Dirac delta. The Einstein equations then imply that
	\begin{equation} \label{surfaceem}
		S_{i j} = -\frac{1}{8\pi} \left( \gamma_{i j} - \gamma h_{i j} \right).
	\end{equation}
	
	\begin{note}
		Any thin shell $\Sigma$ with $\gamma_{i j} \neq 0$ is a distributional solution of the Einstein equations for some matter distribution $S_{i j}$ that can be read off from \eqref{surfaceem}, in the same way that one can `solve' Einstein's equations by writing down a metric and reading off $T_{\mu \nu}$. Any \emph{physical} thin shell should have a prescribed type of matter: a choice of the form of $S_{i j}$.
	\end{note}
		
	The following equivalent coordinate-invariant\footnote{Lanczos only published his result \cite{Lanczos} because he was mis-referenced in a paper which claimed that his approach was dependent on a certain set of coordinates.} definition goes back to Lanczos \cite{Lanczos}:
	\begin{definition} \label{Lanczos}
		Let $\mathcal{U}$ be a neighbourhood in a spacetime $(\mathcal{M},g) $ containing a thin shell $\Sigma$, such that $\mathcal{U}\cap \Sigma \neq \emptyset$. The \underline{surface energy momentum tensor} of $\Sigma$, $S_{\mu \nu}$, is related to the spacetime energy momentum tensor by:
		\begin{equation}
			\int_\mathcal{U} T_{\mu \nu} =	\int_{\mathcal{U}^+} T^+_{\mu \nu} + \int_{\mathcal{U}^-} T^-_{\mu \nu} +  \int_{\Sigma \cap \mathcal{U}} S_{\mu \nu},
		\end{equation}
		where integration over $\mathcal{U}$ is with respect to the volume form of $g$, while integration over $\Sigma$ is with respect to the volume form of the induced metric, $h$.
	\end{definition}
	 
	A thin shell of dust, first considered by Israel \cite{I}, is defined by choosing $S_{i j} = \sigma v_i v_j$, so that the matter `on the thin shell' is dust. In spherical symmetry, with $s$ proper time along $\Sigma$, the induced metric on $\Sigma$ is
	\begin{equation}
		h = -ds^2 + r^2 g_{\mathbb{S}^2},
	\end{equation}
	and \eqref{surfaceem} for dust is equivalent to
	\begin{equation} \label{thinshell}
		[K_{s s}] = -4\pi\sigma, \qquad [K_{A B}] = -4\pi\sigma r^2 (g_{\mathbb{S}^2})_{A B}.
	\end{equation}
	
	We can therefore define a thin shell of dust to be a timelike hypersurface $\Sigma$ such that \eqref{thinshell} holds, for some $\sigma(s)$ defined along $\Sigma$.
	
	\subsection{Piecewise smooth solutions of (E-d)} \label{pwisesection}
	
	We now consider piecewise smooth solutions of the system (E-d) on the quotient manifold $\mathcal{Q}$, and show that they are thin shells of dust in the `upstairs' manfiold, $\mathcal{M} = \mathcal{Q} \times \mathbb{S}^2$.
	
	In this section, we consider spacetimes $(\mathcal{Q}, r, \Omega^2)$, where the quotient manifold is given by
	\begin{equation}
		\mathcal{Q}:= [0,u_1) \times [0,v_1) \subset \R^{1,1}_{u,v},
	\end{equation}
	for $u_1,v_1 > 0$, and the metric is $g_{\mathcal{Q}} = - \Omega^2 du dv$, with $r, \Omega^2 \in C^{0, 1}(\mathcal{Q})$. Let $\Sigma$ be a future-directed $C^2$ timelike curve: given $s_1 > 0$ and  $u_{\Sigma}(s), v_{\Sigma}(s) \in C^1([0,s_1)) \cap C^2((0,s_1))$, with $\dot{u}_{\Sigma}(s), \dot{v}_{\Sigma}(s) > 0$ for $s \in [0,s_1)$, then
	\begin{equation}
		\Sigma := \{ (u_{\Sigma}(s), v_{\Sigma}(s)) : s \in [0,s_1)\} \subset \mathcal{Q}.
	\end{equation}
	\begin{note}
		We immediately drop the notation $u_{\Sigma}(s), v_{\Sigma}(s)$ in place of $u(s), v(s)$, where the parametrisation by $s$ indicates that they are the coordinates of $\Sigma$. Differentiation by $s$ will be denoted by a dot, e.g.~$\dot{u}(s)$, $\dot{v}(s)$.
	\end{note}
	We denote the past-endpoint of $\Sigma$ by $\partial^-\Sigma := (u(0),v(0))$, where by convention we always have $\partial^-\Sigma \in \Sigma$. The future-endpoint is $\partial^+\Sigma := (u(s_1),v(s_1)) \in \overline{\mathcal{Q}}$. We use the definition of endpoints to define the `interior' and `exterior' regions of a timelike curve $\Sigma$:
	\begin{definition} \label{def:Dpm}
		For a connected $C^1$ timelike curve $\Sigma \subset \mathcal{Q}$, we define:  $D(\Sigma,\mathcal{Q}) := J^+(\partial^- \Sigma) \cap I^-(\partial^+ \Sigma)$, so that $D(\Sigma,\mathcal{Q}) \setminus \Sigma \subset \mathcal{Q}$ has two connected components. We define the `exterior' region:
		\begin{equation}
			D^+(\Sigma,\mathcal{Q}) := \{ (u,v) \in D(\Sigma,\mathcal{Q}) : \exists \tilde{v} < v \text{ s.t. } (u,\tilde{v}) \in \Sigma \}
		\end{equation}
		and the `interior' region:
		\begin{equation}
			D^-(\Sigma,\mathcal{Q})  := \{ (u,v) \in D(\Sigma,\mathcal{Q})  : \exists \tilde{u} < u \text{ s.t. } (\tilde{u},v) \in \Sigma \}.
		\end{equation}
	\end{definition}
	\begin{remark}
		Throughout this section, whenever the measure of integration is suppressed, any integral over (some subset of) the quotient manifold is understood to be taken with respect to the volume form of the quotient manifold, for example:
		\begin{equation}
			\int_{\mathcal{Q}} \varphi(u,v) = \int_0^{u_1} \int_0^{v_1} \frac{1}{2} \Omega^2(u,v) \varphi(u,v) dudv. 
		\end{equation}
	\end{remark}
	
	\begin{definition} \label{piecesmooth}
		We have a piecewise smooth distribution of dust $(\rho,\mathbf{u}^\mu)$ on $(\mathcal{Q}, r, \Omega^2)$ with `delta shell' supported on $\Sigma$, given by functions $u(s), v(s) \in C^1([0,s_1)) \cap C^2((0,s_1))$ for $s_1 > 0$:
		\begin{equation}
			\Sigma := \{ (u(s), v(s)) : s \in [0,s_1)\} \subset \mathcal{Q},
		\end{equation}
		if there exist non-negative functions $\mathring{\rho} \in C^{1}_{u,v}(\mathcal{Q} \setminus \Sigma)$, $\mathring{\mathbf{u}}^u, \mathring{\mathbf{u}}^v \in C^1_{u,v}(\text{supp}(\mathring{\rho}) \setminus \Sigma)$ and $\sigma \in C^0([0,s_1)) \cap C^1([0,s_1))$, such that:
		\begin{equation}
			\mathbf{u}(u,v) = \begin{cases}
				\dot{u}(s) \partial_u + \dot{v}(s) \partial_v &: \exists s  \in [0,s_1) \ s.t. \ (u,v) =(u(s),v(s)), \\ \mathring{\mathbf{u}}^u \partial_u + \mathring{\mathbf{u}}^v \partial_v  &: (u,v) \in \text{supp}(\mathring{\rho}) \setminus \Sigma,
			\end{cases}
		\end{equation}
		and for all $\varphi \in C^0_c(\mathcal{Q})$:
		\begin{align}
			\int_\mathcal{Q} \varphi \rho &= \int_{s_-}^{s_+} \varphi(s) \sigma(s) ds + \int_{\mathcal{Q}\setminus \Sigma} \varphi(u,v) \mathring{\rho}(u,v), \\
			\int_\mathcal{Q} \varphi \rho (\mathbf{u}^u)^2 &= \int_{s_-}^{s_+} \varphi(s) \sigma(s) \dot{u}(s)^2 ds + \int_{\mathcal{Q}\setminus \Sigma} \varphi(u,v) \mathring{\rho}(u,v) (\mathring{\mathbf{u}}^u(u,v))^2, \\ \int_\mathcal{Q} \varphi \rho (\mathbf{u}^v)^2 &= \int_{s_-}^{s_+} \varphi(s) \sigma(s) \dot{v}(s)^2 ds + \int_{\mathcal{Q}\setminus \Sigma} \varphi(u,v) \mathring{\rho}(u,v) (\mathring{\mathbf{u}}^v(u,v))^2.
		\end{align}
		We require moreover that $\mathring{\rho}, \mathring{\mathbf{u}}^\mu$ extend to $C^1_{u,v}$ functions on $(D^\pm(\Sigma,\mathcal{Q}) \cup \Sigma) \setminus \partial^- \Sigma$, $(\text{supp}(\mathring{\rho}) \cap (D^\pm(\Sigma,\mathcal{Q}) \cup \Sigma)) \setminus \partial^- \Sigma$ respectively, that $\Omega^2(u,v) \mathbf{u}^u(u,v) \mathbf{u}^v(u,v) = 1$ and $\mathbf{u}^u(u,v), \mathbf{u}^v(u,v) > 0$ for all $(u,v) \in \text{supp}(\mathring{\rho}) \cup \Sigma$ and $\sigma(s) > 0$ for all $s \in (0,s_1)$.
	\end{definition}
	\begin{remark}
		Let $(\rho,\mathbf{u}^\mu)$ be a piecewise smooth distribution of dust as in Definition \ref{piecesmooth}, we have:
		\begin{equation}
			\int_\mathcal{Q} \rho \mathbf{u}^\mu \mathbf{u}^\nu = \int_\Sigma \sigma(s) \mathbf{u}^\mu(s) \mathbf{u}^\nu(s) + \int_{\mathcal{Q}\setminus\Sigma} \mathring{\rho} \mathring{\mathbf{u}}^\mu \mathring{\mathbf{u}}^\nu,
		\end{equation}
		so this is exactly Lanczos' definition \eqref{Lanczos} (where for convenience we have raised the indices), in spherical symmetry: the delta shell has the surface energy momentum tensor of dust. 
	\end{remark}
	\begin{remark}
		The normalisation $\Omega^2 \mathbf{u}^u \mathbf{u}^v = 1$ implies that the parameter $s$ is proper time along the timelike curve $\Sigma$, for all $s \in [0,s_1)$:
		\begin{equation}
			\Omega^2(u(s), v(s)) \dot{u}(s) \dot{v}(s) = 1.
		\end{equation}
	\end{remark}
	We can now define piecewise smooth solutions of (E-d): 
	\begin{definition} \label{pwisesmooth}
		A \underline{piecewise smooth solution of (E-d)}, $(\mathcal{Q}, r, \Omega^2, \rho, \mathbf{u}^\mu)$, with underlying manifold $\mathcal{Q} = [0,u_1) \times [0,v_1) \subset \R^{1 + 1}_{u,v}$, consists of positive functions $r, \Omega^2 \in C^{0,1}(\mathcal{Q})$ and a piecewise smooth distribution of dust $(\rho,\mathbf{u}^\mu)$, as in Definition \ref{piecesmooth}, with delta shell supported on $\Sigma \subset \mathcal{Q}$, a timelike curve given by functions $u(s), v(s) \in C^1([0,s_1)) \cap C^2((0,s_1))$ for $s_1>0$:
		\begin{equation}
			\Sigma := \{ (u(s), v(s)) : s \in [0,s_1)\} \subset \mathcal{Q},
		\end{equation}
		such that $(\mathcal{Q} \setminus \Sigma, r, \Omega^2, \rho, \mathbf{u}^\mu)$ is a $C^1$ solution\footnote{Definition \ref{kcollapse}.} of (E-d), where each function $r, \Omega^2, \rho, \mathbf{u}^\mu$ extends in the respective regularity for a $C^1$ solution to $\Sigma \setminus \partial^-\Sigma$ from $D^\pm(\Sigma,\mathcal{Q}) \setminus \partial^-\Sigma$. Moreover, the functions $(r,\Omega^2, \rho, \mathbf{u}^\mu)$ satisfy the Einstein-dust system in `weak' integrated form: $\forall \varphi \in C^1_{c}(\mathcal{Q})$,
		\begin{align}
			 \int_0^{u_1} \int_0^{v_1} \partial_v r \partial_u \varphi dudv + \int_0^{v_1} \varphi(0,v)\partial_v r(0,v) dv &= \int_\mathcal{Q} \varphi\left(\frac{m}{r^2} - 2\pi r\rho \right), \label{rwaveweakv} \\ \int_0^{u_1} \int_0^{v_1} \partial_u r \partial_v \varphi dudv + \int_0^{u_1} \varphi(u,0)\partial_u r(u,0) du &= \int_\mathcal{Q} \varphi\left(\frac{m}{r^2} - 2\pi r\rho \right), \label{rwaveweaku} \\  \int_0^{u_1} \int_0^{v_1} \partial_v \log\Omega^2 \partial_u \varphi dudv + \int_0^{v_1} \varphi(0,v)\partial_v \log \Omega^2(0,v) dv &= \int_\mathcal{Q} \varphi\left(4\pi\rho - \frac{2m}{r^3}\right), \label{omegawaveweakv} \\ \int_0^{u_1} \int_0^{v_1} \partial_u \log\Omega^2 \partial_v \varphi dudv + \int_0^{u_1} \varphi(u,0)\partial_u \log \Omega^2(u,0) du &= \int_\mathcal{Q} \varphi\left(4\pi\rho - \frac{2m}{r^3}\right), \label{omegawaveweaku} \\ \int_0^{u_1} \int_0^{v_1} \frac{\partial_u r}{\Omega^2} \partial_u \varphi dudv + \int_0^{v_1} \frac{\partial_u r(0,v)}{\Omega^2(0,v)} \varphi(0,v) dv &= \int_\mathcal{Q} 2\pi r \rho (\mathbf{u}^v)^2 \varphi, \label{Rayuweak} \\  \int_0^{u_1} \int_0^{v_1} \frac{\partial_v r}{\Omega^2} \partial_v \varphi dudv + \int_0^{u_1} \frac{\partial_v r(u,0)}{\Omega^2(u,0)} \varphi(u,0) du &= \int_\mathcal{Q} 2\pi r \rho (\mathbf{u}^u)^2 \varphi. \label{Rayvweak}
		\end{align}
	\end{definition}
	\begin{remark}
		We define piecewise smooth solutions only on domains of the form $\mathcal{Q} = [0,u_1) \times [0,v_1)$ to simplify the definition, and because we will use it only in the case of thin shell development and for the \emph{propagation condition}, where this choice is sufficient. For the more general definition of piecewise smooth solutions with finitely many thin shells, see Section \ref{finiteshells}.
	\end{remark}
	\begin{note}
		In the following, we use the notation $[f](s) = f^+(u(s),v(s)) - f^-(u(s),v(s))$, for functions that are continuous with extension to $\Sigma$, where $f^\pm(u(s),v(s))$ is the unique limit to $(u(s),v(s))$ from $D^\pm(\Sigma,\mathcal{Q})$.
	\end{note}
	The following lemma shows the equivalence of piecewise smooth solutions of (E-d) and thin shells of dust:
	\begin{lemma} \label{thinshellequiv}
		Given $(\mathcal{Q}, r, \Omega^2, \rho, \mathbf{u}^\mu)$ a piecewise smooth solution of (E-d), then the integrals \eqref{rwaveweakv}, \eqref{rwaveweaku}, \eqref{omegawaveweaku}, \eqref{omegawaveweakv}, restricted to $D(\Sigma,\mathcal{Q})$ are equivalent to the following jump conditions evaluated on $\Sigma$:
		\begin{equation}
			\dot{u}(s) [\partial_u r](s) = 2\pi \sigma(s) r(s), \qquad \dot{v}(s) [\partial_v r](s) = -2\pi\sigma(s) r(s) \label{prjump}
		\end{equation}
		\begin{equation}
			 \dot{u}(s) [\partial_u \log\Omega^2](s) = -4\pi\sigma(s), \qquad \dot{v}(s) [\partial_v \log\Omega^2](s) = 4\pi\sigma(s), \label{pomegajump}
		\end{equation}
		and \eqref{Rayuweak}, \eqref{Rayvweak} are satisfied by definition of $\mathbf{u}^\mu$: $\mathbf{u}^u(u(s),v(s)) = \dot{u}(s)$, $\mathbf{u}^v(u(s),v(s)) = \dot{v}(s)$ for all $s \in [0,s_1)$. 
		
		The timelike hypersurface $\Sigma \times \mathbb{S}^2$ in the `upstairs' manifold $(\mathcal{M},g)$ is therefore a \underline{thin shell of dust} with respect to the interior/exterior spacetimes $\mathcal{M}^\pm = \mathcal{Q}^\pm \times \mathbb{S}^2$.
	\end{lemma}
	
	\begin{remark}
		The jump conditions \eqref{prjump} and \eqref{pomegajump} imply that
		\begin{equation}
			\dot{u}[\partial_u \log (r\Omega)^2] = 0 = \dot{v} [\partial_v \log (r\Omega)^2].
		\end{equation}
	\end{remark}
	
	\begin{proof}
		By definition of the piecewise smooth solution of (E-d), if $\text{supp} (\varphi) \cap \Sigma = \emptyset$, then we can integrate by parts in the integrals \eqref{rwaveweakv}--\eqref{Rayvweak}, which then vanish by assumption that $(D^\pm(\Sigma,\mathcal{Q}), r, \Omega^2, \rho, \mathbf{u}^\mu)$ is a $C^1$ solution of (E-d).
		
		Now, suppose that $\text{supp}(\varphi) \cap \Sigma \neq \emptyset$. We integrate \eqref{rwaveweakv}--\eqref{Rayvweak} by parts, where we now get boundary terms on $\Sigma$. For example, integrating by parts the left hand side of \eqref{rwaveweakv}:
		\begin{align}
			 \int_0^{u_1} \int_0^{v_1} &\partial_v r \partial_u \varphi dudv + \int_0^{v_1} \varphi(0,v)\partial_v r(0,v) dv = \\ &= \left( \int_{\mathcal{Q}^+} + \int_{\mathcal{Q}^-} \right) \left( \partial_u (\varphi \partial_v r ) - \varphi \partial_u \partial_v r \right) du dv + \int_0^{v_1} \varphi(0,v)\partial_v r(0,v) dv \\ &= \int_{v(0)}^{v(s_1)} [\partial_v r](u(v^{-1}(\tilde{v})),\tilde{v}) \varphi(u(v^{-1}(\tilde{v})),\tilde{v}) d\tilde{v} - \int_{\mathcal{Q}^+} \varphi \partial_u\partial_v r dudv - \int_{\mathcal{Q}^-} \varphi \partial_u\partial_v r dudv \\ &= \int_{0}^{s_1} \dot{v}(s) [\partial_v r](s) \varphi(s) ds - \int_{\mathcal{Q}^+} \varphi \partial_u\partial_v r dudv - \int_{\mathcal{Q}^-} \varphi \partial_u\partial_v r dudv. 
		\end{align}
		Equating with the right hand side in \eqref{rwaveweakv}, the integrals over $\mathcal{Q}^\pm$ cancel by assumption, and we use the definition of the delta shell to get
		\begin{equation}
			\int_{s_-}^{s_+} \varphi(s) (\dot{v}(s)[\partial_v r](s) + 2\pi r(s) \sigma(s)) ds = 0.
		\end{equation}
		Proceeding similarly for the other integrals, we have a weak solution of (E-d) if and only if the jump conditions \eqref{prjump} and \eqref{pomegajump} hold on $\Sigma$ (where the Raychaudhuri equations \eqref{Rayuweak}, \eqref{Rayvweak} are satisfied if the jump conditions \eqref{prjump} hold and $\mathbf{u}^u(u(s), v(s)) = \dot{u}(s)$, $\mathbf{u}^v(u(s), v(s)) = \dot{v}(s)$). Let $n$ be the unit normalised outward pointing normal to $\Sigma$. In spherical symmetry, we have
		\begin{equation} \label{KAB}
			K_{A B} = n(r) r \cdot (g_{\mathbb{S}^2})_{A B}.
		\end{equation}
		In double null coordinates, $n = \dot{v} \partial_v - \dot{u}\partial_u$, so \eqref{prjump} implies that $[n(r)] = -4\pi\sigma r$, so from \eqref{KAB} we have $[K_{A B}] = -4\pi \sigma r^2 (g_{\mathbb{S}^2})_{A B}$. In double null coordinates, we compute:
		\begin{equation}
			K_{s s} = \frac{\ddot{u}}{\dot{u}} + \dot{u}\partial_u \log\Omega^2 = -\frac{\ddot{v}}{\dot{v}} - \dot{v}\partial_v \log\Omega^2, \label{extcurv}
		\end{equation}
		so \eqref{pomegajump} implies that $[K_{s s}] = -4\pi\sigma$, and $\Sigma \times \mathbb{S}^2$ is therefore a thin shell of dust in $(\mathcal{M},g)$.
	\end{proof}
	
	\subsection{The thin shell ODE} \label{thinshellode}
	
	The definition of a thin shell of dust in terms of the `jump' in extrinsic curvature is manifestly coordinate invariant. In this section, we first show, in Proposition \ref{tsodeprop}, that the regularity assumptions in the definition of piecewise smooth solutions (Definition \ref{pwisesmooth}) allow us to define comoving coordinates, in which the jump conditions \eqref{prjump}, \eqref{pomegajump} imply that quantities defined along the thin shell solve an ordinary differential equation: the `thin shell ODE'. 
	
	We then show in Proposition \ref{prop:tsode-eik} that given a smooth solution of the collapsing Einstein-dust system in comoving coordinates $(\mathscr{Q}, r, \rho, m, f)$, and a solution of the thin shell ODE on $\mathscr{Q}$, then after identifying solutions of the eikonal equation on the solution of the thin shell ODE, and transforming to double null coordinates, we recover the jump conditions \eqref{prjump}, \eqref{pomegajump}. This is essential for \emph{constructing} piecewise smooth solutions of (E-d). In Section \ref{detcondsec} we define the determinism condition for a piecewise smooth solution of (E-d).
	
	We first define the following functions, which will make up the thin shell ODE.
	\begin{definition} \label{def:ODE}
		Given $C^\infty$ solutions of the collapsing Einstein-dust system in comoving coordinates, \\ $(\mathscr{Q}^\pm, r^\pm, \rho^\pm, m^\pm, f^\pm)$, we gauge fix the comoving coordinate $\chi^\pm=m^\pm$, and let $T^\pm(r^\pm,m^\pm)$ be the smooth function as in Remark \ref{Tremark} for each solution. We define:
		\begin{align} 
			\mathcal{R}(r,m^+,m^-,\epsilon) &:= - \left(\epsilon^2 - 1 + \frac{m^+ + m^-}{r} + \frac{(m^+ - m^-)^2}{4\epsilon^2 r^2}\right)^\frac{1}{2} \label{def:R} \\ \mathcal{M}^\pm(T^\pm, \rho^\pm, f^\pm; r,m^+,m^-,\epsilon) &:= 4\pi r^2 \rho^\pm(T^\pm(r,m^\pm),m^\pm) \widetilde{\mathcal{X}}^\pm(f^\pm;r,m^+,m^-,\epsilon) \label{def:Mpm} 
		\end{align}
		\begin{equation}
			\mathcal{E}(T^\pm, \rho^\pm, f^\pm; r,m^+,m^-,\epsilon) := \frac{\mathcal{M}^+}{m^+ - m^-}\left( \epsilon - \frac{\epsilon^2 \mathcal{T}^+}{(1+f^+(m^+))^\frac{1}{2}} \right) - \frac{\mathcal{M}^-}{m^+ - m^-}\left( \epsilon - \frac{\epsilon^2 \mathcal{T}^-}{(1+f^-(m^-))^\frac{1}{2}}\right), \label{def:E} 
		\end{equation}
		where
		\begin{align} \label{def:T}
			\mathcal{T}^\pm(f^\pm; r, m^+, m^-,\epsilon) &:= \frac{1 + f^\pm(m^\pm) + \mathcal{R}^2}{\left(\epsilon \mp \frac{m^+ - m^-}{2\epsilon r}\right)(1+f^\pm(m^\pm))^\frac{1}{2} - \mathcal{R} \left( \frac{2m^\pm}{r} + f^\pm(m^\pm) \right)^\frac{1}{2}}, \\ \widetilde{\mathcal{X}}^\pm(f^\pm;r,m^+,m^-,\epsilon) &:= \mathcal{R}(r,m^+,m^-,\epsilon) + \left( \frac{2m^\pm}{r} + f^\pm(m^\pm) \right)^\frac{1}{2} \mathcal{T}^\pm(f^\pm; r, m^+, m^-,\epsilon) \label{def:X0} \\ &= \frac{(1+f^\pm(m^\pm))^\frac{1}{2}(\frac{2m^\pm}{r} + f^\pm(m^\pm) - \mathcal{R}^2)}{\left(\epsilon \mp \frac{m^+ - m^-}{2\epsilon r}\right)(\frac{2m^\pm}{r} + f^\pm(m^\pm))^\frac{1}{2} - \mathcal{R}(1+f^\pm(m^\pm))^\frac{1}{2}}. \label{def:X}
		\end{align}
	\end{definition}
	
	We now show that, given suitable conditions on a piecewise smooth solution of (E-d), the thin shell ODE, which we define, must be satisfied along the thin shell. 
	\begin{proposition} \label{tsodeprop}
		Given a piecewise smooth solution of (E-d), $(\mathcal{Q}, r, \Omega^2,\rho,\mathbf{u}^\mu)$ with thin shell 
		\begin{equation}
			\Sigma := \{ (u(s), v(s)) : s \in [0,s_1) \},
		\end{equation}
		such that $\rho(u,v) > 0$ for all $(u,v) \in D^\pm(\Sigma,\mathcal{Q})$, and $(D^\pm(\Sigma,\mathcal{Q}),r,\Omega^2,\rho,\mathbf{u}^\mu)$ is a \underline{collapsing} $C^1$ solution of (E-d), then: there exist interior and exterior comoving coordinates, $(\tau^\pm(u,v),m^\pm(u,v)) \in C^{2}(D^\pm(\Sigma,\mathcal{Q}))$ (where $m^\pm(u,v)$ is the Hawking mass), extending to $C^2_{u,v}$ functions on $(D^\pm(\Sigma,\mathcal{Q}) \cup \Sigma) \setminus \partial^-\Sigma$, so that $\tau^\pm(s) := \tau^\pm(u(s),v(s)) \in C^2((0,s_1))$, $m^\pm(s) := m^\pm(u(s),v(s)) \in C^2((0,s_1))$. We impose that $m^+(s) > m^-(s)$ for $s \in (0,s_1)$. We define 
		\begin{equation} \label{epsdef}
			\epsilon(s) := \frac{1}{2} n^+(r)(s) + \frac{1}{2} n^-(r)(s),
		\end{equation}
		where $n^\pm$ are the outward-directed unit normals to $\Sigma$ in $D^\pm(\Sigma,\mathcal{Q})$ respectively. The jump conditions \eqref{prjump} imply that
		\begin{equation} \label{eps}
			\epsilon(s) = \frac{m^+(s) - m^-(s)}{4\pi r(s)^2 \sigma(s)},
		\end{equation}
		and therefore $\epsilon(s) > 0$ for $s \in (0,s_1)$. Defining
		\begin{equation}
			\mathscr{Q}^\pm := \{ (\tau(u,v),\chi(u,v)) : (u,v) \in D^\pm(\Sigma,\mathcal{Q}) \},
		\end{equation}
		we suppose that, transforming to comoving coordinates on $\mathscr{Q}^\pm$, $(\mathscr{Q}^\pm, r^\pm, \rho^\pm, m^\pm, f^\pm)$ is a $C^\infty$ collapsing solution of the Einstein-dust system. Assuming moreover that the thin shell is collapsing, $\dot{r}(s) < 0$, then $(r(s),m^+(s),m^-(s),\epsilon(s)) \in C^2((0,s_1))$ is a solution of the \underline{thin shell ODE}, for all $s \in (0,s_1)$:
		\begin{align}
			\dot{r}(s) &= \mathcal{R}(r(s),m^+(s),m^-(s),\epsilon(s)), \label{rdot0} \\ \dot{m}(s) &= \mathcal{M}^\pm(T^\pm, \rho^\pm, f^\pm; r(s),m^+(s),m^-(s),\epsilon(s)), \label{mpmdot0} \\ \dot{\epsilon}(s) &= \mathcal{E}(T^\pm, \rho^\pm, f^\pm; r(s),m^+(s),m^-(s),\epsilon(s)), \label{epsdot0}
		\end{align}
		and we also have
		\begin{equation}
			\dot{\tau}^\pm(s) = \mathcal{T}^\pm(f^\pm; r(s),m^+(s),m^-(s),\epsilon(s)). \label{Tdot0}
		\end{equation}
	\end{proposition}
	
	\begin{remark}
		We'll see that every thin shell formed at shell crossing in the \emph{collapsing} Einsten-dust system is initially collapsing, $\dot{r}(0) < 0$, and remains so for its time of existence. This formulation of the ODE will therefore be sufficient for any thin shell formed at shell crossing in collapsing dust.
	\end{remark}
	
	\begin{proof}
		For the existence of interior and exterior comoving coordinates, note that $D^\pm(\Sigma,\mathcal{Q})$ are simply connected, and $\rho>0$ on $D^\pm(\Sigma,\mathcal{Q})$ so $\mathbf{u}^\mu \in C^1(D^\pm(\Sigma,\mathcal{Q}))$, and the components of the 1-form
		\begin{equation}
			\beta^\pm_\mu := - g_{\mu \nu} (\mathbf{u}^\pm)^\nu
		\end{equation}
		in double null coordinates are $C^{1}_{u,v}$ by definition of the piecewise smooth solution. Therefore, there exists $\tau^\pm \in C^{2}(D^\pm(\Sigma,\mathcal{Q}))$ such that $\beta = d\tau^\pm$. From \eqref{dn:pm} we have $m \in C^{2}(D^\pm(\Sigma,\mathcal{Q}))$ and $\partial_u m < 0$, $\partial_v m > 0$ by the assumption that the $C^1$ solutions of (E-d) are collapsing. Morevoer, $\tau^\pm(u,v), m^\pm(u,v)$ extend to $C^2_{u,v}$ functions on $\Sigma$ from $D^\pm(\Sigma,\mathcal{Q}) \setminus \partial^-\Sigma$ by assumption of the regularity of piecewise smooth solutions, and the explicit expressions for the first derivatives: \eqref{dn:pm} and \eqref{velcomp} with \eqref{uvtau}.
		\begin{note}
			In the following, we suppress the $\pm$ notation: any expression not involving $\pm$ (or $[\cdot]$) is evaluated on $\Sigma$ as a boundary to $D^\pm(\Sigma,\mathcal{Q})$.
		\end{note}
		We first differentiate the relations \eqref{prjump} in $s$ along $\Sigma$. For the $u$ term, we have
		\begin{equation} \label{diffpur}
			\ddot{u} [\partial_u r] + \dot{u}^2 [\partial^2_u r] + \dot{u}\dot{v}[\partial_u \partial_v r] = \frac{d}{ds} (2\pi\sigma r).
		\end{equation}
		Using the wave equation for $r$, \eqref{rwave}, and expanding Raychaudhuri \eqref{Rayu}:
		\begin{equation}
			\partial^2_u r = \partial_u r \partial_u \log\Omega^2 - \frac{\pi r \rho}{(\mathbf{u}^u)^2},
		\end{equation}
		we substitute into \eqref{diffpur} to get
		\begin{equation} \label{dsigu}
			\frac{d}{ds}(2\pi\sigma r) = \frac{\ddot{u}}{\dot{u}} \cdot 2\pi\sigma r + \dot{u}^2 [\partial_u r \cdot\partial_u \log\Omega^2] - \pi r \dot{u}^2 \left[ \frac{\rho}{(\mathbf{u}^u)^2}\right]+ \pi r [\rho] - \frac{[m]}{2r^2}.
		\end{equation}
		Computing similarly for $v$, we have
		\begin{equation} \label{dsigv}
			\frac{d}{ds}(2\pi\sigma r) = \frac{\ddot{v}}{\dot{v}} \cdot 2\pi\sigma r - \dot{v}^2 [\partial_v r \cdot\partial_v \log\Omega^2] + \pi r \dot{v}^2 \left[ \frac{\rho}{(\mathbf{u}^v)^2}\right] - \pi r [\rho] + \frac{[m]}{2r^2},
		\end{equation}
		and summing \eqref{dsigu}, \eqref{dsigv} gives 
		\begin{equation} \label{dsigb}
			\frac{d}{ds}(4\pi \sigma r) = 2\pi\sigma r \left( \frac{\ddot{u}}{u} +  \frac{\ddot{v}}{v}\right) + \dot{u}^2 [\partial_u r \cdot\partial_u \log\Omega^2] - \dot{v}^2 [\partial_v r \cdot\partial_v \log\Omega^2] + \left[ \pi r \rho \left( \frac{\dot{v}^2}{(\mathbf{u}^v)^2} - \frac{\dot{u}^2}{(\mathbf{u}^u)^2}\right)\right].
		\end{equation}
		The middle terms in the right hand side of \eqref{dsigb} are equivalent to: 
		\begin{equation} \label{middle}
			\dot{u}^2 [\partial_u r \cdot\partial_u \log\Omega^2] - \dot{v}^2 [\partial_v r \cdot\partial_v \log\Omega^2] = \left[ \dot{r} \cdot \dot{u} \partial_u \log\Omega^2\right] - \left[ \frac{d}{ds}(\log\Omega^2) \cdot \dot{v} \partial_v r\right] = -4\pi\sigma \dot{r} + 2\pi\sigma r \frac{d}{ds}(\log\Omega^2),
		\end{equation}
		and the last term on the right hand side of \eqref{middle} cancels with the first term in the right hand side of \eqref{dsigb}: from the normalisation $g(\mathbf{u},\mathbf{u}) = -1$ we have $\log\Omega^2 = - \log\dot{u} - \log\dot{v}$ and therefore
		\begin{equation}
			\frac{d}{ds}(\log\Omega^2) = - \frac{\ddot{u}}{u} - \frac{\ddot{v}}{v}.
		\end{equation}
		We now derive an expression for the last term on the right hand side of \eqref{dsigb}. Differentiating $\tau(u(s),v(s))$ in $s$, we get
		\begin{equation} \label{taudot}
			\dot{\tau} = \frac{1}{2} \left( \frac{\dot{u}}{\mathbf{u}^u} + \frac{\dot{v}}{\mathbf{u}^v}\right) > 0
		\end{equation}
		where the inequality follows from the assumption $\mathbf{u}^\mu > 0$ (Definition \ref{piecesmooth}). In double null coordinates, the outward-directed normal to $\Sigma$ is $n^\sharp = \dot{v}\partial_v - \dot{u} \partial_u$, so that
		\begin{equation} \label{ntau}
			n(\tau) = \frac{1}{2} \left(\frac{\dot{v}}{\mathbf{u}^v} - \frac{\dot{u}}{\mathbf{u}^u}\right).
		\end{equation}
		From \eqref{dn:pm} we have
		\begin{align}
			\dot{m} &= 2\pi r^2 \rho (1+f)^\frac{1}{2} \left( \frac{\dot{v}}{\mathbf{u}^v} - \frac{\dot{u}}{\mathbf{u}^u} \right) \label{mdot0} \\ 
			n(m) &= 4 \pi r^2 \rho (1+f)^\frac{1}{2} \dot{\tau},
		\end{align}
		so from \eqref{ntau} we get $n(\tau) = (1+f)^{-\frac{1}{2}} \frac{\dot{m}}{4\pi r^2 \rho}$. Using $\dot{\tau}^2 - n(\tau)^2 = 1$, we get
		\begin{equation} \label{norm}
			\dot{\tau}^2 = 1 + \frac{1}{1+f} \left( \frac{\dot{m}}{4\pi r^2 \rho}\right)^2,
		\end{equation}
		which is simply the normalisation $g(\mathbf{u},\mathbf{u}) = -1$ in comoving coordinates. Using the expression for $n(\tau)$, we get
		\begin{equation} \label{last}
			\frac{\dot{v}^2}{(\mathbf{u}^v)^2} - \frac{\dot{u}^2}{(\mathbf{u}^u)^2} = 4 \dot{\tau} n(\tau) = 4 (1+f)^{-\frac{1}{2}} \frac{\dot{\tau} \dot{m}}{4\pi r^2 \rho}.
		\end{equation}
		Substituting \eqref{middle} and \eqref{last} into \eqref{dsigb}:
		\begin{equation} \label{dsigma}
			\frac{d}{ds}(4\pi\sigma r^2) = \frac{d}{ds} (4\pi\sigma r) r + 4 \pi \sigma \dot{r} r = \left[ \frac{\dot{m} \dot{\tau}}{(1+f)^\frac{1}{2}} \right].
		\end{equation}
		We have, using the expression for $n$ in double null coordinates
		\begin{equation} \label{nsquared}
			n(r)^2 = (\dot{v}\partial_v r - \dot{u} \partial_u r)^2 - (\dot{v}\partial_v r + \dot{u} \partial_u r)^2 + \dot{r}^2 = -4 \dot{u}\dot{v} \partial_u r \partial_v r = 1 - \frac{2m}{r} + \dot{r}^2.
		\end{equation}
		Differentiating $r(\tau,m)$ with respect to $s$: $\dot{r} = \dot{\tau} \partial_\tau r + \dot{m} \partial_m r$, we get
		\begin{equation} \label{mdot}
			\frac{\dot{m}}{4 \pi r^2 \rho} = \dot{r} + \left( \frac{2m}{r} + f \right)^\frac{1}{2} \dot{\tau}.
		\end{equation}
		Using the expressions for $n(\tau)$, $n(m)$, we can express $n^\sharp$ in comoving coordinates:
		\begin{equation} \label{ncom}
			n^\sharp = \frac{\dot{m}}{4\pi r^2 \rho} \frac{\partial_\tau}{(1+f)^\frac{1}{2}} + \dot{\tau}(1+f)^\frac{1}{2} 4\pi r^2 \rho \partial_m,
		\end{equation}
		and therefore derive the following expression for $n(r)$:
		\begin{align} \label{nr}
			(1+f)^\frac{1}{2}n(r) &= \dot{\tau}(1+f) -  \left( \left( \frac{2m}{r} + f\right)^\frac{1}{2} \dot{r} + \left( \frac{2m}{r} + f\right)\dot{\tau} \right) \\ &= \left(1- \frac{2m}{r}\right) \dot{\tau} - \left(\frac{2m}{r} + f\right)^\frac{1}{2}\dot{r}. \label{nrcomoving}
		\end{align}
		We can rearrange \eqref{nr} into an expression for $\dot{\tau}$ in terms of $\dot{r}$, $n(r)$:
		\begin{align} 
			\dot{\tau} &= \frac{\dot{r} \left( \frac{2m}{r} + f\right)^\frac{1}{2} + n(r)(1+f)^\frac{1}{2}}{1 - \frac{2m}{r}}, \label{dtau} \\ &= \frac{1 + f + \dot{r}^2}{n(r) (1+f)^\frac{1}{2} - \dot{r} \left( \frac{2m}{r} + f \right)^\frac{1}{2}}. \label{dtau2}
		\end{align}
		From \eqref{epsdef}, we derive \eqref{eps} using \eqref{prjump} and \eqref{nsquared}. Differentiating \eqref{eps} in $s$ and using \eqref{dsigma}, we get 
		\begin{align} \label{doteps}
			\dot{\epsilon} = \frac{\dot{m}^+}{m^+ - m^-}\left( \epsilon - \frac{\epsilon^2 \dot{\tau}^+}{(1+f^+)^\frac{1}{2}} \right) - \frac{\dot{m}^-}{m^+ - m^-}\left( \epsilon - \frac{\epsilon^2 \dot{\tau}^-}{(1+f^-)^\frac{1}{2}}\right).
		\end{align}
		From the jump condition \eqref{prjump}, and the definition of $n$ in double null coordinates, we get
		\begin{equation}
			[n(r)] = [\dot{v} \partial_v r - \dot{u} \partial_u r] = - 4\pi\sigma r,
		\end{equation}
		which we combine with \eqref{epsdef} to see that
		\begin{equation} \label{npmr}
			n^+(r) = \epsilon - 2\pi\sigma r, \ n^-(r) = \epsilon + 2\pi\sigma r.
		\end{equation}
		Squaring either of \eqref{npmr} and using \eqref{nsquared}, we get
		\begin{equation}  \label{rdot2}
			\dot{r}^2 = \epsilon^2 - 1 + \frac{m^+ + m^-}{r} + \frac{(m^+ - m^-)^2}{4\epsilon^2 r^2}.
		\end{equation}
		and \eqref{rdot0} follows from the assumption that $\dot{r}(s) < 0$. The equality \eqref{Tdot0} then follows from using $\dot{r} = \mathcal{R}$ in \eqref{dtau2}, \eqref{mpmdot0} follows from substituting \eqref{rdot0} and \eqref{Tdot0} into \eqref{mdot}, and \eqref{epsdot0} follows from substituting into \eqref{doteps}.
	\end{proof}
	
	We note that from \eqref{norm}, we can see that the following identity holds, which in terms of the thin shell ODE is equivalent to the normalisation of $s$ to be proper time along the thin shell.
	\begin{lemma}
		The functions $\mathcal{T}^\pm, \widetilde{\mathcal{X}}^\pm$ defined by \eqref{def:T}, \eqref{def:X} satisfy the identity
		\begin{align} \label{id:T}
			(\mathcal{T}^\pm(f^\pm; r, m^+, m^-,\epsilon))^2 &= 1 + \frac{1}{1+f^\pm(m^\pm)} (\widetilde{\mathcal{X}}^\pm(f^\pm; r, m^+, m^-,\epsilon))^2
		\end{align}
	\end{lemma}
	
	We now prove a partial converse of Proposition \ref{tsodeprop}.
	\begin{proposition} \label{prop:tsode-eik}
		Given $C^\infty$ solutions of the collapsing Einstein-dust system, $(\mathscr{Q}^\pm, r^\pm, \rho^\pm, m^\pm, f^\pm)$, let $T^\pm(r,m)$ be the function such that, for all $(\tau,m) \in \mathscr{Q}^\pm$: $T^\pm(r(\tau,m),m) = \tau$. Suppose that $(r(s), m^+(s), m^-(s), \epsilon(s)) \in C^\infty((0,s_1))$ is a solution of the thin shell ODE:
		\begin{align}
			\dot{r}(s) &= \mathcal{R}(r(s),m^+(s),m^-(s),\epsilon(s)), \\ \dot{m}(s) &= \mathcal{M}^\pm(T^\pm, \rho^\pm, f^\pm; r(s),m^+(s),m^-(s),\epsilon(s)), \\ \dot{\epsilon}(s) &= \mathcal{E}(T^\pm, \rho^\pm, f^\pm; r(s),m^+(s),m^-(s),\epsilon(s)).
		\end{align}
		Defining $\tau^\pm(s) := T(r(s),m^\pm(s))$ and
		\begin{equation}
			\Sigma^\pm := \{ (\tau^\pm(s),m^\pm(s)) : s \in (0,s_1)\} \subset \mathscr{Q}^\pm,
		\end{equation}
		suppose that, for open subsets $\mathscr{U}_0 \subset \mathscr{U}_-(\Sigma^-, \mathscr{Q}^-)$, $\mathscr{V}_0 \subset \mathscr{V}_-(\Sigma^+, \mathscr{Q}^+)$, with $\Sigma^- \subset \mathscr{U}_0$, $\Sigma^+ \subset \mathscr{V}_0$, and some $k \in \mathbb{N}+1$, we have $v(\tau,m) \in C^k_{\tau, m}(\mathscr{V}_0), u(\tau,m) \in C^k_{\tau, m}(\mathscr{U}_0)$ solutions of the eikonal equation \eqref{eiku}, \eqref{eikv} on $\mathscr{Q}^\pm$, respectively, with $\partial_\tau u, \partial_\tau v > 0$. Defining:
		\begin{equation}
			u^-(s) := u(\tau^-(s),m^-(s)), \qquad v^+(s) := v(\tau^+(s),m^+(s)),
		\end{equation}
		then there exist unique solutions of the eikonal equation $u(\tau,m) \in C^k_{\tau,m}( \mathscr{U}_+(\Sigma^+, \mathscr{Q}^+))$, $v(\tau,m) \in C^k_{\tau,m}( \mathscr{V}_+(\Sigma^-, \mathscr{Q}^-))$, attaining the data
		\begin{equation}
			u(\tau^+(s),m^+(s)) = u^-(s), \qquad v(\tau^-(s),m^-(s)) = v^+(s),
		\end{equation}
		respectively, and we have $\partial_\tau u, \partial_\tau v > 0$ on the respective domains. Defining also 
		\begin{equation}
			\mathscr{W}^- := \mathscr{U}_0 \cap \mathscr{V}_+( \Sigma^-,  \mathscr{Q}^-), \qquad \mathscr{W}^+ := \mathscr{V}_0 \cap \mathscr{U}_+(\Sigma^+, \mathscr{Q}^+),
		\end{equation}
		and 
		\begin{equation}
			\mathcal{Q}^\pm := \{ (u(\tau,m), v(\tau,m)) : (\tau,m) \in \mathscr{W}^\pm \setminus \Sigma^\pm \}, \qquad \Sigma := \{ (u^-(s),v^+(s)) : s \in (0,s_1) \},
		\end{equation}
		then: transforming to double null coordinates from $\mathscr{W}^\pm \setminus \Sigma^\pm$ to $\mathcal{Q}^\pm$,  $(\mathcal{Q}^\pm, r, \Omega^2, \rho,\mathbf{u}^\mu)$ are collapsing $C^{k-1}$ solutions of (E-d), with  extension in regularity to $\Sigma$. 
		
		Moreover, $r, \Omega^2$ extend to continous functions on $\mathcal{Q} := \mathcal{Q}^- \cup \Sigma \cup \mathcal{Q}^+$: $r, \Omega^2 \in C^{0,1}(\mathcal{Q})$, and for $s \in (0,s_1)$ the jump conditions \eqref{prjump}, \eqref{pomegajump} are satisfied on $\Sigma$ for $u(s) := u^-(s)$, $v(s) := v^+(s)$ and
		\begin{equation} \label{sigma}
			\sigma(s) := \frac{m^+(s) - m^-(s)}{4\pi r(s)^2 \epsilon(s)}.
		\end{equation}
	\end{proposition}
	
	\begin{remark}
		Given the statement of the proposition, for any $s \in (0,s_1)$ there exists $(u_1, v_1) \in I^+(u(s),v(s))$ such that the restriction to $[u(s), u_1) \times [v(s), v_1)$ is a piecewise smooth solution of (E-d), with $(\rho, \mathbf{u}^\mu)$ as in  Definition \ref{piecesmooth}, given by: $\sigma(s), u(s), v(s)$ and the $C^1$ solution of (E-d) on $([u(s), u_1) \times [v(s), v_1)) \cap \mathcal{Q}^\pm$.
	\end{remark}
	
	\begin{proof}
		The existence and uniqueness result for the solutions of the eikonal equation follow from Proposition \ref{eikprop}, and $(\mathcal{Q}^\pm,r,\Omega^2,\rho,\mathbf{u}^\mu)$ being a $C^{k-1}$ solution of (E-d) follows from Proposition \ref{dntransf}. 
		
		For $r^\pm(\tau,\chi)$ the solution of \eqref{Edcollapse} on $\mathscr{Q}^\pm$, we have $r^-(\tau^-(s),m^-(s)) = r^+(\tau^+(s),m^+(s)) = r(s)$, and, after transforming to double null coordinates, we therefore have $r \in C^{0, 1}(\mathcal{Q})$ (since $\tau^\pm(u,v), m^\pm(u,v)$ extend in $C^k$ to $\Sigma$ from $\mathcal{Q}^\pm$). From the boundary conditions for the eikonal equation on $\Sigma^\pm$, with $u(s) := u^-(s)$, $v(s) := v^+(s)$:
		\begin{align}
			\dot{u}(s) \dot{v}(s) &= \partial_\tau u(\tau^\pm(s),m^\pm(s)) \partial_\tau v(\tau^\pm(s),m^\pm(s)) \left( (\dot{\tau}^\pm(s))^2 - \frac{ (\dot{m}^\pm(s))^2 (4 \pi r(s)^2 \rho^\pm(\tau^\pm(s),m^\pm(s)))^{-2}}{1+f^\pm(m^\pm(s))} \right) \\ &= \partial_\tau u(\tau^\pm(s),m^\pm(s)) \cdot \partial_\tau v(\tau^\pm(s),m^\pm(s)),
		\end{align}
		where the second equality follows from \eqref{norm}. Hence, from the transformation to double null coordinates, in particular \eqref{omega}, we see that $\Omega^2 \in C^{0,1}(\mathcal{Q})$. 
		
		We now show that the jump conditions are satisfied. Continuity of $r$ implies that $[\dot{r}]=0$, so that $\dot{u}[\partial_u r](s) = - \dot{v} [\partial_v r](s)$. For $n$ the outward-directed unit normal, we have $n(r) = \dot{v} \partial_v r - \dot{u} \partial_u r$, so that $\dot{v} [\partial_v r] = \frac{1}{2} [n(r)] = - \dot{u} [\partial_u r]$. We compute $n(r)$ in comoving coordinates using the eikonal equation and the thin shell ODE to see that $n^\pm(r) = \epsilon \mp \frac{m^+ - m^-}{2\epsilon r}$, and therefore $\dot{u} [\partial_u r] = 2\pi \sigma r = - \dot{v} [\partial_v r]$, using the definition \eqref{sigma}. It remains to show that the jump conditions for $\log \Omega^2$ hold. We note that $\sigma(s)$ as defined solves the ODE \eqref{dsigma}: we now reverse the steps of deriving the ODE for $4\pi r^2 \sigma$ in the proof of Proposition \ref{tsodeprop}. Differentiating the jump conditions $\dot{u} [\partial_u r] = 2\pi \sigma r = - \dot{v} [\partial_v r]$, we get \eqref{dsigu}, \eqref{dsigv}, which we sum to get \eqref{dsigb}. The last term on the right hand side is then equal to $r^{-1}\frac{d}{ds}(4\pi\sigma r^2)$, and we can rearrange the middle terms as in \eqref{middle} where we use $[\dot{r}] = 0 = [\frac{d}{ds} (\log \Omega^2)]$, and we're left with 
		\begin{equation}
			\frac{d}{ds} (4\pi \sigma r) = \dot{r} \dot{u} [\partial_u \log \Omega^2] + r^{-1}\frac{d}{ds}(4\pi \sigma r^2),
		\end{equation}
		so that $\dot{u} [\partial_u \log \Omega^2] = - 4\pi \sigma = - \dot{v} [\partial_v \log \Omega^2]$.
	\end{proof}
	
	\subsubsection{The determinism condition} \label{detcondsec}
	
	In Section \ref{mdd}, we showed that smooth solutions of Einstein-dust are determined by the evolution of the fluid flow lines (which holds also for $C^{2}$ solutions of the Einstein-dust system in comoving coordinates). We introduce the \emph{determinism condition} which implies that the fluid flow lines emanating from $p \in \Sigma$ must lie in $J^-(p)$. Without such a condition, there would be fundamental non-uniqueness of piecewise smooth solutions attaining (a suitable notion of) initial data.
	\begin{definition} \label{detcond}
		Let $(\mathcal{Q},r,\Omega^2,\rho,\mathbf{u}^\mu)$ be a piecewise smooth solution with thin shell $\Sigma$, given by the functions $u(s), v(s) \in C^1([0,s_1)) \cap C^2((0,s_1))$,
		\begin{equation}
			\Sigma = \{ (u(s),v(s)) : s \in [0,s_1) \}.
		\end{equation}
		The thin shell satisfies the \underline{determinism condition} if, for all $s \in (0,s_1)$ such that $\rho^+(s) > 0$:
		\begin{equation} 
			(\mathbf{u}^v)^+(u(s),v(s)) < \dot{v}(s),
		\end{equation}
		and for all $s \in (s_0,s_1)$ such that $\rho^-(s) > 0$:
		\begin{equation}
			\dot{v}(s) < (\mathbf{u}^v)^-(u(s),v(s)).
		\end{equation}
	\end{definition}
	By continuity of $\Omega^2(u,v)$, and $\Omega^2(u(s),v(s)) \dot{u}(s) \dot{v}(s) = 1$, the conditions are equivalent respectively to $ \dot{u}(s) < (\mathbf{u}^u)^+(u(s),v(s))$ and $(\mathbf{u}^u)^-(u(s),v(s)) < \dot{u}(s)$. 

	\begin{lemma} \label{detcondlemma}
		Let $(\mathcal{Q},r,\Omega^2,\rho,\mathbf{u}^\mu)$ be a piecewise smooth solution with thin shell $\Sigma$ as in the statement of Proposition \ref{tsodeprop}. Then $\Sigma$ satisfies the determinism condition if and only if:
		\begin{equation} \label{detcond-X}
			\widetilde{\mathcal{X}}^+(f^+;r(s),m^+(s),m^-(s),\epsilon(s)) > 0, \qquad \widetilde{\mathcal{X}}^-(f^-;r(s),m^+(s),m^-(s),\epsilon(s)) < 0,
		\end{equation}
		(equivalent to $\dot{m}^-(s) < 0$, $\dot{m}^+(s) > 0$) for all $s \in (0,s_1)$. Moreover, \eqref{detcond-X} is equivalent to:
		\begin{equation} \label{key}
			n^+(r) = \epsilon - \frac{m^+ - m^-}{2\epsilon r} < (1+ f^+)^\frac{1}{2}, \qquad n^-(r) = \epsilon + \frac{m^+ - m^-}{2\epsilon r} > (1+f^-)^\frac{1}{2},
		\end{equation}
		and the bound for $n^-(r)$ is equivalent to
		\begin{equation} \label{rdotneg}
			\dot{r}(s) < - \left( \frac{2m^-(s)}{r(s)} + f^-(s) \right)^\frac{1}{2}.
		\end{equation}
	\end{lemma}
	
	\begin{proof}
		From \eqref{mdot0}, we see that (suppressing the $\pm$ and argument of $\widetilde{\mathcal{X}}$):
		\begin{equation} \label{X-dn}
			\widetilde{\mathcal{X}} = \frac{1}{2} (1+f)^\frac{1}{2} \left( \frac{\dot{v}}{\mathbf{u}^v} - \frac{\dot{u}}{\mathbf{u}^v}\right), 
		\end{equation}
		so \eqref{detcond-X} follows directly from the definition of the determinism condition, and clearly $\dot{m}$ has the corresponding sign directly from \eqref{mdot0}. 
		
		Now, we have $n^-(r)(s) > 0$ for all $s \in (0,s_1)$, since $\epsilon(s) > 0$ and $m^+(s) > m^-(s)$, so the denominator of $\widetilde{\mathscr{X}}^-$, as in \eqref{def:X}, is always positive, so $\widetilde{X}^-<0$ is equivalent to $\dot{r}^2 > \frac{2m^-}{r} + f^-$, and since $\dot{r} < 0$ this is equivalent to \eqref{rdotneg}, as well as $n^-(r)^2 > 1+ f^-$ by \eqref{nsquared}, equivalent to $n^-(r) > (1+f^-)^\frac{1}{2}$. 
		
		For $n^+(r) \geq 0$, we can use the same argument, the denominator of $\widetilde{\mathcal{X}}^+$ as in \eqref{def:X} is positive, and therefore $\widetilde{\mathcal{X}}^+ > 0$ is equivalent to $\dot{r}^2 < \frac{2m^+}{r} + f^+$, and by \eqref{nsquared} we must have $n^+(r) < (1+f^+)^\frac{1}{2}$. If $n^+(r) < 0$, then we must have $\frac{2m^+}{r} > 1$, which follows from e.g.~\eqref{nrcomoving}, and in this case we can express $\widetilde{X}^+$ as:
		\begin{equation}
			\widetilde{\mathcal{X}}^+ = \frac{\left( 1 + f^+ \right)^\frac{1}{2}}{1 - \frac{2m^+}{r}} \left( \dot{r} (1+f^+)^\frac{1}{2} + \left( \frac{2m^+}{r} + f^+\right)^\frac{1}{2} n^+(r) \right) \label{rpdchi1},
		\end{equation}
		which is clearly positive whenever $n^+(r) > 0$, $\frac{2m^+}{r} > 1$ (since $\dot{r} < 0$), so the determinism condition always holds when $n^+(r) < 0 < (1+f^+)^\frac{1}{2}$.
	\end{proof}
	
	\section{Thin shell development at shell crossing} \label{development}

	\subsection{Statement of the development theorem (Theorem \ref{devthm})} \label{statementdevthm}
	
	We now give the precise statement of  the `thin shell development theorem' (Theorem \ref{thm:1} in the introduction):
	\begin{theorem}[Thin shell development] \label{devthm}
		Let $(\mathcal{Q},r,\Omega^2,\rho,\mathbf{u}^\mu)$ be a globally hyperbolic $C^\infty$ collapsing solution of (E-d), with $(u_c,v_c) \in \overline{\mathcal{Q}} \setminus \mathcal{Q}$ a non-degenerate shell crossing singularity\footnote{See Section \ref{firstscsec} for Definitions \ref{firstscsing}, \ref{nondegensc}.}, then by Proposition \ref{regext} there exists $(u_0,v_0) \in I^-((u_c,v_c))$, and $\delta' > 0$ such that (after possibly uniquely extending $(\mathcal{Q},r,\Omega^2,\rho,\mathbf{u}^\mu)$): 
		\begin{equation}
			\mathcal{R}' := ( [u_0,u_c) \times [v_0,v_c + \delta']) \cup ( [u_0,u_c + \delta'] \times [v_0,v_c)) \subset \mathcal{Q}.
		\end{equation}
		\textbf{Existence.} Then there exists $0 < \delta_* \leq \delta'$ and a piecewise smooth solution\footnote{As in Definition \ref{pwisesmooth}.} of (E-d), $(\mathcal{R}_{\delta_*}, \tilde{r}, \tilde{\Omega}^2, \tilde{\rho}, \tilde{\mathbf{u}}^\mu)$, with underlying manifold
		\begin{equation}
			\mathcal{R}_{\delta_*} = [u_0,u_c + \delta_*) \times [v_0,v_c + \delta_*),
		\end{equation}		
		that extends the $C^\infty$ collapsing solution of (E-d): for all $(u,v) \in \mathcal{R}' \cap \mathcal{R}_{\delta_*}$,
		\begin{equation} \label{equivR'}
			\tilde{r}(u,v) = r(u,v), \qquad \tilde{\Omega^2}(u,v) = \Omega^2(u,v), \qquad \tilde{\rho}(u,v) = \rho(u,v), \ \tilde{\mathbf{u}}^u(u,v) = \mathbf{u}^u(u,v),
		\end{equation}
		and contains a thin shell $\Sigma_*$ forming at $(u_c, v_c)$, given by functions $u_*(s), v_*(s) \in C^1([0,\tilde{s}_*)) \cap C^\infty((0,\tilde{s}_*))$:
		\begin{equation} \label{sigmastar}
			\Sigma_* = \{ (u_*(s), v_*(s)) : s \in [0,\tilde{s}_*) \} \subset \mathcal{R}_*, 
		\end{equation}
		with $(u_*(0),v_*(0)) = (u_c,v_c)$. The thin shell $\Sigma_*$ satisfies the determinism condition\footnote{See Definition \ref{detcond}.}, and letting $\mathcal{R}_{\delta_*, f} := (\mathcal{R}_{\delta_*} \setminus \Sigma_*) \cup \{(u_c,v_c)\}$, we have:
		\begin{equation} \label{regthm1}
			\tilde{r}, \tilde{\Omega}^2, \tilde{\rho}, \tilde{\mathbf{u}}^\mu \in C^2_{u,v}( \mathcal{R}_{\delta_*} \setminus \Sigma_*), \qquad \tilde{r}, \tilde{\Omega}^2 \in C^1_{u, v}(\mathcal{R}_{\delta_*, f}), \qquad \log(\tilde{r}\tilde{\Omega})^2 \in C^2_{u, v}(\mathcal{R}_{\delta_*, f}), 
		\end{equation}
		\begin{equation} \label{regthm2}
			\tilde{\mathbf{u}}^\mu \in C^0_{u, v}(\mathcal{R}_{\delta_*, f}), \qquad \lim_{\mathcal{R}_* \setminus \Sigma_* \ni (u,v) \to (u_c,v_c)} \tilde{\rho}(u,v) = \infty. 
		\end{equation}
		\textbf{Uniqueness.} Given any $\delta \in (0,\delta_*]$, and a piecewise smooth solution of (E-d), $(\mathcal{R}_{\delta},\underline{r}, \underline{\Omega}^2, \underline{\rho}, \underline{\mathbf{u}}^\mu)$, with underlying manifold
		\begin{equation}
			\mathcal{R}_{\delta} = [u_0,u_c + \delta) \times [v_0,v_c + \delta),
		\end{equation}
		and thin shell $\underline{\Sigma}$ satisfying the determinism condition, that extends the $C^\infty$ collapsing solution of (E-d): for all $(u,v) \in \mathcal{R}' \cap \mathcal{R}_{\delta}$,
		\begin{equation} \label{equivR'2}
			\underline{r}(u,v) = r(u,v), \qquad \underline{\Omega}^2(u,v) = \Omega^2(u,v), \qquad \underline{\rho}(u,v) = \rho(u,v), \qquad \underline{\mathbf{u}}^\mu(u,v) = \mathbf{u}^\mu(u,v),
		\end{equation}
		then $\underline{\Sigma} = \Sigma_* \cap \mathcal{R}_{\delta}$, and for all $(u,v) \in \mathcal{R}_{\delta}$:
		\begin{equation}
			\underline{r}(u,v) = \tilde{r}(u,v), \qquad \underline{\Omega}^2(u,v) = \tilde{\Omega}^2(u,v), \label{equivuniq1}
		\end{equation}
		and therefore for all $(u,v) \in \mathcal{R}_{\delta} \setminus \underline{\Sigma}$:
		\begin{equation}
			 \underline{\rho}(u,v) = \tilde{\rho}(u,v), \qquad \underline{\mathbf{u}}^\mu(u,v) = \tilde{\mathbf{u}}^\mu(u,v). \label{equivuniq2}
		\end{equation}
	\end{theorem}
	
	We start by showing that the proof of Theorem \ref{devthm} reduces to an existence and uniqueness result for the `thin shell ODE' on a fixed maximal dust development, determined by the prior $C^\infty$ collapsing solution of (E-d), with initial data at a first shell crossing singularity. The ODE result is:
	\begin{proposition} \label{thinshellODEprop}
		Let $(\mathcal{Q},r,\Omega^2,\rho,\mathbf{u}^u)$ be a globally hyperbolic $C^\infty$ collapsing solution of (E-d) as in the statement of Theorem \ref{devthm}, with $\delta' > 0$ such that $\mathcal{R}' \subset \mathcal{Q}$. Let $(\tau(u,v),m(u,v))$ be comoving coordinates on $\mathcal{R}'$ (where $m(u,v)$ is the Hawking mass), gauge fixed by $\tau(u_0,v_0)=0$. Let $m_0 := m(u_c,v_c)$, $r_0 := r(u_c,v_c)$ and $f_0 := f(m_0)$. Define
		\begin{equation} \label{Dprop}
			\mathscr{D} := \{ (\tau(u,v),m(u,v)) : u = u_0, v_0 \leq v \leq v_c + \delta' \text{ or } u_0 \leq u \leq u_c + \delta', v = v_0  \},
		\end{equation}
		and let $(\mathscr{Q}_{MDD}, r, \rho, m ,f)$ be the maximal dust development of the induced data\footnote{See Remark \ref{bifurcatedata}.} on $\mathscr{D}$. Let $T(r,m)$ be the smooth function\footnote{See in Remark \ref{Tremark}.} such that, for all $(\tau,m) \in \mathscr{Q}_{MDD}$: $\tau = T(r(\tau,m),m)$.
		
		\textbf{Existence.} Then there exist numbers $s_*, \Delta_* > 0$, and functions $r_*, m^+_*, m^-_*, \epsilon_* \in C^0([0,s_*)) \cap C^\infty((0,s_*))$, with $\Delta_* = m^+_*(s_*) - m^-_*(s_*)$, such that, defining $\tau^\pm_*(s) := T(r_*(s),m^\pm_*(s))$, the functions satisfy
		\begin{equation} \label{detstat1}
			(\tau^\pm_*(s),m^\pm_*(s)) \in \mathscr{Q}_{MDD} \text{ for all } s \in (0,s_*),
		\end{equation}
		\begin{equation} \label{detstat2}
			\epsilon_*(s) > 0, \qquad \dot{r}_*(s) < 0, \qquad \dot{m}^+_*(s) > 0, \qquad \dot{m}^-_*(s) < 0, \qquad \text{for all } s \in (0,s_*),
		\end{equation}
		\begin{equation} \label{detstat3}
			r_*(0) = r_0, \qquad m^\pm_*(0) = m_0, \qquad \epsilon_*(0) = (1+f_0)^\frac{1}{2},
		\end{equation}
		and $(r_*(s), m^+_*(s), m^-_*(s), \epsilon_*(s))$ solves the \underline{thin shell ODE} for all $s \in (0,s_*)$:
		\begin{align}
			\dot{r}_*(s) &= \mathcal{R}(r_*(s),m^+_*(s),m_*^-(s),\epsilon_*(s)), \label{ODE1} \\ 
			\dot{m}_*^\pm(s) &= \mathcal{M}^\pm(T, \rho, f; r_*(s),m^+_*(s),m_*^-(s),\epsilon_*(s)), \label{ODE2} \\ \dot{\epsilon}_*(s) &= \mathcal{E}(T, \rho, f; r_*(s),m^+_*(s),m_*^-(s),\epsilon_*(s)),  \label{ODE3}
		\end{align}
		
		\textbf{Uniqueness.} Given a number $s_1>0$ and functions $\underline{r}, \underline{m}^+, \underline{m}^-, \underline{\epsilon} \in C^0([0,s_1)) \cap C^2((0,s_1))$, with $\underline{m}^+(s_1) - \underline{m}^-(s_1) \leq \Delta_*$, and satisfying:
		\begin{equation} \label{cond1}
			(T(\underline{r}(s),\underline{m}^\pm(s)),\underline{m}^\pm(s)) \in \mathscr{Q}_{MDD} \text{ for all } s \in (0,s_1),
		\end{equation}
		\begin{equation} \label{cond2}
			\underline{\epsilon}(s) >0, \qquad \dot{\underline{r}}(s) < 0, \qquad \underline{\dot{m}}^+(s) > 0, \qquad \underline{\dot{m}}^-(s) < 0, \qquad \text{for all } s \in (0,s_1),
		\end{equation}
		\begin{equation} \label{cond3}
			\underline{r}(0) = r_0, \qquad \underline{m}^\pm(0) = m_0, \qquad \underline{\epsilon}(0) = (1+f_0)^\frac{1}{2},
		\end{equation}
		and for all $s \in (0,s_1)$:
		\begin{align} 
			\underline{\dot{r}}(s) &= \mathcal{R}(\underline{r}(s),\underline{m}^+(s),\underline{m}^-(s),\underline{\epsilon}(s)), \label{rode} \\ 
			\underline{\dot{m}}^\pm(s) &= \mathcal{M}^\pm(T, \rho,f; \underline{r}(s),\underline{m}^+(s),\underline{m}^-(s),\underline{\epsilon}(s)), \label{mode} \\ \underline{\dot{\epsilon}}(s) &= \mathcal{E}(T, \rho,f; \underline{r}(s),\underline{m}^+(s),\underline{m}^-(s),\underline{\epsilon}(s)), \label{eode}
		\end{align}
		then $s_1 \leq s_*$, and for all $s \in [0,s_1)$:
		\begin{equation}
			\underline{r}(s) = r_*(s), \qquad \underline{m}^\pm(s) = m^\pm_*(s), \qquad \underline{\epsilon}(s) = \epsilon_*(s). 
		\end{equation}
	\end{proposition}

	\subsection{Equivalence of the development theorem (Theorem \ref{devthm}) to an existence and uniqueness result for the thin shell ODE (Proposition \ref{thinshellODEprop})} \label{equiv}
	
	In this section, we prove that Theorem \ref{devthm} follows from Proposition \ref{thinshellODEprop}. We then prove Proposition \ref{thinshellODEprop} in Section \ref{proof}.
	\begin{proposition} \label{equivprop}
		Proposition \ref{thinshellODEprop} implies Theorem \ref{devthm}.
	\end{proposition}
	
	See the first half of Section \ref{introdev} for a brief outline of the proof. 
	
	\begin{proof}
		\textbf{Existence. } We first prove that the existence part of Proposition \ref{thinshellODEprop} implies the existence part of Theorem \ref{devthm}. Indeed, let $(r_*(s),m^+_*(s),m^-_*(s),\epsilon_*(s)) \in C^0([0,s_*]) \cap C^\infty((0,s_*))$ be the solution of the thin shell ODE \eqref{ODE1}, \eqref{ODE2}, \eqref{ODE3}, as given in the statement of Proposition \ref{thinshellODEprop}. We start by choosing the number $\delta_* > 0$ for which we prove Theorem \ref{devthm}.
		
		\textbf{1. Choosing $\delta_*$.} Given the induced data on $\mathscr{D}$ for the eikonal equation, by Proposition \ref{eikprop} there exist unique solutions:
		\begin{equation}
			u(\tau,m) \in C^\infty(\mathscr{U}_+(\mathscr{D}, \mathscr{Q}_{MDD} \cup \mathscr{C})), \ v(\tau,m) \in C^\infty(\mathscr{V}_+(\mathscr{D}, \mathscr{Q}_{MDD} \cup \mathscr{C})),
		\end{equation}
		of \eqref{eiku} and \eqref{eikv} respectively. By Lemma \ref{nondeglemma}, we can choose $\Delta_1 > 0$ sufficiently small such that: $[m_0 - \Delta_1, m_0 + \Delta_1] \subset \mathscr{C}_{\chi}$, and $\tau_c'(m) < 0$ for $m \in [m_0 - \Delta_1, m_0)$, $\tau_c'(m) > 0$ for $m \in (m_0, m_0 + \Delta_1]$. Since $m(u,v) \in C^\infty(\mathcal{R}')$ extends continuously to $\overline{\mathcal{R}'}$, we can choose $0 < \delta_1 \leq \delta'$ such that
		\begin{equation}
			m(u_c, v_c + \delta_1) - m(u_c + \delta_1, v_c) = \Delta_1. 
		\end{equation} 
		We now define $0 < \delta_* \leq \delta'$ to satisfy, given $\Delta_* > 0$ of Proposition \ref{thinshellODEprop}:
		\begin{equation} \label{deltastar}
			m(u_c, v_c + \delta_*) - m(u_c + \delta_*, v_c) = \min\{\Delta_1, \Delta_*\}. 
		\end{equation}
		Let $\tau_{+, \delta_*}(m)$ be the unique outgoing null curve (i.e.~solving \eqref{outnull}) in $\mathscr{Q}_{MDD} \cup \mathscr{C}$ with $\tau_{+, \delta_*}( m(u_c + \delta_*, v_0) ) = \tau( u_c + \delta_*, v_0)$. Then there exists $m_{1} \in (m(u_c + \delta_*, v_0), m_0)$ such that $\tau_{+, \delta_*}(m) \in C^\infty_m( [m(u_c + \delta_*, v_0), m_{1}])$, and $\tau_{+, \delta_*}(m_1) = \tau_c(m_1)$, with $\{ (\tau_{+, \delta_*}(m),m) : m \in [m(u_c + \delta_*, v_0), m_1) \} \subset \mathscr{Q}_{MDD}$. Let $\tau_{\mathscr{D}}(m) \in C^0([m(u_c + \delta', v_0), m(u_0, v_c + \delta')])$ be such that $\mathscr{D} = \{ (\tau_{\mathscr{D}}(m), m) : m \in [m(u_c + \delta', v_0), m(u_0, v_c + \delta')] \}$. We then have
		\begin{align}
			\mathscr{U}_{\delta_*, -} &:= \{ (\tau,m) \in \mathscr{U}(\mathscr{D}, \mathscr{Q}_{MDD} \cup \mathscr{C}) : m < m_0, \ u(\tau, m) < u_c + \delta_* \} \\ &=  \{ (\tau, m) : m \in [m(u_c + \delta_*, v_0), m_1), \tau \in [\tau_{\mathscr{D}} (m), \tau_{+, \delta_*}(m)] \} \cup \{ (\tau, m) : m \in [m_1, m_0), \tau \in [\tau_{\mathscr{D}}, \tau_c(m)) \},
		\end{align}
		analogously, let $\tau_{-, \delta_*}(m)$ be the unique ingoing null curve (solving \eqref{innull}) in $\mathscr{Q}_{MDD} \cup \mathscr{C}$ with $\tau_{-, \delta_*}( m(u_0, v_c + \delta_*) ) = \tau( u_0, v_c + \delta_*)$. There exists $m_0 < m_2 < m(u_0, v_c + \delta_*)$ such that 
		\begin{align}
			\mathscr{V}_{\delta_*, +} &:= \{ (\tau,m) \in \mathscr{V}(\mathscr{D}, \mathscr{Q}_{MDD} \cup \mathscr{C}) : m > m_0, \ v(\tau, m) < v_c + \delta_* \} \\ &= \{ (\tau, m) : m \in (m_2, m(u_0, v_c + \delta_*)], \tau \in [\tau_{\mathscr{D}}(m), \tau_{-, \delta_*}(m)] \} \cup \{ (\tau, m) : m \in (m_0, m_2], \tau \in [\tau_{\mathscr{D}}, \tau_c(m)) \}.
		\end{align}
		There must exist $\tilde{s}_{*, -} > 0$ such that: 
		\begin{equation} \label{nulldom}
			(\tau_*^-(s), m_*^-(s)) \in \mathscr{U}_{\delta_1, -}, \qquad (\tau_*^+(s), m_*^+(s)) \in \mathscr{V}_{\delta_*, +}, 
		\end{equation}
		for all $s \in (0, \tilde{s}_{*, -})$. Now, for all $s \in (0,s_*)$ such that \eqref{nulldom} holds, we have $m^+_*(s) - m^-_*(s) < \Delta_*$ by definition of $\delta_*$, \eqref{deltastar}, and monotonicity of $m$ (in double null coordinates $\partial_u m(u,v) < 0$, $\partial_v m(u,v) > 0$). Hence, there must exist $0 < \tilde{s}_* \leq s_*$ such that \eqref{nulldom} holds for all $s \in (0,\tilde{s}_*)$, but not for $s = \tilde{s}_*$. We can now define, for $s \in [0,s_*)$:
		\begin{equation} \label{ustarvstar}
			u_*(s) := u(\tau^-_*(s), m_*^-(s)), \qquad v_*(s) := v(\tau^+_*(s), m^+_*(s)),
		\end{equation}
		with $u_*(s), v_*(s) \in C^0([0,s_*)) \cap C^\infty((0,\tilde{s}_*))$, $(u_*(s),v_*(s)) \in \mathcal{R}_{\delta_*}$ for $s \in [0,\tilde{s}_*)$, and $\lim_{s \to \tilde{s}_*} (u_*(s), v_*(s)) \notin \mathcal{R}_*$. We define the `exterior' and `interior' of the thin shell respectively as
		\begin{equation}
			\Sigma_ *^\pm := \{ (\tau_*^\pm(s),m_*^\pm(s)) : s \in (0,\tilde{s}_*)\} \subset \mathscr{Q}_{MDD} \cup \{ (\tau_c(m_0),m_0)\},
		\end{equation}
		and define $\Sigma_*$ by \eqref{sigmastar} with \eqref{ustarvstar}. Differentiating \eqref{ustarvstar} with respect to $s$, we see that, using the eikonal equation \eqref{eiku}, \eqref{eikv} respectively:
		\begin{align}
			\dot{u}_*(s) &= \partial_\tau u (\tau^-_*(s), m_*^-(s)) \left[ (\dot{\tau}^-(s))^2 + ((\dot{\tau}^-(s))^2 - 1)^\frac{1}{2} \right], \label{dotustar} \\ \dot{v}_*(s) &= \partial_\tau v (\tau^+_*(s), m_*^+(s)) \left[ (\dot{\tau}^+(s))^2 + ((\dot{\tau}^+(s))^2 - 1)^\frac{1}{2} \right], \label{dotvstar}
		\end{align}
		where we've used $\dot{m}^-(s) < 0$, $\dot{m}^+(s) > 0$ and 
		\begin{equation} \label{comnorm}
			(\dot{\tau}_*^\pm(s))^2 = 1 + \frac{[\dot{m}^\pm_*(s) \partial_m r(\tau^\pm_*(s), m_*^\pm(s))]^2}{1 + f(m_*^\pm(s))}.
		\end{equation}
		We have $\tau^\pm_*(s) \in C^1([0,s_*))$, and the initial data  \eqref{detstat3} implies that $\dot{\tau}^\pm_*(0) = 1$ (using \eqref{def:T}). Proposition \ref{eikprop} implies that $\partial_\tau u$, $\partial_\tau v$ extend continuously to $(\tau_c(m_0),m_0)$ from their respective domains, and therefore \eqref{dotustar}, \eqref{dotvstar} imply that $u_*, v_* \in C^1([0,s_*))$, and $\dot{u}_*(s), \dot{v}_*(s) > 0$ for all $s \in [0,s_*)$ (since $\dot{\tau}^\pm(s) > 1$ for $s \in (0,s_*)$).
		
		\textbf{2. Constructing the piecewise smooth solution on $\mathcal{R}_*$ by identifying double null coordinates on $\Sigma_*^\pm$. } We define the following subsets of $\mathscr{Q}_{MDD}$:
		\begin{equation}
			\mathscr{U}_{*, +} := \mathscr{U}( \Sigma_*^+, \mathscr{Q}_{MDD}) \cap \{ v(\tau,m) < v_*(\tilde{s}_*)\}, \label{Uplus} \qquad \mathscr{V}_{*, -} := \mathscr{V}( \Sigma_*^-, \mathscr{Q}_{MDD}) \cap \{ u(\tau,m) < u_*(\tilde{s}_*)\},
		\end{equation}
		By Proposition \ref{eikprop}, there exists a unique solution of the eikonal equation \eqref{eiku}: $\tilde{u}(\tau,m) \in C^\infty(\mathscr{U}_{*, +})$, given  
		\begin{equation} \label{sigmaplusdata}
			\tilde{u}(\tau^+_*(s), m^+_*(s)) = u_*(s), \text{ for } s \in (0,\tilde{s}_*).
		\end{equation}
		Similarly, there exists a unique solution of \eqref{eikv}, $\tilde{v}(\tau,m) \in C^\infty(\mathscr{V}_{*, -})$, given 
		\begin{equation} \label{sigmaminusdata}
			\tilde{v}(\tau^-_*(s), m^-_*(s)) := v_*(s), \text{ for } s \in (0,\tilde{s}_*).
		\end{equation}
		By Proposition \ref{prop:tsode-eik}, we have $\partial_\tau \tilde{u}, \partial_\tau \tilde{v} > 0$ on $\mathscr{U}_{\star, +}, \mathscr{V}_{*, -}$ respectively, and transforming to double null coordinates on 
		\begin{equation}
			\mathcal{R}^+ := \{ (\tilde{u}(\tau,m), v(\tau,m)) : (\tau,m) \in \mathscr{U}_{*, +}\} \setminus \Sigma_*, \qquad \mathcal{R}^- := \{ (u(\tau,m), \tilde{v}(\tau,m)) : (\tau,m) \in \mathscr{V}_{*, -}\} \setminus \Sigma_*,
		\end{equation}
		as in Proposition \ref{prop:tsode-eik}, $(\mathcal{R}^\pm,\tilde{r},\tilde{\Omega}^2,\tilde{\rho},\tilde{\mathbf{u}}^\mu)$ are $C^\infty$ collapsing solutions of (E-d), with $C^\infty$ extension to $\Sigma_* \setminus \{(u_c,v_c)\}$, $r, \Omega^2$ extend to $C^{0,1}$ functions on $(u_c, u_c + \delta_*) \times (v_c, v_c + \delta_*)$ and the jump conditions \eqref{prjump}, \eqref{pomegajump} are satisfied along $\Sigma_* \setminus \{(u_c,v_c)\}$. In particular, we have for all $s \in (0,s_*)$:
		\begin{equation}
			\dot{u}_*(s) [ \partial_u \log(\tilde{r}\tilde{\Omega})^2 ](s) = 0 = \dot{v}_*(s) [ \partial_v \log(\tilde{r}\tilde{\Omega})^2 ](s). \label{promegajump}
		\end{equation} 
	
		We now show that the $C^\infty$ solutions of (E-d) defined on $\mathcal{R}^\pm$ \emph{extend} in $C^2_{u,v}$ the given $C^\infty$ solution of (E-d) on $\mathcal{R}'$. Indeed, by Proposition \ref{regext} the functions $r, \Omega^2, \rho, \mathbf{u}^\mu \in C^\infty(\mathcal{R}')$, for the given $C^\infty$ collapsing solution of (E-d) on $\mathcal{R}'$, extend in $C^\infty$ to $\overline{\mathcal{R}'} \setminus (u_c,v_c)$. We then extend the functions $\tilde{r}, \tilde{\Omega}^2, \tilde{\rho}, \tilde{\mathbf{u}}^\mu$ to the larger set $\mathcal{R}_{\delta_*}$, by imposing \eqref{equivR'} for $(u,v) \in \overline{\mathcal{R}'} \cap \mathcal{R}_{\delta_*}$. We now show that \eqref{regthm1}, \eqref{regthm2} are satisfied for this definition of $\tilde{r}, \tilde{\Omega}^2, \tilde{\rho}, \tilde{\mathbf{u}}^\mu$. 
		
		Let $\mathscr{R}' := \{ (\tau,m) \in \mathscr{U}(\mathscr{D}, \mathscr{Q}_{MDD} \cup \mathscr{C}) \cap \mathscr{V}(\mathscr{D}, \mathscr{Q}_{MDD} \cup \mathscr{C}) : u(\tau,m) < u_c + \delta_*, \ v(\tau,m) < v_c + \delta_* \}.$ Note in particular that $(\tau_c(m_0), m_0) \in \mathscr{R}'$, as well as the unique future-directed ingoing and outgoing null curves from $(\tau_c(m_0),m_0)$. Defining also,
		\begin{equation}
			\mathscr{R}_{\delta_*} := \mathscr{R}' \cup \mathscr{U}_{*, +} \cup \mathscr{V}_{*,-},
		\end{equation}
		we extend $\tilde{u}(\tau,m), \tilde{v}(\tau,m)$ to functions on $\mathscr{R}_{\delta_*}$, by equating $\tilde{u}(\tau,m) = u(\tau,m)$ for $(\tau,m) \in \mathscr{U}(\mathscr{D}, \mathscr{Q}_{MDD})$, and $\tilde{v}(\tau,m) = v(\tau,m)$ for $(\tau,m) \in \mathscr{V}(\mathscr{D}, \mathscr{Q}_{MDD})$. Then \eqref{regthm1}, \eqref{regthm2} follow if, as defined, $\tilde{u}, \tilde{v} \in C^3_{\tau,m}(\mathscr{R}_{\delta_*})$. (This follows by the transformation to double null coordinates, see Lemma \ref{dntransf} and Proposition \ref{regext}.)
		\begin{lemma} \label{eiklemma}
			$\tilde{u}, \tilde{v} \in C^3_{\tau,m}(\mathscr{R}_{\delta_*})$.
		\end{lemma}
		\begin{proof}
			By definition of $\tilde{u}, \tilde{v}$, we have $\tilde{u}, \tilde{v} \in C^0_{\tau,m}(\mathscr{R}_{\delta_*})$. Moreover, defining, with $\tau_0 := \tau_c(m_0)$,
			\begin{equation}
				\mathscr{CH}^- := \{ (\tau,m) \in \mathscr{R}_{\delta_*} : \tilde{u}(\tau,m) > u_c, \tilde{v}(\tau,m) = v_c \}, \qquad \mathscr{CH}^+ := \{ (\tau,m) \in \mathscr{R}_{\delta_*} : \tilde{u}(\tau,m) = u_c, \tilde{v}(\tau,m) > v_c \},
			\end{equation}
			we have $\tilde{u}, \tilde{v} \in C^\infty_{\tau,m}(\mathscr{R}_{\delta_*} \setminus (\mathscr{CH}^- \cup \{(\tau_0, m_0)\} \cup \mathscr{CH}^+))$, and $\tilde{u} \in C^\infty_{\tau,m}(\mathscr{U}(\mathscr{D}, \mathscr{Q}_{MDD}))$, $\tilde{v} \in C^\infty_{\tau,m}(\mathscr{V}(\mathscr{D}, \mathscr{Q}_{MDD}))$. It  suffices to show that, for $1 \leq k \leq 3$, $\partial_\tau^k \tilde{u}(\tau_*^+(s),m_*^+(s)) \to \partial_\tau^k \tilde{u} (\tau_0, m_0)$, and $\partial_\tau^k \tilde{v}(\tau_*^-(s),m_*^-(s)) \to \partial_\tau^k \tilde{v} (\tau_0, m_0)$, as $s \to 0$. Then $\tilde{u}$ extends in $C^3_{\tau,m}$ to $\mathscr{CH}^+$, and $\tilde{v}$ extends in $C^3_{\tau,m}$ to $\mathscr{CH}^-$, and we'd have $\tilde{u}, \tilde{v} \in C^3_{\tau,m}(\mathscr{R}_{\delta_*})$.
			
			We have, for all $k \in \mathbb{N}$, $\partial_\tau^k \tilde{u}(\tau_*^-(s),m_*^-(s)) \to \partial_\tau^k \tilde{u} (\tau_0, m_0)$ as $s \to 0$. We differentiate the definitions \eqref{ustarvstar} and \eqref{sigmaplusdata} to see that
			\begin{equation} \label{id:ptauu}
				\frac{\partial_\tau \tilde{u}(\tau_*^+(s),m_*^+(s))}{\partial_\tau \tilde{u}(\tau_*^-(s),m_*^-(s))} = \frac{\dot{\tau}_*^-(s) + \left[ (\dot{\tau}_*^-(s))^2 - 1 \right]^\frac{1}{2}}{\dot{\tau}_*^+(s) - \left[ (\dot{\tau}_*^+(s))^2 - 1 \right]^\frac{1}{2}},
			\end{equation}
			where we've used $\dot{m}^-(s) < 0$, $\dot{m}^+(s) > 0$ for $s > 0$ and \eqref{comnorm}. We have $\dot{\tau}^\pm_*(0)=1$, so $\partial_\tau \tilde{u}(\tau_*^+(s),m_*^+(s)) \to \partial_\tau \tilde{u}(\tau_0,m_0)$ as $s \to 0$. Now, using the expression for $\partial_u \log(r\Omega)^2$ in terms of the solution of the eikonal equation, i.e.~combining \eqref{pomega} with \eqref{pr}, and using \eqref{promegajump}, we see that $\partial_\tau^2 \tilde{u}(\tau_*^+(s),m_*^+(s)) \to \partial_\tau^2 \tilde{u}(\tau_0,m_0)$. Differentiating $[ \partial_u \log(\tilde{r} \tilde{\Omega})^2] = 0$ in $s$, and using $[ \partial_u \partial_v \log (\tilde{r} \tilde{\Omega})^2](s) = - \frac{\Omega_*^2(s)}{r_*(s)^3} [\tilde{m}(s)] \to 0$ as $s \to 0$, we see that $[ \partial_u^2 \log(\tilde{r}\tilde{\Omega})^2] \to 0$. From the  expression \eqref{pusqrom}, we get $\partial_\tau^3 \tilde{u}(\tau_*^+(s),m_*^+(s)) \to \partial_\tau^3 \tilde{u}(\tau_0,m_0)$. The argument is analogous to show that $1 \leq k \leq 3$, $\partial_\tau^k \tilde{v}(\tau_*^+(s),m_*^+(s)) \to \partial_\tau^k \tilde{v} (\tau_0, m_0)$. 
		\end{proof}
		
		We've constructed a piecewise smooth solution of (E-d) on $[u_c, u_c + \delta_*) \times [v_c, v_c + \delta_*)$, with thin shell $\Sigma_*$ and $(\rho,\mathbf{u}^\mu)$ as in Definition \ref{piecesmooth}, with $\sigma_*(s) := \frac{m^+_*(s) - m_*^-(s)}{4\pi r_*(s)^2 \epsilon_*(s)}$. The integrated weak formulation of the Einstein equations, \eqref{rwaveweakv}--\eqref{Rayvweak}, hold on the restriction to $[u_c, u_c + \delta_*) \times [v_c, v_c + \delta_*)$ by the jump conditions (extending continuously to $s=0$), and they moreover hold on $(\mathcal{R}_{\delta_*},\tilde{r},\tilde{\Omega}^2,\tilde{\rho},\tilde{\mathbf{u}}^u)$ since $(\mathcal{R}', r,\Omega^2,\rho,\mathbf{u}^u)$ is a $C^\infty$ solution of (E-d). This finishes the proof of existence of the piecewise smooth solution of (E-d). 
			                                                                              
		\textbf{Uniqueness. } We now prove that the uniqueness part of Proposition \ref{thinshellODEprop} implies the uniqueness part of Theorem \ref{devthm}. For any $\delta \in (0,\delta_*]$, let $(\mathcal{R}_\delta, \underline{r}, \underline{\Omega}^2, \underline{\rho}, \underline{\mathbf{u}}^u)$ be any piecewise smooth solution of (E-d), with underlying manifold
		\begin{equation}
			\mathcal{R}_{\delta} = [u_0,u_c + \delta) \times [v_0,v_c + \delta),
		\end{equation}
		and thin shell $\underline{\Sigma}$ satisfying the determinism condition, such that \eqref{equivR'2} holds for all $(u,v) \in \mathcal{R}' \cap \mathcal{R}_{\delta}$.
		
		\textbf{1. Constructing comoving coordinates; the determinism condition implies embedding into $\mathscr{Q}_{MDD}$. } Let $\underline{u}(s),\underline{v}(s) \in C^0([0,s_1]) \cap C^2(0,s_1)$ be functions such that
		\begin{equation}
			\underline{\Sigma} = \{ (\underline{u}(s),\underline{v}(s)) : s \in (0,s_1) \} \subset (u_c, u_c + \delta) \times (v_c, v_c + \delta),
		\end{equation}
		where we must have $(\underline{u}(0),\underline{v}(0)) = (u_c,v_c)$. We consider a neigbourhood of the exterior of the thin shell in $J^+(u_c,v_c)$: choosing $\delta_2 \in (0,\delta)$ such that there exists $s_2 \in (0,s_1)$ with $\underline{v}(s_2) = v_c + \delta_2$, we define
		\begin{equation}
			\mathcal{R}_{\delta_1}^+ := \{ u \geq u_c, v \leq v_c + \delta_2 \} \cap \{ (u,v) : \exists s \in (0,s_2] \text{ s.t. } \underline{u}(s) = u, \ \underline{v}(s) \leq v \}.
		\end{equation}
		We want to show $\rho(u,v) > 0$ for all $(u,v) \in \mathcal{R}_{\delta_1}^+$, and that $\mathcal{R}_{\delta_1}^+ \cap \{ u >u_c \} \subset D_{ac}^+( \{ u= u_c, v \in (v_c, v_c + \delta_2)\} )$. Indeed, define 
		\begin{equation}
			\mathcal{F}_{\delta_2} := D_{ac}^+( \{ u= u_c, v \in (v_c, v_c + \delta_2)\} ),
		\end{equation}
		then we have, by continuity of $\rho$ and uniqueness/existence of fluid flow lines, $\mathcal{F}_{\delta_2} \cap \mathcal{R}_{\delta_1}^+ \neq \emptyset$. Now, we will use the assumed regularity of the piecewise smooth solution, in particular that $\rho$ must extend continuously to $\underline{\Sigma}$ in $\mathcal{R}_{\delta_1}^+$, to show that $\underline{\Sigma} \cap \mathcal{R}_{\delta_1}^+$ must be foliated by flow lines from $\{ u = u_c, v \in (v_c, v_c + \delta_2)\}$. 
		
		We have $\rho(u_c, v_c+\delta_2) > 0$, by continuity and $\rho > 0$ in $\mathcal{R}'$. Suppose that there exists $u_2 \in (u_c, \underline{u}(s_2)]$ such that $\rho(u_2,v_c+\delta_2) = 0$, with $\rho(u,v_c+\delta_2)>0$ for $u \in [u_c,u_2)$. For all $u \in [u_c,u_2)$, there exist unique past-directed fluid flow lines from $(u,v_c+\delta_2)$. By the determinism condition, the fluid flow lines cannot emanate from $\underline{\Sigma}$, so they must intersect $\{ u = u_c, v \geq v_c \}$. Suppose that there exists such a fluid flow line with past limit point $\{ u = u_c, v= v_c\}$. Then by existence and uniqueness of fluid flow lines, we must have $\mathcal{F}_{\delta_2} \cap \{ v = v_c + \delta_2\} \subset \{ u \in (u_c,u_2), v = v_c + \delta_2\}$. For any $\tilde{v} \in (v_c, v_c + \delta_2)$, then $\mathcal{F} : =\mathcal{R}' \cup (D^+_{ac}(\{ u= u_c, v \in (\tilde{v}, v_c + \delta_2)\}) \cap \mathcal{R}_{\delta_1}^+ )$ is simply connected and orientable, so by Lemma \ref{comovinglemma}, and the assumed regularity of piecewise smooth solutions, there exists a unique choice of comoving time coordinate $\tau(u,v) \in C^2(\mathcal{F})$, and $\inf_{(u,v) \in \mathcal{F}} \rho(u,v), \mathbf{u}_*(r) > 0$, so the Hawking mass $m(u,v) \in C^2(\mathcal{F})$ satisfies $\partial_u m(u,v) < 0$, $\partial_v m(u,v) >0$ for all $(u,v) \in \mathcal{F}$. Let $\mathscr{F} := \{ (\tau(u,v),m(u,v)) : (u,v) \in \mathcal{F}\}$. We invert $(\tau(u,v),m(u,v)) \in C^2(\mathcal{F})$: there exists $\underline{u}(\tau,m), \underline{v}(\tau,m) \in C^2_{\tau,m}(\mathscr{F})$ such that
		\begin{equation}
			\underline{u}(\tau(u,v),m(u,v)) = u, \qquad \underline{v}(\tau(u,v),m(u,v)) = v,
		\end{equation}
		and we define
			\begin{align}
			\uu{r}(\tau,m) &:= \underline{r}(\underline{u}(\tau,m), \underline{v}(\tau,m)), \label{rcom} \\ \uu{f}(\tau,m) &:= \underline{\mathbf{u}}_*(\underline{r})^2(\underline{u}(\tau,m), \underline{v}(\tau,m)) - 1, \label{fcom}
		\end{align}
		where $\uu{r}, \uu{f} \in C^2(\mathscr{F})$ are $C^2_{\tau,m}$ solutions of the collapsing Einstein-dust system (since by assumption the dust is collapsing on $\mathcal{R}'$ and $\mathbf{u}(\mathbf{u}(r))<0$ implies it is collapsing on $D^+_{ac}(\mathcal{R}')$): 
		\begin{equation}
			\partial_\tau \uu{r} = - \left( \frac{2m}{\uu{r}} + \uu{f} \right)^\frac{1}{2}, \qquad \partial_\tau \uu{f} = 0. \label{solEd}
		\end{equation}
		By assumption, \eqref{equivR'2} holds for all $(u,v) \in \mathcal{R}_{\delta} \cap \mathcal{R}'$, so, letting $(\mathscr{Q}_{MDD}, r, f)$ be the maximal dust development of the induced data on $\mathscr{D}$, given by \eqref{Dprop}, then we have 
		\begin{equation} \label{equivcom}
			\uu{r}(\tau,m) = r(\tau,m), \qquad \uu{f}(\tau,m) = f(\tau,m),
		\end{equation}
		for all $(\tau,m) \in \mathscr{R}'$. In fact, uniqueness of $C^2$ solutions of \eqref{solEd} implies that \eqref{equivcom} holds for all $(\tau,m) \in \mathscr{F}$, and therefore $\mathscr{F} \subset \mathscr{Q}_{MDD}$. 
		
		Now, for $u \in (u_c,u_2)$, we have $\tau(u,v_c+\delta_2) > \tau(u_c,v_c+\delta_2) > \tau_c(m_0)$, where $m_0 := m(u_c,v_c)$. But, by continuity of $m(u_c,v)$ and $\tau_c(m)$, we can choose $\tilde{v} \in (v_c,v_c+\delta_2)$ such that for $\tilde{m} := m(u_c,\tilde{v})$, we have $\tau_c(\tilde{m}) < \tau(u_c,v_c+\delta_2)$. This is a contradiction to $\sup_{(u,v) \in \mathscr{F}} \rho(u,v) < \infty$, so we must have $\rho(u,v_c+\delta_2) > 0$ for all $u \in [u_c, \underline{u}(s_2)]$. We therefore have $\rho(u,v) > 0$ for all $(u,v) \in \mathcal{R}^+ := \{ u \geq u_c, v < v_c + \delta\} \cap \{ (u,v) : \exists s \in (0,s_1) \text{ s.t. } \underline{u}(s) = u, \underline{v}(s) \leq v\}$. We can therefore uniquely extend $\tau(u,v) \in C^2(\mathcal{R}^+)$, and the Hawking mass satisfies monotonicity on $\mathcal{R}^+$, so in particular, defining 
		\begin{equation}
			\mathscr{R}^+ := \{ (\tau(u,v),m(u,v)) : (u,v) \in \mathcal{R}^+ \},
		\end{equation}
		we can uniquely extend $\underline{u}(\tau,m), \underline{v}(\tau,m) \in C^2_{\tau,m}(\mathscr{R}^+)$. The above argument implies in particular, that for each $s \in (0,s_1)$, the unique past-directed fluid flow line in $\mathcal{R}^+$, which exists by the determinism condition and since $\rho^+(s) > 0$ (defining $+$ quantities on $\underline{\Sigma}$ to be the unique limit from $\mathcal{R}^+$), must intersect $\{ u=u_c, v \in (v_c, v_c + \delta)\}$. Defining \eqref{rcom}, \eqref{fcom} for all $(\tau,m) \in \mathscr{R}^+$, then \eqref{solEd} holds, and $C^2$ uniqueness of solutions of \eqref{solEd} implies that $\mathscr{R}^+ \subset \mathscr{Q}_{MDD}$, and \eqref{equivcom} holds for all $(\tau,m) \in \mathscr{R}^+$. We define 
		\begin{equation}
			\underline{\Sigma}^+ := \{ (\underline{\tau}^+(s),\underline{m}^+(s)) : s\in (0,s_1)\},
		\end{equation}
		where $\underline{\tau}(s), \underline{m}(s)$ are the limits of $\tau(u,v), m(u,v)$ to $\underline{\Sigma}$ from $\mathcal{R}^+$. We can apply Lemma \ref{detcondlemma}, since we have $\mathbf{u}_*(r), \rho > 0$ in $\mathcal{R}^+$, so we have $\underline{m}^+(s) > 0$, for all $s \in (0,s_1)$. We have $\lim_{s \to 0} \underline{m}^+(s) = m_0$, $\lim_{s \to 0} \underline{\tau}^+(s) = \tau_0 := \tau(u_c,v_c)$. 
		
		The analogous argument for the interior of $\underline{\Sigma}$ implies that $\rho > 0$ on $\mathcal{R}^+ \setminus \overline{\Sigma}$, that $\tau(u,v)$ uniquely extends to $\tau(u,v) \in C^2(\mathcal{R}^+ \setminus \overline{\Sigma})$ and, defining
		\begin{equation}
			\mathscr{R}_{\delta} := \{ (\tau(u,v), m(u,v)) : (u,v) \in \mathcal{R}_{\delta} \setminus \underline{\overline{\Sigma}} \},
		\end{equation}
		along with $\underline{\tau}^-(s), \underline{m}^-(s)$ as the limits of $\tau(u,v)$, $m(u,v)$ from the interior of $\underline{
		\Sigma}$, and
		\begin{equation}
			\underline{\Sigma}^- := \{ (\underline{\tau}^-(s),\underline{m}^-(s)) : s\in (0,s_1)\},
		\end{equation}
		then we have $\mathscr{F} := \mathscr{R}_{\delta} \cup \underline{\Sigma}^+ \cup \underline{\Sigma}^-$, and we can extend $\underline{u}(\tau,m), \underline{v}(\tau,m) \in C^2(\mathscr{F})$, and \eqref{equivcom} holds for all $(\tau,m) \in \mathscr{F}$. We have $\underline{m}^-(s) < 0$ for $s \in (0,s_1)$, and $\underline{\tau}^-(s), \underline{m}^-(s)$ extend continuously to $s=0$: $\underline{\tau}^-(0) = \tau_0$, $\underline{m}^-(0) = m_0$. 
		
		\textbf{2. Uniqueness of solutions for eikonal equation implies that $\underline{\Sigma}^\pm$ satisfies the thin shell ODE. } Now, $\underline{u}(\tau,m)$, $\underline{v}(\tau,m)$ are $C^2_{\tau,m}$ solutions of the eikonal equation, \eqref{eiku}, \eqref{eikv} respectively, for $(\tau,m) \in \mathscr{R}_{\delta} \cup \underline{\Sigma}^- \cup \underline{\Sigma}^+ \subset \mathscr{Q}_{MDD}$. Letting $u(\tau,m) \in C^\infty(\mathscr{U}(\mathscr{D},\mathscr{Q}_{MDD}))$, $v(\tau,m) \in C^\infty(\mathscr{V}(\mathscr{D},\mathscr{Q}_{MDD}))$ be the unique solutions of the eikonal equation \eqref{eiku}, \eqref{eikv} respectively, given the induced data on $\mathscr{D}$, then we have $\underline{u}(\tau,m) = u(\tau,m)$ for all $(\tau,m) \in \mathscr{U}(\mathscr{D},\mathscr{Q}_{MDD})) \cap (\mathscr{R}_{\delta} \cup \underline{\Sigma}^- \cup \underline{\Sigma}^+)$ and $\underline{v}(\tau,m) = v(\tau,m)$ for all $(\tau,m) \in \mathscr{V}(\mathscr{D},\mathscr{Q}_{MDD})) \cap (\mathscr{R}_{\delta} \cup \underline{\Sigma}^- \cup \underline{\Sigma}^+)$ by uniqueness. In particular, we have:
		\begin{align}
			u(\underline{\tau}^-(s),\underline{m}^-(s)) = \underline{u}(s) = \underline{u}(\underline{\tau}^+(s), \underline{m}^+(s)),
		\end{align}
		for $s \in (0,s_1)$. By the reverse argument as for existence, we have $\underline{u}(\tau,m) \in C^2(\mathscr{R}_{\delta} \cup \underline{\Sigma}^- \cup \underline{\Sigma}^+)$, so in particular we must have $\frac{\partial_\tau \underline{u}(\underline{\tau}^+(s), \underline{m}^+(s))}{\partial_\tau u(\underline{\tau}^-(s),\underline{m}^-(s))} \to 1$ as $s \to 0$. But, now reversing the proof of Lemma \ref{eiklemma}, we therefore have
		\begin{equation}
			\frac{\underline{\dot{\tau}}^-(s) + ((\dot{\underline{\tau}}^-)^2 - 1)^\frac{1}{2}}{\underline{\dot{\tau}}^+(s) - ((\dot{\underline{\tau}}^+)^2 - 1)^\frac{1}{2}} \to 1 \text{ as } s \to 0,
		\end{equation}
		where $\dot{\underline{\tau}}^\pm(s) > 1$ for $s \in (0,s_1)$. Hence, we have $\dot{\tau}^\pm(s) \to 1$ as $s \to 0$, and therefore $\dot{\underline{r}}(s) \to - \left( \frac{2m_0}{r_0} + f_0 \right)^\frac{1}{2}$. Therefore, there exists $0 < \tilde{s}_1 \leq s_1$ such that $\dot{\underline{r}}(s) < 0$ for $s \in (0,\tilde{s}_1)$. We also have, using the monotonicity of $m(u,v)$ and $\delta \leq \delta_*$: $\underline{m}^+(s) - \underline{m}^-(s) \leq \underline{m}^+(s_1) - \underline{m}^-(s_1) \leq \Delta_*$, for all $s \in [0,s_1]$.
		
		Proposition \ref{tsodeprop} then implies that $(\underline{r}(s), \underline{m}^+(s), \underline{m}^-(s), \underline{\epsilon}(s))$ is a solution of the thin shell ODE, \eqref{ODE1}--\eqref{ODE3}, for $s \in (0,\tilde{s}_1)$, and satisfies \eqref{detstat1}--\eqref{detstat3}, where $\underline{\epsilon}(0) = (1+f_0)^\frac{1}{2}$ follows from \eqref{epsdef} and \eqref{nr}, with
		$\dot{\underline{r}}(0) = - \left( \frac{2m_0}{r_0} + f_0 \right)^\frac{1}{2}$, $\dot{\underline{\tau}}^\pm(0) = 1$. Therefore, the uniqueness statement of Proposition \ref{thinshellODEprop} implies that 
		\begin{equation} \label{equivuniq}
			\underline{r}(s) = r_*(s), \qquad \underline{m}^\pm(s) = m_*^\pm(s), \qquad \underline{\epsilon}(s) = \epsilon_*(s),
		\end{equation}
		for $s \in (0,\tilde{s}_1)$. Now, existence and uniqueness for the thin shell ODE with $s > 0$ implies that \eqref{equivuniq} holds for all $s \in (0,s_1)$. Uniqueness and existence of solutions to the eikonal equation then implies that $\underline{u}(s) = u_*(s), \underline{v}(s) = v_*(s)$ for all $s \in (0,s_1)$, so $\underline{\Sigma} = \Sigma_* \cap \mathcal{R}_{\delta}$, and \eqref{equivuniq1} holds for all $(u,v) \in \mathcal{R}_{\delta}$, and \eqref{equivuniq2} holds for all $(u,v) \in \mathcal{R}_{\delta} \setminus \underline{\Sigma}$.
	\end{proof}

	\subsection{Proof of Proposition \ref{thinshellODEprop}} \label{proof}
	
	For an explanation of the methodology and an outline of the proof, see the second half of Section \ref{introdev}. 
	
	\begin{note}
		Throughout the following sections we fix $\mathscr{Q}_{MDD}$ to be as in the statement of Proposition \ref{thinshellODEprop}, with $r_0, m_0, f_0$ and the functions $T, \rho, f$ fixed accordingly. We therefore suppress the dependence on $T, \rho, f$ in the functions $\mathcal{R}$, $\mathcal{M}^\pm$, $\mathcal{E}$, $\mathcal{T}^\pm$, $\widetilde{\mathcal{X}}^\pm$ (defined in \eqref{def:R}--\eqref{def:X}).
	\end{note}
	
	\subsubsection{ODE system with the determinism condition} \label{ODEdet}
	
	We start by using the determinism condition to reparametrise solutions of the ODE system as in the uniqueness statement in Proposition \ref{thinshellODEprop} to a system parametrised by the jump in Hawking mass. As discussed in Section \ref{introdev}, this will allow us to ultimately reduce the thin shell ODE to a system with a regular singular point.
	
	Since $m_0 \in \mathscr{C}_\chi$, there exists $\Delta_0 > 0$ such that $m(v_0,u_c+\delta') \leq m_0 - \Delta_0 < m_0 < m_0 + \Delta_0 \leq m(u_0,v_c+\delta')$, and in particular $[m_0 - \Delta_0,m_0 + \Delta_0] \subset \mathscr{C}_\chi$, and then for all $m \in [m_0 - \Delta_0,m_0 + \Delta_0]$, $\frac{1}{2} r_0 \leq r \leq r_0$ we have:
	\begin{equation} \label{sim1}
		1 + f(m), \ \frac{2m}{r} + f(m), \ \frac{2}{r_c(m)} + f'(m) \sim 1,
	\end{equation}
	where the last bound holds from Proposition \ref{mddprop}. 
	
	We use double underlined quantities in the following to distinguish the functions from the subsequent functions defined after a change of variables. Indeed, for $\uu{r}, \uu{m}^+, \uu{m}^-, \uu{\epsilon} \in C^0([0,s_1)) \cap C^2((0,s_1))$ solving \eqref{rode}, \eqref{mode}, \eqref{eode} and satisfying \eqref{cond1}, \eqref{cond2}, \eqref{cond3}, we define $\uu{\Delta}(s) := \uu{m}^+(s) - \uu{m}^-(s) \in C^0([0,s_1]) \cap C^2((0,s_1))$, which satisfies $\uu{\Delta}(0) = 0$, $\uu{\Delta}(s),\uu{\Delta}(s) > 0$ for $s \in (0,s_1)$, and we define $\underline{\Delta} := \uu{\Delta}(s_1)$, where we \emph{impose that} $\underline{\Delta} \leq \Delta_0$. We define
	\begin{equation}
		\uu{\eta}(s) := [\uu{\Delta}(s)]^{-1} (m_0 - \uu{m}^-(s)) \in C^2((0,s_1)),
	\end{equation}
	which satisfies $\uu{\eta}(s) \in (0,1)$ for all $s \in (0,s_1)$. The function $\uu{\Delta}(s)$ is invertible for $s \in [0,s_1]$, with $\uu{\Delta}^{-1} \in C^0([0,s_1]) \cap C^2((0,s_1))$. We may therefore define, for $\Delta \in [0,\underline{\Delta})$:
	\begin{equation}
		\underline{r}(\Delta) := \uu{r}(\uu{\Delta}^{-1}(\Delta)), \qquad \underline{\epsilon}(\Delta) := \uu{\epsilon}(\uu{\Delta}^{-1}(\Delta)), \qquad \underline{\eta}(\Delta) := \uu{\eta}(\uu{\Delta}^{-1}(\Delta)), 
	\end{equation}
	with $\underline{r}, \underline{\epsilon} \in C^0([0,\underline{\Delta}_1)) \cap C^2((0,\underline{\Delta}_1))$, $\underline{\eta} \in C^2((0,\underline{\Delta}_1))$, and \eqref{cond3} is equivalent to
	\begin{equation} \label{IC}
		\underline{r}(0) = r_0, \qquad \underline{\epsilon}(0) = (1+f_0)^\frac{1}{2}.
	\end{equation}
	\begin{definition} \label{def:mpm}
		We define for $\Delta,\eta \in \R$:
		\begin{equation} \label{mpm}
			m^\pm(\Delta,\eta) := m_0 + a^\pm(\eta) \Delta, \qquad a^+(\eta) := 1 - \eta, \qquad a^-(\eta) := - \eta,
		\end{equation}
		along with the function $\Psi_0 : \R^4 \to \R^4$ defined by:
		\begin{equation}
			\Psi_0(\Delta,r,\eta,\epsilon) := (r, m^+(\Delta,\eta),m^-(\Delta,\eta),\epsilon).
		\end{equation} 
	\end{definition}
	Note that \eqref{mpm} is defined so that $\underline{m}^\pm(s) = m^\pm(\underline{\Delta}(s), \underline{\eta}(s))$, and by the choice of $\Delta_0 > 0$ and the assumption $\underline{\Delta} \leq \Delta_0$, we have:  $m^-(\Delta,\underline{\eta}(\Delta)) \in (m_0 - \Delta_0,m_0)$, $m^+(\Delta,\underline{\eta}(\Delta)) \in (m_0,m_0 + \Delta_0)$ for all $\Delta \in (0,\underline{\Delta})$. Letting
	\begin{equation}
		\underline{\Psi_0}(\Delta) := \Psi_0(\Delta,\underline{r}(\Delta),\underline{\eta}(\Delta),\underline{\epsilon}(\Delta)),
	\end{equation}
	then $(\underline{r},\underline{\eta},\underline{\epsilon})(\Delta)$ is a solution of the following ODE for $\Delta \in (0,\underline{\Delta})$:
	\begin{align} 
		\frac{d\underline{r}}{d\Delta}(\Delta) &= \frac{\mathcal{R}(\underline{\Psi_0}(\Delta))}{\mathcal{M}^+(\underline{\Psi_0}(\Delta))-\mathcal{M}^-(\underline{\Psi_0}(\Delta))} ,\label{drdDelta}\\ \frac{d(\Delta\underline{\eta}(\Delta))}{d\Delta} &= \frac{-\mathcal{M^-}(\underline{\Psi_0}(\Delta))}{\mathcal{M}^+(\underline{\Psi_0}(\Delta))-\mathcal{M}^-(\underline{\Psi_0}(\Delta))}, \label{detadDelta} \\ \frac{d\underline{\epsilon}}{d\Delta} (\Delta) &= \frac{\underline{\epsilon}(\Delta)\left(1 - \underline{\epsilon}(\Delta) T(\underline{\Psi_0}(\Delta))\right)}{\Delta}, \label{depsdDelta}
	\end{align}
	where
	\begin{equation} \label{Tdef}
		T(\underline{\Psi_0}(\Delta)) := \left( \mathcal{M}^+(\underline{\Psi_0}(\Delta))-\mathcal{M}^-(\underline{\Psi_0}(\Delta))\right)^{-1} \left(\frac{\mathcal{M}^+(\underline{\Psi_0}(\Delta))\mathcal{T}^+(\underline{\Psi_0}(\Delta))}{(1 + f(m^+(\Delta,\underline{\eta}(\Delta))))^{\frac{1}{2}}}-\frac{\mathcal{M}^-(\underline{\Psi_0}(\Delta))\mathcal{T}^-(\underline{\Psi_0}(\Delta))}{(1 + f(m^-(\Delta,\underline{\eta}(\Delta))))^{\frac{1}{2}}} \right),
	\end{equation}
	and \eqref{cond1}, \eqref{cond2} are equivalent to the following conditions for all $\Delta \in (0,\underline{\Delta})$:
	\begin{equation} \label{MDDbd}
		\underline{r}(\Delta) - r_c(m^\pm(\Delta,\underline{\eta}(\Delta))) > 0,
	\end{equation}
	\begin{equation} \label{sign}
		\underline{\epsilon}(\Delta) > 0, \ \mathcal{R}(\underline{\Psi_0}(\Delta)) < 0, \ \underline{r}(\Delta) \leq r_0,
	\end{equation}
	\begin{equation} \label{detcondDelta}
		\widetilde{\mathcal{X}}^+(\underline{\Psi_0}(\Delta)) > 0, \ \widetilde{\mathcal{X}}^-(\underline{\Psi_0}(\Delta)) < 0.
	\end{equation}
	
	\begin{note}
		The conditions \eqref{MDDbd} and \eqref{detcondDelta} imply that $\mathcal{M}^+(\underline{\Psi_0}(\Delta)) > 0$, $\mathcal{M}^-(\underline{\Psi_0}(\Delta)) < 0$ for all $\Delta \in (0,\underline{\Delta})$.
	\end{note}
	
	\begin{remark} \label{rem:ode}
		Throughout the following sections, we'll assume that $\underline{r}, \underline{\epsilon} \in C^0([0,\underline{\Delta})) \cap C^2((0,\underline{\Delta}))$, $\underline{\eta} \in C^2((0,\underline{\Delta}))$ constitute a solution of the ODE \eqref{drdDelta}, \eqref{detadDelta}, \eqref{depsdDelta} satisfying \eqref{IC}, along with \eqref{MDDbd}, \eqref{sign}, \eqref{MDDbd}. We'll successively restrict $0 < \underline{\Delta} \leq \Delta_i$, for $0 < \Delta_{i+1} \leq \Delta_i \leq \Delta_0$, depending on the functions $f, r_c \in C^\infty([m_0 - \Delta_0,m_0+\Delta_0])$ as well as $r_0, m_0, \Delta_0$.
	\end{remark}
	We will expand smooth functions, in particular $f, r_c$, as in the following:
	\begin{note}
		For any smooth function, e.g.~$f \in C^\infty([m_0 - \Delta_0,m_0+\Delta_0])$, we can expand
		\begin{equation}
			f(m) = f_0 + (m-m_0) f_1(m),
		\end{equation}
		where $f_1(m) = \int_0^1 f'(m_0 + s(m-m_0)) ds \in C^\infty([m_0 - \Delta_0, m_0 + \Delta_0])$. We define
		\begin{equation} \label{f1}
			f_1^\pm(\Delta,\eta) := \frac{f(m^\pm(\Delta,\eta)) - f_0}{\Delta} = a^\pm(\eta) f_1(m^\pm(\Delta,\eta)).
		\end{equation}
		Expanding again with $f_{1,0} := f_1(m_0) = f'(m_0)$, we get
		\begin{equation}
			f^\pm(\Delta,\eta) := f(m^\pm(\Delta,\eta)) = f_0 + f_{1, 0} \cdot a^\pm(\eta) \Delta + (a^\pm(\eta) \Delta)^2 f_2^\pm(\Delta,\eta),
		\end{equation}
		where $f_2^\pm(\Delta,\eta) \in C^\infty([0,\Delta_0] \times [0,1])$, given by $f_2^\pm(\Delta,\eta) := f_2(m^\pm(\Delta,\eta))$, $f_2(m) := \int_0^1 f_1'(m_0 + s(m-m_0)) ds \in C^\infty([m_0 - \Delta_0, m_0 + \Delta_0])$, so that $f_2(m_0) = f_1'(m_0) = \frac{1}{2} f''(m_0)$.
	\end{note}
	Now, for $r_c(m) \in C^\infty([m_0-\Delta_0,m_0+\Delta_0])$, we note that by the assumption of Theorem \ref{devthm}, $(u_c,v_c)$ is a \emph{non-degenerate} first shell crossing singularity, and the statement of Lemma \ref{nondeglemma} therefore implies that for some $n \in \mathbb{N}$, $r_c(m)$ has the expansion: 
	\begin{equation} \label{rc:nondeg-exp}
		r_c(m) = r_0 + \frac{r_c^{(2n)}(m_0)}{(2n)!} (m-m_0)^{2n} + r_{2n+1}(m) (m-m_0)^{2n+1}, 
	\end{equation}
	with $r_{2n} := - \frac{r_c^{(2n)}(m_0)}{(2n)!} > 0$, $r_{2n+1}(m) \in C^\infty([m_0-\Delta_0,m_0+\Delta_0])$, and $r_{2n+1}(m_0) = \frac{r_c^{(2n+1)}(m_0)}{(2n+1)!}$. In terms of $(\Delta,\eta)$:
	\begin{equation}
		r^\pm_c(\Delta,\eta) := r_c(m^\pm(\Delta,\eta)) = r_0 - r_{2n} a^\pm(\eta)^{2n} \Delta^{2n} + r_{2n} \cdot r^\pm_{c, 2}(\Delta,\eta) \Delta^{2n+1}, \label{rcpm} 
	\end{equation}
	where
	\begin{equation} \label{rc2}
		r_{c, 2}^\pm(\Delta,\eta) := r_{2n}^{-1} a^\pm(\eta)^{2n+1} r_{2n+1}(m^\pm(\Delta,\eta)),
	\end{equation}
	so we have $r_{c, 2}^\pm(\Delta,\eta) \in C^\infty([0,\Delta_0] \times [0,1])$. 
	
	\begin{definition} \label{phi-alpha-def}
		We define the following functions for $\Delta, \phi, \alpha \in \R$:
		\begin{equation} \label{def:phi-alpha}
			r(\Delta,\phi) := r_0 - r_{2n} \phi \Delta^{2n}, \qquad \epsilon(\Delta,\alpha) := (1+f_0)^\frac{1}{2} (1 + \alpha \Delta),
		\end{equation}
		and for $\theta := (\alpha, \eta, \phi) \in \R^3$ we define the function $\Psi_1: \R^4 \to \R^4$ by:
		\begin{equation}
			\Psi_1(\Delta, \theta) := (r(\Delta,\phi), m^+(\Delta,\eta), m^-(\Delta,\eta), \epsilon(\Delta,\alpha)).
		\end{equation}
	\end{definition}
	\begin{lemma} \label{proof1}
		There exists $0 < \Delta_1 \leq \Delta_0$ depending on the functions $f, r_c$ and the constants $r_0, m_0, \Delta_0$ such that: for any $0 < \underline{\Delta} \leq \Delta_1$, then the functions $\underline{\alpha}, \underline{\phi} \in C^0((0,\underline{\Delta}))$ defined by:
		\begin{equation} \label{under-alpha-phi}
			\epsilon(\Delta, \underline{\alpha}(\Delta)) := \underline{\epsilon}(\Delta), \qquad r(\Delta, \underline{\phi}(\Delta)) := \underline{r}(\Delta),
		\end{equation}
		satisfy for all $\Delta \in (0,\underline{\Delta})$:
		\begin{equation}
			\underline{\alpha}(\Delta) \in [-\alpha_-,\alpha_+], \qquad \underline{\phi}(\Delta) \in [0, 2^{1-2n}],
		\end{equation}
		where $\alpha_\pm > 0$ depend on $f(m)$ and are determined in the proof. Moreover, defining 
		\begin{equation}
			\mathcal{B} := [-\alpha_-,\alpha_+] \times [0,1] \times [0,2^{1-2n}],
		\end{equation}
		then for all $(\Delta,\theta) \in [0,\Delta_1] \times \mathcal{B}$, we have:
		\begin{align} \label{epsbd}
			\frac{1}{2} (1+f_0)^\frac{1}{2} &\leq \ \epsilon(\Delta,\alpha) \leq \frac{3}{2} (1+f_0)^\frac{1}{2}, \\ \frac{r_0}{2} &\leq \ r(\Delta,\phi) \leq r_0, \label{rbd} \\ - \frac{3}{2}\left(\frac{2m_0}{r_0} + f_0 \right)^\frac{1}{2} \leq \ &\mathcal{R}(\Psi_1(\Delta,\theta)) \leq - \frac{1}{2}\left(\frac{2m_0}{r_0} + f_0 \right)^\frac{1}{2}, \label{Rbd}
		\end{align}
		so that the equivalent bounds are satisfied for $\underline{\epsilon}(\Delta), \underline{r}(\Delta), \mathcal{R}(\underline{\Psi_0}(\Delta))$ respectively.
	\end{lemma}
	
	\begin{remark}
		The functions $\underline{r}, \underline{\eta}, \underline{\epsilon}$ need not satisfy the ODE for the statement of the lemma to hold, we only require \eqref{IC} and the conditions \eqref{MDDbd}, \eqref{sign}, \eqref{detcondDelta}.
	\end{remark}
	 
	\begin{proof}
		From \eqref{MDDbd}, \eqref{rcpm} and \eqref{def:phi-alpha}, we have:
		\begin{align}
			r(\Delta,\underline{\phi}(\Delta)) - r_c^+(\Delta,\underline{\eta}(\Delta)) &= \Delta^{2n} r_{2n} ( (1-\underline{\eta}(\Delta))^{2n} - \underline{\phi}(\Delta)  +  r_{c, 2}^+(\Delta,\underline{\eta}(\Delta))\Delta ) > 0, \label{rminrcp} \\ r(\Delta,\underline{\phi}(\Delta)) - r_c^-(\Delta,\underline{\eta}(\Delta)) &= \Delta^{2n} r_{2n} ( \underline{\eta}(\Delta)^{2n} - \underline{\phi}(\Delta) + r_{c, 2}^-(\Delta,\underline{\eta}(\Delta))\Delta) > 0, \label{rminrcm}
		\end{align}
		so that, defining $r_{2,+} := \max_{\Delta \in [0,\Delta_1], \ \eta\in[0,1], \pm}|r_{c, 2}^\pm(\Delta,\eta)|$, we get $(1-\underline{\eta}(\Delta))^{2n} - \underline{\phi}(\Delta), \underline{\eta}(\Delta)^{2n} - \underline{\phi}(\Delta) > - r_{2,+} \Delta$, for all $0 < \Delta \leq \underline{\Delta}$, and therefore
		\begin{equation}
			\underline{\phi}(\Delta) < \min\{\underline{\eta}(\Delta)^{2n},(1-\underline{\eta}(\Delta))^{2n}\} + r_{2,+}\Delta \leq 2^{-2n} +  r_{2,+}\Delta,
		\end{equation}
		We can then choose $0 < \Delta_1 r_{2, +} \leq 2^{-2n}$ so that $\underline{\phi}(\Delta) \leq 2^{1-2n}$ for all $\Delta \in (0,\underline{\Delta}]$, while $\underline{\phi}(\Delta) \geq 0$ follows immediately from \eqref{sign}. We then choose again $\Delta_1 > 0$ possibly smaller so that $r_{2n} 2^{1-2n} \Delta_1^{2n} \leq \frac{1}{2} r_0$ and \eqref{rbd} holds. Next, we use Lemma \ref{detcondlemma} to see that $\widetilde{\mathcal{X}}^-(\underline{\Psi_0}(\Delta)) < 0$ implies, for all $\Delta \in (0,\underline{\Delta})$:
		\begin{equation}
			\left( \underline{\epsilon}(\Delta) + \frac{\Delta}{2\underline{\epsilon}(\Delta) \underline{r}(\Delta)} \right)^2 > 1 + f^-(\Delta,\underline{\eta}(\Delta)),
		\end{equation}
		which is equivalent to, using \eqref{IC}, and defining $f_{0,-} := \min_{m \in [m_0-\Delta_0,m_0+\Delta_0]} f(m)$ (so $1 + f_{0, min} \sim 1$ by \eqref{sim1}) along with $f_{1,+} := \sup_{\Delta \in [0,\Delta_1], \eta \in [0,1]} |f_1^\pm(\Delta,\eta)|$:
		\begin{align} \label{detminus}
			\underline{\epsilon}(\Delta) &> \frac{(1 + f^-(\Delta,\underline{\eta}(\Delta)))^\frac{1}{2}}{2}  \left[1 + \left(1 - \frac{2\Delta}{\underline{r}(\Delta)(1+f^-(\Delta,\underline{\eta}(\Delta)))} \right)^\frac{1}{2} \right], \\ &\geq (1+f_0)^\frac{1}{2} \left(1 - \frac{\Delta f_{1,+}}{1+f_0}\right) \left( 1 - \frac{2\Delta}{r_0 (1+f_{0,-})}\right),
		\end{align}
		and we choose $\Delta_1$ again possibly smaller so that $0 < \Delta_1 \leq \frac{r_0}{4} (1+f_{0,-})$, so we get the following bound on $\underline{\alpha}(\Delta)$ for $\Delta \in (0,\Delta_1)$:
		\begin{equation} \label{alphaminus}
			\underline{\alpha}(\Delta) \geq - \frac{2}{r_0(1+f_{0,-})} - \frac{f_{1,+}}{2(1+f_0)} =: -\alpha_-.
		\end{equation}
		We now choose again $\Delta_1 > 0$ possibly smaller so that $\Delta_1 \leq \frac{1}{2\alpha_-}$ and $\epsilon(\Delta,\alpha) \geq \frac{1}{2}(1+f_0)^\frac{1}{2}$ for $\Delta \in [0,\Delta_1]$, $\alpha \geq -\alpha_-$. Using again Lemma \ref{detcondlemma}, the condition $\widetilde{\mathcal{X}}^+(\underline{\Psi_0}(\Delta)) > 0$ is equivalent to $\underline{\epsilon}(\Delta) - \frac{\Delta}{2\underline{\epsilon}(\Delta) \underline{r}(\Delta)} < (1+f^+(\Delta,\underline{\eta}(\Delta)))^\frac{1}{2}$. If we choose moreover $0 < \Delta_1 \leq \frac{r_0}{4}(1+f_0)$, then we get for $\Delta \in [0,\underline{\Delta}]$:
		\begin{equation}
			\underline{\epsilon}(\Delta)^2 \geq \frac{1}{4}(1+f_0) > \frac{\Delta_1}{r_0} \geq \frac{\Delta}{2\underline{r}(\Delta)},
		\end{equation}
		so that we can square the bound equivalent to the determinism condition to get $\left( \underline{\epsilon} - \frac{\Delta}{2\underline{\epsilon}(\Delta) \underline{r}(\Delta)} \right)^2 < 1+f^+(\Delta,\eta(\Delta))$, and therefore
		\begin{align}
			\underline{\epsilon}(\Delta) &< \frac{(1 + f^+(\Delta,\underline{\eta}(\Delta)))^\frac{1}{2}}{2}  \left[1 + \left(1 + \frac{2\Delta}{\underline{r}(1+f^+(\Delta,\underline{\eta}(\Delta)))} \right)^\frac{1}{2} \right], \\ &\leq (1+f_0)^\frac{1}{2} \left(1 + \frac{\Delta f_{1,+}}{2(1+f_0)}\right) \left(1 + \frac{\Delta}{r_0 (1 + f_{0,-})}\right),
		\end{align}
		and
		\begin{equation} \label{alphaplus}
			\underline{\alpha}(\Delta) \leq \frac{1}{r_0(1 + f_{0,-})} + \frac{f_{1,+}}{2(1+f_0)} =: \alpha_+.
		\end{equation}
		Choosing again $\Delta_1 > 0$ such that $\alpha_+ \Delta_1 \leq \frac{1}{2}$, we get \eqref{epsbd}. Finally, we expand \eqref{def:R} for $(\Delta, \theta) \in [0,\Delta_1] \times \mathcal{B}$ to see that
		\begin{equation}
			\mathcal{R}(\Psi_1(\Delta,\theta))^2 = (1+f_0)(1+\alpha \Delta)^2 - 1 + \frac{2m_0 + (1-2\eta)\Delta}{r(\Delta,\phi)} + \frac{\Delta^2}{4\epsilon(\Delta,\alpha)^2 r(\Delta,\phi)^2}, 
		\end{equation}
		which we can rewrite as
		\begin{equation}
			\mathcal{R}(\Psi_1(\Delta,\theta))^2 = \frac{2m_0}{r_0} + f_0 + \Delta \left( (1+f_0)(2\alpha + \alpha^2 \Delta) + \frac{2m_0 r_1 \phi \Delta^{2n-1}}{r_0 r(\Delta,\phi)} + \frac{1-2\eta}{r(\Delta,\phi)} + \frac{\Delta}{4\epsilon(\Delta,\alpha)^2 r(\Delta,\phi)^2}\right), \label{Rexp}
		\end{equation}
		where we've used
		\begin{equation} \label{rinv}
			r(\Delta,\phi)^{-1} = r_0^{-1} + \frac{r_1 \phi \Delta^{2n}}{r_0 r(\Delta,\phi)},
		\end{equation}
		and \eqref{Rbd} then follows from \eqref{epsbd}, \eqref{rbd} and choosing again $\Delta_1 > 0$ sufficiently small.
	\end{proof}
	
	\subsubsection{Uniqueness of first order terms} \label{firstorder}
	
	In this section, we will show that, after successively choosing $0 < \Delta_3 \leq \Delta_2 \leq \Delta_1$ depending on $f, r_c \in C^\infty([m_0 - \Delta_0,m_0 + \Delta_0])$ and the constants $r_0,m_0$, that solutions of the ODE \eqref{drdDelta}, \eqref{detadDelta}, \eqref{depsdDelta}, which satisfy the conditions \eqref{MDDbd}, \eqref{sign}, \eqref{detcondDelta} for $\Delta \in (0,\Delta_3)$, must have $\underline{\alpha}, \underline{\eta}, \underline{\phi} \in C^0([0,\Delta_3))$, where $\underline{\alpha}(0), \underline{\eta}(0), \underline{\phi}(0)$ are \emph{uniquely determined}. We start by expanding the relevant functions in order to expand \eqref{depsdDelta}.
	
	For $(\Delta,\theta) \in [0,\Delta_1] \times \mathcal{B}$, we define: $\mathcal{R}(\Psi_1(\Delta,\theta))^2 =: \frac{2m^\pm(\Delta,\eta)}{r(\Delta,\phi)} + f^\pm(\Delta,\eta) + \Delta \mathcal{R}^\pm_1(\Delta,\theta)$, so that, using \eqref{rinv}, we have
	\begin{align} \label{R2}
		\mathcal{R}^\pm_1(\Delta,\theta) &:= 2\alpha(1+f_0) \mp \frac{1}{r_0} - f'(m_0) a^\pm(\eta) + \Delta \mathcal{R}^\pm_2(\Delta,\theta), \\ 
		\mathcal{R}^\pm_2(\Delta,\theta) &:= \alpha^2(1+f_0) + \frac{(1+\alpha\Delta)^{-2}}{4r(\Delta,\phi)^2(1+f_0)} \mp \frac{r_1 \phi \Delta^{2n-1}}{r_0 r(\Delta,\phi)} - a^\pm(\eta)^2 f_2(m_0 + a^\pm(\eta) \Delta), \label{R2def}
	\end{align}
	and from \eqref{rbd}, \eqref{epsbd}, we have $\mathcal{R}_1^\pm, \mathcal{R}_2^\pm \in C^\infty([0,\Delta_1] \times \mathcal{B})$. We now consider an expansion for $\widetilde{\mathcal{X}}^\pm(\Psi_0(\Delta,\theta))$, defined in \eqref{def:X}. We define
	\begin{equation}
		\epsilon(\Delta,\alpha) \mp \frac{\Delta}{2\epsilon(\Delta,\alpha)r(\Delta,\phi)} =: (1+f_0)^\frac{1}{2}\left( 1 + \Delta n^\pm_1(\Delta,\theta) \right),
	\end{equation}
	where
	\begin{equation} \label{n1pm}
		n^\pm_1(\Delta,\theta) = \alpha \mp \frac{r(\Delta,\phi)^{-1}}{2(1+f_0)(1+\alpha\Delta)},
	\end{equation}
	and \eqref{epsbd}, \eqref{rbd} imply that $n^\pm_1 \in C^\infty([0,\Delta_1] \times \mathcal{B})$. 
	We now expand $(1+f^\pm(\Delta,\eta))^{-\frac{1}{2}}$, which we'll need in order to expand \eqref{Tdef}, as well as $\widetilde{X}^\pm(\Psi_0(\Delta,\theta))$, where we work from \eqref{def:X} rearranged to:
	\begin{equation} \label{tildeX1}
		\widetilde{\mathcal{X}}^\pm = \left( \frac{2m^\pm}{r} + f^\pm \right)^{-\frac{1}{2}} \frac{\frac{2m^\pm}{r} + f^\pm - \mathcal{R}^2}{n^\pm(r) (1+f^\pm)^{-\frac{1}{2}} - \mathcal{R}(\frac{2m^\pm}{r}+f^\pm)^{-\frac{1}{2}}}.
	\end{equation}
	We define:
	\begin{align} \label{F1}
		(1 + f^\pm(\Delta,\eta))^{-1/2} = (1+f_0)^{-1/2} \left(1 + \frac{\Delta f_1^\pm(\Delta,\eta)}{1+f_0}\right)^{-1/2} =: (1+f_0)^{-1/2} (1 + \Delta F^\pm_1(\Delta,\eta)),
	\end{align}
	where
	\begin{align} \label{F1-2}
		F^\pm_1(\Delta,\eta) &= - \frac{f_{1,0} a^\pm(\eta)}{2(1+f_0)} + \Delta F^\pm_2(\Delta,\eta), \\ F^\pm_2(\Delta,\eta) &:= -\frac{a^\pm(\eta)^2f_2(m^\pm(\Delta,\eta))}{2(1+f_0)} + \frac{3}{8} \left( \frac{f_1^\pm(\Delta,\eta) }{1+f_0}\right)^2 F_{-1/2, 1}\left( \frac{\Delta f_1^\pm(\Delta,\eta)}{1+f_0}\right), \label{F2}
	\end{align}
	where $F_{-1/2, 1}(x)$ is defined\footnote{The functions are defined to satisfy $(1+x)^{\pm1/2} = 1 \pm \frac{x}{2} F_{\pm1/2}(x)$, and $F_{-1/2}(x) = 1 - \frac{3x}{4} F_{-1/2, 1}(x)$.} for $x > -1$:
	\begin{align}
		F_{-1/2, 1}(x) := \frac{1}{3} (F_{1/2}(x))^2 + \frac{2}{3} F_{-1/2}(x) + \frac{x}{6} (1+x)^{-1/2} (F_{1/2}(x))^3, \\ F_{1/2}(x) := 2(1+(1+x)^\frac{1}{2})^{-1}, \ F_{-1/2}(x) := (1+x)^{-\frac{1}{2}}F_{1/2}(x),
	\end{align}
	and $F_{\pm1/2}, F_{-1/2,1} \in C^\infty((-1,\infty))$, so $F_1^\pm, F_2^\pm \in C^\infty([0,\Delta_1] \times \mathcal{B})$ by \eqref{sim1}. Next, we define
	\begin{equation}
		 \frac{\epsilon(\Delta,\alpha) \mp \frac{\Delta}{2\epsilon(\Delta,\alpha)r(\Delta,\phi)}}{(1+f^\pm(\Delta,\eta))^{\frac{1}{2}}}  - \frac{\mathcal{R}(\Psi_0(\Delta,\theta))}{\left(\frac{2m^\pm(\Delta,\eta)}{r(\Delta,\phi)}+f^\pm(\Delta,\eta)\right)^{\frac{1}{2}}} =: 2(1 + \Delta X_d(\Delta,\theta)),
	\end{equation}
	which, after using the definition of $\mathcal{R}_1^\pm(\Delta,\theta)$ to express
	\begin{equation}
		-\mathcal{R}(\Psi_0(\Delta,\theta))\left(\frac{2m^\pm(\Delta,\eta)}{r(\Delta,\phi)} + f^\pm(\Delta,\eta) \right)^{-1/2} = 1 + \frac{\Delta \mathcal{R}^\pm_1(\Delta,\theta) \left(\frac{2m^\pm(\Delta,\eta)}{r(\Delta,\phi)}+f^\pm(\Delta,\eta)\right)^{-\frac{1}{2}}}{\left(\frac{2m^\pm(\Delta,\eta)}{r(\Delta,\phi)}+f^\pm(\Delta,\eta)\right)^{\frac{1}{2}} - \mathcal{R}(\Psi_0(\Delta,\theta))},
	\end{equation}
	we derive
	\begin{equation} \label{Xdpm}
		X_d^\pm(\Delta,\theta) =  \frac{1}{2}\frac{ \mathcal{R}^\pm_1(\Delta,\theta) \left(\frac{2m^\pm(\Delta,\eta)}{r(\Delta,\phi)}+f^\pm(\Delta,\eta)\right)^{-\frac{1}{2}}}{\left(\frac{2m^\pm(\Delta,\eta)}{r(\Delta,\phi)}+f^\pm(\Delta,\eta)\right)^{\frac{1}{2}} - \mathcal{R}(\Psi_0(\Delta,\theta))} + \frac{1}{2} n_1^\pm(\Delta,\theta) + \frac{1}{2}F_1^\pm(\Delta,\theta)(1 + \Delta n_1^\pm(\Delta,\theta)),
	\end{equation}
	and from \eqref{sim1}, \eqref{epsbd}, \eqref{rbd}, \eqref{Rbd} we get $X_d^\pm \in C^\infty([0,\Delta_1] \times \mathcal{B})$. 
	\begin{definition}
		We define $X_{d,+} := \max_{\Delta \in [0,\Delta_1], \theta \in \mathcal{B}} |X_d(\Delta,\theta)|$, and 
		\begin{equation} \label{def:Delta2}
			\Delta_2 := \min\left\{ \Delta_1, \frac{1}{2 X_{d, +}}\right\}.
		\end{equation}
	\end{definition}
	We can finally express \eqref{tildeX1} as:
	\begin{equation} \label{tildeX2}
		\widetilde{\mathcal{X}}^\pm(\Delta,\theta) = \frac{\Delta}{2} \left( \frac{2m^\pm(\Delta,\eta)}{r(\Delta,\phi)} + f^\pm(\Delta,\eta) \right)^{-1/2} \frac{-\mathcal{R}^\pm_1(\Delta,\theta)}{1 + \Delta X_d(\Delta,\theta)} =: \Delta X_0^\pm(\Delta,\theta),
	\end{equation}
	where by \eqref{def:Delta2} (and \eqref{sim1}, \eqref{epsbd}, \eqref{rbd}) we have $X_0^\pm \in C^\infty([0,\Delta_2];\mathcal{B})$. 
	
	For the expansion of $\mathcal{T}^\pm(\Psi_0(\Delta,\theta))$, we note that from \eqref{def:T} we have $\mathcal{T}^\pm(\Psi_0(0,\theta)) = 1$, and so \eqref{id:T} implies that
	\begin{equation}
		\mathcal{T}^\pm(\Psi_0(\Delta,\theta)) = \left(1 + \frac{\widetilde{\mathcal{X}}^\pm(\Psi_0(\Delta,\theta))^2}{1+f^\pm(\Delta,\eta)} \right)^\frac{1}{2},
	\end{equation}
	and so $\mathcal{T}^\pm \circ \Psi_0 \in C^\infty([0,\Delta_2] \times \mathcal{B})$, with $\mathcal{T}^\pm(\Psi_0(\Delta,\theta)) \geq 1$ for all $(\Delta, \theta) \in [0,\Delta_2] \times \mathcal{B}$. We expand 
	\begin{equation} \label{Tpm}
		\mathcal{T}^\pm(\Psi_0(\Delta,\theta)) =: 1 + \Delta^2 \mathcal{T}_2^\pm(\Delta,\theta),
	\end{equation}
	where
	\begin{equation} \label{Tpm2}
		\mathcal{T}^\pm_2(\Delta,\theta) = \frac{X_0^\pm(\Delta,\theta)^2}{(1+f^\pm(\Delta,\eta))(1 + \mathcal{T}^\pm(\Delta,\theta))},
	\end{equation}
	and we again have $\mathcal{T}^\pm_2 \in C^\infty([0,\Delta_2] \times \mathcal{B})$. 

	\begin{lemma} \label{lem:beta}
		We restrict to $\underline{\Delta} \in (0,\Delta_2]$. Let $\underline{\alpha}, \underline{\phi} \in C^2((0,\underline{\Delta}))$ be as defined in \eqref{def:phi-alpha} (along with $\underline{\eta} \in C^2((0,\underline{\Delta}))$). Defining $\beta_0 := \frac{f'(m_0)}{4(1+f_0)}$, and
		\begin{equation} \label{def:beta}
			\underline{\beta}(\Delta) := \underline{\alpha}(\Delta) + 2 \beta_0 \cdot \underline{\eta}(\Delta),
		\end{equation}
		then $\underline{\beta} \in C^0([0,\underline{\Delta}))$, with $\underline{\beta}(0) = \beta_0$. Moreover, defining $\underline{\beta}(\Delta) =: \beta_0 + \underline{\tilde{\beta}}(\Delta)\Delta$, we have $\tilde{\beta}(\Delta) \in [- B_+,B_+]$ for $\Delta \in (0,\underline{\Delta})$, where $B_+$ depends only on $f, r_c$ and $r_0, m_0, \Delta_2$.
	\end{lemma}
	
	\begin{proof}
		We define $\underline{\theta}(\Delta) := (\underline{\alpha}(\Delta), \underline{\eta}(\Delta),\underline{\phi}(\Delta))$, along with
		\begin{equation} \label{def:N}
			N(\Delta,\underline{\theta}(\Delta)) := \frac{-\mathcal{M^-}(\underline{\Psi_0}(\Delta))}{\mathcal{M}^+(\underline{\Psi_0}(\Delta))-\mathcal{M}^-(\underline{\Psi_0}(\Delta))},
		\end{equation}
		where \eqref{MDDbd} and \eqref{detcondDelta} (and \eqref{def:Mpm}) imply that $N(\Delta,\underline{\theta}(\Delta)) \in (0,1)$ for all $\Delta \in (0,\underline{\Delta})$. We now expand $T(\Psi_1(\Delta,\underline{\theta}(\Delta)))$, using \eqref{Tdef}, \eqref{F1}, \eqref{F1-2}, \eqref{Tpm}, \eqref{def:N}:
		\begin{equation} \label{Texp}
			(1+f_0)^\frac{1}{2} T(\underline{\Psi_0}(\Delta)) = 1 + T_1(\Delta, \underline{\theta}(\Delta)) \cdot \Delta + T_2(\Delta,\underline{\theta}(\Delta)) \cdot \Delta^2,
		\end{equation}
		where we've defined (suppressing the $\Delta$ argument in $\underline{\theta}(\Delta)$ for legibility):
		\begin{equation} \label{T1}
			T_1(\Delta, \underline{\theta}) := 2\beta_0 \cdot (\underline{\eta} + N(\Delta,\underline{\theta}) - 1),
		\end{equation}
		and
		\begin{align} \label{T2}
			T_2(\Delta, \underline{\theta}) := \left[ 1-N(\Delta,\underline{\theta})\right] &\cdot \left[F_2^+(\Delta,\underline{\theta}) + \mathcal{T}_2^+(\Delta, \underline{\theta})\cdot(1+f^+(\Delta,\underline{\eta}))^\frac{1}{2}\right] \\ &+ N(\Delta,\underline{\theta})\left[F_2^-(\Delta,\underline{\eta}) + \mathcal{T}_2^-(\Delta, \underline{\theta}) \cdot (1+f^-(\Delta,\underline{\eta}))^\frac{1}{2}\right],
		\end{align}
		We now expand the right hand side of the ODE for $\underline{\epsilon}$, \eqref{depsdDelta}, using \eqref{Texp}:
		\begin{equation} \label{dalphaexp}
			\Delta \frac{d}{d\Delta} (\Delta \underline{\alpha}) = (1+\Delta\underline{\alpha})\left[ 1 - (1+\Delta\underline{\alpha})(1+\Delta T_1(\Delta,\underline{\theta}) + \Delta^2 T_2(\Delta,\underline{\theta})) \right],
		\end{equation}
		which we rearrange to:
		\begin{equation} \label{dalpha}
			\frac{d\underline{\alpha}}{d\Delta} + \frac{2\underline{\alpha}+ T_1(\Delta,\underline{\theta})}{\Delta} = B(\Delta,\underline{\theta}),
		\end{equation}
		where we've defined
		\begin{equation} \label{B}
			B(\Delta,\underline{\theta}) := - \underline{\alpha}^2 - 2\underline{\alpha} T_1(\Delta,\underline{\theta}) - \underline{\alpha}^2 T_1(\Delta,\underline{\theta}) \Delta - T_2(\Delta,\underline{\theta}) \cdot (1+\underline{\alpha} \Delta)^2.
		\end{equation}
		Now, we have $N(\Delta,\underline{\theta}(\Delta)) \in (0,1)$ and $\underline{\theta}(\Delta) \in \mathcal{B}$ for $\Delta \in (0,\underline{\Delta})$, so we can bound
		\begin{equation} \label{Bplus}
			|B(\Delta,\underline{\theta}(\Delta))| \leq 3B_+,
		\end{equation}
		where $B_+ \in (0,\infty)$ depends on $f, r_c$ and $r_0, m_0, \Delta_2$. The ODE for $\underline{\eta}(\Delta)$, \eqref{detadDelta}, is equivalent to: 
		\begin{equation} \label{deta}
			\frac{d\underline{\eta}}{d\Delta} + \frac{\underline{\eta} - N(\Delta, \underline{\theta})}{\Delta} = 0.
		\end{equation}
		Differentiating the definition of $\underline{\beta}(\Delta)$, for $\Delta \in (0,\Delta_2)$, and using \eqref{dalpha}, \eqref{deta}, we get:
		\begin{equation} \label{dbeta}
			\frac{d\underline{\beta}}{d\Delta} + \frac{2(\underline{\beta} - \beta_0)}{\Delta} = B(\Delta,\underline{\theta}),
		\end{equation}
		where we use an integrating factor along with \eqref{Bplus} to get the bound
		\begin{equation} \label{dbeta1}
			- 3 B_+ \Delta^2 \leq \frac{d}{d\Delta}(\Delta^2(\underline{\beta} - \beta_0)) \leq 3B_+ \Delta^2.
		\end{equation}
		We may now show that $\underline{\tilde{\beta}}(\Delta)\in[-B_+,B_+]$ for all $\Delta \in (0,\underline{\Delta})$. Indeed, suppose that for any $\tilde{\Delta} \in (0,\underline{\Delta})$, $\underline{\tilde{\beta}}(\tilde{\Delta}) > B_+$. Integrating \eqref{dbeta1} (using the upper bound for the derivative) from $\tilde{\Delta}$ to $\Delta \in (0,\tilde{\Delta})$,  we get
		\begin{equation}
			\underline{\beta}(\Delta) \geq \beta_0 + \Delta B_+ +  \frac{\tilde{\Delta}^3 (\underline{\tilde{\beta}}(\tilde{\Delta}) - B_+)}{\Delta^2},
		\end{equation}
		which contradicts $\underline{\beta}(\Delta) \lesssim 1$ for $\Delta \in (0,\underline{\Delta})$, as $\Delta \to 0$. The argument is identical if $\tilde{\beta}(\tilde{\Delta}) < -B_+$ for any $\tilde{\Delta} \in (0,\underline{\Delta})$, using the other inequality in \eqref{dbeta1}.
	\end{proof}
	\begin{definition} \label{def:Theta1}
		We define, for $\Delta, \tilde{\beta}, \eta \in \R$:
		\begin{equation} \label{alpha}
			\alpha(\Delta,\tilde{\beta},\eta) := \beta_0(1 - 2\eta) + \tilde{\beta} \Delta,
		\end{equation}
		along with
		\begin{equation}
			\mathcal{B}_1 := [-B_+-1,B_+ +1] \times [0,1] \times [0,2^{1-2n}].
		\end{equation}
		For $\theta_1 := (\tilde{\beta}, \eta, \phi) \in \R^3$ we define the function $\Theta_{1} : \R^4 \to \R^4$ by:
		\begin{equation}
			\Theta_{1}(\Delta, \theta_1) := (\Delta,\alpha(\Delta,\tilde{\beta},\eta),\eta,\phi). 
		\end{equation}
	\end{definition}
	Note that the definition of $\alpha(\Delta,\tilde{\beta},\eta)$ is such that $\underline{\alpha}(\Delta) = \alpha(\Delta,\underline{\tilde{\beta}}(\Delta),\underline{\eta}(\Delta))$. We now expand $\mathcal{M}^\pm$ in $(\Delta,\theta_1)$, in order to obtain a simplified differential equation for $\underline{\eta}$. From \eqref{alphaminus}, \eqref{alphaplus}, we have $\alpha_\pm > 2|\beta_0| + \frac{1}{r_0(1+f_0)} $. In order to use the previous expansions, we need to choose $\Delta_3 > 0$ such that $\alpha(\Delta,\tilde{\beta}.\eta) \in [-\alpha_-,\alpha_+]$ for all $(\Delta,\tilde{\beta},\eta) \in [0,\Delta_3] \times [-B_+,B_+] \times [0,1]$. We'll also choose $\Delta_3 > 0$ in order to control $|\mathcal{R}_1^\pm| \sim 1$, as well as a term in the expression for $\partial_m r$, see the statements after the definition.
	\begin{definition}
		Let $A(r,m)$ be the function defined in \eqref{pchirA} (in Proposition \ref{mddprop}), so that in particular, for $\mathscr{A} := [\frac{1}{2} r_0, r_0] \times [m_0 - \Delta_0, m_0 + \Delta_0]$, we have $A \in C^\infty(\mathscr{A})$. 
		\begin{align}
			A_+ &:= \max_{(r,m) \in \mathscr{A}} |A(r,m)|, \\
			\mathcal{R}_{2, +} &:= \max_{(\Delta,\theta) \in [0,\Delta_2] \times \mathcal{B}, \pm} |\mathcal{R}_2^\pm(\Delta,\theta)|, \\ 
			\Delta_3 &:= \min\left\{\Delta_2, \frac{(1+B_+)^{-1}}{r_0(1+f_0)}, \frac{ \frac{2}{r_0} + f'(m_0)}{8(1+f_0)(1+B_+) + 4\mathcal{R}_{2, +}}, \left[ 2A_+ r_{2n} (2 + r_{2, +} \Delta_2)\right]^{-\frac{1}{2n}} \right\}. \label{def:Delta3}
		\end{align}
	\end{definition} 
	For $\Delta \in [0,\Delta_3]$, $\theta_1 \in \mathcal{B}_1$, we have $\Theta_1(\Delta,\theta_1) \in [0,\Delta_3] \times \mathcal{B}$, and, using \eqref{alpha} in particular:
	\begin{equation} \label{R1beta}
		\mathcal{R}_1^\pm(\Theta_1(\Delta,\theta_1)) = \mp\frac{1}{2}\left( \frac{2}{r_0} + f'(m_0)\right) + \Delta (\mathcal{R}_2^\pm(\Theta_1(\Delta,\theta_1)) + 2(1+f_0) \tilde{\beta}),
	\end{equation}
	so \eqref{def:Delta3} implies, for all $(\Delta, \theta_1) \in [0,\Delta_3] \times \mathcal{B}_1$:
	\begin{equation} \label{R1sign}
		\mathcal{R}_1^+(\Theta_1(\Delta,\theta_1)) < - \frac{1}{4} \left( \frac{2}{r_0} + f'(m_0)\right), \ \mathcal{R}_1^-(\Theta_1(\Delta,\theta_1)) > \frac{1}{4} \left( \frac{2}{r_0} + f'(m_0)\right).
	\end{equation}
	Noting that \eqref{pchirA} takes the following form (where we've equated $\chi=m$ as the comoving coordinate), and we can restrict to $(r,m) \in [\frac{1}{2} r_0, r_0] \times [m_0 - \Delta_3, m_0 + \Delta_3]$:
	\begin{equation} \label{pmr}
		\partial_m r(T(r,m),m) = \frac{1}{2} \frac{\frac{2}{r_c(m)}+ f'(m)}{\frac{2m}{r} + f(m)} (r - r_c(m)) \left[ 1 + (r-r_c(m)) A(r,m) \right],
	\end{equation}
	the last term in \eqref{def:Delta3} implies that, for all $(\Delta,\theta) \in [0,\Delta_3] \times \mathcal{B}$:
	\begin{equation} \label{Abd}
		|A(r(\Delta,\phi),m^\pm(\Delta,\eta)) \cdot (r(\Delta,\phi) - r_c^\pm(\Delta,\eta))| \leq \frac{1}{2}.
	\end{equation}
	We may therefore define, for $(\Delta,\theta) \in [0,\Delta_3] \times \mathcal{B}$:
	\begin{equation} \label{M1pm}
		1 + M^\pm_1(\Delta,\theta) \Delta := \frac{ \left( \frac{2m^\pm(\Delta,\eta)}{r(\Delta,\phi)} + f^\pm(\Delta,\eta)\right)^\frac{1}{2}\left( \frac{2m_0}{r_0} + f_0 \right)^{-\frac{1}{2}} \left( \frac{2}{r_0} + f'(m_0)\right)\left( \frac{2}{r^\pm_c(\Delta,\eta)} + f'(m^\pm(\Delta,\eta)) \right)^{-1}}{\left(1+X^\pm_d(\Delta,\theta) \Delta\right) \left[1 + A(r(\Delta,\phi),m^\pm(\Delta,\eta))(r(\Delta,\phi)-r^\pm_c(\Delta,\eta))\right]},
	\end{equation}
	so that, by \eqref{def:Delta2}, \eqref{def:Delta3} and \eqref{sim1}, \eqref{Abd}, we have $M^\pm_1 \in C^\infty([0,\Delta_3] \times \mathcal{B})$, and $1 + M^\pm_1(\Delta,\theta) \Delta > 0$ for all $(\Delta,\theta) \in [0,\Delta_3] \times \mathcal{B}$. We also define, for $(\Delta,\theta_1) \in [0,\Delta_3] \times \mathcal{B}_1$:
	\begin{equation} \label{X1pm}
		1 + X_1^\pm(\Delta,\theta_1) \Delta := \mp \left( \frac{f'(m_0)}{2} + \frac{1}{r_0} \right)^{-1} \mathcal{R}_1^\pm(\Theta_1(\Delta,\theta_1)),
	\end{equation}
	so $1 + \Delta X_1^\pm(\Delta,\theta_1) > \frac{1}{4}$ for all $(\Delta,\theta_1) \in [0,\Delta_3] \times \mathcal{B}_1$ by \eqref{R1sign}, and from \eqref{R1beta} we get
	\begin{equation}
		X_1^\pm(\Delta,\theta_1) = \mp \left( \frac{f'(m_0)}{2} + \frac{1}{r_0}\right)^{-1} [2(1+f_0)\tilde{\beta} + \mathcal{R}_2^\pm(\Theta_1(\Delta,\theta_1))],
	\end{equation}
	so $X_1^\pm \in C^\infty([0,\Delta_3] \times \mathcal{B}_1)$. We define also: 
	\begin{equation} \label{Y1pm}
		1 + Y_1^\pm(\Delta,\theta_1) \Delta := (1+ X_1^\pm(\Delta,\theta_1)\Delta )(1+ M_1^\pm(\Theta_1(\Delta,\theta)) \Delta),
	\end{equation}
	where $Y_1^\pm \in C^\infty([0,\Delta_3] \times \mathcal{B}_1)$ and $1 + Y_1^\pm(\Delta,\theta_1) \Delta > 0$ for all $(\Delta,\theta_1) \in [0,\Delta_3] \times \mathcal{B}_1$. We finally define for $(\Delta,\theta_1) \in [0,\Delta_3] \times \mathcal{B}_1$:
	\begin{align} \label{def:Npm}
		\mathcal{N}^\pm(\Delta,\theta_1) &:= a^\pm(\eta)^{2n} - \phi + N_1^\pm(\Delta,\theta_1) \Delta := (a^\pm(\eta)^{2n}-\phi + r_{c, 2}^\pm(\Delta,\eta)\Delta) (1 + Y_1^\mp(\Delta,\theta_1) \Delta), \\ D(\Delta,\theta_1) &:= \mathcal{N}^+(\Delta,\theta_1) - \mathcal{N}^-(\Delta,\theta_1) =: (1-\eta)^{2n} - \eta^{2n} + D_1(\Delta,\theta_1) \Delta \label{def:D}
	\end{align}
	so that $\mathcal{N}_1^\pm, N_1^\pm, D, D_1 \in C^\infty([0,\Delta_3] \times \mathcal{B}_1)$. 
	\begin{definition}
		\begin{equation}
			D_{1, +} := \max_{(\Delta,\theta_1) \in [0,\Delta_3] \times \mathcal{B}_1} |D_1(\Delta,\theta_1)|.
		\end{equation}
	\end{definition}
	
	\begin{definition} \label{def:tilde-eta}
		For $\Delta, \tilde{\eta} \in \mathbb{R}$:
		\begin{equation} \label{def:tildeeta}
			\eta(\Delta,\tilde{\eta}) := \frac{1}{2} + \tilde{\eta} \Delta.
		\end{equation}
	\end{definition}
	
	We may now show:
	\begin{lemma} \label{lem:eta}
		We impose that $\underline{\Delta} \in (0,\Delta_3]$. Then $\underline{\eta} \in C^0([0,\underline{\Delta}))$ with $\underline{\eta}(0) = \frac{1}{2}$. Moreover, defining $\underline{\eta}(\Delta) =: \frac{1}{2} + \underline{\tilde{\eta}}(\Delta) \Delta$, we have $\underline{\tilde{\eta}}(\Delta) \in [-N_+, N_+]$ for all $\Delta \in (0,\underline{\Delta})$, where $N_+ := \frac{1}{n} 2^{2n-3}(1 + D_{1, +})$.
	\end{lemma}
	\begin{proof}
		We define $\underline{\theta_1}(\Delta) := (\underline{\tilde{\beta}}(\Delta), \eta(\Delta), \phi(\Delta))$. We may now expand $\mathcal{M}^\pm$, using $\underline{\theta_1}(\Delta) \in \mathcal{B}_1$ for $\Delta \in (0,\underline{\Delta})$, along with \eqref{rminrcm}, \eqref{rminrcp}, and \eqref{tildeX2}, \eqref{pmr}, \eqref{X1pm}, \eqref{Y1pm}, to see that: (where the $\Delta$ argument in $\underline{\eta}$, $\underline{\phi}$ is suppressed for legibility)
		\begin{equation} \label{Mpm}
			\mathcal{M}^\pm(\underline{\Psi_0}(\Delta)) = \frac{\widetilde{X}^\pm(\underline{\Psi_0}(\Delta))}{\partial_m r(T(r(\Delta,\underline{\phi}),m^\pm(\Delta,\underline{\eta})),m^\pm(\Delta,\underline{\eta}))} = \pm \frac{\Delta^{1-2n}}{2r_{2n}} \left( \frac{2m_0}{r_0} + f_0 \right)^\frac{1}{2} \frac{1 + Y_1^\pm(\Delta,\underline{\theta_1}(\Delta))}{a^\pm(\underline{\eta})^{2n} - \underline{\phi} + r_{c, 2}^\pm(\Delta,\underline{\eta})\Delta}.
		\end{equation}
		We rearrange the definition of $N$, \eqref{def:N}, for $\Delta \in (0,\underline{\Delta})$, to see that:
		\begin{equation} \label{N-Npm}
			N(\Theta_1(\Delta,\underline{\theta}_1(\Delta))) = \frac{\mathcal{N}^+(\Delta,\underline{\theta_1}(\Delta))}{\mathcal{N}^+(\Delta,\underline{\theta_1}(\Delta)) + \mathcal{N}^-(\Delta,\underline{\theta_1}(\Delta))},
		\end{equation}
		where we've used the definition \eqref{def:Npm}, and we have $\mathcal{N}^\pm(\Delta,\underline{\theta_1}(\Delta)) > 0$ for all $\Delta \in [0,\Delta_3]$ by \eqref{MDDbd} (and more directly \eqref{rminrcm}, \eqref{rminrcp}) and \eqref{Y1pm}. We observe that, $\Delta \in (0,\underline{\Delta})$ we have $N(\Theta_1(\Delta,\underline{\theta}_1(\Delta))) > \frac{1}{2} \iff D(\Delta,\underline{\theta_1}(\Delta)) > 0$, and analogously $N(\Theta_1(\Delta,\underline{\theta}_1(\Delta))) < \frac{1}{2} \iff D(\Delta,\underline{\theta_1}(\Delta)) < 0$, for $D$ defined in \eqref{def:D}. 
		
		We consider the function, for $x \in [0,1]$ (where $n \in \mathbb{N}$ is fixed by leading order term in $r_c$):
		\begin{equation}
			h(x) := (1-x)^{2n} - x^{2n}.
		\end{equation}
		Differentiating, we bound for all $x \in [0,1]$:
		\begin{equation}
			h'(x) = -2n \left[ (1-x)^{2n-1} + x^{2n-1}\right] \leq -n 2^{3-2n} =: - C_n
		\end{equation}
		Therefore, using \eqref{def:tildeeta}, we have for all $\Delta \in [0,\Delta_3]$, and any $\tilde{\eta} \in (0,\infty)$ such that $\eta(\Delta,\tilde{\eta}) \in [\frac{1}{2},1]$: 
		\begin{equation}
			h(\eta(\Delta,\tilde{\eta}))  \leq - C_n \tilde{\eta} \Delta,
		\end{equation}
		and for any $\tilde{\eta} \in (-\infty,0)$ such that $\eta(\Delta,\tilde{\eta}) \in [0,\frac{1}{2}]$:
		\begin{equation}
			h(\eta(\Delta,\tilde{\eta})) \geq - C_n \tilde{\eta} \Delta.
		\end{equation}
		We note that for all $\Delta \in (0,\tilde{\Delta})$ we have by definition $\underline{\eta}(\Delta) \in (0,1)$. Now, by definition of $N_+$, if $\underline{\tilde{\eta}}(\tilde{\Delta}) > N_+$ for some $\tilde{\Delta} \in (0,\underline{\Delta})$, then 
		\begin{equation}
			D(\tilde{\Delta},\underline{\theta_1}(\tilde{\Delta})) \leq (-C_n N_+ + D_{1,+} )\tilde{\Delta} = - \tilde{\Delta},
		\end{equation}
		and so we have $N(\Theta_1(\tilde{\Delta},\underline{\theta}_1(\tilde{\Delta}))) < \frac{1}{2}$. For any $\tilde{\Delta} \in (0,\underline{\Delta})$, we define the set:
		\begin{equation} \label{A}
			\mathcal{A}(\tilde{\Delta}) := \left\{ x \in (0,\tilde{\Delta}) : \text{ for all } \Delta \in (x,\tilde{\Delta}], \ \underline{\tilde{\eta}}(\Delta) > \left(\frac{\tilde{\Delta}}{\Delta}\right)^2 \cdot N_+ \right\}.
		\end{equation}
		If $\underline{\tilde{\eta}}(\tilde{\Delta}) > N_+$, then $\mathcal{A}(\tilde{\Delta})$ is non-empty by continuity, while $\mathcal{A}(\tilde{\Delta})$ is closed by definition. We now show that $\mathcal{A}(\tilde{\Delta})$ is open. Suppose that $x \in \mathcal{A}(\tilde{\Delta})$, then for all $\Delta \in (x,\tilde{\Delta}]$, the definition \eqref{A} implies that $\underline{\tilde{\eta}}(\Delta) > N_+$, and so $N(\Theta_1(\tilde{\Delta},\underline{\theta}_1(\tilde{\Delta}))) < \frac{1}{2}$. But then from \eqref{deta}, we have:
		\begin{equation}
			\frac{d\underline{\eta}}{d\Delta}(\Delta) = \frac{N(\Theta_1(\Delta,\underline{\theta}_1(\Delta))) - \underline{\eta}(\Delta)}{\Delta} < \frac{1/2 - \underline{\eta}(\Delta)}{\Delta} \implies \frac{d}{d\Delta} \left[ \Delta \left( \underline{\eta}(\Delta) - \frac{1}{2}\right)\right] < 0.
		\end{equation}
		Integrating from $\tilde{\Delta}$ to $x$, we get:
		\begin{equation}
			\underline{\eta}(x) > \frac{1}{2} + \frac{\tilde{\Delta}^2 N_+}{x} \iff \tilde{\underline{\eta}}(x) > \left( \frac{\tilde{\Delta}}{x} \right)^2 N_+,
		\end{equation}
		and therefore by continuity we have $\mathcal{A}(\tilde{\Delta})$ open. But this is clearly a contradiction to $\underline{\eta}(\Delta) \in (0,1)$ for all $\Delta \in (0,\underline{\Delta})$, for sufficiently small $\Delta$, so we must have $\tilde{\underline{\eta}}(\Delta) \leq N_+$ for all $\Delta \in (0,\tilde{\Delta})$. 
		
		If there exists $\tilde{\Delta} \in (0,\underline{\Delta})$ with $\underline{\tilde{\eta}}(\Delta) < -N_+$, then we'd have $D(\tilde{\Delta},\underline{\theta_1}(\tilde{\Delta})) \geq \tilde{\Delta}$, and therefore $N(\Theta_1(\tilde{\Delta},\underline{\theta}_1(\tilde{\Delta}))) > \frac{1}{2}$. An analogous argument shows that we must therefore have $\tilde{\underline{\eta}}(\Delta) \geq - N_+$ for all $\Delta \in (0,\underline{\Delta})$. Hence, we have $\underline{\eta}(\Delta)$ extending continuously to $\Delta=0$, with $\underline{\eta}(0) = \frac{1}{2}$.
 	\end{proof}
 		
	We now consider the ODE for $\underline{\phi}$, for which we need to expand \eqref{drdDelta}. We define, for $(\Delta,\theta) \in [0,\Delta_3] \times \mathcal{B}$:
	\begin{equation} \label{R1}
		1 + \mathcal{R}_1(\Delta,\theta) \Delta := - \left( \frac{2m_0}{r_0} + f_0 \right)^{-\frac{1}{2}} \mathcal{R}(\Delta,\theta),
	\end{equation}
	where \eqref{Rbd} implies that $\frac{1}{2} \leq 1 + \mathcal{R}_1(\Delta,\theta) \Delta \leq \frac{3}{2}$ for all $(\Delta,\theta) \in [0,\Delta_3] \times \mathcal{B}$, and by definition $\mathcal{R}_1(\Delta,\theta) \in C^\infty([0,\Delta_3] \times \mathcal{B})$.
	\begin{definition} \label{def:a+Delta4}
		\begin{equation} \label{anplus}
			a_{n, +} := \max_{x \in [-\frac{1}{2},\frac{1}{2}]} \frac{2^{-2n}}{x}  \left[ (1+x)^{2n} - 1 \right], \qquad M_{1, +} := \max_{(\Delta, \theta) \in [0,\Delta_3] \times \mathcal{B}, \pm} |M_1^\pm(\Delta,\theta)|, \qquad \delta(n) := \frac{n 2^{-2n}}{24+2n},
		\end{equation}
		\begin{equation} \label{def:Delta4}
			\Delta_4 := \min\left\{ \Delta_3, \frac{1}{4(1+N_+)}, \frac{1}{2 M_{1, +}}, \frac{\delta(n)}{r_{2, +} + 2(1 + N_+)a_{n, +}}\right\}.
		\end{equation}
	\end{definition}
	For $(\Delta,\tilde{\eta}) \in [0,\Delta_4] \times [-N_+-1,N_+ + 1]$, we have $\eta(\Delta,\tilde{\eta}) \in [\frac{1}{4}, \frac{3}{4}]$. We now expand the term involving $\underline{\eta}$ in the denominator of \eqref{Mpm}, and use the remaining bounds on $\Delta_4$ from the definition. Indeed, we define, for $(\Delta,\tilde{\eta}) \in [0,\Delta_4] \times [-N_+-1,N_++1]$:
	\begin{equation} \label{def:tilderc}
		\tilde{r}_{c, 2}^\pm(\Delta,\tilde{\eta}) := r_{c, 2}^\pm(\Delta,\eta(\Delta,\tilde{\eta})) + \Delta^{-1} \left[ a^\pm(\eta(\Delta,\tilde{\eta}))^{2n} - 2^{-2n} \right],
	\end{equation}
	which is smooth, $\tilde{r}_{c,2}^\pm \in C^\infty([0,\Delta_4] \times [-N_+-1,N_++1])$, and we can bound for all $(\Delta,\tilde{\eta}) \in [0,\Delta_4] \times [-N_+-1,N_++1]$:
	\begin{equation} \label{tilderbd}
		|\tilde{r}_{c, 2}^\pm(\Delta,\tilde{\eta})|\Delta \leq (r_{2, +} + 2(1 + N_+)a_{n, +}) \Delta_4 \leq \delta(n),
	\end{equation}
	where we've used $|\Delta \tilde{\eta}| \leq \frac{1}{4}$, \eqref{anplus}, and \eqref{def:Delta4}. The bound in terms of $M_{1, +}$ in \eqref{def:Delta4} implies that $1 + M_1^\pm(\Delta,\theta) \Delta > \frac{1}{2}$ for all $(\Delta,\theta) \in [0,\Delta_4] \times \mathcal{B}$, and therefore: for all $(\Delta, \theta_1) \in [0,\Delta_4] \times \mathcal{B}_1$:
	\begin{equation} \label{Y1bd}
		1 + Y_1^\pm(\Delta,\theta_1) \Delta > \frac{1}{8}, 
	\end{equation}
	where we bound $X_1^\pm$ using \eqref{X1pm}. We define:
	\begin{definition} \label{def:tilde-phi}
		$\phi(\Delta,\tilde{\phi}) = \frac{2^{-2n}}{1+2n} + \Delta \tilde{\phi}$.
	\end{definition}
	Using the bounds that follow from the choice of $\Delta_4$, \eqref{def:Delta4}, we can show:
	\begin{lemma} \label{lem:phi}
		We impose that $\underline{\Delta} \in (0,\Delta_4]$. Then $\underline{\phi} \in C^0([0,\underline{\Delta}))$, with $\underline{\phi}(0) = \frac{2^{-2n}}{1+2n} =: \phi_0$. Moreover, defining $\underline{\phi}(\Delta) =: \phi_0 + \underline{\tilde{\phi}}(\Delta) \Delta$, we have $\underline{\tilde{\phi}}(\Delta) \in [- P_+, P_+]$ for all $\Delta \in (0,\underline{\Delta})$, for $P_+$ determined in the proof, which depends on $n, r_0, m_0$ and $f,r_c$. 
	\end{lemma}
	\begin{proof}
		Note that using \eqref{def:tilderc}, we have 
		\begin{equation}
			a^\pm(\underline{\eta}(\Delta)) - \underline{\phi}(\Delta) + r_{c, 2}^\pm(\Delta,\underline{\eta}(\Delta)) = 2^{-2n} - \phi + \tilde{r}_{c,2}^\pm(\Delta,\underline{\tilde{\eta}}(\Delta)) \Delta > 0,
		\end{equation}
		for all $\Delta \in [0,\underline{\Delta}]$, by \eqref{MDDbd}. From \eqref{Mpm} and \eqref{R1}, we have:
		\begin{equation}
			\frac{\mathcal{R}(\underline{\Psi_0}(\Delta))}{\mathcal{M}^+(\underline{\Psi_0}(\Delta)) - \mathcal{M}^-(\underline{\Psi_0}(\Delta))} = -2 r_{2n} \Delta^{2n-1}  \widetilde{P}_0(\Delta, \tilde{\underline{\beta}}(\Delta), \tilde{\underline{\eta}}(\Delta), \underline{\phi}(\Delta)) (1+ \Delta \mathcal{R}_1(\Delta,\underline{\theta}(\Delta))),
		\end{equation}
		where:
		\begin{equation} \label{P0}
			\widetilde{P}_0(\Delta, \tilde{\underline{\beta}}, \tilde{\underline{\eta}}, \underline{\phi}) := \frac{( 2^{-2n} - \underline{\phi} + \tilde{r}_{c, 2}^+(\Delta,\underline{\tilde{\eta}}) \Delta)( 2^{-2n} - \underline{\phi} + \tilde{r}_{c, 2}^-(\Delta,\underline{\tilde{\eta}})\Delta)}{(2^{-2n} - \underline{\phi} + \tilde{r}_{c, 2}^+(\Delta,\underline{\tilde{\eta}}) \Delta)(1+Y_1^-(\Delta,\underline{\theta_1}) \Delta) + (2^{-2n} - \underline{\phi} + \tilde{r}_{c, 2}^-(\Delta,\underline{\tilde{\eta}}) \Delta)(1+Y_1^+(\Delta,\underline{\theta_1})\Delta)},
		\end{equation}
		and so \eqref{drdDelta} becomes:
		\begin{equation} \label{dphi}
			\frac{d\underline{\phi}}{d\Delta} + \frac{2n \underline{\phi}}{\Delta} = \frac{2 (1+ \mathcal{R}_1(\Delta,\underline{\theta})\Delta)}{\Delta} \cdot \widetilde{P}_0(\Delta, \tilde{\underline{\beta}}, \tilde{\underline{\eta}}, \underline{\phi}).
		\end{equation}
		We'll first show that by the definition of $\Delta_4$, \eqref{def:Delta4}, we must have $2^{-2n} - \underline{\phi}(\Delta) \geq \delta(n)$ for all $\Delta \in [0,\Delta_4]$. Suppose that there exists $\tilde{\Delta} \in (0,\underline{\Delta})$ such that $2^{-2n} - \underline{\phi}(\tilde{\Delta}) < \delta(n)$. We then have, using \eqref{tilderbd}:
		\begin{equation} \label{2delta}
			2^{-2n} - \underline{\phi}(\tilde{\Delta}) + \tilde{r}_{c,2}^\pm(\tilde{\Delta},\underline{\tilde{\eta}}(\tilde{\Delta})) \Delta < 2\delta(n),
		\end{equation}
		and we can then bound $\widetilde{P}_0$ using \eqref{Y1bd} and \eqref{2delta}:
		\begin{equation}
			\widetilde{P}_0(\tilde{\Delta}, \tilde{\underline{\beta}}(\tilde{\Delta}), \tilde{\underline{\eta}}(\tilde{\Delta}), \underline{\phi}(\tilde{\Delta})) < 8 \delta(n).
		\end{equation}
		Using the bound for \eqref{R1}, and \eqref{anplus}, we get:
		\begin{equation} \label{dphibd}
			\frac{d\underline{\phi}}{d\Delta}(\tilde{\Delta}) < \frac{24 \delta(n) - 2n \underline{\phi}(\tilde{\Delta})}{\tilde{\Delta}} < \frac{-n 2^{-2n}}{\tilde{\Delta}},
		\end{equation}
		but we then have $2^{-2n} - \underline{\phi}(\Delta) < \delta(n)$ for all $\Delta \in (0,\tilde{\Delta}]$, and so \eqref{dphibd} holds for all $\Delta \in (0,\tilde{\Delta}]$, which leads to a contradiction to $\underline{\phi} \leq 2^{-2n}$, after integrating to sufficiently small $\Delta > 0$. 
		
		Now, let $\mathcal{B}_2 := [-B_+-1,B_++1] \times [-N_+-1,N_++1] \times [0,2^{1-2n} - \delta(n)]$. We may divide through in \eqref{P0} by $2^{-2n} - \underline{\phi} \geq \delta(n)$, which gives:
		\begin{equation} \label{P1}
			\widetilde{P}_0(\Delta, \tilde{\underline{\beta}}, \tilde{\underline{\eta}}, \underline{\phi}) =: \frac{1}{2}\left[ 2^{-2n} - \underline{\phi} +  \widetilde{P}_1(\Delta, \tilde{\underline{\beta}}, \tilde{\underline{\eta}},\underline{\phi}) \Delta \right], 
		\end{equation}
		where $\widetilde{P}_1(\Delta,\tilde{\beta},\tilde{\eta},\phi) \in C^\infty([0,\Delta_4] \times \mathcal{B}_2)$. Using \eqref{P1} in \eqref{dphi}:
		\begin{equation}
			\frac{d\underline{\phi}}{d\Delta} + \frac{(1+2n) \underline{\phi} - 2^{-2n}}{\Delta} = \widetilde{P}_2(\Delta, \tilde{\underline{\beta}}, \tilde{\underline{\eta}}, \underline{\phi}),
		\end{equation}
		where $\widetilde{P}_2(\Delta, \tilde{\beta}, \tilde{\eta}, \phi) := \widetilde{P}_1(\Delta, \beta, \eta, \phi) + 2 \mathcal{R}_1(\Delta,\alpha(\Delta,\tilde{\beta},\tilde{\eta}), \eta(\Delta,\tilde{\eta}),\phi) \widetilde{P}_0(\Delta, \tilde{\beta}, \tilde{\eta}, \phi) \in C^\infty([0,\Delta_4] \times \mathcal{B}_2)$. We define
		\begin{equation}
			P_+ := \frac{1}{2n+2} \max_{(\Delta, \tilde{\beta}, \tilde{\eta}, \underline{\phi}) \in [0,\Delta_4] \times \mathcal{B}_2} |\widetilde{P}_2(\Delta, \tilde{\beta}, \tilde{\eta}, \phi)|,
		\end{equation}
		and it follows (see e.g.~the proof of Lemma \ref{lem:beta}) that we must have $\underline{\tilde{\phi}}(\Delta) \in [-P_+,P_+]$ for all $\Delta \in (0,\underline{\Delta})$, and therefore $\underline{\phi}(\Delta)$ extends continuously to $\Delta=0$, with $\underline{\phi}(0) = \phi_0$. 
	\end{proof}
	
	\subsubsection{Expanding to second order} \label{secondorder}
			
		We now expand all functions in the ODE \eqref{drdDelta}, \eqref{detadDelta}, \eqref{depsdDelta} to second order, so that we can show that the ODE for the second terms has a regular singular point structure, see already Section \ref{contraction}. 
	
		\begin{definition} \label{def:tilde-theta}
			For $\tilde{\theta} := (\tilde{\beta}, \tilde{\eta}, \tilde{\phi}) \in \R^3$, we define the map $\Psi_2 : \R^4 \to \R^4$:
			\begin{align}
				\Theta_2(\Delta,\tilde{\theta}) &:= (\Delta, \alpha(\Delta,\tilde{\beta},\eta(\Delta,\tilde{\eta})), \eta(\Delta,\tilde{\eta}), \phi(\Delta,\tilde{\phi})), \\
				\Psi_2(\Delta,\tilde{\theta}) &:= \Psi_1(\Theta_2(\Delta,\tilde{\theta})).
			\end{align} 
			We define also 
			\begin{equation} \label{tildeB}
				\tilde{\mathcal{B}} := [-B_+-1,B_++1] \times [-N_+-1,N_++1] \times [-P_+-1,P_++1].
			\end{equation}
		\end{definition}
		\begin{remark} \label{rem:firstorder}
			In Section \ref{firstorder}, we've shown: $(\underline{\tilde{\beta}}(\Delta), \underline{\tilde{\eta}}(\Delta), \underline{\tilde{\phi}}(\Delta)) \in [-B_+,B_+] \times [-N_+,N_+] \times [-P_+,P_+] \subset \tilde{\mathcal{B}}$, for all $\Delta \in (0,\underline{\Delta}) \subseteq (0,\Delta_4)$.
		\end{remark}
		We choose $\Delta_5$ so that we control $\phi(\Delta,\tilde{\phi})$ for $(\Delta,\tilde{\theta}) \in \B$:
		\begin{definition}
			\begin{equation} \label{def:Delta5}
				\Delta_5 := \min\left\{ \Delta_4, \frac{(1+P_+)^{-1}}{2^{1+2n}(1+2n)}\right\}.
			\end{equation}
		\end{definition}
		For $(\Delta,\tilde{\phi}) \in [0,\Delta_5] \times [-P_+-1,P_++1]$, we have $|\Delta \tilde{\phi}| \leq \frac{1}{2} \frac{2^{-2n}}{1+2n}$, and therefore $\phi(\Delta,\tilde{\phi}) \in [\frac{1}{2} \frac{2^{-2n}}{1+2n},\frac{3}{2} \frac{2^{-2n}}{1+2n}]$, so we have $\Theta_2(\Delta,\tilde{\theta}) \in [0,\Delta_5] \times \mathcal{B}$ for all $\tilde{\theta} \in \tilde{\mathcal{B}}$. We show moreover that \eqref{def:Delta5} implies that we can control the denominator of \eqref{Mpm} $\sim 1$. Indeed, from \eqref{def:tilderc} we have, for all $(\Delta,\tilde{\theta}) \in [0,\Delta_5] \times \tilde{\mathcal{B}}$:
		\begin{align} \label{denbd}
			a^\pm(\eta(\Delta,\tilde{\eta})) - \phi(\Delta,\tilde{\phi}) + r_{c, 2}^\pm(\Delta,\eta(\Delta,\tilde{\eta})) \Delta &= \frac{n2^{1-2n}}{1+2n} + ( \tilde{r}_{c, 2}^\pm(\Delta,\tilde{\eta}) - \tilde{\phi}) \Delta  \\ &\geq 2^{-1-2n} \left[ \frac{4n}{1+2n} - \frac{2n}{24+2n} -  \frac{1}{1+2n} \right] > 0,
		\end{align}
		where we've used \eqref{tilderbd} and \eqref{def:Delta5}.
		\begin{remark} \label{rem:cond}
			For all $(\Delta,\tilde{\theta}) \in [0,\Delta_5] \times \tilde{\mathcal{B}}$, we have:
			\begin{equation}
				r(\Delta,\phi(\Delta,\tilde{\phi})) - r_c^\pm(\Delta,\eta(\Delta,\tilde{\eta})) > 0,
			\end{equation}
			\begin{equation}
				\widetilde{\mathcal{X}}^+(\Psi_2(\Delta,\tilde{\theta})) > 0, \qquad \widetilde{\mathcal{X}}^-(\Psi_2(\Delta,\tilde{\theta})) < 0,
			\end{equation}
			while $\epsilon(\Delta,\alpha(\Delta,\eta(\Delta,\tilde{\eta})))$, $r(\Delta,\phi(\Delta,\tilde{\phi}))$, $\mathcal{R}(\Psi_2(\Delta,\tilde{\theta}))$ satisfy the respective bounds: \eqref{epsbd}, \eqref{rbd}, \eqref{Rbd}.
		\end{remark}
		\begin{lemma} \label{lem:unique}
			We restrict to $\underline{\Delta} \in (0,\Delta_5]$. We then have $\underline{\tilde{\theta}}(\Delta) := (\underline{\tilde{\beta}}(\Delta), \underline{\tilde{\eta}}(\Delta),\underline{\tilde{\phi}}(\Delta)) \in C^2((0,\underline{\Delta}); \tilde{\mathcal{B}})$, then we can uniquely extend to $\underline{r},\underline{\eta},\underline{\epsilon} \in C^0([0,\Delta_5]) \cap C^2((0,\Delta_5])$ solving \eqref{drdDelta}, \eqref{detadDelta}, \eqref{depsdDelta} and satisfying \eqref{MDDbd}, \eqref{sign}, \eqref{detcondDelta}, with $\underline{\tilde{\theta}} \in C^2((0,\Delta_5])$.
		\end{lemma}
		\begin{proof}
			We've shown in Section \ref{firstorder} that we have $\underline{\tilde{\theta}}(\Delta) \in \tilde{\mathcal{B}}$ for all $\underline{\Delta} \leq \Delta_5$, but then from Remark \eqref{rem:cond} we see that the conditions \eqref{MDDbd}, \eqref{sign}, \eqref{detcondDelta} are satisfied for all $\underline{\Delta} \leq \Delta_5$, and we can then apply Picard-Lindel\"of to uniquely extend any solution of the ODE to $\underline{\Delta} = \Delta_5$.
		\end{proof}
		\begin{remark}
			We may therefore set $\underline{\Delta} = \Delta_5$.
		\end{remark}
		We now expand all functions to the next order, in terms of $(\Delta,\tilde{\theta}) \in [0,\Delta_5] \times \tilde{\mathcal{B}}$, see the following section for the conclusion of the argument based on the expansions in this section.
		
		We start by expanding the relevant terms in order to determine $B$, \eqref{B}, to next order: we need to expand $T(\Psi_2(\Delta,\tilde{\theta}))$ and $\alpha(\Delta,\tilde{\beta},\eta(\Delta,\tilde{\eta}))$. We see from \eqref{alpha} that we have:
		\begin{equation} \label{tilde-alpha}
			\alpha(\Delta,\tilde{\beta},\eta(\Delta,\tilde{\eta})) = \Delta(\tilde{\beta} - 2 \beta_0 \tilde{\eta}) =: \Delta \tilde{\alpha}(\tilde{\beta},\tilde{\eta})
		\end{equation}
		Now, we note that \eqref{denbd} implies that  $\mathcal{N}^\pm(\Delta,\tilde{\beta},\eta(\Delta,\tilde{\eta}),\phi(\Delta,\tilde{\phi})) > 0$ (defined in \eqref{def:Npm}) for all $(\Delta,\tilde{\theta}) \in [0,\Delta_5] \times \tilde{\mathcal{B}}$ , and therefore, for $N$ defined in \eqref{def:N}, \eqref{N-Npm} implies that $N \circ \Theta_2 \in C^\infty([0,\Delta_5] \times \tilde{\mathcal{B}})$. We may now expand $T$, \eqref{Tdef}, as a smooth function in $(\Delta,\tilde{\theta})$. We'll use \eqref{Texp}, which we write in terms of $\tilde{\theta}$ as:
		\begin{equation}
			(1+f_0)^\frac{1}{2} T(\Psi_2(\Delta,\tilde{\theta})) = 1 + T_1(\Theta_2(\Delta,\tilde{\theta})) \cdot \Delta + T_2(\Theta_2(\tilde{\theta})) \cdot \Delta^2,
		\end{equation}
		where we have $T_1 \circ \Theta_2, T_2 \circ \Theta_2 \in C^\infty(\B)$ by definition, \eqref{T1} and \eqref{T2}, and therefore $B \circ \Theta_2 \in C^\infty(\B)$ by definition, \eqref{B}. From \eqref{tilde-alpha} and \eqref{B}, we see that we need only expand $T_2 \circ \Theta_2$. We expand \eqref{F2}:
		\begin{equation}
			F_2^\pm(\Delta,\eta(\Delta,\tilde{\eta})) =: F_{2, 0} + \Delta F_3^\pm(\Delta,\tilde{\eta}),
		\end{equation}
		where $F_{2, 0} := -\frac{f_2(m_0)}{8(1+f_0)} + \frac{3}{8}\left(\frac{f'(m_0)}{2(1+f_0)}\right)^2$, and then $F_3^\pm \in C^\infty([0,\Delta_5] \times [-N_+-1,N_++1])$ by \eqref{sim1}, and we've used $a^\pm(\eta(\Delta,\tilde{\eta})) = \pm\frac{1}{2} - \tilde{\eta}\Delta$. Expanding also $X_0^\pm$, defined in \eqref{tildeX2}, using in particular \eqref{R1beta}:
		\begin{equation}
			X_0^\pm(\Theta_2(\Delta,\tilde{\theta})) = \pm \frac{1}{4} \left( \frac{2m_0}{r_0} + f_0 \right)^{-1/2} \left( \frac{2}{r_0} + f'(m_0) \right) + \Delta X_{0,1}^\pm(\Delta,\tilde{\theta}),
		\end{equation}
		where $X_{0,1}^\pm \in C^\infty(\B)$. We may now expand \eqref{Tpm2}: 
		\begin{equation}
			\mathcal{T}_2^\pm(\Theta_2(\Delta,\tilde{\theta})) =: \mathcal{T}_{2, 0} + \mathcal{T}_3^\pm(\Delta,\tilde{\theta}) \Delta,
		\end{equation}
		where 
		\begin{equation}
			\mathcal{T}_{2,0} = \frac{\left( \frac{2}{r_0}+f'(m_0)\right)^2}{32(1+f_0)\left( \frac{2m_0}{r_0} + f_0 \right)},
		\end{equation}
		and $\mathcal{T}_3^\pm \in C^\infty(\B)$. Then expanding \eqref{T2}, using that the leading order terms in each bracket are independent of $\pm$, we get
		\begin{equation}
			T_2(\Theta_2(\Delta,\tilde{\theta})) =: -3\tilde{\beta}_0 +  T_3(\Delta,\tilde{\theta}) \Delta,
		\end{equation}
		where $\tilde{\beta}_0 = - \frac{1}{3}[F_{2, 0} + \mathcal{T}_{2, 0}(1 + f_0)^{1/2}]$, and $T_3 \in C^\infty(\B)$. We can now expand \eqref{B}:
		\begin{equation} \label{Bexp}
			B(\Theta_2(\Delta,\tilde{\theta})) =: 3\tilde{\beta}_0 + P_1(\Delta,\tilde{\theta}) \Delta,
		\end{equation}
		where $P_1 \in C^\infty(\B)$. We now expand $\mathcal{R}$ to next order: expanding directly from \eqref{Rexp}, we get
		\begin{equation} \label{R2exp}
			\mathcal{R}(\Psi_2(\Delta,\tilde{\theta})) = - \left( \frac{2m_0}{r_0} + f_0\right)^\frac{1}{2} \left(1 + \mathcal{R}_2(\Delta,\tilde{\theta}) \Delta^2 \right),
		\end{equation}
		where $\mathcal{R}_2 \in C^\infty(\B)$. The remaining term to expand is $\mathcal{M}^\pm$. Expanding \eqref{R2def}:
		\begin{equation}
			\mathcal{R}_2^\pm(\Theta_2(\Delta,\tilde{\theta})) =: \frac{1}{4r_0^2(1+f_0)} - \frac{1}{4} f_2(m_0) + \mathcal{R}_3^\pm(\Delta,\tilde{\theta}) \Delta,
		\end{equation}
		with $\mathcal{R}_3^\pm \in C^\infty(\B)$, we can then expand \eqref{X1pm}:
		\begin{equation}
			X_1^\pm(\Delta,\tilde{\beta},\eta(\Delta,\tilde{\eta}),\phi(\Delta,\tilde{\phi})) = X_{1, 0}^\pm(\tilde{\beta}) + X_2^\pm(\Delta,\tilde{\theta}) \Delta,
		\end{equation}
		where
		\begin{equation}
			X_{1,0}^\pm(\tilde{\beta}) = \mp \frac{4(1+f_0)\tilde{\beta} + \frac{1}{2r_0^2(1+f_0)} - \frac{1}{4} f''(m_0)  }{\frac{2}{r_0} + f'(m_0)},
		\end{equation}
		and $X_2^\pm \in C^\infty(\B)$. We now expand the relevant terms for $M_1^\pm$, where in order to expand \eqref{Xdpm} we've also expanded \eqref{n1pm}:
		\begin{align}
			X_d^\pm(\Theta_2(\Delta, \tilde{\theta})) &=: \mp \frac{1}{8} \left( \frac{2}{r_0} + f'(m_0) \right)\left(\frac{1}{1+f_0} + \frac{1}{\frac{2m_0}{r_0} + f_0}\right) + X_{d,  1}^\pm(\Delta,\tilde{\theta}) \Delta, \\
			\left(\frac{\frac{2m^\pm(\Delta,\eta(\Delta,\tilde{\eta}))}{r(\Delta,\phi(\Delta,\tilde{\phi}))} + f^\pm(\Delta,\eta(\Delta,\tilde{\eta}))}{\frac{2m_0}{r_0} + f_0} \right)^\frac{1}{2} &=: 1 \pm \frac{\Delta}{4}\frac{\frac{2}{r_0} + f'(m_0)}{\frac{2m_0}{r_0} + f_0} + R_1(\Delta,\tilde{\theta}) \Delta^2, \\
			\frac{\frac{2}{r_0} + f'(m_0)}{\frac{2}{r(\Delta,\phi(\Delta,\tilde{\phi}))} + f'(m^\pm(\Delta,\eta(\Delta,\tilde{\eta})))} &=: 1 \mp \frac{\Delta f''(m_0)}{2(\frac{2}{r_0} + f'(m_0))} + R_2(\Delta,\tilde{\theta}) \Delta^2,
		\end{align}
		where $X_{d, 1}^\pm, R_1, R_2 \in C^\infty(\B)$, and from \eqref{M1pm} we get:
		\begin{equation}
			M_1^\pm(\Theta_2(\Delta,\tilde{\theta})) =: M_{1,0}^\pm + M_2^\pm(\Delta,\tilde{\theta}) \Delta,
		\end{equation}
		where
		\begin{equation}
			M_{1,0}^\pm := \pm \frac{3}{8}\frac{\frac{2}{r_0} + f'(m_0)}{\frac{2m_0}{r_0} + f_0} \mp \frac{1}{2}\frac{f''(m_0)}{\frac{2}{r_0} + f'(m_0)} \pm \frac{1}{8} \frac{\frac{2}{r_0} + f'(m_0) }{1+f_0},
		\end{equation}
		and $M_2^\pm \in C^\infty(\B)$. We now expand the denominator in \eqref{Mpm}. From \eqref{rc2}, we have
		\begin{equation}
			r_{c, 2}^\pm(\Delta, \eta(\Delta,\tilde{\eta})) =: \pm 2^{-1-2n} r_{2n+1, 0} + r_{c, 3}^\pm(\Delta,\tilde{\eta})\Delta,
		\end{equation}
		where $r_{2n+1, 0} := \frac{r_{2n+1}(m_0)}{r_{2n}}$, and $r_{c, 3}^\pm \in C^\infty([0,\Delta_5] \times [-N_+-1,N_++1])$. Noting also that 
		\begin{align}
			a^\pm(\eta(\Delta,\tilde{\eta}))^{2n} =: 2^{-2n}(1 \mp 4n\tilde{\eta}\Delta + \Delta^2 a_2^\pm(\Delta,\tilde{\eta})),
		\end{align}
		where $a_2^\pm \in C^\infty([0,\Delta_5] \times [-N_+-1,N_++1])$, we can then expand:
		\begin{align}
			d(\Delta,\tilde{\eta},\tilde{\phi}) &:= a^\pm(\eta(\Delta,\tilde{\eta}))^{2n} - \phi(\Delta,\tilde{\phi}) + \Delta r_{c, 2}^\pm(\Delta,\eta(\Delta,\tilde{\eta})) \\ &= 2n\phi_0 + 2^{-2n}\left[ \mp 4n\Bar{\eta}\Delta + \Delta^2 a_2^\pm(\Delta,\Bar{\eta})\right] - \tilde{\phi}\Delta \pm 2^{-1-2n} r_{2n+1,0} \Delta + r_{c, 3}^\pm(\Delta,\tilde{\eta})\Delta^2 \\ &=:2n\phi_0 \left(1 - d_{1}^\pm(\tilde{\eta},\tilde{\phi}) \Delta + d^\pm_2(\Delta,\tilde{\theta})\Delta^2 \right).
		\end{align}
		where
		\begin{equation}
			d_{1}^\pm(\tilde{\eta},\tilde{\phi}) := \frac{\tilde{\phi}}{2n\phi_0} \pm 2(1+2n)\tilde{\eta} \mp \frac{1}{2}r_{2n+1,0}\left( 1 + \frac{1}{2n} \right),
		\end{equation}
		and $d_{2}^\pm \in C^\infty(\B)$. We can now expand $\mathcal{M}^\pm$, starting from \eqref{Mpm}:
		\begin{align}
			\mathcal{M}^\pm(\Psi_2(\Delta,\tilde{\theta})) &= \pm \frac{\Delta^{1-2n}}{4nr_{2n}\phi_0}\left(\frac{2m_0}{r_0} + f_0\right)^\frac{1}{2} \frac{(1 + X_{1, 0}^\pm(\tilde{\beta}) \Delta + X_2^\pm(\Delta,\tilde{\theta}) \Delta^2)(1 + M_{1, 0}^\pm \Delta + M_2^\pm(\Delta,\tilde{\theta}) \Delta^2)}{1 - d_{1}^\pm(\tilde{\eta},\tilde{\phi}) \Delta + d_2^\pm(\Delta,\tilde{\theta}) \Delta^2} \\ &=: \pm \frac{\Delta^{1-2n}}{4nr_{2n}\phi_0} \left(\frac{2m_0}{r_0} + f_0\right)^\frac{1}{2}(1 + \mathcal{M}_{1}^\pm(\tilde{\theta}) \Delta + \mathcal{M}_2^\pm(\Delta,\tilde{\theta}) \Delta^2),
		\end{align}
		where $\mathcal{M}_1^\pm(\tilde{\theta}) = X_{1,0}^\pm(\tilde{\beta}) + d_1^\pm(\tilde{\eta},\tilde{\phi}) + M_{1,0}^\pm$, and $\mathcal{M}_2^\pm \in C^\infty(\B)$. We then obtain the following expression for the difference, $\mathcal{M}^+-\mathcal{M}^-$:
		\begin{equation} \label{MplusMminus}
			\mathcal{M}(\Delta,\tilde{\theta}) := \mathcal{M}^+(\Psi_2(\Delta,\tilde{\theta})) - \mathcal{M}^-(\Psi_2(\Delta,\tilde{\theta})) =: \frac{\Delta^{1-2n}}{2nr_{2n}\phi_0} \left(\frac{2m_0}{r_0} + f_0\right)^\frac{1}{2}(1 + \mathcal{M}_{1}(\tilde{\phi}) \Delta + \mathcal{M}_2(\Delta,\tilde{\theta}) \Delta^2),
		\end{equation}
		where: 
		\begin{equation}
			\mathcal{M}_1(\tilde{\phi}) = \frac{1}{2}( \mathcal{M}_1^+(\tilde{\theta}) + \mathcal{M}_1^-(\tilde{\theta})) = \frac{\tilde{\phi}}{2n\phi_0},
		\end{equation}
		and $\mathcal{M}_2 \in C^\infty(\B)$. We can now expand the right hand side of \eqref{drdDelta}, using \eqref{R2exp} and \eqref{MplusMminus}:
		\begin{equation} \label{RMplusminus}
			\frac{\mathcal{R}(\Psi_2(\Delta,\tilde{\theta}))}{\mathcal{M}^+(\Psi_2(\Delta,\tilde{\theta})) - \mathcal{M}^-(\Psi_2(\Delta,\tilde{\theta}))} =: - r_{2n} \Delta^{2n-1} \left( 2n \phi_0  - \tilde{\phi} + P_3(\Delta,\tilde{\theta}) \Delta^2 \right),
		\end{equation}
		where $P_3 \in C^\infty(\B)$. We finally expand \eqref{N-Npm}:
		\begin{equation} \label{Nexp}
			N(\Theta_2(\Delta,\tilde{\theta})) = \frac{1}{2} \frac{1 + \mathcal{M}_{1}^-(\tilde{\theta}) \Delta + \mathcal{M}_2^-(\Delta,\tilde{\theta}) \Delta^2}{1 + \mathcal{M}_{1}(\tilde{\phi}) \Delta + \mathcal{M}_2(\Delta,\tilde{\theta}) \Delta^2} =: \frac{1}{2} + N_1(\tilde{\beta},\tilde{\eta}) \Delta + N_2(\Delta,\tilde{\theta}) \Delta^2,
		\end{equation}
		where
		\begin{align}
			N_1(\tilde{\beta},\tilde{\eta}) &= \frac{1}{2}(\mathcal{M}_1^-(\tilde{\theta}) - \mathcal{M}_1(\tilde{\phi})) = c_{\beta} \tilde{\beta} - (1+2n) \tilde{\eta} + c_{\eta}, \\ c_{\eta} &:= \frac{1}{2} \frac{\frac{1}{r_0^2(1+f_0)} - f''(m_0)}{\frac{2}{r_0} + f'(m_0)} + \frac{1}{4}\left( 1 + \frac{1}{2n}\right) r_{2n+1,0} + \frac{1}{2}M_{1,0}^-, \\ c_{\beta} &:= \frac{2(1+f_0)}{\frac{2}{r_0} + f'(m_0)} > 0,
		\end{align}
		and $N_2 \in C^\infty(\B)$.
		
		\subsubsection{Contraction mapping for a unique solution} \label{contraction}
		
		We start by showing the equivalence between solutions of the thin shell ODE: \eqref{rode}, \eqref{mode}, \eqref{eode}; satisfying \eqref{cond1}, \eqref{cond2}, \eqref{cond3}, and solutions of the ODE for $\tilde{\theta}$ which we derive using the previous section. We first make some definitions so that we can recover the change of variables $\Delta(s) = m^+(s) - m^-(s) \in C^0([0,s_1]) \cap C^2((0,s_1))$, for $s_1>0$ such that $\Delta(s_1) = \Delta_5$, for any $\tilde{\theta} \in C^2((0,\Delta_5);\tilde{\mathcal{B}})$. Indeed, we define for $(\Delta, \tilde{\theta}) \in \B$: $\widetilde{\mathcal{M}}(\Delta,\tilde{\theta}) := \Delta^{1-2n} [\mathcal{M}(\Delta,\tilde{\theta})]^{-1}$, so $\widetilde{\mathcal{M}} \in C^\infty(\B)$ by \eqref{MplusMminus} and Remark \ref{rem:cond}, and $\widetilde{\mathcal{M}}(\Delta,\tilde{\theta}) > 0$ for all $(\Delta,\tilde{\theta}) \in \B$. We then define, for $\tilde{\theta} \in C^2((0,\Delta_5);\tilde{\mathcal{B}})$:
		\begin{equation}
			I(\Delta;\tilde{\theta}) := \int_0^\Delta [\mathcal{M}^+(\Psi_2(\tilde{\Delta},\tilde{\theta}(\tilde{\Delta}))) - \mathcal{M}^-(\Psi_2(\tilde{\Delta},\tilde{\theta}(\tilde{\Delta})))]^{-1} d\tilde{\Delta} = \Delta^{2n} \int_0^1 t^{2n-1} \widetilde{\mathcal{M}}(t\Delta,\tilde{\theta}(t\Delta)) dt,
		\end{equation}
		and $\widetilde{I}(\Delta,\tilde{\theta}) := [I(\Delta;\tilde{\theta}) ]^{\frac{1}{2n}}$, so that $\widetilde{I}(\Delta;\tilde{\theta}) \in C^0([0,\Delta_5]) \cap C^2((0,\Delta_5))$, and $\frac{d}{d\Delta} \widetilde{I}(\Delta;\tilde{\theta}) > 0$ for all $\Delta \in (0,\Delta_5)$, so by the inverse function theorem we have $\widetilde{I}^{-1}(\cdot;\tilde{\theta}) \in C^0([0,\widetilde{I}(\Delta_5;\tilde{\theta})]) \cap C^2((0,\widetilde{I}(\Delta_5;\tilde{\theta})))$.
		\begin{lemma} \label{lem:finalode}
			Let $\tilde{\theta} = (\tilde{\beta}, \tilde{\eta}, \tilde{\phi}) \in C^2((0,\Delta_5)); \tilde{\mathcal{B}})$. Then, letting $s_1 := I(\Delta_5;\tilde{\theta})$, the functions
			\begin{equation} \label{Psi2}
				(r(s),m^+(s),m^-(s),\epsilon(s)) := \Psi_2(\widetilde{I}^{-1}(s^{\frac{1}{2n}}; \tilde{\theta}), \tilde{\theta}(\widetilde{I}^{-1}(s^{\frac{1}{2n}}; \tilde{\theta}))) \in C^0([0,s_1]) \cap C^2((0,s_1)),
			\end{equation}
			satisfy the conditions \eqref{cond1}, \eqref{cond2}, \eqref{cond3}, with 
			\begin{equation} \label{Delta-s}
				m^+(s) - m^-(s) = \widetilde{I}^{-1}(s^\frac{1}{2n};\tilde{\theta}).
			\end{equation}
			Moreover, $(r(s),m^+(s),m^-(s),\epsilon(s))$ is a solution of the thin shell ODE, \eqref{rode}, \eqref{mode}, \eqref{eode}, if and only if $\tilde{\theta}(\Delta)$ is a solution of the following ODE for $\Delta \in (0,\Delta_5)$:
			\begin{equation} \label{ode:beta}
				\frac{d\tilde{\beta}}{d\Delta} + \frac{3(\tilde{\beta} - \tilde{\beta}_0)}{\Delta} = P_1(\Delta,\tilde{\theta}),
			\end{equation}
			\begin{equation} \label{ode:eta}
				\frac{d\tilde{\eta}}{d\Delta} + \frac{(2n+3)\tilde{\eta} - c_{\beta} \tilde{\beta} - c_{\eta}}{\Delta} = N_2(\Delta,\tilde{\theta}),
			\end{equation}
			\begin{equation} \label{ode:phi}
				\frac{d\tilde{\phi}}{d\Delta} + \frac{(2n+2)\tilde{\phi}}{\Delta} = P_3(\Delta,\tilde{\theta}),
			\end{equation}
			where $P_1, N_2, P_3 \in C^\infty(\B)$ and $\tilde{\beta}_0, c_{\beta}, c_{\eta}$ are as defined in Section \ref{secondorder}.
		\end{lemma}

		\begin{proof}
			The conditions \eqref{cond1}, \eqref{cond2}, \eqref{cond3} hold by Remark \ref{rem:cond}. By definition of $\widetilde{I}(\Delta;\tilde{\theta})$, we have:
			\begin{equation} \label{sDelta}
				\frac{d}{ds} \left[ \widetilde{I}^{-1}(s^{\frac{1}{2n}};\tilde{\theta}) \right] = \mathcal{M}^+(r(s),m^+(s),m^-(s),\epsilon(s)) - \mathcal{M}^-(r(s),m^+(s),m^-(s),\epsilon(s)),
			\end{equation}
			so that if $(r(s),m^+(s),m^-(s),\epsilon(s))$ is a solution of \eqref{rode}, \eqref{mode}, \eqref{eode}, then \eqref{Delta-s} holds by integrating \eqref{sDelta}. The implication that $\tilde{\theta}$ solves \eqref{ode:beta}--\eqref{ode:phi} follows from the change of variables to $\Delta(s) = m^+(s) - m^-(s)$ as in Section \ref{ODEdet}, and then the equivalence of \eqref{drdDelta}--\eqref{depsdDelta} to \eqref{ode:beta}--\eqref{ode:phi}, which follows by direct computation using the expansion in Section \ref{secondorder}. Indeed, \eqref{ode:beta} follows from expanding \eqref{dbeta} (which follows from expanding \eqref{depsdDelta}), and then using \eqref{Bexp}; \eqref{ode:eta} follows by directly expanding \eqref{detadDelta} using \eqref{def:N} and \eqref{Nexp} and \eqref{ode:phi} similarly follows by directly expanding \eqref{drdDelta} using \eqref{RMplusminus}.
			
			For the other direction, we note that the derivation of \eqref{ode:beta}--\eqref{ode:phi} is reversible and \eqref{rode}, \eqref{mode}, \eqref{eode} follow by differentiating \eqref{Psi2}.
		\end{proof}
		
		\begin{lemma} \label{lem:cont}
			Let $\tilde{\theta} \in C^2((0,\Delta_5];\tilde{\mathcal{B}})$ be a solution of the ODE \eqref{ode:beta}, \eqref{ode:eta}, \eqref{ode:phi}, then $\tilde{\theta}$ extends continously to $\Delta=0$: $\tilde{\theta} \in C^0([0,\Delta_5];\tilde{\mathcal{B}})$; with:
			\begin{equation} \label{initval}
				\tilde{\beta}(0) = \tilde{\beta}_0 \in [-B_+,B_+], \ \tilde{\eta}(0) = \frac{c_{\beta}\tilde{\beta}_0 + c_{\eta}}{2n+3} =: \tilde{\eta}_0 \in [-N_+,N_+], \ \tilde{\phi}(0) = 0 \in [-P_+,P_+]. 
			\end{equation}
		\end{lemma}
		
		\begin{proof}
			To see that the functions extend continuously to $\Delta=0$ with the given values, we first show that there is a change of variables so that the system has a regular singular point. Indeed, we define:
			\begin{equation} \label{tildemu}
				\tilde{\mu}(\Delta) := \tilde{\eta}(\Delta) - \frac{c_{\beta}}{2n} \tilde{\beta}(\Delta),
			\end{equation}
			and see from \eqref{ode:beta}, \eqref{ode:eta} that:
			\begin{equation} \label{ode:mu}
				\frac{d\tilde{\mu}}{d\Delta} + \frac{(2n+3)(\tilde{\mu} - \tilde{\mu}_0)}{\Delta} = P_2(\Delta,\tilde{\theta}),
			\end{equation}
			where $P_2 := N_2 - \frac{c_{\beta}}{2n} P_1 \in C^\infty([0,\Delta_5] \times \tilde{\mathcal{B}})$, and
			\begin{equation}
				\tilde{\mu}_0 := \frac{c_{\eta} - \frac{3c_{\beta}}{2n} \tilde{\beta}_0}{2n+3}.
			\end{equation}
			An identical argument to that given in e.g.~the proof of Lemma \ref{lem:beta}, using $P_1, P_2, P_3 \in C^\infty([0,\Delta_5] \times \tilde{\mathcal{B}})$, implies that the functions extend continuously to $\Delta = 0$ with $\tilde{\beta}(0) = \tilde{\beta}_0$, $\tilde{\mu}(0) = \tilde{\mu}_0$ and $\tilde{\phi}(0) = 0$, and therefore $\tilde{\eta}(0) = \tilde{\eta}_0$. 
			
			Now, we must have $\tilde{\theta}_0 := (\tilde{\beta}_0,\tilde{\eta}_0,0) \in [-B_+,B_+] \times [-N_+,N_+] \times [-P_+,P_+]$ by the uniqueness arguments of Section \ref{firstorder}. Indeed, from \eqref{B} and \eqref{Bexp} we must have $|\beta_0| \leq B_+$, and certainly $0 \in [-P_+,P_+]$. To see that $\tilde{\eta}_0 \in [-N_+,N_+]$, observe that $N_1(\tilde{\beta}_0,\tilde{\eta}_0) = 2\tilde{\eta}_0$ and therefore $N(\Theta_2(\Delta,\tilde{\theta}_0)) - \eta(\Delta,\tilde{\eta}_0) = \tilde{\eta}_0 \Delta + N_2(\Delta,\tilde{\theta}_0) \Delta^2$, so the sign of $\frac{d\eta(\Delta,\tilde{\eta}(\Delta))}{d\Delta}$ for sufficiently small $\Delta$ is the same as $\tilde{\eta}_0$, so we must have $\tilde{\eta}_0 \in [-N_+,N_+]$ (and if $\tilde{\eta}_0 = 0$ then the result also holds).
		\end{proof}
		
		We conclude the proof of Proposition \ref{thinshellODEprop} by showing that there is a unique solution of the ODE \eqref{ode:beta}, \eqref{ode:eta}, \eqref{ode:phi} in $C^0([0,\Delta_*]; \tilde{\mathcal{B}}) \cap C^2([0,\Delta_*]; \tilde{\mathcal{B}})$, for some $\Delta_* \in (0,\Delta_5]$, so that the result then follows by Lemmas \ref{lem:unique}, \ref{lem:finalode} and \ref{lem:cont}.
		
		\begin{lemma} \label{lem:contraction}
			There exists $0 < \Delta_* \leq \Delta_5$, depending on $\Delta_0, r_0, m_0$ and the functions $f,r_c$, such that there is a unique $\tilde{\theta}_* \in C^0([0,\Delta_*];\tilde{\mathcal{B}}) \cap C^2((0,\Delta_*);\tilde{\mathcal{B}})$ solving \eqref{ode:beta}, \eqref{ode:eta}, \eqref{ode:phi}, and moreover $\tilde{\theta}_* \in C^\infty([0,\Delta_*];\tilde{\mathcal{B}})$. 
		\end{lemma}
		
		\begin{remark}
			The $s_*$ of Proposition \ref{thinshellODEprop} is then given by $s_* := I(\Delta_*;\tilde{\theta}_*) > 0$.
		\end{remark}

		\begin{proof}
			For $\Delta \in [0,\Delta_5]$, suppose that $\tilde{\theta} \in C^0([0,\Delta_*];\tilde{\mathcal{B}}) \cap C^2((0,\Delta_*);\tilde{\mathcal{B}})$ is a solution of \eqref{ode:beta}, \eqref{ode:eta}, \eqref{ode:phi}. Then $\tilde{\mu} \in C^0([0,\Delta_*];\tilde{\mathcal{B}}) \cap C^2((0,\Delta_*);\tilde{\mathcal{B}})$ (defined by \eqref{tildemu}), and we integrate \eqref{ode:beta}, \eqref{ode:phi}, \eqref{ode:mu} using an integrating factor, so that the system is equivalent to:
			\begin{equation} \label{ode:int}
				\tilde{\beta}(\Delta) = \frac{c_1}{\Delta^3} + T_1(\tilde{\theta})(\Delta), \qquad \tilde{\mu}(\Delta) = \frac{c_2}{\Delta^{2n+3}} + T_2(\tilde{\theta})(\Delta), \qquad \tilde{\phi}(\Delta) = \frac{c_3}{\Delta^{2n+2}} + T_3(\tilde{\theta})(\Delta), 
			\end{equation}
			where $c_i \in \R$ are constants of integration, and the functions $T_i(\tilde{\theta})$ are defined by:
			\begin{align}
				T_1(\tilde{\theta})(\Delta) &:= \tilde{\beta}_0 + \Delta \int_0^1 t^{3} P_1(t\Delta,\tilde{\theta}(t\Delta)) dt, \\ T_2(\tilde{\theta})(\Delta) &:= \tilde{\mu}_0 + \Delta \int_0^1 t^{2n+3} P_2(t\Delta,\tilde{\theta}(t\Delta)) dt, \\ T_3(\tilde{\theta})(\Delta) &:= \Delta \int_0^1 t^{2n+2} P_3(t\Delta,\tilde{\theta}(t\Delta)) dt.
			\end{align}
			We define
			\begin{equation}
				\Gamma_i := \max_{\Delta\in[0,\Delta_5], \tilde{\theta} \in \tilde{\mathcal{B}}}  | P_i(\Delta,\tilde{\theta})|,
			\end{equation}
			so that, now only assuming that $\tilde{\theta} \in C^0([0,\Delta_5];\tilde{\mathcal{B}})$, and letting $d_1 = 3$, $d_2 = 2n+3$, $d_3 = 2n+2$, we have
			\begin{equation} \label{Tbd}
				|T_i(\tilde{\theta})(\Delta) - \tilde{\theta_0}| \leq \frac{\Gamma_i \Delta}{1+d_i},
			\end{equation}
			for each $i \in \{1,2,3\}$. We therefore define: $\Delta_6 := \min\left\{ \Delta_5, \frac{4} {\Gamma_1}, \frac{4n}{c_{\beta} \Gamma_1}, \frac{n+2} {\Gamma_2}, \frac{2n+3} {\Gamma_3}\right\}$, so that from \eqref{initval}, \eqref{tildemu}, and the definition of $\tilde{\mathcal{B}}$, \eqref{tildeB}, we have $\mathbf{T}(\tilde{\theta}) := (T_1(\tilde{\theta}), T_2(\tilde{\theta}) + \frac{c_{\beta}}{2n} T_1(\tilde{\theta}),T_3(\tilde{\theta})) \in C^0([0,\Delta_6];\tilde{\mathcal{B}} )$ for all $\tilde{\theta} \in C^0([0,\Delta_6];\tilde{\mathcal{B}})$. Any $\tilde{\theta} \in C^0([0,\Delta_6];\tilde{\mathcal{B}})$ solving \eqref{ode:int} must therefore have $c_i = 0$ for each $i \in \{1,2,3\}$. 
			
			We now choose $0 < \Delta_* \leq \Delta_6$ such that we can show there is a unique solution $\tilde{\theta} \in C^0([0,\Delta_*]; \tilde{\mathcal{B}})$ solving $\tilde{\theta}(\Delta) = \mathbf{T}(\tilde{\theta})(\Delta)$. We consider the following norm, defined for $\tilde{\theta}_1, \tilde{\theta}_2 \in C^0([0,\Delta_*];\tilde{\mathcal{B}})$, by:
			\begin{equation}
				\| \tilde{\theta}_1 - \tilde{\theta}_2 \| = \sum_i \| (\tilde{\theta}_1)_i - (\tilde{\theta}_2)_i \|_{\infty},
			\end{equation} 
			where $\| \cdot \|_{\infty}$ denotes the supermum norm on $[0,\Delta_*]$, so that $C^0([0,\Delta_*],\tilde{\mathcal{B}})$ is a complete metric space under $\| \cdot \|$. We then define
			\begin{equation}
				\Lambda_{i j} := \max_{\Delta\in[0,\Delta_6], \tilde{\theta} \in \tilde{\mathcal{B}}} \left| \frac{\partial}{\partial\tilde{\theta}_j} P_i(\Delta,\tilde{\theta}) \right|, \qquad \Lambda := \max_{1 \leq i,j \leq 3} \Lambda_{ i j},
			\end{equation}
			so that for all $\tilde{\theta}_1, \tilde{\theta}_2 \in C^0([0,\Delta_*];\tilde{\mathcal{B}})$, $i \in \{1,2,3\}$, we have
			\begin{equation}
				\sup_{\Delta \in [0,\Delta_*]} | P_i(\Delta, \tilde{\theta}_1(\Delta)) - P_i(\Delta, \tilde{\theta}_2(\Delta))| \leq \Lambda \cdot \| \tilde{\theta}_1 - \tilde{\theta}_2 \|,
			\end{equation}
			and therefore
			\begin{equation}
				\| T_i(\tilde{\theta}_1) - T_i(\tilde{\theta}_2)\|_{\infty} \leq \frac{\Lambda \Delta_*}{1 + d_i} \cdot \| \tilde{\theta}_1 - \tilde{\theta}_2 \|,   
			\end{equation}
			so that 
			\begin{equation}
				\| \mathbf{T}(\tilde{\theta}_1) - \mathbf{T}(\tilde{\theta}_2)\| \leq \left( 3 + \frac{c_{\beta}}{2n}\right) \Lambda \Delta_* \cdot \| \tilde{\theta}_1 - \tilde{\theta}_2 \|, 
			\end{equation}
			which implies that for any $\Delta_* < \left[ \left( 3 + \frac{c_{\beta}}{2n}\right) \Lambda\right]^{-1}$, $\mathbf{T} : C^0([0,\Delta_*];\tilde{\mathcal{B}}) \to C^0([0,\Delta_*];\tilde{\mathcal{B}})$ is a contraction map, and therefore it has a unique fixed point $\tilde{\theta}_* \in C^0([0,\Delta_*];\tilde{\mathcal{B}})$. Then differentiating \eqref{ode:int} with $c_i = 0$, and using $T_i(\tilde{\theta}) = (\tilde{\theta}_0)_i + \int_0^{\Delta} \tilde{\Delta}^{d_i} P_i(\tilde{\Delta},\tilde{\theta}(\tilde{\Delta})) d\tilde{\Delta}$, we get $\tilde{\theta}_* \in C^\infty([0,\Delta_*];\tilde{\mathcal{B}})$, since $P_i \in C^\infty([0,\Delta_*] \times \tilde{\mathcal{B}})$.
		\end{proof}
		
		This completes the proof of Theorem \ref{devthm}.
	
	\section{Maximal globally hyperbolic piecewise smooth development (MGHPSD) of generic smooth initial data} \label{maxdev}
	
	In this section, we consider the problem of extending smooth solutions of (E-d) given smooth initial data, as piecewise smooth solutions of (E-d), building up to the precise statement, and proof, of Theorem \ref{MGHD} in the introduction.
	
	In Section \ref{smoothdataEd}, we define the space of initial data that we'll consider: compactly supported `collapsing' dust clouds with vacuum exterior. We then prove a local existence and uniquness result, and define a family of metric space structures ($\metric$ for $k \in \mathbb{N}$) on initial data for which Cauchy stability holds. 
	
	In Section \ref{finiteshells}, we define piecewise smooth solutions of (E-d) with finitely many thin shells (as opposed to restricting to a single thin shell e.g.~as in Section \ref{pwisesection}). We define the \emph{propagation condition}, the assumption of which implies uniqueness of solutions: see the extension principle of Section \ref{extp}. We then give the precise statement of Theorem \ref{MGHD} in Section \ref{sec:thm2}, see Section \ref{MGHDproof} for the proof, in particular Section \ref{outlinethm2} for the outline of the proof. 
	
	\subsection{Smooth initial data for (E-d)} \label{smoothdataEd}
	
	The moduli space of smooth initial data for the Einstein-dust system is most natural in comoving coordinates. We want to pose initial data for (E-d) for a dust cloud with vacuum exterior, consisting of a choice of `dust data' in comoving coordinates, with a `gauge choice' of double null coordinates. We first consider the problem of posing smooth initial data in comoving coordinates including the centre, in Section \ref{initial-data}. We then use this to define initial data for (E-d) on a simultaneous curve, in Section \ref{dataE-d}, containing a `dust cloud' with compactly supported energy density. In Section \ref{localex} we prove a local existence and uniqueness result, where in particular we show that the solution is smooth in a neighbourhood of the intersection of the centre with the initial curve, and that the compactly supported dust cloud has a smooth boundary with the vacuum exterior. Finally, we define a topology on the moduli space of initial data, modulo the gauge choice, and prove a corresponding Cauchy stability result, in Section \ref{top}.
	
	\subsubsection{Smooth solutions of the collapsing Einstein-dust system in comoving coordinates including the centre} \label{initial-data}
	
	This section is essentially an extension of Section \ref{comoving} to include the centre: we consider the problem of posing smooth, compactly supported data for the Einstein-dust system in comoving coordinates, up to the centre.
	
	We choose the initial curve to be `simultaneous', this is convenient in order to understand the moduli space of solutions, as well as the issue of regularity at the centre. Indeed, suppose $(Q,g,\rho,\mathbf{u})$ is a smooth solution of Einstein-dust in spherical symmetry with $\Gamma \neq \emptyset$, and suppose that $\mathcal{D} \subset Q$ is a simultaneous curve with $\mathcal{D} \cap \Gamma \neq \emptyset$, with a choice of comoving coordinate $\mathring{\chi} : \mathcal{D} \to [0,\chi_b]$ such that $r|_{\mathcal{D}}(\mathring{\chi}) = \mathring{\chi}$. We construct $(\tau,\chi)$ comoving coordinates on $(Q \setminus \Gamma) \cap D_{ac}^+(\mathcal{D})$, where we gauge fix $\tau(\mathcal{D}) = 0$ and $\chi|_{\mathcal{D}} = \mathring{\chi}$, and define $\mathring{\mathscr{Q}} := \{ (\tau(p),\chi(p)) : p \in (Q \setminus \Gamma) \cap D_{ac}^+(\mathcal{D}) \} \subset \R_{\tau,\chi}^2 \setminus \{\chi=0\}$. We further assume that $\mathbf{u}(r)(p) < 0, \mathbf{u}_*(r)(p) > 0$ for $p \in \mathcal{D} \setminus \Gamma$, so that the transformation to comoving coordinates defines a $C^\infty$ solution of the collapsing Einstein-dust system away from the centre, $(\mathring{\mathscr{Q}}, r, \rho, m, f)$. 
	
	We now consider whether the solution extends smoothly to the centre in the constructed comoving coordinates. We define $\mathscr{D}_{0, \chi_b} := \{\tau = 0, \chi \in [0, \chi_b]\} \subset Q$, where $r(0,\chi) = \chi$. From \eqref{metric}, we see that the induced metric on $\mathscr{D}_{0, \chi_b}$ as a hypersurface in the `upstairs' manifold is
	\begin{equation}
		h_{\mathscr{D}_{0, \chi_b}} = \frac{d\chi^2}{1 + f(\chi)} + \chi^2 g_{\mathbb{S}^2},
	\end{equation}
	with $\mathbf{u}_*(r)^2\big|_{\mathscr{D}_{0,\chi_b}} = 1+f(\chi) > 0$ for $\chi > 0$, and we must have $f(\chi)$ extending smoothly to $\chi=0$ with $1 + f(0) > 0$. We define comoving coordinates about the centre, satisfying $\chi^2 = x^2 + y^2 + z^2$, so that:
	\begin{equation}
		h_{\mathcal{D}_0} = dx^2 + dy^2 + dz^2 + \frac{1}{\chi^2}\left( 1 - \frac{1}{1+f(\chi)}\right) (xdx + ydy + zdz)^2.
	\end{equation}
	The metric is smooth at the centre if and only if $f(0)=0$ and $f(\chi)$ is a smooth even function (i.e.~$f$ is smooth in $\chi^2$). The initial energy density, $\rho_0(\chi) := \rho(0,\chi)$, is smooth in Cartesian coordinates if and only if it is a smooth even (non-negative) function on $[0,\chi_b]$. The Hawking mass is defined in terms of the initial energy density by:
	\begin{equation} \label{m-com}
		m(\chi) = 4\pi \int_0^{\chi} x^2 \rho_0(x) dx,
	\end{equation}
	where $m(0) = 0$ is necessary for the centre to be regular. We define:
	\begin{equation} \label{tilde}
		m =: \chi^3 \tilde{m}(\chi), \qquad f =: \chi^2 \tilde{f}(\chi), 
	\end{equation}
	where by the considerations above, $\tilde{m}, \tilde{f}$ must be smooth even functions of $\chi$ for smooth initial data at the centre. 
	
	We now define the moduli space of smooth data in comoving coordinates satisfying these requirements, including the collapsing and compactly supported assumptions. Given $\chi_b > 0$, we define the function spaces: 
	\begin{align}
		\mathcal{E}_{\chi_b} &:= \{f \in C^\infty([0,\chi_b]) : f \text{ extends to an even function in } C^\infty([-\chi_b,\chi_b])\}, \\ \mathcal{E}_{\rho, \chi_b} &:= \{f \in \mathcal{E}_{\chi_b} : f(x) > 0 \ \forall x \in [0,\chi_b); \ \forall k \in \mathbb{N}, \ \lim_{\chi \to \chi_b} f^{(k)}(\chi) = 0 \}.
	\end{align}
	\begin{definition} \label{D}
		We define the moduli space of smooth initial data for the collapsing Einstein-dust system posed on $\mathscr{D}_{0, \chi_b}$:		
		\begin{align} \label{data-centre}
			\mathfrak{D}_{\chi_b} := \Big\{&(\tilde{m},\tilde{f}) \in (\mathcal{E}_{\chi_b})^2 : \rho_0 := \frac{1}{4\pi} (3\tilde{m} + \chi \partial_\chi \tilde{m}) \in \mathcal{E}_{\rho, \chi_b}, \\ &\tilde{m}, \ 2\tilde{m}(\chi) + \tilde{f}(\chi), \ 1 + \chi^2 \tilde{f}(\chi) > 0 \ \forall \chi \in [0,\chi_b], \text{ and } \frac{d}{d\chi}(\chi^2 \tilde{f}(\chi))|_{\chi = \chi_b} = 0 \Big\}.
		\end{align}
	\end{definition}
	
	\begin{remark}
		The condition that, for $f(\chi) := \chi^2 \tilde{f}(\chi)$, we have $f'(\chi_b) = 0$, is so that $\tau_s'(\chi_b) > 0$ (which follows from \eqref{tausp}), and therefore shell crossing can only occur in the interior of the support of the dust. 
	\end{remark}
	
	We now define the notion of a $C^\infty$ solution of the Einstein-dust system in comoving coordinates, including the centre: the smooth, even conditions on $\rho$, $f$, are necessary for the solution to be smooth in Cartesian coordinates at the centre.
	\begin{definition} \label{comdustcentre}
		A $C^\infty$ solution of the collapsing Einstein-dust system in comoving coordinates, including the centre, consists of a domain $\mathscr{Q} \subset \R^2_{\tau,\chi \geq 0}$, with $\Gamma := \{ \chi = 0\} \subset \mathscr{Q}$, and a possibly empty exterior boundary $\mathscr{B}_{\chi_b} := \mathscr{Q} \cap \{\chi = \chi_b\}$, $\mathscr{Q} \subset \{ \chi \in [0,\chi_b]\}$, with functions $r, \rho, m, f \in C^\infty_{\tau,\chi}(\mathscr{Q})$ such that: $r(\tau,0) = 0 = m(\tau,0)$, $\rho(\tau,\chi_b) = 0$, and $1+f, \partial_\chi r > 0$ for all $(\tau,\chi) \in \mathscr{Q}$, $\rho > 0$ for all $(\tau,\chi) \in \mathscr{Q} \setminus \mathscr{B}_{\chi_b}$, and $r,m, \frac{2m}{r} + f > 0$ for all $(\tau,\chi) \in \mathscr{Q} \setminus \Gamma$. The functions $r, \rho, m, f$ satisfy \eqref{mtau}, \eqref{rho}, \eqref{Edcollapse} on $\mathscr{Q} \setminus \Gamma$. 
		
		Moreover, on any simultaneous curve up to the centre, i.e.~for fixed $\tau= \tilde{\tau}$ such that $(\tilde{\tau},0) \in \mathscr{Q}$, let $\chi_{\tilde{\tau}}(r)$ be such that $r(\tilde{\tau},\chi_{\tilde{\tau}}(\tilde{r})) = \tilde{r}$, then we require that $\rho_0(\tilde{\tau}, \chi_{\tilde{\tau}}(r)), f(\chi_{\tilde{\tau}}(r))$ are smooth, even functions of $r$, with $\rho_0(\tilde{\tau},0) > 0, f(0)=0$.
	\end{definition}
	
	We can now extend the notion of maximal dust development to initial data sets including the centre: the following proposition extends the result of Proposition \ref{mddprop} to include the centre.
	\begin{proposition} \label{prop:mddcentre}
		Given initial data $(\tilde{m}, \tilde{f}) \in \mathfrak{D}_{\chi_b}$ posed on $\mathscr{D}_{0, \chi_b}$, there exists a unique maximal acoustically globally hyperbolic development:
		\begin{equation}
			(\mathscr{Q}_{MDD}, r, \rho, m, f)(\mathscr{D}_{0, \chi_b}, \tilde{m},\tilde{f}),
		\end{equation}
		attaining the data on $\mathscr{D}_{0, \chi_b}$. The decomposition of the singular boundary as in Proposition \ref{mddprop} holds, with 
		\begin{equation} \label{taus}
			\tau_s(\chi) = \int_0^1 \frac{R^\frac{1}{2}}{(2\tilde{m}(\chi) + \tilde{f}(\chi)R)^\frac{1}{2}} dR \in C^\infty([0,\chi_b]),
		\end{equation}
		an even function, where $\tau_s'(0)=0$ so that $(\tau_s(0),0) \in \mathscr{S}$, and in addition $\tau_s'(\chi_b) > 0$, so that:
		\begin{equation}
			\mathscr{C}_\chi \subset (0,\chi_b).
		\end{equation}
		If $\inf(\mathscr{C}_\chi) = 0$, then: $\lim_{\mathscr{C}_{\chi} \ni \chi \to 0} \tau_c(\chi) = \tau_s(0)$.
	\end{proposition}
	
	\begin{proof}
		We define $R(\tau,\chi) := \chi^{-1} r(\tau,\chi)$, so that $R(0,\chi) = 1$. Given $(\tilde{m},\tilde{f}) \in \mathfrak{D}_{\chi_b}$, with $\rho_0$ defined as in \eqref{data-centre} the energy density is given by
		\begin{equation}
			\rho(\tau,\chi) = \frac{\rho_0(\chi)}{\partial_\chi r(\tau,\chi) R(\tau,\chi)^2},
		\end{equation}
		where $\partial_\chi r = R + \chi \partial_\chi R$, and $R(\tau,\chi)$ is the solution of:
		\begin{equation} \label{tauR}
			\partial_\tau R(\tau,\chi) = - \left( \frac{2\tilde{m}(\chi)}{R(\tau,\chi)} + \tilde{f}(\chi) \right)^{\frac{1}{2}},
		\end{equation}
		with $R(0,\chi) = 1$. Integrating \eqref{tauR}, we have $\tau = T(R(\tau,\chi),\chi)$, where
		\begin{equation} \label{T}
			T(R,\chi) :=  \int_R^1 \left( \frac{2\tilde{m}(\chi)}{\tilde{R}} + \tilde{f}(\chi) \right)^{-\frac{1}{2}} d\tilde{R},
		\end{equation}
		which is well-defined since $2\tilde{m} + \tilde{f} > 0$, $\tilde{m} > 0$. We then have $\tau_s(\chi) = T(0,\chi)$ which is a smooth, even function by the choice of $\tilde{m}, \tilde{f}$; and $R(\tau,\chi) \in (0,1]$ for $\tau \in [0,\tau_s(\chi))$. By the implicit function theorem, since
		\begin{equation}
			\partial_R T(R,\chi) = - \left( \frac{2\tilde{m}(\chi)}{R} + \tilde{f}(\chi) \right)^{-\frac{1}{2}} < 0,
		\end{equation}
		we have that $R(\tau,\chi)$ is smooth for $0 \leq \tau < \tau_s(\chi), 0 \leq \chi \leq \chi_b$. 
		
		We now show that the solution is smooth at the centre in Cartesian coordinates for $R>0$. Differentiating \eqref{tauR} with respect to $\chi$, we see that $\partial_\chi R(\tau,0)$ satisfies the ODE:
		\begin{equation}
			\partial_\tau (\partial_\chi R(\tau,0)) = \left( \frac{2\tilde{m}(0)}{R(\tau,0)} + \tilde{f}(0) \right)^{-\frac{1}{2}} \frac{\tilde{m}(0)}{R(\tau,0)^2} \partial_\chi R(\tau,0),
		\end{equation}
		with $\partial_\chi R(0,0) = 0$, so that $\partial_\chi R(\tau,0) = 0$ for all $\tau \in [0,\tau_s(\chi))$. Similarly, all odd $\partial_\chi$ derivatives of $R$ vanish at the centre: $\partial_\chi^{2k-1} R(\tau,0) = 0$ for all $\tau \in [0,\tau_s(\chi))$, $k \in \mathbb{N}$, so $R(\tau,\chi)$ is a smooth even function in $\chi$ for each $\tau$. The metric in $(\tau,r)$ coordinates is:
		\begin{equation} 
			g = \frac{1}{1+f} \left( \left( -1 + \frac{2m}{r} \right)d\tau^2 + 2 \left( \frac{2m}{r} + f \right)^\frac{1}{2} d\tau dr + dr^2 \right) + r^2 g_{\mathbb{S}^2},
		\end{equation}
		so if we define Cartesian coordinates with respect to $r$: $r^2 = x^2 + y^2 + z^2$, $dr^2 + r^2 g_{\mathbb{S}^2} = dx^2 + dy^2 + dz^2$; then the metric components in these Cartesian coordinates are smooth at the centre, since $\frac{2m(\chi)}{r(\tau,\chi)} = \chi^2 \cdot \frac{2\tilde{m}(\chi)}{R(\tau,\chi)}$ and $\chi^2 \tilde{f}(\chi)$ are smooth even functions vanishing at $\chi = 0$, and for any fixed $\tau$: $\chi_\tau(r) = r \cdot R^{-1}(\tau,\chi_\tau(r))$, so any even function of $\chi$ is an even function of $r$.
		
		From \eqref{tausp} with $m'(\chi_b) = 0 = f'(\chi_b)$ we get
		\begin{equation}
			\tau_s'(\chi_b) = \left( \frac{2m(\chi_b)}{\chi_b} + f(\chi_b) \right)^{-\frac{1}{2}} > 0,
		\end{equation}
		so by definition of $\mathscr{C}_\chi$ in Proposition \ref{mddprop} we get $\mathscr{C}_\chi \subset (0,\chi_b)$. We have $\rho(\tau,0) = \frac{\rho_0(0)}{R(\tau,0)^3}$, so $\{ \tau \in [0,\tau_s(0)), \chi = 0\} \subset \mathscr{Q}_{MDD}$. If there exists $\delta>0$ such that $\tau_s'(\chi) < 0$ for $\chi \in (0,\delta)$, then $\lim_{\chi \to 0} \tau_c(\chi) = \tau_s(0)$, since we have $\tau_c(\chi) < \tau_s(\chi)$ for all $\chi \in \mathscr{C}_\chi$. 
	\end{proof}	
	
	\subsubsection{Smooth initial data for (E-d) on a simultaneous curve} \label{dataE-d}
	
	We consider initial data for (E-d) on a simultaneous curve: data in comoving coordinates, with a gauge choice for double null coordinates. For solutions of (E-d) including the centre, we impose the `boundary condition' for the gauge:
	\begin{equation}
		\Gamma = \{u = v\}.
	\end{equation}
	We now define data on a spacelike curve for (E-d), up to and including the centre (corresponding to asymptotically flat initial data on a spacelike hypersurface homeomorphic to $\R^3$ in the `upstairs' manifold), with compactly supported initial energy density:
	\begin{definition} \label{E-ddata}
		Smooth \emph{collapsing} initial data for (E-d) consists of an initial curve
		\begin{equation}
			\mathcal{D}_0 := \{ (-v,v) : v \in [0,\infty)\},
		\end{equation}
		with $(\tilde{m},\tilde{f}) \in \mathfrak{D}_{\chi_b}$, such that for $v \in [0,v_b]$: 
		\begin{equation}
			m_0(v) := v^3 \tilde{m}(v), \qquad f_0(v) := v^2 \tilde{f}(v),
		\end{equation}
		\begin{equation} \label{E-ddata1}
			r(-v,v) = v, \ m(-v,v) = m_0(v), \qquad \Omega^2(-v,v) = (1+f_0(v))^{-1},
		\end{equation}
		and the curve $\mathcal{D}_0$ is simultaneous: $\mathbf{u}^u(-v,v) = \mathbf{u}^v(-v,v) = \Omega^{-1}(-v,v)$, with $\mathbf{u}_*(r)(-v,v) = (1+f_0(v))^\frac{1}{2}$ and
		\begin{equation} \label{vel-collapse}
			\mathbf{u}(r)(-v,v) = - \left( \frac{2m_0(v)}{v} + f_0(v) \right)^\frac{1}{2}.
		\end{equation}
		For $v \geq \chi_b$, the energy density vanishes:
		\begin{equation} \label{Schw}
			r(-v,v) = v, \qquad m(-v,v) = M := m_0(\chi_b), \qquad \Omega^2(-v,v) = \Omega_0^2(v),
		\end{equation}
		where $\Omega_0^2(v)$ is a gauge choice such that $\Omega^2(-v,v) \in C^\infty_v([0,\infty))$, $\Omega_0^2(v)(1-\frac{2M}{v}) < 1$ for all $v > \chi_b$, and $\mathcal{D}_0$ is asymptotically flat (i.e.~$\Omega_0^2(v) = 1 + \frac{2M}{v} + \mathcal{O}\left(\frac{1}{v^2}\right)$ and the asymptotic behaviour holds for all derivatives).
	\end{definition}
	
	\begin{remark}
		The condition that $\Omega_0^2(v)(1-\frac{2M}{v}) < 1$ holds for all $v > \chi_b$ is so that $\partial_u r(-v,v)$, $\partial_v r(-v,v)$ are well-defined. Note that we have $\partial_v r(-v,v) = 1 + \partial_u r(-v,v)$ for all $v$, so using the definition of the Hawking mass \eqref{m}, along with the assumption \eqref{vel-collapse}, we see that:
		\begin{equation}
			\partial_u r(-v,v) = -\frac{1}{2} - \frac{1}{2} \left[ 1 - \Omega^2(-v,v) \left( 1 - \frac{2m_0(v)}{v} \right)\right]^\frac{1}{2},
		\end{equation} 
		(which conicides with e.g.~\eqref{pr} for $v \leq \chi_b$, using \eqref{E-ddata1}).
	\end{remark}
	
	\begin{definition}
		Given $(\tilde{m}, \tilde{f}) \in \mathfrak{D}_{\chi_b}$, let $\mathfrak{G}_{\chi_b}(\tilde{m},\tilde{f})$ be the set of all $\Omega_0^2 \in C^\infty((\chi_b,\infty))$, satisfying the conditions of Definition \ref{E-ddata}.
	\end{definition}
	
	\subsubsection{Local existence and uniqueness of solutions to (E-d) from smooth initial data} \label{localex}
	
	We now prove a local existence and uniqueness result for solutions of (E-d) with initial data as in Definition \ref{E-ddata}, we first define smooth solutions of (E-d) including the centre:
	\begin{definition} \label{def:E-d-centre}
		A $C^\infty$ solution of (E-d) including the centre consists of a domain $\mathcal{Q} \subset \R^{1+1}_{u,v}$, with $\Gamma = \{ (u,v) \in \mathcal{Q} : u = v \} \neq \emptyset$, and non-negative functions $r, m, \rho \in C^\infty(\mathcal{Q})$, along with positive functions $\Omega^2 \in C^\infty(\mathcal{Q})$, $\mathbf{u}^u, \mathbf{u}^v \in C^\infty(\text{supp}(\rho))$, with $r(v,v) = 0 = m(v,v)$, with $r(u,v) > 0$ for all $(u,v) \in \mathcal{Q} \setminus \Gamma$. The functions satisfy \eqref{rwave}-\eqref{m}.
		
		Moreover, there exists a choice of comoving coordinates in a neighbourhood of $\Gamma \subset \mathcal{Q}$ such that the solution is $C^\infty$ in comoving coordinates at the centre as in Definition \ref{comdustcentre}. 
	\end{definition}
	
	\begin{proposition} \label{vacuumext}
		Given $(\tilde{m},\tilde{f}) \in \mathfrak{D}_{\chi_b}$, and $\Omega_0^2(v) \in \mathfrak{G}_{\chi_b}(\tilde{m},\tilde{f})$, then there exists $\delta(\tilde{m}, \tilde{f}) > 0$ such that $(\mathcal{Q}_\delta, r, \Omega^2, \rho, \mathbf{u}^\mu)$ is the unique $C^\infty$ solution of (E-d) including the centre with underlying manifold
		\begin{equation}
			\mathcal{Q}_\delta = \{ (u,v) : v \geq u, \ 0 \leq u+v < \delta \},
		\end{equation} 
		extending the data $(\mathcal{D}_0,\tilde{m},\tilde{f},\Omega_0^2)$. 
		
		For any globally hyperbolic $\mathcal{Q} \subset \R^{1 + 1}_{u,v}$ with $\mathcal{Q}_\delta \subset \mathcal{Q}$ such that there exists a $C^\infty$ solution of (E-d), $(\mathcal{Q}, r, \Omega^2, \rho, \mathbf{u}^\mu)$, extending the data on $\mathcal{D}_0$, then the solution is unique, and: the `dust cloud' has a smooth boundary with the vacuum exterior: there exists $u_{b, +} > -\chi_b$ and $v_b(u) \in C^\infty([-\chi_b, u_{b, +}))$ such that
		\begin{equation}
			\mathcal{B}_d := \{ (u,v) \in \text{supp}(\rho) : \rho(u,v) = 0\} = \{ (u, v_b(u)): u \in [-\chi_b, u_{b, +}) \},
		\end{equation}
		where $v_b(-\chi_b) = \chi_b, \ v_b'(-\chi_b) = 1$. The vacuum exterior region is given by:
		\begin{equation}
			\mathcal{V} := \{(u,v) : \rho(u,v) = 0\} = \{ u \leq - \chi_b\} \cup \{ u \in (-\chi_b, u_{b, +}), \ v \geq v_b(u)\},
		\end{equation}
		and is isometric to a subset of Schwarschild, with Hawking mass $m(u,v) = M$ determined from initial data by \eqref{Schw}.
	\end{proposition}
	
	\begin{remark}
		For the Einstein-dust system with no symmetry assumptions, see \cite{C-B} for the existence of a unique maximal globally hyperbolic development (MGHD). 
	\end{remark}
	
	\begin{remark}
		The MGHDs of initial data as in Definition \ref{E-ddata} may have infinitely many first shell crossing singularities.
	\end{remark}

	\begin{proof}
		For the existence of a $C^\infty$ solution of (E-d) extending the data on $\mathcal{D}_0$, we construct a solution of the eikonal equation on the MDD of the given data in comoving coordinates, $(\mathscr{Q}_{MDD}, r, \rho, m, f)(\mathscr{D}_{0,\chi_b}, \tilde{m}, \tilde{f})$. Indeed, consider the following data posed for the eikonal equation on $\mathscr{D}_{0, \chi_b} \subset \mathscr{Q}_{MDD}$: $u(0, \chi) = - \chi = - v(0,\chi)$. We have $\partial_\chi u(0, \chi) = -1 = - \partial_\chi v(0,\chi)$, so that from \eqref{eiku}, \eqref{eikv}:
		\begin{equation}
			\partial_\tau u (0,\chi) = (1 + f(\chi))^\frac{1}{2} = \partial_\tau v(0,\chi),
		\end{equation}
		and therefore, after transforming to double null coordinates, the data coincides with that on $\mathcal{D}_0 \cap \{ v \leq \chi_b \}$. There exists a unique smooth solution of the eikonal equation on $\mathscr{U}_+(\mathscr{D}_{0, \chi_b}, \mathscr{Q}_{MDD}) \cap \mathscr{V}(\mathscr{D}_{0, \chi_b}, \mathscr{Q}_{MDD})$, by Proposition \ref{eikprop}. We now show that, with the boundary condition $u(\tau,0) = v(\tau,0)$, the solution of the eikonal equation is smooth in a neighbourhood of $\Gamma \cap \mathscr{D}_{0,\chi_b}$. 
		
		Differentiating the boundary condition, we see that $\partial_\tau^k (u-v)(\tau,0) = 0$. The eikonal equation on $\mathscr{Q}_{MDD}$ is:
		\begin{equation}
			\partial_\tau u = - A(\tau,\chi) \partial_\chi u, \qquad \partial_\tau v = A(\tau,\chi) \partial_\chi v, \label{eik}
		\end{equation}
		where $A(\tau,\chi) = \frac{(1+f(\chi))^\frac{1}{2}}{\partial_\chi r(\tau,\chi)}$ is a smooth, even function of $\chi$ for each $\tau \in [0,\tau_s(\chi))$, and for all $\tau$ derivatives. We can locally solve for $v(\tau,0)$ from initial data, using the method of characteristics, then $u(\tau,0) := v(\tau,0)$ and we can solve for $v(\tau,\chi)$ in a neighbourhood of $\Gamma \cap \mathscr{D}_{0,\chi_b}$. The solution is smooth if and only if $\partial_\tau^k u(0,0) = \partial_\tau^k v(0,0)$ for all $k \in \mathbb{N}$. Differentiating \eqref{eik} with respect to $\tau$, $(k-1)$ times, the difference satisfies:
		\begin{equation}
			\partial_\tau^k (v-u)(0,\chi) = \sum_{j=0}^{k-1} \binom{k-1}{j} \partial_\tau^{k-1-j} A(0,\chi) \cdot \partial_\chi \partial_\tau^j ( u + v)(0,\chi),
		\end{equation}
		and:
		\begin{equation}
			\partial_\chi \partial_\tau^k (u+v)(0,\chi) = \sum_{j=0}^{k-1} \binom{k-1}{j} \left( \partial_\chi \partial_\tau^{k-1-j} A(0,\chi) \cdot \partial_\chi \partial_\tau^j ( v - u)(0,\chi) + \partial_\tau^{k-1-j} A(0,\chi) \cdot \partial_\chi^2 \partial_\tau^j ( v - u)(0,\chi) \right).
		\end{equation}
		We keep commuting derivatives until we reach an expression for $\partial_\tau^k(v-u)(0,\chi)$ in terms of $\chi$ derivatives of $\partial_\tau^k A(0,\chi)$, and $(u+v)(0,\chi) = 0$, $(v-u)(0,\chi) = 2\chi$. Note that, when we set $\chi = 0$, any term with an odd number of $\chi$ derivatives on $\partial_\tau^k A(\tau,\chi)$ must vanish. Using $\partial_\chi \partial_\tau^{k-1-j} A(0,0) = 0$, and $\partial_\chi (u+v)(0,\chi)=0$, we get:
		\begin{equation}
			\partial_\tau^k (v-u)(0,0) = \sum_{j=1}^{k-1} \binom{k-1}{j} \partial_\tau^{k-1-j} A(0,0) \left[ \sum_{i=0}^{j-1} \binom{j-1}{i} \partial_\tau^{j-1-i}A(0,0) \partial_\chi^2 \partial_\tau^i (v-u)(0,0) \right].
		\end{equation}
		Consider $\partial_\chi^{2l} \partial_\tau^k (v-u)(0,0)$ for $l \in \mathbb{N}$: we must have an even number of derivatives of $\partial_\chi \partial_\tau^j(u+v)(0,0)$, because any odd number of derivatives on $\partial_\tau^k A(0,\chi)$ must vanish at $\chi=0$. By the same argument, we see that we can only get terms of the form $\partial_\chi^{2l} \partial_\tau^k (v-u)(0,0)$ in the expression for $\partial_\chi^{2l+1} \partial_\tau^k (u+v)(0,0)$. The only possible non-zero term is $\partial_\chi (v-u)(0,\chi) = 2$, which cannot appear, so we have $\partial_\tau^k (v-u)(0,0) = 0$ for all $k \in \mathbb{N}$. 
		
		The `boundary' in comoving coordinates is $\{ \chi = \chi_b\}$, where from local existence there exists $\delta_b > 0$ such that $\{ \tau \in [0,\delta_b), \ \chi = \chi_b\} \subset \mathscr{Q}_{MDD} $, and the function $r(\tau,\chi_b)$ is smooth for $\tau \in [0,\delta_b)$. We can locally solve the eikonal equation in $\mathscr{Q}_{MDD}$ for $u(\tau,\chi_b)$, $\tau \in [0,\delta_b)$, assuming $\delta_b$ sufficiently small. Given the gauge choice $\Omega_0^2(v)$ for $v \geq \chi_b$, we solve for the boundary as follows. Suppose that we have a smooth choice of double null coordinates in a neighbourhood of the boundary, then \eqref{pr} implies that:
		\begin{equation} \label{Bode1}
			\partial_\tau v(\tau,\chi_b) = \frac{1}{2 \partial_v r(u(\tau,\chi_b),v(\tau,\chi_b))} \left[ (1+f(\chi_b))^\frac{1}{2} - \left( \frac{2M}{r(\tau,\chi_b)} + f(\chi_b) \right)^\frac{1}{2} \right],
		\end{equation}
		where we've used $m(\chi_b) = M$. In the vacuum exterior, Raychaudhuri's equation \eqref{Rayu} with $\rho(u,v) = 0$ implies that 
		\begin{equation}
			(\Omega^{-2} \partial_u r)(u,v) = (\Omega^{-2} \partial_u r)(-v,v),
		\end{equation} 
		and the equation for the Hawking mass, with $m(u,v) = M$, leads to:
		\begin{equation} \label{vr}
			\partial_v r(u,v) = \frac{\frac{2M}{r(u,v)} - 1}{4(\Omega^{-2} \partial_u r)(-v,v)}.
		\end{equation}
		Substituting \eqref{vr} into \eqref{Bode1}, we arrive at an ODE for $v_b(\tau) := v(\tau,\chi_b)$:
		\begin{equation} \label{tauvb}
			\partial_\tau v_b(\tau) = \frac{-2(\Omega^{-2} \partial_u r)(-v_b(\tau),v_b(\tau))}{(1+f(\chi_b))^\frac{1}{2} + (\frac{2M}{r(\tau,\chi_b)} + f(\chi_b))^\frac{1}{2}},
		\end{equation}
		which has a unique local solution, with $v_b(0) = \chi_b$, $v_b'(0) = \mathbf{u}^v(-\chi_b,\chi_b) = (1+f(\chi_b))^{\frac{1}{2}}$. Since the initial data is smooth, we get $v(\tau,\chi)$ is smooth in a neighbourhood of $(0,\chi_b) \in \mathscr{Q}_{MDD}$, with the boundary condition $v(\tau,\chi_b) = v_b(\tau)$.
		
		Combining the local existence results at the centre and exterior boundary satisfying the appropriate boundary conditions, there exists a local smooth solution of the eikonal equation extending the data on $\mathscr{D}_{0,\chi_b}$. Transforming to double null coordinates, there exists $\delta > 0$ such that there is a smooth solution on $\mathcal{Q}_\delta$, which attains the data on $\mathcal{D}_0$. The development can be extended to the transformation to double null coordinates of the extension of the smooth solution of the eikonal equation, satisfying the boundary conditions. The boundary-defining function $v_b(u)$ is determined by defining $\tau_b(u)$ to be the inverse of $u(\tau,\chi)$, and then defining $v_b(u) := v_{b, com}(\tau_b(u))$, where $v_b(-\chi_b) = \chi_b$ and $v_b'(-\chi_b) = 1$ since $\partial_\tau u(0,\chi_b) = \partial_\tau v(0,\chi_b)$.
		
		We now show uniqueness of $C^\infty$ solutions. Given any $C^\infty$ solution of (E-d) extending the data, let $\mathcal{P} := \{ (u,v) \in \mathcal{Q} : \rho(u,v) > 0\}$. We now show that for any $(u,v) \in \mathcal{P}$, the unique past-directed fluid flow line from $\mathcal{P}$ must intersect $\mathcal{D}_0 \cap \{ v < \chi_b \}$. Indeed, let $u_f(t), v_f(t)$ be the solutions of: 
		\begin{equation}
			u_f'(t) = - \mathbf{u}^u(u_f(t),v_f(t)), \qquad v_f'(t) = - \mathbf{u}^v(u_f(t),v_f(t)),
		\end{equation}
		with $u_f(0) = u$, $v_f(0) = v$. By Picard-Lindelo\"f there exists $t_0 > 0$ such that there is a unique solution $u_f, v_f \in C^\infty([0,t_0))$, with $\rho(u_f(t),v_f(t)) > 0$, $u_f(t) + v_f(t) > 0$ for $t \in [0,t_0)$. Now, mass conservation implies that
		\begin{equation}
			\mathbf{u}(\log \rho) = \mathbf{u}^\mu \partial_\mu (\log \rho) = - \nabla_\mu \mathbf{u}^\mu,
		\end{equation}
		so we must have $\rho(u_f(t_0), v_f(t_0)) > 0$, and so if $u_f(t_0) + v_f(t_0) > 0$ then the solution extends. Now in order to see that the flow line must reach $\mathcal{D}_0 \cap \{ v < \chi_b\}$, i.e.~that the solution extends to $t = t_1$ with $u_f(t_1) + v_f(t_1) = 0$, we note that $(u_f(t), v_f(t)) \in J^-(u,v) \subset \mathcal{Q}$ for all $t$, and $\Omega^2 \sim 1$ on $J^-(u,v)$, so there exists $c > 0$ such that:
		\begin{equation}
			\frac{d}{dt} (u_f(t) + v_f(t)) = - (\mathbf{u}^v + \Omega^{-2} (\mathbf{u}^v)^{-1}) \leq - (\mathbf{u}^v + c (\mathbf{u}^v)^{-1}) \leq - 2 c^\frac{1}{2}. 
		\end{equation}
		The subset $\mathcal{P}$ is therefore acoustically globally hyperbolic. Defining comoving coordinates on $D^+_{ac}(\mathcal{D}_0 \cap \{ v \leq \chi_b \})$ satisfying the gauge choice $\tau(-v,v) = 0$, $\chi(-v,v) = v$, then 
		\begin{equation}
			\mathscr{Q} := \{ (\tau(u,v),\chi(u,v)) : (u,v) \in D^+_{ac}(\mathcal{D}_0 \cap \{ v \leq \chi_b \})\} \subset \mathscr{Q}_{MDD},
		\end{equation}
		and the solution in comoving coordinates on $\mathscr{Q}$ coincides with the restriction of the unique solution on the maximal dust development to $\mathscr{Q}$. Now, since $\mathbf{u}^u, \mathbf{u}^v \in C^\infty(\text{supp}(\rho))$, then $\{ \chi = \chi_b\}$ is a fluid flow line with $\rho(\tau, \chi_b) = 0$, and by uniqueness of $C^\infty$ solutions of the eikonal equation, with initial and boundary conditions as in the existence part, the solution is unique. 
	\end{proof}
	
	\subsubsection{Metric space structure on smooth data for (E-d) and Cauchy stability} \label{top}
	
	We consider the moduli space of initial data for (E-d) as in Definition \ref{E-ddata}, modulo the choice of gauge in the exterior vacuum region. By construction, this is the moduli space of the Einstein-dust system in comoving coordinates with initial data on $\mathscr{D}_{0, \chi_b}$:
	\begin{definition} \label{metricspace}
		For $k \in \mathbb{N}$, we define the function $d_k : \mathfrak{D}_{\chi_b} \times \mathfrak{D}_{\chi_b} \to [0, \infty)$: for $(\tilde{m},\tilde{f}),(\underline{\tilde{m}},\underline{\tilde{f}}) \in \mathfrak{D}_{\chi_b}$,
		\begin{equation}
			d_k((\tilde{m},\tilde{f}),(\tilde{m},\tilde{f})) := \| \tilde{m} - \underline{\tilde{m}}\|_{C^{k}([0,\chi_b])} + \| \tilde{f} - \underline{\tilde{f}}\|_{C^{k}([0,\chi_b])},
		\end{equation}
		so that $(\mathfrak{D}_{\chi_b},d_k)$ is a metric space. We define also $d_\infty : \mathfrak{D}_{\chi_b} \times \mathfrak{D}_{\chi_b} \to [0, \infty)$ by:
		\begin{equation}
			d_\infty((\tilde{m},\tilde{f}),(\underline{\tilde{m}},\underline{\tilde{f}})) := \sum_{k=0}^\infty 2^{-k} \frac{d_k((\tilde{m},\tilde{f}),(\underline{\tilde{m}},\underline{\tilde{f}}))}{1 + d_k((\tilde{m},\tilde{f}),(\underline{\tilde{m}},\underline{\tilde{f}}))},
		\end{equation}
		so that $(\mathfrak{D}_{\chi_b},d_\infty)$ is a metric space which induces the $C^\infty$ topology on $\mathfrak{D}_{\chi_b}$.
	\end{definition}
	
	\begin{definition}
		$B_{\chi_b, k}((\tilde{m},\tilde{f}), \delta) := \{ (\underline{\tilde{m}},\underline{\tilde{f}}) \in \mathfrak{D}_{\chi_b} : d_k((\tilde{m},\tilde{f}), (\underline{\tilde{m}},\underline{\tilde{f}})) < \delta \}$.
	\end{definition}
	
	We now prove a Cauchy stability statement for (E-d) with respect to the metric space $(\mathfrak{D}_{\chi_b},d_k)$.
	\begin{proposition} \label{prop:CS}
		Given $(\tilde{m},\tilde{f}) \in \mathfrak{D}_{\chi_b}$, with a choice of `exterior gauge': $\Omega_0^2(v) \in \mathfrak{G}_{\chi_b}(\tilde{m},\tilde{f})$; let $(\mathcal{Q}, r, \Omega^2, \rho, \mathbf{u}^\mu)$ be a globally hyperbolic $C^\infty$ solution of (E-d) attaining the data $(\mathcal{D}_0, \tilde{m}, \tilde{f}, \Omega_0^2)$. Then for any compact globally hyperbolic set $K \subset \mathcal{Q} \subset \R^{1+1}_{u,v}$, $K \cap \mathcal{D}_0 \neq \emptyset$, and any $k \in \mathbb{N}$, $\varepsilon > 0$, there exists $\delta(\tilde{m}, \tilde{f}, \Omega_0^2, K, \varepsilon, k) > 0$ such that, for all $(\underline{\tilde{m}},\underline{\tilde{f}}) \in B_{\chi_b, k}((\tilde{m},\tilde{f}),\delta)$: there exists $\underline{\Omega}_0^2 \in \mathfrak{G}_{\chi_b}(\underline{\tilde{m}}, \underline{\tilde{f}})$ such that $\| \Omega_0^2 - \underline{\Omega_0^2} \|_{C^{k}( K \cap \mathcal{D}_0)} < \varepsilon$, and a globally hyperbolic $C^\infty$ solution of (E-d) attaining the data $(\mathcal{D}_0, \underline{\tilde{m}}, \underline{\tilde{f}}, \underline{\Omega_0^2})$, $(\underline{\mathcal{Q}}, \underline{r}, \underline{\Omega}^2, \underline{\rho}, \underline{\mathbf{u}}^\mu)$, satisfying:
		\begin{equation} \label{CS:dust}
			K \subset \underline{\mathcal{Q}}, \qquad K_d := \text{supp}(\rho) \cap K = \text{supp}(\underline{\rho}) \cap K, 
		\end{equation}
		\begin{equation} 
			K_v := \{ (u,v) \in K : \rho(u,v) = 0\} =  \{ (u,v) \in K : \underline{\rho}(u,v) = 0\}, \label{CS:vacuum}
		\end{equation}
		and:
		\begin{equation}
			\| r- \underline{r}\|_{C^{k+1}(K)}, \qquad \| \Omega^2 - \underline{\Omega}^2 \|_{C^{k}(K)} < \varepsilon, \label{CS:romega} 
		\end{equation}
		\begin{equation}
			 \| r^{-3} m - \underline{r}^{-3} \underline{m} \|_{C^k(K_d)}, \qquad \| \underline{r}^{-2} f - \underline{r}^{-2} \underline{f} \|_{C^k(K_d)}, \qquad \| \rho - \underline{\rho} \|_{C^{k-1}(K_d)}, \qquad \| \mathbf{u}^\mu - \underline{\mathbf{u}}^\mu \|_{C^{k}( K_d)} < \varepsilon. \label{CS:rhou}
		\end{equation}
	\end{proposition}
	
	\begin{proof}
		We'll first show that Cauchy stability holds for $k=1$, where we'll use Cauchy stability in comoving coordinates and Cauchy stability of null curves on smooth solutions in comoving coordinates. 
		
		Let $\tau(u,v), \chi(u,v)$ be comoving coordinates on $K_d \subset \text{supp}\{\rho\}$, gauge fixed by $\tau(-v,v) = 0$, $\chi(-v,v) = v$, so that, transforming to comoving coordinates on $\mathscr{K} := \{ (\tau(u,v), \chi(u,v)) : (u,v) \in K_d \}$, we have a $C^\infty$ solution of the Einstein-dust system in comoving coordinates that is exactly the restriction of $(\mathscr{Q}_{MDD},r,\rho,m,f)(\mathscr{D}_{0, \chi_b}, \tilde{m}, \tilde{f})$ to $\mathscr{K}$. Now, by construction $\mathscr{K}$ is acoustically globally hyperbolic, so there exists $\tau_{\mathscr{K}} \in C^0([0,\chi_b])$, ($\tau_{\mathscr{K}}(\chi) \geq0$) such that $\mathscr{K} = \{ (\tau,\chi) : 0 \leq \chi \leq \chi_b, \ 0 \leq \tau \leq \tau_{\mathscr{K}}(\chi) \}$. We define:
		\begin{definition}
			$F_{null}(f,\partial_\chi r; \tau,\chi) := (1+f(\chi))^{-\frac{1}{2}} \partial_\chi r(\tau,\chi)$.
		\end{definition}
		Since $\mathscr{K}$ is a compact subset of $\mathscr{Q}_{MDD}$, there exists $\delta_{\mathscr{K}} > 0$ sufficiently small (depending on $\tilde{m}, \tilde{f}, \Omega_0^2, K$) such that $\widetilde{\mathscr{K}} := \{ (\tau,\chi) : 0 \leq \tau \leq \tau_{\mathscr{K}}(\chi) + \delta_{\mathscr{K}} \} \subset \mathscr{Q}_{MDD}$, and $F_- := \min_{(\tau,\chi) \in \widetilde{\mathscr{K}}} |F_{null}(f,\partial_\chi r; \tau,\chi)| \geq \frac{1}{2}\min_{(\tau,\chi) \in \mathscr{K}} |F_{null}(f,\partial_\chi r; \tau,\chi)| > 0$. There then exists $\tilde{\delta}_{\mathscr{K}} > 0$ such that for all $\chi \in [0,\chi_b]$:
		\begin{equation} \label{tildedeltaK}
			\mathscr{R}_{\tilde{\delta}_{\mathscr{K}}}(\tau_\mathscr{K}(\chi),\chi) := [ \tau_{\mathscr{K}}(\chi), \tau_{\mathscr{K}}(\chi) + \tilde{\delta}_{\mathscr{K}}] \times \left(  \left[ \chi - (F_-)^{-1}\tilde{\delta}_{\mathscr{K}},  \chi + (F_-)^{-1}\tilde{\delta}_{\mathscr{K}}\right] \cap [0,\chi_b] \right) \subset \widetilde{\mathscr{K}},
		\end{equation}
		and therefore $\mathscr{R}_{\tilde{\delta}_{\mathscr{K}}}(\tau,\chi) \subset \widetilde{\mathscr{K}}$ for all $(\tau,\chi) \in \mathscr{K}$. We now need the following lemmas:
		\begin{lemma}[Cauchy stability in comoving coordinates] \label{lem:CS-comoving}
			Given $k \in \mathbb{N}$, $(\tilde{m},\tilde{f}) \in \mathfrak{D}_{\chi_b}$ with maximal dust development $(\mathscr{Q}_{MDD},r,\rho,m,f)(\mathscr{D}_{0, \chi_b}, \tilde{m},\tilde{f})$,
			and a compact acoustically globally hyperbolic subset $\mathscr{Q} \subset \mathscr{Q}_{MDD}$, then there exists $\varepsilon_{com}(\tilde{m},\tilde{f},\mathscr{Q}) > 0$ such that for all $0 < \varepsilon \leq \varepsilon_{com}$, there exists $\delta_{com}(\tilde{m},\tilde{f},\mathscr{Q},\varepsilon,k) > 0$ such that: for all $(\underline{\tilde{m}},\underline{\tilde{f}}) \in B_{\chi_b, k}((\tilde{m},\tilde{f}),\delta_{com})$ with $(\underline{\mathscr{Q}_{MDD}}, \underline{r}, \underline{\rho}, \underline{m}, \underline{f})(\mathscr{D}_{0, \chi_b}, \underline{\tilde{m}}, \underline{\tilde{f}})$, $\mathscr{Q} \subset \underline{\mathscr{Q}_{MDD}}$, and 
			\begin{equation} \label{com-bd}
				\| \tilde{m} - \underline{\tilde{m}} \|_{C^k(\mathscr{Q})}, \qquad \| \tilde{f} - \underline{\tilde{f}}\|_{C^k(\mathscr{Q})}, \qquad \| R - \underline{R} \|_{C^k(\mathscr{Q})}, \qquad \| \partial_\chi r - \partial_\chi \underline{r} \|_{C^{k-1}(\mathscr{Q})} < \varepsilon.
			\end{equation}
		\end{lemma}
		\begin{proof}
			From \eqref{tauR}, $\underline{R}(\tau,\chi)$ is the solution of the ODE
			\begin{equation}
				\partial_\tau \underline{R}(\tau,\chi)  = - \left( \frac{2\underline{\tilde{m}}(\chi)}{\underline{R}(\tau,\chi)} + \tilde{f}(\chi) \right)^\frac{1}{2},
			\end{equation}
			which we differentiate in $\chi$ to see that $\partial_\chi \underline{R}(\tau,\chi)$ satisfies the ODE:
			\begin{equation} \label{pchiptauR}
				\partial_\tau(\partial_\chi \underline{R}(\tau,\chi)) = -\frac{1}{2} \left( \frac{2\underline{\tilde{m}}(\chi)}{\underline{R}(\tau,\chi)} + \tilde{f}(\chi) \right)^{-\frac{1}{2}} \left( \frac{2\underline{\tilde{m}'}(\chi)}{\underline{R}(\tau,\chi)} + \underline{\tilde{f}}'(\chi) - \frac{2\underline{\tilde{m}}(\chi)}{\underline{R}(\tau,\chi)^2} \partial_\chi \underline{R}(\tau,\chi) \right).
			\end{equation}
			The subset $\mathscr{Q} \subset \mathscr{Q}_{MDD}$ is compact, so $\min_{\mathscr{Q}} R, \partial_\chi r =: 2\varepsilon_{com}(\tilde{m},\tilde{f},\mathscr{Q}) > 0$. Now, since $\partial_\chi r = R + \chi \partial_\chi R$, by Cauchy stability of ODEs there exists, for all $0 < \varepsilon \leq \varepsilon_{com}$, $\delta_{com}(\tilde{m},\tilde{f},\mathscr{Q},\varepsilon,1) > 0$ such that for all $(\underline{\tilde{m}},\underline{\tilde{f}}) \in B_{\chi_b, k}((\tilde{m},\tilde{f}),\delta_{com})$, then $\mathscr{Q} \subset \underline{\mathscr{Q}_{MDD}}$ (i.e.~$\min_{\mathscr{Q}}\underline{R}, \partial_\chi \underline{r} > 0$) and \eqref{com-bd} holds with $k=1$. Differentiating \eqref{pchiptauR} $(k-1)$ times in $\chi$, we see that $\partial^k_\chi R$ satisfies an ODE in $\tau$ for each $k$, depending on $\tilde{m}^{(l)}(\chi), \tilde{k}^{(l)}(\chi)$ for $0 \leq l \leq k$, and therefore Cauchy stability for $k>1$ also follows from Cauchy stability of ODE.
		\end{proof}
		
		In order to state the lemma for Cauchy stability of null curves, we first make the following definition:
		\begin{definition}
			For $\varepsilon \geq 0$, $(\tilde{m},\tilde{f}) \in \mathfrak{D}_{\chi_b}$ with maximal dust development $(\mathscr{Q}_{MDD},r,\rho,m,f)(\mathscr{D}_{0, \chi_b}, \tilde{m},\tilde{f})$, and compact acoustically globally hyperbolic subset $ \mathscr{Q} \subset \mathscr{Q}_{MDD}$, and a point $(\tau_0,\chi_0) \in \mathscr{Q}$, we define:
			\begin{align}
				\chi_{+, \varepsilon}((\tau_0,\chi_0),\mathscr{Q},f,\partial_\chi r) := \max\{ &\chi \in [\chi_0, \chi_b] : \exists \tau_+ \in C^1([\chi_0,\chi]) \text{ s.t. } \tau_+(\chi_0) = \tau_0, \text{ and } \\ &\forall \tilde{\chi} \in [\chi_0,\chi], (\tau_+(\tilde{\chi}) + \varepsilon,\tilde{\chi}) \in \mathscr{Q} \text{ and } \tau_+'(\tilde{\chi}) = F_{null}(f,\partial_\chi r;\tau_+(\tilde{\chi}),\tilde{\chi})  \}, \\ \chi_{-, \varepsilon}((\tau_0,\chi_0),\mathscr{Q},f,\partial_\chi r) := \min\{ &\chi \in [0, \chi_0] : \exists \tau_- \in C^1([\chi,\chi_0]) \text{ s.t. } \tau_-(\chi_0) = \tau_0, \text{ and } \\ &\forall \tilde{\chi} \in [\chi,\chi_0], (\tau_-(\tilde{\chi}) + \varepsilon,\tilde{\chi}) \in \mathscr{Q} \text{ and } \tau_-'(\tilde{\chi}) = -F_{null}(f,\partial_\chi r;\tau_-(\tilde{\chi}),\tilde{\chi})  \},
			\end{align}
			along with the corresponding null curves: let $\tau_{+, \varepsilon}((\tau_0,\chi_0), \mathscr{Q}, f, \partial_\chi r) \in C^1([\chi_0, \chi_{+, \varepsilon}((\tau_0,\chi_0),\mathscr{Q},f,\partial_\chi r)])$, be the solution of $\tau_+'(\tilde{\chi}) = F_{null}(f,\partial_\chi r;\tau_+(\tilde{\chi}),\tilde{\chi})$ with $\tau_+(\chi_0) = \tau_0$, and let $\tau_{-, \varepsilon}((\tau_0,\chi_0), \mathscr{Q}, f, \partial_\chi r) \in C^1([\chi_{-, \varepsilon}((\tau_0,\chi_0),\mathscr{Q},f,\partial_\chi r), \chi_0])$, be the solution of $\tau_-'(\tilde{\chi}) = -F_{null}(f,\partial_\chi r;\tau_-(\tilde{\chi}),\tilde{\chi})$ with $\tau_-(\chi_0) = \tau_0$.
		\end{definition}
	
		\begin{lemma}[Cauchy stability for null curves] \label{lem:CS-nullcurves}
			Given $(\tilde{m},\tilde{f}) \in \mathfrak{D}_{\chi_b}$ with $(\mathscr{Q}_{MDD},r,\rho,m,f)(\mathscr{D}_{0, \chi_b}, \tilde{m},\tilde{f})$, a compact acoustically globally hyperbolic subset $\mathscr{Q} \subset \mathscr{Q}_{MDD}$, and $0 < \varepsilon \leq \varepsilon_{com}(\tilde{m},\tilde{f},\mathscr{Q})$, then there exists $ \delta_{\tau}(\tilde{m},\tilde{f},\mathscr{Q},\varepsilon) > 0$ and $0 < \delta_{null}(\tilde{m},\tilde{f},\mathscr{Q},\varepsilon) \leq \delta_{com}(\tilde{m},\tilde{f},\mathscr{Q},\varepsilon,1)$, such that for all $(\tau_0,\chi_0) \in \mathscr{Q}$, $\tilde{\tau}_0 \in (\max\{\tau_0 - \delta_{\tau}, 0\}, \tau_0 + \delta_{\tau})$, and all $(\underline{\tilde{m}},\underline{\tilde{f}}) \in B_{\chi_b, 1}((\tilde{m},\tilde{f}), \delta_{null})$ with $(\underline{\mathscr{Q}_{MDD}}, \underline{r}, \underline{\rho}, \underline{m}, \underline{f})(\mathscr{D}_{0, \chi_b}, \underline{\tilde{m}}, \underline{\tilde{f}})$:
			\begin{align}
				&\chi_{+, 0}((\tilde{\tau}_0,\chi_0), \mathscr{Q}, \underline{f}, \partial_\chi\underline{r}) \geq \chi_{+, \varepsilon}((\tau_0,\chi_0), \mathscr{Q}, f, \partial_\chi r), \\ &\chi_{-, 0}((\tilde{\tau}_0,\chi_0), \mathscr{Q}, \underline{f}, \partial_\chi\underline{r}) \leq \chi_{-, \varepsilon}((\tau_0,\chi_0), \mathscr{Q}, f, \partial_\chi r),
			\end{align}
			and 
			\begin{align} \label{out-null-bd}
				&\|\tau_{+, 0}((\tilde{\tau}_0,\chi_0), \mathscr{Q}, \underline{f}, \partial_\chi\underline{r}) - \tau_{+, 0}((\tau_0,\chi_0), \mathscr{Q}, f, \partial_\chi r) \|_{C^0([\chi_0, \chi_{+, \varepsilon}((\tau_0,\chi_0), \mathscr{Q}, f, \partial_\chi r)])} < \varepsilon, \\ &\|\tau_{-, 0}((\tilde{\tau}_0,\chi_0), \mathscr{Q}, \underline{f}, \partial_\chi\underline{r}) - \tau_{-, 0}((\tau_0,\chi_0), \mathscr{Q}, f, \partial_\chi r) \|_{C^0([\chi_{-, \varepsilon}((\tau_0,\chi_0), \mathscr{Q}, f, \partial_\chi r), \chi_0])} < \varepsilon, \label{in-null-bd}
			\end{align}
		\end{lemma}
		
		\begin{proof}
			Standard Cauchy stability of ODEs (see Section \ref{cauchystability}), using Cauchy stability in comoving coordinates.
		\end{proof}
		
		We define $u_{\Gamma, +} := \max\{ u : (u,u) \in K_d\}$, $u_{b, +} := \max\{ u : (u,v) \in K_d \cap K_v \}$, $v_{b, +} := \max\{ v : (u,v) \in K_d \cap K_v \}$, and let $v_b(u) \in C^\infty([-\chi_b, u_{b, +}])$ be the boundary-defining function of $K_d$, i.e.~$\mathcal{B} |_{K} := K_d \cap K_v = \{ (u,v_b(u)) : u \in [-\chi_b, u_{b, +}]\}$. Note that by definition (since $K_d$ is bounded to the future by an achronal curve) we have $v_{b, +} \geq u_{\Gamma, +} \geq v_b^{-1}(v_{b,+})$. We define also $\tau_\Gamma(u) := \tau(u,u)$ for $u \in [0, v_{\Gamma, +}]$, and $\tau_b(v) := \tau(v_b^{-1}(v),v)$ for $v \in [\chi_b, v_{b, +}]$.
		
		In order to use Cauchy stability of null curves, we need to take into account the number of `bounces' of null curves between $\Gamma$ and $\mathcal{B}$ in $K_d$. Indeed, we define the sequences $u_{\Gamma, i}, v_{b, j} \in \mathbb{R}$ as follows. Let $u_{\Gamma, 0} := 0$, $v_{b, 0} := \chi_b$ so that $\{(u_{\Gamma, 0}, u_{\Gamma, 0})\} = \Gamma \cap \mathcal{D}_0$, and $\{ (v_b^{-1}(v_{b, 0}), v_{b, 0})\} = \mathcal{B} \cap \mathcal{D}_0$. Then for $n_\Gamma \in \mathbb{N} - 1$ such that for $0 \leq i \leq n_{\Gamma}$:
		\begin{equation}
			\{ u = u_{\Gamma, i} \} \cap K_d = \{ u = u_{\Gamma, i}, v \in [u_{\Gamma, i}, v_b(u_{\Gamma, i})] \},
		\end{equation}
		we define $v_{b, i + 1} := v_b(u_{\Gamma, i})$; for $n_{b} \in \mathbb{N} - 1$ such that for $0 \leq j \leq n_b$:
		\begin{equation}
			\{ v = v_{b, j} \} \cap K_d = \{ v = v_{b, j}, u \in [v_b^{-1}(v_{b, j}), v_{b, j}]\},
		\end{equation}
		we define $u_{\Gamma, j + 1} := v_{b, j}$. We define $N_{\Gamma}(\tilde{m},\tilde{f},K_d), N_{b}(\tilde{m},\tilde{f},K_d) \in \mathbb{N} - 1$ to be the maximal such $n_{\Gamma}, n_b$ respectively (which must satisfy $|N_{\Gamma} - N_b| \leq 1$), which must be finite by consideration of null curves on $\widetilde{\mathscr{K}}$. Indeed, we must have $\tau(u_{\Gamma, i}, v_b(u_{\Gamma, i})) \leq \tau_{\mathscr{K}}(\chi_b)$, $\tau(v_{b, j}, v_{b, j}) \leq \tau_{\mathscr{K}}(0)$, while $\tau(u_{\Gamma, i}, v_b(u_{\Gamma, i})) - \tau(u_{\Gamma, i}, u_{\Gamma, i}) \geq F_- \cdot \chi_b$, $\tau(v_{b, j}, v_{b, j}) - \tau(v_b^{-1}(v_{b, j}), v_{b, j}) \geq F_- \cdot \chi_b$. Maximality of $N_{\Gamma}$, $N_b$ implies that one of the following conditions are satisfied: if $N_{\Gamma} = N_{b} - 1$ then $u_{\Gamma, +} < v_{b, N_\Gamma}$ (and $v_b^{-1}(v_b^+) \geq u_{\Gamma, N_{\Gamma}}$); similarly if $N_b = N_\Gamma - 1$ then $v_b^{-1}(v_b^+) < u_{\Gamma, N_b}$ (and $u_{\Gamma, +} \geq v_{b, N_b}$); if $N_{\Gamma} = N_b$ then $v_b^{-1}(v_{b, +}) < u_{\Gamma, N_{\Gamma}}$, $u_{\Gamma, +} < v_{b, N_b}$. 
	
		Now, we let $N_{null}(\tilde{m},\tilde{f},K_d) := \max\{N_{\Gamma}(\tilde{m},\tilde{f},K_d), N_{b}(\tilde{m},\tilde{f},K_d)\} + 1$. We define $\varepsilon_{null} := \frac{1}{3} \tilde{\delta}_{\mathscr{K}}$, and for $\varepsilon_{null, 0} \in (0,\varepsilon_{null}]$, we define $\varepsilon_{null, 1}(\varepsilon_{null, 0}) := \delta_\tau(\tilde{m},\tilde{f},\widetilde{\mathscr{K}},\varepsilon_{null, 0})$, and:
		\begin{equation} \label{eps-null-i}
			\varepsilon_{null, i}(\varepsilon_{null, 0}) := \delta_\tau(\tilde{m},\tilde{f},\widetilde{\mathscr{K}},\varepsilon_{null, i-1}(\varepsilon_{null, 0})),
		\end{equation}
		for $2 \leq i \leq N_{null}$. Now, for any $(\underline{\tilde{m}},\underline{\tilde{f}}) \in B_{\chi_b, 1}((\tilde{m},\tilde{f}), \delta_{null}(\tilde{m},\tilde{f},\widetilde{\mathscr{K}},\varepsilon_{null, N_{null}}))$, let $\underline{u}(\tau,\chi), \underline{v}(\tau,\chi)$ be solutions of the eikonal equation,  \eqref{eiku} and \eqref{eikv} respectively on $(\underline{\mathscr{Q}_{MDD}}, \underline{r},\underline{\rho},\underline{m},\underline{f})(\mathscr{D}_{0, \chi_b}, \underline{\tilde{m}}, \underline{\tilde{f}})$, restricted to  $\widetilde{\mathscr{K}} \subset \underline{\mathscr{Q}_{MDD}}$, with initial/boundary conditions:
		\begin{equation} \label{under-IBCs}
			\underline{u}(0,\chi) = - \chi = -\underline{v}(0,\chi), \qquad \underline{u}(\tau,0) = \underline{v}(\tau,0), \qquad \underline{v}(\tau,\chi_b) = v_b(\underline{u}(\tau,\chi_b)),
		\end{equation}
		where we moreover restrict to $\underline{v}(\tau,0) \leq v_{\Gamma, +}$, $\underline{u}(\tau,\chi_b) \leq u_{b, +}$. For any $\underline{\tau_\Gamma} \in [0, \tau_{\mathscr{K}}(0) + \delta_{\mathscr{K}}]$, $\underline{\tau_b} \in [0, \tau_{\mathscr{K}}(\chi_b) + \delta_{\mathscr{K}}]$, define $\Gamma(\underline{\tau_\Gamma}) := \{ \chi =0, \tau \in [0, \underline{\tau_\Gamma}]\} \subset \widetilde{\mathscr{K}}$, $\mathscr{B}_{\chi_b}(\underline{\tau_b}) := \{ \chi = \chi_b, \tau \in [0, \underline{\tau_b}]\} \subset \widetilde{\mathscr{K}}$. Then $\underline{u}, \underline{v}$ are uniquely defined on 
		\begin{equation}
			\underline{\mathscr{K}}(\underline{\tau_{\Gamma, +}},\underline{\tau_{b, +}}) := \mathscr{U}_+(\mathscr{D}_{0, \chi_b} \cup \Gamma(\underline{\tau_{\Gamma, +}}), \widetilde{\mathscr{K}}) \cap \mathscr{V}_+(\mathscr{D}_{0, \chi_b} \cup \mathscr{B}_{\chi_b}(\underline{\tau_{b, +}}), \widetilde{\mathscr{K}}),
		\end{equation}
		for $\underline{\tau_{\Gamma, +}}, \underline{\tau_{b, +}} > 0$ such that
		\begin{equation}
			\Gamma(\underline{\tau_{\Gamma, +}}) \subset \mathscr{V}_+(\mathscr{D}_{0, \chi_b} \cup \mathscr{B}_{\chi_b}(\underline{\tau_{b, +}}), \widetilde{\mathscr{K}}), \ \mathscr{B}_{\chi_b}(\underline{\tau_{b, +}}) \subset \mathscr{U}_+(\mathscr{D}_{0, \chi_b} \cup \Gamma(\underline{\tau_{\Gamma, +}}), \widetilde{\mathscr{K}}),
		\end{equation}
		and $u(\underline{\tau_{\Gamma, +}}, 0) \leq u_{\Gamma, +}$, $v(\underline{\tau_{b, +}}, \chi_b) \leq v_{b, +}$, $\underline{\tau_{\Gamma, +}} \leq \tau_{\mathscr{K}}(0) + \delta_{\mathscr{K}}$, $\underline{\tau_{b, +}} \leq \tau_{\mathscr{K}}(\chi_b) + \delta_{\mathscr{K}}$. Now let $\underline{\tau_{\Gamma, +}}, \underline{\tau_{b, +}}$ be the maximal such choice, and let $\underline{\widetilde{\mathscr{K}}} := \underline{\mathscr{K}}(\underline{\tau_{\Gamma, +}}, \underline{\tau_{b, +}})$. We invert $(\underline{u}(\tau,\chi),\underline{v}(\tau,\chi))$ on $\underline{\widetilde{\mathscr{K}}}$, defining a $C^\infty$ collapsing solution of (E-d), $(\underline{K_d},\underline{r}, \underline{\Omega^2}, \underline{\rho}, \underline{\mathbf{u}^\mu})$, where the underlying manifold is 
		\begin{equation}
			\underline{K_d} := \{ (\underline{u}(\tau,\chi), \underline{v}(\tau,\chi)) : (\tau,\chi) \in \underline{\widetilde{\mathscr{K}}}\}.
		\end{equation}
		Let $\underline{\tau}, \underline{\chi} \in C^\infty_{u, v}(\underline{K_d})$ be the inverse of $\underline{u}, \underline{v} \in C^\infty_{\tau, \chi}(\underline{\widetilde{\mathscr{K}}})$. We define $\underline{\tau_{\Gamma}}(u) := \underline{\tau}(u,u)$, and $\underline{\tau_b}(v) := \underline{\tau}(v_b^{-1}(v),v)$. 
		
		Next, we show using the definition \eqref{eps-null-i} that $\underline{u_{\Gamma, +}} := \max\{ u: (u,u) \in \underline{K_d} \} = u_{\Gamma, +}$, and $\underline{v_{b, +}} := \max\{ v : (v_b^{-1}(v),v) \in \underline{K_d} \} = v_{b, +}$, and moreover that
		\begin{equation} \label{delta-bd}
			\| \tau_b - \underline{\tau_b} \|_{C^0([\chi_b,v_{b, +}])}, \| \tau_\Gamma - \underline{\tau_\Gamma} \|_{C^0([0,u_{\Gamma, +}])} < \delta_\tau(\tilde{m},\tilde{f},\widetilde{\mathscr{K}}, \varepsilon_{null, 0}),
		\end{equation}
		so that we can appeal to Cauchy stability for null curves to any $(u,v) \in K_d$. Assume $v_{b, +}, u_{\Gamma, +} > 0$. We show \eqref{delta-bd} by proving the following result by induction: that for all $1 \leq j \leq N_b$, $1 \leq i \leq N_\Gamma$, we have $\underline{v_{b, +}} \geq v_{b, j}$, $\underline{u_{\Gamma, +}} \geq u_{\Gamma, i}$, with
		\begin{equation} \label{inductive-step}
			\| \tau_b - \underline{\tau_b} \|_{C^0([v_{b, j-1}, v_{b, j}])} < \delta_\tau(\tilde{m},\tilde{f},\widetilde{\mathscr{K}}, \varepsilon_{null, N_{null}-j}), \qquad  \| \tau_\Gamma - \underline{\tau_\Gamma} \|_{C^0([u_{\Gamma, i-1}, u_{\Gamma, i}])} < \delta_\tau(\tilde{m},\tilde{f},\widetilde{\mathscr{K}}, \varepsilon_{null, N_{null}-i}),
		\end{equation}
		and then that $\underline{v_{b, +}} = v_{b, +}$, $\underline{u_{\Gamma, +}} = u_{\Gamma, +}$, with
		\begin{equation} \label{final-step}
			\| \tau_b - \underline{\tau_b} \|_{C^0([v_{b, N_b}, v_{b, +}])} < \delta_\tau(\tilde{m},\tilde{f},\widetilde{\mathscr{K}}, \varepsilon_{null, N_{null}-N_b -1}), \qquad \| \tau_\Gamma - \underline{\tau_\Gamma} \|_{C^0([u_{\Gamma, N_\Gamma}, u_{b, +}])} < \delta_\tau(\tilde{m},\tilde{f},\widetilde{\mathscr{K}}, \varepsilon_{null, N_{null}-N_{\Gamma} - 1}). 
		\end{equation}
		We start with the base case of $i = 1 = j$. If $N_b = 0$, then let $\chi_b < v < v_{b, +}$, otherwise let $\chi_b < v \leq v_{b, 1}$. By the choice of $\tilde{\delta}_{\mathscr{K}}$, \eqref{tildedeltaK}, we have $\chi_{+, \varepsilon_{null, 0}}((0,v),\widetilde{\mathscr{K}},f,\partial_\chi r) = \chi_b$, so by Lemma \ref{lem:CS-nullcurves} we have $\chi_{+, 0}((0,v),\widetilde{\mathscr{K}},\underline{f},\partial_\chi \underline{r}) = \chi_b$, and therefore $\underline{v_{b, +}} \geq v$, while $\tau_b(v) = \tau_{+, 0}((0,v), \widetilde{\mathscr{K}}, f, \partial_\chi r)(\chi_b)$, $\underline{\tau_b}(v) := \tau_{+, 0}((0,v), \widetilde{\mathscr{K}}, \underline{f}, \partial_\chi \underline{r})(\chi_b)$, and from Lemma \ref{lem:CS-nullcurves}, in particular \eqref{out-null-bd}, we have $|\tau_b(v) - \underline{\tau_b}(v)| < \varepsilon_{null, N_{null}} (\leq \varepsilon_{null, 0})$. Similarly, if $N_\Gamma = 0$ then let $0 < u < u_{\Gamma, +}$, otherwise let $0 < u < u_{\Gamma, 1}$. Then the analogous argument for ingoing null curves implies that $\underline{u_{\Gamma, +}} \geq u$, and $|\tau_{\Gamma}(u) - \underline{\tau_{\Gamma}}(u)| < \varepsilon_{null, N_{null}} = (\leq \varepsilon_{null, 0})$. 
		
		We now show the inductive step (if one of $N_b, N_\Gamma$ greater than zero). Assume that \eqref{inductive-step} holds for $i,j \leq l < \min\{ N_\Gamma, N_b\}$. Then for any $v \in [v_{b, l}, v_{b, l+1}]$, we have $v_b^{-1}(v) \in [u_{\Gamma, l -1}, u_{\Gamma, l}]$ and therefore by \eqref{inductive-step} for $i,j = l$ we have $|\tau_{\Gamma}(v_b^{-1}(v)) - \underline{\tau_\Gamma}(v_b^{-1}(v))| < \delta_\tau(\tilde{m},\tilde{f},\widetilde{\mathscr{K}}, \varepsilon_{null, N_{null}-l})$, and $\chi_{+, \varepsilon_{null, N_{null}-l}}( (\tau_\Gamma(v_b^{-1}(v)), 0), \widetilde{\mathscr{K}}, f, \partial_\chi r) = \chi_b$ by \eqref{tildedeltaK}, so by  Lemma \ref{lem:CS-nullcurves} we have $\chi_{+, 0}( (\underline{\tau_\Gamma}(v_b^{-1}(v)), 0), \widetilde{\mathscr{K}}, \underline{f}, \partial_\chi \underline{r}) = \chi_b$, and therefore $\underline{v_{b, +}} \geq v$, and moreover $|\tau_b(v) - \underline{\tau_b}(v)| < \varepsilon_{null, N_{null}-l} = \delta_\tau(\tilde{m},\tilde{f},\widetilde{\mathscr{K}}, \varepsilon_{null, N_{null}-l-1})$. The argument is analogous for ingoing null curves to $\Gamma$, and clearly extends to the case where $N_b \neq N_\Gamma$, as well as \eqref{final-step}, and so \eqref{delta-bd} holds. 
		
		We may now show that $K_d \subset \underline{K_d}$, and:
		\begin{equation} \label{tau-chi-bd}
			\| \tau - \underline{\tau}\|_{C^0(K_d)} < 3 \varepsilon_{null, 0}, \qquad \| \chi - \underline{\chi} \|_{C^0(K_d)} < 3 (F_-)^{-1} \varepsilon_{null, 0}.
		\end{equation}
		The restriction of \eqref{tau-chi-bd} to $K_d \cap (\Gamma \cup \mathcal{B})$ holds by \eqref{delta-bd} and $\underline{\chi}(v,v) = 0$, $\underline{\chi}(v_b^{-1}(v),v) = \chi_b$. For $(u,v) \in K_d \setminus (\Gamma \cup \mathcal{B})$, we define $v_-(u) := \min\{ v : (u,v) \in K_d \} \in [0, u_{\Gamma, +}]$, and $u_-(v) := \min\{ u : (u,v) \in K_d\} \in [-\chi_b, v_b^{-1}(v_{b, +})]$. In the following we abbreviate $\tau_+ := \tau_{+, \varepsilon_{null, 0}}((\tau(u,v_-(u)),\chi(u,v_-(u))), \widetilde{\mathscr{K}}, f, \partial_\chi r)$ and $\tau_- := \tau_{-, \varepsilon_{null, 0}}((\tau(u_-(v),v),\chi(u_-(v),v)), \widetilde{\mathscr{K}}, f, \partial_\chi r)$. From \eqref{tildedeltaK}, we have $\mathscr{R}_{\tilde{\delta}_{\mathscr{K}}}(\tau(u,v),\chi(u,v)) \subset \widetilde{\mathscr{K}}$, and we may define $\tilde{\chi}_\pm(u,v)$ to satisfy: 
		\begin{equation} \label{Fminus-bd}
			\chi(u,v) - 3 (F_-)^{-1} \varepsilon_{null, 0} < \tilde{\chi}_-(u,v) < \chi(u,v) < \tilde{\chi}_+(u,v) < \chi(u,v) + 3 (F_-)^{-1} \varepsilon_{null, 0} ,
		\end{equation}
		\begin{equation} \label{tilde-chi-def}
			 \chi_{-, \varepsilon_{null, 0}}((\tau(u_-(v),v),\chi(u_-(v),v)), \widetilde{\mathscr{K}}, f, \partial_\chi r) < \tilde{\chi}_\pm(u,v) < \chi_{+, \varepsilon_{null, 0}}((\tau(u,v_-(u)),\chi(u,v_-(u))), \widetilde{\mathscr{K}}, f, \partial_\chi r)),
		\end{equation}
		and either: $\tau_\pm(\tilde{\chi}_\pm(u,v)) - \tau_\mp(\tilde{\chi}_\pm(u,v)) = 2\varepsilon_{null, 0}$, or $0 < \tau_\pm(\tilde{\chi}_\pm(u,v)) - \tau_\mp(\tilde{\chi}_\pm(u,v)) < 2\varepsilon_{null, 0}$ with either $\tilde{\chi}_+(u,v) = \chi_b$, $\tilde{\chi}_-(u,v) = 0$ and/or $\tau_\mp(\tilde{\chi}_\pm(u,v)) = 0$. (This is well-defined because as long as the null curves don't reach the boundary, $|\tau_\pm'(\chi)| \geq F_-$ implies that we must have $\tau_\pm(\chi) = \tau(u,v) + 3\varepsilon_{null, 0}$ for some $\chi \in [\chi(u,v) - 3(F_-)^{-1} \varepsilon_{null, 0}, \chi(u,v) + 3(F_-)^{-1} \varepsilon_{null, 0}]$, while we can restrict to $\chi$ such that $\tau_\pm(\chi) \leq \tau(u,v) + 2\varepsilon_{null, 0}$, and we must therefore attain $\tau_\pm(\tilde{\chi}_\pm(u,v)) - \tau_\mp(\tilde{\chi}_\pm(u,v)) = 2\varepsilon_{null, 0}$ while satisfying \eqref{tilde-chi-def}.) Now, from \eqref{tilde-chi-def}, and since $|\underline{\tau}(u_-(v),v) - \tau(u_-(v),v)|, |\underline{\tau}(u,v_-(u)) - \tau(u,v_-(u))|  < \delta_\tau(\tilde{m},\tilde{f},\widetilde{\mathscr{K}}, \varepsilon_{null, 0})$ by \eqref{delta-bd} (where the difference vanishes if $(u_-(v),v), (u,v_-(u)) \in \mathcal{D}_0$) and $\chi(u,v_-(u)) = \underline{\chi}(u,v_-(u)), \chi(u_-(v),v) = \underline{\chi}(u_-(v),v)$, Lemma \ref{lem:CS-nullcurves} implies that:
		\begin{align}
			\chi_{+, 0}((\underline{\tau}(u,v_-(u)),\underline{\chi}(u,v_-(u))), \widetilde{\mathscr{K}}, \underline{f}, \partial_\chi \underline{r})) \geq \tilde{\chi}_+(u,v), \\ \chi_{-, 0}((\underline{\tau}(u_-(v),v),\underline{\chi}(u_-(v),v)), \widetilde{\mathscr{K}}, \underline{f}, \partial_\chi \underline{r}) \leq \tilde{\chi}_-(u,v),
		\end{align}
		and moreover, abbreviating
		\begin{equation}
			\underline{\tau_+} := \tau_{+, 0}((\underline{\tau}(u,v_-(u)),\underline{\chi}(u,v_-(u))), \widetilde{\mathscr{K}}, \underline{f}, \partial_\chi \underline{r}), \qquad \underline{\tau_-} := \tau_{-, 0}((\underline{\tau}(u_-(v),v),\underline{\chi}(u_-(v),v)), \widetilde{\mathscr{K}}, \underline{f}, \partial_\chi \underline{r}),
		\end{equation}  
		we have 
		\begin{equation}
			\underline{\tau_\pm}(\tilde{\chi}_\pm(u,v)) - \underline{\tau_\pm}(\tilde{\chi}_\pm(u,v)) > 0,
		\end{equation}
		where in the case that $\tau_\pm(\tilde{\chi}_\pm(u,v)) - \tau_\mp(\tilde{\chi}_\pm(u,v)) = 2\varepsilon_{null, 0}$, this follows immediately from \eqref{out-null-bd}, \eqref{in-null-bd} with $\varepsilon = \varepsilon_{null, 0}$, while if $\tilde{\chi}_+(u,v) = \chi_b$, $\tilde{\chi}_-(u,v) = 0$ then it follows by uniqueness of null curves on $\widetilde{\mathscr{K}}$, and if $\tau_\mp(\tilde{\chi}_\pm(u,v)) = 0$ then it follows immediately from $\tau_\pm(\tilde{\chi}_\pm(u,v)) > 0$. We must therefore have $(u,v) \in K_d$, since the curves $(\underline{\tau_+}(\chi),\chi)$, $(\underline{\tau_-}(\chi),\chi)$ must intersect in $\widetilde{\mathscr{K}}$, and we have $\underline{\chi}(u,v) \in (\tilde{\chi}_-(u,v),\tilde{\chi}_+(u,v))$, and $\underline{\tau}(u,v) \in (\tau(u,v) - 3\varepsilon_{null, 0}, \tau(u,v) + 3 \varepsilon_{null, 0})$, which implies \eqref{tau-chi-bd}.
		
		\textbf{$C^0$ Cauchy stability for $\mathbf{u}^\mu(u,v), \Omega^2(u,v),  \rho(u,v), r^{-3}m(u,v), r^{-2}f(u,v)$.} We integrate \eqref{Lptauu} on $\tau_+$, $\underline{\tau_+}$ (continuing to use the abbreviations from the previous paragraph) in comoving coordinates, in order to bound $\frac{\mathbf{u}^u(u,v)}{\underline{\mathbf{u}}^u(u,v)}$ for $(u,v) \in K_d$. We define $F_1(r,m,f;\tau,\chi) := (1+f(\chi))^{-\frac{1}{2}} \left( \frac{2m(\chi)}{r(\tau,\chi)}+f(\chi) \right)^{-\frac{1}{2}} \left( \frac{2m'(\chi)}{r(\tau,\chi)} + f'(\chi) - \frac{2m(\chi)}{r(\tau,\chi)^2} \partial_\chi r(\tau,\chi) \right)$, so \eqref{Lptauu} implies that:
		\begin{equation} \label{u-int}
			\underline{\mathbf{u}}^u(u,v) = \underline{\mathbf{u}}^u(u,v_-(u)) \exp\left( \int_{\underline{\chi}(u,v_-(u))}^{\underline{\chi}(u,v)} F_1(\underline{r},\underline{m},\underline{f}; \underline{\tau_+}(\chi),\chi) d\chi \right).
		\end{equation}
		We let $F_{2, +} := \max_{(\tau,\chi) \in \widetilde{\mathscr{K}}} |\partial_\tau F_1 (r,m,f;\tau,\chi)|$, and $\underline{F}_{1, +} := \max_{(\tau,\chi) \in \widetilde{\mathscr{K}}} |F_1(\underline{r},\underline{m},\underline{f};\tau,\chi)|$, where $\underline{F}_{1, +}$ can be bounded in terms of $\varepsilon_{com}(\tilde{m},\tilde{f},\widetilde{\mathscr{K}})$ and $\delta_{com}(\tilde{m},\tilde{f},\widetilde{\mathscr{K}},\varepsilon_{null, N_{null}},1)$. For $\chi \in [\chi(u,v_-(u)), \min\{ \chi(u,v), \underline{\chi}(u,v)\}]$, we can bound:
		\begin{align} \label{F1-bd}
			|F_1(r,m,f;\tau_+(\chi),\chi) - F_1(\underline{r},\underline{m},\underline{f};\underline{\tau_+}(\chi),\chi)| &\leq F_{2, +}|\tau_+(\chi) - \underline{\tau}_+(\chi)|  + |F_1(r,m,f;\underline{\tau_+}(\chi),\chi) - F_1(\underline{r},\underline{m},\underline{f};\underline{\tau_+}(\chi),\chi)| \\ &< F_{2,+} \varepsilon_{null, 0} + C_{F_1} \varepsilon_{null, N_{null}} \leq (F_{2,+} + C_{F_1}) \varepsilon_{null, 0},
 		\end{align}
 		where $C_{F_1}$ depends on $\varepsilon_{com}, \delta_{com}$. We can then bound:
 		\begin{equation} \label{exp-F-bd}
 			\left| \exp\left( \int_{\chi(u,v_-(u))}^{\chi(u,v)} F_1(r,m,f; \tau_+(\chi),\chi) d\chi - \int_{\underline{\chi}(u,v_-(u))}^{\underline{\chi}(u,v)} F_1(\underline{r},\underline{m},\underline{f}; \underline{\tau_+}(\chi),\chi) d\chi\right) - 1\right| < C_F \cdot \varepsilon_{null, 0},
 		\end{equation}
 		where $C_F$ depends on $\tilde{m},\tilde{f},\Omega_0^2,K$ (through $F_{2, +}, C_{F_1}, F_-, \tilde{\delta}_{\mathscr{K}}$), and the bound follows by combining the integrals over $[\chi(u,v_-(u)), \min\{ \chi(u,v), \underline{\chi}(u,v)\}]$ and then using \eqref{F1-bd}, and then bounding the integrand by $\underline{F}_{1, +}$ for the integral over $[\min\{ \chi(u,v), \underline{\chi}(u,v)\}, \max\{ \chi(u,v), \underline{\chi}(u,v)\}]$. Using \eqref{u-int} and \eqref{exp-F-bd} we can bound:
 		\begin{align}
 			|\mathbf{u}^u(u,v) - &\underline{\mathbf{u}}^u(u,v)| < \exp(\chi_b \cdot \underline{F}_{1, +}) \cdot |\mathbf{u}^u(u,v_-(u)) - \underline{\mathbf{u}}^u(u,v_-(u))| +  C_F \mathbf{u}^u(u,v) \cdot \varepsilon_{null, 0} \\ &\leq C_{u, 0}(\tilde{m},\tilde{f},\Omega_0^2,K) \cdot |\underline{\mathbf{u}}^u(u,v_-(u)) - \mathbf{u}^u(u,v_-(u))| + C_{u, 1}(\tilde{m},\tilde{f},\Omega_0^2,K) \cdot \varepsilon_{null, 0},  \label{u-u-bd}
 		\end{align}
 		and the analogous bound holds for $|\mathbf{u}^v - \underline{ \mathbf{u}}^v|$ by integrating \eqref{Lptauv}:
 		\begin{equation} \label{u-v-bd}
 			|\mathbf{u}^v(u,v) - \underline{\mathbf{u}}^v(u,v)| < C_{v, 0}(\tilde{m},\tilde{f},\Omega_0^2,K) \cdot |\underline{\mathbf{u}}^v(u_-(v),v) - \mathbf{u}^v(u_-(v),v)| + C_{v, 1}(\tilde{m},\tilde{f},\Omega_0^2,K) \cdot \varepsilon_{null, 0}.
 		\end{equation}
 		We have $|\mathbf{u}^\mu(-v,v) - \underline{\mathbf{u}}^\mu(-v,v)| \leq \varepsilon_{null, 0}$ using the choice of $\delta_{null}$, and we can use \eqref{u-u-bd}, \eqref{u-v-bd} along with the boundary conditions $	\underline{\mathbf{u}}^u(v,v) = \underline{\mathbf{u}}^v(v,v), \underline{\mathbf{u}}^v(u,v_b(u)) = v_b'(u) \underline{\mathbf{u}}^u(u,v_b(u))$, 
 		for $0 \leq v \leq \underline{u_{\Gamma, +}}$, $0 \leq v_b(u) \leq \underline{v_{b, +}}$, to bound
 		\begin{equation}
 			\|\mathbf{u}^\mu - \underline{\mathbf{u}}^\mu\|_{C^0(K_d)} < C_{\mathbf{u}, 0}(\tilde{m},\tilde{f},\Omega_0^2,K) \cdot \varepsilon_{null, 0},
 		\end{equation}
 		so we immediately get the bound $\|\Omega^2 - \Omega_0^2\|_{C^0(K_d)} < C_{\Omega^2, 0}(\tilde{m},\tilde{f},\Omega_0^2,K) \cdot \varepsilon_{null, 0}$.
 		
 		Now, given $\varepsilon_1 > 0$, there exists $0 < \varepsilon_{cont}(\varepsilon_1) \leq \varepsilon_{null}, \varepsilon_1$, depending on the uniform continuity of $\tilde{m}, \tilde{f} \in C^1([0,\chi_b])$ (and bounds for $\tau$ derivatives of $\underline{R}(\tau,\chi)$ which depend on $\tilde{m},\tilde{f} \in C^1([0,\chi_b])$), such that for all $0 < \varepsilon_{null, 0} \leq \varepsilon_{cont}(\varepsilon_1)$ (where we still have $(\underline{\tilde{m}},\underline{\tilde{f}}) \in B_{\chi_b, 1}((\tilde{m},\tilde{f}), \delta_{null}(\tilde{m},\tilde{f},\widetilde{\mathscr{K}},\varepsilon_{null, N_{null}}(\varepsilon_{null, 0})))$), then we use Cauchy stability in comoving coordinates and \eqref{tau-chi-bd} to get:
 		\begin{equation} \label{Czero-bd}
 			\| \rho - \underline{\rho} \|_{C^0(K_d)}, \ \| r^{-3} - \underline{r}^{-3} \underline{m} \|_{C^0(K_d)}, \ \| r^{-2} f - \underline{r}^{-2} \underline{f} \|_{C^0(K_d)} < \varepsilon_1,
 		\end{equation}
 		(where we use $r^{-3} m = R^{-3} \tilde{m}$, $r^{-2} f = R^{-2} \tilde{f}$ in comoving coordinates). 
 		
 		\textbf{$C^1$ Cauchy stability of $\Omega^2$ and conclusion of the proof of $k=1$ Cauchy stability on $K_d$.} We use the wave equation for $\log \Omega^2$, \eqref{Ruv}:
 		\begin{equation} \label{wave:Omega}
 			\partial_u \partial_v \log \Omega^2(u,v) = \Omega^2(u,v) \left( (r^{-3}m)(u,v) - 2\pi \rho(u,v)\right),
 		\end{equation}
 		where \eqref{Czero-bd} implies that $\| \partial_u \partial_v \log \Omega^2 - \partial_u \partial_v \log \underline{\Omega}^2\|_{C^0(K_d)} \leq C_{\Omega^2, 1}(\tilde{m},\tilde{f},\Omega_0^2,K) \cdot \varepsilon_1$. We can then bound $\| \Omega^2 - \underline{\Omega}^2 \|_{C^1(K_d)} \leq C_{\Omega^2, 2}(\tilde{m},\tilde{f},\Omega_0^2,K) \cdot \varepsilon_1$ using $| \partial_\mu \log \Omega^2(-v,v) - \partial_\mu \log \underline{\Omega}^2(-v,v)| \lesssim d_1((\tilde{m},\tilde{f}), (\underline{\tilde{m}},\underline{\tilde{f}}))$ and the boundary conditions: 
 		\begin{equation} \label{OmegaBCs}
 			\partial_u \log \Omega^2(v,v) = \partial_v \log\Omega^2(v,v), \qquad \partial_v \log\Omega^2(u,v_b(u)) = v_b'(u) \partial_u \log\Omega^2(u,v_b(u)) - \frac{v_b''(u)}{v_b'(u)} \mathbf{u}^u(v_b(u),u),
 		\end{equation}
 		for $0 \leq v \leq u_{\Gamma, +}$, $\chi_b \leq v_b(u) \leq v_{b, +}$. We can use \eqref{pr} and \eqref{rwave}, \eqref{Rayu}, \eqref{Rayv} to bound $\|r - \underline{r}\|_{C^2(K_d)} \lesssim \varepsilon_1$, and \eqref{tauwave} to bound $\| \mathbf{u}^\mu - \underline{\mathbf{u}}^\mu \|_{C^1(K_d)} \lesssim \varepsilon_1$. Hence, from \eqref{tauwave} and \eqref{uvtau}, there exists  $(C_{\tau, 2}, C_{\chi,1})(\tilde{m},\tilde{f},\Omega_0^2,K) \in (0,\infty)^2$, such that $\| \tau - \underline{\tau}\|_{C^2(K_d)} < C_{\tau, 2} \cdot \varepsilon_{1}$, and $\| \chi - \underline{\chi} \|_{C^1(K_d)} < C_{\chi, 1} \cdot \varepsilon_1$ (where we only control $\chi(u,v)$ in $C^1$ because one can see from \eqref{inveikv}, \eqref{inveiku} that second derivatives depend on $\partial_\chi^2 r \sim \tilde{m}''(\chi), \tilde{f}''(\chi)$). We can then again use uniform continuity of $\tilde{m}, \tilde{f} \in C^1([0,\chi_b])$, to see that there exists $0 < \varepsilon_{cont, 1}(\varepsilon_1) \leq \varepsilon_{cont}(\varepsilon_1)$ such that for all $0 < \varepsilon_{null, 0} \leq \varepsilon_{cont, 1}(\varepsilon_1)$, we have \eqref{CS:rhou} for $k=1$ for $\varepsilon = \varepsilon_1$.
 		
 		\textbf{$C^1$ Cauchy stability in the vacuum exterior.} From $\| \tau - \underline{\tau}\|_{C^2(K_d)} < C_{\tau, 2} \cdot \varepsilon_{1}$, we have $\| \tau_b - \underline{\tau_b} \|_{C^2([\chi_b,v_{b, +}])} < 3 C_{\tau, 2} \cdot \varepsilon_{dust, 1}$. From \eqref{tauvb}, we get, for $v \in [\chi_b, v_{b, +}]$:
 		\begin{equation}
 			\underline{R_0}(v) := [\underline{\Omega_0^2}(v)]^{-1} \partial_u \underline{r}(-v,v) =  - \frac{1}{2 \underline{\tau_b}'(v)} \left[ (1+\underline{f}(\chi_b))^\frac{1}{2} + \left( \frac{2\underline{m}(\chi_b)}{\underline{r}(\underline{\tau_b}(v),\chi_b)} + \underline{f}(\chi_b) \right)^\frac{1}{2}\right].
 		\end{equation}
 		We can then use the gauge choice $\underline{r}(-v,v) = v$ along with the definition of the Hawking mass \eqref{m} to express
 		\begin{equation}
 			\underline{\Omega_0^2}(v) = \frac{1}{\underline{R_0}(v)} \left[ \frac{\frac{2\underline{m}(\chi_b)}{v} - 1 }{\underline{R_0}(v)} - 1 \right],
 		\end{equation}
 		so we can choose $\varepsilon_{dust, 1} > 0$ sufficiently small depending on $\varepsilon$ such that $\| \Omega_0^2 - \underline{\Omega_0^2} \|_{C^{1}([\chi_b, v_{b, +}])} < \frac{1}{2}\varepsilon$, and we can then choose $\underline{\Omega_0^2}(v)$ for $v > v_{b, +}$ so that $\| \Omega_0^2 - \underline{\Omega_0^2} \|_{C^{1}(K \cap \mathcal{D}_0)} < \varepsilon$, and we can also choose $\varepsilon_{dust, 1} > 0$ sufficiently small such that \eqref{CS:romega} holds with $k=1$.
 		
 		\textbf{$C^k$ Cauchy stability by induction.} The full Cauchy stability result follows by induction, indeed, suppose that Cauchy stability holds for $k\in\mathbb{N}$, and for $\varepsilon_k > 0$ let $\delta_k(\varepsilon_k) := \delta(\tilde{m},\tilde{f},\Omega_0^2,K,\varepsilon_k,k)$. For $(\underline{\tilde{m}},\underline{\tilde{f}}) \in B_{\chi_b, k}((\tilde{m},\tilde{f}),\delta_k(\varepsilon_k))$, we have $\| \tau - \underline{\tau}\|_{C^{k+1}(K_d)} \lesssim \varepsilon_k$, from \eqref{uvtau} and $\| \mathbf{u}^\mu - \underline{\mathbf{u}}^\mu \|_{C^k(K_d)} < \varepsilon_k$. From \eqref{inveikv}, \eqref{inveiku}, we only have $\| \chi - \underline{\chi}\|_{C^{k}(K_d)} \lesssim \varepsilon_k$, since $(k+1)$ derivatives of $\chi(u,v)$ depend on $\partial_\chi^{k+1} r(\tau(u,v),\chi(u,v))$. We can then choose, given $\varepsilon_{k+1}>0$, $0 < \delta(\varepsilon_{k+1}) \leq \delta(\tilde{m},\tilde{f},\Omega_0^2,K,\varepsilon_k,k)$ sufficiently small such that, using Cauchy stability in comoving coordinates, we have, for all $(\underline{\tilde{m}}, \underline{\tilde{f}}) \in B_{\chi_b, k+1}((\tilde{m},\tilde{f}), \delta)$: $\| \tau- \underline{\tau}\|_{C^{k+1}(K_d)}, \| \chi - \underline{\chi}\|_{C^{k +1}(K_d)} < \varepsilon_{k+1}$, and $\| \rho - \underline{\rho} \|_{C^k(K_d)}, \| r^{-3} m - \underline{r}^{-3} \underline{m}\|_{C^{k+1}(K_d)}, \| r^{-2} f - \underline{r}^{-2} \underline{f}\| < \varepsilon_{k+1}$. We then need to extend Cauchy stability of the double null gauge: we use \eqref{wave:Omega}, where we now control the right hand side in $C^k$, so we control all $(k+1)$ mixed derivatives of $\log \Omega^2$ directly by differentiation, while we bound $\| \partial_\mu^{k+1} \log \Omega^2 - \partial_\mu^{k+1} \log \underline{\Omega}^2\|_{C^{0}(K_d)} \lesssim \varepsilon_{k+1}$ by differentiating \eqref{wave:Omega} with $\partial_\mu^{k}$, and then integrating in the other direction, using $| \partial_\mu^{k+1} \log \Omega^2(-v,v) - \partial_v^{k+1} \log \underline{\Omega}^2(-v,v)| \lesssim d_{k+1}((\tilde{m},\tilde{f}),(\underline{\tilde{m}},\underline{\tilde{f}}))$, and the boundary conditions which follow from differentiating \eqref{OmegaBCs}. It then follows from \eqref{tauwave} that we have $\| \mathbf{u}^\mu - \underline{\mathbf{u}}^\mu\|_{C^{k+1}(K_d)} \lesssim \varepsilon_{k+1}$, and \eqref{rwave}, \eqref{Rayu}, \eqref{Rayv} imply that we have $\| r- \underline{r} \|_{C^{k+2}(K_d)} \lesssim \varepsilon_{k+1}$. We then use $\| \tau - \underline{\tau} \|_{C^{k+2}(K_d)} \lesssim \varepsilon_{k+1}$ to bound $\| \Omega_0^2 - \underline{\Omega_0}^2\|_{C^{k+1}([\chi_b, v_{b, +}])} \lesssim \varepsilon_{k+1}$, and it follows that we have $(k+1)$ Cauchy stability in the vacuum exterior.
	\end{proof}
	
	\subsection{Piecewise smooth solutions with finitely many thin shells and the propagation condition} \label{finiteshells}
	
	In Section \ref{pwisesection}, we defined the notion of a piecewise smooth solution of the Einstein-dust system in double null coordinates, containing one thin shell. We now generalise this definition to finitely many thin shells, and define the \emph{propagation condition}: given any point on a thin shell, the solution to the future is locally given by the propagation of a single thin shell. Given these definitions, we give the precise statement of Theorem \ref{MGHD} in Section \ref{sec:thm2}.
	\begin{definition} \label{pwisefinite}
		We say that $(\mathcal{Q},r,\Omega^2,\rho,\rho(\mathbf{u}^\mu)^2)$ is a piecewise smooth solution of (E-d) with finitely many thin shells if $\mathcal{Q} \subset \R^{1+1}_{u,v}$, $r, \Omega^2 \in C^{0,1}(\mathcal{Q})$; there exists $N \in \mathbb{N}$, and functions $u_{i}(s), v_{i}(s) \in C^1([0,s_i)) \cap C^2((0, s_{i}))$ for each $1 \leq i \leq N$, with $\dot{u}_i(s), \dot{v}_i(s) > 0$ for all $0\leq s < s_i$, such that for
		\begin{equation} \label{def:P}
			\Sigma_i := \{ (u_{i}(s), v_{i}(s)) \in \mathcal{Q} : s \in [0,s_{i}) \}, \ \mathcal{P}(\mathcal{Q}) := \mathcal{Q} \setminus \left( \bigcup_{i=1}^N \Sigma_i \right),
		\end{equation}
		there exist non-negative functions $\sigma_i(s) \in C^0([0,s_i)) \cap C^1((0,s_i))$ and $\mathring{\rho} \in C^1(\mathcal{P}(\mathcal{Q}))$, $\mathring{\mathbf{u}}^u, \mathring{\mathbf{u}}^v \in C^1( \text{supp}(\mathring{\rho}) \cap \mathcal{P}(\mathcal{Q}))$ such that: for all $(u,v) \in \mathcal{P}(\mathcal{Q})$, $\rho(u,v) = \mathring{\rho}(u,v)$, and for all $(u,v) \in \text{supp}(\mathring{\rho}) \cap \mathcal{P}(\mathcal{Q})$, $\mathbf{u}^u(u,v) = \mathring{\mathbf{u}}^u(u,v)$, $\mathbf{u}^v(u,v) = \mathring{\mathbf{u}}^v(u,v)$, and for all $\varphi \in C_c(\mathcal{Q})$:
		\begin{align}
			\int_{\mathcal{Q}} \varphi \rho &= \sum_{i=1}^{N} \int_0^{s_i} \sigma_i(s) \varphi(u_i(s),v_i(s)) ds + \int_{\mathcal{P}(\mathcal{Q})} \varphi \mathring{\rho}, \\ \int_{\mathcal{Q}} \varphi \rho (\mathbf{u}^u)^2 &= \sum_{i=1}^{N} \int_0^{s_i} \sigma_i(s) (\dot{u}_i(s))^2 \varphi(u_i(s),v_i(s)) ds + \int_{\mathcal{P}(\mathcal{Q})} \varphi \mathring{\rho} (\mathring{\mathbf{u}}^u)^2, \\ \int_{\mathcal{Q}} \varphi \rho (\mathbf{u}^v)^2 &= \sum_{i=1}^{N} \int_0^{s_i} \sigma_i(s) (\dot{v}_i(s))^2 \varphi(u_i(s),v_i(s)) ds + \int_{\mathcal{P}(\mathcal{Q})} \varphi \mathring{\rho} (\mathring{\mathbf{u}}^v)^2.
		\end{align}
		Moreover, we require that the restriction $(\mathcal{P}(\mathcal{Q}) \setminus \Gamma, r, \Omega^2, \rho, \mathbf{u}^u)$ is a $C^1$ solution of (E-d), where we impose $\Gamma = \{ u= v\}$ and if $(v,v) \in \mathcal{Q}$: $r(v,v) = 0$, $m(v,v) = 0$. For any $1 \leq i \leq N$, and any $(u,v) \in (\Sigma_i \setminus \partial^- \Sigma_i) \setminus (\cup_{j \neq i} \Sigma_j)$, then $r, \Omega^2, \rho, \mathbf{u}^\mu$ extend in the respective regularity for a $C^1$ solution of (E-d) to $(u,v)$ from $D^\pm(\Sigma_i,\mathcal{Q}) \cap \mathcal{P}(\mathcal{Q})$. 
		
		We require that the integrated weak formulation of the Einstein equations, \eqref{rwaveweaku}--\eqref{Rayvweak} holds for all $\varphi \in C_c(\text{int}(\mathcal{Q}))$ (where we replace $\int_0^{u_1} \int_0^{v_1} du dv$ with $\iint_{(u,v) \in \mathcal{Q}} du dv$, and drop the integrals over $\{ u = 0\}, \{v=0\}$).
		
		We impose moreover that the set of formation points of thin shells:
		\begin{equation}
			\partial^- \Sigma := \{ (u_i(0),v_i(0)) : 1 \leq i \leq N\},
		\end{equation}
		must either be first shell crossing singularities (with respect to the local prior $C^1$ solution of (E-d)), or the future limit point of a thin shell:
		\begin{equation}
			\partial^+ \Sigma := \left\{ \lim_{s \to s_i}(u_{i}(s),v_{i}(s)) : 1 \leq i \leq N\right\}.
		\end{equation}
		We finally also impose the convention that $\mathcal{Q}$ is `open to the future' as in Remark \ref{open-future}. 
	\end{definition}
	
	\subsubsection{The propagation condition} \label{propcond}
	
	As discussed in the introduction (Section \ref{sec:int-pwise-ext}), if no additional conditions are imposed, piecewise smooth solutions with finitely many thin shells are fundamentally non-unique. We will restrict to a class of piecewise smooth solutions for which we can show uniqueness. We first define the notion of a solution to the `thin shell propagation' problem:
	\begin{definition} \label{propagation}
		We say that $(\mathcal{Q},r,\Omega^2,\rho,\mathbf{u}^\mu)$, a piecewise smooth solution of (E-d) as in Definition \ref{pwisesmooth}, with $\mathcal{Q} = [u_0, u_1) \times [v_0,v_1)$, is a solution to the `thin shell propagation' problem if: $\Sigma$ satisfies the determinism condition (Definition \ref{detcond}), $\rho, \mathbf{u}^\mu$ extend from $D^\pm(\Sigma,\mathcal{Q})$ (see Definition \ref{def:Dpm}) to continuous functions on $D^\pm(\Sigma,\mathcal{Q}) \cup \Sigma$ (where possibly $\rho$ extends continuously to infinity at $(u_0,v_0)$) and $r, \Omega^2$ extend from $D^\pm(\Sigma,\mathcal{Q})$ to $D^\pm(\Sigma,\mathcal{Q}) \cup \Sigma$ in $C^1$. 
	\end{definition}
	
	\begin{definition} \label{def:propcond}
		A piecewise smooth solution of (E-d) with finitely many thin shells, $(\mathcal{Q},r,\Omega^2,\rho, \rho(\mathbf{u}^\mu)^2)$, satisfies the \underline{propagation condition} if for all $(u_0, v_0) \in \mathcal{Q} \cap \bigcup_{i} \overline{\Sigma_i}$, there exists $(u_1,v_1) \in I^+(u_0,v_0)$ such that the restriction to $[u_0, u_1) \times [v_0, v_1)$ is a solution of the thin shell propagation problem, as in Definition \ref{propagation}.
		
		For a piecewise smooth solution of (E-d) with finitely many thin shells, $(\mathcal{Q},r,\Omega^2,\rho, \rho(\mathbf{u}^\mu)^2)$, satisfying the propagation condition, we can define $\mathbf{u}^\mu$ pointwise on $\mathcal{Q}$, as in e.g.~Definition \ref{piecesmooth}. We therefore refer to any piecewise smooth solution of (E-d) satisfying the propagation condition as $(\mathcal{Q},r,\Omega^2,\rho,\mathbf{u}^\mu)$. 
	\end{definition}
	
	\begin{remark}
		For brevity we will often drop `with finitely many thin shells' and simply refer to piecewise smooth solutions of (E-d) satisfying the propagation condition.
	\end{remark}
	
	\begin{remark}
		The propagation condition implies that for any $p \in \cup_i \Sigma_i$, there exists $q \in I^+(p)$ such that there is only one thin shell in $J^+(p) \cap I^-(q)$. If $p$ is a point of thin shell formation, i.e.~$p \in \partial^- \Sigma$, then there can however be arbitrarily many thin shells in $J^-(p)$. Indeed, we'll see that thin shell \emph{collisions}, where multiple thin shells have the same future limit point, can occur in evolution.
	\end{remark}
	
	\subsubsection{Extension principle for piecewise smooth solutions satisfying the propagation condition} \label{extp}
	
	In this section, we first state a Corollary of thin shell development applicable to $C^1$ solutions of (E-d) (with smooth initial data), and then prove the main extension principle for unique extensions beyond limit points of thin shells, given conditions which we'll later show apply for thin shells formed in generic data, in the proof of Theorem \ref{MGHDprecise}.
	
	Since we only require that away from thin shells in piecewise smooth solutions of (E-d), we have a $C^1$ solution of (E-d), we make the following corollary of Theorem \ref{devthm}, in order to be able to use thin shell development when the double null gauge is only $C^2$ (with respect to smooth comoving coordinates). 
	\begin{corollary}[of Theorem \ref{devthm}] \label{devcor}
		Let $(\mathcal{Q},r,\Omega^2,\rho,\mathbf{u}^\mu)$ be a piecewise smooth solution of (E-d) satisfying the propagation condition, such that for $u_0 < u_1 < u_3$, $v_0 < v_1 < v_3$: 
		\begin{equation}
			\mathcal{R}' := ([u_0, u_1) \times [v_0, v_3]) \cup ([u_0, u_3] \times [v_0, v_1)) \subset \mathcal{P}(\mathcal{Q}),
		\end{equation}
		and $(u_1,v_1)$ a \underline{non-degenerate} first shell crossing singularity. Suppose moreover that $\rho(u,v) > 0$ for all $(u,v) \in \mathcal{R}'$, and letting $(\tau(u,v), m(u,v))$ be comoving coordinates on $\mathcal{R}'$, where $m(u,v)$ is the Hawking mass, that: transforming to $(\tau,m)$ coordinates on
		\begin{equation}
			\mathscr{R}' := \{ (\tau(u,v),m(u,v)) : (u,v) \in \mathcal{R}' \},
		\end{equation}
		$(\mathscr{R}',r,\rho,m,f)$ is a $C^\infty$ solution of the collapsing Einstein-dust system. Then: there exists $u_2 \in (u_1,u_3)$, $v_2 \in (v_1,v_3)$ such that, letting $\mathcal{R}_2 := \mathcal{R}' \cup ([u_1, u_2) \times [v_1, v_2))$, there is a unique extension as a piecewise smooth solution of (E-d) satisfying the propagation condition to $\mathcal{R}_2$, with a thin shell forming as in Theorem \ref{devthm}.
	\end{corollary}
	
	\begin{proof}
		This is an immediate generalisation of Theorem \ref{devthm}. The proof of equivalence to the existence/uniqueness result for the thin shell ODE is now restricted to $C^2$ solutions of the eikonal equation, otherwise the proof is identical. Moreover, from the statement of Theorem \ref{devthm}, the propagation condition is satisfied.
	\end{proof}
		
	\begin{definition}
		For $(\mathcal{Q},r,\Omega^2,\rho,\mathbf{u}^\mu)$ a piecewise smooth solution of (E-d) satisfying the propagation condition, we denote the fluid flow line from a point $(u,v) \in \mathcal{P}(\mathcal{Q})$ by:
		\begin{equation}
			\mathcal{F}(u,v;\mathcal{Q}) := \{ (\tilde{u}(t), \tilde{v}(t)) : t \in [0,t_*)\} \subset \mathcal{P}(\mathcal{Q}),
		\end{equation}
		where $\tilde{u}(0) = u$, $\tilde{v}(0) = v$, and $\tilde{u}(t), \tilde{v}(t) \in C^2([0,t_*))$ solve
		\begin{equation}
			\tilde{u}'(t) = \mathbf{u}^u(\tilde{u}(t),\tilde{v}(t)), \ \tilde{v}'(t) = \mathbf{u}^v(\tilde{u}(t),\tilde{v}(t)).
		\end{equation}
		We can always choose $t_*>0$ such that $\partial^+ \mathcal{F}(u,v;\mathcal{Q}) \notin \mathcal{P}(\mathcal{Q})$.
	\end{definition}	
		
	We now state the extension principle for the case in which $r, \Omega^2$ do not extend in $C^1$ to $(u_1,v_1)$. See Section \ref{sec:int-pwise-ext} in the introduction for a discussion of the result and a brief outline of the proof. 
	\begin{proposition} \label{prop:pwise-extension}
		Let $(\mathcal{Q},r,\Omega^2,\rho,\mathbf{u}^\mu)$ be a globally hyperbolic piecewise smooth solution of (E-d) satisfying the propagation condition, with $(u_1,v_1) \in \partial^+\Sigma$, $(u_0,v_0) \in I^-(u_1,v_1)$ such that
		\begin{equation}
			\mathcal{R}' := ([u_0, u_1] \times [v_0, v_1]) \setminus \{(u_1,v_1)\} \subset \mathcal{Q},
		\end{equation}
		and the functions $r, \Omega^2$ extend continuously to $(u_1,v_1)$, with $r(u_1,v_1), \Omega^2(u_1,v_1) > 0$, and, defining $C := \{ u=u_1, v_0 \leq v < v_1\}$, $\underline{C} := \{ u_0 \leq u < u_1, v =v _1\}$, that $C, \underline{C} \subset \mathcal{P}(\mathcal{Q})$ and $r,\Omega^2$ extend in $C^1$ from $C$, $\underline{C}$ to $(u_1,v_1)$, where we denote e.g.~$(\partial_u r)_1^- := \lim_{v \to v_1} \partial_u r(u_1,v)$, $(\partial_u r)_1^+ := \lim_{u \to u_1} \partial_u r(u,v_1)$ and $[\partial_u r]_1 := (\partial_u r)_1^+ - (\partial_u r)_1^-$. We suppose that the limits satisfy:
		\begin{equation}
			0 > (\partial_u r)_1^+ > (\partial_u r)_1^-, \qquad (\partial_v r)_1^+ < (\partial_v r)_1^-, \qquad m_1^+ > m_1^-, \qquad [\partial_u \log(r\Omega)^2]_1 = 0 = [\partial_v \log(r\Omega)^2]_1. \label{firstbd}
		\end{equation}
		We suppose moreover that: there exists $\tilde{u} \in (u_0, u_1]$ such that $\underline{\widetilde{C}} := \{ \tilde{u} \leq u \leq u_1, v = v_0\} \subset \mathcal{P}(\mathcal{Q})$, with $\rho(u,v) > 0$ for $(u,v) \in \underline{\widetilde{C}}$, $m(\tilde{u}, v_0) = m_1^-$, and $\partial^+ \mathcal{F}(\tilde{u},v_0;\mathcal{Q}) = (u_1,v_1)$, with
		\begin{equation}
			\widetilde{\mathcal{R}}^- := D^-(\mathcal{F}(\tilde{u},v_0;\mathcal{Q}),\mathcal{Q}) \cup \mathcal{F}(\tilde{u},v_0;\mathcal{Q}) \subset \mathcal{P}(\mathcal{Q}),
		\end{equation}
		and, for $(\tau^-(u,v),m(u,v))$ comoving coordinates on $\widetilde{\mathcal{R}}^-$, the induced data on $\mathscr{D}^- := \{ (\tau^-(u,v),m(u,v)) : (u,v) \in \underline{\widetilde{C}}\}$ is smooth initial data for the collapsing Einstein-dust system, with maximal dust development $(\mathscr{Q}_{MDD}^-, r^-,\rho^-,m^-,f^-)$. We then have $\mathbf{u}^\mu$ extending continuously to $(u_1,v_1)$ from $C$, and we require that: 
		\begin{equation}
			\frac{[\partial_u r]_1}{- [\partial_v r]_1} < \frac{(\mathbf{u}^v)^-_1}{(\mathbf{u}^u)^-_1}, \label{mindet}
		\end{equation}
		and that there exists $0 \leq \beta < 1$ such that $\rho(u_1,v) \lesssim (v_1 - v)^{-\beta}$ for all $v \in [v_0,v_1)$. 
		
		We suppose that \underline{either} the same condition holds in the exterior: there exists $\tilde{v} \in (v_0,v_1]$ such that $\widetilde{C} := \{ u=u_0, \tilde{v} \leq v \leq v_1 \} \subset \mathcal{P}(\mathcal{Q})$, $m(u_0,\tilde{v}) = m_1^+$, $\partial^+ \mathcal{F}(u_0,\tilde{v};\mathcal{Q}) = (u_1,v_1)$,
		\begin{equation}
			\widetilde{\mathcal{R}}^+ := D^+(\mathcal{F}(u_0,\tilde{v};\mathcal{Q}),\mathcal{Q}) \cup \mathcal{F}(u_0,\tilde{v};\mathcal{Q}) \subset \mathcal{P}(\mathcal{Q}),
		\end{equation}
		and for $(\tau^+(u,v),m^+(u,v))$ comoving coordinates on $\widetilde{\mathcal{R}}^+$, the induced data on $\mathscr{D}^+ := \{ (\tau^+(u,v),m(u,v)) : (u,v) \in \widetilde{C}\}$ is smooth initial data for the collapsing Einstein-dust system, with maximal dust development $(\mathscr{Q}_{MDD}^+, r^+,\rho^+,m^+,f^+)$, and 
		\begin{equation}
			\frac{(\mathbf{u}^v)^+_1}{(\mathbf{u}^u)^+_1} < \frac{[\partial_u r]_1}{- [\partial_v r]_1}, \label{posdet}
		\end{equation}
		$\rho(u,v_1) \lesssim (u_1 - u)^{-\beta}$ for all $u \in [u_0,u_1)$; \underline{or}
		$\rho(u,v_1) = 0$ for all $u \in [u_0,u_1)$.
		
		Then $(\mathcal{Q}, r, \Omega^2, \rho, \mathbf{u}^\mu)$ can be \emph{uniquely extended} as a piecewise smooth solution satisfying the propagation condition: for some $u_2 > u_1$, $v_2 > v_1$, $[u_0, u_2) \times [v_0, v_2) \subset \mathcal{Q}$, and the restriction to $\mathcal{R}_2 := [u_1, u_2) \times [v_1, v_2)$ is a piecewise smooth solution (as in Definition \ref{pwisesmooth}) with thin shell
		\begin{equation}
			\Sigma := \{ (u(s),v(s)) : s \in [0,s_1)\},
		\end{equation}
		given by $u(s),v(s) \in C^1([0,s_1)) \cap C^2((0,s_1))$ satisfying the determinism condition, and
		\begin{equation}
			\dot{u}(0) = \mathbf{u}_0^u := \Omega(u_1,v_1)^{-1} \left( \frac{-[\partial_v r]_1}{[\partial_u r]_1}\right)^\frac{1}{2}, \qquad \dot{v}(0) = \mathbf{u}_0^v := \Omega(u_1,v_1)^{-1} \left( \frac{[\partial_u r]_1}{- [\partial_v r]_1}\right)^\frac{1}{2}, \label{initvel}
		\end{equation}
		and $D^\pm(\mathcal{R}_2,\Sigma) \subset \mathcal{P}(\mathcal{Q})$ with, defining the comoving coordinate $\tau^\pm(u,v) \in C^2(D^\pm(\mathcal{R}_2,\Sigma))$ (so that $\tau^\pm(u_1,v_1)$ coincides with the respective limit from $\widetilde{\mathcal{R}}^\pm$, and $\tau^+(u,v)$ is only defined if $\rho$ does not vanish on $\underline{C}$), then
		\begin{equation}
			\mathscr{R}_2^\pm := \{ (\tau^\pm(u,v),m^\pm(u,v)) : (u,v) \in D^\pm(\mathcal{R}_2,\Sigma)\} \subset \mathscr{Q}_{MDD}^\pm,
		\end{equation}
		and $(r(s),m^+(s),m^-(s),\epsilon(s)) \in C^0([0,s_1)) \cap C^2((0,s_1))$ (defined for $\Sigma$ as in Proposition \ref{tsodeprop}) is the \emph{unique} solution of the thin shell ODE \eqref{rdot0}--\eqref{epsdot0} with initial data
		\begin{equation} \label{extdata1}
			r(0) = r(u_1,v_1), \qquad m^\pm(0) = m_1^\pm, 
		\end{equation} 
		\begin{equation} \label{extdata2}
			\epsilon(0) = \frac{1}{2} \mathbf{u}_0^v ( (\partial_v r)_1^+ + (\partial_v r)_1^-) - \frac{1}{2} \mathbf{u}_0^u ( (\partial_u r)_1^+ + (\partial_u r)_1^-) > 0, 
		\end{equation}
		where 
		\begin{equation} \label{init-R}
			\mathcal{R}(r(0),m^+(0),m^-(0),\epsilon(0)) = \mathcal{R}_0 := \mathbf{u}_0^u (\partial_u r)^\pm_1 + \mathbf{u}_0^v (\partial_v r)^\pm_1 < 0,
		\end{equation}
		and if $\rho_1^+ > 0$: $\dot{m}^-(s)<0$, $\dot{m}^+(s)>0$, $(\tau^\pm(s),m^\pm(s)) \in \mathscr{Q}_{MDD}^\pm$ for $s \in (0,s_1)$; if $\rho$ vanishes on $\underline{C}$ then $\dot{m}^-(s)<0$, $m^+(s) = m_1^+$, $(\tau^-(s),m^-(s)) \in \mathscr{Q}_{MDD}^-$ for $s \in (0,s_1)$.
	\end{proposition}
	
	\begin{remark}
		The extension principle can be thought of as the generalisation of Theorem \ref{devthm} to the case of formation of a thin shell with positive initial mass, where we require the propagation condition to impose that any solution locally extending to include $(u_1,v_1)$ consists of the propagation of a single thin shell. 
	\end{remark}

	\begin{proof}
		We first show that the assumptions on $C$, $\underline{C}$ imply that $\mathcal{Q}$ uniquely extends to include, for some $\delta>0$ sufficiently small: 
		\begin{equation} \label{step1}
			([u_1, u_1 + \delta) \times [v_0, v_1)) \cup ([u_0, u_1) \times [v_1, v_1 + \delta)) \subset \mathcal{Q},
		\end{equation}
		and moreover that $([u_1, u_1 + \delta) \times [v_1 - \delta, v_1)) \cup [u_1 - \delta, u_1) \times [v_1, v_1+ \delta) \subset \mathcal{P}(\mathcal{Q})$, and that $r, \Omega^2$ extend in $C^1$, $\rho,\mathbf{u}^\mu$ extend in $C^0$ to $(u_1,v_1)$ from the interior/exterior respectively, with the limit coinciding with the limit along $C$, $\underline{C}$ respectively. We can then consider the problem of extending as a piecewise smooth solution satisfying the propagation condition to (a subset of) $[u_1, u_1 + \delta) \times [v_1, v_1 + \delta)$.
		
		We suppose first that $\rho_1^+ > 0$. The assumption that the induced data on $\mathscr{D}^+$ is smooth data for the collapsing Einstein-dust system, along with $\widetilde{\mathcal{R}}^+ \subset \mathcal{P}(\mathcal{Q})$, implies that, transforming to $(\tau^+(u,v),m(u,v))$ comoving coordinates on $\widetilde{\mathscr{R}}^+ := \{ (\tau^+(u,v),m(u,v)) : (u,v) \in \widetilde{\mathcal{R}}^+\}$, we have $(\widetilde{\mathscr{R}}^+, r^+, \rho^+, m^+, f^+) \subset (\mathscr{Q}_{MDD}, r^+, \rho^+, m^+, f^+)$. If $\rho(u,v_1) \to \infty$ as $u \to u_1$, then the `non-degeneracy' condition $\rho(u,v) \lesssim (u_1 - u)^{-\beta}$ implies that $\tau_c^+(m)$ (defined on $(\mathscr{C}_\chi)^+$ for $\mathscr{Q}_{MDD}^+$) is increasing, with $\tau_c^+(m) - \tau_c^+(m_1^+) \sim (m - m_1^+)^{n_+}$ for some $n_+ \in \mathbb{N}$, and $m \geq m_1^+$ sufficiently close to $m_1^+$ (see Lemma \ref{nondeglemma}). Hence, there exists a (unique) outgoing null curve from $(\tau_1^+, m_1^+)$ in $\mathscr{Q}_{MDD}$ (see e.g.~the proof of Proposition \ref{regext}). For $u_0 < u < u_1$ sufficiently close to $u_1$, the flow line $\mathcal{F}(u,v_1; \mathcal{Q})$ is non-empty (since $\underline{C} \subset \mathcal{P}(\mathcal{Q})$ and by assumption there are finitely many \emph{timelike} thin shells in $\mathcal{Q}$), and ($\mathcal{Q}$ can be uniquely extended so that) there exists $\delta >0$ with $[u_1-\delta, u_1) \times [v_1, v_1 + \delta) \subset D^-(\mathcal{F}(u,v_1; \mathcal{Q}), \mathcal{Q})$. Moreover, by construction we have $[u_1-\delta, u_1) \times [v_1, v_1 + \delta) \subset D_{ac}^+(\widetilde{C}) \subset \mathcal{P}(\mathcal{Q})$. Extending $\tau^+(u,v)$ to be defined (uniquely) on $D^+_{ac}(\widetilde{C})$, we define:
		\begin{align} \label{CHdelta}
			\mathscr{CH}_{\delta}^+ &:= \{ (\tau^+(u,v),m(u,v)) : u = u_1, v \in [v_1,v_1+\delta) \} \subset \mathscr{Q}_{MDD}^+ \cup \{ (\tau_1^+, m_1^+)\}, \\
			\underline{\mathscr{N}}^+_{\delta} &:= \{ (\tau^+(u,v),m(u,v)) : v = v_1, u \in [u_1 - \delta,u_1] \} \subset \mathscr{Q}_{MDD}^+.
		\end{align}
		Let $u^+(\tau,m), v^+(\tau,m)$ be the inverse of $(\tau^+(u,v),m(u,v))$ on $\mathscr{R}^+_{ext} := \{ (\tau^+(u,v),m(u,v)) : (u,v) \in D_{ac}^+(\widetilde{C}) \} \subset \mathscr{Q}_{MDD}^+$. Then $u^+(\tau,m), v^+(\tau,m) \in C^2(\mathscr{R}^+_{ext})$, and by Proposition \ref{eikprop}, we have $v^+(\tau,m)$ extending in $C^2_{\tau, m}$ to $\mathscr{CH}_{\delta}^+$, since we have e.g.~$C^2$ induced data for the eikonal equation on the outgoing null curve $\{ (\tau^+(u,v),m(u,v)) : u = u_1 - \delta, v \in [v_1, v_1 + \delta]\}$. Moreover, by the assumption that $\partial_u r$, $\partial_u \log \Omega^2$ extends continuously to $(u_1,v_1)$ along $\underline{C}$, we have $\partial_\tau u^+$, $\partial_\tau^2 u^+$ extending continuously to $(\tau_1^+,m_1^+)$ along $\underline{\mathscr{N}}^+_{\delta}$ (and $\mathbf{u}^\mu$ must therefore extend continuously along $\underline{C}$ to $(u_1,v_1)$). Then $\mathscr{CH}_{\delta}^- \subset \mathscr{U}_+(\underline{\mathscr{N}}^+_{\delta})$ and so $u^+(\tau,\chi)$ extends in $C^2$ to $\mathscr{CH}_{\delta}^-$. The $C^1$ collapsing solution of (E-d) therefore extends in regularity to $\{ u = u_1, v \in (v_1, v_1 + \delta)\}$, and the limit to $(u_1,v_1)$ coincides with limit along $\underline{C}$.
		
		The analogous argument holds in the interior, and by compactness we can choose $\delta > 0$ sufficiently small such that \eqref{step1} holds, with
		\begin{equation}
			[u_1-\delta, u_1) \times [v_1, v_1 + \delta) \subset D_{ac}^+(\widetilde{C}) \subset \mathcal{P}(\mathcal{Q}), \ [u_1, u_1 + \delta) \times [v_1 - \delta, v_1) \subset D_{ac}^+(\underline{\widetilde{C}}) \subset \mathcal{P}(\mathcal{Q}),
		\end{equation}
		and $r,\Omega^2$ extend in $C^1$, $\mathbf{u}^\mu, \rho$ extend in $C^0$ (with possibly $\rho \to \infty$ at $(u_1,v_1)$) from $[u_1, u_1 + \delta) \times [v_1 - \delta, v_1)$ to $\{ u \in [u_1, u_1 + \delta), v = v_1 \}$, and from $[u_1 - \delta, u_1) \times [v_1, v_1 + \delta)$ to $\{ u = u_1, v \in [v_1, v_1 + \delta) \}$. Moreover, defining $\mathscr{CH}_{\delta}^-$ analogously to \eqref{CHdelta}, we have $\mathscr{CH}_{\delta}^\pm \setminus (\tau_1^\pm,m_1^\pm) \subset \mathscr{Q}_{MDD}^\pm$.
		
		In the case of vacuum exterior, the same result holds in the interior, but with $\rho(u,v) = 0$ for all $(u,v) \in [u_1-\delta, u_1) \times [v_1, v_1 + \delta) \subset \mathcal{P}(\mathcal{Q})$ instead, and again $r, \Omega^2$ extending in $C^1$ to $\{ u = u_1, v \in [v_1, v_1 + \delta) \}$ from $[u_1 - \delta, u_1) \times [v_1, v_1 + \delta)$, where by continuity we have $\partial_u r < 0$.
		
		We now make the necessary observations from the statement of the proposition to show existence and uniqueness of solutions to the thin shell ODE, as stated. We will then be able to show existence and uniqueness of piecewise smooth extensions satisfying the propagation condition. From \eqref{mindet} and the definitions \eqref{initvel}, we see that: 
		\begin{equation} \label{detcondzero}
			\frac{\mathbf{u}_0^u}{(\mathbf{u}^u)_1^-} = \left( \frac{- (\mathbf{u}^v)_1^-[\partial_v r]_1}{(\mathbf{u}^u)_1^-[\partial_u r]_1}\right)^\frac{1}{2} > 1, \qquad \frac{\mathbf{u}_0^v}{(\mathbf{u}^v)_1^-} < 1,
		\end{equation}
		and similarly if $\rho_1^+ > 0$ then $\frac{\mathbf{u}_0^u}{(\mathbf{u}^u)_1^+} < 1 < \frac{\mathbf{u}_0^v}{(\mathbf{u}^v)_1^+}$. Then the equality \eqref{init-R} follows from the definition \eqref{def:R} along with \eqref{extdata2}. We now show the inequality, using the equality \eqref{init-R} on the $-$ side. We have $(\partial_u r)_1^- < 0$ by assumption, if $(\partial_v r)_1^- > 0$, then:
		\begin{equation}
			\mathbf{u}(r)^-_1 - \mathcal{R}_0 = ( (\mathbf{u}^u)_1^- - \mathbf{u}_0^u) (\partial_u r)_1^- + ( (\mathbf{u}^v)_1^- - \mathbf{u}_0^v) (\partial_v r)_1^- > 0,
		\end{equation}
		by \eqref{detcondzero}, and so $\mathbf{u}(r)^-_1 < 0$ implies that $\mathcal{R}_0 < 0$. If $(\partial_v r)_1^- < 0$ then we clearly have $\mathcal{R}_0 = \mathbf{u}_0^u (\partial_u r)_1^- + \mathbf{u}_0^v (\partial_v r)_1^- < 0$. We can now show that there exists a unique solution to the thin shell ODE as stated. This follows by Picard-Lindel\"of if $(\tau^\pm(u_1,v_1),m^\pm(u_1,v_1)) \in \mathscr{Q}_{MDD}^\pm$: we extend the result to allow $(\tau^\pm(u_1,v_1),m^\pm(u_1,v_1)) \in \mathscr{C}$, with the non-degeneracy condition as stated.  
		
		\begin{lemma} \label{lem:ode}
			There exists $s_1>0$ such that there is a unique solution $(r(s),m^+(s),m^-(s),\epsilon(s)) \in C^0([0,s_1)) \cap C^2((0,s_1))$ of the thin shell ODE \eqref{rdot0}--\eqref{epsdot0}, with $(\tau^\pm(s),m^\pm(s)) \in \mathscr{Q}_{MDD}^\pm$, $\dot{m}^+(s) > 0$, $\dot{m}^-(s) < 0$ for $s \in (0,s_1)$ ($m^+(s) = m_1^+$ if $\rho_1^+ = 0$), and with initial data
			\begin{equation}
				r(0) = r(u_1,v_1), \qquad m^\pm(0) = m_1^\pm, \qquad \epsilon(0) = \frac{1}{2} \dot{v}(0) ( (\partial_v r)_1^+ + (\partial_v r)_1^-) - \frac{1}{2} \dot{u}(0) ( (\partial_u r)_1^+ + (\partial_u r)_1^-).
			\end{equation} 
		\end{lemma}
		\begin{proof}
			We show there exists a unique solution in the case $\rho_1^+ > 0$. If $\rho_1^+ = 0$ then the proof simplifies with $\dot{m}^+(s) = 0$. 
			
			From the initial data satisfying \eqref{mindet}, \eqref{posdet}, and the choice of initial data \eqref{extdata2}, it follows that $\tilde{\mathcal{X}}^-(r(0),m^+(0),m^-(0),\epsilon(0)) < 0$ and $\tilde{\mathcal{X}}^+(r(0),m^+(0),m^-(0),\epsilon(0)) > 0$. We therefore have, by continuity for sufficiently small $s$: $\dot{m}^\pm(s) \partial_m r(\tau^\pm(s), m^\pm(s)) \sim \pm 1$.
			
			The non-degeneracy conditions for $\rho(u_1,v)$, $\rho(u,v_1)$ imply that, if $m_1^\pm \in (\mathscr{C}_\chi)^\pm$, then there exist $n_\pm \in \mathbb{N}-1$ such that $r_c^\pm(m^\pm(s)) - r(u_1,v_1) \sim |m^\pm(s) - m_1^\pm|^{n_\pm}$ (if the density is bounded then the argument holds with $n_\pm = 0$, ignoring the bounds explicitly involving $r_c^\pm(m)$). We have $\dot{r}(0) \sim -1$, so from \eqref{pchirA} we have $\partial_m r(\tau^\pm(s),m^\pm(s)) \lesssim |m^\pm(s) - m_1^\pm|^{n_\pm}$, and therefore $|\dot{m}^\pm(s)| \gtrsim |m^\pm(s) - m_1^\pm|^{-n_\pm}$, so $|m^\pm(s) - m_1^\pm| \gtrsim s^{\frac{1}{1+n_\pm}}$, and therefore $r_c^\pm(m^\pm(s)) - r(u_1,v_1) \gtrsim s^{\frac{n_\pm}{1+n_\pm}}$. However, $r(u_1,v_1) - r(s) \sim s$, so we in fact have $\partial_m r(\tau^\pm(s),m^\pm(s)) \sim |m^\pm(s) - m_1^\pm|^{n_\pm}$ and $|m^\pm(s) - m_1^\pm| \sim s^{\frac{1}{1+n_\pm}}$, so $\partial_m r(\tau^\pm(s),m^\pm(s)) \sim s^{\frac{n_\pm}{1+n_\pm}}$. We therefore expand:
			\begin{equation}
				r(s) = r_0 - s \tilde{r}(s), \qquad m^\pm(s) = m_1^\pm \pm s^{\frac{1}{1+n_\pm}} \tilde{m}^\pm(s),
			\end{equation}
			where for any solution we have $\tilde{r}(s), \tilde{m}^\pm(s) \sim 1$ in a neighbourhood of $s=0$. Writing \eqref{epsdot0} in the form
			\begin{equation} \label{epsdot2}
				\dot{\epsilon} = \mathcal{M}^+ \mathcal{E}^+ - \mathcal{M}^- \mathcal{E}^-,
			\end{equation}
			we see that $|\epsilon(s) - \epsilon(0)| \lesssim s^{\frac{1}{1+n_{\epsilon}}}$, where $n_{\epsilon} = \max_{\pm} n_\pm$, so expanding $\epsilon(s) = \epsilon(0) + s^\frac{1}{1+n_{\epsilon}} \tilde{\epsilon}(s)$, we have $|\tilde{\epsilon}(s)| \lesssim 1$. Now, we reparametrise by $s =: t^{n}$ where $n := (1+n_+)(1+n_-)$, and we can expand the ODEs for $r, m^\pm$ in the form:
			\begin{equation}
				\frac{d\tilde{r}}{dt} + \frac{n}{t} (\tilde{r} + \dot{r}(0)) = \mathcal{R}_1(t,\tilde{r},\tilde{m}_1^+,\tilde{m}_1^-,\tilde{\epsilon}),
			\end{equation}
			\begin{equation}
				\frac{d \left[(\tilde{m}^\pm)^{1+n_\pm} \right]}{dt} + \frac{n}{t} ((\tilde{m}^\pm)^{1+n_\pm} - \widetilde{\mathcal{M}}_0^\pm) = \widetilde{\mathcal{M}}_1^\pm(t,\tilde{r},\tilde{m}_1^+,\tilde{m}_1^-,\tilde{\epsilon}), 
			\end{equation}
			where $\widetilde{\mathcal{M}}_0^\pm$ is a constant and $\mathcal{R}_1, \widetilde{\mathcal{M}}_1^\pm$ are smooth functions. Now, expanding the right hand side of \eqref{epsdot2}, we get
			\begin{equation} \label{epsode}
				\frac{d}{ds}( \epsilon - \mathcal{E}^+(0)(m^+ - m_1^+) - \mathcal{E}^-(0)(m_1^- - m^-)) = t( \mathcal{M}^+ \mathcal{E}_1^+(t,\tilde{r},\tilde{m}_1^+,\tilde{m}_1^-,\tilde{\epsilon}) - \mathcal{M}^- \mathcal{E}_1^-(t,\tilde{r},\tilde{m}_1^+,\tilde{m}_1^-,\tilde{\epsilon})),
			\end{equation}
			where $\mathcal{E}_1^\pm$ are smooth functions. If $n_\epsilon = n_+ = n_-$ then we define: $\tilde{\gamma}(s) := \tilde{\epsilon}(s) - \mathcal{E}^+(0) \tilde{m}^+ - \mathcal{E}^-(0) \tilde{m}^-$, otherwise if $n_\epsilon = n_\pm$ then we define $\tilde{\gamma}(s) := \tilde{\epsilon}(s) - \mathcal{E}^\pm(0) \tilde{m}^\pm$, and in either case \eqref{epsode} takes the form $\frac{d\tilde{\mu}}{dt} + (1+n_0)\frac{\tilde{\mu}}{t} = G(t, \tilde{r}, \tilde{m}_1^+, \tilde{m}_1^-, \tilde{\gamma})$, where $n_0 := \min_{\pm} n_\pm$, and $G$ is a smooth function.
			
			We've arrived at an ODE system for $(\tilde{r}, (\tilde{m}^+)^{1+n_-}, (\tilde{m}^-)^{1+n_-}, \tilde{\gamma})$ with regular singular point structure, and $\tilde{r}, (\tilde{m}^+)^{1+n_-}, (\tilde{m}^-)^{1+n_-} \sim 1$, $|\tilde{\gamma}| \lesssim 1$ for any solution of the thin shell ODE. A standard contraction map argument, cf.~Section \ref{contraction} shows that there exists a unique solution with $(\tilde{r}, (\tilde{m}^+)^{1+n_-}, (\tilde{m}^-)^{1+n_-}, \tilde{\gamma})$ bounded, hence there exists a unique $(r(s),m^+(s),m^-(s),\epsilon(s)) \in C^0([0,s_1)) \cap C^2((0,s_1))$ solving the thin shell ODE and satisfying the conditions. 
		\end{proof}
		
		We now show that given the existence of a solution to the thin shell ODE as in Lemma \ref{lem:ode}, there exists a piecewise smooth extension satisfying the propagation condition as stated. This is an almost identical argument to that in Proposition \eqref{equivprop}, indeed, suppose that there exists $(r(s),m^+(s),m^-(s),\epsilon(s)) \in C^0([0,s_1)) \cap C^2((0,s_1))$ solving the thin shell ODE in $\mathscr{Q}_{MDD}^\pm$ with initial data \eqref{extdata1}, \eqref{extdata2}, and let, for $0 < s_2 \leq s_1$
		\begin{equation}
			\Sigma^\pm := \{ (\tau^\pm(s),m^\pm(s)) : s \in [0,s_2) \}.
		\end{equation}
		We can choose $s_2, \delta > 0$ sufficiently small such that we have $\Sigma^- \subset \mathscr{U}_{+}(\mathscr{CH}_\delta^-, \mathscr{Q}_{MDD}^- \cup (\tau_1^-,m_1^-))$, $\Sigma^+ \subset \mathscr{U}_{+}(\mathscr{CH}_\delta^+, \mathscr{Q}_{MDD}^+ \cup (\tau_1^+,m_1^+))$. We solve the eikonal equation with boundary conditions on $\Sigma^\pm$ in the usual way, as in Proposition \ref{prop:tsode-eik}, and then the statement (for the necessary extension of regularity at $s=0$) reduces to showing that we have $\partial_\tau^k u(\tau^+(s),m^+(s)) \to \partial_\tau^k u^+(\tau_1^+,m_1^+)$ as $s \to 0$, and similarly for $v(\tau,m)$ in the interior. Note that, transforming to double null coordinates as in Proposition \ref{prop:tsode-eik} in $u > u_1, v > v_1$, the jump conditions \eqref{prjump}, \eqref{pomegajump} hold for $s \in (0,s_2)$. The choice of initial data, in particular \eqref{extdata2}, implies that the jump conditions are satisfied continuously to $s=0$: e.g.~$[\partial_u r](s) = \frac{2\pi r(s) \sigma(s)}{\dot{u}(s)} \to [\partial_u r]_1$ as $s \to 0$. Using also $[\partial_u \log \Omega^2](s) \to [\partial_u \log \Omega^2]_1$, then we do have $\partial_\tau^k u(\tau^+(s),m^+(s)) \to \partial_\tau^k u^+(\tau_1^+,m_1^+)$, for $k \leq 2$, so the constructed solution is $C^2$ across $\{ u = u_1, v \in (v_1,v_1+\delta)\}$, and the analogous argument in the interior implies that it is $C^2$ across $\{ v = v_1, u \in (u_1,u_1+\delta)\}$. 
		
		In the case that $\rho_1^+ = 0$, then we solve eikonal equation as usual in the interior, with $\Sigma^- \subset \mathscr{U}_{+}(\mathscr{CH}_\delta^-, \mathscr{Q}_{MDD}^- \cup (\tau_1^-,m_1^-))$, and $u(s) := u(\tau^-(s),m^-(s))$. We need to solve for $v(s)$ for the thin shell propagating in the vacuum exterior. We note that the following identity holds:
		\begin{equation} \label{id:vdot}
			2 \dot{v}(s) \partial_v r(u(s),v(s)) = \dot{r}(s) + n^+(r)(s),
		\end{equation}
		and we have, using Raychaudhuri \eqref{Rayu} with $\rho=0$:
		\begin{equation}
			\partial_v r (u(s),v(s)) = - \frac{1}{4} \left( 1 - \frac{2m_1^+}{r(s)} \right) [(\Omega^{-2} \partial_u r)(u_1,v(s))]^{-1}, 
		\end{equation} 
		so that \eqref{id:vdot} implies the following ODE for $v(s)$:
		\begin{equation}
			\dot{v}(s) = - \frac{2(\Omega^{-2} \partial_u r)(u_1,v(s))}{ -\dot{r}(s) + n^+(r)(s)},
		\end{equation}
		where $n^+(r)(s) = \epsilon(s) - \frac{m_1^+ - m^-(s)}{2 \epsilon(s) r(s)}$, and we have $-\dot{r} + n^+(r) > 0$ even when $n^+(r) < 0$. Now, one can see from the proof of Proposition \ref{prop:tsode-eik} that the jump conditions \eqref{prjump}, \eqref{pomegajump} hold for a thin shell with vacuum exterior, and are continuous to $s=0$, so the constructed spacetime is $C^2$ across $\{ u = u_1, v \in (v_1,v_1+\delta)\}$ and $\{ v = v_1, u \in (u_1,u_1+\delta)\}$.
		
		To conclude the proof of existence, we note that by definition of thin shells containing their limit points, the weak formulation of the Einstein equation can be split into: an integral over $\mathcal{R}'$ which is satisfied by assumption that it is a piecewise smooth solution, and then an integral over the extension, which satisfies the weak formulation by construction. We've therefore constructed a piecewise smooth extension satisfying the propagation condition.
		
		Suppose that $(\mathcal{Q},r,\Omega^2,\rho,\mathbf{u}^\mu)$ is extended, as a piecewise smooth solution, to include for some $u_2 > u_1$, $v_2 > v_1$: $[u_0, u_2) \times [v_0, v_2)$. We must have $(u_1,v_1) \in \partial^- \Sigma$ by the propagation condition, and therefore there exists $u_1 < u_3 < u_1 + \delta$, $v_1 < v_3 < v_1 + \delta$ such that the restriction to $[u_1, u_3) \times [v_1,v_3) \subset \mathcal{Q}$ is a piecewise smooth solution as in Definition \ref{pwisesmooth} satisfying the determinism condition, with thin shell given by $u(s), v(s) \in C^1([0,s_0)) \cap C^2((0,s_0))$ for some $s_0>0$. We must have first derivatives in the interior and exterior extending continuously to data (by the propagation condition), so we have from \eqref{prjump}:
		\begin{equation} \label{jump0}
			\dot{u}(0) [\partial_u r]_1 = 2\pi r(u_1,v_1) \sigma(0) = - \dot{v}(0) [\partial_v r]_1,
		\end{equation}
		but we also have $\Omega^2(u_1,v_1) \dot{u}(0) \dot{v}(0) = 1$, so \eqref{initvel} must hold. Any such thin shell $\Sigma$ must satisfy \eqref{extdata2} by definition of $\epsilon$, \eqref{epsdef}. Note that we have $\sigma(0) > 0$ by the assumption \eqref{firstbd}, and then \eqref{eps} implies $\epsilon(0) > 0$, again using \eqref{firstbd}. By \eqref{detcondzero}, and the analogous inequality in the exterior if $\rho_1^+ > 0$: the determinism condition, which holds by assumption for $s \in (0,s_1)$, actually holds with extension to $s=0$. By continuity of $\rho$ for solutions satisfying the propagation condition, we assume, by possibly restricting the domain, that $\rho > 0$ in $D^\pm(\Sigma, [u_1, u_3) \times [v_1,v_3))$ (or vanishes if $\rho_1^+ = 0$). The usual argument using the determinism condition implies that all past-directed fluid flow lines from $\Sigma$, in $D^\pm(\Sigma, [u_1, u_3) \times [v_1,v_3))$, must intersect $\mathcal{R}'$, and in particular $D^-(\Sigma, [u_1, u_3) \times [v_1,v_3)) \subset D_{ac}^+(\underline{\widetilde{C}})$, $D^+(\Sigma, [u_1, u_3) \times [v_1,v_3)) \subset D_{ac}^+(\widetilde{C})$. Extending the respective definitions of comoving coordinates, and using uniqueness of $C^2$ solution of the collapsing Einstein-dust system in comoving coordinates, the exterior (again only if $\rho_1^+ > 0$) and interior respectively embed into $\mathscr{Q}_{MDD}^\pm$. Hence, $(D^\pm(\Sigma, [u_1, u_3) \times [v_1,v_3)), r, \Omega^2, \rho, \mathbf{u}^\mu)$ are collapsing $C^1$ solutions of (E-d). Now, since $\dot{r}(0) = \mathcal{R}_0 < 0$ by \eqref{extdata2}, then $(r(s),m^+(s),m^-(s),\epsilon(s)) \in C^0([0,s_1)) \cap C^2((0,s_1))$ as defined for the thin shell (e.g.~as defined in \eqref{tsodeprop}) must be a solution of the thin shell ODE, with initial data \eqref{extdata1}, \eqref{extdata2}. The uniqueness result of Lemma \ref{lem:ode}, along with uniqueness of solutions of the eikonal equation, implies that the piecewise smooth extension satisying the propagation condition is locally unique.
	\end{proof}
	
	\subsubsection{Statement of Theorem \ref{MGHDprecise}} \label{sec:thm2}
	
	We define the following subset of the singular boundary for any piecewise smooth solution of (E-d):
	\begin{definition}
		For any piecewise smooth solution of (E-d) satisfying the propagation condition, $(\mathcal{Q}, r, \Omega^2, \rho, \mathbf{u}^\mu)$, we define the following subset of the first singularities of $\mathcal{Q}$:
		\begin{align}
			\mathcal{S}_{r>0}(\mathcal{Q},r) := \{ (u,v) \in \R^{1+1}_{u,v}:& \ \exists (u_0,v_0) \in I^-(u,v) \text{ s.t. } \mathcal{R}' := (J^+(u_0,v_0) \cap J^-(u,v)) \setminus\{ (u,v)\} \subset \mathcal{Q}, \\ &\text{ and } \inf_{(u,v) \in \mathcal{R}'} r(u,v) > 0 \}
		\end{align}
	\end{definition}
	
	\begin{note}
		We include the parametrisation by initial data for the functions $\tau_s$, $\tau_c$, and the sets $\mathscr{C}_\chi$ when necessary for clarity: for $(\tilde{m},\tilde{f}) \in \mathfrak{D}_{\chi_b}$, we use $\tau_s(\tilde{m},\tilde{f}), \tau_c(\tilde{m},\tilde{f}), \mathscr{C}_\chi(\tilde{m},\tilde{f})$ to denote the functions and set $\tau_s, \tau_c, \mathscr{C}_\chi$ defined for the maximal dust development $(\mathscr{Q}_{MDD},r,\rho,m,f)(\mathscr{D}_{0,\chi_b},\tilde{m},\tilde{f})$, so that e.g.~$\tau_s(\tilde{m},\tilde{f}) \in C^\infty([0,\chi_b])$, $\tau_c(\tilde{m},\tilde{f}) \in C^\infty(\mathscr{C}_\chi(\tilde{m},\tilde{f}))$. (See Propositions \ref{mddprop} and \ref{prop:mddcentre} for the definitions.)
	\end{note}
	
	We are now in a position to state the precise version of Theorem \ref{MGHD} (as stated in `rough' form in the introduction).\footnote{In the statement of the theorem, we follow the example of Kommemi's boundary decomposition in \cite{K}.}
	\begin{theorem} \label{MGHDprecise}
		There exists a subset $\mathfrak{D}_{\chi_b}^* \subset \mathfrak{D}_{\chi_b}$ that is open and dense in the metric space\footnote{By Proposition \ref{prop:CS}, a Cauchy stability result holds on any compact globally hyperbolic $C^\infty$ development of initial data $(\tilde{m},\tilde{f}) \in \mathfrak{D}_{\chi_b}$ with gauge choice $\Omega_0^2 \in \mathfrak{G}_{\chi_b}(\tilde{m},\tilde{f})$, in the metric $d_k$. } $\metric$ for $k \geq 3$, such that for any $(\tilde{m},\tilde{f}) \in \mathfrak{D}_{\chi_b}^*$ and any $\Omega_0^2 \in \mathfrak{G}_{\chi_b}(\tilde{m},\tilde{f})$, then there exists a unique maximal globally hyperbolic piecewise smooth development (MGHPSD) of $(\mathcal{D}_0,\tilde{m},\tilde{f},\Omega_0^2)$, satisfying the propagation condition, $(\mathcal{Q}_{MGHD}, r, \Omega^2, \rho, \mathbf{u}^\mu)$. The MGHPSD satisfies the property:
		\begin{equation}
			\mathcal{S}_{r>0}(\mathcal{Q}_{MGHD}, r) = \emptyset.
		\end{equation}
		\textbf{Uniqueness:} Given any globally hyperbolic piecewise smooth solution of (E-d) satisfying the propagation condition, attaining the data $(\mathcal{D}_0,\tilde{m},\tilde{f},\Omega_0^2)$, $(\mathcal{Q}, \tilde{r}, \tilde{\Omega}^2, \tilde\rho, \tilde{\mathbf{u}}^\mu)$, then $\mathcal{Q} \subseteq \mathcal{Q}_{MGHD}$, and for all $(u,v) \in \mathcal{Q}$:
		\begin{equation}
			r(u,v) = \tilde{r}(u,v), \qquad \Omega^2(u,v) = \tilde{\Omega}^2(u,v).
		\end{equation}
		\textbf{Boundary decomposition of the `dust cloud':} 
		The set $\{ (u,v) \in \mathcal{P}(\mathcal{Q}_{MGHD}) : \rho(u,v) > 0\}$ is bounded in $\R^{1+1}_{u,v}$, and we define\footnote{Closures are taken with respect to the topology on the ambient $\R^{1+1}_{u,v}$. For the notation $\mathcal{P}(\mathcal{Q})$, see \eqref{def:P}.}:
		\begin{align}
			\mathcal{P}_{d}(\mathcal{Q}_{MGHD}) &:= \overline{\{ (u,v) \in \mathcal{P}(\mathcal{Q}_{MGHD}) : \rho(u,v) > 0\}} \cap \mathcal{P}(\mathcal{Q}_{MGHD}), \\ \mathcal{P}_{d+ts}(\mathcal{Q}_{MGHD}) &:= \overline{\mathcal{P}_d(\mathcal{Q}_{MGHD})} \cap \mathcal{Q}_{MGHD}, 
		\end{align}
		and we also define $\mathcal{B}_d$ to be the `boundary of the dust cloud': the set of $(u,v) \in \mathcal{P}_{d+ts}(\mathcal{Q}_{MGHD})$ for which $\rho(\tilde{u},\tilde{v}) = 0$ for all $(\tilde{u}, \tilde{v}) \in \mathcal{Q}_{MGHD}$ with $\tilde{v} > v$, $\tilde{u} < u$.
		
		The restriction to $\mathcal{P}_{d}(\mathcal{Q}_{MGHD})$ is acoustically globally hyperbolic, with acoustic Cauchy surface $\mathcal{D}_0 \cap \{ v \leq \chi_b\}$. The set $\mathcal{P}_{d+ts}(\mathcal{Q}_{MGHD})$ has boundary $\Gamma \cup (\mathcal{D}_0 \cap \{ v \leq \chi_b \}) \cup \mathcal{B}_d$ in the sense of manifold-with-boundary, and boundary induced by the ambient $\R^{1+1}_{u,v}$: $\partial \mathcal{P}_{d+ts}(\mathcal{Q}_{MGHD}) := \overline{\mathcal{P}_{d+ts}(\mathcal{Q}_{MGHD})} \setminus \mathcal{P}_{d+ts}(\mathcal{Q}_{MGHD})$ which can be decomposed as:
		\begin{equation}
			\partial\mathcal{P}_{d+ts}(\mathcal{Q}_{MGHD}) = \mathcal{O} \cup \mathcal{CH}_d \cup \mathcal{S}_d, 
		\end{equation}
		where:
		\begin{itemize}
			\item $\mathcal{O} =: \{ (v_{\mathcal{O}}, v_{\mathcal{O}})\}$ is the unique future limit point of $\Gamma$ in $\R^{1+1}_{u,v}$, to which $r$ extends continuously to zero.
			\item $\mathcal{CH}_d \subset \{ u = v_{\mathcal{O}}\}$ is a half-open connected (possibly empty) outgoing null segment emanating from (but not including) $\mathcal{O}$, to which $r$ extends to a positive function (except possibly at its future endpoint).
			\item $\mathcal{B}_d$ is a half-open connected (possibly piecewise) $C^1$ timelike curve with future limit point $\partial^+\mathcal{B}_d =: \{ (u_{\mathcal{B}_d, +}, v_{\mathcal{B}_d, +})\} \in \mathcal{O} \cup \mathcal{CH}_d \cup \mathcal{S}_d$, and past endpoint $(-\chi_b,\chi_b) \in \mathcal{D}_0$.
			\item $\mathcal{S}_d \subset \{ u < v_{\mathcal{O}}, v \leq v_{\mathcal{B}_d, +} \}$ is a half-open connected (possibly empty) achronal curve, to which $r$ extends continuously to zero. 
		\end{itemize}
		\textbf{Common intersection/limit points of the components of $\partial\mathcal{P}_{d+ts}(\mathcal{Q}_{MGHD})$:}
		\begin{itemize}
			\item $\partial^+\mathcal{B}_d \in \mathcal{O}$ if and only if $\mathcal{S}_d = \emptyset = \mathcal{CH}_d$.
			\item If $\partial^+\mathcal{B}_d \in \mathcal{CH}_d$ then $\mathcal{S}_d = \emptyset$, and $\partial^+\mathcal{B}_d$ is the future endpoint of $\mathcal{CH}_d$. 
			\item If $\mathcal{S}_d \neq \emptyset$ then $\partial^+ \mathcal{B}_d \in \mathcal{S}_d$ is the (`right') endpoint of $\mathcal{S}_d$. 
			\item If $\mathcal{CH}_d = \emptyset \neq \mathcal{S}_d$, then $\mathcal{O}$ is the (`left') limit point of $\mathcal{S}_d$. 
			\item If $\mathcal{CH}_d \neq \emptyset \neq \mathcal{S}_d$, then the future endpoint of $\mathcal{CH}_d$ is the (`left') limit point of $\mathcal{S}_d$. 
		\end{itemize}
		\textbf{Thin shell formation and future limit points:}
		There are $N_{ts} \in \mathbb{N}-1$ thin shells: $\Sigma_i \subset \mathcal{P}_{d+ts}(\mathcal{Q}_{MGHD})$ for $1 \leq i \leq N_{ts}$, with future limit points/past endpoints $\partial^\pm \Sigma_i$, and we denote $\partial^\pm \Sigma = \cup_{i=1}^n \partial^\pm \Sigma_i$. $N_{ts} = 0$ if and only if $\mathscr{C}_\chi(\tilde{m},\tilde{f}) = \emptyset$. If $N_{ts} \in \mathbb{N}$:
		\begin{itemize}
			\item The set $\partial^- \Sigma \setminus \partial^+ \Sigma$ consists of \underline{non-degenerate} first shell crossing singularities.
			\item $(u,v) \in \partial^+ \Sigma \cap \mathcal{P}_{d+ts}(\mathcal{Q}_{MGHD})$ only if $(u,v) \in \partial^- \Sigma_i$ for a \underline{unique} $1 \leq i \leq N_{ts}$, and either: 
			\begin{itemize}
				\item There is a \underline{thin shell collision} at $(u,v)$: there exist $1 \leq i < j \leq N_{ts}$ such that $\partial^+\Sigma_i = \partial^+ \Sigma_j$,
				\item and/or at least one of:
				\begin{equation}
					\lim_{\mathcal{P}(\mathcal{Q}_{MGHD}) \ni (u,\tilde{v}) \to (u,v)} \rho(u,\tilde{v}) = \infty, \qquad \lim_{\mathcal{P}(\mathcal{Q}_{MGHD}) \ni (\tilde{u},v) \to (u,v)} \rho(\tilde{u},v) = \infty,
				\end{equation}
				corresponding to the formation of a \underline{non-degenerate} first shell crossing singularity on a thin shell, either in the interior or exterior of the (possibly colliding) thin shell(s) respectively, or possibly in both the interior and exterior.
			\end{itemize}
		\end{itemize}
		\textbf{Dichotomy of `thin shells collapsing to the centre' and non-empty Cauchy horizons:}\footnote{This dichotomy only holds because of the particular choice of generic subset $\mathfrak{D}_{\chi_b}^*$, where we exclude, for example, \emph{non-generic} initial data such that $\tau_s(\tilde{m},\tilde{f})''(0) = 0$, of which the Oppenheimer--Snyder spacetime \cite{OS} is an example. See the discussion in Section \ref{gen-data}.} 
		\begin{itemize}
			\item $\mathcal{O} \in \partial^+\Sigma$ if and only if $\mathcal{CH}_d = \emptyset$.
			\item If $\tau_s(\tilde{m},\tilde{f})''(0) < 0$, then $\mathcal{O} \in \partial^+\Sigma$
			and moreover, $\sup_{(u,v) \in \Gamma} \rho(u,v) < \infty$. The set $\{ (\tilde{m},\tilde{f}) : \tau_s''(0) < 0\}$ is non-empty and open in $(\mathfrak{D}_{\chi_b},d_k)$ for $k \geq 3$. If $\mathcal{CH}_d \neq \emptyset$, then $\tau_s(\tilde{m},\tilde{f})''(0) > 0$.
		\end{itemize}
		\textbf{Extension of functions to $\partial\mathcal{P}_{d+ts}(\mathcal{Q}_{MGHD})$:} 
		\begin{itemize}
			\item If $\mathcal{S}_d \neq \emptyset$, then $r^\frac{3}{2} \rho$, $r^{-\frac{1}{2}} \mathbf{u}^\mu, (-r \partial_\mu r), r \Omega^2, m$ extend continuously from $\mathcal{P}_d(\mathcal{Q}_{MGHD})$ to positive functions on $\mathcal{S}_d \setminus \partial^+\Sigma$, the connected components of which are $C^1$ spacelike.
			\item If $\mathcal{CH}_d \neq \emptyset$, then $\mathbf{u}^\mu, \Omega^2$ extend continuously as positive functions, and $r$ extends in $C^1$ (with $\partial_v r > 0$ at $\mathcal{O}$), to $\mathcal{O}$ from $\{ v \leq v_{\mathcal{O}}\}$; $\frac{m}{r}, r$ extend continuously to zero at $\mathcal{O}$ from $\mathcal{P}_d(\mathcal{Q}_{MGHD})$, and
			\begin{equation} \label{pvOmega-blowup}
				\lim_{u \to v_{\mathcal{O}}} |\partial_v \log \Omega^2(u, v_{\mathcal{O}})| = \infty.
			\end{equation}
			There is \underline{finite blueshift} on $\{ v= v_{\mathcal{O}}\}$: 
			\begin{equation} \label{blueshift}
				\int_{-v_{\mathcal{O}}}^{v_{\mathcal{O}}} \frac{\frac{2m(u, v_{\mathcal{O}})}{r(u,v_{\mathcal{O}})}}{1 - \frac{2m(u, v_{\mathcal{O}})}{r(u,v_{\mathcal{O}})}} \frac{\partial_u r(u, v_{\mathcal{O}})}{r(u, v_{\mathcal{O}})} du < \infty. 
			\end{equation}
		\end{itemize}
		\textbf{Boundary decomposition of the Penrose diagram of $\mathcal{Q}_{MGHD}$:} Let $\tilde{v}(v) \in C^\infty(\R)$ be such that $\tilde{v}'(v) > 0$ for all $v\in\R$; $\tilde{v}(v) = v$ for $v \leq v_{\mathcal{B}_d, +} + 1$, and $\tilde{v}([v_{\mathcal{B}_d, +} + 1, \infty))$ is relatively compact in $\R$, then $\widetilde{\mathcal{Q}}_{MGHD} := \{ (u,\tilde{v}(v)) : (u,v) \in \mathcal{Q}_{MGHD}\}$ is a bounded subset of $\R^{1+1}_{u,\tilde{v}}$, with boundary $\Gamma \cup \widetilde{\mathcal{D}_0}$ in the sense of manifold-with-boundary\footnote{$\widetilde{\mathcal{D}}_0 := \{ (u,\tilde{v}(v)) : (u,v) \in \mathcal{D}_0\}$.}, and boundary $\partial \widetilde{\mathcal{Q}}_{MGHD}$ induced by the ambient $\R^{1+1}_{u,\tilde{v}}$, with decomposition:
		\begin{equation}
			\partial \widetilde{\mathcal{Q}}_{MGHD} = \mathcal{O} \cup \mathcal{CH} \cup \mathcal{S} \cup i^\Box \cup \mathcal{I}^+ \cup i^0,
		\end{equation}
		where:
		\begin{itemize}
			\item `Spacelike infinity', $i^0$, is the unique limit point of $\widetilde{\mathcal{D}}_0$ in $\partial \widetilde{\mathcal{Q}}_{MGHD}$, to which $r$ extends continuously to $\infty$.
			\item `Future null infinity', $\mathcal{I}^+$, is a (non-empty) connected open (ingoing) null segment emanating from $i^0$, defined by the set of $(u,v) \in \partial \widetilde{\mathcal{Q}}_{MGHD}$ for which $\lim_{\widetilde{\mathcal{Q}}_{MGHD} \ni (u,\tilde{v}) \to (u,v)} r(u,\tilde{v}) = \infty$.
			\item $i^\Box$ is the unique future limit point of $\mathcal{I}^+$.
			\item $\mathcal{CH}, \mathcal{S}$ satisfy the properties of $\mathcal{CH}_d, \mathcal{S}_d$ (without the restriction of $v \leq v_{\mathcal{B}_d, +}$ for $\mathcal{S}$) as defined in the `boundary decomposition of the dust cloud' part, and $\mathcal{S}_d \subseteq \mathcal{S}$, $\mathcal{CH}_d \subseteq \mathcal{CH}$. 
		\end{itemize}
		\textbf{Vacuum exterior of the dust cloud:} The restriction of $(\widetilde{\mathcal{Q}}_{MGHD},r,\Omega^2,\rho,\mathbf{u}^\mu)$ to
		\begin{equation}
			\widetilde{\mathcal{V}} := \{ (u,\tilde{v}) \in \widetilde{\mathcal{Q}}_{MGHD} : \rho(u,v) = 0\} \subset \mathcal{P}(\widetilde{\mathcal{Q}}_{MGHD}),
		\end{equation}  
		is a classical solution of the spherically symmetric Einstein vacuum equations (with the gauge regularity of a $C^1$ solution of (E-d)), which is $C^2$-isometric to a subset of Schwarzschild of mass $M := m(-\chi_b,\chi_b)$, with boundary $\mathcal{B}_d \cup (\widetilde{\mathcal{D}_0} \cap \{ \tilde{v} > \chi_b \})$ in the sense of manifold-with-boundary, and the boundary induced by the ambient $\R^{1+1}_{u,\tilde{v}}$ has the decomposition:
		\begin{equation}
			\partial \widetilde{\mathcal{V}} = (\mathcal{CH} \setminus \mathcal{CH}_d) \cup (\mathcal{S} \setminus \mathcal{S}_d) \cup i^\Box \cup \mathcal{I}^+ \cup i^0.
		\end{equation}
		\begin{itemize}
			\item The functions $r\Omega^2$, $-r\partial_\mu r$ extend continuously to positive functions on $\mathcal{S} \setminus \partial^+\Sigma$, the connected components of which are $C^1$ spacelike.
			\item $\mathcal{CH}\setminus \mathcal{CH}_d \neq \emptyset$ only if: $\mathcal{CH}_d \neq \emptyset, \mathcal{S}_d = \emptyset$ and $r_{\partial^+\mathcal{B}_d} := \lim_{\mathcal{Q}_{MGHD}\ni(u,v) \to \partial^+\mathcal{B}_d} r(u,v) > 0$. 
			\begin{itemize}
				\item If $\mathcal{CH}\setminus \mathcal{CH}_d = \emptyset$, or $\frac{2M}{r_{\partial^+\mathcal{B}_d}} \geq 1$, then $\mathcal{I}^+$ is complete\footnote{As defined in \cite{C4}.}, and $i^\Box = i^+$ is future timelike infinity.
				\item If $\frac{2M}{r_{\partial^+\mathcal{B}_d}} < 1$, then $\mathcal{I}^+$ is not complete and we write $i^\Box = i^{naked}$: $\mathcal{S} = \emptyset$ and $i^{naked}$ is the future endpoint of $\mathcal{CH}$.
			\end{itemize}
			\item If $\frac{2M}{r_{\partial^+\mathcal{B}_d}} \leq 1$ then $\mathcal{BH} := \widetilde{\mathcal{Q}}_{MGHD} \setminus J^-(\mathcal{I}^+) = \emptyset$, and if $\frac{2M}{r_{\partial^+\mathcal{B}_d}} > 1$ (or $\mathcal{CH}\setminus \mathcal{CH}_d = \emptyset$) then $\mathcal{BH} \neq \emptyset$. 
		\end{itemize} 
		\textbf{Violation of cosmic censorship.}
		The set $\{ (\tilde{m},\tilde{f}) \in \mathfrak{D}_{\chi_b}^* : \mathscr{C}_\chi(\tilde{m},\tilde{f}) = \emptyset \} \subset \{ (\tilde{m},\tilde{f}) \in \mathfrak{D}_{\chi_b}^* : \mathcal{CH} \neq \emptyset \}$ is non-empty and open in $\metric$ for $k\geq3$, and from Christodoulou's \cite{C1}, there exists a non-empty, open subset of $\{ (\tilde{m},\tilde{f}) \in \mathfrak{D}_{\chi_b}^* : \mathscr{C}_\chi(\tilde{m},\tilde{f}) = \emptyset \}$ for which: $\frac{2M}{r_{\partial^+\mathcal{B}_d}} < 1$ and $\mathcal{I}^+$ is not complete.
	\end{theorem}

	\subsection{Proof of Theorem \ref{MGHDprecise}} \label{MGHDproof}

	\subsubsection{Outline of the proof of Theorem \ref{MGHDprecise}} \label{outlinethm2}
	
	Here, we outline the proof of Theorem \ref{MGHDprecise}, which consists of the following Sections \ref{genericsection} - \ref{firstsingrzero}. See also Section \ref{intromaxdev} in the introduction for an outline of the important parts of the proof.
	
	By the extension principle of Section \ref{extp}, for any piecewise smooth solution, with thin shells forming at non-degenerate first shell crossing singularities, we can construct global comoving coordinates (continuous only on the subset $\mathcal{P}(\mathcal{Q}) \subset \mathcal{Q}$), for which $\mathcal{P}(\mathcal{Q})$ embeds into the \emph{maximal dust development}, and the thin shells correspond to solutions of the \emph{thin shell ODE} on the MDD. We therefore need to understand the singular boundary of the MDD of a suitable `generic' subset of initial data, and understand the thin shell ODE on the MDD.
	
	We first show in \ref{genericsection} that there exists a generic subset $\mathfrak{D}_{\chi_b}^* \subset \mathfrak{D}_{\chi_b}$, such that for the maximal dust development of data $(\tilde{m}, \tilde{f}) \in \mathfrak{D}_{\chi_b}^*$: any first shell crossing singularity is non-degenerate, and there can only be finitely many first shell crossing singularities, with a corresponding decomposition of the singular boundary of the MDD (Corollary \ref{structuregendata}). 
	
	In Section \ref{ODEMDD} we consider the maximal solutions of the thin shell ODE on a fixed maximal dust development, for data in $\mathfrak{D}_{\chi_b}^*$. We show that solutions of the thin shell ODE exist until the `interior' or `exterior' meet the singular boundary of the MDD, and show that the interior/exterior cannot have limit points in certain subsets of the shell crossing singular boundary (defined in Corollary \ref{structuregendata}). See Proposition \ref{odemddprop}. 
	
	In Section \ref{uniqueMGHPSD}, we use the results of the previous sections to construct the maximal globally hyperbolic piecewise smooth development for generic initial data. The proof proceeds in steps, where at each step we show that the first singularities of the constructed spacetime with $\inf r > 0$ satisfy the extension principle of Section \ref{extp}, then solve for the unique extension as solutions of the thin shell ODE on the MDD (where we then solve the eikonal equation with suitable boundary conditions to construct the corresponding globally hyperbolic piecewise smooth solution). We extend the thin shells maximally, and show that we can apply the extension principle to the limit points if $r>0$ (and any other first singularities with $\inf r > 0$). In particular, we show that we can always uniquely extend beyond \emph{thin shell collisions}, where in particular we show that the determinism condition is always satisfied for a thin shell formed at collision.
	
	 Finally, in Section \ref{firstsingrzero}, we consider the first singularities of the maximal piecewise smooth development for data in $\mathfrak{D}_{\chi_b}^*$, where we use the embedding of $\mathcal{Q}_{MGHD}$ (after constructing comoving coordinates), into a subset of $\mathscr{Q}_{MDD}$ to understand the structure of the boundary of $\mathcal{P}_+(\mathcal{Q}_{MGHD})$.

	\subsubsection{Generic smooth initial data for (E-d)} \label{genericsection}
	
	See Section \ref{gen-data} in the introduction for a discussion of the results in this section. 
	
	Given a smooth function $F(\chi)$, we define
	\begin{equation}
		\mathcal{N}_2(F)(\chi) := |F'(\chi)| + |F''(\chi)|.
	\end{equation} 
	
	\begin{remark}
		In this section, we will often use underlined quantities to distinguish a different initial data set: $(\underline{\tilde{m}}, \underline{\tilde{f}}) \in \mathfrak{D}_{\chi_b}$. We may then notate:
		\begin{equation}
			\underline{\tau_s} := \tau_s(\underline{\tilde{m}},\underline{\tilde{f}}), \qquad \underline{\tau_c}(\chi) := \tau_c(\underline{\tilde{m}},\underline{\tilde{f}}).
		\end{equation}
	\end{remark}
	
	\begin{proposition} \label{genericdata}
		There exists a \emph{generic} subset of initial data $\mathfrak{D}_{\chi_b}^* \subset \mathfrak{D}_{\chi_b}$, open and dense in the metric space $(\mathfrak{D}_{\chi_b}, d_k)$ for $k \geq 3$, such that for $(\tilde{m},\tilde{f}) \in \mathfrak{D}_{\chi_b}^*$:
		\begin{equation}
			\min_{\chi \in [0,\chi_b]} \mathcal{N}_2(\tau_s(\tilde{m},\tilde{f}))(\chi) > 0,
		\end{equation}
		and, if for the maximal dust development $(\mathscr{Q}_{MDD},r,\rho,m,f)(\mathscr{D}_{0,\chi_b},\tilde{m},\tilde{f})$ we have $\mathscr{C}_{\chi} \neq \emptyset$, then:
		\begin{equation}
			\inf_{\chi \in \mathscr{C}_\chi} \mathcal{N}_2(\tau_c(\tilde{m},\tilde{f}))(\chi) > 0.
		\end{equation}
	\end{proposition}
	
	The importance of Proposition \ref{genericdata} is that for $(\tilde{m},\tilde{f}) \in \mathfrak{D}_{\chi_b}^*$, any extrema of the functions $\tau_s, \tau_c$ are either parabolic minima or maxima. We therefore have the following decomposition of the singular boundary of $\mathscr{Q}_{MDD}$ for generic data:
	\begin{corollary} \label{structuregendata}
		For $(\tilde{m},\tilde{f}) \in \mathfrak{D}_{\chi_b}^*$, with maximal dust development $(\mathscr{Q}_{MDD}, r, \rho, m ,f)(\mathscr{D}_{0,\chi_b},\tilde{m},\tilde{f})$ the singular boundary of $\mathscr{Q}_{MDD}$ has the following decomposition: 
		\begin{itemize}
			\item If $\tau_s'(\chi) \geq 0$ for all $\chi \in [0,\chi_b]$, then $\tau_s''(0) > 0$ and $\mathscr{S}_{\chi} = [0,\chi_b]$, $\mathscr{C}_{\chi} = \emptyset$.
			\item Otherwise, $\mathscr{C}_{\chi} \neq \emptyset$, and there exists $N_c \in 2\mathbb{N}$, and real numbers $0 \leq \chi_{i} < \chi_{i+1} < \chi_b$, for $1 \leq i \leq N_c - 1$, such that:
			\begin{equation} \label{decomp1}
				\mathscr{C}_{\chi} = \bigcup_{i=1}^{N_c/2} (\chi_{2i-1}, \chi_{2i}).
			\end{equation}
			\item When $\mathscr{C}_{\chi} \neq \emptyset$, we can further decompose:
			\begin{equation}
				\mathscr{C}_{\chi} = \mathscr{C}_{\chi}^- \cup \partial^- \mathscr{C}_{\chi} \cup \mathscr{C}_{\chi}^+ \cup \partial^+ \mathscr{C}_{\chi},
			\end{equation}
			where $\mathscr{C}_{\chi}^+$, $\mathscr{C}_{\chi}^-$ are a finite union of open intervals, such that:
			\begin{equation}
				\mathscr{C}_{\chi}^\pm = \{ \chi \in \mathscr{C}_{\chi} : \pm \tau_c'(\chi) > 0 \},
			\end{equation}
			and $\partial^-\mathscr{C}_{\chi}$, $\partial^+\mathscr{C}_{\chi}$ are finite sets that correspond to minima and maxima of $\tau_c(\chi)$ respectively:
			\begin{equation}
				\partial^\pm\mathscr{C}_{\chi} = \{ \chi \in \mathscr{C}_{\chi} : \tau_c'(\chi) = 0, \ \pm \tau_c''(\chi) < 0 \}.
			\end{equation}
			The singular boundary $\mathscr{C}$ then decomposes into:
			\begin{equation}
				\mathscr{C} = \mathscr{C}^- \cup \partial^- \mathscr{C} \cup \mathscr{C}^+ \cup \partial^+ \mathscr{C},
			\end{equation}
			where
			\begin{equation}
				\mathscr{C}^\pm = \{ (\tau_c(\chi),\chi) : \chi \in \mathscr{C}_\chi^\pm \}, \qquad \partial^\pm \mathscr{C} = \{ (\tau_c(\chi),\chi) : \chi \in \partial^\pm \mathscr{C}_\chi \}.
			\end{equation}
		\end{itemize}
	\end{corollary}
	
	\begin{remark}
		The corollary implies that any first shell crossing singularity of a smooth solution of (E-d) with data $(\tilde{m}, \tilde{f}) \in \mathfrak{D}_{\chi_b}^*$, must correspond to a minima of $\tau_c(\chi)$ with $\tau_c''(\chi) > 0$. Any such first singularity is therefore \emph{non-degenerate} and by Corollary \ref{devcor} there exists a unique local extension as a piecewise smooth solution of (E-d).
	\end{remark}

	\begin{remark}
		We say that $f$ is a \emph{Morse function} if $\min_{x \in [a,b]} \mathcal{N}_2(f)(x) > 0$ (and more generally $f : \R^k \to \R$ is Morse if all critical points are non-degenerate, see e.g.~\cite{Diff-Top}).
	\end{remark}
	
	For the proof of Proposition \ref{genericdata}, we need the following basic result of Morse theory. 
	\begin{lemma} \label{lem:dense}
		For $a, b \in \R$, $a < b$, the set $\{ f \in C^\infty([a,b]) : \min_{x \in [a,b]} \mathcal{N}_2(f)(x) > 0 \}$ is dense in $C^\infty([a,b])$, in the metric induced by the $C^k$ norm, for any $k \in \mathbb{N}$ (and therefore in the $C^\infty$ topology). 
		
		The result also holds in the case $a=0$ restricted to even functions. 
	\end{lemma}
	
	Note that the density of smooth even Morse functions in the space of smooth even functions is a special case of \cite[Lemma 4.8]{Wasserman}, for completeness we include a simple proof. 
	
	\begin{proof}
		Density of the Morse functions $\{ f \in C^\infty([a,b]) : \min_{x \in [a,b]} \mathcal{N}_2(f)(x) > 0 \}$ in $C^\infty([a,b])$ follows from the following lemma, which is an immediate consequence of Sard's theorem\footnote{Sard's theorem (see \cite[Section 7]{Diff-Top}): the set of critical values of a smooth map between manifolds has measure zero.}:
		\begin{lemma}[Section 7, \cite{Diff-Top}]
			Let $f$ be a smooth function on an open set $U \subset \R^k$, then for almost all $a \in \R^k$, the function $f_a := f + a \cdot x$ is a Morse function on $U$. 
		\end{lemma}
		The restriction to even functions is almost immediate. We use Whitney's theorem for even functions \cite{W}: for all $f \in C^\infty([0,b])$ there exists $g \in C^\infty([0,b^2])$ such that $f(x) = g(x^2)$. Now, from
		\begin{equation}
			f'(x) = 2x g'(x^2), \qquad f''(x) = 2g'(x^2) + 4x^2 g''(x^2),
		\end{equation}
		we see that (since $f'(x) = 0 = f''(x)$ for some $x>0$ if and only if $g'(x^2) = 0 = g''(x^2)$): $f$ is Morse if and only if $g$ is Morse \emph{and} $g'(0) \neq 0$. Using Sard's theorem, $g_a \in C^\infty([0,b^2])$ is Morse with $g'(0) \neq 0$ for almost all $a \in \R$, so the result follows.
	\end{proof}
	
	\begin{proof}[Proof of Proposition \ref{genericdata}]
		
	We first outline the proof, which is composed of Lemmas \ref{Dsgen}--\ref{genericsc}. We define:
	\begin{definition}
		\begin{equation}
			\mathfrak{D}_{\chi_b, \mathscr{C} = \emptyset} := \{ (\tilde{m},\tilde{f}) \in \mathfrak{D}_{\chi_b} : \mathscr{C}_{\chi} = \emptyset\}, \ \mathfrak{D}_{\chi_b, \mathscr{C} \neq \emptyset} := \{ (\tilde{m},\tilde{f}) \in \mathfrak{D}_{\chi_b} : \mathscr{C}_{\chi} \neq \emptyset\}, 
		\end{equation}
		\begin{equation}
			\mathfrak{D}^*_{\chi_b, s} := \{ (\tilde{m},\tilde{f}) \in \mathfrak{D}_{\chi_b} : \min_{\chi \in [0,\chi_b]} \mathcal{N}_2(\tau_s(\tilde{m},\tilde{f}))(\chi) > 0 \}, \qquad \mathfrak{D}_{\chi_b, \mathscr{C} = \emptyset}^* := \{ (\tilde{m},\tilde{f}) \in \mathfrak{D}_{\chi_b, s}^* : \mathscr{C}_{\chi} = \emptyset \},
		\end{equation}
		\begin{equation}
			\mathfrak{D}_{\chi_b, \mathscr{C} \neq \emptyset}^* := \{ (\tilde{m}, \tilde{f}) \in \mathfrak{D}_{\chi_b, \mathscr{C} \neq \emptyset} \cap \mathfrak{D}_{\chi_b, s}^* : \inf_{\chi \in \mathscr{C}_\chi} \mathcal{N}_2(\tau_c(\tilde{m},\tilde{f}))(\chi) > 0 \}.
		\end{equation}
	\end{definition}
	We then have $\mathfrak{D}_{\chi_b} = \mathfrak{D}_{\chi_b, \mathscr{C} = \emptyset} \cup \mathfrak{D}_{\chi_b, \mathscr{C} \neq \emptyset}$, $\mathfrak{D}_{\chi_b}^* = \mathfrak{D}_{\chi_b, \mathscr{C} = \emptyset}^* \cup \mathfrak{D}_{\chi_b, \mathscr{C} \neq \emptyset}^*$. We show in Lemma \ref{Dsgen} that $\mathfrak{D}^*_{\chi_b, s}$ is open and dense in $\mathfrak{D}_{\chi_b}$ (using Lemma \ref{lem:dense}), in the metric space $\metric$ for $k \geq2$, and $\mathfrak{D}_{\chi_b, \mathscr{C} \neq \emptyset}$ is open for $k \geq 1$. For $(\tilde{m},\tilde{f}) \in \mathfrak{D}_{\chi_b, \mathscr{C} \neq \emptyset} \cap \mathfrak{D}^*_{\chi_b, s}$, we have already the finite decomposition \eqref{decomp1}. We note here that since $\mathfrak{D}_{\chi_b, \mathscr{C} \neq \emptyset}$ is open and $\mathfrak{D}^*_{\chi_b, s}$ is dense for $k \geq 2$, then for $k \geq 3$: $\mathfrak{D}_{\chi_b, \mathscr{C} \neq \emptyset}^*$ is dense in $\mathfrak{D}_{\chi_b, \mathscr{C} \neq \emptyset}$ if and only if it is dense in $\mathfrak{D}_{\chi_b, \mathscr{C} \neq \emptyset} \cap \mathfrak{D}^*_{\chi_b, s}$. (We show this in the last lemma in this section, Lemma \ref{genericsc}). 
	
	Before showing genericity for data that has non-empty shell crossing singular boundary to the MDD, we show in Lemma \ref{lem:dense-nosc} that $\mathfrak{D}_{\chi_b, \mathscr{C} = \emptyset}^*$ is open, and dense in $\mathfrak{D}_{\chi_b, \mathscr{C} = \emptyset}$, in $\metric$ for $k \geq 2$. 
	
	In Lemmas \ref{scprop1}, \ref{scprop2}, we use $\mathcal{N}_2(\tau_s) > 0$ for $(\tilde{m},\tilde{f}) \in \mathfrak{D}_{\chi_b, \mathscr{C} \neq \emptyset} \cap \mathfrak{D}^*_{\chi_b, s}$ to control, for $(\underline{\tilde{m}}, \underline{\tilde{f}})$ in a neighbourhood of $(\tilde{m},\tilde{f})$ (for $k\geq2$), the shell crossing boundary near the endpoints of each open interval of $\underline{\mathscr{C}_\chi}$ (see \eqref{decomp1}). This is essential for the proof, in Lemma \ref{genericsc}, that $\mathfrak{D}_{\chi_b, \mathscr{C} \neq \emptyset}^*$ is open for $k \geq 3$. Finally, we show that $\mathfrak{D}_{\chi_b, \mathscr{C} \neq \emptyset}^*$ is dense in $\mathfrak{D}_{\chi_b, \mathscr{C} \neq \emptyset}$, making essential use of Cauchy stability of ODEs to construct initial data sets with $\mathcal{N}_2(\tau_c) > 0$ (and $\mathcal{N}_2(\tau_s) > 0$), arbitrarily close to any $(\tilde{m},\tilde{f}) \in \mathfrak{D}_{\chi_b, \mathscr{C} \neq \emptyset} \cap \mathfrak{D}_{\chi_b, s}^*$. This will conclude the proof.

	\begin{lemma} \label{Dsgen}
		$\mathfrak{D}^*_{\chi_b, s}$ is an open, dense subset of $\mathfrak{D}_{\chi_b}$ in the metric space $(\mathfrak{D}_{\chi_b},d_k)$ for $k \geq 2$, and $\mathfrak{D}_{\chi_b, \mathscr{C} \neq \emptyset}$ is open in $(\mathfrak{D}_{\chi_b},d_k)$ for $k \geq 1$.
	\end{lemma}
	
	\begin{proof}
		From the expression \eqref{taus}, for any $(\tilde{m},\tilde{f}) \in \mathfrak{D}_{\chi_b}$, $k\in\mathbb{N}$, and any $\varepsilon_s > 0$, there exists $\delta_s(\tilde{m},\tilde{f},\varepsilon_s) > 0$ such that, for all $(\underline{\tilde{m}},\underline{\tilde{f}}) \in B_{\chi_b, k}((\tilde{m},\tilde{f}),\delta_s)$, we have
		\begin{equation} \label{taus-bd}
			\| \tau_s - \underline{\tau_s} \|_{C^k([0,\chi_b])} < \varepsilon_s.
		\end{equation}
		Note that $(\tilde{m},\tilde{f}) \in \mathfrak{D}_{\chi_b, \mathscr{C} \neq \emptyset}$, if and only if there exists $\chi \in (0,\chi_b)$ with $\tau_s'(\chi) < 0$, so by \eqref{taus-bd}: $\mathfrak{D}_{\chi_b, \mathscr{C} \neq \emptyset}$ is open in $(\mathfrak{D}_{\chi_b},d_k)$ for $k \geq 1$.
		
		Suppose that $(\tilde{m},\tilde{f}) \in \mathfrak{D}_{
		\chi_b, s}^*$, and choose $\varepsilon_s := \min_{\chi \in [0, \chi_b]} \mathcal{N}_2(\tau_s(\tilde{m},\tilde{f}))(\chi) > 0$, then by \eqref{taus-bd}, for $k \geq 2$ we have $\min_{0, \chi_b} \mathcal{N}_s(\tau_s(\underline{\tilde{m}},\underline{\tilde{f}}))(\chi) > 0$ and therefore $B_{\chi_b, k}((\tilde{m},\tilde{f}),\delta_s(\tilde{m},\tilde{f},\varepsilon_s)) \subset \mathfrak{D}_{\chi_b, s}^*$, so $\mathfrak{D}_{\chi_b, s}^*$ is open for $k \geq 2$. For any $(\tilde{m},\tilde{f}) \in \mathfrak{D}_{\chi_b}$, we can express \eqref{taus} as:
		\begin{equation}
			\tau_s(\tilde{m},\tilde{f}) = (2\tilde{m})^{-\frac{1}{2}} I(z(\tilde{m},\tilde{f})),
		\end{equation}
		where we've defined $I(x) \in C^\infty((-1,\infty))$:
		\begin{equation}
			I(x) := \int_0^1 \frac{R^\frac{1}{2}}{(1+xR)^\frac{1}{2}} dR, \ z(\tilde{m},\tilde{f}) := \frac{\tilde{f}}{2\tilde{m}}.
		\end{equation}
		Now, suppose that $(\tilde{m},\tilde{f}) \in \mathfrak{D}_{\chi_b} \setminus \mathfrak{D}_{\chi_b, s}^*$. We have $\tau_s'(\chi_b) > 0$, so there exists $0 < \chi_1 < \chi_b$ such that $\delta_1 := \min_{\chi \in [\chi_1, \chi_b]} \tau_s'(\chi) > 0$. We choose also $\chi_2 \in (\chi_1, \chi_b)$. We define the smooth cut-off function $h_{\chi_1, \chi_2} \in C^\infty([\chi_1,\chi_2])$:
		\begin{equation} \label{def:h}
			h_{\chi_1, \chi_2}(\chi) := \left( 1 + \exp[\left[ (\chi-\chi_1)^{-1} - (\chi_2 - \chi)^{-1}\right] \right)^{-1}, 
		\end{equation}
		satisfying $h_{\chi_1, \chi_2}(\chi_1) = 0$, $h_{\chi_1, \chi_2}(\chi_2) = 1$, and $h_{\chi_1, \chi_2}^{(l)}(\chi_1) = 0 = h_{\chi_1, \chi_2}^{(l)}(\chi_2)$ for all $l \in \mathbb{N}$. For any $k \in \mathbb{N}$, $\varepsilon > 0$, by Lemma \ref{lem:dense} there exists $\tilde{\tau}_{s, \varepsilon} \in \mathcal{E}_{\chi_b}$ such that $\min_{\chi \in [0,\chi_b]} \mathcal{N}_2(\tilde{\tau}_{s, \varepsilon})(\chi) > 0$, and $\| \tau_s - \tilde{\tau}_{s, \varepsilon}\|_{C^k([0,\chi_b])} < \varepsilon$. We define:
		\begin{equation}
			\tau_{s, \varepsilon}(\chi) := \begin{cases}
				\tilde{\tau}_{s, \varepsilon}(\chi) &: \chi \in [0,\chi_1], \\ \tau_s(\chi) h_{\chi_1, \chi_2}(\chi) + \tilde{\tau}_{s, \varepsilon}(\chi) (1 - h_{\chi_1, \chi_2}(\chi)) &: \chi \in [\chi_1,\chi_2], \\ \tau_s(\chi) &: \chi \in [\chi_2, \chi_b],
			\end{cases}
		\end{equation}
		so that $\tau_{s, \varepsilon} \in \mathcal{E}_{\chi_b}$, and there exists $C(k) \in (0,\infty)$ depending on $h$ such that $\| \tau_{s, \varepsilon} - \tau_s \|_{C^k([0,\chi_b])} < C(k) \cdot \varepsilon$. We restrict to $\varepsilon \leq C(1)^{-1} \delta_1$ so that $\tau_{s, \varepsilon}'(\chi) > 0$ for $\chi \in [\chi_1, \chi_2]$, and therefore $\min_{\chi \in [0,\chi_b]} \mathcal{N}_2(\tau_{s, \varepsilon})(\chi) > 0$. We then define:
		\begin{equation} 
			\tilde{m}_\varepsilon := \frac{1}{2} I(z(\tilde{m},\tilde{f}))^2 (\tau_{s, \varepsilon})^{-2}, \qquad \tilde{f}_\varepsilon := 2\tilde{m}_\varepsilon \cdot z(\tilde{m},\tilde{f}),
		\end{equation}
		so that $z(\tilde{m}_\varepsilon,\tilde{f}_\varepsilon) = z(\tilde{m},\tilde{f})$, 
		$\tau_s(\tilde{m}_\varepsilon, \tilde{f}_\varepsilon) = \tau_{s, \varepsilon}$, and for all $\chi \in [\chi_2, \chi_b]$: $\tilde{m}_{\varepsilon}(\chi) = \tilde{m}(\chi), \tilde{f}_{\varepsilon}(\chi) = \tilde{f}(\chi)$. For any $\delta > 0$, we choose $\varepsilon$ sufficiently small depending on $\tilde{m}, \tilde{f}, \delta, k$ so that $d_k((\tilde{m},\tilde{f}),(\tilde{m}_{\varepsilon},\tilde{f}_{\varepsilon})) < \delta$. Note that $\tau_{s, \varepsilon} \in \mathcal{E}_{\chi_b}$ implies that $\tilde{m}_{\varepsilon}, \tilde{f}_{\varepsilon} \in \mathcal{E}_{\chi_b}$, and for all $\chi \in [0,\chi_b]$ we have $\tilde{m}_{\varepsilon}(\chi) > 0$, $2\tilde{m}_{\varepsilon}(\chi) + \tilde{f}_{\varepsilon}(\chi) = 2\tilde{m}_{\varepsilon}(\chi) \cdot (1 + z(\tilde{m},\tilde{f})(\chi)) > 0$. Let $\delta_d := \min_{\chi \in [0,\chi_2]} \left\{ \frac{4\pi \rho_0(\chi)}{3+ \chi_b}, \ \frac{1 + f(\chi)}{\chi_b^2} \right\} > 0$, and define 
		\begin{equation}
			\rho_{0, \varepsilon} := \frac{1}{4\pi} ( 3\tilde{m}_{\varepsilon} + \chi \partial_\chi \tilde{m}_{\varepsilon}), \qquad f_\varepsilon := \chi^2 \tilde{f}_{\varepsilon}.
		\end{equation}
		For any $\delta \in (0,\delta_d]$, we have:
		\begin{align}
			\| \rho_0 - \rho_{0, \varepsilon} \|_{C^0([0,\chi_2])} &\leq \frac{3 + \chi_b}{4\pi} \| \tilde{m} - \tilde{m}_{\varepsilon} \|_{C^1([0,\chi_2])} < \min_{\chi \in [0,\chi_2]} \rho_0(\chi), \\ \| f - f_{\varepsilon} \|_{C^0([0,\chi_2])} &\leq \chi_b^2 \| \tilde{f} - \tilde{f}_{\varepsilon} \|_{C^0([0,\chi_2])} < \min_{\chi \in [0,\chi_2]} (1 + f(\chi)),
		\end{align}
		so that $\rho_{0,\varepsilon}(\chi), 1 + f_{\varepsilon}(\chi) > 0$ for all $\chi \in [0,\chi_b]$, and therefore $(\tilde{m}_{\varepsilon}, \tilde{f}_{\varepsilon}) \in \mathfrak{D}_{\chi_b, s}^* \cap B_{\chi_b, k}((\tilde{m},\tilde{f}),\delta)$, so $\mathfrak{D}_{\chi_b, s}^*$ is dense in $\mathfrak{D}_{\chi_b}$ for all $k \in \mathbb{N}$.
	\end{proof}
	
	\begin{lemma} \label{lem:dense-nosc}
		$\mathfrak{D}_{\chi_b, \mathscr{C} = \emptyset}^*$ is open in $\mathfrak{D}_{\chi_b}$ and dense in $\mathfrak{D}_{\chi_b, \mathscr{C} = \emptyset}$, in the metric space $\metric$ for $k \geq 2$.
	\end{lemma}
	
	\begin{proof}
		For $(\tilde{m},\tilde{f}) \in \mathfrak{D}_{\chi_b, \mathscr{C} = \emptyset}^*$, $\mathscr{C}_{\chi} = \emptyset$ implies that $\tau_s'(\chi) \geq 0$ and therefore $\tau_s''(0) > 0$. If $\tau_s'(\tilde{\chi}) = 0$ for any $\tilde{\chi} \in (0,\chi_b)$, then we must have $\tau_s''(\tilde{\chi}) > 0$, but this contradicts $\tau_s'(\chi) \geq 0$ by integrating to $\chi < \tilde{\chi}$ sufficiently close, so we must have $\tau_s'(\chi) > 0$ for all $\chi \in (0,\chi_b]$. By continuity, there exists $\chi_0 \in (0,\chi_b)$ such that $\tau_s''(\chi) > 0$ for all $\chi \in [0,\chi_0]$. Letting $\varepsilon_0 := \frac{1}{2} \min\{ \min_{\chi \in [0,\chi_0]}\tau_s''(\chi), \min_{\chi \in [\chi_0,\chi_b]} \tau_s'(\chi) \} > 0$, there exists $\delta_0 > 0$ such that for all $(\underline{\tilde{m}},\underline{\tilde{f}}) \in B_{\chi_b, 2}((\tilde{m},\tilde{f}),\delta_0)$, we have $\| \tau_s - \underline{\tau_s} \|_{C^2([0,\chi_b])} < \varepsilon_0$, and therefore $\underline{\tau_s}'(\chi) > 0$ for all $\chi \in [\chi_0,\chi_b]$, and $\underline{\tau_s}''(\chi) > 0$ for all $\chi \in [0,\chi_0]$ which implies that $\underline{\tau_s}'(\chi) > 0$ for all $\chi \in (0,\chi_b]$ and $\min_{\chi \in [0,\chi_b]} \mathcal{N}_2(\underline{\tau_s})(\chi)$, so we have $(\underline{\tilde{m}},\underline{\tilde{f}}) \in \mathfrak{D}_{\chi_b, \mathscr{C} = \emptyset}^*$ and so $\mathfrak{D}_{\chi_b, \mathscr{C} = \emptyset}^*$ is open.
		
		For any $k \geq 2$, we now show that $\mathfrak{D}_{\chi_b, \mathscr{C} = \emptyset}^*$ is dense in $\mathfrak{D}_{\chi_b, \mathscr{C} = \emptyset}$. Let $(\tilde{m},\tilde{f}) \in \mathfrak{D}_{\chi_b, \mathscr{C} = \emptyset} \setminus \mathfrak{D}_{\chi_b, \mathscr{C} = \emptyset}^*$, where we have $\tau_s''(0) \geq 0$ and $\tau_s'(\chi) \geq 0$ for all $\chi \in (0,\chi_b]$. For $\varepsilon > 0$ we define: $\tilde{\tau}_{s, \varepsilon} := \tau_s + \frac{1}{2} \varepsilon \chi^2 \in \mathcal{E}_{\chi_b}$, so that $\| \tau_s - \tau_{s, \varepsilon} \|_{C^k([0,\chi_b])} \leq \varepsilon \cdot \chi_b( 1+ \frac{1}{2} \chi_b)$. We have $\tilde{\tau}_{s, \varepsilon}''(0) > 0$, $\tilde{\tau}_{s, \varepsilon}'(\chi) > 0$ for all $\chi \in (0,\chi_b]$, so $\min_{\chi \in [0,\chi_b]} \mathcal{N}_2(\tilde{\tau}_{s, \varepsilon})(\chi) > 0$. Density then follows from the identical construction to that in Lemma \ref{Dsgen} for $\tau_{s, \varepsilon}$ and $\tilde{m}_{\varepsilon}, \tilde{f}_{\varepsilon}$ in terms of $\tilde{\tau}_{s, \varepsilon}$, where we have $\tau_{s, \varepsilon}'(\chi) > 0$ for all $\chi \in (0,\chi_b]$ so that $(\tilde{m}_{\varepsilon},\tilde{f}_{\varepsilon}) \in \mathfrak{D}_{\chi_b, \mathscr{C}=\emptyset}^* \cap B_{\chi_b, k}((\tilde{m},\tilde{f}),\delta)$ for sufficiently small $\delta > 0$. 
	\end{proof}
	We now prove the first of two lemmas controlling the shell crossing singular boundary in an open ball at $(\tilde{m},\tilde{f}) \in \mathfrak{D}_{\chi_b, \mathscr{C}\neq \emptyset} \cap \mathfrak{D}_{\chi_b, s}^*$.
	\begin{lemma} \label{scprop1}
		Given $(\tilde{m},\tilde{f}) \in \mathfrak{D}_{\chi_b, \mathscr{C} \neq \emptyset} \cap \mathfrak{D}_{\chi_b, s}^*$ and any $0 < \chi_* \in \overline{\mathscr{C}_{\chi}} \setminus \mathscr{C}_{\chi}$ so that $\tau_s'(\chi_*) = 0$, $|\tau_s''(\chi_*)| > 0$, then there exist $\varepsilon_c(\tilde{m},\tilde{f}, \chi_*), \delta_c(\tilde{m},\tilde{f}, \chi_*) > 0$ such that, for all $(\underline{\tilde{m}},\underline{\tilde{f}}) \in B_{\chi_b, 2}((\tilde{m},\tilde{f}), \delta_c)$, there exists $\underline{\chi}_*(\underline{\tilde{m}},\underline{\tilde{f}}, \chi_*) \in (\chi_* - \varepsilon_c, \chi_* + \varepsilon_c)$ such that $\underline{\tau_s}'(\underline{\chi_*})=0$. Moreover, if $\tau_s''(\chi_*) > 0$, then $\underline{\tau_s}''(\underline{\chi_*})>0$, 
		\begin{equation} \label{scsingbd1}
			\underline{\mathscr{C}_{\chi}} \cap [\chi_* - \varepsilon_c, \chi_* + \varepsilon_c] = [\chi_* - \varepsilon_c, \underline{\chi_*}),
		\end{equation}
		and $\underline{\tau_c}$ extends to a $C^1$ function on $[\chi_* - \varepsilon_c, \underline{\chi_*}]$, with $\min_{\chi \in [\chi_* - \varepsilon_c, \underline{\chi_*}]} \underline{\tau_c}'(\chi) > 0$. Similarly, if $\tau_s''(\chi_*) < 0$, then $\underline{\tau}_s''(\underline{\chi_*})<0$, 
		\begin{equation} \label{scsingbd2}
			\underline{\mathscr{C}_{\chi}} \cap [\chi_* - \varepsilon_c, \chi_* + \varepsilon_c] = (\underline{\chi_*},  \chi_* + \varepsilon_c],
		\end{equation}
		and $\underline{\tau_c}$ extends to a $C^1$ function on $[\underline{\chi_*},\chi_* + \varepsilon_c]$, with $\min_{\chi \in [\underline{\chi_*},\chi_*+\varepsilon_c]} \underline{\tau_c}'(\chi) < 0$. 
		
		Therefore, $\inf_{\chi \in \underline{\mathscr{C}_\chi} \cap [\chi_* - \varepsilon_c, \chi_* + \varepsilon_c]}\mathcal{N}_2(\underline{\tau_c})(\chi) > 0$.
	\end{lemma}
	
	\begin{proof}
		We give the proof for the case $\tau_s''(\chi_*) > 0$, the proof for the negative case is identical up to the change in sign.
		
		We can choose $\varepsilon_{c, 1} > 0$ so that $\tau_s'(\chi) > 0$ for $\chi \in [\chi_* - \varepsilon_{c, 1}, \chi_*) \cup (\chi_*, \chi_* + \varepsilon_{c, 1}]$, and $\frac{3}{4}\tau_s''(\chi_*) < \tau_s''(\chi) < \frac{5}{4} \tau_s''(\chi_*)$ for all $\chi \in [\chi_* - \varepsilon_{c, 1}, \chi_* + \varepsilon_{c, 1}]$. We then define $\varepsilon_{s, 2} := \frac{1}{2} \min\{ \frac{1}{2}\tau_s''(\chi_*), \tau_s'(\chi_* + \varepsilon_{c, 1}), -\tau_s'(\chi_* - \varepsilon_{c, 1}) \} > 0$, so that for all $(\underline{\tilde{m}},\underline{\tilde{f}}) \in B_{\chi_b, 2}((\tilde{m},\tilde{f}),\delta_s(\tilde{m},\tilde{f},\varepsilon_{s, 2}))$ (using \eqref{taus-bd}), we have:
		\begin{equation}
			\underline{\tau_s}'(\chi_* + \varepsilon_{c, 1}) > \frac{1}{2} \varepsilon_{c, 2}, \ \underline{\tau_s}'(\chi_* - \varepsilon_{c, 1}) < -\frac{1}{2} \varepsilon_{c, 2}, \ \underline{\tau_s}''(\chi) \in \left[\frac{1}{2}\tau_s''(\chi_*), \frac{3}{2}\tau_s''(\chi_*) \right] \text{ for all } \chi \in [\chi_*-\varepsilon_{c,1}, \chi_* + \varepsilon_{c, 1}].
		\end{equation}
		Hence, there exists a unique $\underline{\chi_*} \in (\chi_*-\varepsilon_{c, 1}, \chi_* + \varepsilon_{c, 1})$ such that $\underline{\tau_s}'(\underline{\chi_*}) = 0$, and \eqref{scsingbd1} holds, for any $0 < \varepsilon_c \leq \varepsilon_{c, 1}$. We define, for $\chi \in [\chi_*-\varepsilon_{c,1}, \chi_* + \varepsilon_{c, 1}]$:
		\begin{equation} \label{def:taus1}
			\underline{\tau_{s, 1}}(\chi) := (\chi - \underline{\chi_*})^{-1}\underline{\tau_s}'(\chi) = \int_0^1 \underline{\tau_s}''(\underline{\chi_*} + s(\chi-\underline{\chi_*})) ds,
		\end{equation}
		so that $\underline{\tau_{s,1}} \in C^\infty([\chi_*-\varepsilon_{c,1}, \chi_* + \varepsilon_{c, 1}])$, $\underline{\tau_{s, 1}}(\chi_*) = \underline{\tau_s}''(\chi_*) > 0$, and for all  $\chi \in [\chi_*-\varepsilon_{c, 1}, \chi_* + \varepsilon_{c, 1}]$: $\underline{\tau_{s, 1}}(\chi) \in \left[\frac{1}{2}\tau_s''(\chi_*), \frac{3}{2}\tau_s''(\chi_*) \right]$. We now derive an expression for $\underline{r_c}(\chi)$ which we'll use to bound $\underline{\tau_c}'$. We define:
		\begin{equation} \label{Hnought}
			\underline{H_0}(r,\chi) := \int_0^1 \frac{2\underline{m}'(\chi) + \underline{f}'(\chi)sr}{(2\underline{m}(\chi) + \underline{f}(\chi)sr)^\frac{3}{2}} s^\frac{1}{2} ds,
		\end{equation}
		which satisfies, for $\chi \in \underline{\mathscr{C}_{\chi}}$: $\underline{H_0}(\underline{r_c}(\chi),\chi) > 0$, by \eqref{diff-pos}. Using \eqref{H} with $\underline{H_s}(\chi) := \underline{H}(\underline{\tau_s}(\chi),\chi)$ we have:
		\begin{equation} \label{Hexp}
			\underline{H}(\tau,\chi) = \underline{H_s}(\chi) -  \frac{1}{2} \underline{r}(\tau,\chi)^\frac{3}{2} \left(\frac{2\underline{m}(\chi)}{\chi} + \underline{f}(\chi)\right)^\frac{1}{2} \underline{H_0}(\underline{r}(\tau,\chi),\chi),
		\end{equation}
		and using the definition of $\underline{\tau_c}$, $\underline{H}(\underline{\tau_c}(\chi),\chi) = 1$ with \eqref{Hs}, we get:
		\begin{equation} \label{rc}
			\underline{r_c}(\chi)^\frac{3}{2} = 2(\underline{H_s}(\chi)-1)  \left(\frac{2\underline{m}(\chi)}{\chi} +\underline{f}(\chi)\right)^{-\frac{1}{2}} \left[ \underline{H_0}(\underline{r_c}(\chi),\chi) \right]^{-1} = - 2 \underline{\tau_s}'(\chi) \left[ \underline{H_0}(\underline{r_c}(\chi),\chi) \right]^{-1}.
		\end{equation}
		Following the definition of $\underline{\tau_{s,1}}$, we define, for $\chi \in [\chi_* - \varepsilon_{c, 1}, \underline{\chi_*})$: $\underline{r_{c, 1}}(\chi) := (\chi - \underline{\chi_*})^{-1} \underline{r_c}(\chi)^{\frac{3}{2}} \in C^\infty(\underline{\mathscr{C}})$, which from \eqref{rc} satisfies, for $\chi \in [\chi_* - \varepsilon_c, \underline{\chi_*})$:
		\begin{equation} \label{rc1}
			\underline{r_{c, 1}}(\chi) = \underline{H_1}(  \underline{r_c}(\chi) ,\chi), \qquad \underline{H_1}(r,\chi) := -2\underline{\tau_{s, 1}}(\chi) \left[ \underline{H_0}\left( r, \chi\right) \right]^{-1}.
		\end{equation}
		Now, from \eqref{Hnought}, we have $H_{0, *} := H_0(0,\chi_*) = \frac{4}{3} (2m(\chi_*))^{-\frac{3}{2}}m'(\chi_*) > 0$, and there exist $0 < \delta_c \leq \delta_s(\tilde{m},\tilde{f},\varepsilon_{s, 2})$, $0 < \varepsilon_{c,2} \leq \varepsilon_{c, 1}$, $\delta_r > 0$, depending on $\tilde{m}, \tilde{f}, \chi_*$ such that, for all $(\underline{\tilde{m}},\underline{\tilde{f}}) \in  B_{\chi_b, 2}((\tilde{m},\tilde{f}),\delta_c)$, and all $(r,\chi) \in [0,\delta_r] \times [\chi_* - \varepsilon_{c, 2}, \chi_* + \varepsilon_{c, 2}]$: $\frac{1}{2} H_{0, *} \leq \underline{H_0}(r,\chi) \leq \frac{3}{2} H_{0, *}$, and
		\begin{equation} \label{H1-bd}
			r_{c, -} := - \frac{6\tau_s''(\chi_*)}{H_{0, *}} \leq \underline{H_1}(r,\chi) \leq - \frac{2\tau_s''(\chi_*)}{3H_{0, *}} =: r_{c, +}.
		\end{equation}
		We further restrict to $\varepsilon_{c, 3} := \min\{ \varepsilon_{c, 2}, (- r_{c, -})^{-1} \delta_r^{\frac{3}{2}}\}$. For $\chi \in [\chi_* - \varepsilon_{c, 3}, \underline{\chi_*})$ such that $\underline{r_c}(\chi) \leq \delta_r$, then from \eqref{rc1} and \eqref{H1-bd} we have $\underline{r_{c, 1}}(\chi) \in [r_{c, -}, r_{c, +}]$, and therefore by definition of $\underline{r_{c, 1}}, \varepsilon_{c, 3}$ we have $\underline{r_c}(\chi) < \delta_r$. By continuity (with $\underline{r_c}(\underline{\chi_*})= 0$), we have $\underline{r_c}(\chi) \in (0,\delta_r], \underline{r_{c, 1}}(\chi) \in [r_{c, -}, r_{c, +}]$ for all $\chi \in [\chi_* - \varepsilon_{c, 3}, \underline{\chi_*})$. Taking the limit of \eqref{rc1} for $[\chi_* - \varepsilon_{c, 3}, \underline{\chi_*}) \ni \chi \to \underline{\chi_*}$, we see that $\underline{r_{c,1}}(\chi)$ extends continuously to $\chi = \underline{\chi_*}$, with $\underline{r_{c,1}}(\underline{\chi_*}) = \underline{H_1}(0,\underline{\chi_*}) < 0$. From $\underline{r_c}(\chi) := \underline{r}(\underline{\tau_c}(\chi),\chi)$ we have $\frac{d}{d\chi}\left[(\underline{r_c})^\frac{3}{2}\right] = - (2\underline{m}+ \underline{r_c} \underline{f})^{\frac{1}{2}} \underline{\tau_c}'$, and using the definition of $\underline{r_{c, 1}}$ we can express:
		\begin{equation} \label{taucp}
			\underline{\tau_c}'(\chi) = -\frac{2}{3} (2\underline{m}(\chi) + \underline{f}(\chi) \underline{r_c}(\chi))^{-\frac{1}{2}} \cdot \left( \underline{r_{c, 1}}(\chi) + (\chi - \underline{\chi_*}) \underline{r_{c,1}}'(\chi) \right).
		\end{equation}
		Let $H_{r, +} := \max\{  |\partial_r \underline{H_0}(r,\chi)| : (r,\chi) \in [0,\delta_r] \times [\chi_* - \varepsilon_{c, 3}, \chi_* + \varepsilon_{c, 3}] \}$, which depends on $\tilde{m}, \tilde{f}, \delta_c$, and choose: $\varepsilon_{c, 4} := \min\left\{ \varepsilon_{c, 3}, \frac{1}{-r_{c,-}} \left( \frac{3H_{0, *}}{8H_{r, +}}\right)^\frac{2}{3} \right\}$, so that, for $\chi \in [\chi_* - \varepsilon_{c, *}, \underline{\chi_*})$ we have:
		\begin{equation}
			\underline{H_d}(\underline{r_{c,1}}(\chi), \chi) := \underline{H_0}(\underline{r_c}(\chi)) - \frac{2}{3} (\underline{\chi_*} - \chi)^\frac{2}{3} (-\underline{r_{c, 1}}(\chi))^{-\frac{2}{3}} \partial_r \underline{H_0}(\underline{r_c}(\chi),\chi) \geq \frac{1}{4} H_{0, *}.
		\end{equation}
		Differentiating \eqref{rc1}, we get: 
		\begin{equation} \label{rc1p}
			(\chi - \underline{\chi_*}) \underline{r_{c, 1}}'(\chi) = (\underline{\chi_*} - \chi)^{\frac{2}{3}} \underline{H_2}( \underline{r_{c,1}}(\chi), \chi ), 
		\end{equation}
		where we've defined:
		\begin{equation}
			\underline{H_2}( \underline{r_{c,1}}, \chi ) := \underline{H_d}(\underline{r_{c,1}}(\chi), \chi)^{-1} \left[ \frac{2}{3} (-\underline{r_{c,1}})^\frac{5}{3} \partial_r H_0(\underline{r_c},\chi) +(\underline{\chi_*} - \chi)^\frac{1}{3} (2 \underline{\tau_{s, 1}}'(\chi) + \underline{r_{c,1}} \partial_\chi \underline{H_0}(\underline{r_c},\chi)) \right], 
		\end{equation}
		and we can bound $|\underline{H_2}(\underline{r_{c,1}}, \chi)| \leq H_{2, +}$ for all $\chi \in [\chi_* - \varepsilon_{c, 4}, \underline{\chi_*})$, $\underline{r_{c,1}} \in [r_{c,-},r_{c,+}]$, for $H_{2,+} > 0$ depending only on $\tilde{m}, \tilde{f}, \delta_c$ (where we've used that $\partial_\chi \underline{H_0}$ depends on second derivatives of $m,f$). We therefore have $(\chi - \underline{\chi_*}) \underline{r_{c, 1}}'(\chi)$ extending continuously to zero at $\chi = \underline{\chi_*}$ by \eqref{rc1p}, and we choose $\varepsilon_{c} := \min\{ \varepsilon_{c, 4}, \left( \frac{-r_{c, +}}{2H_{2, +}}  \right)^\frac{3}{2} \}$, so that for all $\chi \in [\chi_* - \varepsilon_{c}, \underline{\chi_*})$: $\underline{r_{c, 1}}(\chi) + (\chi - \underline{\chi_*}) \underline{r_{c,1}}'(\chi) \leq \frac{1}{2} r_{c,+} < 0$, and $\underline{\tau_c}'(\chi)$ extends continuously to $\chi = \underline{\chi_*}$, and $\min_{\chi \in [\chi_* - \varepsilon_{c}, \underline{\chi_*}]} \underline{\tau_c}'(\chi) > 0$.
	\end{proof}
	
	We now show that there is a open ball at $(\tilde{m},\tilde{f}) \in \mathfrak{D}_{\chi_b, \mathscr{C}\neq \emptyset} \cap \mathfrak{D}_{\chi_b, s}^*$, for which we can control the shell crossing boundary in a fixed neighbourhood of $\chi = 0$ when $\inf\{\chi \in \mathscr{C}_\chi\} = 0$.
	\begin{lemma} \label{scprop2}
		Given $(\tilde{m},\tilde{f}) \in \mathfrak{D}_{\chi_b, \mathscr{C}\neq \emptyset} \cap \mathfrak{D}_{\chi_b, s}^*$ such that $\tau_s''(0) < 0$, then there exist $\tilde{\varepsilon}_{c}(\tilde{m},\tilde{f}), \tilde{\delta}_{c}(\tilde{m},\tilde{f}) > 0$ such that, for all $(\underline{\tilde{m}},\underline{\tilde{f}}) \in B_{\chi_b, 2}((\tilde{m},\tilde{f}), \tilde{\delta}_{c})$, we have $(0,\tilde{\varepsilon}_{c}] \subset \underline{\mathscr{C}_{\chi}}$, and $\underline{\tau_c}$ extends to a $C^2$ function on $[0,\tilde{\varepsilon}_{c}]$, with $\underline{\tau_c}'(\chi) > 0$ for $\chi \in (0, \varepsilon_{c, 0}]$ and: $\underline{\tau_c}(0) = \underline{\tau_s}(0), \ \underline{\tau_c}'(0) = 0, \ \underline{\tau_c}''(0) = \frac{7}{3} \underline{\tau_s}''(0) < 0$, so that $\inf_{\chi \in (0,\tilde{\varepsilon}_c]} \mathcal{N}_2(\underline{\tau_c})(\chi) > 0$.
	\end{lemma}
	
	\begin{proof}
		We proceed almost identically to the proof of Lemma \ref{scprop1}, where we now rescale the relevant quantities at the centre. Since $\tau_s''(0) < 0$, there exists $\tilde{\varepsilon}_{c, 1} > 0$ such that $\tau_s''(\chi) \in [\frac{5}{4} \tau_s''(0), \frac{3}{4} \tau_s''(0)]$ for all $\chi \in [0,\tilde{\varepsilon}_{c, 1}]$. For all $(\underline{\tilde{m}},\underline{\tilde{f}}) \in B_{\chi_b, 2}((\tilde{m},\tilde{f}),\delta_s(\tilde{m},\tilde{f},-\frac{1}{4}\tau_s''(0)))$, and all $\chi \in [0,\tilde{\varepsilon}_{c,1}]$ (using \eqref{taus-bd}) we have: $\underline{\tau_s}''(\chi) \in [\frac{3}{2} \tau_s''(0), \frac{1}{2} \tau_s''(0)]$, and therefore for all $0 < \varepsilon \leq \tilde{\varepsilon}_{c, 1}$, $(0, \varepsilon] \subset \underline{\mathscr{C}_\chi}$.
		
		We define, for $\chi \in [0,\tilde{\varepsilon}_{c, 1}]$: $\underline{\tau_{s, 2}}(\chi) := \chi^{-1} \underline{\tau_s}'(\chi) = \int_0^1 \underline{\tau_s}''(s \chi) ds$, so that: $\underline{\tau_{s,2}}(\chi) \in [\frac{3}{2} \tau_s''(0), \frac{1}{2} \tau_s''(0)]$, and we define also $\underline{R_c}(\chi) := \underline{R}(\underline{\tau_c}(\chi),\chi) = \chi^{-1} \underline{r_c}(\chi)$. We define, for $\chi \in (0,\tilde{\varepsilon}_{c, 1}]$: $\underline{R_{c, 2}}(\chi) := \chi^{-2} \underline{R_c}^{\frac{3}{2}}$, so that $\underline{R_{c, 2}} \in C^\infty((0,\tilde{\varepsilon}_{c, 1}])$, and we see from \eqref{rc} that $\underline{R_c}$ satisfies:
		\begin{equation} \label{Rc2}
			\underline{R_{c, 2}}(\chi) = \underline{\widetilde{H}_1}(\underline{R_c}(\chi),\chi), \qquad \underline{\widetilde{H}_1}(R,\chi) := - \frac{2\underline{\tau_{s, 2}}(\chi) }{\underline{\widetilde{H}_0}(R,\chi)},
		\end{equation}
		\begin{equation} \label{tildeH0}
			\underline{\widetilde{H}_0}(R,\chi) := \chi^{5/2} \underline{H_0}( \chi R, \chi) =   \int_0^1 \frac{6\underline{\tilde{m}}(\chi) + 2 \underline{\tilde{f}}(\chi) sR +  \chi ( 2 \underline{\tilde{m}}'(\chi) + \underline{\tilde{f}}'(\chi) sR) }{(2\underline{\tilde{m}}(\chi) + \underline{\tilde{f}}(\chi)s R)^\frac{3}{2}} s^\frac{1}{2} ds,
		\end{equation}
		where $\widetilde{H}_0(0,0) = 2(2\tilde{m}(0))^{-\frac{1}{2}} =: \widetilde{H}_{0,0}$. Now, there exists $0 < \tilde{\delta}_c \leq \delta_s(\tilde{m},\tilde{f},-\frac{1}{4}\tau_s''(0)))$ and $0 < \tilde{\varepsilon}_{c, 2} \leq\tilde{\varepsilon}_{c,1}$, $\delta_R > 0$, depending on $\tilde{m}, \tilde{f}$ such that, for all $(\underline{\tilde{m}},\underline{\tilde{f}}) \in  B_{\chi_b, 2}((\tilde{m},\tilde{f}),\tilde{\delta}_c)$, and all $(R,\chi) \in [0,\delta_R] \times [0, \tilde{\varepsilon}_{c, 2}]$: $\frac{1}{2} \widetilde{H}_{0,0} \leq \underline{\widetilde{H}_{0}}(R,\chi) \leq \frac{3}{2} \widetilde{H}_{0,0}$, and 
		\begin{equation} \label{tildeH1-bd}
			R_{c, -} := -\frac{2\tau_s''(0)}{3\widetilde{H}_{0, 0}} \leq \underline{\widetilde{H_1}}(R,\chi) \leq -\frac{6\tau_s''(0)}{\widetilde{H}_{0, 0}} =: R_{c, +}.
		\end{equation}
		We further restrict to $\tilde{\varepsilon}_{c, 3} := \min\left\{ \tilde{\varepsilon}_{c,2}, \delta_R^{\frac{3}{4}}\left( \frac{-6 \tau_s''(0)}{\widetilde{H}_{0,0}}\right)^\frac{1}{2}\right\}$, so that, by continuity (from Proposition \ref{prop:mddcentre}, we have $\underline{R_c}(\chi) \to 0$ as $\chi \to 0$), we have for all $\chi \in (0,\tilde{\varepsilon}_{c, 3})$: $\underline{R_{c,2}}(\chi) \in [R_{c, -}, R_{c, +}], \underline{R_c}(\chi) \in (0,\delta_R]$. Hence, from \eqref{Rc2} we see that $\underline{R}_{c,2}(\chi)$ extends continuously to $\chi = 0$. We can express $\underline{\tau_c}'(\chi)$ as (equivalent to \eqref{taucp}):
		\begin{align}
			\underline{\tau_c}'(\chi) &= - \frac{2}{3} \chi^{-\frac{3}{2}} (2\underline{\tilde{m}}(\chi) + \underline{\tilde{f}}(\chi) R_c(\chi))^{-\frac{1}{2}} \frac{d}{d\chi} \left( \chi^{\frac{7}{2}} \underline{R_{c, 2}}(\chi) \right) \\ &= - \frac{7\chi}{3} (2\underline{\tilde{m}}(\chi) + \underline{\tilde{f}}(\chi) \underline{R_c}(\chi))^{-\frac{1}{2}}  \left( \underline{R_{c, 2}}(\chi) + \frac{2\chi}{7} \underline{R_{c, 2}}'(\chi) \right). \label{taucp-centre}
		\end{align}
		We now derive a bound for $\underline{R}_{c, 1}'$. Indeed, defining $\widetilde{H}_{1, +} := \max\{ |\partial_r \underline{\widetilde{H}_1}| : (R,\chi) \in [0,\delta_R] \times [0, \tilde{\varepsilon}_{c, 3}]\}$ which depends on $\tilde{m}$, $\tilde{f}$, $\tilde{\delta}_c$, we choose: $\tilde{\varepsilon}_{c, 4} := \min\{ \tilde{\varepsilon}_{c, 3}, \left( \frac{3(R_{c, -})^\frac{1}{3}}{4 \widetilde{H}_{1, +}}\right)^\frac{3}{4} \}$, so that for $\chi \in (0,\tilde{\varepsilon}_{c, 4}]$:
		\begin{equation}
			\underline{\widetilde{H}_d}(\underline{R_{c,2}}(\chi),\chi) := 1 - \frac{2}{3} \chi^{\frac{4}{3}}( \underline{R_{c, 2}})^{-\frac{1}{3}} \partial_r \underline{\widetilde{H}_1}(\underline{R_c}(\chi),\chi) \geq \frac{1}{2}.
		\end{equation}
		Differentiating \eqref{Rc2}, we get:
		\begin{equation}
			\underline{R_{c,2}}'(\chi) = \underline{\widetilde{H}_d}(\underline{R_{c,2}}(\chi),\chi)^{-1} \left[ \frac{4}{3} \chi^\frac{1}{3} \underline{R_{c,2}}(\chi)^\frac{2}{3} \partial_r \underline{\widetilde{H}_1}(\underline{R_c}(\chi),\chi) + \partial_\chi \underline{\widetilde{H}_1}(\underline{R_c}(\chi),\chi)\right],
		\end{equation}
		so there exists $\widetilde{H}_{2, +}$ depending only on $\tilde{m},\tilde{f},\tilde{\delta}_c$, such that $|\underline{R_{c,2}}'(\chi)| \leq \widetilde{H}_{2, +}$ for all $\chi \in (0,\tilde{\varepsilon}_{c, 4}]$. We finally choose $\tilde{\varepsilon}_c := \min\{ \tilde{\varepsilon}_{c, 4}, \frac{7 R_{c, -}}{4 \widetilde{H}_{2, +}}\}$, so that for all $\chi \in (0,\tilde{\varepsilon}_c]$: $\underline{R_{c, 2}}(\chi) + \frac{2\chi}{7} \underline{R_{c, 2}}'(\chi) \geq \frac{1}{2} R_{c, -}$ and therefore $\underline{\tau_c}'(\chi) < 0$, extending continuously to zero at $\chi=0$.
		
		Using then that $\underline{R_{c, 2}}''(\chi) \lesssim \chi^{-\frac{2}{3}}$ (and $\underline{R_c}'(\chi) \lesssim \chi^\frac{1}{3}$), from differentiating \eqref{taucp-centre} we see that we must have, for $\chi \to 0$:
		$\underline{\tau_c}''(\chi) \to -\frac{7}{3} (2\underline{\tilde{m}}(0))^{-\frac{1}{2}} \underline{R_{c,2}}(0) = \frac{7}{3}\underline{\tau_s}''(0)$.
	\end{proof}
	
	Using Lemmas \ref{scprop1} and \ref{scprop2}, we show:
	\begin{lemma} \label{genericsc}
		$\mathfrak{D}^*_{\chi_b, \mathscr{C} \neq \emptyset}$ is open in $\mathfrak{D}_{\chi_b}$ and dense in $\mathfrak{D}_{\chi_b, \mathscr{C} \neq \emptyset} \cap \mathfrak{D}_{\chi_b, s}^*$, in the metric space $\metric$ for $k \geq 3$. 
	\end{lemma}
	\begin{proof}
		For $(\tilde{m},\tilde{f}) \in \mathfrak{D}_{\chi_b, \mathscr{C} \neq \emptyset} \cap \mathfrak{D}_{\chi_b, s}^*$ and $k \geq 2$, the Lemmas \ref{Dsgen}, \ref{scprop1}, \ref{scprop2}, imply that there exists $\delta_1(\tilde{m},\tilde{f},k) > 0$ such that, for all $(\underline{\tilde{m}},\underline{\tilde{f}}) \in B_{\chi_b, k}((\tilde{m},\tilde{f}),\delta_1) \subset \mathfrak{D}_{\chi_b, \mathscr{C} \neq \emptyset} \cap \mathfrak{D}_{\chi_b, s}^*$:
		\begin{itemize}
			\item There exists $N_c(\tilde{m},\tilde{f}) \in 2\mathbb{N}$ such that there are $\frac{1}{2} N_c$ connected components of $\underline{\mathscr{C}_\chi}$: there exist real numbers $0 \leq \underline{\chi_{i}} < \underline{\chi_{i+1}} < \chi_b$, for $1 \leq i \leq N_c - 1$, such that:
			\begin{equation} 
				\underline{\mathscr{C}_{\chi}} = \bigcup_{i=1}^{N_c/2} (\underline{\chi_{2i-1}}, \underline{\chi_{2i}}).
			\end{equation}
			\item There exist $0 < \tilde{\chi}_i < \tilde{\chi}_{i+1} < \chi_b$, $1 \leq i \leq 2N_c - 1$, depending only on $(\tilde{m},\tilde{f})$, such that: $[\tilde{\chi}_{2i-1},\tilde{\chi}_{2i}] \subset (\underline{\chi_{2i-1}}, \underline{\chi_{2i}})$ for all $1 \leq i \leq \frac{1}{2}N_c$, and defining $\mathscr{K}_{\chi} := \bigcup_{i=1}^{\frac{1}{2}N_c} [\tilde{\chi}_{2i-1}, \tilde{\chi}_{2i}] $,  $\underline{\mathscr{N}_{\chi}} := \underline{\mathscr{C}_\chi} \setminus \mathscr{K}_\chi$, we have:
			\begin{equation} \label{endpts}
				\inf_{\chi \in \underline{\mathscr{N}_\chi}} \mathcal{N}_2(\underline{\tau_c})(\chi) > 0,
			\end{equation}
			where moreover, for each $1 \leq i \leq \frac{1}{2} N_c$: $\underline{\tau_c}'(\chi) < 0$ for $\chi \in (\underline{\chi_{2i-1}}, \tilde{\chi}_{2i-1}]$, $\underline{\tau_c}'(\chi) > 0$ for $\chi \in [\tilde{\chi}_{2i}, \underline{\chi_{2i}})$.
		\end{itemize}
		In order to show that $\mathfrak{D}^*_{\chi_b, \mathscr{C} \neq \emptyset}$ is open in $(\mathfrak{D}_{\chi_b}, d_k)$ for $k \geq 3$, it remains to show that there exists $0 < \delta_* \leq \delta_1$ such that, for all $ (\underline{\tilde{m}},\underline{\tilde{f}}) \in B_{\chi_b, k}((\tilde{m},\tilde{f}),\delta_*)$, we have $\min_{\chi \in \mathscr{K}_{\chi}} \mathcal{N}_2(\underline{\tau_c})(\chi) > 0$. Note that since $\mathscr{K}_{\chi} \subset \underline{\mathscr{C}_{\chi}}$, we have $\min_{\chi \in \mathscr{K}_\chi} \underline{r_c}(\chi) > 0$, and (from differentiating $\underline{r_c}(\chi) := \underline{r}(\underline{\tau_c}(\chi),\chi)$) we therefore have $\min_{\chi \in \mathscr{K}_{\chi}} \mathcal{N}_2(\underline{\tau_c})(\chi) > 0$ if and only if $\min_{\chi \in \mathscr{K}_{\chi}} \mathcal{N}_2(\underline{r_c})(\chi) > 0$. Letting $\varepsilon_2 := \min_{\chi \in \mathscr{K}_{\chi}} \mathcal{N}_2(r_c)(\chi) > 0$, it is sufficient to show that we can choose $\delta_* > 0$ sufficiently small such that $\| r_c - \underline{r_c} \|_{C^2(\mathscr{K}_{\chi})} < \varepsilon_2$. 
		
		We now reparametrise $H(\tau,\chi)$ in order to directly bound $r_c(\chi)$. Indeed, we define for all $\chi \in [0,\chi_b]$, $0 < r \leq \chi$: $\underline{G}(r,\chi) := \underline{H}(\underline{T}(r,\chi), \chi)$. We can exploit the global properties of $G(r,\chi)$, i.e.~the properties of $H(\tau,\chi)$ used in the proof of Proposition \ref{mddprop}, in order to bound $\| r_c - \underline{r_c}\|_{C^0(\mathscr{K}_\chi)}$. Indeed, let $\varepsilon_G := \min_{\chi \in \mathscr{K}_\chi} \{ \min\{ r_c(\chi), \chi - r_c(\chi)\} \}> 0$, then for any $0 < \varepsilon < \varepsilon_G$, define:
		\begin{equation}
			\delta_G(\varepsilon) := \frac{1}{2} \min\{ \min_{\chi \in \mathscr{K}_\chi} |G(r_c(\chi) \pm \varepsilon/2,\chi) - 1| \} > 0.
		\end{equation} 
		Letting $\mathscr{K}_0 := \{ (r,\chi) : r \in [\frac{1}{2} \varepsilon_G, \chi], \chi \in \mathscr{K}_\chi \}$, there exist constants $C_k(\tilde{m},\tilde{f},\mathscr{K}_0) > 0$ such that $\| G - \underline{G} \|_{C^k(\mathscr{K}_0)} \leq C_k(\tilde{m},\tilde{f},\mathscr{K}_0) \cdot d_{k+1}((\tilde{m},\tilde{f}),(\underline{\tilde{m}},\underline{\tilde{f}}))$. Then, defining $\delta_2(\varepsilon) := \min\{ \delta_1, C_0^{-1} \delta_G(\varepsilon)\}$, we have that for all $ (\underline{\tilde{m}},\underline{\tilde{f}}) \in B_{\chi_b, 2}((\tilde{m},\tilde{f}),\delta_2(\varepsilon))$, and all $\chi \in \mathscr{K}_\chi$: $\underline{G}(r_c(\chi) + \varepsilon/2,\chi) < 1$, $\underline{G}(r_c(\chi) - \varepsilon/2,\chi) > 1$, therefore $\| r_c - \underline{r_c}\|_{C^0(\mathscr{K}_\chi)} < \varepsilon$. 
		
		Note that $\partial_r G(r,\chi) = - \left( \frac{2m}{r} + f \right)^{-\frac{1}{2}} \partial_\tau H(T(r,\chi),\chi)$, so in particular $\partial_r G(r_c(\chi),\chi) < 0$ for all $\chi \in \mathscr{C}_\chi$ and $\min_{\chi \in \mathscr{K}_\chi} |\partial_r G(r_c(\chi),\chi)| > 0$. Therefore, by differentiating the identity $\underline{G}(\underline{r_c}(\chi),\chi)=1$ implicitly, there exists $0 < \delta_* \leq \delta_1$ depending on $\tilde{m},\tilde{f},\mathscr{K}_0, \varepsilon_2$ such that for all $ (\underline{\tilde{m}},\underline{\tilde{f}}) \in B_{\chi_b, 3}((\tilde{m},\tilde{f}),\delta_*)$: $\| r_c - \underline{r_c} \|_{C^2(\mathscr{K}_{\chi})} < \varepsilon_2$. 
		
		 We now show that $\mathfrak{D}_{\chi_b, \mathscr{C} \neq \emptyset}^*$ is dense in $\mathfrak{D}_{\chi_b, \mathscr{C} \neq \emptyset} \cap \mathfrak{D}_{\chi_b, s}^*$ for all $k \geq 2$. Let $(\tilde{m},\tilde{f}) \in (\mathfrak{D}_{\chi_b, \mathscr{C} \neq \emptyset} \cap \mathfrak{D}_{\chi_b, s}^*) \setminus \mathfrak{D}_{\chi_b, \mathscr{C} \neq \emptyset}^*$. We can write $H(\tau_c(\chi),\chi) = 1$, for $\chi \in \mathscr{C}_\chi$, as:
		 \begin{equation} \label{rc-ode}
		 	\frac{1}{2} f'(\chi) = J( \chi, f(\chi), r_{c}(\chi)) := \frac{ \chi^{\frac{1}{2}} \left( 2m(\chi) + f(\chi) \chi \right)^{-\frac{1}{2}} - m'(\chi)\int_{r_{c}(\chi)}^\chi ( 2m(\chi) + f(\chi) \tilde{r})^{-\frac{3}{2}} \tilde{r}^\frac{1}{2} d\tilde{r} }{ \int_{r_c(\chi)}^\chi ( 2m(\chi) + f(\chi) \tilde{r} )^{-\frac{3}{2}} \tilde{r}^\frac{3}{2} d\tilde{r}}.
		 \end{equation}
		 We'll use Cauchy stability of this ODE to prove the density statement. For $1 \leq i \leq \frac{1}{2} N_c(\tilde{m},\tilde{f})$, we choose $\tilde{\chi}_{2i-1, -} \in (\chi_{2i-1},\tilde{\chi}_{2i-1})$, $\tilde{\chi}_{2i, +} \in (\tilde{\chi}_{2i},\chi_{2i})$, and we define $\widetilde{\mathscr{K}}_{\chi} := \cup_{i=1}^{\frac{1}{2} N_c} [\tilde{\chi}_{2i-1, -}, \tilde{\chi}_{2i, +}]$.
		 
		 By Lemma \ref{lem:dense}, for any $\delta > 0$, $k \geq 2$, and each $1 \leq i \leq \frac{1}{2} N_c$, there exists $\tilde{r}_{c, \delta, i} \in C^\infty([\tilde{\chi}_{2i-1, -}, \tilde{\chi}_{2i, +}])$ such that $\min_{\chi \in [\tilde{\chi}_{2i-1, -}, \tilde{\chi}_{2i, +}]} \mathcal{N}_2(\tilde{r}_{c, \delta, i})(\chi) > 0$, and $\| r_c - r_{c, \delta, i}\|_{C^{k-1}([\tilde{\chi}_{2i-1, -}, \tilde{\chi}_{2i, +}])} < \delta$. We then define $r_{c, \delta, i} \in C^\infty([\tilde{\chi}_{2i-1}, \tilde{\chi}_{2i, +}])$, using \eqref{def:h}, by:
		 \begin{equation}
		 	r_{c, \delta, i} := \begin{cases}
		 		r_c(\chi)(1 - h_{\tilde{\chi}_{2i-1, -}, \tilde{\chi}_{2i-1}}(\chi)) + \tilde{r}_{c, \delta, i}(\chi) h_{\tilde{\chi}_{2i-1, -}, \tilde{\chi}_{2i-1}}(\chi) &: \chi \in [\tilde{\chi}_{2i-1, -}, \tilde{\chi}_{2i-1}], \\ \tilde{r}_{c, \delta, i}(\chi) &: \chi \in [\tilde{\chi}_{2i-1}, \tilde{\chi}_{2i, +}],
		 	\end{cases}
		 \end{equation}
		 so that $\| r_c - r_{c, \delta, i}\|_{C^{k-1}([\tilde{\chi}_{2i-1, -}, \tilde{\chi}_{2i, +}])} < C(k,r_c) \cdot \delta$, and we choose $\delta$ sufficiently small depending on $r_c$ such that $r_{c, \delta, i}(\chi) > \frac{1}{2} \min_{\chi \in \widetilde{K}_\chi} r_c(\chi)$, $r_{c, \delta, i}'(\chi) > 0$ for $\chi \in [\tilde{\chi}_{2i-1, -}, \tilde{\chi}_{2i-1}]$ and therefore
		 \begin{equation} \label{min-N2}
		 	\min_{\chi \in [\tilde{\chi}_{2i-1, -}, \tilde{\chi}_{2i, +}]} \mathcal{N}_2(r_{c, \delta, i})(\chi) > 0.
		 \end{equation}
		 Now, $f(\chi)$ is a solution of the ODE \eqref{rc-ode} that exists (at least) for $\chi \in [\tilde{\chi}_{2i-1, -}, \tilde{\chi}_{2i, +}]$. Using the notation of the statement of Propostition \ref{CS}, we choose $\varepsilon_0, \delta_0 > 0$ sufficiently small such that for $0 < \delta \leq \delta_0$: $\widetilde{J}_{\delta, i} \in C^\infty(K([\tilde{\chi}_{2i-1, -}, \tilde{\chi}_{2i, +}],f,\varepsilon_0))$, where for $(\chi, \xi) \in K([\tilde{\chi}_{2i-1, -}, \tilde{\chi}_{2i, +}],f,\varepsilon_0)$, we've defined:
		 \begin{equation}
		 	\widetilde{J}_{\delta, i}(\chi, \xi) := J(\chi, \xi, r_{c,\delta, i}(\chi)),
		 \end{equation}
		 (i.e.~so that $2m(\chi) + \xi \chi, 2m(\chi) + \xi r_{c,\delta, i}(\chi) > 0$ for all $(\chi,\xi) \in K([\tilde{\chi}_{2i-1, -}, \tilde{\chi}_{2i, +}],f,\varepsilon_0)$). By Proposition \ref{CS}, for any $0 < \varepsilon \leq \varepsilon_0$, we can choose $0 < \delta \leq \delta_0$ sufficiently small such that $f_{\varepsilon, i}(\chi) \in C^\infty([\tilde{\chi}_{2i-1, -}, \tilde{\chi}_{2i, +}])$ is the unique solution of:
		 \begin{equation}
		 	\frac{1}{2} f_{\varepsilon, i}'(\chi) = J( \chi, f_{\varepsilon, i}(\chi), r_{c, \delta, i}(\chi)),
		 \end{equation}
		 with initial data $f_{\varepsilon, i}(\tilde{\chi}_{2i-1, -}) = f(\tilde{\chi}_{2i-1, -})$, and satisfies
		 \begin{equation}
		 	\| f - f_{\varepsilon, i} \|_{C^k([\tilde{\chi}_{2i-1, -}, \tilde{\chi}_{2i, +}])} < \varepsilon,
		 \end{equation}
		 where we moreover require that $1 + f_{\varepsilon, i}(\chi)> 0$ (and $2m(\chi) + \chi f_{\varepsilon, i}(\chi) > 0$ follows by the assumption on $\delta_0, \varepsilon_0$), for all $\chi \in [\tilde{\chi}_{2i-1, -}, \tilde{\chi}_{2i, +}]$. We then define $f_{\varepsilon} \in \mathcal{E}_{\chi_b}$ by, for $\chi \in [0,\chi_b] \setminus \widetilde{\mathscr{K}}_\chi$: $f_{\varepsilon}(\chi) = f(\chi)$, and for each $1 \leq i \leq \frac{1}{2} N_c$, $\chi \in [\tilde{\chi}_{2i-1, -}, \tilde{\chi}_{2i, +}]$:
		 \begin{equation}
		 	f_{\varepsilon}(\chi) := \begin{cases}
		 		f_{\varepsilon, i}(\chi) &: \chi \in [\tilde{\chi}_{2i-1, -}, \tilde{\chi}_{2i}], \\ f_{\varepsilon, i}(\chi)(1 - h_{\tilde{\chi}_{2i}, \tilde{\chi}_{2i, +}}(\chi)) + f(\chi) h_{\tilde{\chi}_{2i}, \tilde{\chi}_{2i, +}}(\chi) &: \chi \in [ \tilde{\chi}_{2i},  \tilde{\chi}_{2i, +}],
		 	\end{cases}
		 \end{equation}
		 so that $1 + f_\varepsilon(\chi), 2m(\chi) + \chi f_{\varepsilon, i}(\chi) > 0$ for all $\chi \in [0,\chi_b]$, and therefore, defining $\tilde{f}_{\varepsilon} := \chi^{-2} f_{\varepsilon} \in \mathcal{E}_{\chi_b}$, we have $(\tilde{m},\tilde{f}_{\varepsilon}) \in \mathfrak{D}_{\chi_b}$, while there exists $\widetilde{C}(f, \tilde{\chi}_1, k)$ such that $\| \tilde{f} - \tilde{f}_{\varepsilon} \|_{C^k([0,\chi_b])} < \widetilde{C} \cdot \varepsilon$, and therefore $(\tilde{m},\tilde{f}_\varepsilon) \in B_{\chi_b, k}((\tilde{m},\tilde{f}), \widetilde{C} \varepsilon)$. By definition of $f_{\varepsilon}$, we have, for each $1 \leq i \leq \frac{1}{2} N_c$, and $\chi \in [\tilde{\chi}_{2i-1, -}, \tilde{\chi}_{2i}]$: $r_c(\tilde{m},\tilde{f}_{\varepsilon})(\chi) = r_{c, \delta, i}(\chi)$. For $0 < \varepsilon \leq \widetilde{C}^{-1} \delta_s(\tilde{m},\tilde{f},\min_{\chi \in [0,\chi_b]} \mathcal{N}_2(\tau_s)(\chi))$, then we have $\mathcal{N}_2(\tau_s(\tilde{m},\tilde{f}_{\varepsilon})) > 0$, while by the argument showing that $\mathfrak{D}_{\chi_b, \mathscr{C} \neq \emptyset}^*$ is open, there exists $\tilde{\varepsilon}_{c, *} > 0$ such that for $0 < \varepsilon \leq \tilde{\varepsilon}_{c, *}$, and each $1 \leq i \leq \frac{1}{2} N_c$: $\| r_c(\tilde{m},\tilde{f}) - r_c(\tilde{m},\tilde{f}_{\varepsilon})\|_{C^1([\tilde{\chi}_{2i}, \tilde{\chi}_{2i, +}])} < \min_{\chi \in [\tilde{\chi}_{2i}, \tilde{\chi}_{2i, +}]} (- r_c'(\chi))$, so that $r_c(\tilde{m},\tilde{f}_{\varepsilon})'(\chi) < 0$ for $\chi \in [\tilde{\chi}_{2i}, \tilde{\chi}_{2i, +}]$. Hence, using also \eqref{min-N2}, we have $\min_{\chi \in \widetilde{\mathscr{K}}_\chi} \mathcal{N}_2(r_c(\tilde{m},\tilde{f}_{\varepsilon})) > 0$ and therefore $(\tilde{m},\tilde{f}_{\varepsilon}) \in \mathfrak{D}_{\chi_b, \mathscr{C} \neq \emptyset}^* \cap B_{\chi_b, k}((\tilde{m},\tilde{f}), \widetilde{C} \cdot \varepsilon)$, so $\mathfrak{D}_{\chi_b, \mathscr{C} \neq \emptyset}^*$ is dense in $\mathfrak{D}_{\chi_b, \mathscr{C} \neq \emptyset} \cap \mathfrak{D}_{\chi_b, s}^*$ for $k \geq 2$.
	\end{proof}
	
	This concludes the proof of Proposition \ref{genericdata}.
	
	\end{proof}
	
	\subsubsection{Maximal solution of the thin shell ODE on generic MDDs} \label{ODEMDD}
	
	By the extension principle for piecewise smooth solutions satisfying the propagation condition (Proposition \ref{prop:pwise-extension}), along with the thin shell development result (Theorem \ref{devthm}, Corollary \ref{devcor}), we need to understand the evolution of the thin shell ODE on generic maximal dust developments. In this section, we show that any solution of the thin shell ODE on a fixed maximal dust development extends to reach the singular boundary in finite time, and we restrict the possible components of the singular boundary that it can reach. See Section \ref{sec:int-ode} for a brief overview of this section in the context of the proof of Theorem \ref{MGHDprecise}.
	
	We first clarify the notation for the thin shell ODE on maximal dust developments including the centre.
	\begin{remark} \label{rem:ODE}
		From now on, we consider the thin shell ODE on the maximal dust development of initial data in $\mathfrak{D}_{\chi_b}$ posed on $\mathscr{D}_{0, \chi_b}$, where the comoving coordinate $\chi$ is gauge fixed. We note that we can still parametrise by $m$ instead of $\chi$: indeed, $m(\chi)$ is invertible on $[0,\chi_b]$, with $m^{-1} \in C^0([0,m(\chi_b)]) \cap C^\infty((0,m(\chi_b)))$. We then `abuse notation' by defining e.g.~$f(m) := f(m^{-1}(m))$, $T(r,m) := T(r,m^{-1}(m))$. 
	\end{remark}
	The results of this section are summarised in the following proposition.
	\begin{proposition} \label{odemddprop}
		Given $(\tilde{m},\tilde{f}) \in \mathfrak{D}_{\chi_b}^*$, with maximal dust development $(\mathscr{Q}_{MDD},r,\rho,m,f)(\mathscr{D}_{0, \chi_b}, \tilde{m},\tilde{f})$, and smooth function $T(r,m)$ as in Remarks \ref{Tremark}, \ref{rem:ODE}, along with $r_0 > 0$, $0 < \chi_{0}^- < \chi_{0}^+ <\chi_b$, $\epsilon_0 > 0$ such that, for $m_0^\pm := m(\chi_0^\pm)$, $\tau_0^\pm := T(r_0,m_0^\pm)$:
		\begin{equation}
			(\tau_0^\pm, \chi_0^\pm) \in \mathscr{Q}_{MDD}, \qquad \mathcal{R}(r_0,m_0^+,m_0^-,\epsilon_0) < 0,
		\end{equation}
		\begin{equation}
			\widetilde{	\mathcal{X}}^+(f; r_0, m^+_{0}, m^-_{0}, \epsilon_0) > 0, \qquad \widetilde{	\mathcal{X}}^-(f; r_0, m^+_{0}, m^-_{0}, \epsilon_0) < 0,
		\end{equation}
		then there exists $s_* > 0$ such that the unique solution, $(r(s),m^+(s),m^-(s),\epsilon(s)) \in C^\infty([0,s_*))$, of the `thin shell ODE':
		\begin{align}
			\dot{r}(s) &= \mathcal{R}(r(s),m^+(s),m^-(s),\epsilon(s)), \label{rdot1} \\ \dot{m}^\pm(s) &= \mathcal{M}^\pm(T, \rho, f; r(s),m^+(s),m^-(s),\epsilon(s)), \label{mpmdot1} \\ \dot{\epsilon}(s) &= \mathcal{E}(T, \rho, f; r(s),m^+(s),m^-(s),\epsilon(s)), \label{epsdot1}
		\end{align}
		with initial data
		\begin{equation}
			r(0) = r_0, \qquad m^\pm(0) = m_0^\pm, \qquad \epsilon(0) = \epsilon_0, \label{data1}
		\end{equation}
		satisfies, for all $s \in [0,s_*)$:
		\begin{equation} \label{maxcond1}
			\dot{r}(s) < 0, \qquad \epsilon(s) > 0, \qquad (\tau^\pm(s),\chi^\pm(s)) \in \mathscr{Q}_{MDD}, \qquad \dot{\chi}^+(s) > 0, \qquad \dot{\chi}^-(s) < 0, 
		\end{equation}
		and $(r(s),m^+(s),m^-(s),\epsilon(s))$ extend continuously to $s = s_*$, with $r(s_*) \geq 0$, $0 \leq m^-(s_*) < m^+(s_*) \leq m(\chi_b)$, $\epsilon(s_*) \geq 0$, such that:
		\begin{itemize}
			\item If $r(s_*) = 0$, then $(\tau^+(s_*), \chi^+(s_*)) \in \mathscr{S} \setminus \overline{\mathscr{C}^+}$, and $(\tau^-(s_*),\chi^-(s_*)) \in \Gamma \cup ( \mathscr{S} \setminus \overline{\mathscr{C}^-} )$.
			\item If $r(s_*) > 0$ and $\chi^+(s_*) < \chi_b$, then (at least) one of $(\tau^\pm(s_*),\chi^\pm(s_*)) \in \mathscr{C}$, with
			\begin{equation} 
				(\tau^+(s_*),\chi^+(s_*)) \notin \mathscr{C}^+ \cup \partial^+\mathscr{C}, \qquad (\tau^-(s_*),\chi^-(s_*)) \notin \mathscr{C}^- \cup \partial^+ \mathscr{C}.
			\end{equation}
			\item If $r(s_*) > 0$ and $\chi^+(s_*) = \chi_b$, then there exists $s_{*, 2} > s_*$ such that there is a unique smooth solution of the thin shell ODE for $(r(s),\chi^-(s), \epsilon(s)) \in C^\infty([0,s_{*,2}))$, with $m^+(s) := m(\chi_b)$ for $s \in [s_*, s_{*, 2})$, and if $r(s_{*, 2}) = 0$ then $(\tau^-(s_{*, 2}),\chi^-(s_{*, 2})) \in \Gamma \cup ( \mathscr{S} \setminus \overline{\mathscr{C}^-} )$; if $r(s_{*, 2}) > 0$ then $(\tau^-(s_{*, 2}),\chi^-(s_{*, 2})) \in \mathscr{C}^+ \cup \partial^- \mathscr{C}$.
		\end{itemize}
	\end{proposition}
	
	The proof of Proposition \ref{odemddprop} consists of Proposition \ref{maximal}, Lemma \ref{lem:detbd}, Lemma \ref{lem:thinshellvacuum} for the case that the thin shell reaches the vacuum exterior and finally Lemma \ref{lem:detbdzero} shows that $(\tau^\pm(s_*), \chi^\pm(s_*)) \notin \overline{\mathscr{C}^\pm}$.
	
	By Picard-Lindel\"of there exists $s_1 > 0$ such that there is a unique solution to \eqref{rdot1}, \eqref{mpmdot1}, \eqref{epsdot1} with initial data \eqref{data1} for $s \in [0,s_1)$, satisfing the conditions by \eqref{maxcond1} by continuity. We now show that the conditions \eqref{maxcond1} remain satisfied with extension to $s= s_1$ as long as $(\tau^\pm(s_1),\chi^\pm(s_1)) \in \mathscr{Q}_{MDD}$.
	\begin{proposition} \label{maximal}
		For any $s_1 > 0$ such that there exists $(r(s),m^+(s),m^-(s),\epsilon(s)) \in C^\infty([0,s_1))$ solving \eqref{rdot1}, \eqref{mpmdot1}, \eqref{epsdot1} with initial data \eqref{data1} and such that the conditions \eqref{maxcond1} hold for all $s \in [0,s_1)$, then: the functions $r(s), m^\pm(s), \epsilon(s)$ extend continuously to $s=s_1$ , with $\epsilon(s_1) \geq 0$, $r(s_1) \geq 0$, and if $r(s_1) > 0$, then: $\dot{r}(s), \dot{\tau}^\pm(s)$ also extend continuously, with $\dot{r}(s_1) < 0$ and
		\begin{align}
			 \widetilde{\mathcal{X}}^+(f;r(s_1),m^+(s_1),m^-(s_1),\epsilon(s_1)) = \dot{r}(s_1) + \left( \frac{2m^+(s_1)}{r} + f(\chi^+(s_1))\right)^\frac{1}{2} \dot{\tau}^+(s_1) &> 0, \label{detplus} \\  \widetilde{\mathcal{X}}^-(f;r(s_1),m^+(s_1),m^-(s_1),\epsilon(s_1)) = \dot{r}(s_1) + \left( \frac{2m^-(s_1)}{r} + f(\chi^-(s_1))\right)^\frac{1}{2} \dot{\tau}^-(s_1) &< 0.
			 \label{detmin}
		\end{align} 
		Hence, there exists $s_* \geq s_1$ such that the solution uniquely extends to $(r(s),m^+(s),m^-(s),\epsilon(s)) \in C^\infty([0,s_*))$, \eqref{maxcond1} holds for all $s\in[0,s_*)$, and (at least) one of $(\tau^\pm(s_*), \chi^\pm(s_*)) \in \partial\mathscr{Q}_{MDD}$.
	\end{proposition}
	
	\begin{proof}
		The bounds $\dot{r}(s) < 0$, $\dot{m}^+(s) > 0$, $m^-(s) < 0$ holding for all $s \in [0,s_1)$, along with $r(s) > 0$, $m^-(s) > 0$, $m^+(s) \leq m(\chi_b)$, imply that the functions $r(s), m^\pm(s)$ extend continuously to $s = s_1$. From \eqref{dsigma}, we have $\frac{d}{ds} (r^2 \sigma)(s) > 0$ for all $s \in [0,s_1)$, and therefore $r^2 \sigma(s)$ extends continuously to $s=s_1$ if and only if $\lim_{s \to s_1} r^2 \sigma(s) < \infty$. Note that \eqref{eps} implies that
		\begin{equation} \label{epsplus}
			\epsilon(s) \leq \frac{m(\chi_b)}{4\pi r^2 \sigma(s)} \leq \frac{m(\chi_b)}{4\pi r^2 \sigma(0)} =: \epsilon_+,
		\end{equation}
		so $\epsilon(s)$ extends continuously to $s=s_1$, where if $\lim_{s \to s_1} r^2 \sigma(s) = \infty$ then $\epsilon(s_1) = 0$. 
		
		We now show that if $r(s_1) > 0$, then $\lim_{s \to s_1} r^2 \sigma(s) < \infty$. By the assumption that the determinism condition is satisfied, we have that \eqref{rdotneg} holds for all $s \in [0,s_1)$ from Lemma \ref{detcondlemma}. Note that for any $s \in [0,s_1)$ such that $n^+(r)(s) \geq - (1 + f(\chi^+(s)))^\frac{1}{2}$,
		\begin{equation} \label{rbd2}
			\dot{r}(s) \geq - \left( \frac{2m^+(s)}{r(s)} + f(\chi^+(s)) \right)^\frac{1}{2}.
		\end{equation}
		Now, define $s_n := \sup\{ s \in [0,s_1) : n^+(r)(s) \geq - (1 + f(\chi^+(s)))^\frac{1}{2} \}$. If $s_n = s_1$, then \eqref{rbd2} implies that $\dot{r}(s_1)^2 \leq \frac{2m^+(s_1)}{r(s_1)} + f(\chi^+(s_1))$. From  \eqref{rdot2}, we can bound for all $s \in [0,s_1)$:
		\begin{equation} \label{sigmabd}
			4\pi \sigma r^2(s) = \frac{\Delta}{\epsilon}(s) < 2r(s) ( 1+ \dot{r}^2(s) )^\frac{1}{2},
		\end{equation}
		and therefore $r^2 \sigma(s)$ extends continuously to $s=s_n$. Suppose then that $s_n < s_1$. For $s \in [s_n,s_1)$, we have $n^+(r)(s) < 0$ so in particular $\frac{2m^+(s)}{r(s)} > 1$ (this follows from \eqref{nrcomoving}). Let $\mu^+_n := \frac{2m^+(s_n)}{r(s_n)} - 1 > 0$, then we have $\frac{d}{ds} \left( \frac{2m^+(s)}{r(s)} \right) > 0$ so $\frac{2m^+(s)}{r(s)} > 1 + \mu_n^+$ for all $s \in [s_n,s_1)$. Note also that $n^+(r)(s) > - |\dot{r}|(s)$ for all $s \in [s_n, s_1)$. 
		
		Let $F_- = \min_{\chi \in [0,\chi_b]} (1 + f(\chi)) > 0$, $F_+ = \max_{\chi \in [0,\chi_b]} (1 + f(\chi))$. Using \eqref{dtau}, for $s \in [s_n, s_1)$ we have:
		\begin{equation}
			\dot{\tau}^+(s) < \frac{|\dot{r}(s)|}{\mu^+_n} \left( \left( \frac{2m(\chi_b)}{r(s_1)} + F_+ - 1 \right)^\frac{1}{2} + (F_+)^\frac{1}{2} \right) =: C^+(\mu_n^+, m(\chi_b), r(s_1), F_+) \cdot |\dot{r}(s)|.
		\end{equation}
		For the bound on $\dot{\tau}^-$, we first define
		\begin{equation} \label{def:Vminus}
			V_-(r(s_1)) := \min_{\chi \in [r(s_1),\chi_b]} \left( \frac{2m(\chi)}{\chi} + f(\chi) \right) > 0, 
		\end{equation}
		so that we have $\frac{2m^-(s)}{r(s)} + f(\chi^-(s)) \geq \frac{2m(\chi^-(s))}{\chi^-(s)} + f(\chi^-(s)) \geq V_-(r(s_1))$. Using \eqref{dtau2}, for $s \in [s_n, s_1)$ we have:
		\begin{equation}
			\dot{\tau}^-(s) < \frac{F_+ + \dot{r}^2(s)}{F_- + (V_-(r(s_1)))^\frac{1}{2} |\dot{r}(s)|} \leq C^-(F_-, F_+, V_-(r(s_1))) (1 + |\dot{r}(s)|).
		\end{equation}
		Using $\Delta(s) = m^+(s) - m^-(s)$, with $\dot{\Delta}(s) > 0$ for all $s$ by the determinism condition, we see from \eqref{dsigma} that:
		\begin{equation} \label{sigmabd1}
			\frac{d}{d\Delta} (4\pi \sigma r^2) = \frac{\dot{m}^+ \dot{\tau}^+ - \dot{m}^- \dot{\tau}^-}{\dot{m}^+ - \dot{m}^-} \leq \max\{ \dot{\tau}^\pm \} \leq  \max\{C^\pm\}(1 + |\dot{r}|).
		\end{equation}
		Now, from \eqref{rdot2} along with \eqref{eps}, \eqref{epsplus}, we get
		\begin{equation}
			\dot{r}^2(s) < \epsilon_+^2 + \frac{2 m(\chi_b)}{r(s_1)} + \frac{(4\pi \sigma r^2)^2}{4 r(s_1)^2},
		\end{equation}
		which we use in \eqref{sigmabd1} to get to
		\begin{equation} \label{sigmabd2}
			\frac{d}{d\Delta} (4\pi \sigma r^2) \leq C (1 + 4\pi \sigma r^2),
		\end{equation}
		where $C$ depends on $C^\pm, \epsilon_+, m(\chi_b), r(s_1)$. Integrating \eqref{sigmabd2} for $\Delta \in [\Delta(s_n), \Delta(s))$, for any $s \in (s_n,s_1)$, we get 
		\begin{equation}
			4\pi \sigma r^2(s) \leq (1 + 4\pi r(s_n)^2 \sigma(s_n)) e^{C(\Delta(s) - \Delta(s_n))} \leq (1 + 4\pi r(s_n)^2 \sigma(s_n)) e^{C m(\chi_b)}, 
		\end{equation}
		and therefore $\lim_{s \to s_1} r^2 \sigma(s) < \infty$. From \eqref{rdot0} we see that $r^2\sigma(s)$ extending continuously to $s_1$ implies that $\dot{r}(s)$ extends continuously, and from \eqref{rdotneg} we must have $\dot{r}(s_1) < 0$. Now, $\dot{\tau}^\pm(s)$ extends continuously to $s = s_1$, with $\dot{\tau}^\pm(s_1) \geq 1$ if and only if
		\begin{equation} \label{Tdenom}
			n^\pm(r)(s_1) (1 + f(\chi^\pm(s_1)))^\frac{1}{2} - \dot{r}(s) \left( \frac{2m^\pm(s)}{r(s)} + f(\chi^\pm(s)) \right)^\frac{1}{2} > 0. 
		\end{equation}
		Now, $n^-(r)(s_1) > 0, \dot{r}(s_1) < 0$, and if $n^+(r)(s_1) \geq 0$ then \eqref{Tdenom} holds. If $n^+(r)(s_1) < 0$, then $\frac{2m^+}{r}(s_1) > 1$, and we have 
		\begin{equation}
			\dot{r}^2 \left( \frac{2m^+(s)}{r(s)} + f(\chi^+(s)) \right) - [n^+(r)]^2 (1 + f^+) = \left( \frac{2m^+}{r} - 1 \right) (1 + f^+ + \dot{r}^2) > 0, 
		\end{equation} 
		so \eqref{Tdenom} also holds in this case. 
	
		We now consider whether the determinism condition holds at $s=s_1$. Independently of whether $(\tau^\pm(s_1),\chi^\pm(s_1)) \in \mathscr{Q}_{MDD}$, we show that \eqref{detplus}, \eqref{detmin} hold. We define
		\begin{equation} \label{thetapm}
			\theta^+(s) := (1+f^+(s))^\frac{1}{2} - n^+(r)(s), \qquad \theta^-(s) := n^-(r)(s) - (1+f^-(s))^\frac{1}{2},
		\end{equation}
		and Lemma \ref{detcondlemma} implies that $\theta^+(s), \theta^-(s) > 0$ for $s \in [0,s_1)$, and we have $\theta^\pm \in C^0([0,s_1])$. We now show explicitly that we have $\widetilde{\mathcal{X}}^\pm(s) \sim \pm \theta^\pm(s)$ for $s \in [0,s_1]$, and so if $\theta^+(s_1), \theta^-(s_1) > 0$, then \eqref{detplus}, \eqref{detmin} hold respectively. Indeed, from \eqref{nsquared} and \eqref{thetapm} we have
		\begin{equation}
			\frac{2m^\pm}{r} + f^\pm - \dot{r}^2 = - \theta^\pm \left[ \theta^\pm \mp 2(1+f(\chi^\pm))^\frac{1}{2}\right].
		\end{equation}
		If $n^+(r)(s_1) \leq 0$, then \eqref{detplus} holds, by \eqref{rpdchi1}, so we may assume that $n^+(r)(s_1) > 0$, which from \eqref{thetapm} is equivalent to $\theta^+(s) < (1+f^+(s))^\frac{1}{2}$. In this case, let $s_+ \in (0,s_1)$ such that $n^+(r)(s) > 0$ for all $s \in [s_+,s_1]$, while we set $s_- := 0$.
		
		Substituting \eqref{thetapm} into \eqref{def:X}, we have for $s \in [s_\pm,s_1]$:
		\begin{equation}
			\widetilde{\mathcal{X}}^\pm(f;r(s),m^+(s),m^-(s),\epsilon(s)) := \pm \theta^\pm(s) \widetilde{\mathcal{X}}_0^\pm(s),
		\end{equation}
		where
		\begin{equation}
			 \widetilde{\mathcal{X}}_0^\pm(s) := \frac{(1+f^\pm(s))^\frac{1}{2} \left(2(1+f^\pm(s))^\frac{1}{2}  \mp \theta^\pm(s) \right) }{n^\pm(r)(s)(\frac{2m^\pm(s)}{r(s)} + f^\pm(s))^\frac{1}{2} - \dot{r}(s)(1+f^\pm(s))^\frac{1}{2}},
		\end{equation}
		which satisfies $\widetilde{\mathcal{X}}_0^\pm \in C^0([0,s_1])$ with $\widetilde{\mathcal{X}}_0^\pm(s_1) > 0$. We now derive an ODE for $\theta^\pm(s)$ which we'll use to show that $\theta^\pm(s_1) > 0$. Differentiating the definition of $\theta^\pm(s)$, \eqref{thetapm}:
		\begin{equation} \label{dtheta1}
			\dot{\theta}^\pm(s) = \mp \frac{d}{ds} (n^\pm(r)(s)) \pm \frac{f'(\chi^\pm(s)) \dot{\chi}^\pm(s)}{(1+f(\chi^\pm(s)))^\frac{1}{2}},
		\end{equation}
		and differentiating $n^\pm(r) = \epsilon \mp 2\pi\sigma r$ using \eqref{dsigma} and \eqref{doteps} we have
		\begin{equation} \label{dtheta2}
			\frac{d}{ds} n^\pm(r) = \frac{\dot{m}^+}{4\pi\sigma r^2} \left( 1 - \frac{\dot{\tau}^+ n^\mp(r)}{(1+f^+)^\frac{1}{2}} \right) - \frac{\dot{m}^-}{4\pi\sigma r^2} \left( 1 - \frac{\dot{\tau}^- n^\mp(r)}{(1+f^-)^\frac{1}{2}} \right) \pm 2\pi\sigma\dot{r} =: \mathcal{N}^\pm.
		\end{equation}
		Rearranging \eqref{dtau2}, we have
		\begin{equation}
			1 - \frac{\dot{\tau}^\pm(s)\cdot n^\pm(r)(s)}{(1+f^\pm(s))^\frac{1}{2}} =  \pm \frac{(-\dot{r}(s))\widetilde{\mathcal{X}}_0^\pm(s) \theta^\pm(s)}{1+f(s)},
		\end{equation}
		so that
		\begin{equation}
			\mathcal{N}^\pm(s) = \pm \frac{\dot{m}^\pm(s)}{4\pi\sigma r^2(s)} \left( 1 - \frac{\dot{\tau}^\pm(s) n^\mp(r)(s)}{(1+f^\pm(s))^\frac{1}{2}} \right) \pm \dot{r}(s) \mathcal{N}_0^\pm(s),
		\end{equation}
		where 
		\begin{equation} \label{N0bd}
			\mathcal{N}_0^\pm(s) := 2\pi \sigma(s) +  \frac{\mp \dot{m}^\mp(s) \widetilde{\mathcal{X}}_0^\mp(s) \theta^\mp(s) }{4\pi\sigma r^2(s)(1+f^\mp(s))} \geq 2\pi\sigma(0),
		\end{equation}
		for all $s \in [s_\pm,s_1]$. We now manipulate the other bracketed term in \eqref{dtheta2}:
		\begin{align}
			1 - \frac{\dot{\tau}^\pm n^\mp(r)}{(1+f^\pm)^\frac{1}{2}} &= (1+f^\pm)^{-\frac{1}{2}} \left( ((n^\pm(r) - \dot{\tau}^\pm n^\mp(r)) \pm \theta^\pm \right) = (1+f^\pm)^{-\frac{1}{2}} \left( \mp 4\pi\sigma r + n^\mp(r)(1 - \dot{\tau}^\pm) \pm \theta^\pm \right) \\ &= (1+f^\pm)^{-\frac{1}{2}} \left( \mp 4\pi\sigma r \pm \theta^\pm \mathcal{N}_1^\pm \right), \label{dtheta3}
		\end{align}
		where
		\begin{equation} \label{N1bd}
			\mathcal{N}_1^\pm(s) = 1 \mp n(r)^\mp(s) \frac{\dot{\tau}^\pm(s) - 1}{\theta^\pm(s)} = 1 \mp \frac{n(r)^\mp(s)}{1+f(\chi^\pm(s))} \frac{\theta^\pm(s) (\widetilde{\mathcal{X}}^\pm_0(s))^2 }{ 1+\dot{\tau}^\pm(s)} \leq 1 + C_1^\pm \cdot \theta^\pm(s),
		\end{equation}
		and $C_1^\pm := \max_{s \in [s_\pm,s_1]}\left|\frac{n(r)^\mp(s) (\widetilde{\mathcal{X}}^\pm_0(s))^2 }{(1+f(\chi^\pm(s)))(1+\dot{\tau}^\pm(s))}\right|$. Combining \eqref{dtheta1}, \eqref{dtheta2}, \eqref{dtheta3}, we have for $s \in [s_\pm,s_1)$:
		\begin{align} \label{dthetads}
			\dot{\theta}^\pm(s) = \mp \mathcal{N}^\pm(s) \pm \frac{f'(\chi^\pm(s)) \dot{\chi}^\pm(s)}{2(1+f^\pm(s))^\frac{1}{2}} \geq 2\pi \sigma(0) (V_-(r(s_1)))^\frac{1}{2} + \frac{ \theta^\pm(s) \widetilde{\mathcal{X}}^\pm_0(s)}{2(1+f^\pm(s))^\frac{1}{2}} \Theta^\pm(s),
		\end{align}
		where we've used \eqref{def:Vminus}, to bound $- \dot{r}(s) \geq (V_-(r(s_1)))^\frac{1}{2}$ for all $s \in [0,s_1]$, \eqref{N0bd}, and we've defined for $s \in [s_\pm,s_1)$:
		\begin{equation}
			\Theta^\pm(s) := \frac{1}{\partial_\chi r(\tau^\pm(s),\chi^\pm(s))} \left( \left(\frac{2m'(\chi^\pm(s))}{r(s)} + f'(\chi^\pm(s))\right) - \frac{ m'(\chi^\pm(s))}{2\pi\sigma r^2(s)} \theta^\pm(s) \mathcal{N}_1^\pm(s)\right).
		\end{equation}
		We need a lower bound for $\Theta^\pm(s)$ in order to obtain a lower bound for $\dot{\theta}^\pm(s)$. From \eqref{diff-pos}, we have
		\begin{equation} \label{Cminus}
			C_{-}(r(s_1)) := \min \left\{ \frac{2m'(\chi)}{r_c(\chi)} + f'(\chi) : \chi \in \mathscr{C}_{\chi}, r_c(\chi) \geq r(s_1) \right\} > 0,
		\end{equation}
		and so we can choose $\delta^\pm > 0$ such that if $\theta^\pm(s) \leq \delta^\pm$, then $\frac{ m'(\chi^\pm(s))}{2\pi\sigma r^2(s)} \theta^\pm(s) \mathcal{N}_1^\pm(s) \leq \frac{1}{2} C_{-}(r(s_1))$, using \eqref{N1bd}, and then
		\begin{equation} \label{Thetabd}
			\Theta^\pm(s) \geq \Theta_0(\tau^\pm(s),\chi^\pm(s)) := \frac{1}{\partial_\chi r(\tau^\pm(s),\chi^\pm(s))} \left( \left(\frac{2m'(\chi^\pm(s))}{r(s)} + f'(\chi^\pm(s))\right) - \frac{1}{2} C_{-}(r(s_1)) \right),
		\end{equation}
		where by \eqref{Cminus}, we have:
		\begin{equation}
			\Theta_{0, -} := \inf_{\chi \in [0,\chi_b], \tau \in [0,\tau_c(\chi))} \Theta_0(\tau,\chi) > - \infty.
		\end{equation}
		Now, if $\inf_{s \in [0,s_1)} \theta^\pm(s) < \delta^\pm$, then let $\tilde{s}^\pm \in [s_\pm, s_1)$ such that $\theta^\pm(s) < \delta^\pm$ for all $s \in [\tilde{s}^\pm,s_1)$. Then \eqref{Thetabd} holds for all $s \in [\tilde{s}^\pm,s_1)$. If $\Theta_{0, -} \geq 0$ then we have $\dot{\theta}^\pm(s) > 0$ and we're done, otherwise we bound \eqref{dthetads} for $s \in [\tilde{s}^\pm,s_1)$:
		\begin{equation} \label{dthetabd}
			\dot{\theta}^\pm(s) \geq c - b^\pm \theta^\pm(s),
		\end{equation}
		where $c := 2\pi \sigma(0) (V_-(r(s_1)))^\frac{1}{2} > 0$, and $b^\pm := (-\Theta_{0, -}) \cdot \max_{s \in [\tilde{s}^\pm,s_1)} \frac{ \widetilde{\mathcal{X}}^\pm_0(s)}{2(1+f^\pm(s))^\frac{1}{2}} > 0$. We now choose $0 < a^\pm \leq c$ such that $b^\pm \theta^\pm(\tilde{s}^\pm) > a^\pm$, and we integrate $\dot{\theta}^\pm(s) \geq b^\pm ( \frac{a^\pm}{b^\pm} - \theta^\pm(s))$ over $[\tilde{s}^\pm, s]$ for $s \in [\tilde{s}^\pm, s_1)$ to see that:
		\begin{equation}
			\theta^\pm(s) \geq \frac{a^\pm}{b^\pm} + e^{- b^\pm (s - \tilde{s}^\pm)} \left(\theta(\tilde{s}^\pm) - \frac{a^\pm}{b^\pm} \right) \geq \frac{a^\pm}{b^\pm},
		\end{equation}
		for all $s \in [\tilde{s}^\pm, s_1)$ and therefore $\theta^\pm(s) \geq \frac{a^\pm}{b^\pm} > 0$.
		
		We conclude by noting that if $(\tau^\pm(s_1),\chi^\pm(s_1)) \in \mathscr{Q}_{MDD}$, then $r(s), m^\pm(s), \epsilon(s)$ extend in $C^\infty$ to $s = s_1$, and we have $\dot{r}(s_1) < 0$, $\epsilon(s_1) > 0$, and $\dot{m}^+(s_1) > 0$, $\dot{m}^-(s_1)$, so we can uniquely extend the solution of  \eqref{rdot1}, \eqref{mpmdot1}, \eqref{epsdot1}, satisfying the conditions \eqref{maxcond1}, to $[0,s_2)$ for some $s_2 > s_1$. One can see from \eqref{rdotneg} that for any $r_- > 0$, if $r(s) > r_-$ then there exists $s_3 < \infty$ such that either one of $(\tau^\pm(s_3),\chi^\pm(s_3)) \notin \mathscr{Q}_{MDD}$, or $r(s_3) = r_-$. For $r_- > 0$ sufficiently small, we have from \eqref{def:R} and \eqref{epsplus} that $\dot{r}(s) \lesssim - r^{-1}$, so the solution must reach $\partial \mathscr{Q}_{MDD}$ in finite time.
	\end{proof}
	
	We now use \eqref{detmin}, \eqref{detplus} in particular to show that $(\tau^+(s_*),\chi^+(s_*)) \notin \mathscr{C}^+ \cup \partial^+ \mathscr{C}$, and $(\tau^-(s_*),\chi^-(s_*)) \notin \mathscr{C}^- \cup \partial^+ \mathscr{C}$.
	
	\begin{lemma} \label{lem:detbd}
		Let $(r(s), m^+(s), m^-(s), \epsilon(s)) \in C^\infty([0,s_1))$ as in the statement of Proposition \ref{maximal}. Suppose that there exists $s_2 \in [0,s_1)$ such that $[\chi^+(s_2), \chi^+(s_1)] \subset \mathscr{C}_{\chi}$, (so that $r(s_1) > 0$), and $\tau_c'(\chi) \geq 0$ for $\chi \in [\chi^+(s_2), \chi^+(s_1)]$. Then the following bound holds:
		\begin{equation} \label{chiplusbd}
			\tau_c(\chi^+(s_1)) - \tau^+(s_1) \geq (\tau_c(\chi^+(s_2))-\tau^+(s_2))e^{-C_+(\chi^+(s_1)-\chi^+(s_2))},
		\end{equation}
		for $C_+>0$, so that $(\tau^+(s_1), \chi^+(s_1)) \in \mathscr{Q}_{MDD}$, and therefore $(\tau^+(s_*),\chi^+(s_*)) \notin \mathscr{C}^+ \cup \partial^+ \mathscr{C}$.
		
		Suppose that $\tau_c'(\chi) \leq 0$ for $\chi \in [\chi^-(s_1), \chi^-(s_2)] \subset \mathscr{C}_\chi$, then 
		\begin{equation} \label{chiminusbd}
		\tau_c(\chi^-(s_1)) - \tau^-(s_1) \geq (\tau_c(\chi^-(s_2))-\tau^-(s_2))e^{-C_-(\chi^-(s_2)-\chi^-(s_1))},
		\end{equation}
		for $C_- > 0$, so that $(\tau^-(s_1), \chi^-(s_1)) \in \mathscr{Q}_{MDD}$, and therefore $(\tau^-(s_*),\chi^-(s_*)) \notin \mathscr{C}^- \cup \partial^+ \mathscr{C}$.
	\end{lemma}
	
	\begin{proof}
		We integrate \eqref{dtdchir} from $\tau$ to $\tau_c(\chi)$:
		\begin{equation} \label{intpchir}
			\partial_\chi r (\tau,\chi) = \frac{1}{2} \left(\frac{2m(\chi)}{r(\tau,\chi)} + f(\chi)\right)^\frac{1}{2} \int_{\tau}^{\tau_c(\chi)}\frac{\frac{2m'(\chi)}{r(\tilde{\tau},\chi)} + f'(\chi)}{\frac{2m(\chi)}{r(\tilde{\tau},\chi)} + f(\chi)} d\tilde{\tau},
		\end{equation}
		so that for all $s \in [s_2,s_1)$:
		\begin{equation}  \label{pchirbd}
			\partial_\chi r(\tau^+(s),\chi^+(s)) \lesssim \tau_c(\chi^+(s)) - \tau^+(s).
		\end{equation}
		We have $r(s) \geq r(s_1) > 0$, so from Proposition \ref{maximal} we get $\dot{\chi}^+(s) \partial_\chi r(\tau^+(s), \chi^+(s)) \sim 1$ for $s \in [s_2,s_1]$. We use also that $\dot{\tau}^+(s) \sim 1$ for $s \in [s_2,s_1]$ so that
		\begin{equation} \label{dchidtau}
			\dot{\chi}^+(s) \sim \frac{\dot{\tau}^+(s)}{\partial_\chi r(\tau^+(s),\chi^+(s))}.
		\end{equation}
		Now, using the assumption $\tau_c'(\chi) \geq 0$, we get $\tau_c(\chi^+(s)) \leq \tau_c(\chi^+(s_1))$, which we use in \eqref{pchirbd} to see that $\partial_\chi r(\tau^+(s),\chi^+(s)) \lesssim \tau_c(\chi^+(s_1)) - \tau^+(s)$ from \eqref{pchirbd}, so \eqref{dchidtau} implies that
		\begin{equation} 
			\dot{\chi}^+(s) \gtrsim \frac{\dot{\tau}^+(s)}{\tau_c(\chi^+(s_1)) - \tau^+(s)}.
		\end{equation}
		Integrating over $(s_2, s_1)$, and using again $\tau_c(\chi^+(s_1)) \geq \tau_c(\chi^+(s_2))$, we get \eqref{chiplusbd}. The argument is analogous to obtain \eqref{chiminusbd}.
	\end{proof}
	
	We now prove the following generalisation of the previous lemma to include the case in which $\chi^\pm(s_1) \in \overline{\mathscr{C}_\chi} \setminus \mathscr{C}_\chi$, where we can no longer bound $r \sim 1$.
	
	\begin{lemma} \label{lem:detbdzero}
		Let $(r(s), m^+(s), m^-(s), \epsilon(s)) \in C^\infty([0,s_1))$ as in the statement of Proposition \ref{maximal}. Suppose that there exists $s_2 \in [0,s_1)$ such that $[\chi^+(s_2), \chi^+(s_1)) \subset \mathscr{C}^+_{\chi}$, with $\chi^+(s_1) \in \overline{\mathscr{C}_\chi} \cap \mathscr{S}_\chi$. Then $\tau^+(s_1) < \tau_s(\chi^+(s_1))$, and so $(\tau^+(s_*),\chi^+(s_*)) \notin \overline{\mathscr{C}^+}$.
		
		Similarly if $(\chi^-(s_1), \chi^-(s_2)] \subset \mathscr{C}^-_{\chi}$, with $\chi^-(s_1) \in \overline{\mathscr{C}_\chi} \cap \mathscr{S}_\chi$, then $\tau^-(s_1) < \tau_s(\chi^-(s_1))$, and so $(\tau^-(s_*),\chi^-(s_*)) \notin \overline{\mathscr{C}^-}$.
	\end{lemma}
	
	\begin{remark}
		Note in particular that if $\chi^-(s_1) = 0$, then Lemma \ref{lem:detbdzero} implies that $(\tau^-(s_1),\chi^-(s_1)) \in \Gamma$. 
	\end{remark}
	
	Lemma \ref{lem:detbdzero} implies that for any thin shell with $(\tau^\pm(s_*), \chi^\pm(s_*)) \in \mathscr{S}$, then $\tau_s'(\chi^\pm(s_*)) > 0$, since for $(\tilde{m}, \tilde{f}) \in \mathfrak{D}_{\chi_b}^*$, we have $\tau_s'(\chi) = 0$ only if $\chi \in \partial{\mathscr{C}_\chi}$. 
	
	\begin{proof}
		We first prove the case in which $\chi^\pm(s_1) > 0$, (we \emph{always} have $\chi^+(s_1) > 0$). We have from e.g.~\eqref{id:T}, for both $\pm$: 
		\begin{equation} \label{chitau-ratio}
			\left|\frac{\dot{\chi}(s) \partial_\chi r(\tau(s), \chi(s))}{\dot{\tau}(s)}\right| = (1+f(\chi(s)))^\frac{1}{2} (1 - (\dot{\tau}(s))^{-2})^\frac{1}{2},
		\end{equation}
		and we note that for $r$ sufficiently small, if $\chi^-(s_1) > 0$, then $\dot{\tau}^\pm(s) \gtrsim r(s)^{-\frac{1}{2}}$, so since we have $\pm \widetilde{\mathcal{X}}^\pm \sim \pm 1$ for $r(s) \sim 1$, we must have $(1 - (\dot{\tau}(s))^{-2})^\frac{1}{2} \sim 1$ for all $s \in [0,s_1]$. Now, from \eqref{intpchir}, we get
		\begin{equation}
			\partial_\chi r(\tau,\chi) \lesssim \left( \frac{2m(\chi)}{r(\tau,\chi)} + f(\chi) \right)^\frac{1}{2} (\tau_c(\chi) - \tau), 
		\end{equation}
		so that
		\begin{equation} \label{chidotbds}
			\dot{\chi}^+(s) \gtrsim \dot{\tau}^+(s) \cdot \frac{r(s)^\frac{1}{2}}{\tau_c(\chi^+(s)) - \tau^+(s)}, \ \dot{\chi}^-(s) \lesssim - \dot{\tau}^-(s) \cdot \frac{r(s)^\frac{1}{2}}{\tau_c(\chi^-(s)) - \tau^-(s)},
		\end{equation}
		Now, from \eqref{tausdef} and \eqref{Trchi}, we have
		\begin{equation}
			\tau_s(\chi) - \tau = \int_0^{r(\tau,\chi)} \frac{\tilde{r}^\frac{1}{2} d\tilde{r}}{(2m(\chi) + \tilde{r} f(\chi))^\frac{1}{2}} d\tilde{r}, 
		\end{equation}
		and since $\chi^-(s_1) > 0$, we can bound $2m(\chi^\pm(s)) + \tilde{r} f(\chi^\pm(s)) \sim 1$ for all $s \in [s_2,s_1)$, $\tilde{r} \in [0,r(s)]$ in either case, so that
		\begin{equation}
			\tau_s(\chi^\pm(s)) - \tau^\pm(s) \lesssim r(s)^\frac{3}{2}.
		\end{equation}
		Now, we have by assumption in the $+$ case: $\tau_s'(\chi^+(s_1)) = 0$, $\tau_c'(\chi^+(s_1)) \sim 1$ by Proposition \ref{scprop1}, while in the $-$ case: $\tau_s'(\chi^-(s_1)) = 0$, $\tau_c'(\chi^-(s_1)) \sim -1$. Since $\tau^\pm(s) < \tau_c(\chi^\pm(s))$ in both cases for all $s \in [s_2,s_1)$, we can bound in either case, for $s_2 < s_1$ sufficiently close:
		\begin{equation}
			r(s)^\frac{1}{2} \gtrsim [\tau_s(\chi^\pm(s)) - \tau_c(\chi^\pm(s)) ]^\frac{1}{3} \gtrsim |\chi^\pm(s) - \chi^\pm(s_1)|^\frac{1}{3}.
		\end{equation}
		Using this bound in \eqref{chidotbds}, along with the bounds in the respective cases: $\tau_c(\chi^+(s)) < \tau_s(\chi^+(s_1))$, $\tau_c(\chi^-(s)) < \tau_s(\chi^-(s_1))$, and integrating over $(s_2,s_1)$, we get in both cases:
		\begin{equation}
			\tau_s(\chi^\pm(s_1)) - \tau^\pm(s_1) \geq ( \tau_c(\chi^-(s_2)) - \tau^\pm(s_2)) \exp \left( - C_\pm| \chi^\pm(s_2) - \chi^\pm(s_1)|^\frac{2}{3}\right),
		\end{equation}
		for $C_\pm > 0$, so the result holds. 
		
		We now consider the case in which $\chi^-(s_*) = 0$, which is essentially the same as the first half of the proof but with $r \leftrightarrow R$. We note that in this case we must have $\tau_s''(0) < 0$ (and Proposition \ref{scprop2} implies that $\tau_c''(0) = \frac{7}{3} \tau_s''(0)$).
		
		One can see from $\dot{r}(s)^2 \sim r(s)^{-2}$ that $n^-(r)(s) \sim |\dot{r}(s)|$, and therefore from \eqref{def:T} we have $\dot{\tau}^-(s) \gtrsim r(s)^{-\frac{3}{2}}$ as $r(s) \to 0$, so indeed from \eqref{chitau-ratio} we do have $\dot{\chi}^-(s) \sim - \frac{\dot{\tau}^-}{\partial_\chi r(\tau^-(s),\chi^-(s))}$. Now, from \eqref{intpchir}, expanding in terms of $r(\tau,\chi) = \chi \cdot R(\tau,\chi)$ and $m(\chi) = \chi^3 \tilde{m}(\chi)$, $f(\chi) = \chi^2\tilde{f}(\chi)$, we see that
		\begin{equation}
			\partial_\chi r(\tau^-(s),\chi^-(s)) \lesssim \left( \frac{2\tilde{m}(\chi^-(s))}{R(s)} + \tilde{f}(\chi^-(s))\right)^\frac{1}{2} (\tau_c(\chi^-(s)) - \tau^-(s)),
		\end{equation}
		while using \eqref{taus} we have:
		\begin{equation}
			\tau_s(\chi^-(s)) - \tau^-(s) \lesssim R(s)^\frac{3}{2},
		\end{equation}
		and so we get the bound
		\begin{equation} \label{chibdcentre}
			\dot{\chi}^-(s) \lesssim \frac{(\tau_s(\chi^-(s)) - \tau^-(s))^\frac{1}{3} \dot{\tau}^-(s)}{\tau_c(\chi^-(s)- \tau^-(s))}.
		\end{equation}
		Using $\tau^-(s) < \tau_c(\chi^-(s)) > \tau_s(\chi^-(s))$ for $s \in [s_2,s_1)$, $s_2 < s_1$ sufficiently close, we get $\tau_s(\chi^-(s)) - \tau^-(s) \gtrsim (\chi^-(s))^2$, which we substitute into \eqref{chibdcentre} and integrate over $(s_2,s_1)$ to obtain the bound:
		\begin{equation}
			\tau_s(0) - \tau^-(s_1) \geq ( \tau_c(\chi^-(s_2)) - \tau^-(s_2) ) \exp\left( - C [\chi^-(s_2)]^\frac{1}{3} \right),
		\end{equation}
		which proves the result.
	\end{proof}
	
	\begin{lemma} \label{lem:thinshellvacuum}
		If $r(s_*) > 0$ and $\chi(s_*) = \chi_b$, with $(\tau^-(s_*),\chi^-(s_*)) \in \mathscr{Q}_{MDD}$, then there is a unique continuous extension as a solution of \eqref{rdot1}, \eqref{mpmdot1}, \eqref{epsdot1}, which satisfies: there exists $s_{*, 2} > s_*$ such that, setting $m^+(s) \equiv m(\chi_b)$ for $s \in [s_*, s_{*,2})$,
		\begin{equation}
			(r(s), m^+(s), m^-(s), \epsilon(s)) \in C^\infty([0,s_{*, 2})),
		\end{equation}
		with $\dot{r}(s) < 0$, $\dot{m}^-(s) < 0$, $(\tau^-(s),\chi^-(s)) \in \mathscr{Q}_{MDD}$ for all $s \in [0,s_{*, 2})$, and if $r(s_{*, 2}) = 0$ then:
		\begin{equation}
			(\tau^-(s_{*, 2}),\chi^-(s_{*, 2})) \in \Gamma \cup (\mathscr{S} \setminus \overline{\mathscr{C}^-}),
		\end{equation}
		otherwise:
		\begin{equation}
			(\tau^-(s_{*, 2}),\chi^-(s_{*, 2})) \in \mathscr{C}^+ \cup \partial^- \mathscr{C}.
		\end{equation}
	\end{lemma}
	
	\begin{proof}
		All derivatives of $m^+(s)$ vanish at $s = s_*$ because $\rho$ is compactly supported in $\chi$. We solve for $\Sigma^-$ for $s \geq s_*$ with the induced initial data at $s = s_*$ from the prior solution, with $m^+(s) \equiv m(\chi_b)$ for $s \geq s_*$. Now, $m^+(s)$ is then smooth for all $s$, and so are $\dot{r}(s), \dot{\epsilon}(s), \dot{\chi}^-(s)$, because the only terms that see $\chi^+(s)$ (which does not extend smoothly to $\chi(s) = \chi_b$ for $s \geq s_*$) are proportional to $\dot{m}^+(s)$. 
		 
		The results of Proposition \ref{maximal} and Lemmas \ref{lem:detbd}, \ref{lem:detbdzero} then hold for $\Sigma^-$. 
	\end{proof}
	
	This concludes the proof of Proposition \ref{odemddprop}. 
	
	\subsubsection{Thin shell collisions and the unique MGHPSD} \label{uniqueMGHPSD}
	
	In this section, we construct the unique MGHPSD for generic initial data in Proposition \ref{prop:MGHPSD}, where in particular we show that we can uniquely extend beyond any \emph{thin shell collision} at which $\inf r > 0$. The proof of Proposition \ref{prop:MGHPSD} makes essential use of the results of Sections \ref{genericsection} and \ref{ODEMDD}. See Section \ref{sec:col} in the introduction for an outline of the proof of Proposition \ref{prop:MGHPSD}.
	\begin{proposition} \label{prop:MGHPSD}
		Let $(\tilde{m},\tilde{f}) \in \mathfrak{D}_{\chi_b}^*$, $\Omega_0^2 \in \mathfrak{G}_{\chi_b}(\tilde{m},\tilde{f})$, then there exists a unique globally hyperbolic piecewise smooth solution $(\mathcal{Q}_{MGHD},r,\Omega^2,\rho,\mathbf{u}^\mu)$ satisfying the propagation condition, attaining smooth initial data $(\mathcal{D}_0,\tilde{m},\tilde{f}, \Omega_0^2)$, and for which
		\begin{equation} \label{allsing-rzero}
			\mathcal{S}_{r>0}(\mathcal{Q}_{MGHD},r) = \emptyset.
		\end{equation}
		The subset $\mathcal{P}_d(\mathcal{Q}_{MGHD})$ is acoustically globally hyperbolic, with acoustic Cauchy surface $\mathcal{D}_0 \cap \{ v \leq \chi_b\}$. There are $N_{ts}(\tilde{m},\tilde{f}) \in \mathbb{N} - 1$ thin shells on $\mathcal{Q}_{MGHD}$, all of which lie in the subset $\mathcal{P}_{d+ts}(\mathcal{Q}_{MGHD})$. If $N_{ts} \geq 1$, the thin shells are given by the functions $u_i, v_i \in C^1([0,s_i)) \cap C^2((0,s_i))$, for $s_i > 0$, $1 \leq i \leq N_{ts}$:
		\begin{equation}
			\Sigma_i := \{ (u_i(s), v_i(s)) : s \in [0,s_i) \},
		\end{equation}
		and satisfy the properties stated in Theorem \ref{MGHDprecise} in the `thin shell formation and future limit points' part. For each $1 \leq i \leq N_{ts}$, define 
		\begin{equation}
			s_{d, i} := \sup\{ 0 \leq s < s_i : \rho^+(\tilde{s}) > 0 \text{ for all } \tilde{s} \in (0,s) \}, 
		\end{equation}
		where if $s_{d, i} < s_i$, then $\{ (u_i(s), v_i(s)) : s \in [s_{d, i},s_i) \} \subset \mathcal{B}_d$, and $\mathcal{B}_d$ is the union of $\mathcal{B}_d \cap \mathcal{P}(\mathcal{Q}_{MGHD}) \neq \emptyset$ along with such subsets of thin shells (possibly empty), and $\mathcal{B}_d$ is then a connected piecewise $C^1$ timelike curve in $\mathcal{Q}_{MGHD}$. Moreover, the curve $\mathcal{B}_d$ is \emph{bounded}, $\partial^+ \mathcal{B}_d \in \R^{1+1}_{u,v}$, and therefore $\mathcal{P}_{d+ts}(\mathcal{Q}_{MGHD})$ is bounded.
		
		There exist unique comoving coordinates $\tau,\chi \in C^2(\mathcal{P}_d(\mathcal{Q}_{MGHD}))$, given the gauge choice $\tau(-v,v) = 0$, $\chi(-v,v) = v$ for $v \in [0,\chi_b]$, and $\tau(u,v), \chi(u,v)$ extend continuously to each $\Sigma_i$ from $\mathcal{P}_d(\mathcal{Q}_{MGHD}) \cap D^\pm(\Sigma_i, \mathcal{P}_d(\mathcal{Q}_{MGHD}))$ (where the extension to the exterior is only for $s < s_{d, i}$) to functions $\tau^-_i,\chi^-_i \in C^0([0,s_{i}]) \cap C^2((0,s_i))$ and (only if $s_{d, 1} > 0$) $\tau^+_i, \chi^+_i \in C^0([0,s_{d, i}]) \cap C^2((0,s_{d, i}))$. For the maximal dust development $(\mathscr{Q}_{MDD},r, \rho,m,f)(\mathscr{D}_{0, \chi_b},\tilde{m},\tilde{f})$, we have the inclusion
		\begin{equation}
			\mathscr{Q}_{MGH}(\tilde{m},\tilde{f}) := \{ (\tau(u,v),\chi(u,v)) : (u,v) \in \mathcal{P}_d(\mathcal{Q}_{MGHD})\} \subset \mathscr{Q}_{MDD},
		\end{equation}
		and we moreover define (where $\Sigma_j^+$ is only defined for $1 \leq j \leq N_{ts}$ such that $s_{d,j} > 0$):
		\begin{equation}
			\Sigma_i^- := \{ (\tau_i^-(s),\chi_i^-(s)) : s \in [0,s_i)\}, \qquad \Sigma_j^+ := \{ (\tau_j^+(s),\chi_j^+(s)) : s \in [0,s_{d, j})\},
		\end{equation}
		so that, letting $\mathscr{B}^+(\mathscr{Q}_{MGH})$ be the induced future boundary\footnote{See Section \ref{mdd}.} of $\mathscr{Q}_{MGH}$ by the ambient $\R^{2}_{\tau,\chi}$: 
		\begin{equation}
			\Sigma^-_i, \Sigma^+_j \subset \mathscr{B}^+(\mathscr{Q}_{MGH}).
		\end{equation}
	\end{proposition}
	
	\begin{proof}
		We construct the globally hyperbolic piecewise smooth solution, satisfying the propagation condition, explicitly in steps. At each step, we identify the (finitely many) first singularities in $\mathcal{S}_{r>0}$ of the (piecewise) smooth solution of (E-d), and show that the extension principles of Section \ref{extp} hold for these first singularities. We then extend uniquely as a piecewise smooth solution of (E-d) satisfying the propagation condition, with thin shells forming at the first singularities, and then consider the first singularities of the maximal such solution, restricted to the already formed thin shells. 
		
		The stated properties of $\mathcal{Q}_{MGHD}$ follow directly from the construction.
		
		As `\underline{Step 0}', we consider the maximal globally hyperbolic development (MGHD) of $(\mathcal{D}_0,\tilde{m},\tilde{f},\Omega_0^2)$ as a $C^\infty$ solution of (E-d), and show that $\mathcal{S}_{r>0}$ consists of finitely many non-degenerate first shell crossing singularities, relying critically on Proposition \ref{genericdata}/Corollary \ref{structuregendata}.
		
		For $\tau_{\Gamma, 0}, \tau_{\chi_b, 0} > 0$, we define the following restrictions of $\Gamma = \{\chi=0\}$, $\mathscr{B}_{\chi_b} = \{\chi=\chi_b\}$ as subsets of $\mathscr{Q}_{MDD}(\mathscr{D}_{0, \chi_b},\tilde{m},\tilde{f})$:
		\begin{equation} \label{Gamma-restricted}
			\Gamma(\tau_{\Gamma, 0}) := \{ (\tau,\chi) \in \mathscr{Q}_{MDD} : 0 \leq \tau < \tau_{\Gamma, 0}, \ \chi = 0\}, \ \mathscr{B}_{\chi_b}(\tau_{\chi_b, 0}) = \{ (\tau,\chi) \in \mathscr{Q}_{MDD} :0 \leq \tau < \tau_{\chi_b, 0}, \ \chi = \chi_b \}.
		\end{equation}
		\begin{definition} \label{def:QGH0}
			We then define the following subsets of $\mathscr{Q}_{MDD}$:
			\begin{equation}
				\QGHa := \mathscr{U}_+(\Gamma(\tau_{\Gamma, 0}) \cup \mathscr{D}_{0, \chi_b}, \mathscr{Q}_{MDD}) \cap \mathscr{V}_+(\mathscr{B}_{\chi_b}(\tau_{\chi_b, 0}) \cup \mathscr{D}_{0, \chi_b}, \mathscr{Q}_{MDD}),
			\end{equation}
			for $\tau_{\Gamma,0}, \tau_{\chi_b, 0} > 0$ such that:
			\begin{equation}
				\Gamma(\tau_{\Gamma, 0}) \subset \mathscr{V}_+(\mathscr{B}_{\chi_b, 0} \cup \mathscr{D}_{0, \chi_b}, \mathscr{Q}_{MDD}), \qquad \mathscr{B}_{\chi_b, 0} \subset \mathscr{U}_+(\Gamma_0 \cup \mathscr{D}_{0, \chi_b}, \mathscr{Q}_{MDD}). 
			\end{equation}
		\end{definition}
		If $\overline{\Gamma(\tau_{\Gamma, 0})} \subset \mathscr{V}_+(\mathscr{B}_{\chi_b, 0} \cup \mathscr{D}_{0, \chi_b}, \mathscr{Q}_{MDD})$, then we can extend $	\mathscr{Q}_{GH, 0}(\tau_{\Gamma,0}, \tau_{\chi_b, 0})$ by increasing $\tau_{\Gamma, 0}$, similarly for $\tau_{\chi_b, 0}$. There is then a unique maximal choice (since $\tau$ is bounded on both $\Gamma$ and $\mathscr{B}_{\chi_b}$ in $\mathscr{Q}_{MDD}$), which we refer to as $\mathscr{Q}_{MGH, 0}$. From the boundary decomposition of $\mathscr{Q}_{MDD}$ given by Proposition \ref{mddprop} along with Corollary \ref{structuregendata} for \emph{generic data}, we see that
		\begin{align}
			\overline{\mathscr{U}_+(\Gamma \cup \mathscr{D}_{0, \chi_b}, \mathscr{Q}_{MDD})} \setminus (\mathscr{Q}_{MDD} \cup \mathscr{S}) \subset \partial^-\mathscr{C} \cup \mathscr{C}^-, \label{null-bd1} \\ \overline{\mathscr{V}_+(\Gamma \cup \mathscr{D}_{0, \chi_b}, \mathscr{Q}_{MDD})} \setminus (\mathscr{Q}_{MDD} \cup \mathscr{S}) \subset \partial^-\mathscr{C} \cup \mathscr{C}^+, \label{null-bd2}
		\end{align}
		since $\mathscr{C}^\pm$ are `outgoing/ingoing' \emph{timelike} singular boundaries (from \eqref{outnull}, \eqref{innull}: $\tau_\pm'(\chi) = 0$ if $(\tau_\pm(\chi),\chi) \in \mathscr{C}$). We therefore have  
		\begin{equation} \label{fsing}
			\overline{\mathscr{Q}_{MGH, 0}} \setminus (\mathscr{Q}_{MDD} \cup \mathscr{S}) \subset \partial^-\mathscr{C}.
		\end{equation}
		Now, there exist unique solutions $u(\tau,\chi), v(\tau,\chi) \in C^\infty(\mathscr{Q}_{MGH, 0})$ of the eikonal equation \eqref{eiku}, \eqref{eikv} respectively (extending smoothly to $\overline{\mathscr{Q}_{MGH, 0}} \setminus (\mathscr{Q}_{MDD} \cup \mathscr{S})$), with initial/boundary conditions:
		\begin{equation} \label{eikIBCs}
			u(0,\chi) = -\chi, \qquad v(0,\chi) = \chi, \ u(\tau,0) = v(\tau,0), \qquad \partial_\tau v(\tau,\chi_b) = \frac{-2(\Omega^{-2} \partial_u r)(-v(\tau,\chi_b),v(\tau,\chi_b))}{(1+f(\chi_b))^\frac{1}{2} + (\frac{2M}{r(\tau,\chi_b)} + f(\chi_b))^\frac{1}{2}},
		\end{equation}
		on $\mathscr{D}_{0, \chi_b}, \Gamma_{0}, \mathscr{B}_{\chi_b, 0}$ respectively (see Proposition \ref{vacuumext}). We define
		\begin{equation} \label{Q0}
			\mathcal{Q}_0 := \{ (u(\tau,\chi),v(\tau,\chi)) : (\tau,\chi) \in \mathscr{Q}_{GH, 0} \} \cup \{ (u,v(\tau,\chi_b)) : 0 \leq \tau < \tau_{\chi_b, 0}, -v(\tau,\chi_b) \leq u < u(\tau,\chi_b) \},
		\end{equation}
		so that, transforming to double null from comoving coordinates on the subset corresponding to $\mathscr{Q}_{GH, 0}$, and identifying the unique solution in the vacuum exterior, $(\mathcal{Q}_0, r, \Omega^2, \rho, \mathbf{u}^\mu)$ is the MGHD of $(\mathcal{D}_0,\tilde{m},\tilde{f},\Omega_0^2)$ as a $C^\infty$ solution of (E-d). Now, if $\mathscr{C}_\chi = \emptyset$, then $\mathcal{S}_{r>0}(\mathcal{Q}_0,r) = \emptyset$. Assuming therefore that $\mathscr{C}_\chi \neq \emptyset$, we then have from \eqref{fsing}: for all $(u,v) \in \mathcal{S}_{r>0}(\mathcal{Q}_0,r)$, there exists $\chi \in \partial^-\mathscr{C}_\chi$ such that $u = u(\tau_c(\chi),\chi), v = v(\tau_c(\chi),\chi)$ (since past-directed null curves from $\partial^-\mathscr{C}$ are well-posed), so $|\mathcal{S}_{r>0}(\mathcal{Q}_0,r)| \leq |\partial^-\mathscr{C}_\chi|$. From Corollary \ref{structuregendata}, we have $|\partial^-\mathscr{C}_\chi| \in \mathbb{N}$.
		
		In `\underline{Step 1a}', we construct the unique maximal globally hyperbolic piecewise smooth development of $\mathcal{Q}_0$ (which we'll define as $\mathcal{Q}_1$) satisfying the propagation condition, restricted to thin shells forming at $\mathcal{S}_{r>0}(\mathcal{Q}_0,r)$. In `Step 1b' , we then show that all $(u,v) \in \mathcal{S}_{r>0}(\mathcal{Q}_1,r)$ satisfy the extension principles of Section \ref{extp}: either a first shell crossing singularity, or a \emph{thin shell collision}.
		
		Now, let $n_1 := |\mathcal{S}(\mathcal{Q}_0,r)|$, and for $1 \leq i \leq n_1$ let $\chi_{f,0,i} \in \partial^-\mathscr{C}$, ordering $0 < \chi_{f,0,i} < \chi_{f,0,i+1} < \chi_b$, be such that $\mathcal{S}(\mathcal{Q}_0,r) = \{ (u(\tau_c(\chi_{f,0,i}),\chi_{0,i}),v(\tau_c(\chi_{f,0,i}),\chi_{f,0,i})) : 1 \leq i \leq n_1 \}$. Defining $p_{0, i} := (\tau_c(\chi_{f,0,i}),\chi_{f,0,i})$, we have:
		\begin{equation}
			\mathscr{U}_-(\{p_{0, i}\}, \mathscr{Q}_{MGH, 0} \cup \{p_{0, i}\}) \cap \mathscr{Q}_{MGH, 0} \neq \emptyset \neq \mathscr{V}_-(\{p_{0, i}\}, \mathscr{Q}_{MGH, 0} \cup \{p_{0, i}\}) \cap \mathscr{Q}_{MGH, 0}. \label{fsing0}
		\end{equation}
		Now, let $(r_{0, i}(s),m^+_{0, i}(s),m^-_{0, i}(s),\epsilon_{0, i}(s)) \in C^0([0,s_{*,0, i}]) \cap C^\infty((0,s_{*,0, i}))$, be the \emph{maximal} unique solution of the thin shell ODE \eqref{rdot1}, \eqref{mpmdot1}, \eqref{epsdot1} (by Proposition \ref{thinshellODEprop}) with initial conditions 
		\begin{equation} \label{ts-formation-ICs}
			r_{0, i}(0) = r(\tau_c(\chi_{f,0,i}),\chi_{f,0,i}), \qquad m^\pm_{0, i}(0) = m(\chi_{f,0,i}), \qquad \epsilon_{0, i}(0) = (1+f(\chi_{f,0,i}))^\frac{1}{2},
		\end{equation}
		satisfying $\dot{m}^+(s) > 0$, $\dot{m}^-(s) < 0$, $(\tau^\pm_{0,i}(s),\chi_{0,i}^\pm(s)) \in \mathscr{Q}_{MDD}$ for all $s \in (0,s_{*, i})$ (where $\tau^\pm_{0,i}(s) := T(r_{0, i}(s),m^\pm_{0,i}(s))$, $m^\pm_{0,i}(s) =: m(\chi_{0,i}^\pm(s))$), and $s_{*,0, i} > 0$ is the maximum proper time such that these conditions are satisfied, so by Proposition \ref{odemddprop} we have $(\tau^\pm_{0, i}(s_{*,i}), \chi^\pm_{0, i}(s_{*,0, i})) \notin \mathscr{Q}_{MDD}$. We then define the interior/exterior curves in $\mathscr{Q}_{MDD} \cup \{p_{0, i}\}$ corresponding to the restriction of the solution of the thin shell ODE to some $0 < s_i \leq s_{*,0, i}$:
		\begin{equation} \label{Sigma-ts}
			\Sigma^\pm_{0, i, s_i} := \{ (\tau_{0, i}^\pm(s),\chi^\pm_{0, i}(s)) : s \in [0,s_i) \}.
		\end{equation}
		For simplicity in the following definitions we also label $\Sigma^+_{0,0, s_0} := \Gamma(\tau_{\Gamma, 0} + s_0)$, $\Sigma^-_{0,n_1+1, s_{n_1+1}} := \mathscr{B}_{\chi_b}(\tau_{\chi_b, 0} + s_{n_1+1})$, for $s_0 \leq \tau_s(0)-\tau_{\Gamma, 0}$, $s_{n_1 + 1} \leq \tau_s(\chi_b) - \tau_{\chi_b, 0}$. We define the following subset of $\mathscr{Q}_{MDD} \cup \partial^-\mathscr{C}$, which will correspond to the subset $\mathcal{P}_{d+ts}(\cdot)$ of any globally hyperbolic piecewise smooth extension of $\mathcal{Q}_0$ (restricted to thin shells forming at $\mathcal{S}_{r>0}(\mathcal{Q}_0,r)$): (in order for consistency with Steps 2-n, we define here $p_{0,i}^\pm := p_{0,i}$)
		\begin{definition} \label{def:QGH1}
			For $0 < s_i \leq s_{*,0, i}$ such that for all $1 \leq i \leq n_1+1$:
			\begin{equation} \label{outgoingint}
				\Sigma_{0, i, s_i}^- \subset \mathscr{U}_+(\mathscr{Q}_{MGH,0} \cup \Sigma_{0,i-1,s_{i-1}}^+, \mathscr{Q}_{MDD} \cup \{p_{0, i}^-\}),
			\end{equation}
			and for all $0 \leq i \leq n_1$:
			\begin{equation} \label{ingoingext}
				\Sigma_{0, i, s_i}^+ \subset \mathscr{V}_+(\mathscr{Q}_{MGH,0} \cup \Sigma_{0, i+1,s_{i+1}}^-, \mathscr{Q}_{MDD} \cup \{p_{0, i}^+\}),
			\end{equation}
			then we define, for $\underline{s} := (s_0, \dots, s_{n_1+1})$:
			\begin{equation}
				\mathscr{Q}_{GH, 1}(\underline{s}) := \mathscr{Q}_{MGH, 0} \cup \bigcup_{i=0}^{n_1} \left( \mathscr{U}_+(\Sigma_{0, i, s_{i}}^+ \cup \mathscr{Q}_{MGH, 0}, \mathscr{Q}_{MDD} \cup \{p_{0, i+1}^-\}) \cap \mathscr{V}_+(\Sigma_{0, i+1, s_{i+1}}^- \cup \mathscr{Q}_{MGH, 0}, \mathscr{Q}_{MDD} \cup \{p_{0, i}^+\}) \right).
			\end{equation}
		\end{definition}
		We now show that for any such $\QGHb$, solutions of the eikonal equation satisfying the conditions \eqref{eikIBCs} along with the following boundary conditions for each $1 \leq i \leq n_1$:
		\begin{equation} \label{SigmaBCs}
			u_{0,i}(s) := u(\tau^+_{0,i}(s),\chi^+_{0,i}(s)) = u(\tau^-_{0,i}(s),\chi^-_{0,i}(s)), \qquad v_{0,i}(s) :=  v(\tau^-_{0,i}(s),\chi^-_{0,i}(s)) = v(\tau^+_{0,i}(s),\chi^+_{0,i}(s)),
		\end{equation}
		extend uniquely to $u, v \in C^2(\QGHb)$ (the regularity is in fact at least $C^3(\QGHb)$ by Theorem \ref{devthm}, but we assume only $C^2$ for consistency with the following steps), and we therefore have $u_{0,i}(s), v_{0,i}(s) \in C^1([0,s_i))\cap C^2((0,s_i))$. In order to see this, we first note that for any $(\tau,\chi) \in \QGHb \setminus (\mathscr{Q}_{MGH, 0} \cup \bigcup_{i=0}^{n_1+1} \Sigma^\pm_{0, i ,s_i})$, both past-directed null curves from $(\tau,\chi)$ either meet $\Sigma_{0, i, s_{i}}^\pm$ for some $i$, or $\mathscr{Q}_{MGH, 0}$, by Definition \ref{def:QGH1}.

		It now suffices to show that the past-directed null curves from each $\Sigma_{0, i, s_i}^\pm$ in $\QGHb$ eventually reach $\mathscr{Q}_{MGH, 0}$, after `identifying' $\Sigma_{0, i, s_i}^\pm$ in accordance with \eqref{SigmaBCs}, and/or `reflecting' off $\Sigma_{0, 0, s_0}^+, \Sigma^-_{0, n_1+1, s_{n_1+1}}$, finitely many times. It is straightforward to see this from a continuity argument: fix any $0 \leq i \leq n_1+1$ and consider the subset of $s \in [0,s_i)$ such that the past null curves reach $\mathscr{Q}_{MGH, 0}$. The subset is clearly non-empty by \eqref{fsing0}, and open, and if any null curve `reflects' off both $\Gamma$ and $\mathscr{B}_{\chi_b}$ before returning to $\Sigma_{0, i, s_i}^\pm$, then by monotonicity in $\tau$ on null curves it must return at proper time $s - \delta$ for some $\delta>0$, hence the subset is closed and equal to $[0,s_i)$.
		
		There is a unique maximal $\QGHb$: for any $\Sigma^\pm_{0, i, s_i}$ such that $\overline{\Sigma_{0, i, s_i}^-} \subset \mathscr{U}_+(\mathscr{Q}_{MGH,0} \cup \Sigma_{0, i-1,s_{i-1}}^+, \mathscr{Q}_{MDD} \cup \{p_i\}) \cap \mathscr{Q}_{MDD}$ and $\overline{\Sigma_{0, i, s_i}^+} \subset \mathscr{V}_+(\mathscr{Q}_{MGH,0} \cup \Sigma_{0, i+1,s_{i+1}}^-, \mathscr{Q}_{MDD} \cup \{p_i\}) \cap \mathscr{Q}_{MDD}$, then we can increase $s_i$, which is necessarily less than $s_{*, i}$. For any two choices of $\underline{s}$, choose the maximum $s_i$ from each, and this clearly satisfies the conditions of the definition, hence there is a unique maximal choice, $\underline{s} = \underline{s}_{1}$, and we define $\mathscr{Q}_{MGH, 1} := \mathscr{Q}_{GH, 1}(\underline{s}_{1})$. For each $1 \leq i \leq n_1$, then at least one of the following holds: either $(s_1)_i = s_{*,i}$ (so that one of, or both, $(\tau_{0, i}^\pm((s_1)_i),\chi_{0, i}^\pm((s_1)_i)) \in \partial\mathscr{Q}_{MDD}$), and/or one of (or both) $(\tau_{0, i}^-((s_1)_i),\chi_{0, i}^-((s_1)_i)) \notin \mathscr{U}_+(\mathscr{Q}_{MGH,0} \cup \Sigma_{0, i-1,(s_1)_{i-1}}^+, \mathscr{Q}_{MDD} \cup \{p_{0, i}^-\})$, $(\tau_{0, i}^+((s_1)_i),\chi_{0, i}^+((s_1)_i)) \notin \mathscr{V}_+(\mathscr{Q}_{MGH,0} \cup \Sigma_{0, i+1,(s_1)_{i+1}}^-, \mathscr{Q}_{MDD} \cup \{p_{0,i}^+\})$.
		
		We now change to double null coordinates on $\mathscr{Q}_{MGH, 1}$, given the unique solutions of the eikonal equation $u, v \in C^2(\mathscr{Q}_{MGH, 1})$, satisfying \eqref{eikIBCs} and \eqref{SigmaBCs}. Indeed, we define $\widetilde{\mathcal{Q}}_1 := \{ (u(\tau,\chi),v(\tau,\chi)) : (\tau,\chi) \in \mathscr{Q}_{MGH, 1} \setminus (\bigcup_{i=1}^{n_1} \Sigma^\pm_{0, i, (s_{1})_i})\}$, and for $1 \leq i \leq n_1$: $\Sigma_{0, i} := \{ (u_i(s),v_i(s)) : 0 \leq s \leq (s_{1})_i \}$, and finally $\mathcal{Q}_1 := \widetilde{\mathcal{Q}}_1 \cup \bigcup_{i=1}^{n_1} \Sigma_{0, i} \cup \{ (u,v(\tau,\chi_b)) : \tau_{\chi_b, 0} \leq \tau < \tau_{\chi_b, 0} + (s_{1})_{n_1 + 1}, -v(\tau,\chi_b) \leq u < u(\tau,\chi_b) \}$. Now, by Proposition \ref{prop:tsode-eik} and Proposition \ref{equivprop}, transforming to double null coordinates on $\widetilde{\mathcal{Q}}_1$ determines a piecewise smooth solution of (E-d): $(\mathcal{Q}_1,r,\Omega^2,\rho,\mathbf{u}^\mu)$ which satisfies the propagation condition. Moreover, any piecewise smooth extension, satisfying the propagation condition, of $\mathcal{Q}_0$ to a globally hyperbolic subset of $\mathcal{Q}_1$ must coincide with the restriction of $(\mathcal{Q}_1,r,\Omega^2,\rho,\mathbf{u}^\mu)$ to that subset, by the extension principle of Section \ref{extp}, and uniqueness of $C^1$ solutions of (E-d).
		
		\underline{Step 1b}: we now consider $\mathcal{S}_{r>0}(\mathcal{Q}_1,r)$. If it's empty, then we're done, otherwise: by the argument used at the end of `Step 0', for any $(u_*,v_*) \in \mathcal{S}_{r>0}(\mathcal{Q}_1,r) \setminus \partial^+\Sigma$, there must exist $(\tau_c(\chi_*),\chi_*) \in \partial^-\mathscr{C}$ such that $u_* = u(\tau_c(\chi_*),\chi_*)$, $v_* = v(\tau_c(\chi_*),\chi_*)$. Any $(u_*,v_*) \in \mathcal{S}_{r>0}(\mathcal{Q}_1,r) \cap \partial^+\Sigma$ corresponds to a `thin shell collision', which we define as:
		\begin{definition} \label{def:collision}
			We say a first singularity $(u_*,v_*) \in \mathcal{S}_{r>0}(\mathcal{Q}_1,r)$ is a \underline{thin shell collision} if: there exist $1 \leq j_- \leq j_+ \leq n_1 + 1$ such that, for $q_{i}^\pm := (\tau^\pm_{0, i}((s_1)_{i})), \chi^\pm_{0, i}((s_1)_{i}))$, $q_{n_1 + 1}^- := (\tau_{\chi_b, 0} + (s_1)_{n_1 + 1}, \chi_b)$: $q_{i}^+ = q_{i+1}^-$ for all $j_- \leq i < j_+$, and 
			\begin{equation} \label{firstsing-collision-U}
				\mathscr{U}_-(\{q_{j_-}^-\}, \mathscr{Q}_{MGH, 1}) \cap \mathscr{Q}_{MGH, 1} \neq \emptyset, 
			\end{equation}
			and if $j_+ < n_1 + 1$:
			\begin{equation} \label{firstsing-collision-V}
				\mathscr{V}_-(\{q_{j_+}^+\}, \mathscr{Q}_{MGH, 1}) \cap \mathscr{Q}_{MGH, 1} \neq \emptyset,
			\end{equation}
			and then $u(q_{j_-}) = u_*$, $v(q_{j_+}) = v_*$.
		\end{definition}
		Any limit points of $\Sigma_{0, i, (s_1)_i}^\pm$ not satisfying the conditions of Definition \ref{def:collision} cannot correspond to a first singularity of $\mathcal{Q}_1$. Indeed, suppose that for $i \in \{ 1, \dots, n_1\}$ we have: $r_{0, i}((s_1)_i) > 0$, with
		\begin{equation} \label{notfirstsing}
			\mathscr{U}_-(\{q_{i}^-\}, \mathscr{Q}_{MGH, 1} \cup \{q_{i}^-\}) \cap \mathscr{Q}_{MGH, 1} = \emptyset, \ q_i^- \neq q_{i-1}^+,
		\end{equation}
		and let $u_{0, i, *} := \lim_{s \to (s_1)_i} u_{0, i}(s)$, $v_{0, i, *} := \lim_{s \to (s_1)_i} v_{0, i}(s)$. Then we must have $\sup\{ v \in \mathbb{R} : (u_{0, i, *},v) \in \mathcal{Q}_1 \} < v_{0, i, *}$ (if the set is non-empty), otherwise by considering the outgoing null curve in $\mathscr{Q}_{MGH, 1}$, we contradict \eqref{notfirstsing}. The argument is analogous with $- \leftrightarrow +$ (note that we've included in Definition \ref{def:collision} the possibility of a `collision' with the boundary of the support of the dust).\footnote{Note that a `collision' with $\Gamma$ cannot be in $\mathcal{S}_{r > 0}(\mathcal{Q}_1,r)$ because we must then have $r_{0, i}((s_1)_i) = 0$.}
		\begin{remark}
			We include in Definition \ref{def:collision} the case of a single thin shell with limit point at a first singularity with $\inf r > 0$, which will correspond to (at least one of) $q_i^\pm \in \partial^-\mathscr{C}$: the formation of a first shell crossing singularity on a thin shell. 
		\end{remark}
		Suppose that $(u_*,v_*) \in \mathcal{S}_{r>0}(\mathcal{Q}_1,r)$ corresponds to a thin shell collision, we'll now show that $(u_*,v_*)$ satisfies the assumptions of the extension principle for piecewise smooth solutions satisfying the propagation condition, Proposition \ref{prop:pwise-extension}. We immediately satisfy the conditions of coinciding with a subset of a maximal dust development in the interior and exterior, we now show that we satisfy the assumptions on extension of regularity of $r, \Omega^2$ (we can see already that $r(u,v)$ must extend continuously to $r_* := r(q_i^\pm)$). 
		
		We first observe that for all $j_- \leq i \leq j_+$, we must have $q_i^\pm \in \mathscr{Q}_{MDD} \cup \partial^-\mathscr{C}$. This follows from Proposition \ref{odemddprop}, we have for all $j_- \leq i \leq j_+$ (where necessarily $r_* := r(q_i^\pm) > 0$): $q_i^- \notin \mathscr{C}^- \cup \partial^+ \mathscr{C}$, $q_i^+ \notin \mathscr{C}^+ \cup \partial^+ \mathscr{C}$, so for $j_- \leq i < j_+$ we have $q_i^+ = q_{i+1}^- \in \mathscr{Q}_{MDD} \cup \partial^-\mathscr{C}$. By consideration of the null curve ODE \eqref{outnull}, \eqref{innull}, and the (timelike) singular boundaries $\mathscr{C}^\pm$, as observed in Step 0, we have (for all $1 \leq i \leq n_1+1$): $\mathscr{U}_+(\mathscr{Q}_{MGH, 0} \cup \Sigma_{i, (s_1)_i}^+, \mathscr{Q}_{MDD} \cup \mathscr{C}) \setminus (\mathscr{Q}_{MDD} \cup \mathscr{S}) \subset \partial^-\mathscr{C} \cup \mathscr{C}^-$ and $\mathscr{V}_+(\mathscr{Q}_{MGH, 0} \cup \Sigma_{i, (s_1)_i}^-, \mathscr{Q}_{MDD} \cup \mathscr{C}) \setminus (\mathscr{Q}_{MDD} \cup \mathscr{S}) \subset \partial^-\mathscr{C} \cup \mathscr{C}^+$, so we must also have $q_{j_-}^-, q_{j_+}^+ \in \partial^-\mathscr{C}$. 
		
		By Proposition \ref{eikprop} and \eqref{firstsing-collision-U}, \eqref{firstsing-collision-V} we certainly have $u(\tau,\chi)$ extending in $C^2_{\tau,\chi}$ to $q_{j_-}^- \in \mathscr{Q}_{MDD} \cup \partial^-\mathscr{C}$ from $\mathscr{Q}_{MGH, 1}$, and $v(\tau,\chi)$ extending in $C^2_{\tau,\chi}$ to $q_{j_+}^+ \in \mathscr{Q}_{MDD} \cup \partial^-\mathscr{C}$ from $\mathscr{Q}_{MGH, 1}$ (where if $j_+ = n_1+1$ then $v(\tau,\chi)$ extends in $C^2$ along $\{\chi=\chi_b\}$ by solving \eqref{eikIBCs}). From Proposition \ref{maximal}, we have $\dot{\tau}^\pm_{0, i} \in C^0([0, (s_1)_i])$, so from differentiating \eqref{SigmaBCs} and using the eikonal equation \eqref{eiku}, \eqref{eikv}, we see that $\dot{u}_{0, i}(s), \dot{v}_{0, i}(s)$ extend continuously to $s = (s_1)_i$. Using also that $\sigma_i(s)$ extends continuously to $s = (s_1)_i$, we see that the jump conditions \eqref{prjump}, \eqref{pomegajump} extend continuously to $s = (s_1)_i$, and therefore, for $k \in \{1,2\}$: $\partial_\tau^k u(\tau^\pm_{0, i}(s),\chi^\pm_{0, i}(s))$, $\partial_\tau^k v(\tau^\pm_{0, i}(s),\chi^\pm_{0, i}(s))$ extend continuously to $s = (s_1)_i$ for each $j_- \leq i \leq j_+$, and $\partial_\tau^k u(q_i^+) = \partial_\tau^k u(q_{i+1}^-)$, $\partial_\tau^k v(q_i^+) = \partial_\tau^k v(q_{i+1}^-)$ for each $j_- \leq i < j_+$. To see that $\Omega^2$ extends continuously to $(u_*,v_*)$, one can compute directly the limiting value in terms of the eikonal equation and limiting values of $\dot{\tau}^\pm_{0, i}(s)$, or simply note that $\Omega^2$ is continuous across each thin shell and we have $|\partial_u \log \Omega^2|, |\partial_v \log \Omega^2| \sim 1$ by \eqref{pomega}. In particular, denoting e.g.~$[\partial_u r]_i := \lim_{s \to (s_1)_i}  \left[ (\partial_u r(u_{0, i}(s),v_{0, i}(s)))^+ - (\partial_u r(u_{0, i}(s),v_{0, i}(s)))^- \right]$, along with $[\partial_u r]_* := \lim_{s \to (s_1)_{j_+}} (\partial_u r(u_{0, i}(s),v_{0, i}(s)))^+ -  \lim_{s \to (s_1)_{j_-}} (\partial_u r(u_{0, i}(s),v_{0, i}(s)))^-$, then we have
		\begin{equation} \label{deriv-sum}
			[\partial_u r]_* = \sum_{i=j_-}^{j_+} [\partial_u r]_i > 0, \qquad [\partial_v r]_* = \sum_{i=j_-}^{j_+} [\partial_v r]_i < 0, \qquad [\partial_u \log(r\Omega)^2]_* = 0 = [\partial_v \log(r\Omega)^2]_*.
		\end{equation} 
		Continuity of derivatives of the eikonal equation on $\mathscr{Q}_{MGH, 1}$, with extension to $q_{j_\pm}^\pm$, (or continuity of derivatives in vacuum if $j_+ = n_1+1$) implies that the limits of derivatives along $\Sigma_{j_\pm, (s_1)_{j_\pm}}^\pm$ coincide with the limits along the respective null curves, as in the requirement of Proposition \ref{prop:pwise-extension} (and $\partial_u r < 0$ by Raychaudhuri \eqref{Rayu} from initial data). It remains to show that \eqref{mindet}, \eqref{posdet} hold. We denote: $\nu_i := \frac{[\partial_u r]_i}{-[\partial_v r]_i} > 0$, $\nu_* := \frac{[\partial_u r]_*}{-[\partial_v r]_*} > 0$, $\nu_{d, i}^\pm := \frac{\partial_\tau v(q_i^\pm)}{\partial_\tau u(q_i^\pm)} > 0$, $\nu_{d, *}^- := \nu_{d, j_-}^-$, $\nu_{d, *}^+ := \nu_{d, j_+}^+$, so we need to show that \eqref{mindet}, \eqref{posdet} hold, which are equivalent to: $\nu_{d, *}^+ < \nu_* < \nu_{d, *}^-$. For each $j_- \leq i \leq j_+$, we have, first using the jump conditions \eqref{prjump}, then the eikonal equation \eqref{eiku} and the continuous extension to $(s_1)_i$ of $\dot{u}(s)$ and finally \eqref{detplus}, \eqref{detmin} from Proposition \ref{maximal}:
		\begin{align}
			\left( \frac{\nu_{d, i}^+}{\nu_i}\right)^\frac{1}{2} &= \frac{\dot{u}_i((s_1)_i)}{\partial_\tau u(q_i^+)} = \left( 1 + \frac{\widetilde{\mathcal{X}}_i^+((s_1)_i)^2}{(1+f(\chi_i^+((s_1)_i))))} \right)^\frac{1}{2} - \frac{\widetilde{\mathcal{X}}_i^+((s_1)_i)}{(1+f(\chi_i^+((s_1)_i)))^\frac{1}{2}} < 1, \label{nu-bd+} \\
			\left( \frac{\nu_{d, i}^-}{\nu_i}\right)^\frac{1}{2} &= \frac{\dot{u}_i((s_1)_i)}{\partial_\tau u(q_i^-)} = \left( 1 + \frac{\widetilde{\mathcal{X}}_i^-((s_1)_i)^2}{(1+f(\chi_i^-((s_1)_i)))} \right)^\frac{1}{2} - \frac{\widetilde{\mathcal{X}}_i^-((s_1)_i)}{(1+f(\chi_i^-((s_1)_i)))^\frac{1}{2}} > 1, \label{nu-bd-}
		\end{align}
		so that $\nu_{d, i}^+ < \nu_i < \nu_{d, i}^-$, and therefore for all $j_- < i$: $\nu_{j_+} < \nu_i$; and all $i < j_+$: $\nu_{i} < \nu_{j_+}$ (since $\nu_{d, i}^+ = \nu_{d, i+1}^-$). Then, using \eqref{deriv-sum}, we see from $[\partial_u r]_* = \sum_{i=j_-}^{j_+} ( - \nu_i [\partial_v r]_i )$ that, if $j_- < j_+$: $- \nu_{j_+} [\partial_v r]_* < [\partial_u r]_* < -\nu_{j_-} [\partial_v r]_*$, and therefore $\nu_{j_+} < \nu_* < \nu_{j_-}$, so we always have $\nu_{d,*}^+ < \nu_* < \nu_{d, *}^-$. Hence, the extension principle holds for all (finitely many) $(u,v) \in \mathcal{S}_{r > 0}(\mathcal{Q}_1,r)$.
		
		\underline{Steps 2-$N$.} We now extend as a globally hyperbolic piecewise smooth solution, satisfying the propagation condition, restricted to thin shells forming at $\mathcal{S}_{r>0}(\mathcal{Q}_1, r)$, and already formed thin shells that can be extended in $C^2$ beyond the `Cauchy horizons' of the first singularities in $\mathcal{S}_{r>0}(\mathcal{Q}_1, r)$. This part, and the rest of Step 2, will be completely analogous for all possible Steps $N > 2$.
		
		We now define the subset of thin shells that do not have a limit point at a first singularity and can be continued in $C^2$ for any piecewise smooth extension of $\mathcal{Q}_1$. For any $\Sigma_{0, i} \subset \mathcal{Q}_1$ with $r_{0, i}((s_1)_i) > 0$, but $(u_{0, i}((s_1)_i), v_{0, i}((s_1)_i)) \notin \mathcal{S}(\mathcal{Q}_1, r)$, there must exist $(u_*,v_*) \in \mathcal{S}_{r>0}(\mathcal{Q}_1,r)$ such that one of $u_{0, i}((s_1)_i) = u_*$ or $v_{0, i}((s_1)_i) = v_*$. If $(\tau_{0, i}^\pm((s_1)_i), \chi_{0, i}^\pm((s_1)_i)) \in \mathscr{Q}_{MDD}$, then $u,v \in C^2(\mathscr{Q}_{MGH, 1})$ extend in $C^2$ to the interior/exterior limit points, and the corresponding solution of the thin shell ODE extends smoothly to $s = (s_1)_i$. If $\mathcal{S}(\mathcal{Q}_1,r) = \emptyset$, then we're done, otheriwse: let $n_2 \in \mathbb{N}$ be the sum of $|\mathcal{S}(\mathcal{Q}_1,r)|$ and the number of `continuing thin shells'. We define $0 < \chi_{f, 1, i}^- \leq \chi_{f, 1, i} \leq \chi_b$ for $1 \leq i \leq n_2$, satisfying $\chi_{f, 1, i}^+ < \chi_{f, 1, i+1}^-$ for $1 \leq i \leq n_2 - 1$, along with $p_{1, i}^\pm = (\tau_{f, 1, i}^\pm, \chi_{f, 1, i}^\pm) \in \mathscr{Q}_{MDD} \cup \partial^-\mathscr{C}$, to correspond to the points of formation (or continuation) of thin shells: for all $1 \leq i \leq n_2$, $p_{1, i}^\pm \in \overline{\mathscr{Q}_{MGH, 1}}$, if $\chi_{f, 1, i}^- = \chi_{f, 1, i}^+ =: \chi_{f, 1, i}$ then $p_{1, i}^- := p_{1, i}^+ := (\tau_c(\chi_{f, 1, i}), \chi_{f, 1, i}) \in \partial^-\mathscr{C}$; if $\chi_{f, 1, i}^- < \chi_{f, 1, i}^+$, then either there exists a thin shell collision $(u_*, v_*) \in \mathcal{S}_{r>0}(\mathcal{Q}_1,r)$ such that, using the notation of Definition \ref{def:collision}: $p_{1, i}^\pm := q_{j_\pm}^\pm$, or there exists $j \in \mathbb{N}$ such that $\Sigma_{0, j}$ is a continuing thin shell, and $p_{1, i}^\pm := (\tau_{0, j}^\pm((s_1)_j), \chi_{0, j}^\pm((s_1)_j))$. We now define the necessary limits in order to state the initial conditions for thin shells as in Proposition \ref{prop:pwise-extension}. In either case of a first singularity or thin shell continuation, we use the unique $C^2$ extension of $u,v \in C^2(\mathscr{Q})$ to $p_{1,i}^\pm$, and define $u_{f, 1, i} := u(p_{1, i}^\pm)$, $v_{f, 1, i} := v(p_{1, i}^\pm)$, and (now considering functions defined on $\mathcal{Q}_1$) we define
		\begin{equation}
			(\partial_\mu r)_{1, i}^- := \lim_{\mathcal{P}(\mathcal{Q}_1)\ni(u_{f, 1, i},v) \to (u_{f, 1, i},v_{f, 1, i})} \partial_\mu r(u_{f, 1, i},v), \qquad (\partial_\mu r)_{1, i}^+ := \lim_{\mathcal{P}(\mathcal{Q}_1)\ni(u,v_{f, 1, i}) \to (u_{f, 1, i},v_{f, 1, i})} \partial_\mu r(u,v_{f, 1, i}),
		\end{equation}
		\begin{equation}
			\dot{u}_{f, 1, i} := \Omega(u_{f, 1, i},v_{f, 1, i})^{-1} \left( \frac{-[\partial_v r]_{1, i}}{[\partial_u r]_{1, i}}\right)^\frac{1}{2}, \qquad \dot{v}_{f, 1, i} = \Omega(u_{f, 1, i},v_{f, 1, i})^{-1} \left( \frac{[\partial_u r]_{1, i}}{- [\partial_v r]_{1, i}}\right)^\frac{1}{2}, 
		\end{equation}
		and
		\begin{equation} \label{eps-extp}
			\epsilon_{f, 1, i} := \frac{1}{2} \dot{v}_{f, 1, i} ( (\partial_v r)_{1,i}^+ + (\partial_v r)_{1,i}^-) - \frac{1}{2} \dot{u}_{f, 1, i} ( (\partial_u r)_{1,i}^+ + (\partial_u r)_{1,i}^-) > 0.
		\end{equation}
		If $\chi_{f, 1, i}^- = \chi_{f, 1, i}^+$, then we define $\Sigma_{1,i,s_i}$ for $0 < s_i \leq s_{*, 1, i}$ exactly as in Step 1a, in terms of the solution of the thin shell ODE with initial conditions \eqref{ts-formation-ICs} and the definition \eqref{Sigma-ts}, with $\{0,i\} \leftrightarrow \{1, i\}$. If $\chi_{f, 1, i}^- < \chi_{f, 1, i}^+$, then we define $r_{1, i}, m_{1, i}^+(s), m_{1, i}^-, \epsilon_{1, i}(s) \in C^0([0,s_{*, 1, i}]) \cap C^2((0,s_{*, 1, i}))$ to be the maximal unique solution of the thin shell ODE \eqref{rdot1}, \eqref{mpmdot1}, \eqref{epsdot1} (by Proposition \ref{thinshellODEprop}) with initial conditions (as in Proposition \ref{prop:pwise-extension}):
		\begin{equation} \label{init-extp}
			r_{1, i}(0) = r(u_{f, 1, i}, v_{f, 1, i}), \qquad m^\pm_{1, i}(0) = m(\chi_{f,1,i}^\pm), \qquad \epsilon_{1, i}(0) = \epsilon_{f, 1, i},
		\end{equation}
		and again define $\Sigma_{1, i}^\pm$ as in the definition \eqref{Sigma-ts} with $\{0,1 \} \leftrightarrow \{1, i\}$. 
		
		If the `innermost' thin shell on the previous step has not `collided' with the centre, i.e.~$\chi_{0, 1}^-((s_1)_1) > 0$, then we again define $\Sigma_{1, 0, s_0}^+ := \Gamma(\tau_{\Gamma,0} + (s_1)_0 + s_0)$ for $s_0 \leq \tau_s(0) - \tau_{\Gamma, 0} - (s_1)_0$. Similarly, if the `exteriormost' thin shell on the previous step has not `collided' with $\{ \chi = \chi_b\}$, i.e.~$\chi_{0, n_1}^+((s_1)_{n_1}) < \chi_b$, then we define $\Sigma_{1, n_2 + 1,s_{n_2+1}} := \mathscr{B}_{\chi_b}(\tau_{\chi_b,0} + (s_1)_{n_1+1} + s_{n_2+1})$, for $s_{n_2+1} \leq \tau_s(\chi_b) - \tau_{\chi_b,0} - (s_1)_{n_1+1}$. We then apply Definition \ref{def:QGH1} for the respective range of $i$ depending on the possible collisions with boundaries to define $\mathscr{Q}_{GH, 2}(\underline{s})$, where we replace $0 \leftrightarrow 1$, $1 \leftrightarrow 2$ in the respective places in the definition. In the case that $\chi_{0, n_1}^+((s_1)_{n_1}) = \chi_b$, then the boundary condition for $v(\tau,\chi)$ on $\Sigma^-_{1, n_2}$ is determined by the ODE:
		\begin{equation} \label{eikBC-vacuum}
			\dot{v}_{n_2}(s) = - \frac{2(\Omega^{-2} \partial_u r)(-v_{n_2}(s),v_{n_2}(s))}{ -\dot{r}_{n_2}(s) + \left(\epsilon_{n_2}(s) - \frac{M - m^-_{n_2}(s)}{2\epsilon_{n_2}(s)r_{n_2}(s)} \right)},
		\end{equation}
		as derived in Proposition \ref{prop:pwise-extension} (and we've used Raychaudhuri \eqref{Rayu} to determine $\Omega^{-2} \partial_ur$ from initial data). There then exists a unique solution of the eikonal equation satisfying \eqref{eikIBCs}, and \eqref{eikBC-vacuum} if necessary, defining $u,v \in C^2(\mathscr{Q}_{GH, 2}(\underline{s}))$, where we've used Corollary \ref{devcor} and Proposition \ref{prop:pwise-extension} to for the $C^2$ extension to $\mathscr{Q}_{GH, 2}$ from $\mathscr{Q}_{MGH, 1}$. We can again find a maximal such $\mathscr{Q}_{GH, 2}$, and the possible first singularities in $\mathcal{S}_{r>0}(\mathcal{Q}_2, r)$ are exactly as for $\mathcal{S}_{r>0}(\mathcal{Q}_1, r)$. We continue until after Step $N_f$, we have $\mathcal{S}_{r>0}(\mathcal{Q}_{N_f},r) = 0$. We can bound $N_f \in \mathbb{N}$ by initial data: we have $|\partial^-\mathscr{C}| = n_c \in \mathbb{N}$, so only $n_c$ thin shells can form. There are then $n_c$ possible collisions (including the `collision' with the vacuum boundary), so the number of total number of steps must satisfy $N_f \leq 2n_c$. We then define $(\mathcal{Q}_{MGHD},r,\Omega^2,\rho,\mathbf{u}^\mu) := (\mathcal{Q}_{N_f},r,\Omega^2,\rho,\mathbf{u}^\mu)$, and $\mathscr{Q}_{MGH} := \mathscr{Q}_{MGH, N_f}$, and we can choose a labelling $\Sigma_i$ for each thin shell defined on $\mathcal{Q}_{MGHD}$ (where we can identify thin shells that have been `continued' at a step). 
		
		We note that the vacuum boundary $\mathcal{B}_d$ of $\mathcal{P}_{d+ts}(\mathcal{Q}_{MGHD})$ consists of the smooth dust boundary initially, then possibly finitely many thin shells, succesively colliding with the boundary: $\mathcal{B}_d$ is therefore possibly only piecewise $C^1$ timelike. If $\mathcal{B}_d$ consists only of the smooth dust boundary, with boundary condition as in \eqref{eikIBCs}, then as $r(\tau,\chi_b) \to 0$ for $\tau \to \tau_s(\chi_b)$, we have $\partial_\tau v(\tau,\chi_b) \sim r(\tau,\chi_b)^\frac{1}{2}$ and so we must have $v_{\mathcal{B}_d, +} < \infty$. In the case of $\mathcal{B}_d$ consisting of a thin shell with $r(s) \to 0$ then from \eqref{eikBC-vacuum} we again see that (where the thin shell has $r(s)\to0$ in finite proper time), $\dot{v}_{n_N}(s) \to 0$, so we again have $v_{\mathcal{B}_d, +} < \infty$. The set $\mathcal{P}_{d+ts}(\mathcal{Q}_{MGHD}) \subset \R^{1+1}_{u,v}$ is therefore bounded. 
	\end{proof}
	
	\subsubsection{Boundary of the Penrose diagram of the MGHPSD} \label{firstsingrzero}
	
	We now show that the decomposition of the boundary of the `dust cloud' for any MGHPSD of generic data is as stated in Theorem \ref{MGHDprecise}, along with the relevant limits of $r, \Omega^2, \rho, \mathbf{u}^\mu$ at each component of the boundary. 
	
	We determine the nature of the boundary by considering null curves on $\mathscr{Q}_{MDD}$, in particular the boundary of the `physical' subset of the maximal dust development, $\mathscr{Q}_{MGH}$, as defined in the statement of Proposition \ref{prop:MGHPSD}.
	
	\begin{lemma} \label{lem:O-limit-thinshell}
		For $(\tilde{m},\tilde{f}) \in \mathfrak{D}_{\chi_b}^*$, and $\Omega_0^2 \in \mathfrak{G}_{\chi_b}(\tilde{m},\tilde{f})$, let $u,v \in C^2(\mathscr{Q}_{MGH}(\tilde{m},\tilde{f}))$ be the unique solution of the eikonal equation with initial/boundary conditions as in Proposition \ref{prop:MGHPSD}. Then: 
		\begin{equation} \label{def:vO}
			v_{\mathcal{O}} := \sup\{ v(\tau,0) : (\tau,0) \in \mathscr{Q}_{MGH}\} < \infty.
		\end{equation}
		If $N_{ts}(\tilde{m},\tilde{f}) \geq 1$, then $\mathcal{O} \in \partial^+\Sigma$ if and only if there exists $1 \leq i \leq N_{ts}$ such that (for $\chi_i^-$ as defined in Proposition \ref{prop:MGHPSD}): $\chi_i^-(s_i) = 0$ (which implies that $\tau_i^-(s_i) \leq \tau_s(0)$).
		
		If $\tau_s(\tilde{m},\tilde{f})''(0) < 0$, then $N_{ts}(\tilde{m},\tilde{f}) \geq 1$ and there must exist $1 \leq i \leq N_{ts}$ such that $\chi_i^-(s_i) = 0$, and moreover $\tau_i^-(s_i) < \tau_s(0)$, so that $\sup_{v < v_{\mathcal{O}}} \rho(v,v) < \infty$. 
	\end{lemma}
	
	\begin{remark}
		The set $\partial^+\Sigma$ is well-defined because $\dot{u}_i(s), \dot{v}_-(s) > 0$ for each thin shell, and $\mathcal{P}_{d+ts}(\mathcal{Q}_{MGHD})$ is bounded by Proposition \ref{prop:MGHPSD}. 
	\end{remark}
	
	\begin{remark} \label{rem:taus-pos}
		It follows from Lemma \ref{lem:O-limit-thinshell}, that if $\mathcal{O} \notin \partial^+\Sigma$, then we must have $\tau_s''(0) > 0$, since $|\tau_s(\tilde{m},\tilde{f})''(0)| > 0$ for all $(\tilde{m},\tilde{f}) \in \mathfrak{D}_{\chi_b}^*$ by definition in Proposition \ref{genericdata}. We'll show in Proposition \ref{prop:NS} that we therefore have $\mathcal{CH}_d \neq \emptyset$.
	\end{remark}
	
	\begin{proof}
		The equality \eqref{def:vO} follows immediately from the definition of $\mathcal{O}$ in the statement of Theorem \ref{MGHDprecise}, and by the construction of $\mathcal{Q}_{MGHD}$ in terms of the unique solution of the eikonal equation on $\mathscr{Q}_{MGH}$, in Proposition \ref{prop:MGHPSD}, where moreover $\partial_\tau v(\tau,0) > 0$ so the limit exists, and the solutions of the eikonal equation are \emph{bounded}, again by Proposition \ref{prop:MGHPSD}, so $v_{\mathcal{O}} < \infty$.
		
		If $\chi_i^-(s_i) = 0$, then we must have $\tau_i^-(s_i) \leq \tau_s(0)$ since  $\Sigma_i^-$ must be the `innermost' thin shell of (at least) one of the steps of the construction of Proposition \ref{prop:MGHPSD}, and therefore there exists $\tau_{\Gamma,+} \leq \tau_s(0)$ such that: 
		\begin{equation}
			\Sigma_{i, s_i}^- \subset \mathscr{U}_+(\Gamma(\tau_{\Gamma,+} \cup \mathscr{D}_{0,\chi_b}),\mathscr{Q}_{MDD} \cup \{ p_i^- \}), 
		\end{equation}
		where $p_i^- = (\tau^-(0),\chi^-(0))$, and we've used the notation \eqref{Gamma-restricted}. The subset of $\mathscr{U}_+(\Gamma(\tau_{\Gamma,+} \cup \mathscr{D}_{0,\chi_b}),\mathscr{Q}_{MDD} \cup \{ p_i^- \})$ to which $\chi$ extends continuously to zero is exactly $\Gamma(\tau_{\Gamma,+}$ (by e.g.~\eqref{outnull}). Now, $\Gamma(\tau_{i}^i(s_i))$ and $\Sigma^-_{i,s_i}$ are timelike disjoint curves in $\mathscr{Q}_{MDD}$ with the same limit point. The functions $u, v \in C^1([0,s_i)) \cap C^2((0,s_i))$ are monotone increasing, $\dot{u}(s), \dot{v}(s) > 0$, and therefore the limit $s \to s_i$ converges. It then follows from the existence of future-directed outgoing/ingoing null curves from $\Gamma(\tau_{i}^i(s_i)), \Sigma^-_{i,s_i}$, which must intersect $\Gamma(\tau_{i}^i(s_i)), \Sigma^-_{i,s_i}$ respectively, that $\lim_{s \to s_i} u(s) = \lim_{s \to s_i} v(s) = v_{\mathcal{O}}$.  
		
		If $\tau_s''(0) < 0$, then by Proposition \ref{genericdata} (and Lemma \ref{scprop2}), there exists $\chi_{*} > 0$ such that $(0,\chi_*] \subset \mathscr{C}_\chi$, $\tau_c'(\chi_*) = 0$, and $\tau_c'(\chi) < 0$ for all $\chi \in (0,\chi_*]$. Either there exists a thin shell forming at $\mathscr{Q}_{MGH}$, or there exists a thin shell $\Sigma_i$ and $0 < s < s_i$ with $\chi_i^-(s) = \chi_*$, $\tau_i^-(s) < \tau_c(\chi_*)$. In either case, the thin shell may collide arbitrarily many times, but by Proposition \ref{maximal}, there must exist $\Sigma_j^-$ such that $\chi_j^-(s_j) = 0$, and $(\tau^-(s_i),\chi^-(s_i)) \notin \overline{\mathscr{C}^-}$, so we must have $\tau_j^-(s_j) < \tau_s(0)$.
	\end{proof}
	
	In order to understand the structure of $\mathcal{S}_d$, we show that the null curve ODE is well-posed with extension to $\overline{\mathscr{Q}_{MGH}} \cap \mathscr{S}$.
	\begin{lemma} \label{lem:eik-S}
		For $(\tilde{m},\tilde{f}) \in \mathfrak{D}_{\chi_b}^*$, with $\mathscr{Q}_{MGH}(\tilde{m},\tilde{f})$ as in the proof of Proposition \ref{prop:MGHPSD}, then for all $\chi_0 \in (0,\chi_b]$ with $\{ (\tau,\chi_0) : \tau \in [0,\tau_s(\chi_0)) \} \subset \mathscr{Q}_{MGH}$: there exists $\delta > 0$ such that $\tau_s'(\chi) > 0$ for $\chi \in [\chi_0 - \delta, \chi_0 + \delta]$, and for $(\tau,\chi) \in \mathscr{X}_{\chi_0, \delta} := \{ (\tau,\chi): \chi \in [\chi_0 - \delta, \chi_0 + \delta], \tau \in [0,\tau_s(\chi)) \} \subset \mathscr{Q}_{MDD}$: $\partial_\chi r(\tau,\chi) \sim r(\tau,\chi)^{-\frac{1}{2}}$, and there exist unique past-directed ingoing and outgoing null curves from $(\tau_s(\chi_0),\chi_0) \in \overline{\mathscr{X}_{\chi_0, \delta}} \subset \R^2_{\tau,\chi}$. 
		
		The solutions of the eikonal equation $u(\tau,\chi), v(\tau,\chi)$ on $\mathscr{Q}_{MGH}$ (with initial/boundary conditions as in Proposition \ref{prop:MGHPSD}) therefore extend continuously to $\overline{\mathscr{Q}_{MGH}} \cap \mathscr{S}$ (if non-empty). 
	\end{lemma}
	
	\begin{proof}
		For any $\chi_0 \in (0,\chi_b]$, if $\{ (\tau,\chi_0) : \tau \in [0,\tau_s(\chi_0)) \} \subset \mathscr{Q}_{MDD}(\mathscr{D}_{0,\chi_b}, \tilde{m},\tilde{f})$ then we must have $\tau_s'(\chi_0) \geq 0$, by Proposition \ref{mddprop}. From Proposition \ref{genericdata}, we can have $\tau_s'(\chi_0) = 0$ only if $\chi_0 \in \overline{\mathscr{C}_{\chi}} \setminus \mathscr{C}_\chi$ (since then we'd have $|\tau_s''(\chi_0)| > 0$), but by Proposition \ref{odemddprop}, and the construction of $\mathscr{Q}_{MGH}$ in Proposition \ref{prop:MGHPSD}, we have $\overline{\mathscr{Q}_{MGH}} \cap \overline{\mathscr{C}^\pm} = \emptyset$, and therefore $\tau_s'(\chi_0) > 0$.
		
		We now show that $\partial_\chi r(\tau,\chi) \sim r(\tau,\chi)^{-\frac{1}{2}}$ for $(\tau,\chi) \in \mathscr{X}_{\chi_0, \delta}$. From \eqref{pchir} along with \eqref{Hs}, \eqref{Hexp}, \eqref{tildeH0}, we have: 
		\begin{equation} \label{pchir-exp}
			\partial_\chi r(\tau,\chi) = R(\tau,\chi)^{-\frac{1}{2}} A_0(R(\tau,\chi),\chi), \qquad A_0(R,\chi) := (2\tilde{m}(\chi) + \tilde{f}(\chi) R)^\frac{1}{2} \left( \chi \tau_s'(\chi) + \frac{1}{2} R^\frac{3}{2} \widetilde{H}_0(R,\chi) \right),
		\end{equation}
		where $\widetilde{H}_0(R,\chi)$ is as defined in \eqref{tildeH0} (removing the underlines), and we have:
		\begin{equation} \label{tildeH0-2}
			\widetilde{H}_0(R,\chi) =: 2(2\tilde{m}(\chi) + \tilde{f}(\chi) R)^{-\frac{1}{2}} + \chi^2 \widetilde{H}_2(R,\chi),
		\end{equation}
		where 
		\begin{equation} \label{tildeH2}
			\widetilde{H}_2(R,\chi) := \chi^{-1} \int_0^1 \frac{2\tilde{m}'(\chi) + \tilde{f}'(\chi) s R}{(2\tilde{m}(\chi) + \tilde{f}(\chi) sR)^\frac{3}{2}} s^\frac{1}{2} ds,
		\end{equation}
		so $\widetilde{H}_2(R(\tau,\chi),\chi) \in C^\infty(\overline{\mathscr{Q}_{MDD}})$ (since $\tilde{m}, \tilde{f}$ are even functions). The function $\widetilde{H}_0(R(\tau,\chi),\chi)$ extends smoothly to a \emph{positive} function on $\mathscr{S}$: $\widetilde{H}_0(0,\chi) = 2(2\tilde{m}(\chi))^{-\frac{1}{2}} + \frac{4\chi}{3} \tilde{m}'(\chi) (2\tilde{m}(\chi))^{-\frac{3}{2}} > 0$ for all $\chi \in \mathscr{S}_{\chi}$. We then have $A_0(R(\tau,\chi),\chi) \in C^\infty(\mathscr{Q}_{MDD})$, extending in $C^1$ to $\mathscr{S}$, with $A_0(0,\chi) = (2\tilde{m})^\frac{1}{2} \chi \tau_s'(\chi)$. For $\chi \in [\chi_0 - \delta, \chi_0 + \delta]$, we have $\tau_s'(\chi) \sim 1$, $\chi \sim 1$, so $A_0(0,\chi) \sim 1$, and therefore $\partial_\chi r(\tau,\chi) \sim R(\tau,\chi)^{-\frac{1}{2}} \sim r(\tau,\chi)^{-\frac{1}{2}}$. 
		
		In order to show that there exists unique past-directed null curves from $(\tau_s(\chi_0),\chi_0)$, we write the null curve ODEs \eqref{outnull}, \eqref{innull} from $(\tau_s(\chi_0),\chi_0)$ in terms of $r_\pm(\chi) := r(\tau_\pm(\chi),\chi)$, which satisfies $r_\pm(\chi_0) = 0$ and:
		\begin{equation}\label{rnull-ode}
			r_\pm'(\chi) = \mp (r_\pm(\chi))^{-1} \frac{A_0(\chi^{-1} r_\pm(\chi),\chi)}{(1+f(\chi))^\frac{1}{2}} \left[ (2m(\chi) + r_\pm(\chi) f(\chi))^\frac{1}{2} \mp r_\mp(\chi)^\frac{1}{2}(1+f(\chi))^\frac{1}{2}\right] =: \frac{1}{8} (r_\pm(\chi))^{-1} F_{0, \pm}(r_\pm(\chi),\chi).
		\end{equation}
		We then define $y_\pm(x) := r_\pm(\chi_0 \mp x^4)^2$, and $Y_\pm(x,y) := \mp x^3 F_{0, \pm}(y^\frac{1}{2}, \chi_0 \mp x^4)$, so that \eqref{rnull-ode} is equivalent to, for $x \in [0,\delta^{\frac{1}{4}}]$: 
		\begin{equation} \label{y-ode}
			y_\pm'(x) = Y_\pm(x,y_\pm(x)) \text{ with } y_\pm(0) = 0.
		\end{equation}
		Note that we have $\widetilde{Y}_{\pm}(x,y) := x^{-3} Y_\pm(x,y)$ extending continuously to $y=0$ for all $x \in [0,\delta_4]$, and from \eqref{rnull-ode}, $A_0(0,\chi) \sim 1$, we see that $\widetilde{Y}_\pm(x,0) > 0$. We can therefore choose $y_{0,+} > 0$ such that: $\{ (T(y^\frac{1}{2},\chi_0 \mp x^4), \chi_0 \mp x^4) : y \in (0,y_{0, +}], x \in [0,\delta^{\frac{1}{4}}]\} \subset \mathscr{X}_{\delta_0, \delta}$ and moreover $\widetilde{Y}_\pm(x,y) \sim 1$ for $(x,y) \in [0,\delta^{\frac{1}{4}}] \times [0,y_{0, +}]$. Hence, there exist constants $0 < C_- < C_+$ such that for any solution $y_\pm(x)$ of \eqref{y-ode} with $(x,y_\pm(x)) \in [0,\delta^{\frac{1}{4}}] \times [0,y_{0, +}]$: $C_- x^4 \leq y_\pm(x) \leq C_+ x^4$. We therefore define, for $ 0 < x_{0,+} \leq \delta^\frac{1}{4}$ such that $C_+ x_{0,+}^4 \leq y_{0, +}$: $\mathcal{Y}(x_{0, +}) := \{ y \in C^0([0,x_{0,+}]) : C_- x^4 \leq y(x) \leq C_+ x^4 \}$. We have $|\partial_r F_{0, \pm}(r,\chi)| \lesssim r^{-\frac{1}{2}}$, so for $y \in \mathcal{Y}(x_{0, +})$: $|\partial_y Y_\pm(x,y)| \lesssim x^3 y^{-\frac{3}{4}} \lesssim 1$ and we can therefore apply Picard-Lindel\"of to show that there exist unique solutions $y_\pm \in \mathcal{Y}(x_{0, +})$ to \eqref{y-ode}. 
		
		Continuous extension of solutions of the eikonal equation to $\overline{\mathscr{Q}_{MGH}} \cap \mathscr{S}$ follows from continuous dependence in $\chi_0$ of the unique solutions $y_\pm(x)$. 
	\end{proof}
	
	Using Lemma \ref{lem:eik-S}, we prove the boundary decomposition of Theorem \ref{MGHDprecise} for $(\tilde{m},\tilde{f}) \in \mathfrak{D}_{\chi_b}^*$ such that $\mathcal{O} \in \partial^+\Sigma$, as defined in Lemma \ref{lem:O-limit-thinshell}.
	\begin{proposition} \label{prop:O-in-limit-ts}
		For any $(\tilde{m},\tilde{f}) \in \mathfrak{D}_{\chi_b}^*$, $\Omega_0^2 \in \mathfrak{G}_{\chi_b}(\tilde{m},\tilde{f})$ such that for $\mathcal{Q}_{MGHD}$ as defined in Proposition \ref{prop:MGHPSD}: $\mathcal{O} \in \partial^+\Sigma$, then $\mathcal{P}_{d+ts}(\mathcal{Q}_{MGHD})$ has boundary decomposition as stated in Theorem \ref{MGHDprecise}, with $\mathcal{CH}_d = \emptyset$. If $\mathcal{S}_d \neq \emptyset$ (equivalent to $\partial^+ \mathcal{B}_d \neq \mathcal{O}$), then the continuous extension of functions to $\mathcal{S}_d$ as stated in Theorem \ref{MGHDprecise} holds, and the connected components of $\mathcal{S}_d \setminus \partial^+\Sigma$ are $C^1$ spacelike.
	\end{proposition}
	
	\begin{proof}
		From the construction of $\mathscr{Q}_{MGH}$ in Proposition \ref{prop:MGHPSD}, $\mathscr{S} \cap \overline{\mathscr{Q}_{MGH}}$ is a finite disjoint union of: points  $\{(\tau_s(\chi_{s, i}),\chi_{s, i})\}$ for $\chi_{s, i} \in [0,\chi_b]$, including $\chi_{s, 1} := 0$, which correspond to the limit point of the interior and exterior of distinct thin shells, i.e.~a `thin shell collision at the boundary' (or `collision' with the centre in the case $\chi_{s,1} = 0$, or possibly $\{ \chi = \chi_b\}$); intervals $\{(\tau_s(\chi),\chi) : \chi_{s, j} \leq \chi \leq \chi_{s, j+1} \}$, where either $\chi_j = 0$ or $(\tau_s(\chi_{s, j}),\chi_{s, j})$ is the limit point of the exterior of a thin shell, and either $\{ \chi= \chi_b\}$ or $(\tau_s(\chi_{s, j+1}),\chi_{s, j+1})$ is the limit point of the exterior of a thin shell. By the same argument as in Lemma \ref{lem:O-limit-thinshell} for thin shells `colliding' with the centre, for any disjoint points in $\mathscr{S} \cap \overline{\mathscr{Q}_{MGH}}$, the limit of $(u_i(s),v_i(s))$ for the corresponding colliding thin shells must coincide. 
		
		Using Lemma \ref{lem:eik-S}, we define $\mathcal{S}_d := \{ (u(\tau_s(\chi),\chi), v(\tau_s(\chi),\chi)) : (\tau_s(\chi),\chi) \in \mathscr{S} \cap \overline{\mathscr{Q}_{MGH}} \} \setminus \mathcal{O}$, which is certainly connected by the previous paragraph, and is half-open with `left' limit point $\mathcal{O}$, `right' limit point $\partial^+\mathcal{B}_d$, which may be a thin shell or the smooth dust boundary. By definition, $\mathcal{S}$ must be achronal by the well-posedness of the null curve ODE on $\mathscr{S} \cap \overline{\mathscr{Q}_{MGH}}$ (see Lemma \ref{lem:eik-S}), and we must also therefore have $\mathcal{CH} = \emptyset$. 
		
		If $\mathcal{S}_d = \emptyset$, then there must exist a thin shell $\Sigma_i$ with limit point $\mathcal{O}$ such that it's exterior (in $\mathscr{Q}_{MDD}$) has $\chi^+_i(s_{d, i}) = \chi_b$, so we do have $\partial^+ \mathcal{B}_d \in \mathcal{O}$.
		
		By the continuous extension of $u,v \in C^2(\mathscr{Q}_{MGH})$, we must have $\tau(u,v), \chi(u,v)$ extending continuously to $\mathcal{S}_d \setminus \partial^+\Sigma$ from $\mathcal{P}_d(\mathcal{Q}_{MGHD})$ (where $\tau,\chi$ are as defined on $\mathcal{P}_d(\mathcal{Q}_{MGHD})$ in the statement of Proposition \ref{prop:MGHPSD}).
		
		We now show the continuous extension of the relevant functions to $\mathcal{S}_d \setminus \partial^+\Sigma$. We write \eqref{tauwave} as a transport equation for $\log(\mathbf{u}^u)$, dividing by $\partial_v r$ using \eqref{pr}, and where the right hand side is understood as functions of the comoving coordinates $(\tau(u,v),\chi(u,v))$:
		\begin{equation}
			\frac{\partial_v \log(\mathbf{u}^u)}{\partial_v r} = \left[ \frac{m}{r^2} - (\partial_\chi r)^{-1} \left(\frac{2\partial_\chi m}{r} + \partial_\chi f\right)\right] \left[ \frac{2m}{r} + f - (1+f)^\frac{1}{2}\left(\frac{2m}{r} + f\right)^\frac{1}{2} \right]^{-1} =: \frac{1}{2r} + r^{-\frac{1}{2}} W_{(u)}(r, \chi),
		\end{equation}
		where for any $\chi \in (0,\chi_b)$ with $\tau_s'(\chi) > 0$, there exists $\delta(\tilde{m},\tilde{f},\chi) > 0$ such that $W_{(u)} \in C^0([0,\delta] \times [\chi - \delta, \chi+ \delta])$. (Here, we've used \eqref{pchir-exp}). We therefore have 
		\begin{equation} \label{u-transp}
			\partial_v \log(r^{-\frac{1}{2}}\mathbf{u}^u) = r^{-\frac{1}{2}} W_{(u)}(r,\chi) \partial_v r, \qquad \partial_u \log(r^{-\frac{1}{2}}\mathbf{u}^v) = r^{-\frac{1}{2}} W_{(v)}(r,\chi) \partial_u r,
		\end{equation}
		where the equation for $\mathbf{u}^v$ follows analogously, with $W_{(v)}$ satisfying the same continuity property as $W_{(u)}$. For any $(u,v) \in \mathcal{S}_d \setminus \partial^+\Sigma$, integrating \eqref{u-transp}, in a neighbourhood of $(u,v)$ in $\mathcal{P}_d(\mathcal{Q}_{MGHD})$, we see that $r^{-\frac{1}{2}} \mathbf{u}^\mu$ extend continuously to $\mathcal{S}_d \setminus \partial^+\Sigma$.
		
		It then follows from \eqref{pr} and \eqref{fluidvel} respectively that $-r^{-1} \partial_\mu r, r\Omega^2$ extend continuously to positive functions on $\mathcal{S}_d \setminus \partial^+\Sigma$ from $\mathcal{P}(\mathcal{Q}_{MGHD})$, while $r^{\frac{3}{2}}\rho$ extends continuously by \eqref{rho} with \eqref{pchir-exp}.
		
		To see that the connected components of $\mathcal{S}_d \setminus \partial^+\Sigma$ are $C^1$ spacelike, we change coordinates to consider the function $u(r,v)$, which satisfies:
		\begin{equation}
			\partial_r u(r,v) = \frac{1}{\partial_u r(u(r,v),v)}, \qquad \partial_v u(r,v) = - \frac{\partial_v r(u(r,v),v)}{\partial_u r(u(r,v),v)}.
		\end{equation}
		For $(u_0,v_0) \in \mathcal{S}_d \setminus \partial^+\Sigma$, we define $u_{\mathcal{S}}(v) := u(0,v)$ in a neighbourhood of $v_0$. For sufficiently small $r$, the derivative 
		\begin{equation}
			\partial_v u(r,v) = - \frac{r^{-\frac{1}{2}} \mathbf{u}^u(u(r,v),v)}{r^{-\frac{1}{2}}\mathbf{u}^v(u(r,v),v)} \cdot \frac{1 - r^\frac{1}{2} \left( \frac{1+f}{2m + rf}\right)^\frac{1}{2}}{1 + r^\frac{1}{2} \left( \frac{1+f}{2m + rf}\right)^\frac{1}{2}},
		\end{equation}
		which extends continuously to $r=0$ in a neighbourhood of $v_0$, so $u_{\mathcal{S}}(v)$ is $C^1$ at $v = v_0$, with $u_{\mathcal{S}}'(v) < 0$, so the normal to the connected components of $\mathcal{S}_d \setminus \partial^+\Sigma$ is timelike.
	\end{proof}
	
	We now demonstrate the boundary decomposition of $\mathcal{P}_{d+ts}(\mathcal{Q}_{MGHD})$ in the remaining case, $\mathcal{O} \notin \partial^+\Sigma$.
	
	\begin{proposition} \label{prop:NS}
		For any $(\tilde{m},\tilde{f}) \in \mathfrak{D}_{\chi_b}^*$, $\Omega_0^2 \in \mathfrak{G}_{\chi_b}(\tilde{m},\tilde{f})$ such that for $\mathcal{Q}_{MGHD}$ as defined in Proposition \ref{prop:MGHPSD}: $\mathcal{O} \notin \partial^+\Sigma$, then $\mathcal{P}_{d+ts}(\mathcal{Q}_{MGHD})$ has boundary decomposition as stated in Theorem \ref{MGHDprecise}, with $\mathcal{CH}_d \neq \emptyset$, and the stated continuous extension/blow up of functions at $\mathcal{O}$.
	\end{proposition}
	
	\begin{proof}
		We first show that if $\tau_s''(0) > 0$, then there exists a unique future-directed outgoing null curve from $(\tau_s(0),0)$ in $\mathscr{Q}_{MDD}$, a subset of which must lie in $\overline{\mathscr{Q}_{MGH}}$. We define, for $(\tau,\chi) \in \mathscr{Q}_{MDD}$: $T_\pm(\tau_\pm(\chi),\chi) := \pm \frac{\partial_\chi r (\tau_\pm(\chi),\chi)}{(1+f(\chi))^\frac{1}{2}}$, so that the null curve ODEs \eqref{outnull}, \eqref{innull} are $\tau_\pm'(\chi) = T_\pm(\tau_\pm(\chi),\chi)$. 
		
		Since $\tau_s''(0) > 0$, there exists $c > 0$ and $\delta_1 > 0$ such that: $\tau_s''(\chi) > 0$ and $\tau_s(\chi) - c \chi^2 > \tau_0 + c \cdot \chi^2$ for all $\chi \in (0,\delta_1]$, and we define, for any $0 < \delta \leq \delta_1$: $\mathscr{Q}_0(\delta) := \{ (\tau,\chi) : \chi \in [0,\delta], 0 \leq \tau \leq \tau_s(\chi) -c \chi^2 \} \subset \mathscr{Q}_{MDD} \cup \{ (\tau_s(0), 0)\}$. We now bound $T_\pm(\tau,\chi), \partial_\tau T_\pm(\tau,\chi)$ for $(\tau,\chi) \in \mathscr{Q}_0(\delta)$. From \eqref{taus} and \eqref{T}, we have for $(\tau,\chi) \in \mathscr{Q}_{MDD}$:
		\begin{equation} \label{R}
			\tau_s(\chi) - \tau = \int_0^{R(\tau,\chi)} \frac{\tilde{R}^\frac{1}{2} d\tilde{R}}{(2\tilde{m}(\chi) + \tilde{f}(\chi) \tilde{R})^{\frac{1}{2}}},
		\end{equation}
		where $2\tilde{m}(\chi) + \tilde{f}(\chi) R \sim 1$ for all $\chi \in [0,\chi_b]$, $R \in [0,1]$, so for $(\tau,\chi) \in \mathscr{Q}_0$ we have: $R(\tau,\chi) \sim (\tau_s(\chi) - \tau)^\frac{2}{3} \sim \chi^\frac{4}{3}$. Now, from \eqref{tildeH0-2} we see that $\tilde{H}_0(R,0) = 2(2\tilde{m}(0) + \tilde{f}(0) R)^\frac{1}{2} > 0$ for $R \in [0,1]$, and we can choose $0 < \delta_2 \leq \delta_1$ sufficiently small such that for all $0 < \delta \leq \delta_2$ and $(\tau,\chi) \in \mathscr{Q}_0(\delta)$: $\tilde{H}_0(R(\tau,\chi),\chi) \sim 1$, so from \eqref{pchir-exp} we see that $A_0(R(\tau,\chi),\chi) \sim \chi^2$ and $\partial_\chi r(\tau,\chi) \sim \chi^\frac{4}{3}$ (and therefore $|T_\pm(\tau,\chi)| \sim \chi^\frac{4}{3}$). We have:
		\begin{equation}
			\partial_\tau T_\pm(\tau,\chi) = \mp (1+f(\chi))^{-\frac{1}{2}} \partial_\chi \left( \frac{2m(\chi)}{r(\tau,\chi)} + f(\chi)\right)^\frac{1}{2},
		\end{equation}
		where using $R, \partial_\chi r \sim \chi^{\frac{4}{3}}$ we get $|\partial_\tau T_\pm| \lesssim \chi^{-\frac{2}{3}}$. In order to apply Picard-Lindel\"of, we make the change of variables $\chi = u^3$, defining $t_\pm(u) := \tau_\pm(u^3)$, $\widetilde{T}_\pm(t,u) := T_\pm(t, u^3)$, $\widetilde{\mathscr{Q}}_0(\delta) := \{ (t,u) : u \in [0,\delta^\frac{1}{3}], 0 \leq t \leq \tau_s(u^3) -c u^6 \}$ and $M_{\delta} := \{ t_+ \in C^0([0,\delta^\frac{1}{3}]) : \tau_s(0) \leq t_+ \leq \tau_s(u^3) - c u^6 \}$. The ODE $\tau_\pm'(\chi) = T_\pm(\tau_\pm(\chi),\chi)$ with $\tau_\pm(0) = \tau_s(0)$ is then equivalent to: 
		\begin{equation} \label{nullode-centre}
			t_\pm'(u) = 3u^2 \widetilde{T}_\pm(t_\pm(u),u), \qquad t_\pm(0) = \tau_s(0). 
		\end{equation}
		By definition of $\widetilde{T}_\pm$, we have, for $(t,u) \in \widetilde{\mathscr{Q}}_0(\delta)$: $|u^2\widetilde{T}_\pm(t,u)| \sim u^6$, $|u^2 \partial_t \widetilde{T}_\pm(t,u)| \lesssim 1$. Integrating \eqref{nullode-centre} from $u=0$, any solution must satisfy (for $u \in [0,\delta^\frac{1}{3}]$):
		\begin{equation} \label{int-nullode-centre}
			t_\pm(u) = \mathbf{T}_\pm(t_\pm)(u) := \tau_s(0) + 3 u^7 \int_0^1 s^2 \widetilde{T}_\pm( t_\pm(su), su) ds. 
		\end{equation}
		By the bounds on $\widetilde{T}_+$, there exists $0 < \delta_3 \leq \delta_2$ such that for all $0 < \delta \leq \delta_3$, and all $t_+ \in M_{\delta}$: $\mathbf{T}_+(t_+) \in M_{\delta}$, and we can apply Picard-Lindel\"of to see that there is a unique solution in $M_{\delta_3}$, which is moreover smooth in $u$, and satisfies $t_+(u) - \tau_s(0) \sim u^7$.
		
		We show also that there is a unique past-directed null curve from $(\tau_s(0),0)$. Any solution $t_-(u)$ of \eqref{nullode-centre} must have $t_-(u) \leq \tau_s(0)$ (so $R(t_-(u),u^3) \gtrsim u^4$), and therefore $|\widetilde{T}_-(t_-(u),u)| \lesssim 1$, but then from \eqref{int-nullode-centre}: $\tau_s(0) - t_-(u) \lesssim u^3$, so $R(t_-(u),u^3) \lesssim u^2$ and from \eqref{pchir-exp}: $- \widetilde{T}_-(t_-(u),u) \sim \partial_\chi r(t_-(u),u^3) \lesssim u$. Iterating again, we see that $\tau_s(0) - t_-(u) \lesssim u^4$, so $R(t_-(u),u^3) \lesssim u^\frac{8}{3}$ and $- \widetilde{T}_-(t_-(u),u) \sim \partial_\chi r(t_-(u),u^3) \lesssim u^2$. Again, $\tau_s(0) - t_-(u) \lesssim u^5$, so $R(t_-(u),u^3) \lesssim u^\frac{10}{3}$ and $- \widetilde{T}_-(t_-(u),u) \sim \partial_\chi r(t_-(u),u^3) \lesssim u^3$. Finally, we see that $\tau_s(0) - t_-(u) \lesssim u^6$, so $R(t_-(u),u^3) \sim u^4$, and $- \widetilde{T}_-(t_-(u),u) \sim \partial_\chi r(t_-(u),u^3) \sim u^6$, $\tau_s(0) - t_-(u) \sim u^7$. One can then apply Picard-Lindel\"of using the same bounds as for $\widetilde{T}_+$ to see that there is a unique solution. 
		
		Now, the solutions of the eikonal equation $u,v \in C^2(\mathscr{Q}_{MGH})$, as defined in Proposition \ref{prop:MGHPSD}, extend continuously to $\overline{\mathscr{Q}_{MGH}}$. By the existence of the outgoing null curve from $(\tau_s(0),0)$, we define: $\mathcal{CH}_d := (\partial \mathcal{P}_{d+ts}(\mathcal{Q}_{MGHD} \cap \{ v= v_{\mathcal{O}}\}) \setminus \mathcal{O}$, which is non-empty and clearly satisfies the stated properties in Theorem \ref{MGHDprecise}, by consideration of the outgoing null curve on $\mathscr{Q}_{MDD}$ (with boundary conditions at the limit points of thin shells in $\overline{\mathscr{Q}_{MGH}}$ as for the eikonal equation). If the null curve meets $\mathscr{S}$, then the properties of $\mathcal{S}$ are satisfied by the argument in Proposition \ref{prop:O-in-limit-ts}.
		
		We have, in comoving coordinates, $\frac{m}{r}, r$ extending to continuous functions on $\mathscr{Q}_0(\delta) \cup \{(\tau_s(0), 0)\}$, vanishing at $(\tau_s(0), 0)$, so we must have, in double null coordinates, $\frac{m}{r} (u,v), r(u,v)$ extending continuously to zero at $\mathcal{O}$. For the continuous extension of $\mathbf{u}^\mu$, we integrate the transport equations for first derivatives of the eikonal equation, i.e.~\eqref{Lptauu}, \eqref{Lptauv}, on $\mathscr{Q}_{MGH}$. Let $\tau_{-, v}(\chi)$ be any ingoing null curve extending to $\chi = 0$ with $v(\tau_{-, v}(\chi),\chi) = v \leq v_{\mathcal{O}}$. We set $\tau = \tau_{-, v}(\chi)$ in \eqref{Lptauv} to see that:
		\begin{equation} \label{tau-v-transp}
			\frac{d}{d\chi}\left[ \left(\log(\partial_\tau v) +\frac{\chi}{R^\frac{1}{2}}\left(\frac{2\tilde{m} + \tilde{f} R}{1+f} \right)^\frac{1}{2} \right)(\tau_{-, v}(\chi),\chi)\right] = V_0(R(\tau_{-, v}(\chi),\chi),\chi),
		\end{equation}
		where we've defined (using \eqref{pchir-exp}):
		\begin{equation} \label{def:V0}
			V_0(R,\chi) := \frac{\chi f'(\chi)( 2\tilde{m}(\chi) + \tilde{f}(\chi) R)^\frac{1}{2} }{2R^\frac{1}{2}(1+f(\chi))^{\frac{3}{2}}} - \frac{\chi \tilde{m}(\chi) A_0(R,\chi)}{R^\frac{5}{2}(1+f(\chi))}.
		\end{equation}
		We have, by the ODE analysis, $R(\tau_{-, v_{\mathcal{O}}}(\chi),\chi) \sim \chi^\frac{4}{3}$, so for $v = v_{\mathcal{O}}$ the term inside the derivative in \eqref{tau-v-transp} is like $\sim \chi^{\frac{1}{3}}$, and $V_0(R(\tau_{-, v_{\mathcal{O}}}(\chi),\chi),\chi) \sim \chi^{-\frac{1}{3}}$, so $\mathbf{u}^v$ extends continuously to $\mathcal{O}$ along $\{v=v_{\mathcal{O}}\}$. For $v < v_{\mathcal{O}}$, we can bound $R(\tau_{-, v}(\chi),\chi) > R(\tau_{-, v_{\mathcal{O}}}(\chi),\chi)$, and therefore we can bound the terms in \eqref{tau-v-transp} by the respective bounds for $v = v_{\mathcal{O}}$, and therefore $\mathbf{u}^v(u,v)$ extends continuously to $\Gamma \cup \mathcal{O}$ for $v \leq v_{\mathcal{O}}$. Using the boundary condition $\mathbf{u}^u(v,v) = \mathbf{u}^v(v,v)$ for $v < v_{\mathcal{O}}$, we use the analogous integral to \eqref{tau-v-transp} for $\partial_\tau u$ on outgoing null curves, where the terms scale as for the ingoing case, and we can again bound $R(\tau_{+, u}(\chi),\chi) > R(\tau_{+, v_{\mathcal{O}}}(\chi),\chi)$ for $u < v_{\mathcal{O}}$. It follows that $\mathbf{u}^u$ extends continuously to $\mathcal{O}$ from $v \leq v_{\mathcal{O}}$, and therefore so does $\Omega^2$ from \eqref{fluidvel}, and $\partial_\mu r$ from \eqref{pr}, where $\partial_v r(v_{\mathcal{O}},v_{\mathcal{O}}) > 0$ since $\frac{m}{r}$ extends to zero.
		
		In order to show \eqref{pvOmega-blowup}, we integrate $\partial_u(\partial_v \log\Omega^2)$ on $\{v=v_{\mathcal{O}}\}$. We first show that $\frac{m}{r^3} - 2\pi \rho \sim R^{-3}$ on $\{ v =v_{\mathcal{O}}\}$, for sufficiently small $\chi$. One can see from \eqref{m-com}, \eqref{tilde} that $\tilde{m}(0) = \frac{4\pi \rho_0(0)}{3}$, and therefore $\frac{m}{r^3} - 2\pi \rho \sim R^{-3} \sim R^{-4} (2 \partial_\chi r - 3R)$. We see from \eqref{R} and $\tau_{-, v_{\mathcal{O}}}(\chi) - \tau_s(0) \sim \chi^\frac{7}{3}$ that: $R(\tau_{-, v_{\mathcal{O}}}(\chi),\chi)^\frac{3}{2} = \frac{3(2\tilde{m}(0))^\frac{1}{2}}{4} \tau_s''(0) \chi^2 + O(\chi^\frac{7}{3})$, while from \eqref{pchir} we see that $(R^\frac{1}{2}\partial_\chi r)(\tau_{-, v_{\mathcal{O}}}(\chi),\chi) = \frac{7(2\tilde{m}_0)^\frac{1}{2}}{4}\chi^2 \tau_s''(0) + O(\chi^\frac{7}{3})$, and so we do have $2 \partial_\chi r - 3R \sim R$ and therefore $\frac{m}{r^3} - 2\pi \rho \sim \chi^{-4}$. We then use $\partial_u \chi \sim - \chi^{-\frac{4}{3}}$ to see that $\frac{\partial_u(\partial_v \log\Omega)^2}{\partial_u \chi} \sim - \chi^{-\frac{8}{3}}$ and therefore \eqref{pvOmega-blowup} holds.
		
		For finite blueshift, we change variables in the integrand \eqref{blueshift} to $\chi$ using $\partial_u r/\partial_u \chi \sim \chi^\frac{4}{3}$, and from $m/r^2 \sim \chi^{-\frac{5}{3}}$ we see that the integral is finite.
	\end{proof}	
	
	We conclude the proof of Theorem \ref{MGHDprecise} by noting that the stated decomposition of the vacuum exterior follows from standard properties of the Schwarzschild spacetime, for some $M>0$ determined by initial data, with gauge determined by initial data and the gauge data on the dust cloud boundary (which is determined by the respective ODE for smooth boundary or a thin shell boundary, see the proof of Proposition \ref{prop:MGHPSD}). The continuous extension of $-r\partial_\mu r$, $r\Omega^2$ follow from the rearranging the definition of the Hawking mass \eqref{m} into an identity for either $\partial_u r$ or $\partial_v r$ in terms of $r, m$ and $\Omega^{-2} \partial_v r$ or $\Omega^{-2} \partial_u r$ respectively. To show that the connected components of $\mathcal{S} \setminus \partial^+ \Sigma$ are $C^1$ spacelike, it remains to show that if $\mathcal{B}_d^+ \notin \partial^+\Sigma$, then $\mathcal{S}_d \cup (\mathcal{S} \setminus \mathcal{S}_d)$ is $C^1$ spacelike. This follows from the continuous extension of $\Omega^{-2} \partial_u r$, $\Omega^{-2} \partial_v r$ to $\mathcal{S}$ in a neighbourhood of $\partial^+\mathcal{B}_d$ (where we use Raychaudhuri's equations \eqref{Rayu}, \eqref{Rayv} in the vacuum region).
	
	\section{Appendix} 
	
	\subsection{Cauchy stability for ODEs} \label{cauchystability}
	
	The following proposition and proof is adapted from \cite{Ryan}.
	\begin{proposition} \label{CS}
		Let $T_0 > 0$, $U \subseteq \R$ an open set, and let $F : [0,T_0] \times U \to \R$ be a smooth function. Suppose that $y \in C^\infty([0,T_0])$ is a solution of the ODE
		\begin{equation}
			\frac{dy}{dt} = F(t,y(t)),
		\end{equation}
		with $y(0) = y_0$ and $y(t) \in U$ for $t \in [0,T_0]$. For $\varepsilon, T > 0$, we define
		\begin{equation}
			K([0,T],y,\varepsilon) := \bigcup_{t \in [0,T]} \{ t \} \times \{ x \in \mathbb{R} : |x - y(t)| \leq \varepsilon\}.
		\end{equation}
		Then, for any $\varepsilon > 0$, $k \in \mathbb{N} - 1$, there exist $\delta_0, \delta > 0$ depending on $(y, F, T_0, k, \varepsilon)$ such that for any smooth function $\widetilde{F} : K([0,T_0],y, \varepsilon) \to \R$ satisfying
		\begin{equation} \label{F-bd}
			\| F - \widetilde{F} \|_{C^{k}(K([0,T_0],y,\varepsilon))} < \delta,
		\end{equation}
		and $\tilde{y}_0 \in \R$ satisfying $|\tilde{y}_0 - y_0| < \delta_0$: there exists $\tilde{y} \in C^\infty([0,T_0])$, the unique solution of $\frac{d\tilde{y}}{dt} = \tilde{F}(t,\tilde{y}(t))$ with $\tilde{y}(0) = \tilde{y}_0$, which moreover satisfies
		\begin{equation} \label{y-bd}
			\| y - \tilde{y} \|_{C^{k+1}([0,T_0])} < \varepsilon.
		\end{equation}
	\end{proposition}
	
	\begin{proof}
		Let $0 < \varepsilon \leq \varepsilon$. Define for $M \geq 1$: 
		\begin{equation}
			A(\delta_0,M) := \left\{ T_* \in [0,T_0] : \tilde{y}(t) \text{ exists on } [0,T_*] \text{ and } \|y - \tilde{y}\|_{C^0([0,T_*])} \leq 2\delta_0 e^{MT_*} < \varepsilon_0 \right\},
		\end{equation}
		which for $\delta_0 < \frac{\varepsilon_0}{2}e^{-MT_0}$ is non-empty by local existence and continuity, and closed by definition. Now, for $T_* > 0$, we suppose that $\tilde{y}(t)$ exists for $t \in [0,T_*]$, and define $\overline{y} := y - \tilde{y} \in C^\infty([0,T_*])$, which is a solution of:
		\begin{equation} \label{overline-y}
			\frac{d\overline{y}}{dt} = F(t,y(t)) - \tilde{F}(t,\tilde{y}(t)),
		\end{equation}
		which we integrate for $t \in [0,T_*]$:
		\begin{align}
			|\overline{y}(t) - \overline{y}(0)| &\leq \int_0^t |F(t,y(t)) - F(t,\tilde{y}(t))| dt +  \int_0^t |F(t,\tilde{y}(t)) - \tilde{F}(t,\tilde{y}(t))| dt \\ &< \max_{(t,z) \in K([0,T_0],y,\varepsilon)} |\partial_z F(t',z)| \cdot \int_0^t 2\delta_0 e^{M t'} dt' +  \int_0^t \delta dt.
		\end{align}
		We choose $M = 2\max_{(t,z) \in K([0,T_0],y,\varepsilon)} |\partial_z F(t',z)|$, so that:
		\begin{equation}
			|\overline{y}(t) - \overline{y}(0)| < \delta_0 (e^{Mt} - 1) + \delta t.
		\end{equation}
		Using the triangle inequality, and choosing $\delta = \frac{\delta_0}{2T_0}$: 
		\begin{equation}
			\|y - \tilde{y}\|_{C^0([0,T_*])} = \| \overline{y} \|_{C^0([0,T_*])} < \delta_0 (1/2 + e^{M T_*}) \leq \frac{3\delta_0}{2} e^{MT_*}.
		\end{equation}
		The set $A(\delta_0,M)$ is therefore open for this choice of $\delta_0, \delta, M$, so $A(\delta_0,M) = [0,T_0]$ and $\|y - \tilde{y}\|_{C^0([0,T_*])} < \varepsilon_0$. We bound directly from \eqref{overline-y}:
		\begin{equation}
			\left\| \frac{dy}{dt} - \frac{d\tilde{y}}{dt} \right\|_{C^0([0,T_0])} \leq \delta + \frac{1}{2} M \varepsilon_0,
		\end{equation}
		so the result for $k=0$ follows from choosing $\delta, \varepsilon_0$ sufficiently small depending on $\varepsilon$ and $M$. The general result for $k \in \mathbb{N}$ follows from differentiating \eqref{overline-y} and using \eqref{F-bd} to bound the difference of derivatives, choosing $\delta, \delta_0$ successively smaller depending on $k$, and $y, F$. 
	\end{proof}
	
	\bibliographystyle{alpha}
	\bibliography{ref}
	
\end{document}